\documentclass[10pt]{report}
\usepackage{graphics}
\def\GR{General Relativity}

\def\beq{\begin{equation}}
\def\bfg{\begin{figure}}

\def\efg{\end{figure}}
\def\eeq{\end{equation}}
\def\bea{\begin{eqnarray}}
\def\eea{\end{eqnarray}}

\def\bh{black hole}
\def\BH{{\mathrm{BH}}} 
\def\mink{\mathrm{Mink}}
\def\rind{\mathrm{Rindler}}
\def\GB{\mathrm{Gauss-Bonnet}}
\def\PRD{Phys. Rev. D}
\def\PRL{Phys.Rev. Lett.}
\def\Tr{\mathrm{Tr}}
\def\ins{{\mathrm{inside}}}
\def\outs{{\mathrm{outside}}}
\def\tot{{\mathrm{total}}}
\def\rad{{\mathrm{rad}}}
\def\mm{\mathrm{mm}}
\newcommand\rlpartial{\stackrel{\leftrightarrow}{\partial}}

\newcommand\Sch{Schwarzschild}
\newcommand\ee{entanglement entropy}

\newcommand\text{Tr_{ext}}

\newcommand\gmn{g_{\mu \nu}}  
\def\Paris{Par\'{\i}s}
\def\Perez{P\'erez}
\def\Gunther{G\"unther}
\def\Schutzhold{Sch\"utzhold}
\def\Lofstedt{L\"{o}fstedt}

\def\Ruggeberg{R\"uggeberg}
\def\ng{ n_{\mathrm {gas}} }
\def\ngi{ n_{\mathrm {gas}}^{\mathrm{in}} }
\def\ngo{ n_{\mathrm {gas}}^{\mathrm{out}} }
\def\nl{n_{\mathrm{liquid}}}

\def\max{\hbox{max}}
\def\min{\hbox{min}}

\def\inn{\mathrm{in}}
\def\out{\mathrm{out}}
\def\half{\frac{\textstyle 1}{\textstyle 2}}
\def\quarter{\frac{\textstyle 1}{\textstyle 4}}
\def\observed{{\mathrm{observed}}}
\def\inside{{\mathrm{inside}}}
\def\outside{{\mathrm{outside}}}
\def\liquid{{\mathrm{liquid}}}
\def\gas{{\mathrm{gas}}}
\def\ni{n_\inn}
\def\no{n_\out}
\def\oi{\omega_\inn}
\def\oo{\omega_\out}
\def\Ni{ {\cal N}_\inn}

\def\half{{\textstyle{1\over2}}}
\def\quarter{{\textstyle{1\over4}}}
\def\Box{\kern0.5pt{\lower0.1pt\vbox{\hrule height.5pt width 6.8pt
    \hbox{\vrule width.5pt height6pt \kern6pt \vrule width.3pt}
    \hrule height.3pt width 6.8pt} }\kern1.5pt}
\def\pl{\mathrm{Pl}}
\def\im{{\mathrm i}}
\def\a{{\mathbf a}}
\def\d{{\mathrm d}}
\def\g{{\mathrm g}}
\def\m{{\mathrm m}}
\def\x{{\mathbf x}}

\def\k{{\mathbf k}}
\def\v{{\mathbf v}}
\def\E{{\mathbf E}}
\def\B{{\mathbf B}}
\def\uu{{\mathbf u}}
\def\n{{\mathbf n}}
\def\c{{\mathrm c}} 
\def\cs{c_{\mathrm s}}

\def\L{{\cal L}}
\def\I{{\cal I}}
\def\H{{\cal H}}

\def\I{{\cal I}}
\def\L{{\cal L}}

\def\VSL{{$\chi$VSL}}
\def\etal{\emph{et al.}}
\def\etc{\emph{etc}}
\def\ie{\emph{i.e.}}
\def\eg{\emph{e.g.}}
\def\beq{\begin{equation}}
\def\eeq{\end{equation}}
\def\beqn{\begin{eqnarray}}
\def\eeqn{\end{eqnarray}}

\def\cp{c_{\mathrm{photon}}}

\def\cn{c_{\mathrm{gravity}}}
\def\cg{c_{\mathrm{gravity}}}
\def\gem{g_{\mathrm{em}}}

\def\matter{{\mathrm{matter}}}

\def\ie{{\em i.e.\/}}
\def\eg{{\em e.g.\/}}
\def\etc{{\em etc.\/}}
\def\etal{{\em et al.\/}}
\def\eff{\mathrm{eff}}
\def\Painleve{Painlev\'e}
\def\aa{{\cal A}}

\def\L{{\cal L}}
\begin{document}
%%%%%%%%%%%%%%%%%%%%%%%%%%%%%%%%%%%%%%%%%%%%%%%%%%%%%%%%%%%%%%%%%%
\pagestyle{empty}

\begin{center}
\vspace*{2cm}
  
{\Large \bf QUANTUM VACUUM EFFECTS IN GRAVITATIONAL FIELDS: 
                                     THEORY AND DETECTABILITY}

\vspace*{3cm}
\large
Thesis submitted for the degree of\\
``Doctor Philosophi\ae"

\vspace*{3cm}

Stefano Liberati\\
International School for Advanced Studies\\
Via Beirut 2-4, 34014 Trieste, Italy.\\
E-mail: {\sl liberati@sissa.it}\\  
\vspace*{2cm}
June 2000

\end{center}
%\date{\today}
%%%%%%%%%%%%%%%%%%%%%%%%%%%%%%%%%%%%%%%%%%%%%%%%%%%%%%%%%%%%%%%%%
\newpage
%%%%%%%%%%%%%%%%%%%%%%%%%%%%%%%%%%%%%%%%%%%%%%%%%%%%%%%%%%%%%%%%
\baselineskip 18pt

\begin{center}

{\sc  QUANTUM VACUUM EFFECTS IN GRAVITATIONAL FIELDS:\\ 
THEORY AND DETECTABILITY}

Stefano Liberati --- {\em Ph.D. Thesis}

\underline{Supervisor: D.W.~Sciama}

\underline{Co-Supervisors: M.~Visser and J.C.~Miller}

{\bf Abstract}
\end{center}
\baselineskip 18pt This thesis is devoted to the study of quantum
vacuum effects in the presence of strong gravitational fields. We
shall see how the quantum vacuum interacts with black hole geometries
and how it can play an important role in the interpretation of the
gravitational entropy. In this respect particular attention will be
given to the peculiar role of the extremal \bh\ solutions.  From this
branch of our research we shall try to collect some important hints
about the relation between quantum gravity theories and the
semiclassical results. After these investigations we shall move our
attention toward possible experimental tests of particle creation from
the quantum vacuum which is an indirect confirmation of the Hawking
effect. This aim will lead us to study acoustic geometries and their
way of ``simulating'' \GR\ structures, such as horizons and black
holes.  We shall study the stability of these structures and the
problems related to setting up experimental detection of phonon
Hawking flux from acoustic horizons.  This research will naturally
lead us to propose a new model for explaining the emission of light in
the phenomenon of Sonoluminescence, based on the dynamical Casimir
effect. Possible experimental tests of this proposal will be
discussed. In this way we shall set up one of the few available models
of quantum vacuum radiation amenable to observational test within the
next few years.  After this journey in the condensed matter world we
shall move to the other arena where our efforts to test the effects of
the quantum vacuum in gravitational fields can find a positive
solution in the future: the high energy phenomena in the early
universe.  We shall concentrate our attention on inflation and its
possible alternatives for solving the cosmological puzzles. This will
lead us to propose a new way to reheat the universe after inflation
via pure gravitational effects.  We shall finally show how some known
phenomena related to the vacuum polarization in the Casimir effects,
can naturally suggest new ways to replace (or at least improve) the
inflationary scenario.

\newpage

%%%%%%%%%%%%%%%%%%%%%%%%%%%%%%%%%%%%%%%%%%%%%%%%%%%%%%%%%%%%%%%%%%%%%%%%%%%

\baselineskip 12pt
\parskip 1ex
\pagenumbering{roman}
\tableofcontents

%%%%%%%%%%%%%%%%%%%%%%%%%%%%%%%%%%%%%%%%%%%%%%%%%%%%%%%%%%%%%%%%%%%%%%%%%%%%%%

\newpage
\pagestyle{headings}
\baselineskip 15pt

%%%%%%%%%%%%%%%%%%%%%%%%%%%%%%%%%%%%%%%%%%%%%%%%%%%%%%%%%%%%%%%%%%%%%%%%%%%%%%

%%%%%%%%%%%%%%%%%%%%%%%%%%%%%%%%%%%%%%%%%%%%%%%%%%%%%%%%%%%%%%%%%%%%%%%%%%%%
%S.Liberati Ph.D. Notations
%%%%%%%%%%%%%%%%%%%%%%%%%%%%%%%%%%%%%%%%%%%%%%%%%%%%%%%%%%%%%%%%%%%%%%%%%%%%
\chapter*{Notation}
%\pagestyle{myheadings}
%\pagestyle{empty}
%\markright{\large \sc Notation}
%\pagenumbering{roman}
\label{chap:not}
%%%%%%%%%%%%%%%%%%%%%%%%%%%%%%%%%%%%%%%%%%%%%%%%%%%%%%%%%%%%%%%%%%%%%%%%%%%%

Unless otherwise stated we shall use units for which $\hbar=c=1$. We
shall make explicit the dependence on the fundamental constants in the
most important formul\ae. The Boltzmann constant is $k_{\mathrm B}$ and
the gravitational constant will be denoted as $G_{\mathrm N}$.  The
Greek indices take values $0\dots 3$ while Latin indices denote
spatial directions and range over $1\dots 3$.

The wide range of problems treated in this thesis has made it
impossible to use the same metric signature through all of the work.
Chapters~\ref{chap:1} and~\ref{chap:3b} use the signature common in
the literature of quantum field theory $(+,-,-,-)$ with the Minkowski
metric given by $\eta_{\mu\nu}={\mathrm{diag}}(1,-1,-1,-1)$.  In
chapters~\ref{chap:2},~\ref{chap:3a} and~\ref{chap:4} we use the
signature commonly used by general relativists $(-,+,+,+)$ with the
Minkowski metric given by
$\eta_{\mu\nu}={\mathrm{diag}}(-1,+1,+1,+1)$.

These notations are consistent with standard reference books in the
subject, which have been used as references for the review parts of
this work.  For purely general relativistic issues we have mainly used
references~\cite{MTW,HawEll73,waldGR}, for quantum field theory in
external/gravitational fields, the principal sources
are~\cite{Birrell-Davies,moste,MostTru}.  For the review on black hole
thermodynamics we have mainly used~\cite{waldGR,wald,FrNo}.

The following special symbols and abbreviations are used throughout

\begin{center}
\noindent
\begin{tabular}{|c|c|}
\hline
$*$ & complex conjugate\\
$\dagger$ or h.c. & Hermitian conjugate\\
$\partial_{\mu}$ or $\frac{\partial}{\partial x^{\mu}}$ or $\ _{,\mu}$ &
Partial derivative\\
$\nabla_{\mu}$ or $\ _{;\mu}$ & Covariant derivative\\ 
$\Box\equiv g^{\mu\nu}\nabla_{\mu}\nabla_{\nu}$ & D'Alambertian operator\\ 
$\Re$ ($\Im$) & Real (imaginary) part\\
$\Tr$ & Trace\\
$[A,B]=AB-BA$ & Commutator\\
$\{A,B\}=AB+BA$ & Anti-commutator\\
$\kappa$ & surface gravity\\
$\kappa_{\mathrm N}=8\pi G_{\mathrm N}$ & rescaled Newton constant\\
$\rho$ & mass density\\
$\varepsilon$ & energy density\\
$r_{\mathrm h}$ & event horizon radius\\
$\g_{\oplus}$ & Earth gravity acceleration\\
\hline
\end{tabular}
\end{center}
%
%%%%%%%%%%%%%%%%%%%%%%%%%%%%%%%%%%%%%%%%%%%%%%%%%%%%%%%%%%%%%%%%%%%%%%%%%%%%

%%%%%%%%%%%%%%%%%%%%%%%%%%%%%%%%%%%%%%%%%%%%%%%%%%%%%%%%%%%%%%%%%%%%%%%%%%%%
%S.Liberati Ph.D. Intro: Introduction
%%%%%%%%%%%%%%%%%%%%%%%%%%%%%%%%%%%%%%%%%%%%%%%%%%%%%%%%%%%%%%%%%%%%%%%%%%%%
\chapter*{Introduction}
\pagestyle{myheadings}
\markright{\large \sc Introduction}
\pagenumbering{arabic}
\addcontentsline{toc}{chapter}{Introduction}
\label{chap:intro}
%%%%%%%%%%%%%%%%%%%%%%%%%%%%%%%%%%%%%%%%%%%%%%%%%%%%%%%%%%%%%%%%%%%%%%%%%%%%
\vspace*{0.5cm}
\rightline{\it For most of human history we have searched for our
               place in the cosmos.} 
\rightline{\it Who are we?  What are we? }
\rightline{\it We find that we inhabit an insignificant planet of a hum-drum 
               star lost in a galaxy}
\rightline{\it tucked away in some forgotten corner of a universe in which 
               there are far more galaxies than people.}
\rightline{\it We make our world significant by the
               courage of our questions, and by the depth of our answers.} 
\vspace*{0.5cm} \rightline{\sf Carl Sagan}
%%%%%%%%%%%%%%%%%%%%%%%%%%%%%%%%%%%%%%%%%%%%%%%%%%%%%%%%%%%%%%%%%%%%%%%%%
\vspace*{0.5cm}
\hrulefill
\vspace*{0.5cm}
%%%%%%%%%%%%%%%%%%%%%%%%%%%%%%%%%%%%%%%%%%%%%%%%%%%%%%%%%%%%%%%%%%%%%%%%%%%
%\subsection*{Motivations}
%\addcontentsline{toc}{section}{Motivations}
%\label{sec:motiv}
%%%%%%%%%%%%%%%%%%%%%%%%%%%%%%%%%%%%%%%%%%%%%%%%%%%%%%%%%%%%%%%%%%%%%%%%%%%%

Among the fundamental forces of nature, gravity still stands in a very
particular role. Although the electromagnetic and weak interactions
have been successfully unified in the Glashow--Weinberg--Salam model,
and the strong force is successfully described with a similar quantum
theory, we still lack a quantum description of the gravitational
interaction.

The past century has seen a large number of formidable theoretical
attacks on the problem of quantum gravity, nevertheless all of these
approaches have so far failed (or at the very least proved
inconclusive).  The reason for such a failure could be just due to the
lack of imagination of scientists, but it is beyond doubt that when
compared, for example, with quantum electrodynamics, the construction of
a quantum theory of gravity turns out to be extremely complicated.

The difficulties on the way to quantum gravity are of different kinds.
First of all, the detection of quantum gravitational effects is by
itself extremely difficult due to the weakness of the gravitational
interaction. In addition one encounters technical problems in
quantizing gravity resulting from basic, and peculiar, properties of
\GR\ such as the non-linearity of the Einstein equations and the
invariance of the theory under the group of diffeomorphisms.

Finally the fact that gravity couples via a dimensional coupling
constant $G_{\mathrm N}$ makes the theory intrinsically
non-renormalizable.  For some time it was believed that supergravity
theories might overcome this problem, but detailed calculations, and
the fact that they are now viewed as effective theories induced from a
more fundamental superstring theory, has led to the conclusion that
they also suffer with the same problem. Nevertheless one should stress
that non-renormalizability of a theory does not necessarily correspond
to a loss of meaning.

Nowadays non-renormalizability can be seen as a natural feature of a
theory for which the action is not fundamental but arises as an
effective action in some energy limit. The Fermi four-fermion model of
weak interactions is certainly a non-renormalizable theory but
nevertheless it can still be useful in giving meaningful predictions
at energies well below those of the $W^{\pm},Z^{0}$ gauge bosons.

In the case of gravity, one may ask when the gravitational interaction
can no longer be treated classically. A general dimensional argument
is that this happens when the gravitational (Einstein--Hilbert) action
is of the same order as the quantum of action $\hbar$
\begin{equation}
  \label{eq:quantgrcond}
  S_{\mathrm{grav}}=\frac{c^3}{16\pi G_{\mathrm N}}\int R \sqrt{g}\; 
                    \d x \approx \hbar
\end{equation}
If $L$ is the typical length scale of the spacetime, one has that the
above equality holds for
\begin{equation}
  \label{eq:lplanck}
  L\approx \sqrt{\frac{\hbar G_{\mathrm N}}{c^3}}\equiv L_{\pl}=10^{-33}
           \,\mbox{cm}
\end{equation}
$L_{\pl}$ is called the Planck length and it corresponds to an energy
$E_{\pl}=c\hbar/L_{\pl}=\sqrt{\hbar c^5/G_{\mathrm N}}= 10^{19}
\,GeV$.  Given the fact that the heaviest particles which we are now
able to produce are ``just'' of the order of
$TeV=10^{3}\,GeV=10^{-16}\,E_{\pl}$ (and that the top end of the
cosmic ray spectrum is at about $10^{13}\,GeV=10^{-6}\,E_{\pl}$) it is
clear that there is a wide range of energies for which matter can be
described quantistically and gravity classically.

This implies that we can limit ourselves to considering theories where
quantum fields are quantized in curved backgrounds and where we at
most consider the linearized gravitational field (gravitons). In this
way the first step of the theory of quantum gravity is naturally
the theory of quantum fields (gravitons included) in curved spaces.

Since its first years, this branch of research has focussed
particularly on the central role of the quantum vacuum. It was soon
discovered that zero point modes of quantum fields are not only
influenced by the geometry but are also able to influence gravity in
an important way. The theories of black hole evaporation and inflation
are nowadays outstanding examples of this.

In this thesis we shall try to present a panoramic view of the general
theory of vacuum effects in strong fields, paying special attention to
the role of gravity. Our approach will be to focus on different sides
of the same physical framework trying to gain new ideas and deeper
understanding by a process of developing cross-connections between
apparently different physical problems. Sometimes we shall try to
learn lessons from our models to then be applied for getting further
insight into other different physical phenomena. On other occasions we
shall seek experimental tests of theoretical predictions of
semiclassical quantum gravity by looking for their analogs in
condensed matter physics.  Finally we shall also try to gain a deeper
understanding of the nature of gravity by considering the possibility
of explaining some of its paradoxes by using different paradigms
borrowed from our general experience about the dynamics of the quantum
vacuum.

Obviously this work is not going to be conclusive but we hope that the
reader will be able to see some of the subtle links that connect the
theory of semiclassical gravity to a much wider theoretical
construction based on the peculiar nature of the quantum vacuum.  It
is these links which we shall be trying to use for developing new
perspectives in this field of research.
 
%%%%%%%%%%%%%%%%%%%%%%%%%%%%%%%%%%%%%%%%%%%%%%%%%%%%%%%%%%%%%%%%%%%%%%%%%%%%
\subsection*{Plan of the work}
%\addcontentsline{toc}{section}{Plan of the work}
\label{sec:plan}
%%%%%%%%%%%%%%%%%%%%%%%%%%%%%%%%%%%%%%%%%%%%%%%%%%%%%%%%%%%%%%%%%%%%%%%%%%%%

This work is divided into five main chapters. Chapter~\ref{chap:1} is
devoted to presenting a general view of the main problems and basic
ideas of vacuum effects in the presence of external fields with
special attention being focussed on the case of gravity.

In chapter~\ref{chap:2} we move on to the study of quantum black
holes. The thermodynamical behaviour of these objects in the presence
of the quantum vacuum will be described and investigated. We shall try
to develop the analogy between \bh\ thermodynamics and the Casimir
effect and then we shall investigate the relationship between black
hole entropy and the global topology of spacetime. Finally we shall
study the nature of extremal black holes.

The following two chapters, chapter~\ref{chap:3a} and
chapter~\ref{chap:3b}, can be seen as two possibly interconnected
parallel lines of research, having the common prospect of possibly
reproducing some of the most important aspects of semiclassical
gravity. The underlying philosophy of these chapters is to investigate
the possible use of condensed matter techniques or phenomena to
generate laboratory analogs of the phenomenon of particle creation
from the quantum vacuum which plays a crucial role in semiclassical
gravity.

We shall see in chapter~\ref{chap:3a} that it may be possible to build
up fluid dynamical analogs of the event horizons of general
relativity.  This research is interesting at different levels. It
could in fact give us the possibility to reproduce in the laboratory
the most important prediction of semiclassical quantum gravity, the
Hawking--Unruh effect. Moreover it can provide us, on the theoretical
side, with a deeper understanding of the possible interpretation of
General Relativity as an {\em effective theory} of gravity.

In chapter~\ref{chap:3b} we pursue an approach which is the reverse of
that in the previous chapter. Instead of trying to build up a
condensed matter model reproducing some semiclassical gravity effects,
we take a well-known, but unexplained, phenomenon and propose a model
based on the production of particles from the quantum vacuum for
explaining it. The phenomenon discussed is {\em Sonoluminescence}: the
emission of visible photons from a pulsating bubble of gas in water.

In chapter~\ref{chap:4} we finally move our attention from the
realm of condensed matter physics to that of cosmology. This is
another promising regime for testing our knowledge of the effects of
the quantum vacuum in the presence of strong fields and, in particular,
of gravity. Actually cosmology and astrophysics is the
only place where we can hope to see the effects of strong gravitational
fields in action. We shall discuss some ways in which the quantum
vacuum can influence gravitation and be influenced by it, and in
particular we study in detail the post-inflationary stage of
preheating. At the end of the chapter we discuss the possibility
of providing alternatives to the inflationary paradigm and again we
shall show that some quantum vacuum effects can play a prominent role
also in this case.

This thesis collects results which have been produced in collaboration
with several people and which are published in the following papers
(listed following the order of appearance in this work)
\cite{BL95,BL97,LP97,LRS99,LSV00,Qed0,PRL,SnPr,Qed1,Qed2,2Gamma,BL98,VSL}.
The research in~\cite{Liberati:1995za} and~\cite{Visser:1998ua} will
not be presented here because these papers concern issues which are
too distant from the main line of this thesis.

\vspace*{0.5cm} 
\rightline{\it Stefano Liberati} 
\leftline{\it Trieste, Italy} 
\leftline{\it April 2000}

%%%%%%%%%%%%%%%%%%%%%%%%%%%%%%%%%%%%%%%%%%%%%%%%%%%%%%%%%%%%%%%%%%%%%%%%%%%%
%S.Liberati Ph.D. Thesis - Introduction: the bridge toward quantum gravity
%%%%%%%%%%%%%%%%%%%%%%%%%%%%%%%%%%%%%%%%%%%%%%%%%%%%%%%%%%%%%%%%%%%%%%%%%%%%
\pagestyle{headings}
\chapter[{Quantum vacuum and Gravitation}]{Quantum vacuum and Gravitation}
%\pagenumbering{arabic}
\label{chap:1}
%signature (+,-,-,-)  like in GMM, MostTru, Birrell-Davies 
%%%%%%%%%%%%%%%%%%%%%%%%%%%%%%%%%%%%%%%%%%%%%%%%%%%%%%%%%%%%%%%%%%%%%%%%%%%%

\vspace*{0.5cm} 
\rightline{\it If the doors of perception were cleansed} 
\rightline{\it everything would appear as it is,}
\rightline{\it infinite}
\vspace*{0.5cm} \rightline{\sf William Blake}

\vspace*{1cm}
\rightline{\it There is a concept which corrupts and upsets all others.}
\rightline{\it I refer not to Evil, whose limited realm is that of ethics;}
\rightline{\it I refer to the infinite.}
\vspace*{0.5cm} \rightline{\sf Jorge Luis Borges}

\newpage
%%%%%%%%%%%%%%%%%%%%%%%%%%%%%%%%%%%%%%%%%%%%%%%%%%%%%%%%%%%%%%%%%%%%%%%%%%%%

This chapter is an introduction to the issue of the vacuum effects in
strong fields. We shall review the basic theory of the quantum vacuum
and its application in the presence of external fields. We shall deal with
both static and dynamical phenomena and, in connection with the latter,
we shall discuss the phenomenon of particle creation from the quantum
vacuum due to a strong, time-varying external field.  Although most of
the chapter is devoted to non gravitational effects, we shall see how
most of the concepts introduced here are necessary tools for
understanding the physics of a quantum vacuum in gravitational fields.  In
particular, at the end of this chapter we shall show how the
application of such a framework has led to fundamental results in modern
theoretical astrophysics. 

%%%%%%%%%%%%%%%%%%%%%%%%%%%%%%%%%%%%%%%%%%%%%%%%%%%%%%%%%%%%%%%%%%%%%%%%%%%
\section[{The nature of the quantum vacuum}]
{The nature of the quantum vacuum} 
\label{sec:qv}
%%%%%%%%%%%%%%%%%%%%%%%%%%%%%%%%%%%%%%%%%%%%%%%%%%%%%%%%%%%%%%%%%%%%%%%%%%%

According to Aristotle {\em vacuum} is ``$\tau\grave{o}\ 
\kappa\epsilon\nu\acute{o}\nu$'', ``the empty''.  The same Latin
world which we now use, ``vacuum'', refers to the absence of anything,
to a ``space bereft of body''. But what actually is this body?

This apparently easy question has received different answers at
different times and these answers often rely on subtle distinctions
like, for example, that between matter and what contains it.  It is
interesting to note that in the last two thousand years both of these
concepts have undergone a continuous (and unfinished) evolution in
their meaning.

What indeed {\em is} matter? The common sense reply (and the one which
our ancestors would have given) is that matter is the real substance
of which objects are made, and that {\em mass} is the concept that
quantifies the amount of matter in a body. But it is easy to see how
this answer has deeply evolved in time. The famous Einstein formula
$E=mc^2$ has definitely broken any barrier between matter and energy
and nowadays physics assumes the existence (at least in a relative
sense) of objects which are never directly subject to our observation.

The concept of space has also changed dramatically. If initially the
bodies were located in a Euclidean space, which had its definition in
a set of positioning laws, with Newton, Lorentz and Einstein this
concept has now evolved. Whereas for Newton ``absolute space'' was
something that ``in its own nature, without relation to anything
external, remains always similar and immovable'', soon it was
recognized that this sort of ``stage'', completely independent of
matter, was actually only a metaphysical category given the fact that
no physical reality can be associated with it. As a consequence 
mechanics in absolute space and time was replaced in practice by the
use of preferred inertial systems e.g. the one defined in terms of the
``fixed stars''. 

The development of the theory of electromagnetism led later on to the
concept of a special ubiquitous medium, the ``ether'', in which the
electromagnetic waves could propagate, and with Lorentz this ether was
described as ``the embodiment of a space absolutely at rest''.  This
concept was also soon rejected when the consequences of Einstein's
theory of Special Relativity were fully understood.  Nevertheless
Einstein was aware of the fact that his theory did not at all imply
the rejection of concepts like empty space; instead he stressed that
the main consequence of Special Relativity regarding the ether was to
force the discarding of the last property that Lorentz left to it, the
immobility.  The ether can exist but it must be deprived of any a
priori mechanical property. This is actually the main feature that the
vacuum (the new name given to the discredited ether) has now in \GR .
As Einstein himself said ``the ether of \GR\ is a medium which by
itself is devoid of any mechanical and kinematical property but at the
same time determines the mechanical (and electromagnetic) processes''.

In this spirit we now see, via the Einstein equations, not only that
the distribution of matter-energy constrains the spacetime itself but
also that solutions described by a vacuum (a null stress-energy tensor)
are endowed with a complex geometrical structure. In a certain sense
we can see now that the synthesis between matter and space, contained
and container, is actually achieved in the modern concept of the vacuum
which is both. 

Still this evolution in the meaning of ``the empty'' is possibly far
from being ended. The development of quantum theory has taught us
that the vacuum is not just a passive canvas on which action takes place.
It is indeed the most important actor.
In modern quantum theories (e.g. string theory) the vacuum
has assumed a central role to such an extent that the identification
of the vacuum state is the central problem which these have to solve.
Particles and matter are merely excitations of the fundamental
(vacuum) state and in this sense are just secondary objects.

The phenomena which we are going to discuss are all manifestations of this
novel active role of the vacuum in modern physics. We shall see how
the vacuum
can indeed manifest itself and how it is sensitive to external
conditions. As a starting point for our investigation we shall review
some basic aspects of the nature of the quantum vacuum in quantum
field theory (QFT).

%--------------------------------------------------------------------------
\subsection[{Canonical quantization}]
{Canonical quantization} 
\label{subsec:canqua}
%--------------------------------------------------------------------------

Let us consider the standard procedure for second quantization of a
scalar field $\phi(t,\x)$ in Minkowski spacetime. The basic
steps required for the quantization are

\begin{itemize}
\item Define the Lagrangian or equivalently the equations of motions
  of the field
  \begin{eqnarray}
    \label{eq:scalar}
    & \L=\half\left(\eta^{\mu\nu}\partial_{\mu}\phi\,\partial_{\nu}\phi-
         m^2\phi^{2}\right)\\
     \label{eq:KG}
    & \Box \phi+m^2\phi=0
  \end{eqnarray}
\item Define a scalar ({\em aka} inner) product
\begin{equation}
(\phi_1,\phi_2) =
 \im \int_{\Sigma_t} \phi_1^* \;
 \stackrel{\leftrightarrow}{\partial}_{t}  \;
 \phi_2\: \d^3x,
 \label{eq:inpro}
\end{equation}  
where $\Sigma_t$ denotes a spacelike hyperplane of simultaneity at
instant $t$. The value of the inner product has the property of being
independent of the choice of $\Sigma_t$.
\item Find a set of solutions of the field equations
  $\{\phi^{(+)}_{\k}(t,\x);\phi^{(-)}_{\k}(t,\x)\}$ which is complete and
  orthonormal with respect to the scalar product just defined
  $(\phi^{(+)}_{\k},\phi^{(-)}_{\k})=0$. The superscript $(\pm)$ here
    denotes solutions with positive and negative energy with respect
    to $t$.
\item Perform a Fourier expansion of the field using the above set
  \begin{equation}
    \label{eq:expa}
       \phi=\sum_{\k} \left [ 
                          a_{\k}\phi^{(+)}_{\k}(t,\x)+
                          a^{\dagger}_{\k}\phi^{(-)}_{\k}    
                  \right]
  \end{equation}
  The sum is appropriate for discrete momenta; for continuous ones, an
  integration is assumed.
\item The quantization of the field is carried out by imposing the
  canonical commutation relations
  \begin{equation}
    \left.
     \begin{array}{lll}
    \left[\phi(t,\x),\dot{\phi}(t,\x')\right] &=&
     i\hbar c^2 \delta(\x-\x')\\
     &&\nonumber\\
    \left[\phi(t,\x),\phi(t,\x')\right] &=&
     [\dot{\phi}(t,\x),\dot{\phi}(t,\x')]=0
     \end{array}
    \right \}
    \label{eq:comrel}     
  \end{equation}
  and these imply the commutation relations for the
  $a_{\k},a^{\dagger}_{\k}$ coefficients. 
  \begin{equation}
    \left.
     \begin{array}{lll}
    \left[ a_{\k}, a^{\dagger}_{\k'} \right] &=& \delta_{\k\k'} \\
    \left[ a_{\k},a_{\k'} \right] &=& 
    \left[ a^{\dagger}_{\k},a^{\dagger}_{\k'} \right]=0
     \end{array}
    \right \}
    \label{eq:comrela}
  \end{equation}
\item In the Heisenberg picture, the quantum states span a Hilbert
  space. A convenient basis in this Hilbert space is the so-called
  Fock representation.  A multiparticle state $|n_{\k}\rangle$ can be
  constructed from the special state $|0\rangle$ by the application
  the coefficients defined above.  The coefficients
  $a_{\k},a^{\dagger}_{\k}$ become, via the relations
  (\ref{eq:comrela}), operators respectively of destruction and
  creation of one particle of momentum $\k$.
  \begin{eqnarray}
    \label{eq:credes}
   & a^{\dagger}_{\k}\,|n_{\k}\rangle = \sqrt{n+1}\, |(n+1)_{\k}\rangle \\
   & a_{\k}\, |n_{\k}\rangle = \sqrt{n}\,|(n-1)_{\k}\rangle
  \end{eqnarray}
  the vacuum state $|0\rangle$ is defined as the state that is
  annihilated by the destruction operator for any $\k$
\begin{equation}
  \label{eq:vacuum}
  a_{\k}|0\rangle\equiv 0 \qquad \forall \k
\end{equation}
\item If we now consider the bilinear operator $N_{\k}\equiv
  a_{\k}^{\dagger}a_{\k}$ then its expectation value on the vacuum
  state and on a multiparticle state are respectively
  \begin{eqnarray}
    \label{eq:partnum}
   & \langle 0|N_{\k}|0\rangle = 0 \qquad \forall \k \\
   & \langle n |N_{\k}|n\rangle = n_{\k}
  \end{eqnarray}
  Thus the expectation value of the operator $N_{k}$ tells us the
  number of particles with momentum $\k$ in a given state. The vacuum
  state is in this sense the only state which has no particles for any
  value of $\k$.
\end{itemize}
The above considerations are all based on the definition of some
complete orthonormal set of classical solutions
$\phi^{(+)}_{\k}(t,\x);\phi^{(-)}_{\k}(t,\x)$. In the absence of
external fields and in Minkowski space the eigenfunctions of the
translation in time operator $\partial/\partial t$, which is the
generator of the Poincar\'e group, form a privileged set of such
solutions (the set associated with inertial observers). In this case
the $\phi^{(+)}$ and $\phi^{(-)}$ have a clear meaning of positive and
negative frequency solutions and the vacuum state $|0 \rangle$ defined
via Eq.\ (\ref{eq:vacuum}) is invariant under transformations of the
Poincar\'e group. So the procedure for constructing the Fock space
turns out to be completely unambiguous.

%--------------------------------------------------------------------------
\subsection[{Canonical quantization in external fields}]
{Canonical quantization in external fields}
\label{subsec:qvext}
%--------------------------------------------------------------------------

The situation is completely different when the quantization has to be
carried out in the presence of an external field. By ``external'' we
mean that the field, which interacts with the one which we want to
quantize, is introduced at a classical level and is not itself a
dynamical variable. We shall see that the role of external field can
be played by very different objects such as electromagnetic and
gravitational fields, some geometrical boundary of the spacetime or
some non trivial topology of the manifold over which the quantization
is performed. Different phenomenologies are also going to be
encountered depending on whether the external field is stationary or
not. If it is stationary, one generically expects vacuum polarization
effects, while if it is not stationary, there can be particle emission
from the quantum vacuum.

In the presence of an external field the translation invariance in
time or space is broken and it is hence impossible to uniquely
define an orthonormal basis. Different complete orthonormal sets of
solutions become equally valid but they lead to different inequivalent
vacua. The basic point is that in this situation the Poincar\'e group
is no longer a symmetry group of the spacetime (e.g. because this is
curved or characterized by boundaries) and hence the vacuum state $|0
\rangle$ defined via Eq.\ (\ref{eq:vacuum}) is now generally dependent
on the basis used for performing the quantization.

In the case that the external field is a gravitational one, the best
thing that one can hope to do is to look for timelike Killing vectors to
use for properly defining the decomposition into positive and negative
energy frequency modes. In some simple cases the global structure of
the spacetime is asymptotically Minkowskian and hence it is possible
to have a natural reference vacuum to use in defining the particle
content of the other vacua allowed by the equivalent sets of
orthonormal bases. We shall come back to this issue later on.

In the next section we shall illustrate a useful technique which allows
inequivalent vacua to be related via a conceptually easy formalism.
This
technique, generally referred to as a ``Bogoliubov transformation'', will
be used several times in the body of this thesis.

%--------------------------------------------------------------------------
\subsubsection[{Bogoliubov transformations}]
{Bogoliubov transformation}
\label{subsubsec:bog}
%--------------------------------------------------------------------------

We can start by considering the case of a scalar field which is
quantized in the presence of an external field (which can be
gravitational or not, static or dynamically changing in time).
Imagine that the problem admits at least two distinct complete
orthonormal sets of modes as solutions of the equations of motion of
the field.  In what follows we shall see similar situations in the case
of time varying external fields or in the case of a static
spacetime which admits two different global Killing vectors associated
with isometries in time (and hence two distinct ways to define positive
frequency modes).

In any case, the above statement implies that it is possible to
decompose the field $\phi$ in two different ways
\begin{eqnarray}
  \phi(t,\x)&=& \sum_{i} 
                 \left(
   a_{i}\varphi_{i}(t,\x)+a^{\dagger}_{i}\varphi^{*}_{i}(t,\x)
                 \right)\label{eq:twodec1}\\
  \phi(t,\x)&=& \sum_{j} 
                 \left(
   b_{j}\psi_{j}(t,\x)+b^{\dagger}_{j}\psi^{*}_{j}(t,\x)
                 \right)\label{eq:twodec2}
\end{eqnarray}
where $i$ and $j$ schematically represent the sets of quantities
necessary to label the modes. 

These two decompositions will correspond to two inequivalent vacua
defined respectively as
\begin{eqnarray}
  \label{eq:ineqvac1}
  |0 \rangle_{\varphi} &\qquad& a_{i}|0\rangle_{\varphi}\equiv 0\\
  \label{eq:ineqvac2}
  |0 \rangle_{\psi}    &\qquad& b_{j}|0\rangle_{\psi}   \equiv 0
\end{eqnarray}
The completeness of both sets allows one to be expanded as a function
of the other
\begin{eqnarray}
  \varphi_{i} &=& \sum_{j}\left(
      \alpha_{ij}\psi_{j}+\beta_{ij}\psi^{*}_{j}\right)\label{eq:bgtr}\\
  \psi_{j} &=& \sum_{i} \left(
      \alpha^{*}_{ij}\varphi_{i}-\beta_{ij}\varphi^{*}_{i}\right)
   \label{eq:bgtr2}
\end{eqnarray}
These relations are called Bogoliubov transformations and the $\alpha$ and
$\beta$ coefficients are called Bogoliubov coefficients.  
The latter are easily calculated via the inner product (\ref{eq:inpro}) as
\begin{equation}
  \label{eq:bgco}
  \alpha_{ij}=-(\psi_{i},\varphi_{j}), \qquad 
   \beta_{ij}=(\varphi^{*}_{i},\psi_{j})  
\end{equation}
From the equivalence of the mode expansions and using (\ref{eq:bgtr})
together with the orthogonality of the modes, it is possible to relate the
creation and destruction coefficients of the two bases:
\begin{eqnarray}
  a_{i}&=&\sum_{j}\left(
   \alpha^{*}_{ij}b_{j}-\beta^{*}_{ij}b^{\dagger}_{j} \right)
  \label{eq:bogop}\\
   &&\nonumber\\
  b_{j}&=&\sum_{i}\left(
   \alpha_{ji}a_{i}+\beta^{*}_{ji}a^{\dagger}_{i}\right)
\end{eqnarray}
From the above relations it is easy to show that the Bogoliubov
coefficients must satisfy the following properties
\begin{eqnarray}
\label{eq:unit}
  \sum_{k}
\left(\alpha_{ik}\alpha^{*}_{jk}-\beta_{ik}\beta^{*}_{jk}\right)
 &=&\delta_{ij}\\
\label{eq:ort}  
  \sum_{k}
\left(\alpha_{ik}\beta_{jk}-\beta_{ik}\alpha_{jk}\right)
 &=& 0
\end{eqnarray}
Now it is clear, from the relation (\ref{eq:bogop}) between the
destruction operators in the two bases, that the two vacuum states
associated with the two choices of the modes $\varphi_{i}$ and
$\psi_{j}$, are different only if $\beta_{ij}\neq 0$.  In fact in this
case one finds that the vacuum state $|0\rangle_{\psi}$ will not be
annihilated by the destruction operator of the $\varphi$ related basis
\begin{equation}
  \label{eq:novac}
  a_{i}|0\rangle_{\psi}=\sum_{j}\beta^{*}_{ij}|1_{j}\rangle_{\psi}\neq 0
\end{equation}
Actually, if we look at the expectation value of the operator,
$N_{i}=a^{\dagger}_{i}a_{i}$, for the number of $\varphi_{i}$-mode particles
in the vacuum state of the $\psi_{j}$ modes, we get
\begin{equation}
  \label{eq:nvac}
  \langle 0 | N_{i} | 0 \rangle_{\psi} =\sum_{j} |\beta_{ji}|^2
\end{equation}
This is equivalent to saying that the vacuum of the $\psi$ modes contains
a non null number of particles of the $\varphi$ mode.

It is then clear that the notion of a particle becomes ambiguous in these
situations as a consequence of the ambiguity in the definition of the
vacuum state. To obtain a more objective qualification of the state of
the field one should refer to objects which are covariant locally
defined quantities such as the expectation value of the stress energy
tensor. 
 
In this sense it is important to recall that the stress energy tensor (SET)
of the field can be cast in a form where an explicit dependence on
$\beta^2$ appears. Nevertheless one can see the it will generally
depend also on ``interference'' terms of the kind $\alpha\beta$~\cite{FD77}.
This is symptomatic of the looseness of the relation between particles
and energy-momentum. We shall directly experience this feature in the
next chapter when we discuss particle creation by collapsing
extremal black holes in section~\ref{sec:exbh}.

Another important point which we want to stress here is that the
Bogoliubov coefficients are just telling us the relation between
vacuum states.  In the case in which the two distinct sets of normal
modes correspond to two asymptotic states in time (for example in the
case in which the external field is static, then changes its value,
then is again static) this relation is interpreted as particle
creation.  If the $\psi$ modes are appropriate at early times and the
$\varphi$ ones are appropriate at late times, then the relation
(\ref{eq:nvac}) can be interpreted as saying that the
$|0\rangle_{\psi}$ has evolved into the $|0\rangle_{\varphi}$ plus
some particles.

Although the Bogoliubov transformation technique has often been used for
describing particle creation by nonstationary external fields it
should be stressed that the information which one can actually get from
the
value of the coefficients is limited. The relation (\ref{eq:nvac})
should more correctly be interpreted in the sense of a ``potential
possibility'', it tells us that the $\psi$ state is potentially
equivalent to the $\varphi$ state plus some $\varphi$-particles. It
does not give {\em a priori} any information about the actual
timing of the emission (when the particles are effectively produced).
This information requires the study of other quantities such as the
stress-energy tensor. 

\begin{center}
  \setlength{\fboxsep}{0.5 cm} 
   \framebox{\parbox[t]{14cm}{
%------------------------------------------------------------------------      
       {\bf Comment}: The inner product (\ref{eq:inpro}) is by
       construction independent of the chosen $t=\mbox{constant}$
       hypersurface.  This implies that in the case considered above,
       when one has only two sets of basis states, also the Bogoliubov
       coefficients (\ref{eq:bgco}) will be time independent. It is
       nevertheless possible to have more complex situations where one
       has more, time dependent, bases of states. In these cases
       one has to build up the Bogoliubov coefficients by taking the
       inner product between some initial basis and a time-dependent
       ``instantaneous'' one. It is clear that in such circumstances
       the Bogoliubov coefficients will indeed depend on time.  We
       shall discuss the instantaneous basis formalism in section
       \ref{subsec:ppEM}.
%------------------------------------------------------------------------
}}
\end{center}
%--------------------------------------------------------------------------
\subsection[{The quantum vacuum}]
{The quantum vacuum}
\label{subsec:nqv}
%--------------------------------------------------------------------------

The concept of the vacuum being a dynamical object endowed with an
autonomous existence is clearly shown in the framework discussed above.  
In fact it is a postulate of QFT that, when a measurement
of a physical quantity is performed, the quantum system is compelled
to occupy an eigenstate of the operator corresponding to the physical
quantity concerned. Since non commuting operators do not share common
eigenstates, this implies that if a system is in an eigenstate of a
given operator then, in general, it will not posses the properties
described by other operators which do not commute with the given one.

We have seen that the vacuum state is an eigenstate with zero
eigenvalue of the particle number operator $N_{\k}$ so in this state
it makes no sense to discuss the properties described by operators
which do not commute with $N_{\k}$ such as the field operator or the
current density.  Moreover, there cannot exist {\em any state} which
is an eigenstate of both $N_{\k}$ and any of the other non commuting
operators.  In the same way that in quantum mechanics it is impossible
for a particle to have zero values of both the coordinate and the
momentum, also in QFT it is impossible to find a state in which there
are simultaneously no photons {\em and} no quantized electromagnetic
field (virtual photons) or electrons {\em and} no electron-positron
current. There is never a truly ``empty'' vacuum state.

A crucial feature of the quantum vacuum, which is important for our
future discussion, is the fact that it is not only ubiquitous but at the
same time it appears to be endowed with an infinite energy.  If we
consider
the stress energy tensor of a free scalar field in flat space
\begin{equation}
  \label{eq:set}
  T_{\mu\nu}=\partial_{\mu}\phi\,\partial_{\nu}\phi
   -\half \eta_{\mu\nu}\eta^{\lambda\delta}
     \partial_{\lambda} \phi \,\partial_{\delta}\phi
     +\half m^2 \phi^2\eta_{\mu\nu}
\end{equation}
then it is easy to calculate the Hamiltonian and the Momentum and
express them in terms of the creation and destruction operators (using
the expansion of the field)
\begin{eqnarray}
  \label{eq:H}
  {\cal H} &=& \int_{t} {T_{00}\d^3 x=\sum_{\k}\omega_{\k}
   \left( a^{\dagger}_{\k}a_{\k}+\frac{1}{2} \right)}\\
  \label{eq:P} 
{\cal P} &=& \int_{t}{T_{0i}\d^3 x=\sum_{\k}a^{\dagger}_{\k}a_{\k}\, k_{i}}
\end{eqnarray}
We can now look at the vacuum expectation values of these operators.
Although the momentum gives the expected vanishing value
\begin{equation}
  \label{eq:momev}
  \langle 0| {\cal P}|0\rangle=0
\end{equation}
the expectation value of the Hamiltonian, the vacuum energy, clearly
shows a divergence
\begin{equation}
  \label{eq:enev}
   \langle 0| {\cal H}|0\rangle={1\over 2} \sum_{\k}\omega_{\k}=\infty
\end{equation}
This divergence of the vacuum energy can be shown to be generic;
nevertheless in Minkowski spacetime it can be removed by the
introduction of ``normal ordering''. This is generally denoted by ::
and demands that in any product of creation and annihilation
operators, the latter should be on the right of the former. Implicitly
this is equivalent to setting the energy of the Minkowski vacuum to
zero, the only justification for this assumption being ``outside'' of
QFT actually in the fact that Minkowski vacuum does not appear to
gravitate --- flat spacetime is a solution of the Einstein equations
where the expectation value of the stress energy tensor is zero
everywhere.

In the case of quantization in curved spacetimes or in the presence of
external fields or boundaries, that is in all the cases where the
central role of the Poincar\'e group breaks down, we do not have a
general way to specify a privileged vacuum and hence a normal ordering
procedure. Although this can generically lead to an intrinsic
ambiguity of the canonical quantization in external fields it is again
true that in a large class of cases an asymptotic symmetry group
exists which allows meaningful construction of a reference vacuum for
{\em the specific physical problem} which one wants to solve.

The relevance of the identification of such a reference background is
again important because it allows ``rescaling'' of the energy of the
(divergent) vacuum in the presence of some external field $\lambda$ by
subtracting the (divergent) energy of the asymptotic one (where the
external field has an asymptotic, possibly vanishing, value
$\lambda_{0}$). 
\begin{equation}
  \label{eq:subtr}
  E_{\mathrm{phys}}=E[\lambda]-E[\lambda_{0}]
\end{equation}
One may wonder why we could not always try to rescale the vacuum
energy in such a way as to make it zero. The point is that also in this
case we would not be allowed to discard the existence of the vacuum, in
fact the expectation value of the fluctuations of the fields would still be
non-zero in the vacuum.
\begin{equation}
  \label{eq:fluct}
  \Delta(\phi^2(t,\x))=\langle 0 | \phi^2(t,\x) | 0 \rangle -
   \langle 0 |\phi(t,\x)|0\rangle ^2
\end{equation}

Of course the above mentioned removal of divergences generally implies
the use of regularization procedures and renormalizations of the
stress energy tensor.  In these procedures there are inherent pitfalls
and difficulties. Most of these are mainly related to the fact that
while the renormalization procedure should obviously be covariant and
gauge invariant, at the same time the presence of external conditions
require the use of both a definite coordinate system and a specific
gauge for the external field.  We shall not deal with these technical
issues here and direct the reader to standard
textbooks~\cite{Birrell-Davies,moste,MostTru}. We shall see this
framework in action several times in this thesis.

As a related remark we would like to make clear why such a heuristic
procedure as the one described above, is generically able to remove
the divergences from the vacuum energies. The basic point here is that
the divergences of the kind shown in Eq.\ (\ref{eq:enev}) are clearly
of ultraviolet nature. They came from the behaviour of the quantum
field theory at extremely short scales. On the other hand the external
fields generally determine a change in the vacuum structure at a {\em
  global} level. This implies a universal nature for the vacuum
divergences and gives a concrete explanation for the cancellation of
the divergent part of the vacuum energies.

Now that we have sketched how there can be finite differences in
vacuum energy it is natural to ask if such energies are indeed
physically relevant and observable. In the next section we shall discuss
some examples in which the ``external field'' is static and lead just
to a change in the vacuum energy and we shall see how these energy shifts
actually lead to observable forces. After this we shall discuss the
influence of nonstationary external fields and their ability to lead
to particle creation from the quantum vacuum. At the end, we shall deal
specifically with vacuum effects in gravitational fields.

\begin{center}
  \setlength{\fboxsep}{0.5 cm} 
   \framebox{\parbox[t]{14cm}{
%------------------------------------------------------------------------      
       {\bf Comment:} Although one can be tempted to link concepts
       like vacuum energy shifts and particle creation respectively to
       effects in static and dynamical external fields, it has to
       be stressed that such a sharp distinction cannot exist.  For
       example a static electric field can create electron-positron
       pairs and a nonstationary electromagnetic field can polarize
       the vacuum without relevant particle production.  The very
       concept of a distinction between vacuum polarization and particle
       production can be subtle and needs to be handled with
       caution.
%------------------------------------------------------------------------
}}
\end{center}  
%%%%%%%%%%%%%%%%%%%%%%%%%%%%%%%%%%%%%%%%%%%%%%%%%%%%%%%%%%%%%%%%%%%%%%%%%%%
\section[{Vacuum effects in static external fields}]
{Vacuum effects in static external fields}
\label{sec:VEEFS}
%%%%%%%%%%%%%%%%%%%%%%%%%%%%%%%%%%%%%%%%%%%%%%%%%%%%%%%%%%%%%%%%%%%%%%%%%%%

In this section we consider a special class of quantum vacuum
effects induced by static external fields. This last phrasing is
generally used in the literature to cover a wide range of
possibilities.  It can be used for vacuum effects in the presence of
boundaries but also for those induced by quantization on spaces with
non trivial topologies (in flat as well as in curved backgrounds). Of
course there are also cases in which the external field is truly a
classical field, for example it can be a scalar, electromagnetic or
gravitational field. 

Here we shall give a brief presentation of only the first two kinds of
effect (those induced by boundaries and topology).  Particular
attention will be given to the case of the gravitational field in the
last section of this chapter where we shall treat static as well as
dynamical cases.

%-------------------------------------------------------------------------
\subsection[{The Casimir effect}]
{The Casimir effect} 
\label{subsec:casimir}
%--------------------------------------------------------------------------

\begin{figure}[hbt]
  \vbox{\hfil \scalebox{0.30}{{\includegraphics{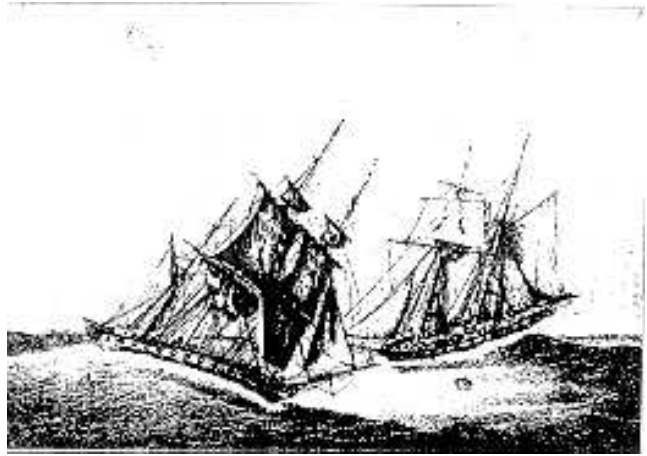}}} \hfil }
  \bigskip
\caption[Maritime Casimir effect]{
%----------------------------------------------------------------------------
  From Causse\`e ``The Mariners Album'', two ships heavily rolling in
  a situation where there are still long waves but no wind.
%---------------------------------------------------------------------------
} \label{f:maritime}
\end{figure}

In 19th century P.C. Causse\`e described in his {\em ``L' Album du
  Marin''} a mysterious phenomenon which was the cause of maritime
disasters~\cite{Caussee,Boer}. Figure \ref{f:maritime} shows the
situation which he called ``Calme avec grosse houle'': no wind but
still with a big swell running. In this situation he stressed that if
two ships end up lying parallel at a close distance then often ``une
certaine force attractive'' was appearing, pulling the two ships
towards each other and possibly leading to a collision.

%==============================================================================
%=============================================================================
%
This phenomenon was for many years considered as just another
superstition of sailors because it was far from clear what force should
be at work in these situations.  It is only recently that this effect
was given the name of the ``Maritime Casimir Effect''~\cite{Boer}.

We shall not discuss here in detail the explanation proposed
in~\cite{Boer} but limit ourselves to a heuristic explanation that
will be useful as an introduction to the real Casimir
effect~\footnote{
%-----------------------------------------------------------------------
  We shall keep the discussion discursive because an actual
  calculation will be redundant with the one which we shall present for
  the Casimir effect in the next section.
%-----------------------------------------------------------------------
}.

Following the discussion of the previous section we can start by
looking at the configuration characterized by just open sea.  In this
case the background is characterized by arbitrarily long waves. Let us
concentrate on the direction orthogonal to the alignment of the boats.
This is the only direction, say the $x$ direction, which is affected
by the location of the ships.

The waves in the open sea can have wavelengths $\lambda$ covering the
whole range of $x$, and so their wave number $k=2\pi/\lambda$ runs
over a continuous range of values.

When the ships are introduced, the $x$ direction automatically
acquires a scale of length and the waves between the two boats are now
obliged to have wave numbers which are integer multiples of a
fundamental mode inversely proportional to the distance $a$ between
the ships, $k_{n}=\pi n/a$.

The energy per mode will obviously be changed, nevertheless the total
energy will in both cases be infinite (being given by the sum over the
modes). What indeed can be expected is that the regularized total
energy in the case of the two ships is less than the corresponding one
for the open sea. In fact just a discrete (although still infinite)
set of waves will be allowed leading to a negative result when
performing the subtraction (\ref{eq:subtr}).

A negative energy density then automatically leads to an attractive
force between the two ships. Although this maritime Casimir effect can
be see as a nice macroscopic example of the quantum effect which we
are going to treat in detail, it should be stressed that it cannot be
more than an analogy; actually it is impossible that for two aligned
boats in an open sea there could be an accurate description of their
attraction in the above terms: other effects of many sorts will
influence the behaviour of the ships and make any interpretation of
the observations obscure. All in all, we are not sure that Mr.
Causse\`e was not telling us one of the numerous mariners stories ...

\subsubsection{A 1+1 version of the Casimir effect} 
 
In 1948 the Dutch physicist H.B.G Casimir predicted that two parallel,
neutral, conducting plates in a vacuum environment should attract each
other by a very weak force per unit area that varies as the inverse
forth power of their separation~\cite{Casimir},
\begin{equation}
  \label{eq:Fcas}
  F=-\frac{\pi^2}{240}\frac{\hbar c}{a^4}
\end{equation}
This effect was experimentally confirmed in the Philips
laboratories~\cite{ExCas} a decade later (see
also~\cite{Lamoreaux:1997wh,Lamoreaux:1999nw} for more recent
results). For plates of area $1\ {\rm cm}^2$ and separation of about
half a micron the force was $\approx 0.2 \cdot 10^{-5}$ N in agreement
with the theory.

To be concrete we shall now show a simplified calculation of the
Casimir effect made using a massless scalar field and working in 1+1
dimensions~\footnote{ 
%----------------------------------------------------------------------  
  A more detailed discussion can be found in various textbooks and
  reports~\cite{moste,MostTru,DW,Casrep}, in particular for an
  original derivation based on pure dimensional analysis see DeWitt
  in~\cite{DeWitt79}.
%--------------------------------------------------------------------------
}.

The 1+1 Klein-Gordon equation takes the form
\begin{equation}
  \label{eq:2dkg}
   \Box \phi(t,x)=0
\end{equation}
We shall assume the internal product and canonical commutation
relations as given in equations (\ref{eq:inpro}) and
(\ref{eq:comrel}) and a standard quantization procedure as previously
discussed in section~\ref{subsec:canqua}. Generically the solutions of
the equation of motion can be expressed in the form of traveling waves:
\begin{eqnarray}
  \label{eq:vacwav}
 & \phi_{k}^{(\pm)}(t,x)=
 \frac{\textstyle 1}{\textstyle \sqrt{4\pi\omega}}\exp[\pm \im(\omega t-k x)]\\
 &\nonumber \\
 & \omega^2=k^2, \qquad -\infty<k<+\infty \nonumber
\end{eqnarray}
We can now consider the one dimensional analogue of the Casimir
effect, that is a boundary in the $x$ direction that imposes Dirichlet
boundary conditions on the field.
\begin{equation}
  \label{eq:cond}
  \phi(t,0)=\phi(t,a)=0
\end{equation}
In this case we have again that there is a discrete set of allowed
wavenumbers and that the wavefunctions are of the form
\begin{eqnarray}
  \label{eq:vacwav2}
  & \phi_{k}^{(\pm)}(t,x)=\frac{\textstyle 1}
    {\textstyle \sqrt{a\omega_{n}}}\exp[\pm \im(\omega_{n} t)] \sin{k_{n} x}\\
  & \nonumber \\
  & \omega^2=k^2_{n}, \qquad k_{n}=\pi n/a \quad n=1,2,\dots \nonumber
\end{eqnarray}
From the above forms of the normal modes it is clear that the creation
and destruction coefficients will be different in the two
configurations (free and bounded space) and hence the vacuum states,
$|0\rangle_{\mathrm{Mink}}$ and $|0\rangle$, will differ as well.

Using the fact that the Hamiltonian of the field has the form
(\ref{eq:H}) we can find the expectation value of the field energy in
the vacuum state for the two configurations (free and bounded)
\begin{eqnarray}
  \label{eq:mincasivev}
  E_{\mathrm{free}}&=&\langle 0|{\cal H}|0\rangle_{\mink}=
  \half 
  \int_{-\infty}^{+\infty}{ \frac{\textstyle \d k}{\textstyle 2\pi} |k| L } \\
   && \nonumber \\
  \label{eq:boundcasivev}
  E_{\mathrm{bound}}&=&
  \langle 0|{\cal H}|0\rangle=
  \half
  \sum_{n=1}^{\infty}{ \frac{\textstyle \pi n}{\textstyle a} }
\end{eqnarray}
where we have introduced in (\ref{eq:mincasivev}) a normalization length 
$ L\to \infty $.

Both of the above quantities are divergent so, in order to compute them
and correctly execute the ``Casimir subtraction'' (\ref{eq:subtr}) we
should adopt some regularization scheme.

This can be done using an exponential cutoff of the kind
$\exp{(-\alpha \omega)}$ and by looking in the Minkowskian case at the
energy in a region of length $a$. To compute this last quantity
we shall look at the energy density $E_{\mathrm{free}}/L$ times $a$.
The results are then
\begin{eqnarray}
  \label{eq:numval}
  E_{\mathrm{free}}^{a}&=&\frac{E_{\mathrm{free}}\cdot a}{L}=
      \half\int_{-\infty}^{\infty}{ \d k |k|a\,\exp{(-\alpha|k|)}}=
      \frac{\textstyle a}{\textstyle 2\pi\alpha^2}\\
  &&\nonumber\\
  E_{\mathrm{bound}}&=&
   \half\sum_{n=1}^{\infty}\frac{\textstyle \pi n}{\textstyle a}
   \exp{(-\alpha\pi n/a)}=
   \frac{\textstyle a}{2\pi\alpha^2}-\frac{\textstyle \pi}{24 a}+O(\alpha^2)
\end{eqnarray}
So we see that in the limit $\alpha\to\infty$ the subtraction
$E_{\mathrm{bound}}-E_{\mathrm{free}}^{a}$ gives a finite quantity
$E_{\mathrm{Casimir}}=-\pi/(24 a )$ to which corresponds to an attractive
force 
\begin{equation}
  \label{eq:Fcasim}
F_{\mathrm{Casimir}}=-\frac{\partial E_{\mathrm{Casimir}}}{\partial a} 
                    =-\frac{\pi}{24a^2}   
\end{equation}
The true Casimir calculation, performed in three spatial dimensions
and with the electromagnetic field, has the same functional form. It
actually differs from the one found above in the value of the
numerical prefactor and in the power of the typical length scale (a
clear consequence of the different number of dimensions).

The calculation just sketched has been performed for different
geometrical configurations and notably it turns out that the value and
even the sign of the Casimir energy is a non trivial function of the
chosen geometry. In particular it is interesting to note that the
Casimir energy in a cubic cavity is negative but that in a spherical
box turns out to be positive.

There are some important points that have to be stressed before
proceeding. 
\begin{enumerate}
\item The Casimir effect could be interpreted as a manifestation of van
  der Waals forces of molecular attraction. However the force
  (\ref{eq:Fcas}) has the property of being absolutely independent of
  the details of the material forming the conducting plates. This is a
  crucial feature which establishes the definitive universal nature of
  the Casimir effect. The independence from the microscopic structure
  of the plates proves that the effect is just a byproduct of the
  nature of the quantum vacuum and of the global structure of the
  manifold.
\item The fact that for different geometries also positive energy
  densities can be obtained implies that the heuristic interpretation,
  that the presence of the boundaries ``takes away'' some modes and
  hence leads to a decrement of the total energy density with respect
  to the unbound vacuum, is wrong. What actually happens is that the
  number of modes allowed between the (ideal) plates is still
  infinite, what changes is the distribution of the vacuum field
  modes. What we see as the appearance of a repulsive or attractive
  vacuum pressure is an effect of this (geometry dependent)
  redistribution of modes~\cite{Hush96}.
\item A common terminology used for describing the shift in the
  vacuum energy appearing in static Casimir effects is that of
  ``vacuum polarization''.  The vacuum is described as a sort of
  dielectric material in which at small distances virtual
  particle-antiparticle pairs are present, analogously to the bounded
  charges in dielectrics.  The presence of boundaries or external
  forces ``polarizes'' the vacuum by distorting the virtual
  particle-antiparticle pairs. If the force is strong enough it can
  eventually break the pairs and ``the bound charges of the vacuum
  dielectric'' become free.
  
  Although this has turned out to be a useful concept, at the same time
  it should be stressed that it is strictly speaking incorrect.  The
  vacuum state is an eigenstate of the number operator which is a
  global operator (because it is defined by the integral of the
  particle density over the
  whole 3-space). On the other hand, the
  field operator, the current density and the other typical operators
  describing the presence of particles are local, so there is a sort of
  complementarity between the observation of a (global) vacuum state
  and (local) particle-antiparticle pairs.
  
  In the following we shall use this standard terminology but keeping in
  mind the fact that it is a misuse of language.
\end{enumerate}

Finally I would like to point out an important feature common to a
wide class of cases of vacuum polarization in external fields.  We can
return to our example of $1+1$ Casimir effect.  Consider the
energy density which in this case is simply given by
Eq.(\ref{eq:boundcasivev}) divided by the interval length $a$.
\begin{equation}
   \label{eq:en-den}
   \varepsilon=\langle 0|T_{00}|0 \rangle =\frac{1}{2a}\sum_{n=1}^{\infty}
   \omega_{n}\, , \qquad \omega_{n}=\frac{\pi n}{a}
\end{equation}
In order to compute it we can adopt a slightly different procedure for
making the Casimir subtraction. It is in fact sometimes very useful in
two dimensional problems to use (especially for problems much more
complicated that the one at hand) the so called Abel--Plana formula
which is generically
\begin{equation}
  \label{eq:ab-pl}
  \sum_{n=0}^{\infty}F(n)-\int_{0}^{\infty}F(\sigma)\d \sigma 
   =\frac{F(0)}{2}+\im \int_{0}^{\infty} \d \sigma 
   \frac{F(\im\sigma)-F(-\im\sigma)}{\exp{(2\pi \sigma)}-1}
  \equiv {\mathrm{reg}}\sum_{0}^{\infty}F(n)
\end{equation}
where $F(z)$ is an analytic function at integer points and $\sigma$ is
a dimensionless variable. This formula is a powerful tool in
calculating spectra because the exponentially fast convergence of the
integral in $F(\pm \im \sigma)$ removes the need for explicitly inserting
a
cut-off.

In our specific case, $F(n)=\omega_{n}$, and so Eq.(\ref{eq:ab-pl})
applied to the former energy density takes the form
\begin{equation}
  \label{eq:ab-pl-spec}
  \varepsilon_{\rm ren}=-\frac{1}{\pi}\int_{0}^{\infty}
   \frac{\omega\d \omega}{\exp{(2a\omega)}-1}
\end{equation}
Although it may appear surprising, we have found that the spectral
density of the Casimir energy of a scalar field on a line interval
does indeed coincide, apart from the sign, with a thermal spectral
density at temperature $T=1/a$.

The appearance of the above ``thermality'' in the static Casimir
effect is not limited to the case of external fields implemented via
boundary conditions and it is not linked to the use of the Abel--Plana
formula (which is merely a computational tool). As we shall see, it
can appear also when quantization is performed in the presence of a
gravitational field (in the case of spacetimes with Killing horizons).
We shall try to give further insight on this point later on in this
chapter.

Before considering other examples of vacuum polarization we shall now
devote some words to an interesting effect which is related to the
Casimir one.

%-----------------------------------------------------------------------------
\subsubsection{The Scharnhorst effect}
%-----------------------------------------------------------------------------

In 1990 K.~Scharnhorst and G.~Barton~\cite{Sch90,Bar90,SchBar93}
showed that the propagation of light between two parallel plates is
anomalous and indeed photons propagating in directions orthonormal to
the plates appear to travel at a speed $c_{\bot}$ which exceeds the
speed of light $c$. The propagation of photons parallel to the Casimir
plates is instead at speed $c_{\|}=c$.

The above results where found starting from the Maxwell Lagrangian
modified via a non-linear term stemming from the energy shifts in the
Dirac sea induced by the electromagnetic fields. The 
action obtained in this way is the well known Euler--Heisenberg one which
in its real part
takes the form ($m_{e}$ is the electron mass)
\begin{equation}
  \label{eq:EuHei}
  \L=\frac{\E^2-\B^2}{8\pi}+
     \frac{\alpha^2}{2^3 \cdot 3^2 \cdot 5 \pi^{2} m_{e}^4}
     \left[ \left(\E^2-\B^2\right)^2+7\left(\E\cdot\B \right)\right] 
\end{equation}
One can see here that the correction is proportional to the square of
the fine structure constant and so it is a one-loop effect.

The above action in particular also describes the scattering of light
by light and can be read as giving to the vacuum an effective,
intensity-proportional, refractive index $n=1+\Delta n$. The
Scharnhorst effect can be seen as a consequence of the fact that in the
Casimir effect the intensity of the zero-point modes is less than in
unbounded space and hence leads to a proportionate drop in the
effective refractive index of the vacuum ($\Delta n<0$).

It should be stressed that the above results are frequency independent
for $\omega\ll m$ and so the phase, group and signal velocities coincide.
Moreover using the Kramers--Kronig dispersion relation
\begin{equation}
  \label{eq:KrKr}
  \Re \left\{n(\omega)\right\}=\Re \left\{n(\infty)\right\}+\frac{2}{\pi} 
  \int_{0}^{\infty}\d \omega'\; 
  \frac{\omega'\Im \left\{n(\omega')\right\}}
       {{\omega'}^{2}-\omega^{2}}
\end{equation}
and assuming that the vacuum always acts a passive medium ($\Im\left\{
n(\omega)\right\}\geq 0$), it has been shown~\cite{SchBar93} that the
effective refractive index of the vacuum at high frequencies must be
smaller than that at low frequencies, and so
\begin{equation}
  \label{eq:krkr2}
  1>n_{\bot}(0)\geq n(\infty)
\end{equation}

Although this anomalous propagation is too small to be experimentally
detected (the corrections to the speed of light are of order
$\alpha^2/(m d)^4$, where $d$ is the distance between the plates) it
is nevertheless an important point that a vacuum which
is polarized by an external field can actually behave as a dispersive
medium with refractive index $n(\omega)$ less than one.  In fact
similar effects have been discovered in the case of vacuum polarization
in gravitational fields~\cite{DH80,DS94,DS96,Shore96}.

This realization has important consequences and deep implications
(e.g.~about the meaning of Lorentz invariance and the possible
appearance of causal pathologies~\cite{DN98}) which are beyond the
scope of the present discussion. We just limit ourselves to noting
the fact that these issues can be dealt with very efficiently in a
geometric formalism where the modification of the photon propagation
can be ``encoded'' in an effective photon metric
$g_{\mu\nu}^{\eff}$~\cite{DG98,Novello99,DeLorenci:2000yh}.  In particular,
in~\cite{Novello99} it is shown that in a general non linear theory of
electrodynamics, this effective metric can be related to the Minkowski
metric via the renormalized stress-energy tensor, in the form
\begin{equation}
  \label{eq:minkoptmetr}
  g^{\eff}_{\mu\nu}=A\eta_{\mu\nu}+B \langle T_{\mu\nu}\rangle
\end{equation}
where $A$ and $B$ are constants depending on the specific form of the
Lagrangian.  

We shall see in chapter~\ref{chap:4} (section~\ref{sec:chi-vsl}) how this
relation generalizes in curved spacetimes and what its possible role
can be in modern cosmology. For the moment, we shall return to our
general discussion of vacuum effects in static external fields.

%----------------------------------------------------------------------
\subsection[{Other cases of vacuum polarization}]
{Other cases of vacuum polarization}
%----------------------------------------------------------------------

The Casimir effect which we have been discussing is just a special
case of a more general class of phenomena leading to vacuum
polarization via some special boundary conditions that makes the
quantization manifold different from the Minkowskian one. Generally
speaking, this class is divided into effects due to physical
boundaries (which are strictly the {\em Casimir effects}) and those
due to nontrivial topology of the spacetime (which are generally
called {\em topological Casimir effects}).

We have briefly discussed some cases of the first class: in general
they are all variations on the main theme of the Casimir effect. They
are all in Minkowski space with some boundary structure, all that
changes is the kind of field quantized, the geometry of the boundaries
(parallel plates, cubes, spheres, ellipsoids etc.) and the number of
spatial dimensions.

In the second class of phenomena, the quantization is performed in
flat spacetimes (or curved ones, as we shall see later) endowed with a
non trivial topology (that is with a topology different from the
standard $R\times R^3$ of Minkowski spacetime). The topology is
generally reflected in periodic or anti-periodic conditions on the
fields.  Again the trivial example of the quantization of a scalar
field on the interval $(0,a)$ turns out to be useful.

Let us consider the case of periodic boundary conditions:
\begin{eqnarray}
  \label{eq:periodic}
  & \phi(t,0) = \phi(t,a)\\
  & \partial_{x}\phi(t,0)=\partial_{x}\phi(t,a)
\end{eqnarray}
This time, in the points $x=0,a$ we are allowed to have also non-zero
modes, and hence we shall have, as possible solutions, the one in
Eq.(\ref{eq:vacwav2}) and another one with $\cos k_{n}x$ replacing the
sine. Hence, we get a doubled number of modes and a different spectrum:
\begin{eqnarray}
  \label{eq:vacwav3}
  & \phi_{k}^{(\pm)}(t,x)=\frac{\textstyle 1}
    {\sqrt{\textstyle 2a\omega_{n}}}\exp[\pm \im(\omega_{n} t-k_{n} x)]\\
  & \nonumber \\
  & \omega^2=k^2_{n}, \qquad k_{n}=2\pi n/a \quad -\infty<n<\infty \nonumber
\end{eqnarray}
The unbounded space solutions will still be those given by
Eq.(\ref{eq:vacwav}). Following the same procedure as before, the
result is now $E=-\pi/6a$~\cite{MostTru}.  Noticeably if one choose
instead anti-periodic boundary conditions 
\begin{eqnarray}
  \label{eq:period}
  & \phi(t,0) =- \phi(t,a)\\
  & \partial_{x}\phi(t,0)=\partial_{x}\phi(t,a)
\end{eqnarray}
then the result changes again to $E=\pi/12a$.  This can be seen as
another proof of the fact that what matters for the determination of
the Casimir energy is a redistribution of the field modes and that it
is not correct to think that the boundary conditions reduce the number of
allowed modes and hence lead to an energy shift. 

The topological Casimir effect has been studied also in more complex
topologies and in a larger number of dimensions
(see~\cite{moste,MostTru,Casrep} for a comprehensive review). A common
feature is that in four dimensions the energy density is generally
inversely proportional to the fourth power of the typical
compactification scale of the manifold\footnote{
%--------------------------------------------------------------------------  
  This dependence can be explained by purely dimensional arguments;
  actually it easy to show that in $(d+1)$ dimensions the energy
  density should behaves as $\epsilon=\hbar c/L^{(d+1)}$ where $L$ is
  the
  typical length scale of the problem.}
%-------------------------------------------------------------------------
 $\epsilon\propto 1/a^4$.
Moreover it is interesting to note that these cases also often show an
effective temperature of the sort which we discussed previously; e.g. in
the
case of the periodic boundary conditions on a string this is again
inversely proportional to $a$. 

Given the special role that the emergence of the effective thermality
has in the case of vacuum effects in strong gravitational fields, we
shall now try to gain further understanding about the nature of this
phenomenon.

%--------------------------------------------------------------------------
\subsection[{Effective vacuum temperature}] 
{Effective vacuum temperature}
\label{subsec:evt}
%--------------------------------------------------------------------------

We have seen how in a number of cases the spectral density, describing
the vacuum polarization of some quantum fields in manifolds which
differ from Minkowski spacetime, is formally coincident with a thermal
one. Given that the effective thermality of the vacuum emerges in very
different physical problems, the question arises of whether it is
possible to give a general treatment to the problem. In particular one
would like to link the presence of a vacuum temperature to some
general feature and if, possible, to find such a temperature
$T_{\eff}$ without necessarily computing explicitly the energy density
of the vacuum polarization.  Such a treatment indeed exists and it can
be instructive for us to review it here
(see~\cite{Birrell-Davies,moste,MostTru} and reference therein).

The basic idea is that the effective thermality is a feature of the
vacuum polarization which is induced by the external field. This
external field (which can be a real field or some sort of boundary
condition) makes the global properties of the manifold under
consideration different from those of reference background (e.g.
unbounded Minkowski spacetime). 

Hence, it is natural to assume that also $T_{\eff}$ should depend on the
same global properties of the manifolds (boundaries, topologies,
curvature etc.) and so information about it should be encoded in
some other object, apart from the stress energy tensor, which is
sensitive to the global domain properties.  Good candidates for such an
object are obviously the Green functions of the field
$G(x,x')$~\footnote{ Here $x=(t,\x)$ is a four-vector}.  Let us work
with the Wightman Green functions,
\begin{eqnarray}
  \label{eq:hadam}
  G^{+}(x,x') &=& \langle 0 |\phi(x)\phi(x')|0\rangle,\\
  G^{-}(x,x') &=& \langle 0 |\phi(x')\phi(x)|0\rangle,\\
  G(x,x') = \langle 0 |\{\phi(x)\phi(x')\}|\rangle
    &=& G^{+}(x,x')+G^{-}(x,x')
\end{eqnarray}
where $|0\rangle$ is the vacuum state of the manifold under consideration and
$G(x,x')$ is the Hadamard Green function.

What we can do then is to construct a tangent bundle to our manifold
by constructing in every point $x$ a tangent Minkowski space. In this
space we can consider the thermal Green functions at a given
temperature $T_{\eff}=(\beta k_{\mathrm B})^{-1}$. These can be built by
making an ensemble average of (\ref{eq:hadam})
\begin{equation}
  \label{eq:termgf}
  G^{\mink}_{\beta}(\phi(x),\phi(x'))=
   \frac{ \Tr \; e^{-\beta \H} \{\phi(x)\phi(x')\} }{\Tr \; e^{-\beta \H}}
\end{equation}
where $\H$ is the field Hamiltonian. In the limit $\beta\to \infty$,
$G^{\mink}_{\beta}$ coincides with the vacuum Green function in Minkowski
space $G^{\mink}$ (the same happens for $G^{\mink}_{\pm}$).

From the universality of the local divergences one can assume that the
leading singularity in $G(x,x')$ for $x\to x'$ is the same as one
encounters for the thermal Green function in the tangent Minkowski
space. If in the limit $x\to x'$ the Green functions of our
manifolds coincide with the thermal ones of the tangent space for some
$T_{\eff}$ then we can say that the vacuum $|0\rangle$ has an
effective thermality at temperature $T_{\eff}$.  So we shall have to
look at the condition
\begin{equation}
  \label{eq:condterm}
  \lim_{x'\to x}\left[G(x,x')-G^{\mink}_{\beta}(x,x')\right]=0
\end{equation}

A first very important property of thermal Green functions (for zero
chemical potentials) is that
\begin{equation}
  \label{eq:imtim}
  G^{\pm}_{\beta}(t,\x;\,t'\x')=
     G^{\mp}_{\beta}(t+\im\beta,\x;\,t',\x')
\end{equation}
This is a direct consequence of the thermal average exponents in
(\ref{eq:termgf}) being equal to the Heisenberg evolution
operators~\footnote{
%------------------------------------------------------------------------ 
  Indeed all that one needs to deduce (\ref{eq:imtim}) is the definition
  (\ref{eq:termgf}) and the Heisenberg equations of motion
\begin{equation}
  \label{eq:Heom}
   \phi(t,\x)=e^{\im \H(t-t_{0})}\phi(t_{0},\x)e^{-\im \H(t-t_{0})}
\end{equation}
%------------------------------------------------------------------------
See~\cite{Birrell-Davies} for further details.}.

The above property allows one to find a very important relationship  
between
the thermal Green functions and the zero
temperature Green functions~\cite{Birrell-Davies} 
\begin{equation}
  \label{eq:imtimsum}
  G_{\beta}(t,\x;\,t'\x')=\sum_{n=-\infty}^{+\infty} 
                                G(t+\im n\beta,\x;\,t',\x')
\end{equation}
that is, the thermal Green function can be written as an infinite
imaginary-time image sum of the corresponding zero-temperature Green
function.

This relation is not only useful for following the present
discussion but also has an intrinsic value for itself. In fact it is the
substantial proof that periodicity in the imaginary time $\tau=it$
automatically implies thermality of the Green function in the standard
time.  This feature is a crucial step, as we shall see in the next
chapter, in order to derive thermality for black holes by purely
geometrical considerations.

Turning back to our problem, we can now try to explicitly compute
Eq.(\ref{eq:condterm}). We first note that the term for $n=0$ in the
sum of Eq.(\ref{eq:imtimsum}) corresponds to the Minkowskian Green
function at zero temperature which is known to have the form
\begin{equation}
G^{\mink}(x,x')=-\frac{1}{2\pi^2 d(x,x')} 
\end{equation}
where $d(x,x')=(x^{\mu}-{x^{\mu}}')(x_{\mu}-{x'}_{\mu})$ is
the interval between $x$ and $x'$ (for the sake of simplicity, I shall
drop the addition of the imaginary term $\im 0$ in the denominator above,
which is
needed to remove the pole at $x=x'$).

Using this we can perform the summation in Eq.(\ref{eq:imtimsum})
separating the $n=0$ term and taking the limit for $x \to x'$ 
\begin{equation}
  \label{eq:subtrGr}
   \lim_{x\to x'}G^{\mink}_{\beta}(x,x')=
          G^{\mink}(x,x')+\frac{1}{6\beta^2}  
\end{equation}
We can now use this expression in the condition (\ref{eq:condterm}) and
get in this way an explicit definition for the vacuum effective
temperature
\begin{equation}
  \label{eq:Teff}
  k_{\mathrm B}T_{\eff}=\frac{1}{\beta}=\sqrt{6}\lim_{x'\to x}
       \left[G(x,x')-G^{\mink}(x,x')\right]^{1/2}
\end{equation}

A final remark is in order before closing this section. The effective
temperature which we have discussed so far is a property related to the
spectral distribution of the vacuum state of a field which in some
cases can have a Planck-like appearance.

In general, real thermality is not just a property of two point Green
functions, but more deeply involves conditions on the structure of the
$n$th-order correlation functions.  In a general case, the vacuum
state is still a non-thermal state even in the presence of boundaries or
non
trivial topologies.  Hence it will in general not admit a
non-trivial density matrix.  In some other cases, such as in spacetimes
with Killing horizons, a non-trivial density matrix arises as a
consequence of the information loss associated with the presence of
the horizon. In these cases the effective thermality is ``promoted''
to a real thermal interpretation.  We shall come back to these topics
again in the following parts of this thesis.

%%%%%%%%%%%%%%%%%%%%%%%%%%%%%%%%%%%%%%%%%%%%%%%%%%%%%%%%%%%%%%%%%%%%%%%%%
\section[{Vacuum effects in dynamical external fields}]
{Vacuum effects in dynamical external fields}
\label{sec:dyncas}
%%%%%%%%%%%%%%%%%%%%%%%%%%%%%%%%%%%%%%%%%%%%%%%%%%%%%%%%%%%%%%%%%%%%%%%%%

In the previous section we have always assumed that the boundary
condition or field acting on the quantum vacuum was unchanging in
time. If the field is non-stationary, this dramatically changes
the physics described above, in particular rapid changes in the
external field (or boundary) can lead to the important phenomenon of
particle production from the quantum vacuum. The variation in time of
the external field perturbs the zero point modes of the vacuum and
drives the production of particles. These are generically produced
in pairs because of conservation of momentum and also of
other quantum numbers (e.g. in the case of production of fermions, the
pairs will actually be particle-antiparticle pairs).

Given that the energy is provided by the time variation of the
external field, then it should be expected that the intensity and
efficiency of this phenomenon is mainly determined by the rapidity of
the changes in time (for a fixed coupling between the quantized
field and the external one).

This phenomenon, sometimes called the ``dynamical Casimir effect'', has a
wide variety of applications. In particular we shall see that in the
case where the time varying external field is a gravitational one, its
manifestation has changed our understanding of the relation between
\GR\ and the quantum world.  Unfortunately we are still lacking
experimental observations of particle creation from the quantum
vacuum, and part of the work presented in this thesis is devoted to
the search for such observational tests.

%--------------------------------------------------------------------------
\subsection[{Particle production in a time-varying external field}]
{Particle production in a time-varying external field}
\label{subsec:ppEM}
%--------------------------------------------------------------------------

To keep the discussion analytically tractable we shall again take the
very simple case of a real scalar field described by the standard
equations of motion in flat space. This will interact with another
scalar field which will be described by a (scalar) external potential
$U$, which contributes to the Lagrangian density (\ref{eq:scalar}) a
term $U\phi^2$. We shall further assume $U$ to be a function of time
but homogeneous in space $U=U(t)$. The equation of motion
(\ref{eq:KG}) then takes the form
\begin{equation}
  \label{eq:eomscalar}
  \Box \phi(t,\x)+\left[m^2+U(t)\right]\phi(t,\x)=0
\end{equation}
We can imagine that our field is again quantized in a spacetime with
zero or periodic (antiperiodic) boundary conditions such as the cases
considered before. The homogeneity of the potential then assures
that the spatial part of the solutions of the equations of motion is
unchanged and that a generic eigenfunction can be factorized as
\begin{equation}
  \label{eq:factoriz}
    \phi_{n}(t,\x)=N_{n}g_{n}(t)\chi_{n}(\x)  
\end{equation}
where $N$ is a normalization factor and $\chi_{n}(x)$ has the form of
the space dependent part of the specific solutions (\ref{eq:vacwav2})
or (\ref{eq:vacwav3}). The above form allows decoupling of the equation
for the time dependent parts of the wave functions which takes the
form of an harmonic oscillator with variable frequency
\begin{equation}
  \label{eq:timedep}
  \ddot{g}_{n}+\omega^{2}_{n}(t)g_{n}=0 \qquad 
                                   \omega_{n}^{2}=m^2+|k|^{2}_{n}+U(t)
\end{equation}
We shall now see how from this equation one can gain information about
the creation of particles. To see what happens as $U$ varies in time we
can consider two limiting cases: the first is the so called {\em
  sudden limit}, the second one is called the {\em adiabatic limit}.

%-----------------------------------%
\subsubsection{Sudden limit}
%-----------------------------------%

Let us consider the case of a sharp impulse described by a short time
jump in $V(t)$. To make the things easier we can model this jump via a
delta function.
\begin{equation}
  \label{eq:jump}
  U(t)=\nu \delta(t)
\end{equation}
At all times except $t=0$ the solution of (\ref{eq:timedep}) has the
standard form
\begin{equation}
  \label{eq:standf}
  g_{n}(t)=\g^{(+)}_{n}\exp(\im\omega_{n}t)+\g^{(-)}_{n}\exp(-\im\omega_{n}t)
\end{equation}
where $\g^{(\pm)}$ are constant factors and
$\omega^{2}_{n}=k^{2}_{n}+m^2$.

The basic observation is that the derivatives of $\dot{g}_{n}$ differ
as $t\to\pm 0$ leading to a discontinuity in the first
derivative~\cite{MostTru}
\begin{equation}
  \label{eq:discont}
   \dot{g}_{n}(0^{+})-\dot{g}_{n}(0^{-})+\nu g_{n}(0)=0
\end{equation}
We can now suppose that for $t<0$ the function $g_{n}$ was
characterized by purely positive-frequency modes so that initially
$\g^{(+)}_{n}|_{t<0}=1,\g^{(-)}_{n}|_{t<0}=0$. Then we can easily see
that after the jump the discontinuity leads to coefficients
\begin{equation}
  \label{eq:discon2}
    \g^{(+)}_{n}|_{t>0}=1-\nu/2\im\omega_{n}\quad 
    \g^{(-)}_{n}|_{t>0}=\nu/2\im\omega_{n}  
\end{equation}
This shows that the variation in time of the potential leads to the
appearance of negative frequency modes which are associated with
creation operators and hence with particles.  Actually a treatment via
a Bogoliubov transformation between the solutions before and after the
jump confirms this interpretation of the mode mixing. Indeed the
coefficients used $\g^{(\pm)}_{n}|_{t>0}$ are nothing other than the
Bogoliubov coefficients $\beta,\alpha$ and the square of the absolute
value of $\g^{(\pm)}_{n}|_{t>0}$ leads to the particle number density,
$\rho_{n}=|\g^{(-)}_{n}|_{t>0}|^{2}$.

%-----------------------------------%
\subsubsection{Adiabatic limit}
%-----------------------------------%

We can now consider the opposite limit of slow variation of the
potential $U(t)$. In this case Eq.~(\ref{eq:timedep}) is governed by a
slowly varying frequency for which $\gamma=\dot{\omega}/{\omega}\ll 1$
and one can apply the quasiclassical approximation of quantum
mechanics. In this approximation the mixing of positive and negative
frequency modes is exponentially suppressed. We can then expect that
particle creation will be exponentially small in $\gamma$ as well.  We
shall encounter this behaviour and discuss it in detail in
chapter~\ref{chap:3b}.

%-----------------------------------%
\subsubsection{General cases}
%-----------------------------------%

In general we shall not be in either of the above regimes and in that
case a detailed discussion of the dynamics of particle creation is
a much more complicated problem. 

An important fact is the presence or not of stationary regimes for the
external potential. In fact in these cases we can define in a standard
way the quantum states asymptotically in the past and in the future
and then define a Bogoliubov transformation that relates them and in
which is encoded all of the information about the way in which the
``in'' states go into the ``out'' ones. Even in this framework there
are few known cases where an analytical treatment is possible. We
refer the reader to~\cite{Birrell-Davies} for some examples; in this
category can also be cast the calculations based on the Bogoliubov
transformations that will be shown in this thesis.

In the most general case in which the physical problem does not allow
a meaningful definition of asymptotic states, then a much more
complicated approach is required. 

Concretely one has to determine somehow an instantaneous basis of
solutions for the equations of motion of the field. This is necessary
in order to define a set of time dependent operators $b^{(\pm)}$ in
terms of which the Hamiltonian of the system can be diagonalized at
all times
\begin{equation}
  \label{eq:hdi}
  \H(t)=\sum_{n}\omega_{n}(t)b_{n}^{(+)}(t)b^{(-)}_{n}(t)
\end{equation}
The $b_{n}^{(\pm)}(t)$ coefficients can be treated as instantaneous
creation and destruction operators for the field.  This obviously
implies also a definition of a time dependent vacuum state
$b_{n}|0\rangle\equiv 0$. Now we can imagine that the variation in
time of the potential $V(t)$ is switched on at some time $t_{0}$ at
which the vacuum state is known. One can then relate the above
$b_{n}^{(\pm)}(t)$ operators to those corresponding to the
wavefunctions at $t_{0}$, say $a_{n'}^{(\pm)}$, via some Bogoliubov
transformations. The instantaneous particle spectrum will be given
again by the Bogoliubov coefficient $\beta_{n,n'}$ squared.

The main difficulty in this procedure is the
determination of the instantaneous basis. 
If one decomposes the field in the standard way:
\begin{equation}
  \label{eq:tddec}
  \phi(t,\x)=
  \sum_{n}\left[b_{n}\varphi(t,\x)+b^{\dagger}_{n}\varphi^{*}_{n}(t,\x)\right]
\end{equation}
then the commonly used ``trick'' is to expand the mode function at a
given instant in time as
\begin{equation}
  \label{eq:mfist}
  \varphi_{n}(t,\x)=\sum_{k}Q_{nk}(t)u_{k}(t,\x)
\end{equation}
where $u_{k}(t,\x)$ is the mode function corresponding to the
stationary case while $Q(t)$ includes all of our ignorance about
what happens to the wavefunction as a consequence of the nonstationarity
of the problem.  Inserting the above expression into the equations of
motion allows an equation for $Q(t)$ to be found which one should then try
to solve at least perturbatively.

A very instructive case is that of a massless scalar field $\phi$ quantized
in flat space between two walls at which Dirichlet boundary conditions are
implemented. We can imagine that one of the walls is at $x=0$ and the
other is at $L_{0}$ for $t\leq 0$ and then starts to move with law
$L(t)$ for $0<t<T$. In this case equations (\ref{eq:tddec}) and (\ref{eq:mfist})
hold with
\begin{equation}
  \label{eq:mfist2}
  u_{k}(t,\x)=\sqrt{\frac{2}{L(t)}}\sin{\omega_{k}(t)x}
\end{equation}
where $\omega_{k}(t)=k\pi/L(t)$.

From the standard wave equation $\Box \phi=0$ then one gets for $Q(t)$
\begin{equation}
  \label{eq:eqqq}
  \ddot{Q}_{nk}+\omega^2_{k}(t)Q_{nk}=2 \lambda  \sum_{j}g_{kj}\dot{Q}_{nj}+
           \dot{\lambda}\sum_{j}g_{kj}Q_{nj}+\dot{\lambda}^2
              \sum_{j,l}g_{jk}g_{jl}Q_{nl}
\end{equation}
where $\lambda=\dot{L}/L$ and
\begin{equation}
  \label{eq:gkj}
  g_{kj}=\left\{
      \begin{array}{cr}
          (-1)^{(k-j)}\frac{2kj}{j^{2}-k^{2}} & \quad (j\neq k)\\
                         0                    & \quad (j  =  k)
      \end{array}
          \right.
\end{equation}

It is clear that the form of equation (\ref{eq:eqqq}) does not
generically allow an easy solution. There is nevertheless a class of
interesting problems where the boundary moves with harmonic motion,
$L(t)=L_{0}[1+\epsilon\sin(\Omega t)]$. In the case of small
displacements, $\epsilon \ll 1$ it is in fact possible to
solve (\ref{eq:eqqq}) perturbatively in
$\epsilon$~\cite{Klaw94a,Klaw94b,Klaw95,Dodo96,LJR96,JJPS97}.

The special cases where the equations of motion of the form (\ref{eq:timedep})
or (\ref{eq:mfist2}) admit a periodic time dependent frequency, all
lead to the general phenomenon of {\em parametric resonance}.

%------------------------------------------------------------------------
\subsubsection{Parametric Resonance}
%-------------------------------------------------------------------------

To be concrete, we can consider the case in which the $U(t)$ is an
electric field oscillating in time and aligned along the $x_{3}$ axis.
\begin{equation}
  \label{eq:oscil}
  A_{3}=(E_{0}/K_{0})\cos K_{0}t
\end{equation}
and scalar particles are created in a plane perpendicular to the
direction of the external field. In this case Eq.(\ref{eq:timedep})
takes the form
\begin{equation}
  \label{eq:timedep2}
   \ddot{g}_{n}+\omega^{2}_{n}(t)g_{n}=0 \qquad 
                    \omega_{n}^{2}=m^2+k_{\bot}^{2}+e^2 A^{2}_{3}
\end{equation}
where $k_{\bot}=k^{2}_{1}+k^{2}_{2}$, $k_{3}=0$ because of the
perpendicularity condition, and $e$ is the coupling constant.

If we redefine the variables by setting $z=K_{0}t-\pi/2$:
\begin{equation}
  \label{eq:redef}
  A = k_{0}^{-2}(m^2+k^{2}_{\bot})+2q, \quad 
  q = \frac{e^2 E_{0}^{2}}{4 k_{0}^{2}}. 
\end{equation}
Eq.(\ref{eq:timedep2}) takes the form:
\begin{equation}
  \label{eq:mathieu}
  \frac{\d^2 g_{n}}{\d z^2}+\left(A-2q \cos 2z \right)g_{n}=0
\end{equation}
This equation is the very well known {\em Mathieu equation}.  The
special feature of the solutions of such a system is that they show a
resonance band structure, determined by the values of $A$ and $q$. 
%==============================================================================
\begin{figure}[hbt]
  \vbox{\hfil \scalebox{0.60}{{\includegraphics{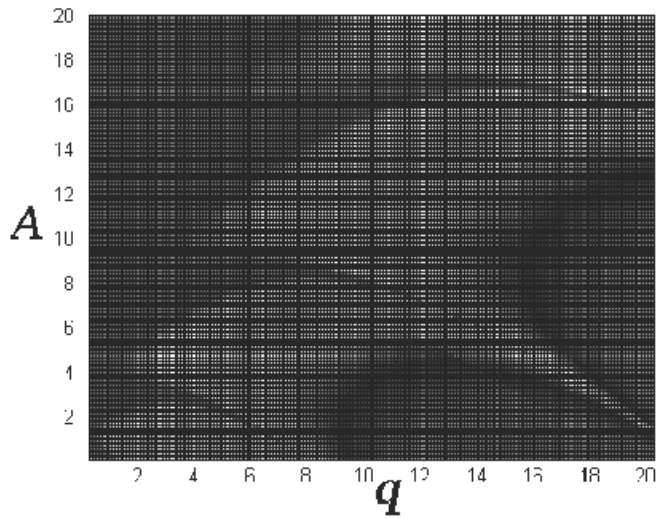}}} \hfil }
  \bigskip
\caption[Mathieu function band structure]{
%----------------------------------------------------------------------------
  Band structure of the Mathieu function (\ref{eq:mathieu}). The
  clearer bands denote the instability regions. The
    author wishes to thank F.~Tamburini for providing the picture and,
    together with B.~Bassett, for allowing the use of it.
%---------------------------------------------------------------------------
} \label{f:mathieu}
\end{figure}
%=============================================================================
%
In correspondence with some of these bands, called ``instability bands'', 
some modes $g_{n}$ can be exponentially amplified
\begin{equation}
  \label{eq:expampl}
  g=\exp(\mu_{N}z)u(z,\sigma) 
\end{equation}
where $\sigma$ is the value of $p_{\bot}^2+m^2$ for which the above
solution is computed. The interval $\sigma \in [-\pi/2,0]$ determines
the width of the instability band. The factor $\mu$ is the so called
{\em Floquet index} for which there is no general expression. Nevertheless
a general form of $\mu^{(N)}$ in the $N$-th instability band in the case
of small $q$ ($q\ll 2 N^{3/2}$) and positive $A$, can be given~\cite{NN74}
\begin{equation}
 \label{eq:floq}
  \mu^{(N)}= -\frac{1}{2N}\frac{\sin 2\sigma}{[2^{N-1}(N-1)!]^2} q^{N}
\end{equation}
In this case it is also possible estimate the spectral density of the
created $\phi$ particles as~\cite{NN74}
\begin{equation}
  \label{eq:spdensh}
  n(\k_{N})=\frac{\sinh^2(2\pi \mu^{(N)})}{\sin^2 2\sigma}
\end{equation}
Note that for massive particles the number of the first instability
zone which gives a contribution to the exponential growth is $N\approx
m/K_{0}$.  Unfortunately for $\pi$-mesons this number leads to the
conclusion that a time which is a huge multiple of the basic period
$T=2\pi/K_{0}$ should pass in order to obtain an observable
effect~\cite{moste}. We shall see in chapter \ref{chap:4} that
perhaps it is not on earth that we should seek for detection of
parametric resonance.

%-------------------------------------------------------------------------
\subsection[{Moving mirrors}]{Moving mirrors}
%-------------------------------------------------------------------------
  
As a last example of particle creation in non-stationary external
fields we now turn our attention to a class of problems which has had
a special role in the study of particle creation by incipient black
holes, that is to particle creation by moving
mirrors~\cite{FD76,FD77} (see also~\cite{Birrell-Davies,moste,MostTru}
for further references).

We can again consider a massless scalar field and to further simplify
the discussion we shall consider it in a two dimensional flat space time.
The action of the moving mirror is modelled by zero boundary
conditions imposed at the moving boundary $x=f(t)$.
\begin{equation}
  \label{eq:movmirbc}
  \phi(t,f(t))=0
\end{equation}
For concreteness we can assume the mirror to be static until $t=0$ and
then to start moving with $|\dot{f}|<1$ (in such a way that the
world-line of the mirror is always timelike).

Being in two dimensions, it is convenient to work in null coordinates
\begin{equation}
  \label{eq:nullcoo}
  u=t-x,\qquad v=t+x
\end{equation}
In this way the equation of motion takes the form
\begin{equation}
  \label{eq:eomnc}
  \partial^2\phi/\partial u\partial v=0
\end{equation}
As usual we have first of all to define a proper basis to be used for
building the ``reference vacuum''. In this case a natural reference
system is given by the zone where the mirror is at rest. From the
above expression it is easy to see that any function of just $u$ or
$v$ is a solution and a complete orthonormal system is
\begin{equation}
  \label{eq:rest}
  \varphi_{\omega}^{(\pm)}(t,\x)=\frac{\pm \im}{2\sqrt{\pi\omega}}
     [\exp(\pm \im\omega u)-\exp(\pm \im\omega v)]
\end{equation}
For $t>0$ the boundary condition (\ref{eq:movmirbc}) has to be
satisfied at a time-dependent point. This implies that the orthonormal
system is now of the form
\begin{equation}
  \label{eq:mov}
  \phi^{\pm}_{\omega}(t,\x)=\frac{\pm \im}{2\sqrt{\pi \omega}}
    [\exp(\pm \im\omega p(u))-\exp(\pm \im\omega v)] 
\end{equation}
where $p(u)$ is given by
\begin{equation}
  \label{eq:pu}
  p(u)=2t_{u}-u; \qquad t_{u}-u=f(t_{u})
\end{equation}
where $t_{u}$ is the time at which the line $u=t-x$ intersects the
mirror. When the mirror is static then $t_{u}=u$ and $p(u)=u$ and so
the function (\ref{eq:mov}) coincides with (\ref{eq:mov}). The
situation changes for $t>0$; in fact the function $p(u)$ acquires a
much more complicated form as a consequence of the reflection of the
incoming (left moving) modes $\exp(\pm i\omega v)$ by the moving
mirror.
 
This implies that the vacuum state defined via the destruction
operators associated to the modes (\ref{eq:rest}) will not in general
be a vacuum state for the modes (\ref{eq:mov}) and hence particle
production should be expected. The complicated exponential of $p(u)$
can be seen as representing a Doppler distortion which excites the
modes of the field and causes particles to appear. Physically this can
be described as a flux of particles which are created by the moving
mirror and which stream out along $u=\mbox{const}$ null rays.

The spectrum of these particles can then be determined via the
Bogoliubov transformations relating the modes (\ref{eq:rest}) and
(\ref{eq:mov}). Regarding the renormalized energy density
$\varepsilon=\langle T_{00}\rangle$ associated with the emission of
particles, this can be found as the difference between the appropriate
integrals over modes for the regions $t\geq0$ and $t=0$. The final
(finite) result can be shown to be~\cite{FD77}:
\begin{equation}
  \label{eq:setmm}
    \varepsilon(u)=-\frac{1}{24\pi}
     \left[\frac{p^{'''}}{p'}-\frac{3}{2}
       \left(\frac{p^{''}}{p'}\right)^{2}
     \right]  
\end{equation}
where the quantity appearing in brackets is called the Schwarzian
derivative.

It is interesting to note that the condition $\varepsilon(u)=0$ is
satisfied not only in the trivial case of uniform motion but also in
the one associated with constant proper acceleration.  We shall
discuss further this last case in the next chapter in relation to
particle creation by extremal incipient black holes.

Moreover, it should be stressed that also the above special
cases can give a non-zero energy density in the case where the
motion of the mirror
is of the kind considered before, that is characterized by an
early static phase and then by a sudden motion from a given instant in
time. In fact this leads to a discontinuity in the derivatives of
$p(u)$ localized at the value of $u$ (or t) at which the motion
started. This discontinuity hence generically gives a delta
contribution localized at the onset of the motion of the mirror.
Physically
this means that there is a burst of radiation emitted at that point.

Finally we want to close this paragraph by considering a special example
of mirror motion, that is the one associated with the asymptotic
trajectory
\begin{equation}
  p(u)=B-Ae^{-\kappa(u+B)}
\end{equation}
with $A,B,\kappa$ being some generally chosen constants.  We see that this
corresponds to a mirror that accelerates non-uniformly and has a
world-line which asymptotically approaches the null ray $v=B$.  In this
case we are not interested in those modes which have $v>B$ because
they will never be reflected by the mirror. The spectrum of the
particles created by the mirror can be computed via Bogoliubov
transformations between the appropriate basis for $t<0$,
Eq.~(\ref{eq:rest}), and that for $t\geq 0$ Eq.~(\ref{eq:mov}).
The Bogoliubov coefficients take the form~\cite{FD77}
\begin{eqnarray}
  \label{eq:bogmovmir}\left.
 \begin{array}{l}
  \alpha_{\omega\omega'}\\
  \beta_{\omega\omega'}
 \end{array}
  \right\} &=& \mp (2\pi)^{-1}(\omega\omega')^{-1/2} 
          [\omega']^{\pm \im\omega/\kappa}\Gamma(1\mp \im\omega/\kappa)\\
          && \qquad e^{\pm \pi\omega/2\kappa}
           e^{\pm \im\omega(\kappa^{-1}\ln A-B)-\im \omega' B}\nonumber
\end{eqnarray}
From the above coefficients it follows that the particle are created
with a Bose-Einstein distribution
\begin{equation}
  \label{eq:bogosquare}
   |\beta_{\omega\omega'}|^{2}=
    \frac{1}{2\pi\kappa\omega'} \left(\frac{1}{e^{\omega/k_{\mathrm B}T}-1}\right)
\end{equation}
with $k_{\mathrm B}T=\kappa/2\pi$. 

Note that the particle spectrum
\begin{equation}
  \label{eq:partspect}
  N_{\omega}=\int^{\infty}_{0}|\beta_{\omega'\omega}|^{2}\d\omega'
\end{equation}
diverges logarithmically. This is a generic feature of this sort of
calculation and is due to the fact that if the mirror keeps
accelerating for an infinite time, an infinite number of particles for
each mode will accumulate. This unphysical feature can be avoided by
using a finite wave packet in place of the plane waves which we used.

As we said, the example just discussed has a special importance because  
it enters as a part of the demonstration that a
collapsing body emits particles with a thermal distribution.  It can
be natural to ask how it is possible that a state which is a
vacuum one (characterized only by zero point fluctuations) can
suddenly evolve into a thermal-like one. To gain a further
understanding of this emergence of thermal distributions in particle
creation from the quantum vacuum we shall now devote some attention
to the quantum properties of the state generated by the dynamical
action of external fields.

%--------------------------------------------------------------------------
\subsection[{Squeezed States and Thermality}]
{Squeezed States and Thermality}
\label{subsec:squeezed}
%--------------------------------------------------------------------------

The clearest understanding of the aforementioned puzzle can be given
in terms of {\em squeezing}. The squeezed vacuum is a particular
distortion of the the quantum electrodynamic one. Actually the basic
characteristic of squeezed states is to have extremely low variance
for some quantum variable and correspondingly (due to the uncertainty
principle) very high variance for the conjugate variable~\footnote{
%--------------------------------------------------------------------------  
  Here lower and higher variance is intended with respect to the
  value one would expect on the basis of the equipartition theorem.
%--------------------------------------------------------------------------
  }. 

Squeezed states are of particular interest for our discussion because
they are the general byproduct of particle creation from the
quantum vacuum. In particular the vacuum In state is a squeezed vacuum
state in the out-Fock space~\footnote{For a demonstration of this
  in the case of a one-mode squeezed state see~\cite{Ford97}.}.

A two-mode squeezed-state is defined by
\begin{equation} 
|\zeta_{ab}\rangle =e^{- \zeta (a^{\dagger}
b^{\dagger} - b a)} |0_{a},0_{b} \rangle, 
\end{equation} 
where $\zeta$ is (for our purposes) a real parameter though more
generally it can be chosen to be complex~\cite{bk}. In quantum optics
a two-mode squeezed-state is typically associated with a so-called
non-degenerate parametric amplifier (one of the two photons is called
the ``signal'' and the other is called the ``idler'' \cite{bk,bk2,yupo}).
Consider the
operator algebra
\begin{equation}
[a,a^{\dagger}]=1=[b,b^{\dagger}],
\qquad
[a,b]=0=[a^{\dagger},b^{\dagger}],
\end{equation}
and the corresponding vacua 
\begin{equation}
|0_a \rangle :\ a |0_a \rangle =0, 
\qquad
|0_b \rangle :\ b |0_b \rangle =0. 
\end{equation}
The two-mode squeezed vacuum is the state $| \zeta \rangle \equiv
|0(\zeta) \rangle $ annihilated by the operators
\begin{eqnarray}
A (\zeta)&=&\cosh (\zeta) \; a -\sinh (\zeta) \; b^{\dagger},
\\
B (\zeta)&=&\cosh (\zeta) \; b-\sinh (\zeta) \; a^{\dagger}.
\end{eqnarray}
A characteristic of two-mode squeezed-states is that if we measure
only one photon and ``trace away'' the second, a thermal density
matrix is obtained~\cite{bk,bk2,yupo}.  Indeed, if $O_a$ represents an
observable relative to one mode (say mode ``a'') its expectation value
in the squeezed vacuum is given by
\begin{equation}
\langle\zeta_{ab}|O_a|\zeta_{ab}\rangle=\frac{1}{\cosh^2 (\zeta)}
\sum_{n=0}^{\infty} 
[\tanh(\zeta)]^{2 n} \langle n_{a}|O_a|n_{a} \rangle.    
\label{E:termo}
\end{equation}
In particular, if we consider $O_a=N_a$, the number operator in mode
$a$, the above reduces to
\begin{equation}
\langle\zeta_{ab}|N_{a}|\zeta_{ab}\rangle=\sinh^2 (\zeta).
\end{equation}
These formulae have a strong formal analogy with thermofield dynamics
(TFD) \cite{TU,ume}, where a doubling of the physical Hilbert
space of states is invoked in order to be able to rewrite the usual
Gibbs (mixed state) thermal average of an observable as an expectation
value with respect to a temperature dependent ``vacuum'' state (the
thermofield vacuum, a pure state). In the TFD approach, a trace over
the unphysical (fictitious) states of the fictitious Hilbert space
gives rise to thermal averages for physical observables completely
analogous to the one in (\ref{E:termo}) {\em except} that we must make
the following identification
\begin{equation}
\tanh(\zeta)= \exp\left(-\frac{1}{2} 
               \frac{\hbar \omega}{k_{\mathrm B} T}\right),
\end{equation}
where $\omega$ is the mode frequency and $T$ is the temperature.  We
note that the above identification implies that the squeezing
parameter $\zeta$ in TFD is $\omega$-dependent in a very special way.

The formal analogy with TFD allows us to conclude that, if we measure
only one photon mode, the two-mode squeezed-state acts as a
thermofield vacuum and the single-mode expectation values acquire a
thermal character corresponding to a ``temperature'' $T_{\mathrm
{squeezing}}$ related with the squeezing parameter $\zeta$ by
\cite{yupo}
\begin{equation}
k_{\mathrm B}\; T_{\mathrm{squeezing}} =
\frac{\hbar \; \omega_i}{2 \log(\coth(\zeta))},
\end{equation}
where the index $i=a,b$ indicates the signal mode or the idler mode
respectively; note that ``signal'' and ``idler'' modes can have
different effective temperatures (in general $\omega_{\rm signal} \neq
\omega_{\rm idler}$)~\cite{yupo}.

It is interesting to note that the squeezing parameter is indeed
linked to the Bogoliubov coefficients via a relation which in the case
of a diagonal Bogoliubov transformation takes the form
\begin{equation}
  \label{eq:bogsq}
  \tanh^2(\zeta)=\frac{|\beta|^2}{|\alpha|^2}
\end{equation}
This in an indication that in special cases where the Bogoliubov
coefficients take an exponential form then a thermal behaviour should
be expected. Noticeably this is exactly the case of the coefficients
(\ref{eq:bogmovmir}) of the ``asymptotic moving mirror''. 

Finally we wish to emphasize the following points:
\begin{enumerate}

\item Squeezed-mode ``effective thermality'' is {\em only} an artifact
  due to the particular type of measurement being made. There is no
  real physical thermal photon distribution in the electromagnetic
  field. However, the complete analogy with TFD implies that no
  measurement involving only single photons can reveal any discrepancy
  with respect to real thermal behaviour. 

\item A similar thermofield dynamics scheme is also often used in the
  case of black hole thermodynamics and in the Unruh effect
  \cite{Israel76}. We stress that in TFD as applied to black
  holes and the Unruh effect there is a physical obstruction to the
  measurement of both ``squeezed photons'' because they live in
  spacelike separated regions of the spacetime. Thermality associated
  with event horizons is ``real'' in that for an observer measuring
  a thermal particle spectrum from a single particle detection in his
  external portion of the Kruskal diagram the other particle of the
  pair is unobservable. This physical hindrance to measuring {\em
    both} of the photons is obviously not present in the case of quantum
  optics measurements so that, by measuring both photons of the pair
  it is (in principle) possible to find strong correlations that are
  absent in the true thermal case.
\item Squeezed states of light have been created in the laboratory
  by non-linear optics techniques~\cite{WKHW86}. Essentially in these
  cases a non-linear medium subject to a time-varying classical
  electromagnetic field, behaves like a material with a time dependent
  dielectric function. As a consequence of this, photon modes
  propagating through the medium undergo a mode mixing that is
  perceived as particle creation. The observation of such an effect
  can be considered as a first indirect test of the general theory which 
  we are dealing with.
\end{enumerate}

%%%%%%%%%%%%%%%%%%%%%%%%%%%%%%%%%%%%%%%%%%%%%%%%%%%%%%%%%%%%%%%%%%%%%%%%%
\section[{Vacuum effects in gravitational fields}] 
{Vacuum Effects in gravitational fields} 
\label{sec:vegf}
%%%%%%%%%%%%%%%%%%%%%%%%%%%%%%%%%%%%%%%%%%%%%%%%%%%%%%%%%%%%%%%%%%%%%%%%%

The theory of quantum vacuum effects in gravitational fields has some
special features which make the problem much harder to solve than for
the cases in flat space which we have seen so far.  The first problem
which one encounters is that the gravitational fields, unlike the
electromagnetic or scalar ones, cannot be described by an external
potential $V$, but are instead described by the metric of a curved
Riemannian spacetime. In the particular case of a scalar field we can
see from the generalization of equations~(\ref{eq:scalar},\ref{eq:KG})
\begin{eqnarray}
    \label{eq:cscalar}
    & \L=\half\left(\nabla_{\mu}\phi\nabla^{\nu}\phi-
         m^2\phi^{2}-\xi R\phi^2 \right)\\
     \label{eq:cKG}
    & \Box \phi+m^2\phi+\xi R\phi=0\\
    & \mbox{where}\quad \Box=g^{\mu\nu}\nabla_{\mu}\nabla_{\nu}=
       \frac{\textstyle 1}{\sqrt{\textstyle -g}} \partial_{\mu}
        \left( \sqrt{-g}\, g^{\mu\nu}\partial_{\nu}\right)\nonumber
\end{eqnarray}
that gravity couples to the field via the metric and a direct coupling
term $-\xi R\phi^2$~\footnote{
%------------------------------------------------------------------------  
  The case $\xi=0$ is said to be ``minimal coupling'' while that with
  $\xi=1/6$ is called ``conformal coupling'' and it can in fact be
  shown~\cite{Birrell-Davies} that for this value the massless field
  equations are invariant under conformal transformations
  $g_{\mu\nu}(x)\to\bar{g}_{\mu\nu}(x)=\Omega^{2}(x)g_{\mu\nu}(x)$.
  Terms of the kind $-\xi R\phi^2$ are possible as well for fields
  of higher spins.
%----------------------------------------------------------------------
}.

As we said in the Introduction, the main consequence of this is to
make the quantization of the fields extremely subtle and
unconventional from the very beginning. One suddenly finds oneself
handling non-linear equations with variable coefficients and problems
related to possible non-trivial structures like event horizons and
singularities.

In the particular case of the procedure of second quantization
sketched in section \ref{subsec:canqua} one finds two correlated
problems: the construction of the Hilbert space of quantized field
states and obtaining finite quantities for the physical observables.

The first problem involves the correct definition of the vacuum state
and then the interpretation of the quantized field in terms of
particles. As we have seen, the vacuum state of second quantization is
by construction Poincar\'e invariant but the presence of external
fields generally prevents the Poicar\'e group being a symmetry group of
the spacetime. In the cases considered so far this was only partially
a problem because it was always possible to define as a reference vacuum
the Minkowski one and to describe vacuum polarization as well as particle
creation with respect to it. In curved spaces it is not always meaningful to
take Minkowski as the reference vacuum. 

The second problem is strictly related to the former one. 
In fact the definition of an expectation value implies as a precondition the
construction of a state space. It has to be stressed that in the case of
gravitation, this issue acquires a central role because in the
special case of the stress-energy tensor the correct definition of
the expectation value determines the right hand side of the
Einstein equations and acts as the source for the geometry~\footnote{
%%%%%%%%%%%%%%%%%%%%%%%%%%%%%%%%%%%%%%%%%%%%%%%%%%%%%%%%%%%%%%%%%%%%%%%%%%%%%  
  Here, we assume that the semiclassical Einstein equations are simply
  given by $G_{\mu\nu}=8\pi\langle \!T_{\mu\nu}\!\rangle_{\rm ren}$ \cite{DW}.
  This, however, might not be a good approximation when $\phi$ is in a
  state with strong correlations (see, e.g., \cite{someth} and
  references therein).
%%%%%%%%%%%%%%%%%%%%%%%%%%%%%%%%%%%%%%%%%%%%%%%%%%%%%%%%%%%%%%%%%%%%%%%%%%%%%
  }.  

Actually, different conditions on the stress-energy tensor have in
the past been a powerful tool for various theorems in \GR .  We just want
to recall that the famous singularity theorems of Hawking and
Penrose~\cite{HawEll73} as well as some the laws of black hole
thermodynamics which we shall discuss in the following chapter, are
based on such conditions. We refer to the book by Visser~\cite{LorWorm} for
further insight into modern applications of them and limit ourselves
here to a brief summary.  

%-------------------------------------------------------------------
\subsubsection{Energy conditions}
%---------------------------------------------------------------------

Given the fact that we shall need to apply these conditions also in a
cosmological framework, we shall specify them in the particular cases
in which the SET has the form ``Type I'' as defined
in~\cite{HawEll73}:
\begin{equation}
 T^{\mu\nu}= 
  \left[ \matrix{\rho &  0  &  0  &  0  \cr
                   0  & p_1 &  0  &  0  \cr                  
                   0  &  0  & p_2 &  0  \cr
                   0  &  0  &  0  & p_3 }\right].  
  \label{eq:t1set}             
\end{equation}
where $\rho$ is the mass density and the $p_j$ are the three principal
pressures. In the case that $p_{1}=p_{2}=p_{3}$ this reduces to
the perfect fluid SET often used in cosmology.  
\begin{description}
\item[Null energy condition] The NEC asserts that for any null vector
  $k^{\mu}$
\begin{equation}
 \label{eq:nec}
  T_{\mu\nu}k^{\mu}k^{\nu}\geq 0
\end{equation}
In the case of a SET of the form (\ref{eq:t1set}) this involves
\begin{equation}
\label{eq:necpf}
\rho+p_{i}\geq 0 \quad \forall i
\end{equation}
\item[Weak energy condition] The WEC states that for any timelike
  vector $v^{\mu}$
\begin{equation}
 \label{eq:wec}
  T_{\mu\nu}v^{\mu}v^{\nu}\geq 0
\end{equation}
Physically one can interpret $T_{\mu\nu}v^{\mu}v^{\nu}$ as the density
of energy measured by any timelike observer with four-velocity
$v^{\mu}$: the WEC requires this quantity to be positive.  In terms of
principal pressures this gives
\begin{equation}
 \label{eq:wecpf}
  \rho\geq 0 \quad \mbox{and} \quad \rho+p_{i}\geq 0 \quad \forall i 
\end{equation}
By continuity, the WEC implies the NEC.
\item[Strong energy condition] The SEC asserts that for any timelike
  vector $v^{\mu}$ the following inequality holds
\begin{equation}
 \label{eq:sec}
  \left( T_{\mu\nu}-\frac{T}{2}\right)v^{\mu}v^{\nu}\geq 0
\end{equation}
where $T$ is the trace of the SET.

In terms of the special SET (\ref{eq:t1set}) the SEC reads
\begin{equation}
 \label{eq:secpf}
  \rho+p_{i}\geq 0 \quad \mbox{and} \quad \rho+\sum_{i}p_{i}\geq 0 \quad 
   \forall i
\end{equation}
The SEC implies the NEC (again by continuity) but not necessarily the WEC.
\item[Dominant energy condition] The DEC states that for any
timelike vector $v^{\mu}$
\begin{equation}
 \label{eq:dec}
  T_{\mu\nu}v^{\mu}v^{\nu}\geq 0 \quad \mbox{and} 
     \quad T_{\mu\nu}v^{\nu}\:\mbox{is not spacelike}
\end{equation}
This implies that not only should the locally observed energy density
be positive but also the energy flux should be timelike or null. The DEC
implies the WEC and hence also the NEC but in general does not force
the SEC. In the case of a SET of the form (\ref{eq:t1set}) one gets
\begin{equation}
 \label{eq:decpf}
  \rho\geq 0 \quad \mbox{and} \quad \: p_{i}\in [-\rho,+\rho] \quad \forall i 
\end{equation}

Apart from the above discussed ``point-wise'' EC there are other ones
which are the ``averaged'' versions (along some null or timelike
curves) of those just described. We refer to~\cite{LorWorm} for
further details.
\end{description}
It is interesting to note that the Casimir effect (resulting from
boundaries or topological) is a typical example in which {\em all} of
the above conditions are violated. We shall see in the next chapter
that this aspect of quantum vacuum effects is especially important
when discussing \bh\ thermodynamics.

Due to the central role of the SET in semiclassical gravity, an
immense effort has been directed to issues related to its
renormalization in the last thirty years and different techniques
(analytical as well as numerical) have been developed for dealing with
the connected problems. It is beyond the scope of this thesis to give
a comprehensive description of all of these issues; we shall limit
ourselves
to an overview, quoting where necessary possible sources for
in-depth analyses.

%--------------------------------------------------------------------------
\subsection[{Renormalization of the stress energy tensor}]
{Renormalization of the stress energy tensor}
\label{subsec:setren}
%--------------------------------------------------------------------------

We have seen how for flat spaces in the presence of boundaries the
standard
procedure for renormalizing the stress energy tensor is generally
based on the regularization of the ultraviolet divergent quantities
and then to the explicit or implicit subtraction of the corresponding
term in free space. The universality of the local divergences in flat
space assures the finiteness of the remaining quantities after the
cut-off removal.

As said previously, in the presence of gravity this whole procedure needs
a deep
revision.  In particular the SET will be sensitive to the curvature of
spacetime and in general will present some additional divergent terms with
respect to the standard one of Minkowski. A simple example of such a
failure of the standard Casimir subtraction is given by the
renormalization of the stress energy tensor of a minimally coupled
scalar field in a flat Friedmann--Lema\^{\i}tre--Robertson--Walker (FLRW)
spacetime described by the metric
\begin{equation}
  \label{eq:conffrw}
   \d s^2=C(\eta)(\d \eta^2-\d\x^2)
\end{equation}
where $\d \eta=\d t/a(t)$ and $a(t)$ is the standard scale factor.  We
shall assume that the scale factor follows the law $C(\eta)=\cos^2
A\eta$.  In this case the mode solutions $u_{\k}$ can easily be found
and hence a Fock space can be constructed. From the vacuum state
defined in this way $|0\rangle_{\rm FLRW}$ one can define the
expectation value of the SET of the field. Introducing into the
integral for the evaluation of the energy density an appropriate
cut-off factor $\exp[-\alpha(\omega^2+A^2)^{1/2}]$ (with $\alpha$ and
$A$ some constants) one gets~\cite{Birrell-Davies}
\begin{equation}
  \label{eq:setexfrw}
   \langle 0|T^{0}_{0}|0\rangle_{\rm FLRW}=
    \frac{48/\alpha^4+(D^2(\eta)-8A^2)/\alpha^2+A^2(D^2(\eta)/2-A^2)\ln\alpha}
     {32\pi^2 C^2(\eta)}+O(\alpha^{0})  
\end{equation}
where $D(\eta)=C^{-1}\partial C/\partial\eta$. We see from the above
expression there are three divergent terms in the limit $\alpha\to 0$.
Normally one would expect that these are exactly those appearing also
in the case of quantization in Minkowski spacetime.  The latter is
recovered in the limit of $C=1$ and $D=A=0$, but we see that in this
case only the first term in equation (\ref{eq:setexfrw}) remains. Hence
the subtraction of the SET of Minkowski would leave the other
two divergent terms: we cannot cure the infinities in FLRW flat space
simply discarding a Minkowski-like term!

This example shows how the whole issue of renormalization becomes
extremely delicate in curved spacetimes. Obviously one is committed to
finding some extra criteria for the SET in order to uniquely fix a
renormalization procedure. This ``axiomatic'' approach to the problem
was pursued in the 1970s by Christensen~\cite{Chris75},
Wald~\cite{Wald77} and others, and led to the identification of a
few basic properties which one should require for a well defined SET.
These can be summarized in a few points:
\begin{enumerate}
\item Covariant conservation: this is required for consistency given
  that the expectation value of the SET has to appear on the
  right-hand side of the Einstein equations.
\item Causality: this implies that the value of $\langle
  T_{\mu\nu}\rangle$ at a point $x$ should depend only on changes in
  the metric structure which took place in the causal past of $x$.
\item Standard forms for the ``off-diagonal'' elements: this is based
  on the fact that off-diagonal elements of the SET are known to be
  finite for orthonormal states and so the value of these elements should
  be the usual, formal, one.
\item Good limit to Minkowski spacetime
\end{enumerate}
Under these conditions it can be proved that $\langle
T_{\mu\nu}\rangle$ is unique to within a local conserved tensor.

Obviously fixing such a set of criteria is not equal to finding the
desired
expectation value. An alternative approach can instead be to treat the
calculation of $\langle T_{\mu\nu}\rangle$ as part of a more general
problem of solving a dynamical theory of gravity and matter. The basic
idea is that one can add appropriate counterterms to the gravitational
action in such a way as to remove the divergent parts by redefinition of
the coupling constants.  A superficial account of the idea can be
given in the path integral approach to quantization.

We shall then consider a theory where the total action is
$S=S_{\g}+S_{\m}$, that is the sum of the matter action $S_{\m}$ and of
the gravitational Einstein--Hilbert action (ignoring for the moment
boundary terms)
\begin{equation}
  \label{eq:EHact}
  S_{\g}=\int \frac{1}{16\pi G_{\mathrm N}}\sqrt{-g} (R-2\Lambda) \d^{n}x
\end{equation}
The classical Einstein equations with a cosmological constant are 
\begin{equation}
  \label{eq:claseineq}
  R_{\mu\nu}-\half R g_{\mu\nu}+\Lambda g_{\mu\nu}=-8\pi G_{\mathrm N}T_{\mu\nu}
\end{equation}
and are given by the condition:
\begin{equation}
  \label{eq:einscond}
  \frac{2}{\sqrt{-g}}\frac{\delta S}{\delta g^{\mu\nu}}=0
\end{equation}
where the variations of the gravitational and matter actions
with respect to the metric tensor determine respectively the left hand side
and right hand side of Eq.(\ref{eq:claseineq}).

In the quantum theory, the generalization of Eq.(\ref{eq:einscond}) is
obtained by using the {\em effective action} for quantum matter $W$,
which by definition satisfies the equation
\begin{equation}
  \label{eq:einscondquant}
  \frac{2}{\sqrt{-g}}\frac{\delta W}{\delta g^{\mu\nu}}=
    \langle T_{\mu\nu}\rangle
\end{equation}
We shall then have the aim to solve a theory now described by the
semiclassical Einstein equations
\begin{equation}
  \label{eq:semiclaseineq}
  R_{\mu\nu}-\half R g_{\mu\nu}+\Lambda g_{\mu\nu}=
 -8\pi G_{\mathrm N}\langle T_{\mu\nu}\rangle
\end{equation}
Notably it can be shown~\cite{Birrell-Davies} that the effective action
can easily be related to the Feynman propagator
$G_{\rm F}(x,x')=\langle {\rm out} |T\left(\phi(x)\phi(x')\right)|{\rm in}
\rangle$.
The $T(\cdots)$ symbol here indicates the time ordered product.
\begin{eqnarray}
  W &=&-\half \Tr \left[\/\ln \left(-G_{\rm F}\right)\right]  
         \label{eq:trln}\\
    &=&-\half \int \d^{n}x \sqrt{-g(x)} 
            \langle x| \left(-G_{\rm F}\right)|x \rangle  
         \label{eq:trln2}\\
    &=&-\half \int \d^n x \sqrt{-g(x)}  
        \lim_{x'\to x } \int^{\infty}_{m^2}\d m^2 G^{\mathrm{DS}}_{\mathrm{F}}(x,x')  
         \label{eq:trln3}\\    
    &=&-\half \int^{\infty}_{m^2}\d m^2 
                \int \d^n x \sqrt{-g(x)} G^{\mathrm{DS}}_{\mathrm{F}}(x,x) 
         \label{eq:trln4}
\end{eqnarray}
where we have used the DeWitt--Schwinger representation $G_{\mathrm
  F}^{\rm DS}$ of the Feynman Green function~\cite{Birrell-Davies}
and $m^2$ is the mass of the matter particles. The integral over
$\d^{n}x$ in (\ref{eq:trln4}) can be shown to be precisely the
expression for the {\em one-loop} Feynman diagrams and so $W$ is called
the {\em one-loop effective action}.  Moreover having taken the limit
of $x'\to x$ the quantity obtained is purely local.

We can now define an effective Lagrangian $L_{\eff}$
\begin{equation}
  \label{eq:efflagra}
  W= \int \sqrt{-g} L_{\eff}(x)\d^{n}x
\end{equation}
and find, via Eq.(\ref{eq:trln3}), that this depends on $G^{\rm DS}_{\rm F}$ as
\begin{equation}
  \label{eq:ellef}
  L_{\eff}(x)=\lim_{x'\to x}\int^{\infty}_{m^2}\d m^2 G^{\rm DS}_{\rm F}(x,x')
\end{equation}
The important point is that this effective Lagrangian
diverges in the limit $x'\to x$ and in four dimensions the potentially
divergent terms can be isolated in the Schwinger--DeWitt expansion of
(\ref{eq:ellef})
\begin{equation}
  \label{eq:ldiv}
  L_{\mathrm{div}}=-\lim_{x'\to x}
                     \frac{ \sqrt{\Delta(x,x')} } {32\pi^2}
  \int_{0}^{\infty}  \frac{\d s} {s^{3}} e^{-\im (m^2 s-\sigma/2s)}
   \left[a_{0}(x,x')+\im s a_{1}(x,x')+(\im s)^2 a_{2}(x,x') \right]
\end{equation}
where $\sigma$ is half of the square proper distance between $x$ and $x'$,
$s$ is a fictitious integration variable, and $\Delta(x,x')=-\det
\left[\partial_{\mu}\partial_{\nu'}\sigma(x,x')\right]/
\sqrt{\left[g(x)g(x')\right]}$ is the van Vleck determinant.

By construction these divergences are the same ones which afflict $\langle
T_{\mu\nu} \rangle$ and a careful investigation of the $a$
coefficients shows that these have a completely geometrical nature
and are built with the (local) curvature tensor $R_{\mu\nu\sigma\tau}$
and its contractions~\footnote{
%--------------------------------------------------------------------  
  The same will not be true for the finite parts of $L_{\eff}$ which
  will instead probe the large scale structure of the spacetime.  In a
  certain sense this fact is another manifestation of the typical
  locality of the ultraviolet divergences which we are dealing with.}.
%--------------------------------------------------------------------
In the limit $x' \to x$, and in the general case of a non-minimally
coupled scalar field, these take the form
\begin{eqnarray}
  & a_{0}(x)=1 \nonumber\\
  & a_{1}(x)=\left(\frac{\textstyle 1}{6}-\xi\right) R \label{eq:sdwcoeff}\\
  & a_{2}(x)=\frac{\textstyle 1}{180}
                 R_{\alpha\beta\gamma\delta}R^{\alpha\beta\gamma\delta}-
             \frac{\textstyle 1}{180}
                 R_{\alpha\beta}R^{\alpha\beta}-
             \frac{\textstyle 1}{6}
                 \left(\frac{\textstyle 1}{5}-\xi\right)\Box R+
             \frac{\textstyle 1}{2}
                 \left(\frac{\textstyle 1}{6}-\xi\right)^2 R^2.
               \nonumber        
\end{eqnarray}
The purely geometrical nature of $L_{\mathrm{div}}$ is the key point
that allows it to be regarded as a contribution to the gravitational part
of the action rather than to the matter part. The general idea is then to
consider a new gravitational action which will be given by the sum of
$S_{\g}$ and the action built up from $L_{\mathrm{div}}$.  Using
(\ref{eq:sdwcoeff}) and putting together (\ref{eq:EHact}) with
(\ref{eq:ldiv}) one obtains~\cite{Birrell-Davies}
\begin{equation}
  \label{eq:lgravtot}
  L_{g}=-\left( A+\frac{\Lambda}{8\pi G_{\mathrm N}}\right)
        +\left( B+\frac{1}{16\pi G_{\mathrm N}}\right) R-a_{2}(x) C
\end{equation}
where $A$, $B$ and $C$ are three divergent coefficients depending on the
particle mass. 

We now see that the first term can be interpreted as a renormalization
of the cosmological constant $\Lambda$. The second can be absorbed by
renormalization of the Newton constant $G^{\mathrm{ren}}_{\mathrm
  N}=G_{\mathrm N}/(1+16 \pi G_{\mathrm N}B)$.  The third term is not
present in the standard Einstein--Hilbert Lagrangian.  This implies
that in order to ``renormalize'' it we should add a term to the
original Einstein--Hilbert gravitational action showing the same
functional dependence on the geometrical tensors appearing in $a_{2}$.
Of course the sum of this new term and the corresponding one coming
from $L_{\mathrm{div}}$ can be set equal to zero in order to recover
standard \GR , it is nevertheless important to see that QFT apparently
indicates that terms involving higher derivatives of the metric should
be expected a priori (this is indeed a result also confirmed by string
theory).

After this brief digression about the problems related to
renormalization of the stress energy tensor in curved spacetimes we
can now turn to the important issue of particle creation by
gravitational fields.

%-------------------------------------------------------------------------
\subsection[{Particle interpretation and creation in gravitational fields}]
{Particle interpretation and creation in gravitational fields}
%-------------------------------------------------------------------------

As we have previously discussed, the main effect of the presence of a
gravitational field is to introduce an intrinsic ambiguity into the
construction of the Fock space. This ambiguity is mainly connected
with the absence of a unique separation of the field operator into
positive and negative frequency parts. We have seen that a general
consequence of this mode mixing (or parametric amplification of the
quantized field oscillators) is the creation of particles from the
quantum vacuum.

Although formally correct, the above statements do not tell us really
{\em what is} creating particles, that is in what way the
gravitational field acts on the quantum vacuum. In order to provide an
answer for this, we will make a brief heuristic investigation.

We can start by taking a quantum field $\varphi(x)$ on a general curved
space characterized, in the neighbourhood of some point, $\x$, by a
typical curvature radius $\rho$. This will be related to the Riemann tensor
by
\begin{equation}
  \label{eq:curv}
  R_{\mu\nu\sigma\tau}R^{\mu\nu\sigma\tau}\approx \rho^{-4}
\end{equation}
It is always possible to introduce at $\x$ a coordinate system which
will be locally Gaussian up to a distance of order $\rho$. In this
coordinate system it will then be possible to construct a complete set of
solutions of the wave equations of the field which for frequencies
$\omega\gg\rho^{-1}$ will be, with exponential accuracy, divided into
positive $\varphi_{n}^{(+)}$ and negative frequency parts
$\varphi_{n}^{(+)}$ via the standard relation of Minkowski space
\begin{equation}
  \label{eq:minkposneg}
  \partial_{0}\varphi^{(\pm)}_{n}(x)=\pm\im \omega_{n}\varphi^{(\pm)}_{n}(x)
\end{equation}
On the other hand, for frequencies of order $\omega_{n}\leq \rho^{-1}$
such a division will not be valid because those modes will start
testing the non Euclidean geometry of the manifold. We shall see later
on that in some special cases there is a way to generalize
Eq.(\ref{eq:minkposneg}).

If we now want to discuss the condition required for particle creation
we can try to apply an analogy with quantum electrodynamics and
qualitatively describe it as a ``breakdown of vacuum loops''. Already
at this level we find an important difference between gravity and
other external fields.  Gravity attracts all particles in the same way
and so only tidal forces can ``break'' particle-antiparticle pairs. A
static gravitational field can then polarize the vacuum only if it is
non-uniform and indeed the presence of an event horizon is required
for actually obtaining particle creation. This particle creation will
automatically lead to a non static geometry so, strictly speaking,
particle creation by a static gravitational field is never truly
realized.

A characteristic distance between the particles of a virtual pair is the
Compton wavelength $\lambda_{\mathrm C}=\hbar/m c$. The relative (tidally
generated) acceleration of a pair of virtual particles is given
by the geodesic deviation equation~\cite{HawEll73}
\begin{equation}
  \label{eq:geoddev}
  \frac{\d^2 \sigma^{\mu}}{\d s^2}=
   R^{\mu}{}_{\nu\delta\gamma}u^{\nu}\sigma^{\delta}u^{\gamma}
\end{equation}
where $u^{\mu}$ is the 4-velocity vector of one of the particles of a
pair and $\sigma^{\mu}$ is a spacelike vector connecting the two
virtual particles which, from what we have just said, has norm
$\sigma_{\mu}\sigma^{\mu}\approx\lambda_{\mathrm C}$.

The condition for the breakdown of the virtual loop and the formation
of two ``real'' particles is now that the tidal forces are able to do 
work at a distance of order $\lambda_{\mathrm C}$ that would exceed $2m$.
Setting $u^{0}=1$, $u^{i}=0$, $\sigma^{0}=0$ and $|\sigma^{i}|\approx
\lambda_{\mathrm C}$ one then gets the condition
\begin{equation}
  \label{eq:condtidal}
  \left| R^{i}{}_{0j0}\right|\geq \lambda_{\mathrm C}^{-2}
\end{equation}
So we obtain that in order to get significant particle creation, the
curvature of the spacetime must be at least of order of the inverse
Compton wavelength. 

In the case of a massless field the above condition is not directly
applicable.  Nevertheless we have just seen that the curvature effects
are important just for modes with frequencies $\omega_{n}\leq
\rho^{-1}$.  Then one should expect that most of the particle creation
will happen at frequencies of order $\omega\sim\rho^{-1}$~\footnote{
%-----------------------------------------------------------------------  
  Alternatively one can apply a line of reasoning similar to the one
  used above by making use of the de~Broglie wavelength
  $\lambda_{\mathrm{dB}}=\hbar/p$ ($p$ is the particle momentum). The
  condition for particle creation again takes the form
  \begin{equation}
    \label{eq:condtidal2}
    \left| R^{i}{}_{0j0}\right|\geq \lambda_{\mathrm{dB}}^{-2}
  \end{equation}
  %
%-----------------------------------------------------------------------  
}. So particle creation will again be relevant only in the presence of
large curvatures of the spacetime.

In a strong gravitational field, when the curvature is $\rho^{-1}\gg
\lambda_{\mathrm C}$, the particles will mainly be created as
ultrarelativistic ones with energies of order $\hbar \omega\gg mc^2$.
Consequently in four dimensions the stress energy tensor of quantum
matter will be
\begin{equation}
  \label{eq:setord}
  \left| T_{\mu\nu}\right|\sim\int^{1/\rho}\omega^{3}\d\omega\sim\rho^{-4}
\end{equation}
It is interesting to note that this result driven by dimensional
arguments has also been confirmed by accurate calculations of the
vacuum polarization in strong gravitational fields.

After this qualitative but instructive discussion about particle creation
we can now return our attention to the problem of suitably defining
particles in curved spacetimes. 

We saw how in the most general case, particles could only be uniquely
defined for wavelengths which are short in comparison with the
spacetime curvature radius.  There are nevertheless special spaces
where Eq.(\ref{eq:minkposneg}) can be suitably generalized; these
spaces are those which admit some preferred reference system generally
associated with a Killing vector field.

A very special and easy to handle situation is that in which the
spacetime is static, that is when it is possible to define a global
timelike Killing vector field $\xi^{\mu}$ which is orthogonal to some
family of hypersurfaces $\left\{\Sigma\right\}$. In this case one can
take as the time coordinate an arbitrary parameter $t$ along the
integral curves of $\xi^{i}$ and consider the family $\left\{
    \Sigma\right\}$ as hypersurfaces at $t$=constant. In this way the
metric can always take the form:
\begin{equation}
  \label{eq:staticmetr}
  \d s^2=g_{00}\d t^2-g_{ij}\d x^{i}\d x^{j}
\end{equation}
with $g_{00}$ and $g_{ij}$ depending only on space coordinates.

Now by definition $\xi$ will be the generator of the symmetry under
time translations and so it will be the behaviour of $\xi$ which
determines the positivity or negativity of the particle frequencies.
It is then easy to see that the natural generalization of equation
(\ref{eq:minkposneg}) will take the form
\begin{equation}
  \label{eq:staticposneg}
  \L_{\xi}\varphi^{(\pm)}_{n}(x)=\pm \im \omega_{n}\varphi_{n}^{(\pm)}(x)
\end{equation}
that is, it will now be the Lie derivative along the direction of the
Killing field that will be used to define the signs of the frequencies. A
straightforward consequence of the globality of the Killing vector is
that the vacuum state, defined using the mode decomposition given by
the above equation, is stable and hence particle creation will not
occur.

A similar derivation (but with more pitfalls and complications) can be
made also in the case of non static but stationary spacetimes (when a
global timelike Killing vector exists but is not orthogonal to any
family of spacelike hypersurfaces). In this case one can again choose
coordinates where the components of the metric will not be time
dependent but now $g_{0i}\neq 0$.

Nevertheless static or stationary spacetimes are not necessarily
protected against particle creation from the quantum vacuum. If the metric
components have coordinate singularities somewhere (such as, for example,
on the event horizon in \Sch\ spacetime) the above discussion is
not applicable.  In fact, in these cases the Killing vector ceases to be
timelike in some regions of the manifold and hence it is impossible to
use it for giving a global description of the vacuum state. 

It is exactly this loss of universality of the vacuum state that leads,
in spacetimes with Killing horizons, to the production of particles that
goes under the general name of Hawking--Unruh effects.

%--------------------------------------------------------------------------
\subsection[{Hawking radiation for general spacetimes with Killing horizons}]
{Hawking radiation for general spacetimes with Killing horizons}
\label{subsec:hawrad}
%--------------------------------------------------------------------------

In this final part of this introductory chapter we shall rapidly
review the basic aspects of particle creation in some spacetimes
with a Killing horizon.  The basic idea is that in these cases the
presence of the event horizon (that is a set of fixed points of
the Killing vector associated with symmetries under time translations)
prevents any unique definition of a vacuum state. Moreover we shall
see that generically the event horizon leads to a exponential redshift
of the outgoing modes which is reflected in an exponential behaviour of
the Bogoliubov coefficient. As we have seen in the previous section
about squeezed states and thermality, this implies a thermal behaviour
of the spectrum of produced particles.

We shall now review some well known examples of particle production in
the presence of event horizons. Given the fact that these cases are now
discussed in detail in standard textbooks (see for
example~\cite{Birrell-Davies,moste}) we shall give the basic facts
here and refer the reader to the literature for further insight.

As a first example we shall start with the so called {\em Unruh
  effect}.  This is an effect appearing in Minkowski spacetime but it
is nevertheless the simplest example of the effects related to the
presence of a Killing horizon and hence it is often used as an
introduction to the more complex cases in curved spacetimes.

%----------------------------------------------------------------
\subsubsection{The Unruh effect}
%----------------------------------------------------------------

The Unruh effect takes place in Minkowski spacetime when one considers
the response of a detector which is uniformly accelerated.  What
actually happens is that the standard vacuum state of Minkowski
appears as a thermal state to the uniformly accelerated observer.
Reintroducing for completeness the fundamental constants $\hbar$ and
$c$, the temperature of such a state is given by
\begin{equation}
  \label{eq:tunruh}
  T=\frac{\hbar a}{2\pi ck_{\mathrm B}}
\end{equation}
where $a$ is the (uniform) acceleration.
In order to see how this result arises we can start from the Minkowski
spacetime looking at it in the standard coordinate system $(t,\x)$ and
in that one appropriate for a uniformly accelerated observer
$(\eta,\mathbf{\xi})$.  Given the fact that the angular variables have
no influence, we can choose to work in two dimensions.

We shall work in null coordinates calling $(\bar{u},\bar{v})$ the ones
for the Minkowski observer (an observer at rest in Minkowski
spacetime), and $(u,v)$ the ones for the Rindler observer (this is a
frequently used name for an observer in uniform accelerated motion in
Minkowski). These coordinates are defined as
\begin{equation}
  \label{eq:nullcoord}
   \mbox{Minkowski}\Rightarrow \left\{
   \begin{array}[h]{lll}
    \bar{u} &=& t-x\\
    \bar{v} &=& t+x    
   \end{array}
   \right. \quad 
   \mbox{Rindler}\Rightarrow \left\{
   \begin{array}[h]{lll}
    u &=& \eta-\xi\\
    v &=& \eta+\xi   
   \end{array}
   \right.
\end{equation}
The relations between the two sets of coordinates are
\begin{equation}
  \label{eq:relatcoo}
  \left\{ 
    \begin{array}[h]{lll}
     t &=& a^{-1} e^{a\xi}\sinh a\eta\\
     x &=& a^{-1} e^{a\xi}\cosh a\eta
    \end{array}
  \right. \qquad
  \left\{
    \begin{array}[h]{lll}
     \bar{u} &=& -a^{-1}e^{-au}\\
     \bar{v} &=& a^{-1}e^{-av}
    \end{array}\right.
\end{equation}
The above transformations map the standard two dimensional Minkowski
metric $\d s^2=\d \bar{u}\d \bar{v}=\d t^2-\d x^2$ into a new,
conformally related, one $\d s^2=\exp(2a\xi)\,\d u\d v =\exp(2a\xi)(\d
\eta^2-\d\xi^2)$. 
 
The most important fact is that the new coordinate system $(\eta,\xi)$
does not cover the whole of Minkowski spacetime but just a quadrant of it
where the condition $x>|t|$ is fulfilled. This implies that only a
portion of Minkowski is accessible to the uniformly accelerated
observer. This portion, also called the {\em Rindler wedge}, is
delimited by the lines $\bar{u}=0$ and $\bar{v}=0$. This can easily be
understood by looking at the behaviour of the $\xi=$constant lines which
are the hyperbolae $x^2-t^2=[\exp(a\xi)/a]^2$. This also shows that the
proper acceleration of the observers is
\begin{equation}
  \label{eq:accel}
   a_{\mathrm{proper}}=a\cdot\exp(-a\xi)
\end{equation}
Obviously, a wedge for $x<|t|$ can also be constructed by simply taking
the mirror image with respect to the $t$ axis of the just defined Rindler
wedge.  This is obtained by reversing the signs on the right hand sides
of the transformations for $t$ and $\bar{v}$.  The conformal diagram
below shows the global structure of the spacetime as experienced by
the two sets of observers.

%==============================================================================
\begin{figure}[hbt]
  \vbox{\hfil \scalebox{0.90}
   {{\includegraphics{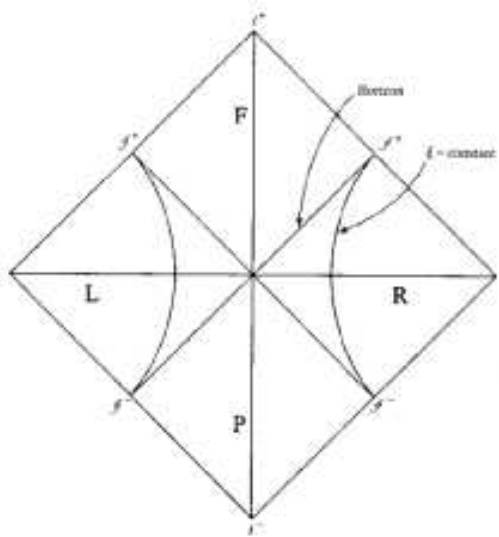}}} 
        \hfil }
  \bigskip
\caption[Penrose diagram of Minkowski spacetime]{
%----------------------------------------------------------------------------
  The conformal ({\em aka} Penrose) diagram of the Minkowski spacetime
  and of the Rindler wedges (the regions labelled with $L$ and $R$).
  Reproduced from~\cite{Birrell-Davies}.
%---------------------------------------------------------------------------
} \label{f:rindl}
\end{figure}
%=============================================================================
%

Given the two bases, we can  perform the mode decomposition for a
scalar field satisfying the trivial wave equation $\Box \phi=0$.  In
Minkowski null coordinates $(\bar{u},\bar{v})$ one gets the standard
orthonormal mode solutions
\begin{eqnarray}
  \label{eq:standsol}
  \bar{u}_{k} &=& \frac{1}{\sqrt{4\pi\omega}}e^{\im kx-\im \omega t}\\
  \omega &=& |k|, \qquad -\infty<k<+\infty \nonumber 
\end{eqnarray}
These modes have positive frequency with respect to the standard global
Killing vector of Minkowski $\partial_{t}$
\begin{equation}
  \label{eq:posmod}
  \L_{\partial_{t}}\bar{u}_{k}=-\im \omega \bar{u}_{k}
\end{equation}
In the case of the Rindler observers, we shall have that the wave
equation is unchanged thanks to its conformal invariance but we shall
have to take into account that two sets of modes (one for the right
wedge and one of the left wedge) are now necessary.  Actually it is
easy to see that the future-pointing timelike Killing vector field is
$+\partial_{\eta}$ in the right wedge and $-\partial_{\eta}$ in the
left wedge and so two different definitions of positive frequency will
exist in the two wedges
\begin{eqnarray}
  \label{eq:posmodrind}
  R:& \quad & \L_{+\partial_{\eta}}{}^{R}u_{k}=-\im \omega {}^{R}u_{k}\quad 
\mbox{and} \quad =0 \quad \mbox{in the left wedge}\\
  \label{eq:posmodrind2}
  L:& \quad & \L_{-\partial_{\eta}}{}^{L}u_{k}=-\im \omega {}^{L}u_{k}\quad 
\mbox{and} \quad =0 \quad \mbox{in the right wedge}
\end{eqnarray}
where the superscripts $R$ and $L$ identify the right and left wedge
modes respectively.  So we shall have two bases each of which is
complete in its respective wedge but not complete over the whole
manifold
\begin{eqnarray}
  \label{eq:sorthsol}
  {}^{R}u_{k} &=& \frac{1}{\sqrt{4\pi\omega}}e^{\im k\xi-\im \omega \eta}\\
  \label{eq:sorthsol2}
  {}^{L}u_{k} &=& \frac{1}{\sqrt{4\pi\omega}}e^{\im k\xi+\im \omega \eta}\\
  \omega &=& |k|, \qquad -\infty<k<+\infty \nonumber 
\end{eqnarray}
with ${}^{R}u_{k}=0$ in $L$ and ${}^{L}u_{k}=0$ in $R$.

At this point is easy to see that the field will admit two different
mode decompositions
\begin{eqnarray}
  \label{eq:moddecrin}
  \phi &=& \sum_{k=-\infty}^{+\infty}{
        \left(a_{k}\bar{u}_{k}+a^{\dagger}\bar{u}^{*}_{k}\right)}\\
&&\nonumber\\
  \phi &=& \sum_{k=-\infty}^{+\infty}{
        \left({}^{L}b_{k}{}^{L}u_{k}+{}^{L}b^{\dagger}_{k}{}^{L}u^{*}_{k}
             +{}^{R}b_{k}{}^{R}u_{k}+{}^{R}b^{\dagger}_{k}{}^{R}u^{*}_{k}
        \right)}
 \label{eq:moddecrin2}
\end{eqnarray}
The Minkowski vacuum will then be defined as
$a_{k}|0\rangle_{\mink}\equiv 0$ and the Rindler vacuum will be
defined as
${}^{L}b_{k}|0\rangle_{L}={}^{R}b_{k}|0\rangle_{R}\equiv 0$.  These two
vacua are clearly inequivalent due to the different mode structure but
to actually see how many Rindler particles are present in the Minkowski
vacuum one has to find the Bogoliubov transformations relating the two
bases~\cite{Birrell-Davies,moste}. The net result is that
\begin{equation}
  \label{eq:rindres}
  \langle 0|{}^{(L,R)}b^{\dagger}_{k}{}^{(L,R)}b_{k}|0\rangle_{\mink}=
 \frac{e^{-\pi\omega/a}}{[2\sinh(\pi\omega/a)]}=\frac{1}{e^{2\pi\omega/a}-1}
\end{equation}
and so we see that the Minkowski vacuum appears to the Rindler observer as
a thermal bath at a temperature $T=a/2\pi k_{\mathrm B}$.

Before continuing, some remarks are in order

\begin{itemize}
\item A general remark that should always be taken into account is
  that in order to state that ``an observer sees a given vacuum as''
  one cannot generally rely just on the results from the Bogoliubov
  coefficients. As we said in section \ref{subsubsec:bog}, the
  presence of typically quantum interference terms in the expression
  of the SET as a function of the Bogoliubov coefficients makes the
  above statement unsafe if just based on non-null $\beta$
  coefficients.  One should always base this sort of statement on more
  reliable tools such as analysis of particle
  detectors~\cite{Birrell-Davies}.
  
  Nevertheless in the special cases in which one gets a thermal nature
  for the emitted radiation, the quantum correlations are lost and the
  interference terms vanish. It is then possible to limit the analysis
  to the Bogoliubov transformations.
\item It should be stressed that no real quanta are created. The total
  SET is covariant under coordinate transformation and so if it is equal
  to zero in Minkowski (as it is set by definition), it should be zero
  also for accelerated observers.  In fact it can be shown that
  the expectation value of the SET in the (Rindler) vacuum state of
  the accelerated observer is non zero but corresponds to a vacuum
  polarization which is equivalent to {\em subtracting} from the Minkowski
  vacuum a thermal bath at the Unruh
  temperature~\cite{CD7778,Sciama:1981hr}. This contribution exactly
  cancels out the other one coming from the fact that a thermal bath of
  real particles is actually experienced from the Rindler observer. So
  in each wedge
  \begin{equation}
    \label{eq:setrindl}
    \langle 0|T_{\mu\nu}|0\rangle_{\mink}=0=
     \langle 0|T_{\mu\nu}|0\rangle_{\rind}+\mbox{Thermal bath}
  \end{equation}
  Physically this tells us that we can regard the non-equivalence of
  the Minkowski and Rindler vacua as an example of vacuum
  polarization.  
\item The thermal distribution can be seen as an effect of the
  apparent horizon at $\bar{u}=\bar{v}=0$ experienced by the Rindler
  observer.  This apparent horizon is similar to the event horizon of
  a black hole although it has a less objective reality: it is
  dependent on the state of motion of some observers and is not
  observed, for example, by Minkowski observers. 
  
  Moreover this apparent horizon can be globally defined for observers
  who have undergone a uniform acceleration for an infinite time. In
  the case of uniform acceleration for a finite time, the thermal
  spectrum is replaced by a more complicated and general distribution.
\end{itemize} 

Having developed a good understanding about the effects linked to the
presence of Killing horizons in flat space, we can now move to the
case of particle creation around black holes.

%-------------------------------------------------------------------
\subsubsection{Hawking Radiation}
%-------------------------------------------------------------------

\begin{figure}[hbt]
  \vbox{\hfil \scalebox{0.70}
   {{\includegraphics{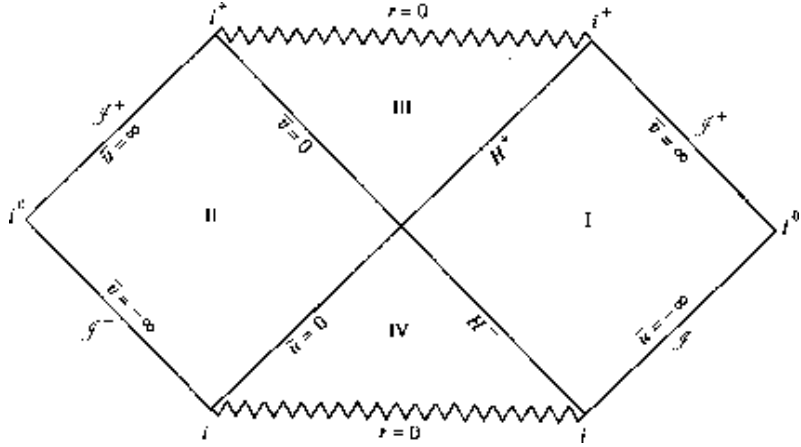}}} 
        \hfil }
  \bigskip
\caption[Penrose diagram of \Sch\ spacetime]{
%----------------------------------------------------------------------------
  Conformal diagram of \Sch\ spacetime (picture reproduced
  from~\cite{Birrell-Davies}).
%---------------------------------------------------------------------------
} \label{f:sch}
\end{figure}

In order to take full advantage of the above results we can
concentrate on the case of a two-dimensional \Sch\ black hole.
The metric in this case is given by
\begin{equation}
  \label{eq:sch2d}
  \d s^2=\left(1-\frac{2M}{r}\right)\d t^2-
         \left( 1-\frac{2M}{r}\right)^{-1}\d r^2
\end{equation}
We can start by passing, as we did before, to the convenient null
coordinates
\begin{equation}
  \label{eq:schnull}
  \left. 
    \begin{array}{lll}
      u &=& t-r-2M\ln \left| \frac{\textstyle r}{2M}-1\right|\\
        & &\nonumber\\
      v &=& t+r+2M\ln \left| \frac{\textstyle r}{2M}-1\right|
    \end{array}\right\}
\end{equation}
In these null coordinates the \bh\ metric takes the form
\begin{equation}
  \label{eq:nullsch}
  \d s^2=\left(1-\frac{2M}{r}\right)\d u \d v
\end{equation}
Together with this null coordinate system, which is appropriate for static
observers at infinity, we shall consider another set of coordinates, the so
called Kruskal coordinates defined as
\begin{equation}
  \label{eq:krusnull}
  \left. 
    \begin{array}{lll}
      \bar{u} &=& -\kappa^{-1}e^{-\kappa u}\\
      \bar{v} &=& \kappa^{-1} e^{\kappa v}
    \end{array}\right\}
\end{equation}
where $\kappa$ is the surface gravity of the \bh\ and in our specific
case is $1/4M$.  The Kruskal coordinates are appropriate for observers
freely falling into the \bh\ (but still outside it) and we
can see that the metric is perfectly well defined also on the event
horizon in this system
\begin{equation}
  \label{eq:nullschmetr}
  \d s^2=\left(\frac{2M}{r}\right)e^{-r/2M} \d \bar{u} \d \bar{v}
\end{equation}
These coordinates allow us to build up the so called {\em maximal
  analytic extension} of the \bh\ spacetime to which can be associated
the conformal diagram shown in figure \ref{f:sch}.
%
%==============================================================================
%=============================================================================
%
The regularity of the Kruskal coordinates also has an important
consequence when we look at the two basis modes for the usual wave
equation $\Box \phi=0$. In fact these bases will be either
proportional to $e^{-\im \omega u},e^{-\im \omega v}$ in the \Sch\ 
observer's system or to $e^{-\im \omega \bar{u}},e^{-\im \omega
  \bar{v}}$ in the Kruskal one. The regular behaviour of the Kruskal
coordinates on the horizon then implies that while the first set above
will oscillate infinitely rapidly near to the horizon, the second one
will be regular over the whole manifold.

This different behaviour of the two mode bases is reminiscent of what
we saw in the Rindler case where the Rindler system of coordinates is
indeed badly defined at the (apparent) horizon and does not cover the
whole of Minkowski spacetime.  Actually if we now confront the form of the
transformation above with those relating the null coordinates of the
Minkowski and Rindler observers, equation (\ref{eq:relatcoo}), it is easy to
see that there is a close relation between the Kruskal observers and
the Minkowskian ones and between \Sch\ and Rindler ones. In this case
the role of the acceleration is played by the surface gravity of the
black hole.

Also in this case we see that the introduction of two separate sets of
\Sch\ modes will be needed in order to expand the scalar field on the
whole manifold. One set will cover the asymptotically flat zone {\bf
  I} of the conformal diagram \ref{f:sch} the other covers the
specular zone {\bf II}.

So, in strict analogy with what we did in the Rindler case,
we get the two mode expansion in Kruskal and \Sch\ coordinates as
\begin{eqnarray}
  \label{eq:modderin}
  \phi &=& \sum_{k=-\infty}^{+\infty}{
        \left(a_{k}\bar{u}_{k}+a^{\dagger}\bar{u}^{*}_{k}\right)}\\
&&\nonumber\\
  \phi &=& \sum_{k=-\infty}^{+\infty}{
        \left({}^{II}b_{k}{}^{II}u_{k}+{}^{II}b^{\dagger}_{k}{}^{II}u^{*}_{k}
             +{}^{I}b_{k}{}^{I}u_{k}+{}^{I}b^{\dagger}_{k}{}^{I}u^{*}_{k}
        \right)}
 \label{eq:modderin2}
\end{eqnarray}
We shall now have two vacuum states $a_{k}|0\rangle_{\rm S}=0$ and
${}^{(I,II)}b_{k}|0\rangle_{\rm K}=0$ to relate via Bogoliubov
transformations and we can again expect a thermal distribution  
\begin{equation}
  \label{eq:schres}
  \langle 0|{}^{(I,II)}b^{\dagger}_{k}{}^{(I,II)}b_{k}|0\rangle_{\rm K}=
   \frac{e^{-\pi\omega/\kappa}}{[2\sinh(\pi\omega/\kappa)]}=
    \frac{1}{e^{2\pi\omega/\kappa}-1}
\end{equation}
So we see that the Kruskal vacuum appears to the \Sch\ observers at infinity
as a thermal bath at temperature $T=\kappa/2\pi k_{\mathrm B}$.

It is interesting to see that the relation between the two vacua can
be written in the form~\cite{Birrell-Davies}
\begin{eqnarray}
  \label{eq:vacrelsch}
  |0\rangle_{\rm K}&=&\exp\left\{ {\textstyle \sum}_{k}
    \left[-\ln\cosh (\zeta_{\omega})+\tanh(\zeta_{\omega})
    {}^{I}b^{\dagger}_{k} {}^{II}b^{\dagger}_{k}\right]\right\}
        |0\rangle_{\rm S}
\nonumber\\
\nonumber\\
               &=& {\textstyle \prod}_{k} 
    \left(\cosh(\zeta_{\omega})\right)^{-1}
                   {\textstyle \sum }^{\infty}_{n_{k}=0}
    e^{-n_{k}\pi\omega/\kappa}|{}^{I}n_{k}\rangle |{}^{II}n_{k}\rangle
\end{eqnarray}
where we have defined $\exp(-\pi\omega/\kappa)\equiv
\tanh(\zeta_{\omega})$, and in the last line we have introduced states
with ${}^{I}n_{k}$ particles in region {\bf I} and ${}^{II}n_{k}$
particles in region {\bf II}. We now see a behaviour which is in
strict relation with what we said about squeezed states and
thermality. In fact if we consider the measurement of an observable
$\hat{A}$ by an observer ``living'' in region {\bf I} we find that the
states of region {\bf II} will simply factor out
\begin{eqnarray}
  \label{eq:schter}
  \langle 0|\hat{A}|0\rangle_{\rm K} &=& \sum_{n_{k}}\prod_{k}
   \langle {}^{I}n_{k}| \hat{A} |{}^{I}n_{k}\rangle 
    \exp(-2n_{k}\pi\omega/\kappa)\times \left[1-\exp(-2\pi\omega/\kappa)\right]
\nonumber\\
   &=& \Tr\left(\hat{A}\rho\right)   
\end{eqnarray}
where, setting $E_{n}=n\omega$ and $\beta=2\pi/\kappa$, $\rho$ can be
regarded as density matrix corresponding to a thermal average
\begin{equation}
  \label{eq:densmat}
  \rho=\sum_{n}\prod_{k}\frac{e^{\beta E_{n}}}
                             {\sum_{m=0}^{\infty}e^{-\beta E_{m}}}
        |{}^{II}n_{k}\rangle  \langle {}^{II}n_{k}|
\end{equation}
In the particular case where $\hat{A}$ is the particle number operator
then we can easily see that from (\ref{eq:schter}) we get the
expected thermal distribution.

The emergence of thermality in \bh\ is again a particular case of the
more general property of squeezed states. Nevertheless, as we said,
the fact that for each pair of particles created from the vacuum, one
gets lost for ever into the black hole is an indication that in this
special case the distinction between the squeezed state and thermal
one disappears.

Before ending this section and this chapter we should stress some
important issues related to the Hawking effect.
\begin{itemize}
\item In the derivation sketched above, the modes that are associated
  with purely outgoing particles with respect to static observers at
  ${\cal I}^{+}$ are $\exp(-i\omega u)$.  These can be traced back in
  time and appear as outgoing from ${\cal H}^{-}$ where the
  appropriate coordinates are those of Kruskal.  Now the Kruskal
  transformations (\ref{eq:krusnull}) imply that any time interval
  $\Delta u$ on $\I^+$ is stretched on ${\cal H}^{-}$, $\Delta
  \bar{u}=-\kappa^{-1}\exp(-\kappa \Delta u)$. Conversely the
  frequencies will get a blueshift factor $\exp(\kappa u)$ when one
  traces the modes back from $\I^{+}$ to $\H^{-}$.  This in particular
  implies that the frequencies of the outgoing modes on $\I^{+}$ in
  proximity of the future event horizon $\H^{+}$ (that is for $u\to
  \infty$) are going to be infinitely blueshifted on $\H^{-}$.
  Actually before arriving at the horizon some modes will get
  frequencies larger than the Planck one $\omega_{pl}=2\pi/t_{pl}$ and so
  the semiclassical approximation should break down. Apparently
  Hawking radiation should be dominated by high frequency modes that would
  lead to a failure of the whole framework used to derive the
  effect.  This is called the {\em problem of transplanckian
    frequencies}. We shall see in chapter~\ref{chap:3a} how such an
  unphysical feature can be interpreted and possibly circumvented.
  
\item We have just seen a very easy example of the Hawking effect in
  two dimensions and for a conformally invariant wave equation.  In
  four dimensions the calculations are much more difficult and in
  particular if we write the generic solution as
  \begin{equation}
    \label{eq:colsol}
    \phi=\frac{R_{\omega l}(r)}{r} Y_{lm}(\theta,\varphi) e^{-\im \omega t}
  \end{equation}
 then from the radial
  part of the standard wave equation $\Box\phi=0$ one gets
  \begin{equation}
   \label{eq:radialeq}
   \frac{\d^{2}R_{\omega l}}{\d {r^{*}}^2}+\left\{
    \omega^{2}-\left[l(l+1)r^{-2}+2 M r^{-3}\right]\left[1-2 M r^{-1}\right]
                                        \right\}R_{\omega l}=0
  \end{equation}
  where $r^{*}=r+2M\ln|(r/2M)-1|$ is the so called ``tortoise''
  coordinate.
  
  Analytical solutions of these equations in terms of simple functions
  are not available.  This has led to elaborate analytical
  approximations~\cite{Page82,BO85,BOP86} as well to numerical techniques
  (see~\cite{Mattvac} and reference therein).  
  
\item The potential terms appearing in square brackets in
  Eq.(\ref{eq:radialeq}) lead to backscattering of the incoming waves
  by the gravitational field. As a consequence of this the spectrum of
  particles reaching future infinity will not be purely thermal but
  will show an overall grey-body factor.
  
\item The vacuum states $|0\rangle_{\rm S}$ and $|0\rangle_{\rm K}$ are not
  the only ones which can be defined on the eternal \bh\ manifold.
  Different initial conditions on the incoming/outgoing modes will
  generically correspond to different vacuum states. This subtle
  issue is not important for the remaining discussion in this thesis
  and we refer the reader to fundamental papers on the subject
  (see e.g.~\cite{Sciama:1981hr}) and standard
  textbooks~\cite{Birrell-Davies,moste,FrNo}.
  
\item We have considered the case of an eternal \bh\ but also the more
  realistic case of particle creation from gravitational collapse can be
  studied~\cite{Birrell-Davies,moste,FrNo}. We shall consider in the
  next chapter an example of gravitational collapse which will apply
  the standard technique for computing particle production.  It
  should be stressed that the aforementioned ambiguity about the
  construction of vacuum states in the eternal case is absent in the
  case of collapsing bodies.
\end{itemize}

With these remarks on the Hawking effect we close this introductory
chapter concerning the effects of the quantum vacuum in strong fields.
We have seen that the new role that vacuum has in modern physics
automatically leads to the prediction of interesting effects such as
polarization and particle production from the vacuum. Although we have
experimental tests of the static Casimir effect we still lack any test
of particle production from nonstationary external fields, the
phenomenon which in the case of gravity takes the name of Hawking
radiation.

The next step in our investigation will now be the discussion of the
consequences of the results described above. In particular we shall
concentrate our attention on black hole thermodynamics because of its
role as a bridge towards quantum gravity and we shall need to make full
use of all of the concepts learned so far.

%%%%%%%%%%%%%%%%%%%%%%%%%%%%%%%%%%%%%%%%%%%%%%%%%%%%%%%%%%%%%%%%%%%%%%%%%%%%
%S.Liberati Ph.D. Thesis - Chapter 2: Quantum Black Holes
%%%%%%%%%%%%%%%%%%%%%%%%%%%%%%%%%%%%%%%%%%%%%%%%%%%%%%%%%%%%%%%%%%%%%%%%%%%%
\chapter[{Quantum Black Holes}]{Quantum Black Holes}
\label{chap:2}
%Lorentzian signature (-,+,+,+) -> Euclidean signature (+,+,+,+) 
%%%%%%%%%%%%%%%%%%%%%%%%%%%%%%%%%%%%%%%%%%%%%%%%%%%%%%%%%%%%%%%%%%%%%%%%%%%%

\vspace*{0.5cm}
\rightline{\it As far as the laws of mathematics refer to reality} 
\rightline{\it they are not certain;} 
\rightline{\it and as far as they are certain,} 
\rightline{\it they do not refer to reality.}
\vspace*{0.5cm} \rightline{\sf Albert Einstein}

\newpage

In this chapter we shall deal with the quantum aspects of black hole
physics, that is with the most famous application of the theory of the
quantum vacuum in gravitational fields. We shall review the basic
notions of this fascinating field of research and discuss its
importance as the key to disclosing the basic properties that a
quantum theory of gravity should have. We shall focus our attention on
the interpretation of black hole entropy and study its relation with
the global structure of spacetime.  From this study we shall move
towards an interpretative proposal for black hole thermodynamics
based on the zero point modes of the quantum fields. The aim of such a
proposal is to unify the Casimir effect and the early Sakharov idea
of induced gravity. We shall finally discuss in detail the special
role that extremal solutions have. The formation of extremal black
holes from gravitational collapse will then be studied to give us
further insight into this issue.  At the end of the chapter we shall
present some speculations about the framework emerging from the
research in this field.

%%%%%%%%%%%%%%%%%%%%%%%%%%%%%%%%%%%%%%%%%%%%%%%%%%%%%%%%%%%%%%%%%%%%%%%%%%
\section[{Black Hole Thermodynamics}]{Black Hole Thermodynamics}
\label{BHT}
%%%%%%%%%%%%%%%%%%%%%%%%%%%%%%%%%%%%%%%%%%%%%%%%%%%%%%%%%%%%%%%%%%%%%%%%%%

Although Hawking's discovery of quantum radiation from \bh s in 1975
\cite{Haw75} was completely unexpected by most of the experts of that
time, it should be stressed that the new paradigm, of a
thermodynamical behaviour of black holes, was already starting to
emerge a few years before.

The first understanding that classical black hole physics was
outrageously in conflict with our basic thermodynamical notions is
generally ascribed to Wheeler. In particular a black hole, by
engulfing an object endowed with a given entropy $\sigma$, would be
able to reduce the total entropy of the universe and in this way would
lead to a violation of the second law of thermodynamics.

This sort of observation led Bekenstein~\cite{Bek72,Bek73} to
postulate that a black hole should be endowed with a proper entropy.
Given the fact that the area of a black hole never decreases in a
classical process~\cite{Haw71,Haw72} then he could argue that this
quantity plays a role similar to entropy.

From this starting consideration the research on \bh\ physics
achieved a series of theoretical results that brought an elegant
and impressive formulation of some General Relativity theorems as
thermodynamical laws. The full awareness of the existence of a sort
of ``black hole thermodynamics'' then found its systematic statement
in a seminal paper by Bardeen, Carter and Hawking~\cite{BCH73}.

What was found was that for each classical thermodynamical law it
is possible to formulate a corresponding one for \bh s. These results
are collected in the table below.
\begin{center}
\noindent
\begin{tabular}{|c|c|c|}\hline 
Law & Thermodynamics & Black Holes \\
\hline \hline 
Zeroth & T is constant throughout & 
The surface gravity $\kappa$ is\\ 
& bodies at thermal equilibrium & constant on the event horizon\\ 
&& of a stationary black hole \\ 
\hline
First & $\d E=T\d s+\mbox{work terms}$ & The mass of a black hole\\ 
&& is related to its area $A$,\\ 
&& surface gravity $\kappa$ and\\ 
&& angular momentum $J$ by\\ 
&& the equation $ \d M={{\kappa }\over {2 \pi}}\d A+\Omega \d J$\\
&& where $\Omega$ is the angular velocity of the \bh \\
\hline
Second & $\delta S\geq 0$ in any process &  $\delta A\geq 0$ in any
processes\\ 
&& satisfying the WEC\\
\hline
Third & It is impossible to achieve & It is impossible to achieve \\
& the $T=0$ state by a & the $\kappa=0$ state\\ 
& physical process in a & (aka {\em extremal state})\\ 
& finite number of steps & by a physical process in a\\ 
&& finite number of steps\\
\hline
\end{tabular}
\end{center}
Although these results were very impressive, the new paradigm became
fully consistent only with the discovery of Hawking radiation by the
application of quantum field theory in curved space. In fact it is
impossible to define a temperature for classical \bh s, thus it is
strictly not even reasonable to talk about entropy in this case. As
discussed before, Hawking proved that, due to the polarization of the
vacuum in the vicinity of the \bh\ horizon, there is a thermal flux of
radiation flowing out towards infinity. By the first law and the
Hawking temperature 
\begin{equation}
T_{\rm H}=\frac{\hbar \kappa}{2\pi c k_{\rm B}}
 \label{eq:T-Haw}
\end{equation}
one finds the entropy of a \bh\ to be one quarter
of its area~\cite{Haw75}
\begin{equation}
S_{\BH}=\frac{A k_{\mathrm B}}{4 L_{\rm Pl}^{2}}  
 \label{eq:Bek-Haw}
\end{equation}
This quantity is the so-called {\em Bekenstein--Hawking entropy}.
Sometimes it also takes the name of {\em gravitational entropy} and we
shall use both terminologies in the rest of this work.

At this point it is perhaps worthwhile to say something more about the
different status of the above quoted thermodynamical laws.  In fact
these do not share the same status or the same robustness.
\begin{itemize}

\item The zeroth law is a firm theorem of classical general
  relativity (which assumes the validity of the DEC to constrain the
  matter behaviour).
  
\item The first law is also a general statement of energy
  conservation when a system incorporating a black hole switches from
  one stationary state to another, it can be rigorously proved at
  least for all the non-extremal black hole systems.
  
\item The second law, in its classical formulation, follows from the
  so-called Hawking area theorem \cite{Haw71,Haw72}. As noted
  previously, this is a theorem of General Relativity which assumes
  the WEC.  Notably the discovery of quantum radiation from black
  holes led also to a revision of the classical statement of the
  second law.  In fact, if a black hole can evaporate due to Hawking
  radiation, then it is possible to reduce its mass and hence its
  area, i.e. its entropy. The point is that WEC is {\em explicitly
    violated} by Hawking radiation (and in general by all processes of
  quantum particle creation from vacuum).
  
  Already before Hawking's discovery, Bekenstein
  \cite{Bek72,Bek73b,Bek74} proposed that the {\em generalized
    entropy}, defined as the sum of the entropy of the \bh\ and that
  of radiation and matter outside the horizon,
  $$
  \tilde{S}=S_{\BH}+S_{\matter}
  $$
  is an always increasing quantity
  $$
  \Delta\tilde{S}=\Delta S_{\BH}+\Delta S_{\matter}\geq 0
  $$
  Later, analytical as well numerical results \cite{Zur82,Page83}
  corroborated this analogy which is now generally stated as the {\em
    generalized second law}. No counter-examples, able to violate this
  general formulation of the second law, have so far been found.
  Nevertheless it is important to stress the intrinsically conjectural
  nature of the above relation and its intrinsic assumption of the
  similar nature of gravitational entropy and quantum matter entropy.
  
\item The third law stands on a much less firm basis than the other
  ones.  First of all it is important to stress that even the standard
  third law of thermodynamics admits at least two different
  formulations both due to Nernst.
  
  The first statement of the third law, {\em aka} the Nernst law, states
  that the zero temperature states of a system are isentropic. If
  $\Delta S$ is the difference between the entropies of two states of
  a thermodynamical system then $\lim_{T\rightarrow 0^{+}}\Delta S=0$.
  This implies that the entropy of all of the states at absolute zero
  temperature does not depend on any other macroscopic parameter, it
  is a constant. {\em Planck's postulate} assumes that this constant is
  strictly zero.
  
  The second statement, {\em aka} the Unattainability law, states the
  impossibility of reaching the absolute zero temperature by a finite
  number of thermodynamical processes.
  
  Although it {\em is generally true} that the above two formulations of the
  third law are equivalent, this {\em is not always true} as stated
  even in some textbooks of thermodynamics \cite{Land}. The black hole
  case is, in this sense, a typical example where the Nernst
  formulation is violated but the unattainability one holds (this was
  first recognized by Davies in \cite{Davies77}). In fact the third
  law as stated above is exactly a formulation, for black holes, of
  the unattainability law. Although in \cite{BCH73} this had just the
  form of an analogy it was later on reformulated and rigorously
  proved by Israel \cite{Israel86}.
  Nowadays the general statement is:
  {\em A non-extremal black hole cannot become extremal at a finite
    advanced time in any continuous process in which the stress-energy
    tensor of accelerated matter stays bounded and satisfies the WEC
    in the neighborhood of the outer apparent horizon.}

\begin{center}
  \setlength{\fboxsep}{0.5 cm} 
   \framebox{\parbox[t]{14cm}{
%------------------------------------------------------------------------    
  The fact that the Nernst formulation does not follow from the
  Unattainability one is also a major cause for being extremely
  suspicious about some common statements, in the existing literature,
  on the failure of the Planck postulate in \bh\ thermodynamics. In
  fact it is generally stated that since the area of extremal black 
  holes is not zero then the entropy as well cannot be zero. This simple
  deduction assumes that the Bekenstein--Hawking formula holds also in the
  extremal limit. This would be safe if a) there are no multiple
  branches in the thermodynamic configuration space, b) there is no
  discontinuity in thermodynamic properties of the system near to
  absolute zero. Both of these assumptions can in principle be violated
  if the Unattainability formulation does not imply the Nernst one.
  This kind of analysis has been carried out for \bh\ thermodynamics
  in~\cite{BM00}.  We shall indeed see, later on in this chapter, how
  semiclassical calculations seem to actually suggest the break down of
  the area law in the extremal limit.
%-------------------------------------------------------------------------
}}
\end{center}
\end{itemize}
As a final remark it should be said that although the laws of black
hole thermodynamics are actually derived as theorems of \GR\ (and so are
substantially encoded at a classical level), the notion of temperature
for a \bh\ emerges just as a quantum effect on the curved background.
How could \GR\ ``be aware'' of the role of quantum fields? How deep
is the relation between the Hawking radiation and the thermodynamical
properties of black holes? It should be stressed that this appears nowadays
as one of the the key points for the whole understanding of the subject.

We shall see later that concepts like gravitational entropy and
Hawking radiation are not necessarily linked and that the
Bekenstein-Hawing formula appears as a general behaviour that emerges
in a whole class of theories of gravitation and is largely independent
of whatever quantum theory of gravity is assumed. 
Apart from this issue other ones arose rapidly after Hawking's
discovery and are still now the subject of investigation.  We cannot
present all of these here, so we shall limit our discussion to the two
other fundamental points that have polarized research in this
field in the last twenty years.

The first of these issues deals with the role of Hawking radiation in the
final destiny of black holes and is often quoted as the puzzle of
Information loss. Information loss is related to the fact that
\bh\ evaporation by means of emission of a thermal particle
spectrum, appears to produce a destruction of quantum information by
converting a pure state into a mixed one. Hawking suggested that this
should be a starting point for a generalization of quantum mechanics
in which non-unitary evolution of quantum states in the presence of strong
gravitational fields is allowed~\cite{Haw76,Haw82}.

In disagreement with Hawking's point of view, several attempts have
been made in order to circumvent this apparent break down of standard
quantum mechanics. We shall limit ourselves here to a brief sketch of
the different proposals (see~\cite{FrNo,Page93} for a comprehensive review)

\begin{itemize}
\item{\bf Information is released via Hawking radiation.}  Although this
  may appear as a natural solution, it implies the discovery of
  some mechanism by which the information hidden inside the black hole
  can be fully ``encoded'' on the horizon and then taken away by the
  radiation.
  
\item{\bf Information is released at the end of evaporation.}  The
  idea of a final ``burst of information'' assumes that a large amount
  of energy is still available at the end of the evaporation in order
  to allow a huge transmission of information. Unfortunately it appears
  very difficult to sustain this assumption.
  
\item{\bf The information is simply ``stored'' in a remnant which is
    the final outcome of the black hole evaporation}.  In this
  framework the information of an arbitrarily massive black hole
  should always be storable in a Planck-size object which, in order to
  do so, should be endowed with an extremely large number of internal
  states.
  
\item{\bf The information escapes to ``baby universes''.}  The black
  hole formation does not lead to the presence of a singularity inside
  the horizon. The collapsing body will instead form a
  closed baby universe to which the information is passed.
  
\item{\bf Quantum Hair.}  The no hair theorem is not true at a quantum
  level and some black hole hair can give partial information about
  the matter that has formed the black hole.

\end{itemize} 

The second issue is related to the statistical origin of the
gravitational entropy and this will be the subject of the next
section.

%%%%%%%%%%%%%%%%%%%%%%%%%%%%%%%%%%%%%%%%%%%%%%%%%%%%%%%%%%%%%%%%%%%%%%%%%%
\section[{Alternative derivations of the \bh\ entropy}]{Alternative
derivations of the \bh\ entropy}
\label{sec:altbh}
%%%%%%%%%%%%%%%%%%%%%%%%%%%%%%%%%%%%%%%%%%%%%%%%%%%%%%%%%%%%%%%%%%%%%%%%%%

After the initial discovery of \bh\ thermodynamics, other techniques
were developed and found capable of obtaining the same results from rather
different approaches. We shall briefly discuss these alternative
formulations of black hole thermodynamics.

%%%%%%%%%%%%%%%%%%%%%%%%%%%%%%%%%%%%%%%%%%%%%%%%%%%%%%%%%%%%%%%%%%%%%%%%%%
\subsection[{Euclidean derivation of \bh\ Thermodynamics}]
{Euclidean derivation of \bh\ Thermodynamics}
\label{sec:EPI}
%%%%%%%%%%%%%%%%%%%%%%%%%%%%%%%%%%%%%%%%%%%%%%%%%%%%%%%%%%%%%%%%%%%%%%%%%%

Semiclassical Euclidean quantum gravity techniques play a key role in
the investigation of the thermodynamics of black holes.  We shall here
summarize the path integral approach procedure following the steps
first delineated by Gibbons and Hawking \cite{GH77}.

Given the classical Einstein--Hilbert action for gravity and the action
of classical matter fields, one formulates the Euclidean path integral
by means of a Wick rotation $t\rightarrow -\im \tau$~\footnote{
%%%%%%%%%%%%%%%%%%%%%%%%%%%%%%%%%%%%%%%%%%%%%%%%%%%%%%%%%%%%%%%%%%%%%%%%
  Working in the Euclidean signature is a standard technique in QFT in
  flat backgrounds. It has the advantages of making the path integral
  ``weighted'' by a real factor and in general transforms the equation
  of motion of free fields from hyperbolic to elliptical, a feature
  that often leads to just one Green function satisfying the boundary
  conditions
%%%%%%%%%%%%%%%%%%%%%%%%%%%%%%%%%%%%%%%%%%%%%%%%%%%%%%%%%%%%%%%%%%%%%%%%
.}.

In particular if one considers the Einstein--Hilbert action plus a
matter contribution in the generating functional of the Euclidean
theory, one finds that the dominant contribution to the Euclidean path
integral is given by gravitational instantons (i.e. non-singular
solutions of the Euclidean Einstein equations).  

In spacetimes with event horizons, metrics extremizing the Euclidean
action are gravitational instantons only after removal of the conical
singularity at the horizon~\cite{GH77}.  A period must therefore be
fixed in the imaginary time, $\tau \rightarrow \tau+\beta$, which
becomes a sort of angular coordinate.

To be concrete we can take the standard \Sch\ metric in Euclidean signature:
\beq
\d s^2_{\mathrm E}=\left(1-\frac{r_{\mathrm h}}{r}\right)\d\tau^2
+\left(1-\frac{r_{\mathrm h}}{r}\right)^{-1}\d r^{2}+r^2\d\Omega^{2}
\eeq
This solution still shows singularities at $r=r_{\mathrm h}=2M$ and $r=0$.  If
we now define a new radial variable, $\rho=2\sqrt{r-r_{\mathrm h}}$, the
metric acquires the form
\beq
\d s^2_{\mathrm E}= \left( \frac{\rho^{2}}{4}+r_{\mathrm h}  \right)^{-1}
\frac{\rho^{2}}{4}\d{\tau}^2 +\left( \frac{\rho^{2}}{4}+r_{\mathrm h}
\right)\d \rho^{2}+ \left( \frac{\rho^{2}}{4}+r_{\mathrm h}\right)\d
\Omega^{2}
\eeq
For $\rho\rightarrow 0$ one then gets:
\beq
\d s^2_{\mathrm E}= 
 \frac{1}{r_{\mathrm h}} \frac{\rho^{2}}{4}\d{\tau}^2 +r_{\mathrm h}\d
  \rho^{2}+ r_{\mathrm h}^{2}\d \Omega^{2}
\eeq
If one now defines $\bar{\rho}\equiv \rho\sqrt{r_{\mathrm h}}$ and
$\bar{\tau}\equiv \tau/(2r_{\mathrm h})$ the metric takes the so
called conical form:
\beq
\d s^2_{\mathrm E}= \bar{\rho}^{2}\d{\bar{\tau}}^2 +\d \bar{\rho}^{2}+
r_{\mathrm h}^{2}\d \Omega^{2}
\eeq
Smoothness at $\bar{\rho}=0$ requires the previously mentioned
periodicity in the imaginary time. The required period $\beta$
corresponds exactly to the inverse Hawking--Unruh temperature
$\beta_{\mathrm H}=2\pi/\kappa$ because of the relation between the
periodicity of the Euclidean Green's functions and the thermal
character of the corresponding Green's functions in Lorentzian
signature.  Hence, in these cases, the effective action is truly a
free energy function $\beta F$.  In such a way thermodynamics appears
as a requirement of consistency of quantum field theory on spacetimes
with Killing horizons, and in this sense we shall define such a
thermodynamics as ``intrinsic".

It is important to stress that the situation changes dramatically in
the case of extremal black holes. In fact the general form of the
metric is now
\beq
\d s^2_{\mathrm E}=\left(1-\frac{r_{\mathrm h}}{r}\right)^{2}\d\tau^2
+\left(1-\frac{r_{\mathrm h}}{r}\right)^{-2}\d r^{2}+r^2\d\Omega^{2}
\eeq
Performing the same transformations as before this can be cast in the form:
\beq \d s^2_{\mathrm E}= \left( \frac{\rho^{2}}{4}+r_{\mathrm h} \right)^{-2}
\left(\frac{\rho^{2}}{4}\right)^{2}\d{\tau}^2 + \left(
  \frac{\rho^{2}}{4}+r_{\mathrm h}\right)^2 \frac{4}{\rho^2}\d \rho^{2}+
\left( \frac{\rho^{2}}{4}+r_{\mathrm h}\right)\d \Omega^{2} \eeq
which in the $\rho\rightarrow 0$ limit gives
\beq
 \d s^2_{\mathrm E}= 
  \left(\frac{1}{r_{\mathrm h}}\right)^{2} \frac{\rho^{4}}{16}\d{\tau}^2 +
   \frac{4 r^{2}_{h}}{\rho^{2}} \d \rho^{2}+ r_{\mathrm h}^{2}\d \Omega^{2}
\eeq
It is now easy to see that in this case passing to a system of
coordinates equivalent to the previous one given by $\bar{\tau}$ and
$\bar{\rho}$ does not give rise to any ``conical-like'' metric.  This
implies that, for extremal black holes, there is no way to fix the
period of the imaginary time and hence the intrinsic temperature.  We
shall see later on, in section \ref{sec:topol}, how these
considerations are strictly related with the emergence of a
gravitational entropy in spacetimes with event horizons.

Coming back to the issue of the Euclidean formulation of \bh\ 
thermodynamics we can now consider the partition function for
solutions where a temperature can be fixed in the way described above.

The partition function in a canonical ensemble $Z$ can be
written in the case that one includes also the gravitational field
\begin{equation}
Z ={\rm{Tr}}\; \exp{[-(\beta \H)]} 
=\int {\cal D} \left[\phi, g_{\mu\nu}\right] 
  \exp{\left[-I\left(\phi, g_{\mu\nu}\right)\right]},
\end{equation}
Using the fact that a black hole solution is an extremum of the
(Euclidean) gravitational action, at the tree level of the
semiclassical expansion one then obtains
\begin{equation}
Z \sim \exp{[-I_{\rm E}]}\, ,\qquad
I_{\rm E}={{1}\over{16 \pi}} 
\int_{M}\left[\left(-R+2\Lambda)+L_{\m}\right)\right]
 +{{1}\over{8\pi}}
\int_{\partial{M}} \left[ K \right], 
\label{ie}
\end{equation}
where $I_{\rm E}$ is the on-shell Euclidean action and $[K]=K-K_{0}$ is
the difference between the extrinsic curvature of the manifold and
that of a reference background.  The Euclidean action is related to
the free energy $F$ as $I_{\rm E}=\beta_{\mathrm H}F$ and the corresponding
entropy is
\begin{equation}
S\equiv \beta^2\frac{\partial F}{\partial \beta}
\end{equation}
\begin{center}
  \setlength{\fboxsep}{0.5 cm} 
   \framebox{\parbox[t]{14cm}{
%------------------------------------------------------------------------      
       {\bf Comment:} It has to be stressed here that the procedure
       just presented has some subtleties that are well studied in the
       literature. One of these is that the \Sch\ solution is a saddle
       point of the Euclidean effective action corresponding to an
       unstable solution. This is related to the fact that any black
       hole in vacuum is a thermally unstable object (as can be seen
       by the negative value of its heat capacity $C_{\BH}=-8 \pi
       M^2$); but of course any self-gravitating system shows such a
       behavior, due to the attractive nature of gravity. The problem
       of performing a thermodynamical analysis of black holes,
       considering a grand-canonical ensemble, has been studied
       thoroughly by York \cite{Yo}, who suggested considering a black
       hole in a box.  Such a choice automatically stabilizes the \bh\ 
       and enables one to perform further semiclassical calculations.
       One can also consider the higher-order corrections to the
       action.  Unfortunately neither one-loop graviton contributions
       \cite{GPJ} nor matter ones \cite{BL95} seem to be able to
       stabilize the black hole, since they are small in comparison
       with the tree-level term, at least in the regime of negligible
       back reaction.
%------------------------------------------------------------------------
}}
\end{center}

An interesting feature of the Euclidean path integral approach, is to
show a link between the thermal properties of the black hole
spacetimes and their global, topological structure. This has led
Hawking and collaborators \cite{HHR94,HH96} to suggest that black hole
entropy has a topological origin. We shall come back to this issue later
on
in section \ref{sec:topol}.

%%%%%%%%%%%%%%%%%%%%%%%%%%%%%%%%%%%%%%%%%%%%%%%%%%%%%%%%%%%%%%%%%%%%%%%%%%
\subsection[{Derivation of \bh\ entropy as a Noether Charge}]
{Derivation of \bh\ entropy as a Noether Charge}
%%%%%%%%%%%%%%%%%%%%%%%%%%%%%%%%%%%%%%%%%%%%%%%%%%%%%%%%%%%%%%%%%%%%%%%%%%

Among alternative derivations of black hole thermodynamics, the so
called {\em Noether charge approach} has attracted growing attention
because of the central role that it appears to give to the symmetry
under diffeomorphisms and the ``localization'' of the \bh\ entropy on
the event horizon. Both of these features have become part of modern
interpretations of black hole thermodynamics as we shall see in the
next section.

Wald and collaborators~\cite{Wal93,IW94} demonstrated that the first
law of black hole thermodynamics can be reformulated in any
diffeomorphism-invariant theory in the presence of bifurcate Killing
horizons~\footnote{
%%%%%%%%%%%%%%%%%%%%%%%%%%%%%%%%%%%%%%%%%%%%%%%%%%%%%%%%%%%%%%%%%%%%%%%%%%%
  A Killing horizon is a null hypersurface whose null generators are
  orbits of a Killing vector field. In General Relativity it has been
  demonstrated that the event horizon of a stationary \bh\ is always a
  Killing horizon.  If the generators of the horizon are geodesically
  complete to the past (and the surface gravity of the \bh\ is
  different from zero) then it contains a 2-dimensional (in four
  dimensional spacetimes) space-like cross section ${\cal B}$ on which the
  Killing vector is null.  $\cal{B}$, called ``surface of bifurcation", is
  a fixed point for the Killing flow and on it the Killing vector
  vanishes. $\cal{B}$ lies at the intersection of the two hypersurfaces
  (past and future) forming the complete horizon. A bifurcate Killing
  horizon is a horizon with a ``surface of bifurcation" $\cal{B}$.
%%%%%%%%%%%%%%%%%%%%%%%%%%%%%%%%%%%%%%%%%%%%%%%%%%%%%%%%%%%%%%%%%%%%%%%%%
  }. From this reformulation one gets that the gravitational entropy
can be identified with the Noether charge associated with the
diffeomorphism invariance of the theory~\cite{Wal93,IW94}.

The basic scheme for the derivation of the first law can be summarized
in a few steps. Let us start with a Lagrangian density ${\cal
  L}(x, \phi, \phi_{,\alpha}, \phi_{,\alpha\beta},...)$
invariant under diffeomorphism transformations $x^{\mu}\rightarrow
x^{\mu} +\xi^{\mu}(x)$.  Here $\phi$ is a complete set of dynamical
variables (metric included).

The Noether current density associated with this symmetry of the
theory, say ${\cal J}^{\alpha}(\phi,\xi)$ is by definition conserved.
This implies that it can be written as:
\beq
\label{eq:noepot}
 {\cal J}^{\alpha}(\phi,\xi)={{\cal N}^{\alpha\beta}}_{,\beta}\qquad
{\cal N}^{\alpha\beta}={\cal N}^{[\alpha\beta]}
 \eeq
When the metric is a dynamical variable it is convenient to work with
ordinary scalars and tensors instead of  densities:
\beq
 L=(-g)^{-1/2}{\cal L} \;\;\; J^{\alpha}=(-g)^{-1/2}{\cal J}^{\alpha}
 \;\;\; N^{\alpha\beta}=(-g)^{-1/2}{\cal N}^{\alpha\beta}
\eeq
In this notation Eq.(\ref{eq:noepot}) takes the form
\begin{equation}
  \label{eq:noepot2}
  J^{\alpha}=N^{\alpha\beta}_{;\beta}
\end{equation}
The integral of $N^{\alpha\beta}$ over a closed two-dimensional
surface $\sigma$ is called the {\em Noether charge} of $\sigma$
relative to $\xi^{\alpha}$.  It is a local function of the dynamical
fields and is linear in $\xi^{\alpha}$ and in its derivatives.
\beq
 N(\sigma,\xi)=\int_{\sigma}N^{\alpha\beta}\d\sigma_{\alpha\beta}
\eeq
We now consider a stationary, axisymmetric, asymptotically flat
spacetime with a bifurcate Killing horizon generated by the Killing
vector $\chi^{\alpha}=\xi^{\alpha}_{(t)}+\Omega^{\mathrm
  h}\xi_{(\phi)}^{\alpha}$. We can reproduce all of the above steps in
the case in which $\xi^{\alpha}=\chi^{\alpha}$ and $\Sigma$ is chosen
to be the three-hypersurface that extends from asymptotic infinity up
to the bifurcation surface.

Considering the variation of the Noether charge at the horizon
resulting from the variation of the fields $\delta \phi$ from the
background solution (keeping fixed $\chi^{\alpha}$), one finds that it
is equal to the difference between the variations of the mass and the
angular momentum of the system measured at infinity in a way which
strictly resembles the first law:
\beq
\delta N(\sigma_{\mathrm h},\chi)=\delta M-\Omega^{\mathrm h}\delta J
\label{eq:nfl}
\eeq
Actually it is not difficult to show that, for black holes which are
vacuum solutions of the Einstein equations, the above formula is
exactly the first law of black hole thermodynamics which we saw
before~\cite{IW94}. In fact for \GR\ one has
\begin{eqnarray}
  \label{corresp}
   L&=&\frac{1}{8\pi}R\nonumber\\
   J^{\alpha}&=&\frac{1}{8\pi} 
    \left[\chi^{[\alpha;\beta]}_{;\beta}+G^{\alpha}_{\beta}\chi^{\beta}\right]
\end{eqnarray}
where $G_{\alpha\beta}$ is the Einstein tensor. For vacuum solutions
this tensor vanishes and then, using Eq.(\ref{eq:noepot2}) it easy to
recognize that
\begin{equation}
  \label{eq:relat}
      N^{\alpha\beta}=\frac{1}{8\pi} \chi^{[\alpha;\beta]}
\end{equation}
Moreover from the fact that $\chi^{\alpha}$ is the generator of the
event horizon it follows that:
\begin{equation}
  \label{eq:noecha}
  N(\sigma_{\mathrm h},\chi)=\frac{1}{8\pi}\int_{\sigma}\d \sigma_{\alpha\beta}
   \chi^{[\alpha;\beta]}=\frac{1}{8\pi}\kappa A
\end{equation}
So equation (\ref{eq:nfl}) takes exactly the standard form of the
first law with $S=A/4$.

The most remarkable fact is that this relation is valid for any
covariant theory and can then be interpreted as a generalization of
the Einsteinian formula. If a diffeomorphism invariant theory does not
contain higher than second order derivatives of the dynamical
variables, the Noether charge $N(\sigma,\chi)$ can always be reduced
to a linear function of $\chi^{\alpha}$ and its first derivative.
Since the Killing vector is, by definition, null on the bifurcation
surface, $\chi^{\alpha}=0$, then $N^{\alpha\beta}\sim
\chi^{\alpha\beta}$.

Finally to obtain a value of the Noether charge independent of the
normalization of the Killing vector one can simply divide
$N(\sigma,\chi)$ by the surface gravity $\kappa$. The normalized
quantity ${\tilde N}(\sigma,\chi)$ is then a purely local
quantity dependent on the geometry and on the dynamical fields at the
horizon. Using it in the analogue of the first law leads to a natural
generalization of the gravitational entropy
\beq
 S=\frac{2\pi}{\kappa}N(\sigma,\chi)= \frac{2\pi}{\kappa}
 \int_{\sigma}N^{\alpha\beta}\d\sigma_{\alpha\beta}
\eeq

To conclude we just stress that the Noether charge approach has several
advantages:

\begin{itemize}
  
\item The strict locality and the geometrical nature of the normalized
  Noether charge make it insensitive to any ambiguity in the
  definition of the Lagrangian (for example to the addition of a total
  divergence).
  
\item The method does not require any ``Euclideanization'', a step
  which is still not fully understood in QFT in curved spaces (see
  \cite{IW95} for a discussion of the link between Euclidean
  methods and Noether charge approach).
  
\item Although the original Wald derivation relied on the
  presence of a bifurcation surface, nevertheless the entropy can be
  calculated by performing the integral over any arbitrary slice of
  the horizon of a stationary black hole \cite{JKM94}.

\end{itemize} 

To conclude the section we call attention to a
crucial fact which is evident in this derivation of the gravitational
entropy.  It is notable that the black hole entropy can be so
intimately related to the symmetry under diffeomorphisms of the
theory, and that such a clean geometrical local nature (localized
exactly on the event horizon) can be attached to it. Is there a deep
link between symmetries and gravitational entropy? Is the Noether
charge approach suggesting a localization on the horizon of the
entropy? We shall discuss these issues again in the following section
where we shall give a panoramic view of proposals advanced for
explaining the origin of gravitational entropy.
 
%%%%%%%%%%%%%%%%%%%%%%%%%%%%%%%%%%%%%%%%%%%%%%%%%%%%%%%%%%%%%%%%%%%%%%%%%%
\section[{Explanations of Black Hole entropy}]
{Explanations of Black Hole entropy}
%%%%%%%%%%%%%%%%%%%%%%%%%%%%%%%%%%%%%%%%%%%%%%%%%%%%%%%%%%%%%%%%%%%%%%%%%%

As often in the history of science, the discovery of such a
complex structure such as black hole thermodynamics opened a whole set
of new questions to which complete answers are still lacking.

The problem of finding a dynamical origin for \bh\ entropy is the effort to
achieve a statistical mechanics explanation of it. This means giving
an interpretation of horizon thermodynamics in a ``familiar'' way, in
the sense that it could be seen as related to dynamical degrees of
freedom associated with \bh\ structure.

It is interesting to note that the terminology ``\bh\ structure'',
used above, is necessarily imprecise given the fact that there is still
quite a heated debate in the literature about the location
of these supposed degrees of freedom giving rise to the
gravitational entropy.  The enumeration of all of the proposals for
explaining black hole entropy is out of the scope of this thesis. In
what follows we shall try to present the main present tendencies.  In
order to make such a classification easier to follow we shall
accumulate different proposals, grouping them in three general classes.

%%%%%%%%%%%%%%%%%%%%%%%%%%%%%%%%%%%%%%%%%%%%%%%%%%%%%%%%%%%%%%%%%%%%%%%%%%
\subsection[{Information based Approaches}] 
{Information based Approaches} 
%%%%%%%%%%%%%%%%%%%%%%%%%%%%%%%%%%%%%%%%%%%%%%%%%%%%%%%%%%%%%%%%%%%%%%%%%%

The first class of explanations, at least in time, for the \bh\ entropy
is that based on the ``{\em sum over possibilities}''. The basic idea
is that when a black hole forms we retain just minimal information
about the initial state (for example the star) which produced it. The
information about the collapsed matter is ``cut off'' by a strong
gravitational field and the black hole ``forgets'' everything about
its initial state except for the charges associated with long range
forces: mass $M$, electrical charge $Q$ and angular momentum $J$~\footnote{
%%%%%%%%%%%%%%%%%%%%%%%%%%%%%%%%%%%%%%%%%%%%%%%%%%%%%%%%%%%%%%%%%
  This is strictly true for solutions of the Einstein-Maxwell equations.
  For black hole solutions in unified field theories other conserved
  charges ({\em aka} hair) can emerge.
%%%%%%%%%%%%%%%%%%%%%%%%%%%%%%%%%%%%%%%%%%%%%%%%%%%%%%%%%%%%%%%%%
}. This is the so-called {\em No Hair Theorem}~\cite{price72}.
In close analogy with standard thermodynamics it is then possible to
associate an entropy with the black hole by counting all of the possible
microscopic configurations of the initial system which would lead to
the same black hole (that is to the same values of the macroscopic
parameters $M,Q,J$). In this case the {\em informational entropy}
({\em aka} von Neumann entropy) would be
\begin{equation}
S_{I}=-\sum_{n}p_{n}\ln p_{n}
\end{equation}
where $p_{n}$ are the probabilities of different initial states.

This general framework includes the early works of Bekenstein
\cite{Bek74} and of Zurek and Thorne \cite{ZT85}. In this proposal the
black hole entropy is related to the ``{\em logarithm of the number of
  quantum-mechanically distinct ways that the hole could have been
  made}''.  In this sense one should also note the proposal by York
\cite{Yo83} who tried to identify the dynamical degrees of freedom at
the origin of \bh\ entropy with the quasi-normal modes excited by the
quantum evaporation process. Since the logarithm of the number of
thermally excited modes at a given time is much less than $M^2$ (in
the \Sch\ solution $S=A/4=4\pi M^2$) York proposed to define the
entropy as the logarithm of the number of distinct excitation states
of these modes in the process of black hole evaporation. More recently
Bekenstein and Mukhanov \cite{BM95} proposed a slightly different way
to explain the gravitational entropy in the context of a model where
the mass (and hence the area) of the black hole is assumed to be
quantized. In this sense one considers the \bh\ in a similar way to an
atomic system, where the absorption/emission of quanta would
correspond to the transition from some energy level to another. The
entropy can then be related to the degeneracy $g(n)$ of the level $n$
in which the black hole is at a given moment: $S=\ln g(n)$.  We stress
the difference between this approach and the Zurek--Thorne one. While
in the latter, the different initial states of matter (which could
have led to the final \bh ) are counted, in the Bekenstein--Mukhanov
approach one counts different trajectories in the space of black hole
parameters.

%%%%%%%%%%%%%%%%%%%%%%%%%%%%%%%%%%%%%%%%%%%%%%%%%%%%%%%%%%%%%%%%%%%%%%%%%%
\subsection[{Quantum Vacuum Based Approaches}]
{Quantum Vacuum Based Approaches}
\label{subsec:qvba}
%%%%%%%%%%%%%%%%%%%%%%%%%%%%%%%%%%%%%%%%%%%%%%%%%%%%%%%%%%%%%%%%%%%%%%%%%%

The gravitational field of a black hole induces strong changes in the
structure of the quantum vacuum surrounding it. As discussed in the
Introduction, even in flat space the vacuum can be seen as a physical
medium characterized by the zero-point fluctuations of the physical
fields.  For eternal black holes the vacuum appears to be polarized by
the \bh\ in such a way as to form a thermal atmosphere. It is then
tempting to relate the modification in the vacuum structure around the
black hole with the vacuum fluctuations around it.

In this direction there was an early paper by Gerlach \cite{Ger76}
where the \bh\ entropy was related to the logarithm of the total
number of vacuum fluctuation modes responsible for the emission of the
black body radiation. Later, `t Hooft \cite{Hooft85} proposed a toy
model, {\em aka} the ``brick wall model'', in which it was assumed
that field modes vanish, encountering an effective boundary, on the
Planck length scale from the event horizon. It is then possible to
consider the thermodynamics associated with the fields restricted to
the exterior of this boundary.  The idea of `t Hooft was to identify
the \bh\ entropy with those of the fields left outside the
horizon/boundary in thermal equilibrium at the Hawking temperature. An
easy calculation yields the formula:
\begin{equation}
S\sim \alpha \frac{A}{\lambda^2}
\end{equation}
where $A$ is the surface area of the \bh , $\lambda$ is the proper
distance of the ``brick wall'' from the horizon. For $\lambda\sim
L_{\pl}$ the above entropy is of the same order of magnitude 
as that of Bekenstein--Hawking.

A development of this model was subsequently proposed by
Srednicki~\cite{Sr93}, Bombelli, Koul, Lee, Sorkin~\cite{BKLS86},
Frolov, Novikov~\cite{FN93} and other works. The \bh\ entropy would be
generated by dynamical degrees of freedom, excited at a certain time,
associated with the matter in the \bh\ interior near to the horizon through
non-causal correlation (EPR) with external matter. The ignorance of an
observer outside the \bh\ about the modes inside the horizon is
associated with a so-called {\em entanglement entropy} which is
identified with the Bekenstein--Hawking one.

%%%%%%%%%%%%%%%%%%%%%%%%%%%%%%%%%%%%%%%%%%%%%%%%%%%%%%%%%%%%%%%%%%%%%
\subsubsection{Entanglement Entropy}
\label{ssubsec:eeba}
%%%%%%%%%%%%%%%%%%%%%%%%%%%%%%%%%%%%%%%%%%%%%%%%%%%%%%%%%%%%%%%%%%%%%%

Let us consider a global Hilbert space ${\mathbf H}$ composed of two
uncorrelated ones ${\mathbf H}_{1}$, ${\mathbf H}_{2}$.
\begin{equation}
{\mathbf H}={\mathbf H}_{1} \otimes {\mathbf H}_{2}
\end{equation}
A general state on ${\mathbf H}$ can be described as a linear superposition
of states on the two Hilbert spaces
\begin{equation}
| \psi \rangle = \sum_{a,b} \psi(a,b) |a \rangle | b\rangle
\label{eq:gs}
\end{equation}
One can define a global density matrix
\begin{equation}
\rho(a,a',b,b')=\psi(a,b) \psi^{*} (a',b')
\end{equation}
The reduced density matrix for the subsystem $a$ is given by
\begin{equation}
\rho(a,a')=\sum_{b=b'}\psi(a,b) \psi^{*} (a',b')
\label{eq:rd}
\end{equation}
Note that even if the general state (\ref{eq:gs}) defined on the
global Hilbert space is a pure one, the form of the reduced density
matrix (\ref{eq:rd}) shows that the corresponding state defined only
in one subspace is a mixed one.  This corresponds to an information
loss that can be properly described by von Neumann entropy (which is
zero for a pure state)
\begin{equation}
S_{1}=-\Tr(\rho_{a} \ln \rho_{a})
\end{equation}
Let us now discuss the same issue in the \bh\ case.  We can consider a
stationary \bh\ and define ${\hat{\rho}\,}^{\tot}$, the density matrix
describing, in Heisenberg representation, the initial state of quantum
matter propagating on its background.  For an external observer the
system consists of two parts: the \bh\ and the radiation outside it.
By defining a spacelike hypersurface we can consider quantum modes of
radiation at a given time so that they can be separated into external
ones and ones internal to the \bh .

The state for external radiation is obtainable from
${\hat{\rho}\,}^{\tot}$ by tracing on all of the states of matter
inside the event horizon and so inaccessible to the external observer
\beq
{\hat{\rho}\,}^{\rad}=\Tr^{\ins}{\hat{\rho}\,}^{\tot} 
\eeq
For an isolated \bh\ alone this density matrix would describe its Hawking
radiation at infinity. We can also define the density matrix for the
\bh\ state
\beq
{\hat \rho\,}^{\BH}=\Tr^{\outs}{\hat{\rho}}^{\tot} 
\eeq 
where one now performs the trace on the external degrees of freedom. From
this matrix it is possible to find the related von Neumann entropy
\beq
S^{\BH}=-\Tr^{\ins}(\hat{\rho}^{\BH} \ln \hat{\rho}^{\BH}) 
\label{Enta} 
\eeq
This is exactly what is called ``entanglement entropy''. It is
important to note that this definition is invariant in the sense that
independent changes of definitions of the vacuum for the ``external'' and
``internal'' states do not change the value of $S^{\BH}$~\cite{BL95}.

Calculations of entanglement entropy were performed by various authors
in a wide range of situations, in flat spaces as well as in curved
ones.  There are three common points that appear as proper
characteristics of \ee :
\begin{enumerate}
\item Entanglement entropy is not a never-decreasing quantity. If the
  internal region is reduced by shrinking the division surface, the
  \ee\ goes to zero in the limit that all of the spacetime is again
  available to the external observer.
\item Entanglement entropy is always proportional to the area of the ``division
  surface'' (the event horizon in the \bh\ case).
\item Entanglement entropy is always divergent on the division surface due
  to the presence of modes of arbitrarily high frequency near to the
  horizon.
  The divergence form is general and independent of the kind of field.
\end{enumerate}
The first point is a deep problem, casting some shadows on what the
computation of entanglement entropy is effectively probing about \bh\ 
physics. Actually only in the Thermofield Dynamics framework which we
introduced in section \ref{subsec:squeezed}, does it appear possible to
safely identify the \ee\ with the \bh\ one~\cite{TU,Israel76}.

Although the second point was initially one of the motivations for the
identification of entanglement entropy with the Bekenstein--Hawking one, it
was soon realized that such a behaviour is {\em too general} and
appears to be valid even in flat space~\cite{Sr93}. It is commonly
felt that the special form of the Einstein equation and
diffeomorphism invariance is at the root of \bh\ 
thermodynamics~\cite{Jacobson:1995ab,Visser:1998yu}. How can one
reconcile this with a general form of \ee\ insensitive to the
dynamical content of the theory?

The third point is an open subject. The divergence of the \ee\ has
sometimes been regulated by introducing a Planckian size cut-off at
the horizon (and in this way getting an ``evolved version'' of the
brick wall approach)~\cite{BFZ95,BL95,MkIs98}. In other cases it has
led to a renormalization procedure for the Newton Gravitational
constant but, in the absence of a renormalizable theory of quantum
gravity, this has had a limited development so far~\cite{SU94,CW95}. As an
alternative to these procedures one can conjecture a framework where
the Newton constant is determined by the matter fields~\cite{Jac94}.
This framework is that of {\em Induced Gravity} and it will be the
next candidate for explaining the \bh\ entropy which we shall
consider.

%%%%%%%%%%%%%%%%%%%%%%%%%%%%%%%%%%%%%%%%%%%%%%%%%%%%%%%%%%%%%%%%%%%%%%%%
\subsubsection{Induced Gravity and black Hole entropy}
%%%%%%%%%%%%%%%%%%%%%%%%%%%%%%%%%%%%%%%%%%%%%%%%%%%%%%%%%%%%%%%%%%%%%%%

According to Sakharov's ideas~\cite{Sak,sakte}, the Einstein--Hilbert
gravitational action is induced by vacuum fluctuations of quantum
matter fields and it represents a type of elastic resistance of the
spacetime to being curved (with elastic constant $G_{\mathrm N}$). The
qualitative basis for this statement~\cite{adler} is the fact that the
Einstein--Hilbert action density is given by the Ricci scalar $R$
times a huge constant (of the order of the square of the Planck mass):
curvature development requires a large action penalty~\cite{adler} to
be paid, that is there is a sort of ``elastic resistance to curvature
deformations''.

The fact that, according to Sakharov, a long--wavelength expansion of
quantum matter fields in curved spacetime contains zero point
divergent terms proportional to the curvature invariants, suggests that
zero point fluctuations induce the gravitational action.
``Induction'' means that no tree level action is considered: quantum
matter fields generate it at a quantum level. Gravitational
interaction in this picture becomes a residual interaction~\cite{MTW}
of a more fundamental one existing at high energy scales (Planck mass).

There are different ways to implement such a fundamental
theory~\cite{adln}. Generically it is assumed that the fundamental
theory is defined by an action $ I [ \phi_{i},g_{\mu\nu} ] $
describing the dynamics of some fields on a curved background
described by the metric $g_{\mu\nu}$. As stated, initially no
gravitational action is associated with this metric which is not a
dynamical field. The gravitational action arises then from the average
over the fundamental fields (and it is in this sense an ``effective
action'')
\begin{equation}
\exp \left( -W \left[ g_{\mu\nu} \right] \right)=
\int {\cal D} \left[\phi_{i}\right] \;
\exp \left( -I \left[ \phi_{i},g_{\mu\nu} \right]\right)
\end{equation}
It appears clear then that in this framework it is natural to relate
gravitational entropy to the statistical mechanics of those high mass
fields that induce gravity in the low energy limit of the theory.

This approach has been developed in the past four years by Frolov,
Fursaev and Zel'nikov
\cite{Frolov:1997aj,Frolov:1997up,Frolov:1997xd,Frolov:1998ea,Frolov:2000gy}
who proposed a model based on some scalar boson fields and fermion
ones.  They showed that it is possible to build up an induced gravity
model where the (induced) cosmological constant is zero and the
(induced) Newton constant is finite. A surprising condition for this
to be possible is the non-minimal coupling between the scalar fields and
curvature.  The statistical entropy $S_{SM}$ is computed again in the
entanglement fashion but it should be stressed that this time it
counts the degrees of freedom of the heavy (unobservable) fields
instead of the light (observable) fields. The former can be
excited just in a narrow region (of Planck size) near to the horizon.

In principle, since the gravitational action is induced by these heavy
constituents, the statistical entropy associated with the latter
should coincide with the Bekenstein--Hawking one.  Nevertheless it is
interesting to note that the above equality is found not to be exact.
It is actually necessary to subtract a quantity ${\cal Q}$ localized
on the event horizon from the (divergent) statistical mechanical
entropy of the heavy fields in order to have equality with the
(finite) classical Bekenstein--Hawking entropy.  This quantity can be
identified with the Noether charge associated with the invariance
under diffeomorphisms of the non-minimally coupled fields. This seems
to indicate some sort of link between the induced gravity approach and
the Wald Noether charge proposal reviewed previously. The nature of
this link is still unclear but it is intriguing that it appears to be
linked to the underlying diffeomorphism symmetry assumed for the theory
in the low energy limit. These elements will appear later on in
section \ref{subsec:symm} where we shall discuss the most recent class
of proposals for explaining black hole entropy.

%%%%%%%%%%%%%%%INDUCED %%%%%%%%%%%%%%%%%%%%%%%%%%%%%%%%%%%%%%%%%%%%%%%%%%
\subsection[{A Casimir approach to black hole thermodynamics}]
{A Casimir approach to black hole thermodynamics}
\label{subsec:bhcas}
%%%%%%%%%%%%%%%%%%%%%%%%%%%%%%%%%%%%%%%%%%%%%%%%%%%%%%%%%%%%%%%%%%%%%%%%%

We shall now described another vacuum based framework --- developed by
the author in collaboration with F. Belgiorno~\cite{BL97} --- in which
a close analogy with the Casimir effect is shown in a natural way by
focusing attention on the dynamics of quantum vacuum fluctuations in
curved spacetime. We stress that this approach has mainly a didactic
aim and is so far speculative. In particular it is interesting because
it shows how, in an induced gravity approach, the distribution of
zero-point modes associated with the fields located ouside a black
hole, can be deformed in such a way to induce a non-trivial
thermodynamical behaviour of the spacetime. In a certain sense we
shall interpret the black hole thermal properties as a special example of
the effective thermality of vacuum we discussed in
section~\ref{subsec:evt}.

%-----------------------------------------------------------------%
\subsubsection{Gravitational action subtraction}
%------------------------------------------------------------------%

The Euclidean path integral approach described schematically in
section~(\ref{sec:EPI}), contains a step which is not definitively
understood, namely the ``reference'' action subtraction for the
gravitational tree level contribution.  

The action on shell consists of the usual Einstein--Hilbert action
(together with a surface term related to the extrinsic curvature) and
a Minkowskian subtraction term (the ``reference'' action). The latter
is introduced by requiring that in flat spacetime the gravitational
action should be zero and it is necessary in order to obtain a finite
value when evaluated on shell. In the following section we will
further analyze this topic.

Hawking and Horowitz~\cite{HH96} have developed this subtraction
scheme for the case of non-compact geometries. They considered the
Lorentzian gravitational action for a metric $g$ and matter fields
$\phi$ on a manifold $M$:
\beq
 I(g,\phi)=\int_{M} \left [{{R}\over{16\pi}}+L_{\m}(g,\phi) \right ]+
 {1\over{8\pi}}\oint_{\partial M} K.
\label{action}
\eeq
The surface term is needed to give rise to the correct equations under
the constraint of fixed induced metric and matter fields on the
boundary $\partial M$.  The action is not well-defined for
non-compact geometries: one has in this case to choose a rather
arbitrary background $g_{0}, \phi_{0}$. Indeed Hawking and Horowitz
chose a static background solution of the field equations.  Their
definition of the physical action is then: 
\beq 
 I_{\mathrm{phys}}(g,\phi)\equiv I(g,\phi)-I(g_{0},\phi_{0}).
 \label{acfi}
\eeq
The physical action for the background is thus zero.  Further, it is
finite for a class of fields $(g,\phi)$ asymptotically equal to
$(g_{0},\phi_{0})$. For asymptotically flat metrics the background is
$(g_{0},\phi_{0})\equiv (\eta, 0)$; the action so obtained is then
equal to that of Gibbons and Hawking:
\begin{equation}
 I_{\mathrm{phys}}(g,\phi)=\int_{M} \left [ 
 {{R}\over{16\pi}}+L_{\m}(g,\phi) \right ]
 +{1\over{8\pi}} \oint_{\partial M} (K-K_{0}).
\label{giha}
\end{equation}
The last term is just the so-called ``Minkowskian subtraction'':
$K_{0}$ is the trace of the extrinsic curvature of the boundary of the
background spacetime.  The above subtraction could be physically
interpreted by requiring that it should represent that of a background
contribution with respect to which a physical effect is measured.

There is a nontrivial point to be stressed about (\ref{acfi}): it 
is implicitly assumed that the boundary metrics $h$ on $\partial M$ 
induced by $g_{0}$ and $g$ are the same. In general it is not 
possible to induce a 3--metric $h$ from a given 4--metric $g_{0}$; 
the same problem arises for the induction of a generic $h$ by 
flat space (See Hawking in~\cite{Eincen}). In the case where the 
asymptotic behaviour of the 4--metrics $g$ and $g_{0}$ is the same, one can 
assume that the 3--metrics, say $h$ and $h_{0}$, induced respectively by 
$g$ and $g_{0}$, become asymptotically equal~\cite{Eincen}.
More generally the requirement to get the same boundary induced 
metric with $g_{0}$ and $g$ can be thought of as a physical constraint 
on the choice of a reference background for a given spacetime.

So far we have argued that the subtraction procedure is a fundamental step 
in the path-integral formulation of semiclassical quantum gravity. In what 
follows, we will recall some well--known facts about the Casimir 
effect \cite{Birrell-Davies,moste,Casrep}, in order to suggest a formal 
similarity between Casimir subtraction and the above gravitational 
action subtraction.

%-----------------------------------%
\subsubsection{Casimir subtraction}
%-----------------------------------%

We start by recalling the problem of two parallel infinite conducting
plates which we met in section \ref{subsec:casimir}; we saw that the energy
density is obtained by means of the subtraction of the zero--point
mode energy in the absence of the plates from the zero--point mode energy
in the presence of the two plates.  One can in general formally define
the Casimir energy as follows~\cite{Casrep}:
\beq
 E_{\mathrm{Cas}}[\partial M]=E_{0}[\partial M]-E_{0}[0]
\label{caster}
\eeq
where $E_{0}$ is the zero--point energy and $\partial M$ is a 
boundary\footnote{
%%%%%%%%%%%%%%%%%%%%%%%%%%%%%%%%%%%%%%%%%%%%%%%%%%%%%%%%%%%%%%%%%%%%%%%%%
  In eq. (\ref{caster}) and in the following analogous equations
  concerning the Casimir effect, a regularization of the right hand
  side terms is understood.
%%%%%%%%%%%%%%%%%%%%%%%%%%%%%%%%%%%%%%%%%%%%%%%%%%%%%%%%%%%%%%%%%%%%%%%%%
}. 

Boundary conditions in the Casimir effect can be considered as
idealizations of real conditions in which matter configurations or
external forces act on a field. The most general formula for the
vacuum energy is~\cite{Casrep} 
\beq
 E_{\mathrm{Cas}}[\mbox{\boldmath $\lambda$}]=E_{0}[\mbox{\boldmath
 $\lambda$}]-E_{0}[\mbox{\boldmath $\lambda_{0}$}]
\label{casgen}
\eeq
where $\mbox{\boldmath $\lambda$}$ is a set of suitable parameters
characterizing the given configuration (for example boundaries, external
fields, nontrivial topology), and $\mbox{\boldmath $\lambda_{0}$}$ is
the same set for the configuration with respect to which the effect has to be
measured.  In the case that $\mbox{\boldmath $\lambda$}$ represents an
external field $\bf A$, the vacuum energy distortion induced by
switching on the external field is given by: 
\beq 
 E_{\mathrm{Cas}}[{\bf A}]=E_{0}[{\bf A}]-E_{0}[0].
 \label{casfie}
\eeq
One can also take into account the finite temperature Casimir
effects~\cite{plut,dowken}: in this case matter fields are not in their
vacuum state, there are real quanta excited which are statistically
distributed according to the Gibbs canonical ensemble.  The Casimir free
energy is:
\beq
 F_{\mathrm{Cas}}[\beta,\mbox{\boldmath $\lambda$}]
 =F[\beta,\mbox{\boldmath $\lambda$}]-F[\beta,\mbox{\boldmath $\lambda_{0}$}].
\label{casfree} 
\eeq
The zero--point contribution~\cite{plut} to the finite temperature
effective action is simply proportional to $\beta$, so it does not
contribute to the thermodynamics because $S=(\beta
\partial_{\beta}-1)\beta F$.

The formal analogy of (\ref{acfi}) with, for example, (\ref{casfree})
consists not just in the fact that in both cases there is a reference
background to be subtracted in order to get the physical result. Indeed
it can be strengthened by making the obvious substitutions,
$\gmn$ in place of {\boldmath $\lambda$} and $\eta_{\mu\nu}$ in place
of {\boldmath $\lambda_{0}$}. Moreover we should take into account
that the Euclidean action $I$ and the gravitational free energy are
easily related via $\beta$: $I=\beta F$.

We stress that there are still substantial differences between
(\ref{acfi}) and (\ref{casfree}) due to the fact that in (\ref{casfree})
the field $A$ is external whereas in (\ref{acfi}) the field $\gmn$ is
the dynamical field itself; moreover, a deeper link of (\ref{acfi})
with the Casimir effect would require a quantum field whose zero point
modes are distorted by spacetime curvature.  Note that in this
case one could naively invoke a Casimir effect with respect to the background
spacetime $(M_{0},g_{0})$:
\begin{equation}
 F_{\mathrm{Casimir}}[\beta,g]_{M}=F[\beta,g]_{M}-F[\beta,g_{0}]_{M_{0}}.
\label{casgra}
\end{equation}
The above relation is purely formal and requires static manifolds
$(M,g)$, $(M_{0},g_{0})$. For zero-temperature the idea underlying the
Casimir effect, as seen above, is to compare vacuum energies in two
physically distinct configurations. If the gravitational field plays
the role of an external field, one can {\em a priori} compare
backgrounds with different manifolds, topology and metric structure.
The non triviality found in defining a gravitational Casimir effect in
a meaningful way can be easily understood, for example, in terms of
the related problem of choice of the vacuum state for quantum
fields~\cite{Birrell-Davies}. Moreover, in the presence of a physical
boundary, the subtraction (\ref{casgra}) is ill-defined in general
because the same embedding problems exist as for (\ref{acfi}).
Despite these problems, we assume that it is possible to give a
physical meaning to (\ref{casgra})~\footnote{
%%%%%%%%%%%%%%%%%%%%%%%%%%%%%%%%%%%%%%%%%%%%%%%%%%%%%%%%%%%%%%%%%%%%%%%%%
  It is at least well-known how to do this in the case of static
  spacetimes with fixed metrics and topology such as $R\times M^3$ where
  the spatial sections $M^3$ are Clifford-Klein space forms of flat,
  spherical or hyperbolic 3-spaces $(M^3=R^3/ \Gamma,S^3/ \Gamma, H^3/
  \Gamma)$ \cite{DB,Ish,GB} and where $\Gamma$ is the group of deck
  transformations for the given space~\cite{Wolf}. We shall discuss
  these cases in chapter~\ref{chap:4}.
%%%%%%%%%%%%%%%%%%%%%%%%%%%%%%%%%%%%%%%%%%%%%%%%%%%%%%%%%%%%%%%%%%%%%%%%% 
}.
In this case the aforementioned Sakharov Induced Gravity represents a
conceptual framework in which the analogy can be strongly
substantiated.

%-------------------------------------------------------%
\subsubsection{Induced gravity and Casimir subtraction}
%-------------------------------------------------------%

From what we have seen earlier it is clear that the induced
gravitational action should be given by the difference between the
quantum effective zero-point action for the matter fields in the
presence of the spacetime curvature and the effective action when the
curvature is zero\footnote{ 
%%%%%%%%%%%%%%%%%%%%%%%%%%%%%%%%%%%%%%%%%%%%%%%%%%%%%%%%%%%%%%%%%%%%%%%%
  This subtraction actually appears in one of Sakharov's seminal
  papers~\cite{sakte}.
%%%%%%%%%%%%%%%%%%%%%%%%%%%%%%%%%%%%%%%%%%%%%%%%%%%%%%%%%%%%%%%%%%%%%%%%
} i.e.
\beq
 I_{\mathrm{Induced}}=I_{\matter} [R]-I_{\matter} [0].
\label{saka}
\eeq
The field $g_{\mu \nu}$ actually appears, in this low energy regime, as an
external field and not as a dynamical one. Then the induced gravity
framework allows us to identify the Min\-kow\-skian subtraction as a
Casimir subtraction.  

We note that there is a boundary term in (\ref{acfi}) that is
necessary in order to implement a Casimir interpretation of the
subtraction and that is missing in the original idea of Sakharov.  But
if the manifold has a boundary it is natural to take into account its
effects on vacuum polarization. 

This implies that in a renormalization scheme for quantum field theory
in curved spacetime it is necessary to introduce suitable boundary
terms in the gravitational action in order to get rid of surface
divergences~\cite{KCD,Birrell-Davies}, so there should be boundary
terms in the induced gravitational action~\cite{denardo}.  Anyway it
is still unknown whether by taking into account boundary terms and
suitable boundary conditions it is possible to produce a
self--consistent theory of induced gravity\footnote{
%%%%%%%%%%%%%%%%%%%%%%%%%%%%%%%%%%%%%%%%%%%%%%%%%%%%%%%%%%%%%%%%%%%%%%%
  For a wider discussion of this point see also~\cite{barv}.
%%%%%%%%%%%%%%%%%%%%%%%%%%%%%%%%%%%%%%%%%%%%%%%%%%%%%%%%%%%%%%%%%%%%%%%
}.
The choice of the boundary conditions should be constrained in such a
way as to get an induced gravity action with a boundary term as in the
Hawking approach. Here we shall limit ourselves to a discussion of the
case of a scalar field and to the divergent part of the effective
action.

In a curved manifold $M$ with a smooth boundary $\partial M$, the
zero--point effective action for a scalar field depends on the
curvature and is divergent:
\beq
 \Gamma [\phi=0,\gmn]=\Gamma [\gmn].
\label{zercur}
\eeq
The zero-point effective action (\ref{zercur}) is composed of
divergent terms which, if $D$ is the dimension of $M$, correspond to
the first $l \leq D/2$ ($l=0,1/2,...$) coefficients, $c_{l}$, in the
heat kernel expansion~\cite{KCD,dowken}. In our case $D=4$ and $l \leq
2$.  For a smooth boundary, the coefficients $c_{l}$ can be expressed
as a volume part plus a boundary part:
\beq
 c_{l}=a_{l}+b_{l}.
\eeq 
The $b_{l}$ depend on the boundary geometry and on the boundary
conditions. The $a_{l}$ coefficients vanish if $l$ is half integral and
for integral values they are equal to the already discussed
Schwinger-DeWitt coefficients (sometimes called Minakshisundaram
coefficients) for the manifold $M$ without boundaries.

If there is a classical gravitational action (not induced but
fundamental), the divergent terms in (\ref{zercur}) can be
renormalized~\cite{KCD} by means of suitable gravitational
counter-terms: that is by re-absorbing the divergences into the bare
gravitational constants appearing in the action for the gravitational
field:
\beq
 S_{\mathrm{ren}}[\gmn]=S_{\mathrm{ext}}[\gmn]+\Gamma_{\mathrm{div}} [\gmn].
\label{reno}
\eeq
In an induced gravity framework, there is no classical (tree level)
term like $S_{\mathrm{ext}}[\gmn]$ to be renormalized and so there should exist
a dynamical cut-off which makes finite also the divergent terms. These
terms give rise to the gravitational (effective) action, so we can
call $\Gamma_{\mathrm{div}} [\gmn]$ ``the gravitational part'' of the
effective action.

We are mainly interested in the case of the \Sch\ black hole:
(\ref{acfi}) could then be interpreted as a Casimir free energy
contribution relative to the matter field zero-point modes.  We
consider the one-loop divergent contribution for a massive scalar
field enclosed in a sphere with radius $r=R_{\mathrm{box}}$.

We choose the boundary condition by looking at the structure of the
boundary terms.  For consistency, one would get in particular the
Einstein--Hilbert action term (\ref{acfi}) at the same order in the
heat kernel expansion, so we direct our attention to the boundary term
$b_{1}$.  It is possible to get the right form for the integrated
coefficient~\cite{denardo} 
\beq 
\int_{\partial M} K 
\eeq 
both for Dirichelet and Neumann boundary conditions.  The first one is
selected on the physical grounds that for a sufficiently large box
(infinite in the limit) the field should be zero on the boundary.

Of course in order to get ordinary gravitational dynamics,
i.e.~General Relativity, it is necessary that the couplings and the
mass of the fundamental theory fulfill suitable renormalization
constraints.  In particular, although generally
\beq 
a_{1}=A\int_{M} R \qquad b_{1}=B \int_{\partial M} K 
\eeq 
in order to get the gravitational sector of the action (\ref{giha})
one has to further impose that the constants $A$ and $B$ fulfill the
special condition $A=B/2$. 

In our conjecture, the gravitational part of the free energy
(\ref{acfi}) becomes a Casimir free energy contribution arising
because of zero--point modes.  What does this mean from a physical
point of view?  The most naive answer to this question is that black
hole (equilibrium) thermodynamics becomes a thermal physics of quantum
fluctuations that are initially in a spherically symmetric spacetime
and then are thermally distorted by the formation of a black hole.  In
this view it seems that there is implicitly an idea of a ``physical
process'' underlying the integral version of the laws of black hole
thermodynamics, i.e. one has to take into account that the vacuum
\Sch\ solution has been generated by a gravitational collapse
drastically deforming the quantum field vacuum.

Also in the general case, we conjecture the same Casimir
interpretation for the subtraction relative to the zero-point
terms (gravitational Lagrangian) in the effective action.  Outside the
induced gravity framework, if one naively considers a generalized
Casimir effect for quantum fields, for example on the \Sch\ background
with respect to flat space, one gets a zero-point contribution to be
renormalized in the gravitational action and, for consistency with
(\ref{casgra}), the gravitational action has to follow the Casimir
subtraction scheme.  What one would miss in this case is a microscopic
interpretation of the tree level gravitational contribution.

%----------------------------------------------------------%
\subsubsection{Boltzmann interpretation in induced gravity}
%----------------------------------------------------------%

In the following, we will pursue the standard Boltzmann interpretation
for black hole entropy and we will try to justify the fact that black
hole entropy can be explained in terms of the statistical mechanical
degrees of freedom by assuming that these are associated with the
zero-point fluctuations of quantum matter fields.

Our attempt consists of finding a link with Casimir physics suggested
by the subtraction procedure in black hole thermodynamics. The
subtraction in our view takes into account a physical process of
adiabatic vacuum energy distortion and in a real collapse we expect
non-adiabatic contributions.  To be more specific about this point, we
briefly summarize the framework which we choose. We have a finite
temperature Casimir scheme in which we subtract from the zero-point
Euclidean action, evaluated on the Euclidean section of a \Sch\ black
hole, the zero-point Euclidean action evaluated for a flat metric on
the same manifold (that is on a variety with {\em exactly the same}
topology $S^2 \times R^2$ (which is not the one of Euclideanized
Minkowski spacetime). In both cases, a periodicity is required in
the Euclidean time, with the period being given by the inverse of the Hawking
temperature.

This means that the Hartle--Hawking state is selected for the black hole,
whereas a mixed state at the Hawking temperature is selected for the
flat spacetime contribution.  We stress again that the manifold (and
so the topology) is the same in both of the above terms. Only the metric
changes.  A posteriori, we can interpret our induced gravity
framework Gibbons--Hawking prescription as the prescription for a
``metric Casimir effect'' on the same manifold (i.e. a Casimir effect in
which the role of the external field is played by the metric).

In both cases, to $r=r_{\mathrm h}$ corresponds a condition of ``no
boundary'':
in the \Sch\ case because the horizon is not a boundary of the
Euclidean manifold (it is a regular point); in the flat space case
because we do not want to change the topology and so we require a
condition of no boundary.

In this way the black hole entropy becomes a Casimir entropy: an
entropy associated with a thermal contribution of zero-point modes.
Note that in the framework of standard statistical mechanics,
zero-point modes cannot contribute to the entropy, their contribution
to the effective action being proportional to $\beta$. However, there
is no real contradiction: in the case of the horizon's thermodynamics,
the subtracted gravitational action is not proportional to $\beta$ but
to $\beta^{2}$ and so zero-point fluctuations do contribute to the
entropy.  Note that this conclusion is independent of the induced
gravity framework: matter fields give a thermal zero-point
contribution (that has to be renormalized in the gravitational action
outside of Sakharov's viewpoint).

About the prescription which one has to follow in order to actually
compute
such an entropy, our framework implies a conceptually easy
prescription.  The gravitational free energy is now the free energy
for the matter field zero-point modes. Once one calculates the latter
one can find the corresponding entropy by applying the usual formula
$\beta F=\beta E-S$.  In the \Sch\ case the internal energy $E$ is the
black hole mass $M$.

Unfortunately the lack of a definite high energy theory from which to
``induce'' the gravitational action, precludes any definitive
conclusion on the effective correctness of such an assumption.

%%%%%%%%%%%%%%%%%%%%%%%%%%%%%%%%%%%%%%%%%%%%%%%%%%%%%%%%%%%%%%%%%%%%%%%%%%
\subsection[{Symmetry based approaches}]
{Symmetry based approaches}
\label{subsec:symm}
%%%%%%%%%%%%%%%%%%%%%%%%%%%%%%%%%%%%%%%%%%%%%%%%%%%%%%%%%%%%%%%%%%%%%%%%%%

We have already seen, in our discussion of the Noether charge approach
proposed by Wald, that it is possible to view the black hole entropy as
something emerging from very general symmetries of the the action.
This starting point has recently been developed in the context of
superstring theories and has led to the first successful attempts to
achieve a statistical explanation of black hole
entropy~\cite{Pol95,SV96,Horo97,Mal97,MSW97,A97}.

Although such results are certainly a major advance in our understanding
of the origin of gravitational entropy it appears nevertheless that
the root of this success has to lie in a general behaviour of {\em
  any} quantum gravity theory when the low energy limit leads to
spacetimes with horizon structures. This point of view recently found
its formalization in Carlip's approach~\cite{Carlip99I,Carlip99II}.

%%%%%%%%%%%%%%%%%%%%%%%%%%%%%%%%%%%%%%%%%%%%%%%%%%%%%%%%%%%%%%%%%%%%%%%%%%
\subsubsection{String theories}
%%%%%%%%%%%%%%%%%%%%%%%%%%%%%%%%%%%%%%%%%%%%%%%%%%%%%%%%%%%%%%%%%%%%%%%%%%

The role of the quantum behaviour of black holes has often been
compared with the problem of the hydrogen atom in the early years of
the last century. The latter led to the revolutionary ideas of
quantum mechanics and it is a general belief that the former will
similarly lead to the development of a theory of quantum gravity. It
is then obvious that one of the main tests of any quantum gravity
theory would be a successful explanation of black hole thermodynamics.
Although this has not been fully achieved in any of the theories of quantum
gravity yet to hand, it is nevertheless true that great progress has been
made in the theory of {\em superstrings}.

According to superstring theory all particles can be interpreted
as excitations of fundamental one dimensional objects (the {\em
  strings}) which can be open or closed, and in the latter case have
periodic or anti-periodic matching conditions. They span a
two-dimensional surface (called the {\em world-sheet}) inside an
initially arbitrary $d$ dimensional space called the {\em target
  space}.  The world-sheet is described by specifying a vector
$\sigma^{\alpha}=(\sigma,\tau)$ which defines a point on the string,
$\sigma$ at a given time $\tau$. A point of the world-sheet in the
$d$-dimensional target space is described by an embedded coordinate
$X^{\mu}(\sigma,\tau)$ with $\mu=0,...,d-1$. On the world-sheet a
metric element $h_{\alpha\beta}$ can be introduced.  

String theory is regulated by two crucial parameters: the string scale
$\sqrt{\alpha'}$ (${\alpha'}$ is also known as {\em inverse string
  tension}) and the {\em string coupling constant} $g_{\mathrm s}$.
The first quantity sets the typical scale of the string so, for
example, in the limit of $\alpha'$ going to zero the strings become
point-like objects and the theory converges to a quantum field theory
(based on particle interactions).  The string coupling regulates the
perturbative order in the quantization of the string action which
classically takes the general form:
\begin{eqnarray}
S_{\mathrm{string}} &=& 
 \frac{1}{2\pi\alpha'} \int \d^2\sigma  \left\{  \sqrt{h}\;
  \left[  h^{\alpha\beta}\; g_{\mu\nu}(X)  \partial_{\alpha}X^{\mu}
   \partial_{\beta}X^{\nu}+T(X) \right]+
    \epsilon^{\alpha\beta} \partial_{\alpha}X^{\mu}
     \partial_{\beta}X^{\nu}B_{\mu\nu} \right\}\nonumber\\ 
 && + \frac{1}{4\pi}\int \d^2\sigma \sqrt{h}\, R\Phi(X).
\end{eqnarray}
From the world sheet point of view the $g_{\mu\nu}$, $T$, $B_{\mu\nu}$
and $\Phi$ fields (the {\em graviton, tachyon, antisymmetric tensor}
and the {\em dilaton}) are coupling constants. From the target space
they are instead dynamical fields. The tachyon field is a
``pathology'' of the theory that it is removed if {\em supersymmetry}
(SUSY) is imposed (and hence for any bosonic field a fermionic partner
is provided).  The quantization of the above action has so far been
successful only in very special target spaces
%($g_{\mu\nu}=\eta_{\mu\nu}+\delta_{\mu\nu}$), 
and then only in a perturbative fashion\footnote{Although more complex
  approaches have also been attempted, for example in the case of the
  sigma-model, these are still intrinsically perturbative and ordinary
  four-dimensional gravity is typically treated in the weak-field
  approximation.}.

Noticeably, while the general covariance of the world-sheet (invariance  
under two-dimensional re\-par\-amet\-riza\-tion) leads to the
conservation of the world-sheet energy momentum tensor, the conformal
invariance\footnote{
%%%%%%%%%%%%%%%%%%%%%%%%%%%%%%%%%%%%%%%%%%%%%%%%%%%%%%%%%%%%%%%%%%%%%
  As already said in the previous chapter, a conformal transformation
  is generically obtained by a rescaling of the metric
  $g_{\mu\nu}\rightarrow \tilde{g}_{\mu\nu}=\Omega^2(x)g_{\mu\nu}$.
  This sort of transformations preserves the ratio between the length
  of the infinitesimal vectors, $x^{\mu},y^{\nu}$ applied at the same
  point and also the angle between them
  $x_{\mu}g^{\mu\nu}y_{\nu}/\sqrt{x^{2}y^{2}}$
%%%%%%%%%%%%%%%%%%%%%%%%%%%%%%%%%%%%%%%%%%%%%%%%%%%%%%%%%%%%%%%%%%%%%
  .}  of the action implies the nullity of the $\beta$ functions of
the theory and leads, in the limit of vanishing $\alpha'$, to a set of
equations of motion for the string massless modes. These can be seen
as obtained from an {\em effective action} which is just a
generalization of the standard Einstein--Hilbert action of \GR\ where
the Newton constant is a function of both of the string theory
fundamental parameters $G_{\mathrm N}\sim g_{\mathrm s}^{2}\alpha'$. 

In addition to the standard Einstein--Hilbert term, which is linear in
the Ricci scalar $R$, there is also present an infinite series of
corrections which are higher powers of the curvature multiplied by
powers of $\alpha'$.  Obviously this implies that when curvatures
approach the inverse of the squared string scale this low energy approximation
breaks down.  This effective action will ``live'' in the target space
and will have a dimension which is fixed by the requirement of
conformal invariance.  For bosonic strings it is $26$, for
superstrings it is $10$. 

The emergence of four dimensional gravity is expected to come from a
process of compactification of the extra six dimensions~\cite{GSW}. It
has to be said that this large scale/low energy regime of the theory,
even after compactification to four dimensions, is not coincident with
``standard'' Einstein gravity.  In fact what one actually finds is
a {\em Supergravity} (SUGRA) theory: a theory of gravity plus other
bosonic as well as fermionic fields related by supersymmetry\footnote{
%%%%%%%%%%%%%%%%%%%%%%%%%%%%%%%%%%%%%%%%%%%%%%%%%%%%%%%%%%%%%%%%%%%%%%
  For example $N=8$ supergravity in four dimensions is the low energy limit
  of a so-called type $II$ string theory compactified on a six-torus
%%%%%%%%%%%%%%%%%%%%%%%%%%%%%%%%%%%%%%%%%%%%%%%%%%%%%%%%%%%%%%%%%%%%%%
}. Moreover, since they are a class of effective
theories, the SUGRA models are not renormalizable (just like \GR )
and can be trusted just for a weak coupling constant (tree level or at
most few-loops expansions).  Now let us come back to the full
superstring theory and see the basic mechanism by which string theory
should give a statistical explanation to \bh\ entrtopy.

Besides the massless fields which we talked about, there are, when one
quantizes a string in flat spacetime, an infinite tower of massive
string excitations. These can be enumerated by an integer $N$ and are
endowed with masses $M_{\mathrm s}^2=N/\alpha'$ (in the limit of zero
string coupling).  These states have a huge degeneracy of order $\exp
N^{1/2}$ for large $N$, so the ``informational entropy'' associated
with these states is $S_{\mathrm s}\sim \sqrt{N}$. The entropy of the
string is then proportional to the string mass (in string units).

This last point can seem to be an unavoidable obstacle if one tries to
explain the \bh\ entropy with the string one. In fact we know that the
Bekenstein--Hawking formula relates the gravitational entropy to the
{\em area} of the black hole and hence to its mass squared, not just
to its mass.

The key point for the resolution of this apparent paradox is to
consider a limiting situation when a black hole solution starts to be
described as some quantum state. This happens when the string coupling
constant goes to zero, in fact in this case the \Sch\ radius,
$G_{\mathrm N}M$, being proportional to the Newton constant shrinks
and becomes of the order of the string scale. At this point the
geometric description of \GR\ breaks down (as Hydrodynamics breaks
down at atomic scales) and one has to deal with pure string states.
So, when the \Sch\ radius $r_{0}$ of the \bh\ is of order of the
string scale $\sqrt{\alpha'}$, it is reasonable to equate the mass of
the string state $M_{\mathrm s}=\sqrt{N/\alpha'}$ to that of the black hole
$M_{\BH}=r_{0}/2G_{\mathrm N}$

It is easy to see that this implies that $\sqrt{\alpha'}/G_{\mathrm
  N}\sim \sqrt{N/\alpha'}$. From the relation $G_{\mathrm N}\sim
g_{\mathrm s}^{2}\alpha'$ one then obtains that this is verified if
$g_{\mathrm s}\sim N^{-1/4}$. But the latter expression actually
implies that the \bh\ entropy is $S\sim r^2_{0}/G_{\mathrm N}= \alpha'/
g^{2}_{\mathrm s}\alpha'= \sqrt{N}$. So the \bh\ entropy comes out to
be proportional to $\sqrt{N}$, the same as the statistical entropy
associated to the degeneracy of the string states.

We have then found a qualitative argument to relate the standard
gravitational entropy $S_{\BH}$ to a purely statistical one
$S_{\mathrm s}$.  Moreover the fact that one requires just $g_{\mathrm
  s}\sim N^{-1/4}$ is an indication that the ``transition'' from a
string state to a black hole one occurs at still small string
couplings (and hence perturbative results are still trustable).
Notably the above results also hold in higher dimensions.  In fact the
entropy of any \Sch\ \bh\ can be written as $S_{\BH}\sim r_{0}M_{\BH}$
so whenever $r_{0}\sim \sqrt{\alpha'}$ the entropy is exactly proportional to
the mass in string units as happens for a free string.

Although the above arguments are quite impressive, it should be stressed
that a definitive agreement, up to the factor $1/4G_{\mathrm N}$ can be
found just in some special cases.  In fact, it is the extremely
powerful tool of supersymmetry that allows such cases to be found.

We saw that the equality of the black hole entropy and the string one
can generally be obtained just for special values of the string
coupling constant. This implies that the identification is valid just
at given energy scales. There are nevertheless black holes for which
the entropy is {\em not} a function of the Newton constant and hence
is not dependent on $g_{\mathrm s}$. These are the so-called
extremally charged black holes which we already met in relation to the
third law of black hole thermodynamics. For these black holes the
relation $G_{\mathrm N}M^2=Q^2$ holds and the Bekenstein--Hawking
entropy would be:
\begin{equation}
  S_{\BH}=\frac{A}{4G_{\mathrm N}}=\frac{\pi}{4}Q^2
 \label{exen}
\end{equation}

We have seen that SUGRA theories, obtained as effective actions from
superstrings, generically exist in a ten-dimensional space.  The
interesting fact is that SUGRA theories generically admit solitonic
(non-perturbative) solutions that correspond to extended objects of
p-spatial dimensions called {\em p-branes} which are the carriers of the
multiple electric charges present in a SUGRA theory (a $0$\/-brane would
be a particle like, for example, an electron). Suitable combinations
of these objects allow, after the compactification process, for them
to be identified as higher dimensional generalizations of the standard
extremal black hole solutions in four or five dimensions. These
solutions have the peculiar property of being solitons which belong to
a very special class of states saturating the so-called
Bogomol'nyi--Prasad--Sommerfield (BPS) bound between mass and charge
$M\geq \gamma {\cal Z}$ (where $\gamma$ is a theory-dependent constant
and ${\cal Z}$ is the so-called {\em central charge} of a given
representation of the SUSY algebra).  These states will henceforth be
called BPS-states.

They have the property that if a state is BPS semiclassically it will
still be BPS at any value of the coupling constant. In fact in a
supersymmetric theory representations of the algebras do not change
under a smooth variation of the coupling constants.  Moreover if the
SUSY is strong enough even the mass $M$ of the state will not
renormalize and neither will the state be allowed to decay.  The
number of allowed BPS-states is a topological invariant of the {\em
  moduli space}, that is of the space of parameters defining a
solution, and depends only on discrete parameters of the theory (for
example the SUSY charges).  So the former number is independent of the
string coupling constant and hence for these states both $S_{\BH}$ and
$S_{\mathrm s}$ are independent of the coupling regime.

Using this property one can then start with a black hole solution
(which is a typical non-perturbative solution of the supergravity
theory) and start decreasing the string coupling until one can perform
the count of all the BPS states at weak coupling endowed with the same
total charge.  Note that in this case there is no issue about where to
match the black hole mass with that of the perturbative string state:
in both of the regimes the mass is completely fixed by the charge.

This appears to be so easy that one may wonder why it was not done
well before the mid nineties. Actually there was an inherent
difficulty related to the fact that for a long time there was no
description in string theory of the (BPS) objects corresponding to the
low-energy $p$\/-branes with which supersymmetric black holes are
constructed.  Basically it was not known how to describe in the weak
coupling limit the ``solitons'' of string theory.

The identification of these objects, called {\em Dp-branes}~\footnote{
%%%%%%%%%%%%%%%%%%%%%%%%%%%%%%%%%%%%%%%%%%%%%%%%%%%%%%%%%%%%%%%%%%%%%%
  $D$ here stands for the Dirichelet boundary condition imposed on
  open strings on these extended objects
%%%%%%%%%%%%%%%%%%%%%%%%%%%%%%%%%%%%%%%%%%%%%%%%%%%%%%%%%%%%%%%%%%%%%%
  .} is due to Polchinski~\cite{Pol95}.  He found that a $Dp$\/-brane has
a mass proportional to $1/g_{\mathrm s}$, so at weak coupling they are
very massive and hence non perturbative.  Nonetheless, since the
Newton constant $G_{\rm N}\sim g_{\mathrm s}^{2}$, one gets that the
gravitational field produced by the $Dp$\/-brane is extremely weak and a
flat spacetime description is allowed.

The quantized object to which an extremal black hole with nonzero
surface area would correspond (in the limit of small $\alpha'$ and
after compactification) is in this case built with a collection of
these charged $Dp$\/-branes and the open strings interacting with
them.  So in order to actuate the programme described above one has to
quantize the $Dp$\/-brane state and then count the number of
excitations of these solitons. This has been done and the number of
BPS states at weak coupling turned out to be equal to the exponential
of the Bekenstein--Hawking entropy of the black hole at strong
coupling~\cite{SV96}.

Although this result was first obtained for a five dimensional
extremal black hole \cite{SV96}, it has been generalized to the case
of extremal rotating \bh s and to that of slightly non-extremal ones.
These results were also re-obtained in the four dimensional case.

This ``corpus'' of results, although impressive is far from being
conclusive. In fact there are still numerous issues which are unclear and
which cast some doubt on the possibility of further generalizing
these results to the most interesting cases such as the \Sch\ and Kerr \bh
s.  We shall limit ourselves to a brief summary here.

\begin{itemize}
  
\item The role of SUSY: we have seen that the supersymmetry of the
  theory appears to be a crucial step for building up the BPS states
  and hence giving a coupling-independent meaning to the identification
  of the gravitational entropy with the string one. Is it possible to
  to generalize this framework when SUSY is broken?

\item In string theory there are always several charges. Black holes
  with just a single charge are not allowed due to the fact that their
  horizon becomes singular in the extremal limit. So, for example, the
  Reissner--Nordstr\"om solution is present in string theory but its
  charge $Z$ is always a function of the several charge parameters
  present in the theory.
  
\item We have seen that in SUGRA calculations what one actually does
  is to calculate the {\em area} of the black hole as a function of
  the charge parameters that extremize the theory in the moduli space.
  After this, one generally {\em imposes} a quarter of this area to be
  the gravitational entropy. This is {\em a priori} unjustified
  because (as explained in Sect.(\ref{BHT})) we are not sure at all
  that the Bekenstein--Hawking law holds semiclassically in the
  extremal case.  The fact that this identification is justified by
  the high energy regime results from $Dp$\/-branes, seems to imply that
  there may be something non trivial going on in the semiclassical
  limit. If the semiclassical dichotomy between extremal and
  non-extremal black holes is confirmed we shall have to assume that
  the stability of the BPS states is not enough for preserving the
  invariance of the statistical entropy. In particular one can imagine
  that, as the statistics for a fermion gas can drastically change due
  to phase transitions (for example in the Cooper pair phenomenon), in
  the same way statistics of string excitations can dramatically
  change in the low energy limit leading to a drastic reduction of the
  actual degeneracy of a state.
  
\item The string calculations always imply a unitary evolution.  It is
  still unclear how from such a behavior can emerge the non-unitary
  evolution associated with black hole evaporation. In relation to
  this point we note the interesting proposal made by
  Amati~\cite{Amati97, Amati:1999ia, Amati:1999fv} about the emergence
  of semiclassical black hole structure from extended string
  solutions.

\end{itemize}

%%%%%%%%%%%%%%%%%%%%%%%%%%%%%%%%%%%%%%%%%%%%%%%%%%%%%%%%%%%%%%%%%%%%%%%%%%
\subsubsection{Carlip's approach}
%%%%%%%%%%%%%%%%%%%%%%%%%%%%%%%%%%%%%%%%%%%%%%%%%%%%%%%%%%%%%%%%%%%%%%%%%%

We have just seen how the results of string theory lead to a
statistical entropy in agreement with the Bekenstein--Hawking one even
if they are performed in completely different regimes and from rather
different microscopic theories. Moreover recent results from the so-called
``quantum geometry'' approach seem to arrive at quite similar
results~\cite{Ashtekar:1998yu}.  In one sense this agreement is not too
surprising because any consistent quantum theory of gravity should be
able to give the semiclassical results in the appropriate limit.
Nevertheless the entropy, as a measure of the density of quantum
states, is a typical quantum object. The link to the semiclassical
quantities does not explain {\em why} the density of quantum states
behaves in such a special way.

It is then natural to ask if such a ``generality of the
Bekenstein--Hawking entropy'' can be understood in a general way,
fixing our attention just on the symmetries of the low energy limit of
the full quantum theory and on the global structure of spacetimes with
horizons.  Carlip's proposal~\cite{Carlip99I,Carlip99II} is an attempt
in this direction and it is substantially based on the idea that
classical symmetries can determine the density of the quantum states.
Although this statement is generally not satisfied, nevertheless it
holds for a large set of theories, that is for the conformal field
theories in two dimensions.

To see this let us consider a conformal field theory on the complex
plane. The infinitesimal diffeomorphism transformation takes the
form:
\begin{eqnarray}
 z &\rightarrow& z+\epsilon f(z)\\
 \bar{z} &\rightarrow& \bar{z}+\bar{\epsilon} f(z)  
\end{eqnarray}
Actually the total conformal group can be decomposed into two independent
ones in $z$ and $\bar{z}$ respectively and so we can work with just one
kind of coordinate. If we now take a basis $f_{n}(z)=z^{n}$ of
holomorphic functions and consider the corresponding algebra of
generators $L_{n}$ of the conformal group (the so-called ``Virasoro
algebra''), we get:
\begin{equation}
 [L_{n},L_{m}]=(n-m)\,L_{n+m}+\frac{\cal C}{12} n\,(n^2-1)\delta_{n,m}  
 \nonumber
 \label{eq:Viralg}
\end{equation}
The constant ${\cal C}$ here is the previously mentioned {\em central
charge} of the theory (in this case the one associated with the
Virasoro algebra), its non-nullity shows the emergence of a
conformal anomaly~\footnote{
%%%%%%%%%%%%%%%%%%%%%%%%%%%%%%%%%%%%%%%%%%%%%%%%%%%%%%%%%%%%%%%%%%%%%%%%%%%%
  An {\em anomaly} in QFT is the breakdown of a classical symmetry of
  a theory after quantization.
%%%%%%%%%%%%%%%%%%%%%%%%%%%%%%%%%%%%%%%%%%%%%%%%%%%%%%%%%%%%%%%%%%%%%%%%%%%%
}.  

As Cardy first showed~\cite{Cardy86}, the central charge ${\cal C}$ is
nearly enough to determine the asymptotic behaviour of the density of
states.  Let us consider the Virasoro generator $L_{0}$ and denote by
$\Delta_{0}$ its lowest eigenvalue (which is identifiable with the
``energy'' of the ground state).  If $\rho(\Delta)$ is the density of
eigenstates of $L_{0}$ with eigenvalue $\Delta$ then for large
$\Delta$ the following relation (also known as the Cardy's formula)
holds
\begin{equation}
 \rho(\Delta) \sim 
 \exp \left\{2\pi\sqrt{\frac{{\cal C}_{\eff}\,\Delta}{6}}\right\}
 \label{eq:cardy}
\end{equation}
where ${\cal C}_{\eff}={\cal C}-24\Delta_{0}$.

A typical example of how this count over states can be related to black
hole entropy is the example of (2+1)-dimensional gravity~\cite{BH86} with a
negative cosmological constant $\Lambda=-1/\ell^2$. For configurations
which are asymptotically anti-de Sitter, the algebra of
diffeomorphism acquires a surface term at infinity, and the induced
algebra at the boundary becomes a pair of Virasoro algebras with
central charges:
\begin{equation}
 {\cal C}=\bar{\cal C}=\frac{3\ell}{2 G_{\mathrm N}}
\end{equation}
Strominger~\cite{Stro98} made the observation that if one takes
the eigenvalues of $L_{0}$ and $\bar{L}_{0}$ that correspond to a
black hole, sets $M=(L_{0}+\bar{L}_{0})/\ell$ and $J=L_{0}-\bar{L}_{0}$,
and takes $\Delta_{0}=0$, then the Cardy formula gives back the
standard Bekenstein--Hawking entropy for the black hole.

This example is just suggestive of what one should look for in the
case of four dimensional black holes. In fact it uses very specific
features of gravity in two dimensions.
Moreover the Virasoro algebra
which it refers to is localized on the two dimensional boundary of the
three dimensional AdS spacetime, so it is in a certain sense
independent of whether the configuration at finite radius is a black
hole or a star.

Carlip's proposal is to concentrate attention on the event horizon
as the peculiar feature of black hole spacetimes. Indeed the obvious
generalization would be to look at the Poisson algebra of generators
of diffeomorphism and see whether an appropriate subgroup of
transformations acquires a central charge on the near-horizon
geometry of the $r-t$ plane.  We shall demonstrate, in the following
section \ref{sec:topol}, that the role of this two dimensional region
is indeed crucial in determining the gravitational entropy.  In fact
we shall see that the Bekenstein--Hawking formula can always be
generalized
to $S=\chi A/8$, where $\chi$ is the Euler characteristic of the four
dimensional manifold. Noticeably the non nullity of the Euler
characteristic is intimately linked to the non-triviality of the
$r-\tau$ surface topology (in Euclidean signature) which is, for
non-extremal black holes, equal to $S^{2}$.

This programme has been pursued in refs.~\cite{Carlip99I,Carlip99II}.
Although conclusive answers have not been obtained yet, the results
appear to be encouraging. The main problem at the moment seems to be
the uncertainty about how to impose boundary conditions at the event
horizon. Moreover aside from technical issues there is also a basic
impossibility in this approach of describing any non-equilibrium
thermodynamical issue. In fact if one has to fix boundary conditions on
an event horizon then this has in a certain sense to fix the horizon
itself. This is not too dramatic for what regards the explanation of
black hole entropy (which is intrinsically an equilibrium
thermodynamics concept) but of course it is a strong limitation if one
wants to achieve a full understanding of the quantum dynamics of a
black hole. In any case this proposal appears to share these sort of
problems with most of the above mentioned approaches\footnote{
%%%%%%%%%%%%%%%%%%%%%%%%%%%%%%%%%%%%%%%%%%%%%%%%%%%%%%%%%%%%%%%%%%%%
  For a thoughtful discussion about the non-equilibrium thermodynamics
  of black holes, see the seminal paper by Sciama~\cite{Sciama:1978ry} and
  following works by Sciama and collaborators~\cite{SciCan, Sciama:1981hr}
  and Hu and collaborators~\cite{Campos:1999xq}.
%%%%%%%%%%%%%%%%%%%%%%%%%%%%%%%%%%%%%%%%%%%%%%%%%%%%%%%%%%%%%%%%%%%
}.

After this survey of the proposals for explaining black hole
thermodynamics we shall now consider three different investigations
with which we shall try to probe some interesting points in this
field. The contents of the next three sections are original. Section
\ref{sec:topol} will present work done in collaboration with
G.~Pollifrone~\cite{LP97}. Section~\ref{sec:exbh} is an investigation
of incipient extremal black holes done in collaboration with
T.~Rothman and S.~Sonego~\cite{LRS99}.

%%%%%%%%%%%%%%%%%%%%%%%%%%%%%%%%% TOPOL %%%%%%%%%%%%%%%%%%%%%%%%%%%%%%
\section[{Gravitational entropy and global structure}]
{Gravitational entropy and global structure}
\label{sec:topol}
%%%%%%%%%%%%%%%%%%%%%%%%%%%%%%%%%%%%%%%%%%%%%%%%%%%%%%%%%%%%%%%%%%%%%%
 
We have just seen that most of the explanations of black hole entropy
give the horizon a central role. Both the vacuum based approaches
and the symmetry based ones have to rely heavily on the presence
of this special structure in order to justify the ``anomalous behaviour''
found for spacetimes with event horizons~\footnote{
%%%%%%%%%%%%%%%%%%%%%%%%%%%%%%%%%%%%%%%%%%%%%%%%%%%%%%%%%%%%%%%%%%%%%%%
  The thermodynamical features described for the black holes can be
  found (with some non trivial differences) in all spacetimes
  endowed with event horizons such as, for example, de Sitter and
  Rindler spacetimes.
%%%%%%%%%%%%%%%%%%%%%%%%%%%%%%%%%%%%%%%%%%%%%%%%%%%%%%%%%%%%%%%%%%%%%%%
}.

A related interesting result can be found in refs.~\cite{HHR94,HH96}.
There it has been shown that the Bekenstein--Hawking law, $S=A/4$, for
\bh\ entropy fails for extremal black holes. These objects were
already considered ``peculiar'', since their metric does not show any
conical structure near to the event horizon, so that no conical
singularity removal is required. Again this seems to imply a special
role for the two-dimensional surface $r-\tau$ in the vicinity of the
horizon.

In this section we shall prove that the Euler characteristic and
gravitational entropy can be related in the same way in almost all
known gravitational instantons endowed with event horizons. The non
triviality of the Euler characteristics of these ``intrinsically
thermal'' spacetimes is strictly related to the nature of the manifold
near to the event horizon.

In particular, we shall show that the Euler characteristic and entropy
have the same dependence on the boundaries of the manifold and we will
relate them by a general formula.  This formulation extends to a wide
class of instantons, and in particular also to the Kerr metric, the
known results~\cite{GHI,GK,BTZ,TC} about such a dependence.

Finally, it is important to stress that in order to obtain this result
one has to consider not only the manifold $M$, associated with the
Euclidean section describing the instanton, but also the related
manifold $V$, which is bounded by the sets of fixed points of the
Killing vector, associated with isometries in the imaginary time.
This will imply boundary contributions also for cosmological, compact
solutions.

%----------------------------------------------------------------------
\subsection[{Euler characteristic and manifold structure}]
{Euler characteristic and manifold structure}
\label{subsec:E-char}
%----------------------------------------------------------------------

The Gauss--Bonnet theorem proves that it is possible to obtain
the Euler characteristic of a closed Riemannian manifold $M^n$ without
boundary from the volume integral of the four-dimensional curvature: 
\begin{equation}
 S_{\GB}={{1}\over{32 \pi^{2}}} \int_{M} \varepsilon_{abcd}
 R^{ab}\wedge R^{cd} ,
\label{egb}
\end{equation}
where the curvature two-form ${R^a}_b$ is defined by the spin 
connection one-forms ${\omega^a}_{b}$ as 
\begin{equation}
 {R^a}_{b}=d{\omega^{a}}_{b}+{\omega^a}_{c}\wedge
 {\omega^c}_{b}.
\label{Rab}
\end{equation}
In a closed  Riemannian manifold $M^n$, Chern \cite{CHI,CHII} 
has defined  
the Gauss--Bonnet differential $n$-form $\Omega$ (with $n$ even) 
\begin{equation}
 \Omega={{(-1)^{n/2}}\over{2^{n}\pi^{n/2}(\frac{n}{2})!}} \varepsilon_{a_{1}
 \ldots a_{n/2}} R^{a_{1} a_{2}} \wedge \ldots \wedge
 R^{a_{n} a_{n+1}}, 
\label{gaussint}
\end{equation}
and has then shown that $\Omega$ can be defined in a manifold
$M^{2n-1}$ formed by the unit tangent vectors of $M^n$. In such a way
$\Omega$ can be expressed as the exterior derivative of a differential
$(n - 1)$-form in $M^{2n-1}$.
\begin{equation}
 \Omega=-\d\Pi.
\end{equation}
In this way the original integral of $\Omega$ over $M^{n}$ can be
performed over a submanifold $V^{n}$ of $M^{2n-1}$. This
$n$-dimensional submanifold is obtained as the image in $M^{2n-1}$ of
a continuous unit tangent vector field defined over $M^n$ with some
isolated singular points.  By applying Stokes' theorem one thus gets
\begin{equation}
 S^{\mathrm {volume}}_{\mathrm {GB}}=\int_{M^{n}}\Omega=\int_{V^{n}}\Omega=
 \int_{\partial V^{n}}\Pi.
\end{equation}
Since the boundary of $V^n$ corresponds exactly to the singular points
of the continuous unit tangent vector field defined over $M^n$, and
bearing in mind that the sum of the indices of a vector field is equal
to the Euler characteristic, one finds that the integral of $\Pi$ over
the boundary of $V^n$ is equal to the Euler number $\chi$.  For
manifolds with a boundary, this formula can be generalized \cite{EGH}:
\begin{eqnarray}
 S_{\GB}&=&S^{\mathrm {volume}}_{\mathrm {GB}}
        +S^{\mathrm {boundary}}_{\mathrm {GB}}
 \nonumber\\
 &=&\int_{M^{n}}\Omega-
 \int_{\partial M^{n}}\Pi=\int_{\partial V^{n}}\Pi-
 \int_{\partial M^{n}}\Pi .
\label{Sgb}
\end{eqnarray}
Thus, the Euler characteristic of a manifold $M^{n}$ vanishes when its
boundary coincides with that of the submanifold $V^{n}$ of $M^{2n-1}$.

The four-manifolds which we shall consider can have a boundary formed
by two disconnected hypersurfaces, say $\partial M^{n} =(r_{\mathrm
  in},r_{\mathrm out})$.  As far as $V^n$ is concerned, the above
quoted unit tangent vector field coincides (again for the cases
considered here) with the time-like Killing vector field
$\partial/\partial\tau$.  Hence the boundary will be the set of
fixed-points for such a vector field. The event horizon is always such
a set; then the boundaries of $V^n$ will be at $r_{\mathrm h}$ and
possibly at one of the actual boundaries of $M^{n}$ which, for sake of
simplicity, we shall assume to be at $r_{\mathrm out}$.

%---------------------------------------------------------------------------
\subsection[{Entropy for manifolds with a boundary}]
{Entropy for manifolds with a boundary}
\label{subsec:Eboun}
%---------------------------------------------------------------------------

In the framework of Euclidean quantum gravity and following the
definition of gravitational entropy adopted in ref.~\cite{KOP}, we
consider a thermodynamical system with conserved charges $C_{i}$ and
relative potentials $\mu_{i}$, and we then work in a grand-canonical
ensemble.  The grand-partition function $Z$, the free energy $W$ and
the entropy $S$ are:
\begin{equation}
 Z ={\rm{Tr}}\; \exp{[-(\beta \H -\mu_{i} C_{i})]} 
 =\exp{[-W]},
\end{equation}
\begin{equation}
 W =E-TS-\mu_{i} C_{i},
\end{equation}
\begin{equation}
 S=\beta(E-\mu_{i}C_{i})+\ln Z,
\label{gce}
\end{equation}
respectively.

We now evaluate separately the two terms appearing on the right-hand
side of Eq.~(\ref{gce}).  At the tree level of the semiclassical
expansion for a manifold $\cal{M}$ one has:
\begin{eqnarray}
 Z& \sim &\exp{[-I_{\mathrm E}]}\nonumber\\
 I_{\mathrm E}&=&{{1}\over{16 \pi}} 
 \int_{\cal M}[(-R+2\Lambda)+L_{\m})]+{{1}\over{8\pi}}
 \int_{\partial{\cal M}} \left[ K \right], 
\end{eqnarray}
where $I_{\mathrm E}$ is the on-shell Euclidean action and $[K]=K-K_{0}$ is
the difference between the extrinsic curvature of the manifold and
that of a reference background.  

In order to compute $Z$ and $I_{\mathrm E}$ it is important to
correctly take into account the boundaries of the Euclidean manifold
$M^4$ which in black hole spacetimes, after the compactification of
the imaginary time, has just a boundary at infinity.  So the logatithm
of the grand-partition function is given by
\begin{equation}
\ln Z=-\left. I_{\rm E}\right|^{\infty}=-\left. I_{\rm E}\right|_{\partial M}
\label{eq:lze}
\end{equation}
To obtain $\beta(E-\mu_{i}C_{i})$ one can consider
the probability of transition between two hypersurfaces at $\tau$
equals constant (where $\tau=i t$), say $\tau_{1}$ and $\tau_{2}$.  In
the presence of conserved charges one gets~\cite{KOP}:
\begin{equation}
\langle \tau_{1} | \tau_{2} \rangle =\exp{[-(\tau_{2}-\tau_{1})
(E-\mu_{i} C_{i})]}
\approx \exp{\left[-{I_{\mathrm E}}\right]_{\partial V}}.
\label{ener}
\end{equation}
The last equality in this equation is explained by the fact that
a hypersurface at $\tau={\mathrm{const}}$ has a
boundary corresponding to the sets of fixed points for 
the Killing vector $\partial/\partial \tau$. 
Hence its boundary coincides with that of $V^{n}$.
\begin{equation}
  \label{eq:becd}
  \beta(E-\mu_{i}C_{i})=\left. I_{\rm E}\right|^{\infty}_{r_{\rm h}}
                        =\left. I_{\rm E}\right|_{\partial V}
\end{equation}
Remarkably the above equations \ref{eq:lze} and \ref{eq:becd} lead to
the conclusion that it is just the boundary term of the Euclidean
action which contributes to the gravitational entropy: the bulk part
of the entropy always cancels also for metrics that are not
Ricci-flat.  The entropy then depends on boundary values of the
extrinsic curvature only. Thus, one obtains
\begin{equation}
 S=\beta(E-\mu_{i} C_{i}) +\ln{Z}
 ={{1}\over{8 \pi}} \left (\int_{\partial V}
 [K] - \int_{\partial M} [K ] \right). 
\label{entr}
\end{equation}
The analogy between Eq. (\ref{entr}) and Eq. 
(\ref{Sgb}) is self-evident.
For the boundaries 
of $V$ and $M$, the same considerations as at the end 
of section \ref{subsec:E-char} hold.

%--------------------------------------------------------------------------
\subsection[{Gravitational entropy and Euler characteristic for spherically 
symmetric metrics}]
{Gravitational entropy and Euler characteristic for spherically 
symmetric metrics}
\label{subsec:Echisf}
%--------------------------------------------------------------------------

We shall now prove, for a given class of Euclidean
spherically symmetric metrics, a general relation between
gravitational entropy and the Euler characteristic. 

%--------------------------------------%
\subsubsection{Euler characteristic}
%--------------------------------------%

In this section we compute the Euler characteristic for 
Euclidean spherically symmetric metrics of the form
\begin{equation}
 \d s^{2}=e^{2U(r)}\d t^{2}+e^{-2U(r)}\d r^{2}+R^{2}(r)\d^{2}\Omega.
\label{metr}
\end{equation}
The associated spin connections are
\begin{eqnarray}
 \omega^{01}&=&{{1}\over{2}} \left( e^{2U}\right)^{\prime} \d t,\qquad
 \omega^{21}=e^{U}R^{\prime}\d\theta, \nonumber\\
 \omega^{31}&=&e^{U}R^{\prime}\sin \theta \d \phi, \qquad
 \omega^{32}=\cos \theta \d \phi ,
\label{spc}
\end{eqnarray}
and the Gauss--Bonnet action takes the form \cite{GK} 
\begin{eqnarray}
 S^{\mathrm {volume}}_{\mathrm{GB}}&=&
 {{1}\over{32 \pi^{2}}} \int_{M} \varepsilon_{abcd}
 R^{ab} \wedge R^{cd}={{1}\over{4 \pi^{2}}} \int_{V}
 d(\omega^{01} \wedge R^{23})
 \nonumber\\
 &=&{{1}\over{4 \pi^{2}}}
 \int_{\partial V} \omega^{01} \wedge R^{23} . 
\label{gbvo}
\end{eqnarray}
The boundary term is \cite{GK,EGH}
\begin{eqnarray}
 S^{\mathrm {boundary}}_{\mathrm{GB}}&=&
 -{{1}\over{32 \pi^{2}}} \int_{\partial M}
 \varepsilon_{abcd} (2\theta^{ab} \wedge R^{cd}-{{4}\over{3}} 
 \theta^{ab} \wedge \theta^{a}_{e} \wedge \theta^{eb})
 \nonumber\\
 &=&-{{1}\over{4 \pi^{2}}} \int_{\partial M}
 \omega^{01} \wedge R^{23}.
\label{gbbo}
\end{eqnarray}
Combining Eqs. (\ref{gbvo}) and (\ref{gbbo}) one eventually gets
\begin{eqnarray}
\label{GBSS}
 S_{\GB}&=&S^{\mathrm{volume}}_{\mathrm{GB}}+
 S^{\mathrm{boundary}}_{\mathrm{GB}}\cr
 &&\cr
 &=&{{1}\over{4 \pi^{2}}} \left ( \int_{\partial V}-\int_{\partial M} \right )
 \omega^{01} \wedge R^{23}.
\end{eqnarray}
where for the metrics (\ref{metr}) 
\begin{eqnarray}
 R^{23}&=&\d\omega^{23}+\omega^{21} \wedge \omega^{13}=(1-
 e^{2U}(R^{\prime})^{2}) \d\Omega \nonumber\\
 \omega^{01} \wedge R^{23}&=&{{1}\over{2}} \left( e^{2U} \right)^{\prime}
 \left[1-e^{2U}(R^{\prime})^{2}\right] \d\Omega\;  \d t ,
\end{eqnarray}
and $d\Omega \equiv \sin\theta \d\theta \d\phi$ is the solid angle.
As already said, we perform our calculations on
Riemannian manifolds
with compactification of imaginary time, $0 \leq\tau
\leq\beta$, which is the 
generalization of the conical singularity removal condition for
the metrics under consideration.
It is easy to see that this corresponds to choosing
\footnote{
%%%%%%%%%%%%%%%%%%%%%%%%%%%%%%%%%%%%%%%%%%%%%%%%%%%%%%%%%%%%%%%%%%%%%%%%%  
  Note that condition (\ref{b}) gives an infinite range of time (no
  period) for extremal \bh\ metrics (i.e. $\left.
    (e^{2U})^{\prime}\right|_{r=r_{\mathrm h}} =0$). This leaves open
  the question of knowing whether the period of imaginary time remains
  unfixed or whether it has to be infinite, in correspondence with a
  zero temperature~\cite{HH96}.
%%%%%%%%%%%%%%%%%%%%%%%%%%%%%%%%%%%%%%%%%%%%%%%%%%%%%%%%%%%%%%%%%%%%%%%%%
}
\begin{equation}
 \beta=4 \pi \left[\left(e^{2U}\right)^{\prime}_{r=r_{\mathrm h}}\right]^{-1}.
\label{b}
\end{equation}
By expressing Eq. (\ref{GBSS}) as a function of the actual boundaries,
which are $\partial {V^4}=(r_{\mathrm h},r_{\out})$ and $\partial {M^4}=(r_{
  \inn},r_{\out})$, one gets
\beq S_{\GB}=2\left[1-\left(e^{U} R^{\prime}\right)^{2}
\right]_{r_{\mathrm h}} -\frac{ \left[ \left(e^{2U}\right)^{\prime}
  \right]_{r=r_{\mathrm h}} } { \left[ (e^{2U})^{\prime} (1-(e^{U}
    R^{\prime})^{2} )\right]_{r_{\inn}}}.
\label{Sgbg}
\eeq
We can also rewrite Eq. (\ref{Sgbg}) in a more suitable form for our
next purposes:
\begin{equation}
 \chi={{\beta}\over{2\pi}} \left[\left(2 U^{\prime} e^{2U}\right)
  \left(1-e^{2U}{R^{\prime}}^{2}\right)\right]^{r_{\mathrm h}}_{r_{\inn}}, 
\label{chib}
\end{equation}
expressing the Euler characteristic as a function of the inverse
temperature $\beta$.

%-----------------------------%
\subsubsection{Entropy}
%-----------------------------%

For the metrics (\ref{metr}) one can obviously use the general formula 
(\ref{entr}). It is well known~\cite{KOP} that one can write 
\begin{equation}
 [K ]=\int_{\partial M} [\omega^{\mu} n_{\mu}],
\label{Ko}
\end{equation}
where for the metrics (\ref{metr}) under investigation, $\omega^{\nu}$
and $n_{\nu}$ are
\begin{eqnarray}
 \omega^{\mu}&=&\left(0,-2e^{2U}\left(\partial_{r} U+
 2 \partial_{r} \ln R \right),
 -{{2 \cot \theta}\over {r^{2}}}, 0 \right), \nonumber \\
 n_{\mu}&=&\left( 0, {{1}\over{\sqrt{g^{11}}}},0,0 \right), 
\label{om}
\end{eqnarray}
and they lead to
\begin{equation}
 \omega^{\mu}n_{\mu}=\omega^{1}n_{1}=-2
 e^{U}\left (\partial_{r} U+2\partial_{r} \ln R \right).
\label{omc}
\end{equation}
By subtracting from Eq. (\ref{omc}) the corresponding flat metric term
\begin{equation}
\label{flat}
 \d s^{2}=\d t^{2}+\d r^{2}+r^{2} \d\Omega^{2},
\end{equation}
one obtains
\begin{eqnarray}
 {\omega^{\mu}_{0}}&=&\left(0,-{4\over r},-
 {{2\cot \theta} \over{r^{2}}},0 \right),\nonumber \\
 {n^{0}_{\mu}}&=&(0,1,0,0) ,
\label{os}
\end{eqnarray}
and so
\beq
 [\omega^{\mu}n_{\mu}]=\omega^{\mu}n_{\mu}-
 \omega_{0}^{\mu}n^{0}_{\mu}\nonumber
 =-2e^{U}(\partial_{r}U+2\partial_{r}\ln R)+{{4}\over{r}} .
 \label{of}
\eeq
Performing  the integration of Eq. (\ref{entr}) for a 
spherically symmetric metric, and 
writing explicitly the dependence on boundaries, one gets
\begin{equation}
 S=-\left. {{\beta R}\over{2}} \left[
 (U^{\prime}R+2R^{\prime})e^{U}-{{2R}\over{r}} \right] e^{U}
 \right|^{r_{\inn}}_{r_{\mathrm h}}.
\label{eg}
\end{equation}

%------------------------------------%
\subsubsection{Entropy and Topology}
%------------------------------------%

We can now prove that a relation between the gravitational entropy and
the Euler characteristic can be found for the general case under
consideration.  For the moment we shall consider only asymptotically
flat solutions with one bifurcate event horizon (more general
solutions will be discussed afterwards). So we can expect to have a
boundary at infinity, $r_{\out}=\infty$ while the inner boundary of
$M$ is generally missing since the horizon, after removal of the
conical singularity, becomes a regular point of the
manifold~\footnote{
%------------------------------------------------------------------------  
  The fact that in extremal geometries such a conical structure is not
  present --- as discussed in section~\ref{sec:EPI} --- can be seen as
  a consequence of the fact that the horizon in these cases is
  infinitely far away along spacelike directions (actually in
  Euclidean signature this is so along all of the possible
  directions): any free falling observer will take an infinite proper
  time to fall into the black hole. In this case the horizon can
    be then treated as an inner boundary of $M^4$~\cite{HHR94}; we
    shall soon see the consequences of this.}.
%--------------------------------------------------------------------
So for asymptotically flat solutions with no inner boundary one gets
\begin{eqnarray}
 A &=&4 \pi R^2 (r_{\mathrm h})\label{sphAr}\\
 S &=& \left.{{\beta R}\over{2}} \left[(U'R+2R')e^{U}-
 {{2R}\over{r}}\right]e^{U}\right|_{r=r_{\mathrm h}}\label{sgene}\\
 \chi &=&\left.{{\beta}\over{2 \pi}}(2U' e^{2U})(1-e^{2U}R'^{2})
 \right|_{r=r_{\mathrm h}}\label{chigene}
\end{eqnarray}
hence one can relate $S$ and $\chi$ by their common dependence on $\beta$
\begin{equation}
 S=\left.\frac{\pi \chi R}{ \left( 2U' e^{2U} \right) 
 \left(1-e^{2U}R'^{2}\right)}
 \left[(U'R+2R')e^{2U}-{{2R}\over{r}}e^{U}\right]\right|_{r=r_{\mathrm h}}.
\label{gen1}
\end{equation}
By definition, one has $\left. e^{2U}\right|_{r=r_{\mathrm h}}=0$, and Eq.
(\ref{gen1}) then gives
\begin{equation}
 S = \pi \chi R(r_{\mathrm h}) 
      \left. \left [ (e^{2U})' \right]^{-1} \right|_{r=r_{\mathrm h}}
   ={{\pi \chi R^{2}(r_{\mathrm h})}\over{2}}=
     {{\chi A}\over{8}} .
\label{SC}
\end{equation}
Some remarks on Eq.~(\ref{SC}) are in order.  

One can wonder why, from the fact that $\left.
  e^{2U}\right|_{r=r_{\mathrm h}}=0$, it does not automatically follow
(see equations (\ref{sgene}) and (\ref{chigene})) that $S=0$ and
$\chi=0$.  The key point here is that for $r\to r_{\mathrm h}$ one
should expect $U'(r)\to\infty$ in such a way as to cancel out with
$e^{2U}$. In fact we can generically write
$U=\half\ln\left[f(r)\right]$ with $f(r_{\mathrm h})=0$. In this case
one gets
\begin{equation}
  \label{eq:uprime}
  U'\cdot e^{2U}={1\over 2} {f'(r)\over f(r)}\cdot f(r)={1\over 2}f'(r)
\end{equation}
So in the above formulae $S=\chi=0$ only if $f'(r_{\mathrm h})=0$, but
this is exactly the condition that leads to zero temperature (infinite
period $\beta$) in equation (\ref{b}), which is the condition for
obtaining extremality.

Actually Eq.~(\ref{entr}) has been obtained in a grand-canonical
ensemble so this formula {\em a priori} is valid only for instantons
endowed with non-zero intrinsic temperature. This could not then be
extended to extremal solutions.  Nevertheless, in analogy
with~\cite{HHR94}, we can work with an arbitrary (un-fixed) $\beta$,
this is all that we need to arrive at equation (\ref{SC}).  We can
then conjecture that Eq. (\ref{SC}) is the general formula, which can
be applied to all of the known cases of instantons with horizons. The
lack of intrinsic thermodynamics is simply deducible from Eq.
(\ref{SC}) by considerations about the topology of the manifold.

In fact the crucial point is that the boundaries of $V$ and $M$ always
coincide for extremal black holes and this automatically leads to the
result $\chi=S=0$ for this class of solutions~\footnote{
%%%%%%%%%%%%%%%%%%%%%%%%%%%%%%%%%%%%%%%%%%%%%%%%%%%%%%%%%%%%%%%%%%%%%%%%%%%%%
  This can also be interpreted as a change in the global structure of
  the $r-\tau$ plane and hence in the topology of the manifold.
  The latter is in fact equal to $S^{1}\times R\times S^2$ for
  extremal solutions instead of the standard $S^2\times R^2$ for
  non-extremal black holes.
%%%%%%%%%%%%%%%%%%%%%%%%%%%%%%%%%%%%%%%%%%%%%%%%%%%%%%%%%%%%%%%%%%%%%%%%%%%%%%
  }. So not only does the topology of the $r-\tau$ plane appear to
determine the gravitational entropy, but we now clearly see how a change
in the topological nature of the horizon is at the basis of the
different thermodynamical behaviour of extremal black holes.  The
last conclusion is also in agreement with various other
semiclassical calculations (see~\cite{KL99} for a comprehensive review).

We shall now consider both black hole and cosmological solutions and
show how relation (\ref{SC}) is implemented in these cases.  As far as
the cosmological solutions are concerned, they are compact, and
therefore $\partial M=0$. Instead, the boundary of $V^{n}$ is only at
the horizon which now is also the maximal radius for the space; hence
the formulae for entropy and Euler characteristic are still
applicable, setting $r_{\out}= 0$ and taking into account a reversal
of sign due to the fact that in these instantons the horizon is an
external boundary of $V^n$. Given that these are just special cases of
the general procedure described before, we shall just give the basic
formulae and refer to the original paper~\cite{LP97} for detailed
derivation.

%--------------------------------------------------------------------%
\subsubsection{\Sch\ instanton: topology $R^2 \times S^2$, $\chi=2$}
%--------------------------------------------------------------------%

We first consider the \Sch\ \bh . In this case the elements of the
general metric (\ref{metr}) are:
\begin{eqnarray}
 e^{2U}&=&(1-2M/r),\nonumber\\
 R&=&r
\end{eqnarray}
and the relations (\ref{sgene} and \ref{chigene}) take the form
\begin{eqnarray}
 S&=&\frac{\beta r_{\mathrm h}}{4},\label{Ss}\\
 \chi &=& \beta r_{\mathrm h} {{1}\over {2\pi r^{2}_{\mathrm h}}}.
\label{Cs}
\end{eqnarray}
Now, combining Eqs. (\ref{Ss}) and (\ref{Cs}), one 
obtains
\begin{equation}
 S={{\pi}\over{2} } \chi r^{2}_{\mathrm h}={{\chi A}\over{8}}.
\end{equation}

%----------------------------------------------------------------------------%
\subsubsection{Dilaton U(1) black holes: topology $R^2 \times S^2$, $\chi=2$}
%----------------------------------------------------------------------------%

The dilaton $U(1)$ \bh\ solutions can be parametrized by a parameter
$a$ with range $0 \leq a \leq 1$ (where $a=0$ corresponds to the
Reissner--Nordstr\"om \bh). One has
\begin{eqnarray}
 e^{2U}&=&\left (1-{{r_{+}}\over{r}} \right )\left (1-{{r_{-}}
 \over{r}} \right )^{{{1-a^{2}}\over{1+a^{2}}}}\nonumber \\
 R&=&r \left (1-{{r_{-}}\over{r}} \right )^{{{a^{2}}\over{1+a^{2}}}}
 \nonumber \\
 M&=&{{r_{+}}\over{2}}+{{1-a^{2}}\over{1+a^{2}}}{{r_{-}}\over{2}}
 \nonumber \\
 Q^{2}&=&{{r_{+}r_{-}}\over{1+a^{2}}} 
 \nonumber \\
 r_{\mathrm h}&=&r_{+}.
\end{eqnarray}
In this case,
\begin{eqnarray}
 S&=&\frac{\beta r_{\mathrm h}}{4},\label{Sd}\\
 \chi &=& \beta r_{\mathrm h} {{1}\over {2\pi R^{2}_{\mathrm h}}}.
\label{Cd}
\end{eqnarray}
Taking into account that for this class of solutions $A=4\pi
R^{2}_{\mathrm h}$, it is again easy to see from Eqs. (\ref{Sd}) and
(\ref{Cd}), that
\begin{equation}
S={{\pi}\over{2} } \chi R^{2}_{\mathrm h}={{\chi A}\over{8}}.
\end{equation} 

%--------------------------------------------------------------%
\subsubsection{de Sitter instanton:  topology $S^4$, $\chi=2$}
%--------------------------------------------------------------%

In the de Sitter cosmological case, we can show that the relation in
Eq.~(\ref{SC}) is due to the boundary structure of the manifold
(horizons and ``real" boundaries) and not to the presence of a black
hole.  There is now only a cosmological horizon and no proper boundary
for $M$, and the topology of the de Sitter instanton is a four-sphere.
One has
\begin{eqnarray}
 e^{2U}&=&\left(1- {{\Lambda}\over{3}} r^{2}\right)
 \nonumber \\
 R&=&r \nonumber \\
 r_{\Lambda}&=&\sqrt{{3}\over{\Lambda}}
 \nonumber \\
 A &=& \frac{12\pi}{\Lambda}
 \nonumber \\
 \beta &=& 2\pi \sqrt{\frac{3}{\Lambda}}
\end{eqnarray}
where $r_{\Lambda}$ and $A$ are respectively the radius and area of
the cosmological horizon.  For this sort of compact manifold no
Minkowskian subtraction is needed; hence, by using Eq. (\ref{omc}) one
straightforwardly gets
\begin{equation}
S={{1}\over{8 \pi}} \int_{\partial V} K .
\label{cosmentr}
\end{equation} 
From Eq. (\ref{of}) it is easy to find 
\beq
\omega^{\mu}n_{\mu}=2\left({{r \Lambda}\over {3}}
{{1}\over{\left[1-\left({{r^{2}\Lambda}\over{3}}\right ) \right]^{1/2}}}
-{{2}\over{r}}
\left[1-\left({{r^{2}\Lambda}\over{3}}\right ) \right]^{1/2} 
\right)
\eeq
Hence, bearing in mind Eq. (\ref{Ko}), one obtains
\begin{equation}
S={{1}\over{16 \pi}} \int_{r_{\Lambda}} 
\omega^{\mu}n_{\mu}
e^{U}\; r^2 \sin{\theta}\; \d\theta\; \d\tau \; \d\phi = 
\frac{\beta^{2}}{4 \pi}.
\label{SdeS}
\end{equation}
By using Eq. (\ref{chib}) with $r_{\inn}=0$ 
and $r_{\mathrm h}=r_{\Lambda}$, the Euler 
characteristic is 
\begin{equation}
\chi=\frac{\beta^{2} \Lambda}{6 \pi^{2}}.
\label{CdS}
\end{equation}
Then, combining Eqs. (\ref{SdeS}) and (\ref{CdS}), it is easy to
check
that Eq. (\ref{SC}) also holds in the de Sitter case.

%---------------------------------------------------------------------%
\subsubsection{Nariai instanton: topology $S^2 \times S^2$, $\chi=4$}
%---------------------------------------------------------------------%

The Nariai instanton is the only non-singular solution of the
Euclidean vacuum Einstein equation for a given mass $M$ and
cosmological constant $\Lambda$.  It can be regarded as the limiting
case of the \Sch --de Sitter solution when one equates the surface
gravity of the \bh\ to that of the cosmological horizon in order to
remove all conical singularities.  This might seem meaningless since,
in \Sch --de Sitter coordinates, the Euclidean section appears to
shrink to zero (the \bh\ and cosmological horizons coincide). However
this is due just to an inappropriate choice of coordinates and by
making an appropriate change of coordinates \cite{GP,HB}, the volume
of the Euclidean section is well defined and non vanishing.  In this
coordinate system, one still deals with a spherically symmetric
metric, and the vierbein forms are
\begin{eqnarray}
 e^{0}&=& {1 \over \sqrt \Lambda}\; \sin\xi  \d\psi, \qquad
 e^{1}={1 \over \sqrt \Lambda}\;\d\xi,\\
 e^{2}&=&{1 \over \sqrt \Lambda}\;\d\theta, \qquad
 e^{3}={1 \over \sqrt \Lambda} \;\sin\theta \d\phi.
\end{eqnarray}
Also, one has
\begin{eqnarray}
 R &=& \Lambda^{-1/2} ,
 \nonumber\\
 A &=& {4\pi \over \Lambda},
 \nonumber\\
 \beta &=& \frac{2 \pi}{\sqrt {\Lambda}}.
\end{eqnarray}
The ranges of integration are $0\leq\psi\leq\beta\sqrt{\Lambda}$,
$0\leq\xi\leq\pi$, $0\leq\theta\leq2\pi$, $0\leq\phi\leq2\pi$.  The
extremes of $\xi$ correspond to the cosmological horizon and to the
\bh\ horizon \cite{HB}.  It is worth noting that the period of the
imaginary time, $\psi$, is $\beta\sqrt{\Lambda}$, instead of the usual
$\beta$.  This is due to the normalization of the time-like Killing
vector which one is forced to choose in this spacetime\footnote{
%%%%%%%%%%%%%%%%%%%%%%%%%%%%%%%%%%%%%%%%%%%%%%%%%%%%%%%%%%%%%%%%%%%%%%%%%  
  For a wider discussion of this point, see the Appendix of
  Ref.~\cite{HB}.
%%%%%%%%%%%%%%%%%%%%%%%%%%%%%%%%%%%%%%%%%%%%%%%%%%%%%%%%%%%%%%%%%%%%%%%%%
}.  

The form of the Nariai metric does not enable us to apply Eq.
(\ref{chib}) and so we compute the Euler characteristic from the
beginning. We obtain
\begin{equation}
 \label{chinar}
  S_{\mathrm GB}={1\over 4\pi^2} \int_{0}^{\pi} \sin\xi \d\xi 
  \int_{0}^{\beta\sqrt{\Lambda}}\d\psi \int_{0}^{\pi} \sin\theta \d\theta 
   \int_{0}^{2\pi}\d\phi 
   =\frac{2\beta\sqrt{\Lambda}}{\pi} .
 \nonumber
\end{equation}
By substituting for $\beta$, one can check that Eq. (\ref{chinar})
gives the correct result.  In fact, the Nariai instanton has topology
$S^{2} \times S^{2}$; hence its Euler number, bearing in mind the
product formula, is $\chi=2\times2=4$.

The entropy can be easily calculated from Eq. (\ref{cosmentr}).
In this case the extrinsic curvature is given by
\begin{equation}
 K=-\sqrt{\Lambda}\;\frac{\cos \xi}{\sin\xi},
\end{equation} 
and one obtains
\beq
 S=-\frac{1}{8\pi}\int_{0}^{2\pi}\d\phi\int_{0}^{\pi}\sin\theta \d\theta 
 \int_{0}^{\beta\sqrt{\Lambda}} \left[ 
 \frac{\sqrt{\Lambda} \cos \xi}{\Lambda^{3/2}}\right]_{0}^{\pi} 
 \d\psi
 =\frac{\beta}{\sqrt{\Lambda}}. 
\label{snar}
\eeq
It is now easy to check that the combination of Eqs. (\ref{chinar})
and (\ref{snar}) gives Eq. (\ref{SC}).  Remarkably, this implies that
Eq. (\ref{SC}) cannot be cast in the form
\begin{equation}
S=\left(\frac{\chi}{2}\right)^{\alpha}\frac{A}{4},
\label{pro}
\end{equation}
where $\alpha$ could in principle be any positive constant. Since Eq.
(\ref{SC}) holds also for the Nariai instanton, for which $\chi/2\neq
1$, then $\alpha$ must be fixed to 1. 

As a further generalization, we shall now see how these results can be
extended to the case of axisymmetric metrics.

%%%%%%%%%%%%%%%%%%%%%%%%%%%%%%%%%%%%%%%%%%%%%%%%%%%%%%%%%%%%%%%%%%%%%%%%
\subsection[{Kerr metric}]{Kerr metric}
\label{subsec:Kerr}
%%%%%%%%%%%%%%%%%%%%%%%%%%%%%%%%%%%%%%%%%%%%%%%%%%%%%%%%%%%%%%%%%%%%%%%%

The Kerr solution describes the stationary axisymmetric
asymptotically flat gravitational field outside a rotating \bh\ with
mass $M$ and angular momentum $J$. The Kerr \bh\ can also be viewed as
the final state of a collapsing star, uniquely determined by its mass
and rate of rotation.  Moreover, its thermodynamical behaviour is very
different from \Sch\ or Reissner--N\"ordstrom \bh s, because of its
much more complicated causal structure~\footnote{
%%%%%%%%%%%%%%%%%%%%%%%%%%%%%%%%%%%%%%%%%%%%%%%%%%%%%%%%%%%%%%%%%%%%%%%  
  For instance, Wald pointed out that in a Kerr \bh\ it is not
  possible to mimic the Unruh--Rindler case to explain its thermal
  behaviour~\cite{wald}.
%%%%%%%%%%%%%%%%%%%%%%%%%%%%%%%%%%%%%%%%%%%%%%%%%%%%%%%%%%%%%%%%%%%%%%
  }. Hence studying it is of great interest for understanding the
properties of astrophysical objects, as well as for checking any
conjecture about the thermodynamical properties of \bh s.

In terms of Boyer--Lindquist coordinates, the Euclidean Kerr metric 
is \cite{Oneill}
\begin{equation}
\label{kermet}
 \d s^{2} = {\Delta \over \rho^{2}}
 \left[\d t-a{\sin^{2}\!\theta}\d \varphi\right]^{2}
 + {\rho^{2} \over \Delta}\d r^{2}
 +\rho^{2}\d \theta^{2} 
 +{\sin^{2}\theta \over \rho^{2}}\Bigr[\left(
 r^{2}+a^{2}\right)\d \varphi -a \d t\Bigr]^{2},
\end{equation}
where
\begin{eqnarray}
 \rho &= &r^{2} + a^{2}\cos^{2}\theta, \nonumber\\ 
 \Delta &=& r^{2}-2Mr +a^{2}.
\end{eqnarray}
Here $a$ is the angular momentum per unit mass as measured from
infinity (which vanishes in the \Sch\ limit) and $\Delta$ is the Kerr
horizon function. The roots of the horizon function $\Delta$
correspond to two null-like surfaces at
\begin{equation}
 r_{\pm}=M \pm \sqrt{M^2-a^2},
\end{equation}
where $r_{+}$ is the Kerr \bh\ event horizon and $r_{-}$ is the Cauchy
horizon around the ring singularity at $\rho=0$.  The area and the
\bh\ angular velocity are respectively
\begin{eqnarray}
 A &=& 4\pi(r_{+}^{2}+a^{2}),\\
\label{AO}
\Omega &=& \frac{a}{(r^{2}_{+}+a^{2})},
\label{AO2}
\end{eqnarray}
In this case, the lack of spherical symmetry forces us to use the
general form of the Gauss--Bonnet integral Eq.\ (\ref{egb}).  From the
Kerr metric one can calculate the spin connections $\omega_{ab}$ and
the Ricci tensor (\ref{Rab}). Using the nilpotency of the exterior
derivative operator $\d$, the Gauss--Bonnet action in Eq.\ (\ref{Sgb})
takes the form
\begin{equation}
\label{kerraction}
 S_{\mathrm{GB}}=
 -{{1}\over{4 \pi^{2}}}\int{\left(\omega^{01} \wedge \d\omega^{23}
 +\omega^{02}\wedge \omega^{21}\wedge\omega^{23}
 +\omega^{03}\wedge \omega^{31}\wedge\omega^{23}
 +\omega^{02}\wedge \d\omega^{31} \right)_{r_{\mathrm h}}}, 
\end{equation}
where $\d\omega^{31}$ can be expressed in terms of a suitable
combination (wedge product) of the type $e^{i}\wedge e^{j}$, and
$r_{\mathrm h}$ is the radius of the Kerr horizon (i.e. the larger
positive root of $\Delta=0$).

At this stage some remarks are in order.  In the Euclidean
path-integral approach, the Kerr solution is an instanton (i.e. a
non-singular solution of the Euclidean action) only after the
identification of the points $(\tau,r,\theta,\varphi)$ and
$(\tau+2\pi\kappa_{1}^{-1},r,\theta,\varphi+2\pi
\kappa_{1}^{-1}\kappa_{2})$ \cite{GH77}, where $\kappa_{1}=\kappa$ is
the surface gravity of the \bh\ and $\kappa_{2}=\pm \Omega$.  With
this identification, the Euclidean section has topology $R^2\times
S^2$ and $\chi=2$. The condition of a periodic isometry group implies
$\kappa_{2}/\kappa_{1}=q$ \cite{GHI}, where $q \in Q$ is a rational
number. By using this relation, it is easy to see that the periods
are:
\begin{eqnarray}
 \beta_{\tau} &=& 2\pi \kappa_{1} = 4\pi 
 \frac{Mr_{ h}}{\sqrt{(M^2-a^2)}},\nonumber\\
 \beta_{\varphi} &= & 2\pi \frac{\kappa_{2} }{\kappa_{1}}= 2\pi q,
\label{betak}
\end{eqnarray}
If one were to set $q \neq 1$, Eq. (\ref{entr}) for the \bh\ entropy
would acquire a factor $q$, but this spurious factor would be absorbed
into the change of the period of $\varphi$ implying a redefinition of
the \bh\ area (\ref{AO}), which would become $A=4\pi q(r_{\rm
  h}^{2}+a^2)$.  Therefore one still expects $S=A/4$, and the fixing
of $q=1$ will not bring about a loss of generality.  Moreover in this
way the area will be the ``physical'' one, as written in Eq.
(\ref{AO}). Hence the Euler number is
\begin{eqnarray}
 \chi &=&
 \frac{M{r_{\rm h}}({r_{\rm h}}-M)}{4\pi^{2}}
 {\int_{0}^{\beta}}\d\tau
 {\int_{0}^{2\pi}}
 {\d\varphi} {\int_{0}^{\pi}}
  {{\left(r_{\rm h}^{2}-3a^{4}\cos^{4}\!\theta\right)} 
  \over {\left(r_{\mathrm h}^{2}+a^{2}\cos^{2}\!\theta\right)^{3}}}
  \sin\theta \d\theta \nonumber\\
  &=&{2 \over \pi} \beta (r_{\mathrm h}-M) {{Mr_{\mathrm h}} 
  \over  {\left(r_{\mathrm h}^{2}+a^{2}\right)^{2}}}.
\label{eq7}
\end{eqnarray}
Bearing in mind Eq. (\ref{betak}) 
and that $(r_{\mathrm h}^{2}+a^{2})= 2Mr_{\mathrm h}$, one eventually gets  
\begin{equation}
 \chi =8 {M^{2}r_{\mathrm h}^{2}\over 
 \left(r_{\mathrm h}^{2}+a^{2}\right)^{2}}= 2.
\end{equation}

As far as the entropy is concerned, we here follow the procedure outlined 
in Sec. \ref{subsec:asympt} and \ref{subsec:cosmcc}. From Eq. (\ref{entr}), 
writing $\omega^{\mu}$ as 
\begin{equation}
 \omega^{\mu}= -{2 \over \sqrt{g}}
 \left({\partial {\sqrt g}\over \partial x^{\nu}}\right)
 g^{\mu\nu} -{\partial g_{\nu\mu}\over \partial x^{\nu}},
\end{equation} 
and bearing in mind that the Kerr determinant is
\begin{equation}
 {\sqrt g} = \rho^{2} \sin\theta ,
\end{equation}
one finds
\begin{eqnarray}
 \omega^{\mu}&=& \left(0,-2{r\Delta\over \rho^{4}},
 -{2(r-M)\over \rho^{2}}, -2{\cot\theta\over \rho^{2}},0\right),
 \nonumber\\
 n_{\mu} &=&\left(0,{\rho \over {\sqrt\Delta}},0,0\right).
\end{eqnarray}
By subtracting the flat Minkowskian term $\omega^\mu$ (see Eq.~(\ref{os}))
one easily obtains
\begin{equation}
 \left[\omega^{\mu}n_{\mu}\right]
 =-{2 \over \rho {\sqrt\Delta}} \left(
 {r\Delta \over \rho^{2}} + r-M\right) 
 + {4\over r}.
\end{equation}
One can then evaluate the Kerr \bh\ entropy:
\begin{eqnarray} 
 S&=&-{1\over 16\pi} \int_{0}^{\beta} \d\tau
 \int_{0}^{\pi}\d\varphi\int_{0}^{2\pi}\d\theta
 \rho{\sqrt \Delta}\sin\theta \:\:\cdot\cr
 &&\left[-{2 \over \rho {\sqrt\Delta}} \left(
 {r\Delta \over \rho^{2}} + r-M\right)
 + {4\over r}\right]_{r_{\mathrm h}}= {\beta \over 2} (r_{\mathrm h} -M),
\label{equ19}
\end{eqnarray}
Thus, combining Eqs. (\ref{eq7}) and (\ref{equ19}), one has
\begin{equation}
 S = {\pi \over 4}
     {\left(r_{\mathrm h}^{2}+a^{2}\right)^{2} \over Mr_{\mathrm h}}\chi 
   = {1\over 2}\pi \left(r_{\mathrm h}^{2}+a^{2}\right) \chi=
 {A \over 8}\chi.
\end{equation}  

%-------------------------------------------------------------------%
\subsection[{Discussion}]{Discussion}
%-------------------------------------------------------------------%

This investigation into the link between topology and gravitational
entropy leads us to two interesting conclusions. The first one is
related to the fact that we have found a very wide class of instantons
for which the same relation between entropy, Euler characteristic and
horizon area holds. This relation is telling us that intrinsic
thermodynamical behaviour is common to most of the spacetimes with non
trivial topological structure in the $\tau-r$ plane.

The fact that the Euler characteristic is the only non-trivial
topological invariant for the instantons which we have considered
suggests that the formula which we have found can eventually be
generalized for solutions with other non null invariants. It would be
interesting to consider spacetimes characterized by horizons with non
spherical topology. In \GR\ event horizons are guaranteed to have an
$S^2$ topology~\cite{Haw72,CW94} and so this proposal would require
the study of solutions of more general theories of gravitation. We
leave this as a possible future research field.

The second interesting issue is that this alternative derivation and
generalization of the Bekenstein--Hawking entropy seems to imply that
the entropy of extremal solutions is actually zero.  So semiclassical
extremal black holes would apparently satisfy Planck's postulate
($S=0$). This is actually not exact. In fact, we saw that the Nernst
formulation of the third law is violated in black hole
thermodynamics. Indeed, even if extremal black holes had a vanishing
entropy, zero is not the value to which the entropy of nearly extremal
black holes tends in the $\kappa\to 0$ limit. 

This lack of a good limit hints at the existence of a discontinuity in
thermodynamical behaviour between non-extremal black holes and
extremal ones~\cite{PSSTW91,GM97,KL99,BM00}. In some sense this seems
to imply that one object is not the zero temperature limit of the
other\footnote{
%%%%%%%%%%%%%%%%%%%%%%%%%%%%%%%%%%%%%%%%%%%%%%%%%%%%%%%%%%%%%%%%%%%%%%%%%%%%%%%
  This is apparently in conflict with string theory results. As we said
  in section \ref{subsec:symm}, this could be explained via some
  non-trivial issue hidden in the procedure for obtaining the
  semiclassical limit. Unfortunately such an issue is still far from
  being understood and we shall not treat it further here.
%%%%%%%%%%%%%%%%%%%%%%%%%%%%%%%%%%%%%%%%%%%%%%%%%%%%%%%%%%%%%%%%%%%%%%%%%%%%%%%
  }. Our calculation and those cited above have so far dealt with
eternal black holes. It is thus unclear whether the thermodynamic
discontinuity just mentioned applies to the case of black holes formed
by collapse. 

For this reason we shall now examine particle production by an
``incipient'' Reissner--Nordstr\"om (RN) black hole: A spherically
symmetric collapsing charged body whose exterior metric is
RN~\footnote{
%%%%%%%%%%%%%%%%%%%%%%%%%%%%%%%%%%%%%%%%%%%%%%%%%%%%%%%%%%%%%%%%%%%%%%%%%%%
  In doing this we shall not address the issue of actually
  constructing solutions of the Einstein equations that describe the
  collapse of charged configurations, because some simple solutions of
  this kind can already be found in the
  literature~\cite{Bou73,FH79,P83}. For the moment we assume that a model can
  be found in which collapse leads to a black hole with $Q^2=M^2$.
%%%%%%%%%%%%%%%%%%%%%%%%%%%%%%%%%%%%%%%%%%%%%%%%%%%%%%%%%%%%%%%%%%%%%%%%%%%
  }. This investigation will lead us to the conclusion that incipient
extremal black holes are not thermal objects and that the notion of
zero temperature is ill-defined for them. We shall see that, as a
consequence of this result, one may need to go to a full semiclassical
theory of gravity, including backreaction, in order to make sense of
the third law of black hole thermodynamics.

%%%%%%%%%%%%%%%%% EXTREMAL %%%%%%%%%%%%%%%%%%%%%%%%%%%%%%%%%%%%%%%%
\section[{Incipient extremal black holes}]
{Incipient Extremal Black Holes}
\label{sec:exbh}
%%%%%%%%%%%%%%%%%%%%%%%%%%%%%%%%%%%%%%%%%%%%%%%%%%%%%%%%%%%%%%%%%%%

We shall start by approaching the problem in the standard fashion,
that is by modelling the collapse by a mirror moving in
two-dimensional Minkowski spacetime \cite{FD77}. The spectra resulting
from the mirror's worldline will then be the same as that of the black
hole, up to gray-body factors due to the nontrivial metric
coefficients of RN spacetime and to the different dimensionality.

In order to pursue this approach the first problem which one has to
face is to determine the appropriate worldline to use for the mirror.
This requires finding a set of coordinates that are regular on the
event horizon, that is Kruskal-like coordinates for an extremal
Reissner--Nordstr\"om black hole.

%-----------------------------------------------------------------------------
\subsection[{Kruskal-like coordinates for the extremal RN solution}]
{Kruskal-like coordinates for the extremal RN solution}
\label{subsec:Kr}
%-----------------------------------------------------------------------------

Several textbooks in \GR\ (see, for example, Refs.\ 
\cite{MTW,HawEll73}) imply that Carter \cite{Carter66} found the
maximal analytical extension of RN spacetime for $Q^2=M^2$. In fact he
made a very ingenious qualitative analysis without actually providing
an analog of the Kruskal coordinates for the extremal case.
Nevertheless, for our analysis it is essential to have such a
coordinate transformation. For this reason we are going to retrace the
steps leading to the maximal analytic extension of RN, paying close
attention to the difference between the non extremal and extremal
situations.

The first step in the procedure is to define the so-called
``tortoise'' coordinate, which is then used to construct the Kruskal
coordinates. We start with the usual form of the RN geometry,
\begin{equation}
 {\rm d}s^2 = -\left(1- \frac{2M}{r} + \frac{Q^2}{r^2}\right){\rm d}t^2
  + \left(1- \frac{2M}{r} + \frac{Q^2}{r^2}\right)^{-1}{\rm d}r^2+r^2\, 
  {\rm d}\Omega^2\;,
\end{equation}
where ${\rm d}\Omega^2$ is the metric on the unit sphere.
The tortoise coordinate $r_*(Q,M)$ is given by
\begin{equation}
 r_*(Q,M) = \int
 \frac{{\rm d}r}{\left(1-2M/r+Q^2/r^2\right)}\;.
\label{tort(QM)}
\end{equation}
Carrying out the integration yields, for the non extremal
case,
\begin{equation}
 r_*(Q,M) = r + \frac{1}{2\sqrt{M^2 - Q^2}}
 \left(r_{+}^2\ln(r-r_{+})-r_{-}^2 \ln(r - r_{-})\right) + \mbox{const},
\label{inttort(QM)}
\end{equation}
where as usual $r_{\pm}= M \pm \sqrt{M^2-Q^2}$.

Now, if we set $Q^2 = M^2$ in Eq.\ (\ref{tort(QM)}) {\it before\/}
integrating, we find the ``extremal'' $r_*$:
\begin{equation}
 r_*(M,M) = r + 2M\left(\ln(r-M) - \frac{M}{2(r-M)}\right)+ \mbox{const}.
\label{inttort(MM)}
\end{equation}
Note that the coordinate $r_*(M,M)$ diverges only at $r = M$, but
setting $Q^2= M^2$ in $r_*(Q,M)$ appears to yield the indeterminate
form $0/0$. However, if we let $Q^2 = M^2(1-\epsilon^2)$, with
$\epsilon \ll 1$, and work to first order in $\epsilon$, it is
straightforward to show that Eq.\ (\ref{inttort(QM)}) does reduce to
Eq.\ (\ref{inttort(MM)}).  Therefore $r_*$ is continuous even for the
extremal case.

Unfortunately, the Kruskal transformation itself breaks down at that
point. The Kruskal transformation is
\begin{equation}
 \left.
  \begin{array}{lll} 
   {\cal U} = -{\rm e}^{-\kappa u}
   & \Leftrightarrow & {\displaystyle u=
   -{1\over\kappa}\ln({\cal -U})}\\
   {\cal V} = {\rm e}^{\kappa v} & \Leftrightarrow &
   {\displaystyle v = {1\over\kappa}\ln {\cal V}}
  \end{array} 
 \right\}\;,
\label{Kruskal}
\end{equation}
where
\begin{equation}
 \left.
  \begin{array}{l}
   u = t - r_*\\ v = t+r_*
  \end{array}
 \right\} 
\label{EF}
\end{equation}
are the retarded and advanced Eddington--Finkelstein coordinates,
respectively, and $\kappa$ is the surface gravity. The latter is
defined as
\begin{equation}
 \kappa = \lim_{r\to r_+}{1\over 2}{{\rm d} \over {\rm d}r}
 \left(1-{2M\over r} + {Q^2\over r^2}\right)=
 {\sqrt{M^2-Q^2}\over r_+^2}\;,
\end{equation}
and vanishes for $Q^2 = M^2$. Therefore the Kruskal coordinates $\cal
U$ and $\cal V$ become constant for any value of $u$ and $v$ and so
the transformation (\ref{Kruskal}) becomes ill-defined at that point.

We are nonetheless able to remedy this situation. Note that the
Eddington-Finkelstein coordinates are constructed by adding or
subtracting $r_*$ to $t$, as in equations (\ref{EF}) above. Now, for the
extremal case, $r_*$ is given by Eq.\ (\ref{inttort(MM)}), which has
the extra pole $M^2/(r-M)$ with respect to the strictly logarithmic
dependence of the Schwarzschild and non extremal RN cases (compare Eq.
(\ref{inttort(QM)})). The simplest thing to do is define a function
\begin{equation}
\psi(\xi)= 4M\left(\ln\xi - \frac{M}{2\xi}\right)
\label{psi}
\end{equation}
and guess that a suitable generalization of the Kruskal transformation
is
\begin{equation}
 \left.
  \begin{array}{l}
   u = -\psi(-{\cal U})\\ v = \psi({\cal V})
  \end{array}
 \right\}\;.
\label{Gen}
\end{equation}
Note that $\psi'(\xi)=4M/\xi+2M^2/\xi^2
> 0$, always, and so $\psi$ is monotonic; therefore
(\ref{Gen}) is a well-defined coordinate transformation.
Note also that
\begin{equation}
r_*(M,M) = r+\frac{1}{2}\psi (r-M) ,
\end{equation}
which means that near to the horizon\footnote{Hereafter, for
two functions $f$ and $g$, we use the notation $f\sim g$ to
mean $\lim f/g=1$ in some asymptotic regime.}
\begin{equation}
r_*(M,M) \sim {1\over 2}\,\psi (r-M)\;. \label{r*(r-M)}
\end{equation}

We can give our choice of $\psi$ added motivation by noting
that near to the horizon Eq.\ (\ref{inttort(QM)}) gives
\begin{equation}
r_*(Q,M)\sim {1\over 2\kappa}\,\ln(r-r_+)\;.
\label{boia}
\end{equation}
Thus we see that the function $\kappa^{-1}\ln(\cdots)$ which appears
in the transformation (\ref{Kruskal}) from Kruskal to the
Eddington--Finkelstein coordinates, is just twice the one which gives
a singular contribution to $r_\ast(Q,M)$ at $r=r_+$. Our extension
(\ref{Gen}) is therefore analogous to the Kruskal transformation
(\ref{Kruskal}): we choose $\psi$ as the part of $r_\ast$ that is
singular at $r=r_+$, a procedure that should work in other, similar
situations.

For (\ref{Gen}) to be a good coordinate extension, the new coordinates
$\cal U$ and $\cal V$ must be regular on the event horizon, ${\cal
  H}^{+}$. This will be the case if after the coordinate
transformation the metric is singular only at $r = 0$. Fortunately it
is possible to explicitly show~\cite{LRS99} that this is verified for
(\ref{Gen}) and consequently, ${\cal U}$ and ${\cal V}$ are good
Kruskal-like coordinates.

Notice that the coordinates $u$ and $v$ defined by the transformation
(\ref{Kruskal}) do not tend to those given by (\ref{Gen}) as $Q^2\to
M^2$. This is related to the fact that the maximal analytic extensions
of RN spacetime are qualitatively different in the two cases
\cite{HawEll73}, and is further evidence of the discontinuous
behaviour mentioned before.

%-----------------------------------------------------------------------------
\subsection[{Asymptotic worldlines}]
{Asymptotic worldlines}
\label{subsec:asympt}
%-----------------------------------------------------------------------------
 
With the result of the previous section in hand we are now able to
construct late-time asymptotic solutions for the incipient extremal
black hole. Our goal is to find an equation for the centre of the
collapsing star (in the coordinates $u$ and $v$) that is valid at late
times.  Equation (\ref{EF}) gives $u$ and $v$ outside the collapsing
star, but the centre of the star is, of course, in the interior. We
must therefore extend $u$ and $v$ into the interior. Since $u$ and $v$
are null coordinates, representing outgoing and ingoing light rays
respectively, the extension can be accomplished almost trivially by
associating any event in the interior of the star with the $u$ and $v$
values of the light rays that intersect at this event.

\begin{figure}[hbt]
  \vbox{\hfil \scalebox{0.50}{{\includegraphics{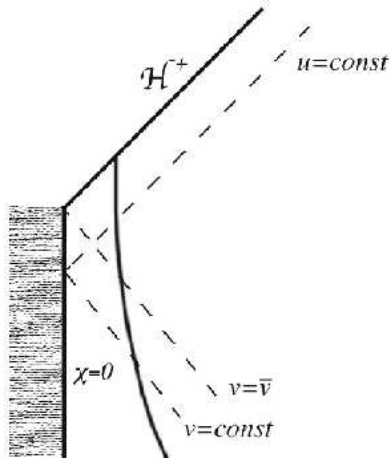}}} \hfil }
  \bigskip
\caption[Gravitational collapse in null coordinates]{
%----------------------------------------------------------------------------
  A representation of gravitational collapse in null coordinates. The
  portion of spacetime beyond the event horizon ${\cal H}^+$ is not
  shown.
%---------------------------------------------------------------------------
} \label{star}
\end{figure}

The most general form of the metric for the interior of a spherically
symmetric star can be written as
\begin{equation}
{\rm d}s^2 = \gamma(\tau,\chi)^2(-{\rm d}\tau^2 + {\rm
d}\chi^2) + \rho(\tau,\chi)^2\,{\rm d}\Omega^2\;,
\end{equation}
where $\gamma$ and $\rho$ are functions that can be chosen to be
regular on the horizon. From the coordinates $\tau$ and $\chi$ we can
construct interior null coordinates $U = \tau - \chi$ and $V = \tau +
\chi$, which will also be regular on the horizon. The centre of the
star can be taken at $\chi = 0$, in which case $V = U$ and ${\rm d}V =
{\rm d}U$ there (see Fig.\ \ref{star}).
%
%==============================================================================
%=============================================================================
%
Because the Kruskal coordinates $\cal U$ and $\cal V$ are regular
everywhere, they can be matched to $U$ and $V$. In particular, if two
nearby outgoing rays differ by ${\rm d}U$ inside the star, then they
will also differ outside by ${\rm d}U=\beta({\cal U}) {\rm d}{\cal
  U}$, with $\beta$ being a regular function. By the same token, since
$V$ and $v$ are regular everywhere, we have ${\rm d}V = \zeta(v) {\rm
  d}v$, where $\zeta$ is another regular function. In fact, if we
consider the last ray $v=\bar v$ that passes through the centre of the
star before the formation of the horizon, then to first order ${\rm
  d}V = \zeta (\bar v){\rm d}v$, where $\zeta (\bar v)$ is now
constant.

We can write near the horizon
\begin{equation}
 {\rm d}U = \beta(0)\frac{{\rm d}{\cal U}}{{\rm d}u}{\rm d}u\;.
\end{equation}
Since for the centre of the star ${\rm d}U = {\rm d}V =
\zeta(\bar v){\rm d}v$, this immediately integrates to
\begin{equation}
 \zeta(\bar v)(v - {\bar v}) = \beta (0){\cal U}(u) =
         -\beta (0) \psi^{-1}\left(-u\right)
        \sim -2 \beta (0)\frac{M^2}{u}\;.
\end{equation}
The last approximation follows from Eq.\ (\ref{psi}) where $\xi\sim
\psi^{-1}(-2M^2/\xi)$ near the horizon.

Thus the late-time worldline for the centre of the star is, finally,
represented by the equation~\footnote{
%%%%%%%%%%%%%%%%%%%%%%%%%%%%%%%%%%%%%%%%%%%%%%%%%%%%%%%%%%%%%%%%%%%%%%%%%%%  
  This result was also obtained by Vanzo \cite{vanzo97} for a
  collapsing extremal thin shell, but without considering a coordinate
  extension. Our method is completely general and shows that Eq.\ 
  (\ref{traj}) follows only from the kinematics of collapse and the
  fact that the external geometry is the extremal RN one.
%%%%%%%%%%%%%%%%%%%%%%%%%%%%%%%%%%%%%%%%%%%%%%%%%%%%%%%%%%%%%%%%%%%%%%%%%%%
}
\begin{equation}
 v \sim {\bar v} - \frac{A}{u}\;,\quad u\to +\infty\;,
\label{traj}
\end{equation}
where $A = 2\beta (0)M^2/\zeta(\bar v)$ is a positive
constant that depends on the details of the internal metric
and consequently on the dynamics of collapse.

We first note that the worldline (\ref{traj}) differs from the one
resulting from the collapse of a nonextremal object, which would be of
the form (see for example\ \cite{FD77,Birrell-Davies,saa})
\begin{equation}
 v\sim {\bar v}-B{\rm e}^{-\kappa u}\;,\quad u\to +\infty\;.
\label{nextraj}\end{equation}
One immediately wonders, then, if our result can be recovered in the
case of nonextremal black holes by simply going to a higher order
approximation for the asymptotic worldline of the centre of the
collapsing star. It is easy to see that this is not the case. Recall
that in the Kruskal coordinates ${\cal U}$ and ${\cal V}$, the horizon
is located at ${\cal U} = 0$. Suppose that the worldline of the centre of
the
star crosses the horizon at some ${\cal V} = \overline{\cal V}$. Let
us expand ${\cal V({\cal U})}$ in a Taylor series around ${\cal U} =
0$ such that ${\cal V} = \overline{\cal V} +\alpha_1\,{\cal
  U}+\alpha_2\,{\cal U}^2$.  The term $\alpha_1\,{\cal U}\propto {\rm
  e}^{-\kappa u}$ is the usual one found for the thermal case and
$\alpha_2\,{\cal U}^2$ is the correction. However, ${\cal U}^2 \propto
{\rm e}^{-2\kappa u}$ and so this term is also a constant for extremal
incipient black holes. In fact the corrections are constant to arbitrary
order. The extremal worldline in no sense, therefore, represents a
limit of the nonextremal case but implies a real discontinuity in the
asymptotic behaviour of the collapsing object.

Equations (\ref{traj}) and (\ref{nextraj}) contain the constants $A$
and $B$, which are determined by the dynamics of collapse. In the
non extremal case, it is known that no measurement performed at late
times can be used to infer the value of $B$, thus enforcing the
no-hair theorems. In particular, the spectrum of Hawking radiation
depends only on the surface gravity $\kappa$. It is natural to ask
whether a similar statement holds true also for extremal black holes.
This point will be analyzed in the following sections.

%-----------------------------------------------------------------------------
\subsection[{Bogoliubov coefficients}]
{Bogoliubov coefficients}
\label{subsec:exbogo}
%-----------------------------------------------------------------------------

Let us now consider a test quantum field in the spacetime of an
incipient extremal RN black hole. For the sake of simplicity, and
without loss of generality, we can restrict our analysis to the case
of a hermitian, massless scalar field $\phi$. Instead of dealing with
a proper black hole, we consider a two-dimensional Minkowski
spacetime
with a timelike boundary \cite{FD77,Birrell-Davies}. This spacetime is
described by null coordinates $(u,v)$ and the equation governing the
boundary is the same that describes the worldline of the centre of the
star, say $v=p(u)$~\footnote{
%%%%%%%%%%%%%%%%%%%%%%%%%%%%%%%%%%%%%%%%%%%%%%%%%%%%%%%%%%%%%%%%%%%%%%%%%%%%%  
  In Refs.\ \cite{FD77,Birrell-Davies,WF99} the function $p$ is defined
  somewhat differently. For a generic shape $x=z(t)$ of the boundary,
  one first defines a quantity $\tau_u$ through the implicit relation
  $\tau_{u}-z(\tau_{u})=u$. Then, the function is defined as
  $p(u)=2\tau_{u}-u$, which is exactly the phase of the outgoing
  component of the In modes, and $v=p(u)$ is just the equation for
  the worldline of the boundary.
%%%%%%%%%%%%%%%%%%%%%%%%%%%%%%%%%%%%%%%%%%%%%%%%%%%%%%%%%%%%%%%%%%%%%%%%%%%
}.  
At the centre of the star the ingoing modes of $\phi$ become outgoing,
and vice versa; this translates into the requirement that at the
spacetime boundary there is perfect reflection, or that
$\phi(u,p(u))\equiv 0$. Hence, the idea of a ``mirror": the timelike
boundary in Minkowski spacetime is traced out by a one-dimensional
moving mirror for the field $\phi$.

In general, for a worldline $v=p(u)$ one has ${\rm
  d}\tau=\sqrt{p'(u)}\,{\rm d}u$, where $\tau$ is the proper time
along the worldline. From this and the fact that the acceleration for
the trajectory in two-dimensional Minkowski spacetime is $a={1\over
  2}\sqrt{p''(u)^2/ p'(u)^3}\ $, one can easily check that Eqs.
(\ref{traj}) and (\ref{nextraj}) yield $a^2=1/A$ and $a^2=\kappa {\rm
  e}^{\kappa u}/(4B)$, respectively. Thus we see that an incipient
extremal black hole is modeled at late times by a uniformly
accelerated mirror; for nonextremal black holes the acceleration of
the mirror increases exponentially with time. In both cases the
worldline of the mirror has a null asymptote $v=\bar{v}$ in the
future, while it starts from the timelike past infinity $i^-$ at
$t=-\infty$.

Without loss of generality, one can assume that the mirror
is static for $t<0$. A suitable worldline is then
\begin{equation}
 p(u)=u\Theta(-u)+f(u)\Theta(u)\;, \label{p(u)}
\end{equation}
where $\Theta$ is the step function, defined as
\begin{equation}
 \Theta(\xi)=
 \left\{
  \begin{array}{l}
    1\quad\mbox{if}\quad\xi\geq 0\;,\\
    0\quad\mbox{if}\quad\xi <0\;,
  \end{array}
 \right.
\end{equation}
and $f(u)$ is a function with the asymptotic form
(\ref{traj}). In order for the worldline to be $C^1$, $f(u)$
must be such that $f(0)=0$ and $f'(0)=1$ . To simplify
calculations, it is convenient to choose $f(u)$ to be hyperbolic
at all times~\cite{Birrell-Davies}, i.e.,
\begin{equation}
 f(u)=\sqrt{A}-{A\over u+\sqrt{A}}\;, \label{f(u)}
\end{equation}
which coincides with the function on the right hand side of Eq.
(\ref{traj}), up to a (physically irrelevant) translation of the
origin of the coordinates.

Due to the motion of the mirror, one expects that the In and Out
vacuum states will differ, leading to particle production whose
spectrum depends on the function $p(u)$. In our case, because the
worldline of the mirror has a null asymptote $v=\bar{v}$ in the future but
no
asymptotes in the past, the explicit forms of the relevant In and Out
modes for $\phi$ are easily shown to be
\begin{equation}
 \phi_\omega^{\rm (in)}(u,v)={{\rm i}\over\sqrt{4\pi\omega}}
 \left({\rm e}^{-{\rm i}\omega v}-{\rm e}^{-{\rm i}\omega
 p(u)}\right) \label{in}
\end{equation}
and
\begin{equation}
 \phi_\omega^{\rm (out)}(u,v)={{\rm i}\over\sqrt{4\pi\omega}}
 \left({\rm e}^{-{\rm i}\omega u}-
 \Theta\left(\bar{v}-v\right){\rm e}^{-{\rm i}\omega q(v)}\right)\;,
\label{out}
\end{equation}
where $q(v)=p^{-1}(v)$ and $\omega>0$. The spectrum of particles
created in such a scenario is known although, to our knowledge, no
one has pointed out the correspondence with the formation of extremal
black holes. However, since the result is something of a textbook
case, we here merely summarize the main steps; for details, see for example\ 
Ref.\ \cite{Birrell-Davies}, p\ 109.

The In and Out states of $\phi$ can be related by the
Bogoliubov coefficients:
\begin{equation}
 \alpha_{\omega\omega'}=\left(\phi_\omega^{\rm (out)},
 \phi_{\omega'}^{\rm (in)}\right)=-{\rm i}\int_0^{+\infty}{\rm d}x
 \left[\phi_\omega^{\rm (out)}(u,v)
 \rlpartial_t\phi_{\omega'}^{{\rm (in)}}(u,v)^\ast\right]_{t=0}\;; 
\label{alphaa}
\end{equation}
\begin{equation}
 \beta_{\omega\omega'}=-\left(\phi_\omega^{\rm (out)},
 \phi_{\omega'}^{\rm (in)\ast}\right)=
 {\rm i}\int_0^{+\infty}{\rm
 d}x\left[\phi_\omega^{\rm
 (out)}(u,v)\rlpartial_t\phi_{\omega'}^{\rm
 (in)}(u,v)\right]_{t=0}\;. \label{betaa}
\end{equation}
There has been some discussion in the literature \cite{Grove86} about
whether the calculation of the Bogoliubov coefficients by Fulling and
Davies \cite{FD77,Birrell-Davies} is correct. We find that their
approximations are valid in the asymptotic regime of interest to us.

The spectrum of created particles is given by the expectation value of
the ``out quanta'' contained in the In state, $\langle
0,\mbox{in}|N_\omega^{\rm (out)}|0,\mbox{in}\rangle$. In terms of the
Bogoliubov coefficients, this spectrum is
\begin{equation}
 \langle N_\omega\rangle=\int_0^{+\infty}{\rm d}\omega'\,
 |\beta_{\omega\omega'}|^2\;,
\label{spectrum}
\end{equation}
where $\langle N_\omega\rangle$ is a shorthand for $ \langle
0,\mbox{in}|N_\omega^{\rm (out)}|0,\mbox{in}\rangle$.

With the choice (\ref{f(u)}), one can compute the Bogoliubov coefficients
which are appropriate in the asymptotic regime $t\to +\infty$.
Performing the integrals in Eqs.\ (\ref{alphaa}) and (\ref{betaa})
gives \cite{Birrell-Davies}
\begin{equation}
 \alpha_{\omega\omega'}\approx {\rm i}\frac{\sqrt A}{\pi}{\rm e}^{-{\rm i}
 {\sqrt A}(\omega+\omega')} K_1(2{\rm i}(A\omega\omega')^{1/2})\;,
\label{alpha}
\end{equation}
\begin{equation}
\beta_{\omega\omega'}\approx \frac{\sqrt A}{\pi}{\rm e}^{{\rm
i}{\sqrt A}(\omega-\omega')}
                K_1(2(A\omega\omega')^{1/2})\;,
\label{beta}
\end{equation}
where $K_1$ is a modified Bessel function, shown in Fig.\ 2.  For
argument $z$, $K_1(z)\sim 1/z$ for $z\to 0$, and
$K_1(z)\sim\sqrt{\pi/(2z)}\,{\rm e}^{-z}$ when $z\to +\infty$
\cite{as}.

%==============================================================================
\begin{figure}[hbt]
\vbox{ \hfil \scalebox{0.60}{{\includegraphics{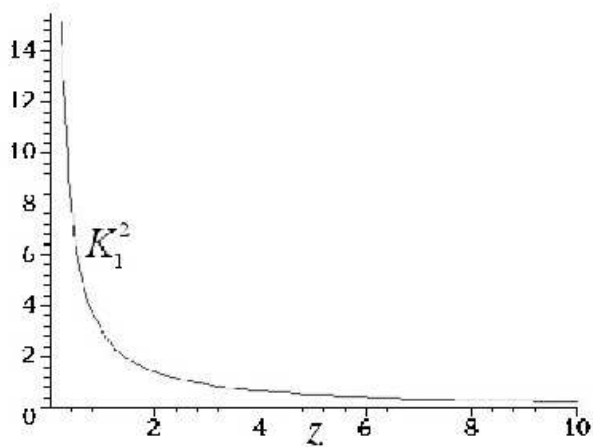}}}
\hfil }
\bigskip
\caption[Modified Bessel function $K_1(z)$ squared]{
%------------------------------
Plot of the modified Bessel function $K_1(z)$, squared.
%------------------------------
} \label{bogol}
\end{figure}
%=============================================================================

We emphasize that Eqs.\ (\ref{alpha}) and (\ref{beta}) {\em do not\/}
correspond to a full evaluation of the integrals in Eqs.\ 
(\ref{alphaa}) and (\ref{betaa}), but only take into account the
contribution for $x\approx\sqrt{A}$, i.e., from the mirror worldline
at $u\to +\infty$. This is the only part of the Bogoliubov
coefficients that can be related to particle creation by an incipient
black hole, because any other contribution corresponds to particles
created much earlier, and depends therefore on the arbitrary choice of
$p(u)$ in the non-asymptotic regime.
Clearly, since $\langle N_\omega\rangle\neq 0$, there is particle
creation by the incipient extremal RN black hole~\footnote{ 
%%%%%%%%%%%%%%%%%%%%%%%%%%%%%%%%%%%%%%%%%%%%%%%%%%%%%%%%%%%%%%%%%%%%%%%%%%%  
  This result is only apparently in contradiction with the analysis
  performed in Ref.\ \cite{vanzo97}, where it is claimed that there is
  no emission of neutral scalar particles. In fact, such a conclusion
  was derived for a massive field in the ultra-relativistic limit, and
  agrees with the exponential behaviour of $K_1$ at large values of
  $\omega$.
%%%%%%%%%%%%%%%%%%%%%%%%%%%%%%%%%%%%%%%%%%%%%%%%%%%%%%%%%%%%%%%%%%%%%%%%%%%
}.

Due to the $1/(\omega\omega')$ in the asymptotic form of
$|\beta_{\omega\omega'}|^2$, the spectrum (\ref{spectrum}) diverges at
low frequencies. The divergence in $\omega'$ has the same origin as
the one that appears in the case of non extremal black holes, where
\begin{equation}
 |\beta_{\omega\omega'}|^2
 =\frac{1}{2\pi\omega'}
 \left(
 \frac{1}{{\rm e}^{2\pi\omega/\kappa}-1}
 \right)\;.
\label{thermal}
\end{equation}
These Bogoliubov coefficients also contain a logarithmic divergence in
$\omega'$, which is due to the evaluation of the mode functions at
$u=+\infty$ and for that reason can be interpreted as an accumulation
of an infinite number of particles after an infinite time. The
divergence can be removed, however, as Hawking suggested
\cite{Hawking75} by the use of wave packets instead of plane-wave In
states; this has the effect of introducing a frequency cutoff. 
%The remaining, Planckian factor in Eq.\ (\ref{thermal}) is well behaved in
%$\omega$.

Contrary to what happens in the non extremal case, $\langle
N_\omega\rangle$ is not a Planckian distribution and therefore the
spectrum of created particles is nonthermal.  Thus, the notion of
temperature is undefined. This result supports the view that an
extremal black hole is not the zero temperature limit of a non extremal
one. However, it would be premature to base such conclusions only on
the basis of Eq.\ (\ref{spectrum}), because the Bogoliubov
coefficients tell us only that particles are created {\em at some
  time\/} in the late stages of collapse, which does not necessarily
mean that such creation takes place at a steady rate. 

Noticeably equations (\ref{spectrum}) and (\ref{beta}) also indicate
that an incipient extremal RN black hole creates particles with a
spectrum that depends on the constant $A$. These results immediately
raise two problems.

First, since particle creation leads to black hole evaporation, it
seems that (some version of) the cosmic censorship conjecture could be
violated. Indeed, emission of neutral scalar particles implies a
decrease in $M$, while $Q$ remains constant; evidently, a transition
to a naked singularity ($Q^2>M^2$) should take place. 

Second, the dependence of the spectrum on $A$, which in turn depends
on the details of collapse, raises the possibility of getting
information about the collapsing object through measurements performed
at late times, a contradiction of the no-hair theorems.  

In order to clarify these issues, in the next two sections we shall
refine our conclusions through an analysis of the stress-energy tensor
of the quantum field.

%-----------------------------------------------------------------------------
\subsection[{Preservation of cosmic censorship}]
{Preservation of cosmic censorship}
\label{subsec:cosmcc}
%-----------------------------------------------------------------------------

We consider here the first of the problems mentioned above.
The luminosity of the black hole, the rate of change of $M$, is given
by the flux of created particles at infinity, or the $T_{uu}$
component of the stress-energy tensor.

We saw in chapter~\ref{chap:1} that the expectation value of
$T_{uu}$ for the case of a moving boundary in two-dimensional
Minkowski spacetime is given by the Schwartzian derivative of $p(u)$
\begin{equation}
 \langle :\!T_{uu}\!:\rangle =
 \frac{1}{4\pi}\left(\frac{1}{4}\left(\frac{p''}{p'}\right)^2
  -{1\over 6}\frac{p'''}{p'}\right)\;.
\label{<:T:>}
\end{equation}
where the primes denote derivatives with respect to $u$.

Inserting the form (\ref{p(u)}) of $p$, with $f$ given by
Eq.\ (\ref{f(u)}), into Eq.\ (\ref{<:T:>}), one gets
\begin{equation}
 \langle :\!T_{uu}\!:\rangle={1\over 24\pi\sqrt{A}}\, \delta(u)\;.
\end{equation}
Thus, the only non-vanishing contribution to $\langle
:\!T_{uu}\!:\rangle$ is due to the transition from uniform to
hyperbolic motion that takes place at $t=0$. For the discussion of
incipient black holes, only the behaviour for $u\to +\infty$ is
relevant, and so this feature is uninteresting. On the other hand, in
the hyperbolic regime $\langle :\!T_{uu}\!:\rangle$ vanishes
identically. (It is also straightforward to check from Eq.\ 
(\ref{<:T:>}) that, conversely, a hyperbolic worldline is the only one
with nonzero acceleration that leads to $\langle
:\!T_{uu}\!:\rangle=0$.)

This result shows that the flux due to an incipient extremal black
hole vanishes asymptotically at late times.  Consequently, extremal
black holes do not lose mass~\footnote{
%%%%%%%%%%%%%%%%%%%%%%%%%%%%%%%%%%%%%%%%%%%%%%%%%%%%%%%%%%%%%%%%%%%%%%%%%%%%%  
  Here, we assume that luminosity is simply related to
  $\langle:\!T_{uu}\!: \rangle$, which amounts to assuming the
  validity of the semiclassical field equation $G_{\mu\nu}=8\pi\langle
  :\!T_{\mu\nu}\!:\rangle$ \cite{DW}.  However, as we said in
  chapter~\ref{chap:1}, this might not be a good approximation when
  $\phi$ is in a state with strong correlations. 
%%%%%%%%%%%%%%%%%%%%%%%%%%%%%%%%%%%%%%%%%%%%%%%%%%%%%%%%%%%%%%%%%%%%%%%%%%%%%
  }, and cosmic censorship is preserved. However, the nonzero value of
$\beta_{\omega\omega'}$ clearly shows that there {\em is\/} particle
creation during collapse. Cosmic censorship has apparently been
rescued only at the price of introducing a paradox, namely: particles
are created {\em and\/} their flux has zero expectation value. How can
these two statements be simultaneously true?

This puzzling situation has been extensively discussed in the context
of particle emission from a uniformly accelerating mirror
\cite{FD77,Birrell-Davies,FD76}. Fulling and Davies \cite{FD77} explain the
net zero energy flux in the presence of nonzero
$\beta_{\omega\omega'}$ by a special cancellation of the created modes
via quantum interference, which is due to contributions from the
coefficients $\alpha_{\omega\omega'}$. In section~\ref{subsec:detect} we analyze
this issue further by examining the response function of an ideal
detector.

%-----------------------------------------------------------------------------
\subsection[{Preservation of the no-hair theorems}]
{Preservation of the no-hair theorems}
\label{subsec:nohair}
%-----------------------------------------------------------------------------

We now turn to the second of the problems mentioned earlier: given
that the spectrum contains the constant $A$, do extremal black holes
violate the no-hair theorems? The result $\langle
:\!T_{uu}\!:\rangle=0$ suggests an escape --- in spite of the nonzero
value of $\langle N_\omega\rangle$, no radiation is actually detected.
However, this resolution raises new questions. If no radiation is
detected, how can one claim that the black hole emits anything at all?
Is the radiation observable? How should one then interpret $\langle
N_\omega\rangle$?

It is premature to claim that no radiation is detected only on the
basis of $\langle :\!T_{uu}\!:\rangle=0$, because there could be other
nonvanishing observables from which one might infer the presence of
quanta. A straightforward calculation shows that the expectation
values of $T_{vv}$ and $T_{uv}$ are also zero. However, let us examine
the variance $\Delta T_{uu}$ of the flux. Wu and Ford \cite{WF99} have
also recently given the following expression for $\langle
:\!T^{2}_{uu}\!:\rangle$ in the case of a minimally coupled, massless
scalar field in two-dimensional Minkowski spacetime with a timelike
boundary described by the equation $v=p(u)$:
\begin{equation}
 \langle :\!T^2_{uu}\!:\rangle= \frac{1}{\left(4\pi\right)^2}
 \left(-\frac{4 p'^{2}}{(v-p(u))^4}+
 \frac{3}{16}\left(\frac{p''}{p'}\right)^{4} -
 \frac{1}{4}\frac{p'''}{p'}\left({p''\over p'}\right)^2+
 {1\over 12}\left({p'''\over p'}\right)^2 \right)\;.
\label{varT}
\end{equation}
If one ignores the so-called cross terms \cite{WF99}, this
coincides with the variance $\Delta T_{uu}$ (because in our
case $\langle :\!T_{uu}\!:\rangle=0$). With $p$ given by
Eqs.\ (\ref{p(u)}) and (\ref{f(u)}), Eq.\ (\ref{varT})
gives, for $u>0$,
\begin{equation}
 \langle :\!T^2_{uu}\!:\rangle=-{A^2\over 4\pi^2
 \left(A+\left(v-\bar{v}\right)u\right)^4}\sim - {A^2\over
 4\pi^2\left(v-\bar{v}\right)^4u^4}\;. \label{powlaw}
\end{equation}
Thus, in spite of the fact that the expectation value of the flux
vanishes identically, its statistical dispersion does not, but its
value becomes smaller and smaller and tends to zero in the limit $u\to
+\infty$. Hence, although one could in principle infer the value of
the constant $A$ by measuring the quantity $\Delta T_{uu}$ at late
times, such measurements will become more and more difficult as
$\Delta T_{uu}$ decreases according to Eq.\ (\ref{powlaw}). This
damping is of course reminiscent of the familiar damping of
perturbations, which prevents one from detecting by late-time
measurements the details of an object that collapses into a black hole
\cite{price72}. Therefore, monitoring $\Delta T_{uu}$ does not lead to a
violation of the no-hair theorems, because no trace of $A$ will
survive in the limit $u\to +\infty$.

This discussion shows only that no violation of the no-hair theorems
can be detected by measuring the variance in the energy flux. The
possibility remains that other types of measurement could allow one to
find out the value of $A$.  If, however, $\Delta T_{\mu\nu}\to 0$ for
$u\to +\infty$, then the random variable $T_{\mu\nu}$ must tend to its
expectation value, i.e., to zero. This means that, asymptotically, the
properties of the field are those of the vacuum state. Consequently,
all local observables will tend to their vacuum value.

Although extremal black holes obey the no-hair theorems, the {\em way}
in which cosmic baldness is enforced differs from the non extremal
situation. Consider again the variance of the flux. Inserting the
function $p$ for non extremal incipient black holes (see Eq.\ 
(\ref{nextraj})) into Eqs.\ (\ref{<:T:>}) and (\ref{varT}), one gets
$\langle :\!T_{uu}\!:\rangle=\kappa^2/(48\pi)$ and
\begin{equation}
 \langle :\!T_{uu}^2\!:\rangle={1\over (4\pi)^2}\left(
 {\kappa^4\over 48}-
 {4\kappa^2B^2{\rm e}^{-2\kappa u}\over\left(
 v-\bar{v}+B{\rm e}^{-\kappa u}\right)^4}\right) \sim
 {\kappa^4\over 768\pi^2}-
 {\kappa^2B^2{\rm e}^{-2\kappa u}\over
 4\pi^2\left(v-\bar{v}\right)^4}\;.
\label{explaw}
\end{equation}
Contrary to the extremal case, the ``non extremal" variance $\Delta
T_{uu}$ tends not to zero as $u\to +\infty$, but to the value
$\kappa^2\sqrt{2}/(48\pi)$, which corresponds to thermal emission;
this is sufficient to guarantee that no information about the details
of collapse is conveyed.  Furthermore, the approach to this value is
exponentially fast, while for the extremal configuration the decay
obeys only a power law.

%-----------------------------------------------------------------------------
\subsection[{Detecting radiation from a uniformly accelerated mirror}]
{Detecting radiation from a uniformly accelerated mirror}
\label{subsec:detect}
%-----------------------------------------------------------------------------

In Sec.\ \ref{subsec:cosmcc} we mentioned the apparently paradoxical
situation in which nonzero particle production (as shown by
non-vanishing Bogoliubov coefficients $\beta_{\omega\omega'}$) is
accompanied by zero energy flux (vanishing expectation value of the
stress-energy-momentum tensor.) Discussions about such issues are
often phrased in terms of ideal detectors \cite{Birrell-Davies,DeWitt79}.
Although our previous arguments are based solely on the behaviour of
the stress-energy-momentum tensor, we can gain some additional insight
into the ``paradox'' by considering the response of a simple monopole
detector on a geodesic worldline $v=u+2x_0$, with $x_0=\mbox{const}$,
in two-dimensional Minkowski spacetime.

We are interested in computing the detector response
function per unit time, defined as
\begin{equation}
 {\cal R}(E)=\lim_{T\to +\infty} {1\over 2T}\int_{-T}^{T}{\rm d}
 \tau\int_{-T}^{T} {\rm d}\tau'\,
 \Theta(E){\rm e}^{-{\rm i}E(\tau
 - \tau')}D^+(u(\tau),v(\tau);u(\tau'),v(\tau'))\;,
\label{response}
\end{equation}
where $D^+$ is the Wightman function of the scalar field in the In
vacuum, $u(\tau)=\tau-x_0$, $v(\tau)=\tau+x_0$, and $E$ is the
excitation energy of the detector. (Note that $E\geq 0$, which is
automatically enforced by the presence of the step function
$\Theta(E)$ on the right hand side of Eq.\ (\ref{response}).) In terms
of the In modes, $D^+$ has the form
\begin{equation}
 D^+(u,v;u',v')=\int_{-\infty}^{+\infty}{\rm d}\omega\,
 \Theta(\omega)\phi_\omega^{\rm (in)}(u,v) \phi_\omega^{\rm
 (in)}(u',v')^\ast\;,
\label{orco}
\end{equation}
where we have extended the integration range to $-\infty$, by
introducing the step function $\Theta(\omega)$.

Since the definition of ${\cal R}(E)$ involves an integration over
time from $-\infty$ to $+\infty$, in the case of a mirror worldline of
the type (\ref{p(u)}) it will get contributions corresponding to the
nonzero flux (such as, for example, the one at $u=0$ when $f(u)$ is given
by
Eq.\ (\ref{f(u)})). These we regard as spurious, because we are really
interested in clarifying the relationship between zero flux and
nonzero spectrum in the hyperbolic regime. For this reason, let us
consider a mirror worldline which is hyperbolic at all times, say
$p(u)=-A/u$ for $u>0$, for which there can be no such spurious
contributions to ${\cal R}(E)$.

The worldline $p(u)=-A/u$ has a null asymptote in the past, thus 
\begin{equation} 
 \phi_\omega^{\rm (in)}(u,v)={{\rm i}\over\sqrt{4\pi\omega}} 
 \left({\rm e}^{-{\rm i}\omega v}- \Theta(u)
 {\rm e}^{-{\rm i}\omega p(u)}\right)\;.
\label{mode}
\end{equation}
The former expression does not form a complete basis in the presence
of a past null asymptote but it takes into account only those outgoing
modes which have been reflected by the moving mirror. 
On substituting Eq.\ (\ref{mode}) into Eq.\ (\ref{orco}), we have
\begin{equation}
 D^+(u,v;u',v')=F_1(v,v')+F_2(u,v')+F_3(v,u')+F_4(u,u')\;,
\end{equation}
where:
\begin{equation}
 F_1(v,v')={1\over 4\pi}\int_{-\infty}^{+\infty}{\rm d}\omega\, 
 {\Theta(\omega)\over |\omega|}\,{\rm e}^{-{\rm i}\omega (v-v')}\;;
\end{equation}
\begin{equation}
 F_2(u,v')=-{1\over 4\pi}
 \Theta(u)\int_{-\infty}^{+\infty}{\rm d}
 \omega\,{\Theta(\omega)\over |\omega|}\,
 {\rm e}^{{\rm i}\omega (v'-p(u))}\;;
\end{equation}
\begin{equation}
 F_3(v,u')=-{1\over 4\pi}
 \Theta(u')\int_{-\infty}^{+\infty}{\rm d}
 \omega\,{\Theta(\omega)\over |\omega|}\,
{\rm e}^{-{\rm i}\omega (v-p(u'))}\;;
\end{equation}
\begin{equation}
 F_4(u,u')=
 {1\over4\pi}\Theta(u)\Theta(u')\int_{-\infty}^{+\infty} {\rm d}\omega\,
 {\Theta(\omega)\over |\omega|}\, {\rm e}^{-{\rm i}
 \omega (p(u)-p(u'))}\;.
\end{equation}
Correspondingly, ${\cal R}(E)$ can be split into four parts:
${\cal R}(E)={\cal R}_1(E)+{\cal R}_2(E)+{\cal R}_3(E)+{\cal
R}_4(E)$.

The terms ${\cal R}_1(E)$, ${\cal R}_2(E)$, and ${\cal
R}_3(E)$ can be computed straightforwardly, by using the
formal identities
\begin{equation}
 \lim_{T\to +\infty}\int_{-T}^T{\rm d}\tau\,{\rm e}^{{\rm i}\xi\tau}
 =2\pi\delta(\xi)
\end{equation}
and
\begin{equation}
 {1\over |E|}\,\Theta(E)\Theta(-E)=2\delta(E)\;,
\label{gnarf}
\end{equation}
the latter being easily established by considering the sequence of
functions $(| E | +\epsilon)^{-1}
\Theta(E+\epsilon)\Theta(-E+\epsilon)$ in the limit $\epsilon\to 0$.
We get ${\cal R}_1(E)=-2{\cal R}_2(E)=-2{\cal R}_3(E)= \delta(E)$, and so
the first three contributions to ${\cal R}(E)$ sum to zero.

The computation of ${\cal R}_4(E)$ is cleaner if one works in
dimensionless variables, such as $\widetilde{E}=\sqrt{A}\,E$,
$\tilde{\omega}=\sqrt{A}\,\omega$, $\tilde{\tau}=\tau/\sqrt{A}$. The
identity
\begin{equation}
\int_0^{+\infty}{\rm d}\tilde{\omega}\,{{\rm e}^{-{\rm
i}\tilde{\omega} (\xi-{\rm
i}0)}\over\tilde{\omega}}=-\ln\left(\xi-{\rm
i}0\right)+I-{\rm i}\,{\pi\over 2}\;,
\end{equation}
where $I$ is the divergent quantity
\begin{equation}
I=\int_0^{+\infty}{\rm d}
\tilde{\omega}
\,
{\cos\tilde{\omega}
\over
\tilde{\omega}}\;,
\end{equation}
together with the properties of the logarithm, allows us to
write
\begin{eqnarray}
&& \int_{-\infty}^{+\infty}{\rm d}\tilde{\omega}\,
 {\Theta(\tilde{\omega})\over |\tilde{\omega}|} \,
 \exp\left(-{\rm i}\tilde{\omega}{\tilde{\tau}-\tilde{\tau}'
 \over\left(\tilde{\tau}-\tilde{x}_0\right)\left(\tilde{\tau}'
 -\tilde{x}_0\right)}\right)=\int_{-\infty}^{+\infty}{\rm d}
 \tilde{\omega}\,{\Theta(\tilde{\omega})\over
 |\tilde{\omega}|}\, {\rm e}^{-{\rm i}\tilde{\omega}
 \left(\tilde{\tau} -\tilde{\tau}'\right)}
\nonumber\\
 &&-\int_{-\infty}^{+\infty} {\rm d}\tilde{\omega}\, 
  {\Theta(\tilde{\omega})\over |\tilde{\omega}|} \, 
 {\rm e}^{-{\rm i}\tilde{\omega}\left(\tilde{\tau}-
 \tilde{x}_0\right)}-\int_{-\infty}^{+\infty}{\rm d}
 \tilde{\omega}\,{\Theta(\tilde{\omega})\over
 |\tilde{\omega}|}\, {\rm e}^{{\rm i}\tilde{\omega}
 \left(\tilde{\tau}'- \tilde{x}_0\right)}+2I\;.
\end{eqnarray}
In this expression we have replaced one of the quantities $I-{\rm i}
\pi/2$ with its complex conjugate by simultaneously changing the sign
in one of the exponents.  This manipulation is allowed by the fact
that, since $\tilde{\tau}-\tilde{x}_0$ and $\tilde{\tau}'-\tilde{x}_0$
can never become negative in $F_4$, their logarithms are always real.
Note that the resulting expression agrees with the property of ${\cal
  R}(E)$ being a real quantity.

We can thus write ${\cal R}_4(E)={\cal R}_{41}(E)+{\cal
  R}_{42}(E)+{\cal R}_{43}(E)+{\cal R}_{44}(E)$. Using the formal
relations
\begin{equation}
 \lim_{\widetilde{T}\to +\infty}
 \int_{\tilde{x}_0}^{\widetilde{T}}{\rm d}\tilde{\tau}\,
 {\rm e}^{{\rm i}\xi\tilde{\tau} }=\pi\delta(\xi)+{\rm i}
 {\rm e}^{ {\rm i}\,\xi\tilde{x}_{0} }\,{\mathcal P}\left(1
 \over\xi\right)\;,
\end{equation}
and
\begin{equation}
 \lim_{\widetilde{T}\to +\infty}{1\over\widetilde{T}}
 \int_{\tilde{x}_0}^{\widetilde{T}}{\rm d}\tilde{\tau}\, {\rm e}^{{\rm i}
 \xi\tilde{\tau}}=\Theta(\xi)\Theta(-\xi)\;,
\end{equation}
together with Eq.\ (\ref{gnarf}), we get ${\cal
  R}_{41}(E)=\delta(E)/4$, ${\cal R}_{42}(E)+{\cal
  R}_{43}(E)=-\delta(E)/2$, and ${\cal R}_{44}(E)=I\delta(E)/4$.
Finally, since $I$ is divergent we can write
\begin{equation}
 {\cal R}(E)={I\over 4}\,\delta(E)\;.
\label{F}\end{equation}
Thus, we have essentially a delta function peaked at zero energy. Now,
${\cal R}(E)$ is related to a quantum mechanical probability and so
this result means that, for any value $E>0$ of the energy, no matter
how small, the detector has probability 1 of making a transition of
amplitude smaller that $E$ and probability 0 of detecting particles of
higher energy. (Of course, this does not mean that it will {\em never\/}
make transitions with $E>0$; only that these take place with probability
0.) The reason for this behaviour is evidently the divergence in the
spectrum as $\omega\to 0$. Of course, the detector does not gain
energy during such a ``detection'' --- in fact, one can say that there
is no detection at all. This is compatible with the zero value of the
flux and moreover it is also in agreement with what should be
expected from the Bogoliubov coefficient (\ref{beta}) which in the low
frequency limit loses any dependence
on $A$ 
\begin{equation}
  \label{eq:betalow}
  \lim_{\omega\to 0}|\beta|^2\approx\frac{1}{(4\pi)^2}
   \frac{1}{\omega}{\omega'}.
\end{equation}
Hence one might be tempted to call the particles emitted by a mirror
in hyperbolic motion ``phantom radiation'': because only arbitrarily
soft particles would be registered by the detector with any nonzero
probability, there would be no chance of determining the spectrum
$\langle N_\omega\rangle$.  The question arises therefore, of whether
there is any way to screen our detector from this overwhelming flux of
soft quanta.

One might think of acting on the selectivity of the detector by using a
two level system that requires at least a minimal energy to switch.
Unfortunately, the detection of the infinite tail of soft quanta
corresponds to the ``transition'' from the ground state to the ground
state, and there is obviously no way to forbid this process. The
detector cannot be forbidden not to switch!

Therefore the analysis of the response function of a detector also
seems to prove that the radiation from uniformly accelerated mirrors
(and extremal incipient black holes) is in some sense like the apple
in Dante's purgatory: We can see it with our mind but we shall never
have it in our hands ...

%-----------------------------------------------------------------------------
\subsection[{Conclusions}]
{Conclusions}
\label{subsec:concl}
%-----------------------------------------------------------------------------

We have just seen how, in a subtle way indeed, cosmic censorship
and the no hair theorem are preserved in the formation of an extremal
black hole.  In fact, although they appear to emit particles, closer
scrutiny reveals that the flux of emitted radiation vanishes
identically, and in the limit $t\to +\infty$ any measurement of local
observables gives results indistinguishable from those in the vacuum
state. This is not incompatible with a nonzero spectrum, which is not
a local quantity and tells us only that particles are created at some
time during collapse (not necessarily at $t=+\infty$). Thus, extremal
black holes are not pathological in this respect.

However, we have seen that there are several clearly defined senses in
which non extremal and extremal black holes differ.  In fact the
conclusions which we can derive from the above results are even more
radical. Not only do they appear to involve a fundamental difference in
the
behaviour of the radiation emitted by non-extremal and extremal black
holes but also they converge towards a picture where our understanding of
the third law of black hole thermodynamics is heavily transformed.
 
A first set of conclusions is related to the nature of the quantum
radiation from the incipient extremal black holes.  Information lost
to an external observer depends on the rate at which the statistical
dispersion of the flux approaches its value for $t\to +\infty$. In the
non extremal case, the dispersion goes to zero exponentially fast, Eq.\ 
(\ref{explaw}), whereas for an incipient extremal black hole it
follows a slower power law, given by Eq.\ (\ref{powlaw}). More
importantly, for $Q^2\to M^2$, Eq.\ (\ref{powlaw}) is not the limit of
Eq.\ (\ref{explaw})\footnote{
%%%%%%%%%%%%%%%%%%%%%%%%%%%%%%%%%%%%%%%%%%%%%%%%%%%%%%%%%%%%%%%%%%%%%%%%%%%%%
  Since $A$ and $B$ do not depend on $u$ by definition, the only case
  that admits a continuous limit is the one in which $A=0$. This
  cannot happen, because it would correspond to a null worldline for
  the centre of the star. Another apparent possibility, that $B\propto
  1/\kappa$ so that $\kappa B$ is constant in the limit $\kappa\to 0$,
  is not viable, because the right hand sides of Eqs.\ (\ref{powlaw})
  and (\ref{explaw}) would still have different functional dependences
  on $u$.
%%%%%%%%%%%%%%%%%%%%%%%%%%%%%%%%%%%%%%%%%%%%%%%%%%%%%%%%%%%%%%%%%%%%%%%%%%%%%
  }. One cannot therefore, consider quantum emission by an incipient
extremal black hole to be the limiting case of emission by a
non extremal black hole.  

In particular, although at $t=+\infty$ a black hole with $Q^2=M^2$ is
totally quiescent it would be incorrect to consider it as the
thermodynamic limit of a non extremal black hole, that is, an object at
zero temperature. Indeed, the quantum radiation emitted by an
incipient extremal black hole is not characterized by a temperature at
any time during collapse. Whereas incipient non extremal black holes
have a well defined thermodynamics, this is not true for extremal
holes, and they should be considered as belonging to a different
class. This result suggests that any calculations that implicitly rely
on a smooth limit in thermodynamic quantities at $Q^2 = M^2$ are
suspect, if not incorrect. Our conclusions, of course, are just
pertinent to incipient black holes; extending them to eternal black
holes seems plausible, but requires care. (Even at the classical
level, eternal black holes must be regarded as fundamentally different
from those deriving from collapse, because the global structure of
spacetime differs in the two cases.)

Moreover it should be stressed that remarkably our calculations lead to
the conclusion that {\em extremal black holes are substantially impossible
to obtain in a semiclassical approach}. We have seen in Sec.\ 
\ref{subsec:cosmcc} that if $Q^2=M^2$ at the onset of the hyperbolic
worldline (\ref{traj}), it will remain so and the cosmic censorship
conjecture is preserved. However, the fact that mass loss from an
incipient extremal black hole is zero \emph{only} in the late
hyperbolic stage seems to imply that an enormous fine tuning is
required in order to produce an extremal object by means of
gravitational collapse. In fact, an object that is extremal from the
start of its collapse might be unstable with respect to transition
to a configuration with $Q^2>M^2$. Such a transition would be
triggered by quantum emission in the early phases of collapse, when
$p(u)$ has not yet assumed its hyperbolic form. This raises the
question of how, in the presence of quantum radiation, the formation of a
naked singularity is prevented (for example, by the emission of charged
particles) and the cosmic censorship conjecture preserved.

A second set of conclusions has a more strictly thermodynamical
relevance.  We have seen in section \ref{sec:topol} how the extremal
solution appears to be ``totally disconnected'' from the
thermodynamical behaviour of the non-extremal one: the different
topology puts them in different categories and does not allow any well
defined limit between the two. Adding these facts to our results,
one may conjecture that the unattainability formulation of the third
law for black hole thermodynamics is actually implemented in an
unexpected way: {\em the temperature of a black hole cannot be reduced
  to zero simply because no black hole zero temperature state exists}.

This conjecture, although it might appear very radical, it is not as
dramatic as it might seem. In standard thermodynamics the absence of
zero-temperature states is not a pathology. Indeed a well-defined
thermodynamics can be defined without them.

It is alternatively possible that to avoid the above conclusion one
should use a more sophisticated theory.  Noticeably the divergence of the
particle spectrum $\langle N_\omega\rangle$ is reminiscent of the
infrared catastrophe typical of QED, which manifests itself, for
instance, in the process of Bremsstrahlung (see, for example, Ref.\ 
\cite{mandl}, pp.\ 165--171).  However, the infrared divergence in the
Bremsstrahlung cross section produces no observable effect, because it
is canceled by analogous terms coming from radiative corrections.
(Thanks to the Bloch--Nordsieck theorem, this cancellation is effective
to all orders of perturbation theory.) One may well wonder whether the
$\omega=0$ singularity in our spectrum is similarly fictitious and
could thus be removed by analogous techniques.

For the mirror this is possible, in principle, if one allows momentum
transfer from the field $\phi$ to the mirror, although such a
calculation is beyond the scope of the present investigation. (See
Ref.\ \cite{par95} for a model that includes recoil.) However,
whatever the answer to the mirror problem might be, it does not seem
that one could transfer it in any straightforward way to the case of
an incipient black hole. Indeed, taking recoil into account would
amount to admitting that backreaction {\em is\/} important and that
the test-field approximation is never valid. Thus, the whole subject
would have to be reconsidered within an entirely different framework.

Nevertheless a recent paper by Anderson, Hiscock and
Taylor~\cite{AHT00} should be quoted where it is demonstrated that for
static RN geometries, zero-temperature black holes cannot exist if one
considers spacetime perturbations due to the back-reaction of quantum
fields.  This seems to imply that even the inclusion of back-reaction
is insufficient to impart thermal properties to extremal black holes.

Apparently then, extremal black holes are either physical states which
are outside of black hole thermodynamics or they represent solutions
in which the external field approach to the semiclassical theory of
gravity breaks down. In both cases they would be dramatically
different objects from their non-extremal counterparts.

%%%%%%%%%%%%%%%%%%%%%%%%%%%%%%%%%%%%%%%%%%%%%%%%%%%%%%%%%%%%%%%%%%%%%%%%%%%%
% S.Liberati Ph.D. Chapter 3: Toward experimental semiclassical gravity
%%%%%%%%%%%%%%%%%%%%%%%%%%%%%%%%%%%%%%%%%%%%%%%%%%%%%%%%%%%%%%%%%%%%%%%%%%%%
\chapter[{Acoustic horizons}]
{Toward experimental semiclassical gravity: Acoustic horizons}
\label{chap:3a}
%%%%%%%%%%%%%%%%%%%%%%%%%%%%%%%%%%%%%%%%%%%%%%%%%%%%%%%%%%%%%%%%%%%%%%%%%%%%

\vspace*{0.5cm} \rightline{\it L'acqua che tocchi de' fiumi \`e
  l'ultima di quella che and\`o} \rightline{\it e la prima di quella
  che viene.} \rightline{\it Cos\`{\i}\ il tempo
  presente.\footnote{The water that you touch in the rivers is the
    last which left and the first which arrives. So it is the present
    time.}}
\vspace*{0.5cm} \rightline{\sf Leonardo da Vinci}

%%%%%%%%%%%%%%%%%%%%%%%%%%%%%%%%%%%%%%%%%%%%%%%%%%%%%%%%%%%%%%%%%%%%%%%%%%%%%

\newpage

This chapter is the first of two in which we shall deal with possible
experimental tests of particle production from the quantum vacuum.
Here we shall study a hydrodynamical system which can closely resemble
\GR\ and which can lead to indirect tests of Hawking radiation. In
particular we shall devote special attention to the actual
realizability of fluid configurations which would be able to simulate
event horizons and hence fluxes of Hawking (phonon) radiation.  In the
subsequent chapter we shall present a tentative explanation of the
observed emitted radiation of {\em Sonoluminescence} as a
manifestation of a dynamical Casimir effect.

%%%%%%%%%%%%%%%%%%%%%%%%%%%%%%%%%%%%%%%%%%%%%%%%%%%%%%%%%%%%%%%%%%%%%%%%%%%
\section[{Seeking tests of the dynamical Casimir effect}]
{Seeking tests of the dynamical Casimir effect} 
\label{sec:tdce}
%%%%%%%%%%%%%%%%%%%%%%%%%%%%%%%%%%%%%%%%%%%%%%%%%%%%%%%%%%%%%%%%%%%%%%%%%%%

It is often true that in theoretical physics the elegance and clarity
of a formula is judged by its property of relying on a few fundamental
constants of nature and basic numbers.  In this sense the formula
describing the temperature in the Hawking--Unruh effect is in an
extremely impressive combination of simplicity and
``interdisciplinarity''
\beq
T=\frac{\hbar}{2\pi c k_{\mathrm B}}\cdot a
\label{eq:THU}
\eeq
In this formula we find together a geometrical constant $\pi$, the
Planck constant of the quantum world, the Boltzmann constant of
thermodynamical laws and the speed of light of Special Relativity. The
acceleration $a$ appearing above is itself a combination of
fundamental quantities in the case of a black hole.  In fact it is
then equal to the surface gravity $\kappa$ which is, in the \Sch\ 
solution, equal to
\beq
a=\kappa=\frac{c^{4}}{4G_{\mathrm N}M}
\eeq
So even the Newton constant of gravitation enters into the count.

This peculiar concentration of fundamental constants is much more
than a curiosity. It is indeed a sign of the fact that the effects
under considerations must arise from some very simple common
background and that this background should be identified in a region
of investigation where quantum effects, thermodynamics and fundamental
forces meet.

This is a broad field and hence different lines of investigation can
be pursued. The realization of this is indeed at the basis of the
concrete possibility for us to test this class of effects and to
improve our understanding of their role in semiclassical quantum
gravity.

In fact it is easy to realize that no ``direct'' experimental test of
the Hawking--Unruh effect will probably be obtainable in the near
future. In the {\em cgs} system Eq.\ (\ref{eq:THU}) takes the form
\begin{equation}
 T=4\times 10^{-23} \frac{\mathrm K s^2}{\mathrm{cm}} \cdot a
\label{eq:THUnum}
\end{equation}
This implies that at least an acceleration of order $10^{23}\g_{\oplus}$
is required for obtaining a thermal spectrum at only one Kelvin!

Similarly, for a black hole to have a Hawking temperature of the order
of one Kelvin requires it to have a mass no greater than
$M_{\mathrm{critical}}\approx 10^{15} g$, that is a mass of the order
of a common mountain (giving a hadron-sized black hole). Although it
is indeed plausible that such objects might form in the very early
universe, it is extremely unlikely that we shall ``meet'' them in the
near future.

It is then legitimate to ask if we can test these effects indirectly
by concentrating our attention on other possible areas where the
production of particles from the quantum vacuum by a time varying
external field can come into action.  In a certain sense we can aim to
test the paradigm at the basis of the Hawking--Unruh effect instead of
the effect itself.

We shall see how this research will be not just a useful
phenomenological study, but also an extremely powerful tool for
investigating some fundamental questions about the deeper nature of
the Hawking--Unruh effect and its role as a bridge towards quantum
gravity.

%%%%%%%%%%%%%%%%%%%%%%%%%%%%%%%%%%%%%%%%%%%%%%%%%%%%%%%%%%%%%%%%%%%%%%%%%%%
\section[{Acoustic Manifolds}]{Acoustic Manifolds} 
\label{sec:acmn}
%%%%%%%%%%%%%%%%%%%%%%%%%%%%%%%%%%%%%%%%%%%%%%%%%%%%%%%%%%%%%%%%%%%%%%%%%%%

It is really true that sometimes deep and fundamental physical answers
can be gained by apparently very simple and basic investigations. In
this sense few other subjects are comparable to the study of the
acoustic manifolds. As we shall see, the answer to the apparently
trivial problem of describing the propagation of acoustic disturbances
in a non-homogeneous flowing fluid has revealed a deep analogy with
Lorentzian differential geometry
\cite{Unruh81,Jacobson91,Jacobson93,Visser93,Unruh94,Visser98,Visser:1998qn}.
This analogy is at the root of the possibility to conceive experiments
to finally test semiclassical quantum gravity effects such as the
Hawking--Unruh one.  However, let us take a step back and pursue an
{\em ab initio} discussion of the subject.

As a first approach to the problem we can start with a hydrodynamical
system characterized by a few fundamental quantities: a density
$\rho$, a velocity field $\v$ and a pressure $p$.  This system is
characterized by the standard equations of fluid dynamics, that is
\begin{itemize}
\item The continuity equation
\begin{equation}
{\partial\rho\over\partial t} 
+ \vec\nabla\cdot\left(\rho\, \v \right)= 0,
\label{E:continuity}
\end{equation}
\item and the Euler equation
\begin{equation}
\rho\;\a={\mathbf f},
\label{E:euler0}
\end{equation}
where $\a$ is the fluid acceleration,
\begin{equation}
\a={\partial\v\over\partial t} +
(\v \cdot \vec\nabla) \v,
\label{E:acc}
\end{equation}
and ${\mathbf f}$ stands for the force density, the sum of all forces
acting on the fluid per unit volume.
\end{itemize}
We shall assume that all of the external forces are gradient-derived
(possibly time-dependent) body forces, which for simplicity we collect
together in a generic term $-\rho\vec\nabla\Phi$. In addition to the
external forces, ${\mathbf f}$ contains a contribution from the
pressure of the fluid and, possibly, a term coming from the kinematic
viscosity $\nu$~\footnote{For simplicity we are here assuming a null
  bulk viscosity. This is sometimes called the ``Stokes assumption''
  (see~\cite{LSV00} for further discussion).}.  Thus, equation
(\ref{E:euler0}) takes the Navier--Stokes form
\begin{equation}
\rho\left({\partial\v\over\partial t} +
(\v \cdot \vec\nabla) \v\right)
=- \vec\nabla p - \rho\;\vec\nabla\Phi +
\rho \,\nu\,
\left(\nabla^2 \v +\frac{1}{3}\vec\nabla (\vec\nabla \cdot \v ) \right).
\label{E:euler}
\end{equation}
In our derivation we are going to make a number of technical
assumptions in order to have an analytically tractable system.
\begin{enumerate}
\item The first assumption is that we have a vorticity-free flow, {\em
    i.e.\/}, that $\vec\nabla\times\v={\mathbf 0}$. This condition is
  generally fulfilled by the superfluid components of physical
  superfluids.  It also implies the possibility to completely specify
  the velocity of the fluid via a scalar field $\psi$ defined as
  $\v=\vec\nabla\psi$.

\item The second assumption is that the fluid has a barotropic
  equation of state, that is, the density $\rho$ is a function only of
  the pressure $p$, so
\begin{equation}
\rho = \rho(p).
\end{equation}
We shall consequently define the speed of sound as
\begin{equation}
\cs^2 = {\d p\over\d\rho}.
\end{equation}
This assumption is crucial because thanks to it the continuity and
Euler equations form a complete set of equations --- if $p$ is not
just a function of $\rho$ one would need an energy equation as well.
\item A third assumption, often made in the existing literature on
  acoustic geometries, is that of a viscosity-free flow. Although this
  is quite a realistic condition for superfluids~\footnote{
%---------------------------------------------------------------------
    Superfluids are often considered the most promising candidates to
    actually reproduce in laboratory the sort of physics we are going
    to describe. See section~\ref{sec:achor} for further details.},
%---------------------------------------------------------------------
  we shall find that viscosity can in general play a prominent role.
  For the moment we shall take it to be zero but we shall come back to
  the issue later.
\end{enumerate}

The system equations (\ref{E:continuity}) and (\ref{E:euler}) can be
shown to be closed and hence we can always find an exact solution of
them $[\rho_{0}(t,\x),p_{0}(t,\x),\psi_{0}(t,\x)]$.  If we want to
describe the propagation of some acoustic perturbations, we can use
such a solution as a background and ask how linearized fluctuations of
it behave. We can then write
\begin{eqnarray}
\rho(t,\x) &=& \rho_0(t,\x) + \epsilon \; \rho_1(t,\x) + \cdots,
\\
p(t,\x) &=& p_0(t,\x) + \epsilon \; p_1(t,\x) + \cdots,
\\
\psi(t,\x) &=& \psi_0(t,\x) + \epsilon \; \psi_1(t,\x) + \cdots.
\end{eqnarray}
Inserting these forms into the equations of motion (\ref{E:continuity})
and (\ref{E:euler}), one finds that the equations for these fluctuations
are
\begin{eqnarray}
{\partial \rho_1 \over \partial t} + 
\nabla \cdot \big( \rho_1 \nabla \psi_0 + \rho_0 \nabla \psi_1 \big)
&=& 0,
\\
\rho_0 \left( {\partial \psi_1 \over \partial t} + 
\nabla\psi_0 \cdot \nabla \psi_1 \right) &=& p_1,
\\
p_1 &=& \cs^2 \rho_1.
\end{eqnarray} 
The same information contained in these three first-order partial
differential equations can be assembled into one second-order partial
differential equation which takes the (apparently more unpleasant) form
\begin{equation}
\label{E:physical}
{\partial\over\partial t} 
\left[ \frac{\rho_0}{\cs^{2}} 
\left( {\partial\psi_1\over\partial t} + {\v}_0 \cdot \nabla \psi_1
\right)
\right]
= \nabla \cdot \left[ 
\rho_0 \nabla \psi_1 - \frac{{\v}_0 \rho_0}{\cs^{2}} 
\left( 
{\partial \psi_1 \over \partial t} + {\v}_0 \cdot \nabla \psi_1 
\right)
\right].
\end{equation}
This is a partial differential equation just in $\psi_{1}(t,x)$ with
coefficients depending on the background field around which we are
studying the perturbations. Once the equation is solved in $\psi_{1}$
the corresponding values of $\rho_{1}$ and $p_{1}$ follow from the
linearized Euler equations and from the equation of state.

So far there is nothing new or strange, but now a surprise appears.
If we introduce four-dimensional coordinates in the usual way
$x^{\mu}\equiv (t,\x)$ and a $4\times 4$ matrix 
\begin{equation}
\label{gcontr}
 g^{\mu\nu}(t,\x) \equiv 
 {1\over \rho_0 \cs}
 \left[ \matrix{-1&\vdots&-{\rm v}_0^j\cr
           \cdots\cdots&\cdot&\cdots\cdots\cdots\cdots\cr
-{\rm v}_0^i&\vdots&(\cs^2 \delta^{ij} -{\rm v}_0^i {\rm v}_0^j )\cr }\right] 
\end{equation}
Then the rather formidable-looking second-order partial
differential equation for $\psi_1$ can be very simply written as
\begin{equation}
\label{E:EOM}
 {1\over\sqrt{-g}} \;
 {\partial \over \partial x^\mu} \left( 
 \sqrt{-g} \; g^{\mu\nu} \; {\partial \over \partial x^\nu} \psi_1 
 \right) = 0.
\end{equation}
where we have defined
\begin{equation}
 g = \left[\det\left(g^{\mu\nu}\right)\right]^{-1}.
\end{equation}
It is easy to see that the differential operator appearing in
Eq. (\ref{E:EOM}) it is just the d'Alembertian built with the inverse
metric $g^{\mu\nu}(t,\x)$. Therefore Eq. (\ref{E:EOM}) is {\em exactly}
the
equation of motion of a minimally coupled scalar field propagating in
a spacetime with inverse metric $g^{\mu\nu}(t,\x)$ !

It is interesting to note that Eq.~(\ref{E:physical}) does not uniquely
fix the equation of motion of $\psi_{1}$. Actually it identifies a
class of conformally related solutions. So we could, in principle,
take a metric which is conformally related to those given in
Eq.~(\ref{gcontr}), the price to pay for this would be a more
complicated equation for $\psi_{1}$ which would have the same
d'Alembertian and some non-minimal coupling.
\begin{center}
  \setlength{\fboxsep}{0.5 cm} 
   \framebox{\parbox[t]{14cm}{
%------------------------------------------------------------------------      
       {\bf Comment:} One may wonder whether this ``geometrical
       interpretation'' of the propagation of acoustic perturbations,
       has just a formal value or is indeed a true built-in property
       of hydrodynamics. A first check of this can be to ask whether
       the equations of motion of $\v_{1}$, $\rho_{1}$ and $p_{1}$ are
       all of the form $\Box X+\cdots$ where $X$ is any one of the
       above quantities and $\Box$ is the d'Alembertian appearing in
       Eq.(\ref{E:EOM}) and defined via the metric (\ref{eq:acmetr}).
       It is actually possible to check that this is the case. This
       equality of the d'Alembertians appearing in the equations of
       motion of all of the physical perturbations, implies that all
       investigations which are based on the study of the Green
       functions of the fields (such as, for example, the treatment of
       Hawking radiation) can be performed using the ``trivial''
       equation for $\psi_{1}$. The basic results can be considered
       valid also for the (easier to detect) perturbations in $\rho$.
%------------------------------------------------------------------------
}}
\end{center}
Let us summarise the situation. We have started with a classical
hydrodynamical system in flat space and have found that the equations
describing the propagation of perturbations can be cast in a form
which is manifestly the propagation of a scalar field in a
$(3+1)$--dimensional Lorentzian (pseudo-Riemannian) geometry described
by what we can call an {\em acoustic metric} $g_{\mu\nu}(t,\x)$.
\begin{equation}
 g_{\mu\nu}(t,\x) \equiv 
 {\rho_0 \over \cs}
 \left[ \matrix{-(\cs^2-{\rm v}_{0}^{2})& \vdots &-{\rm v}_0^j\cr
                \cdots\cdots\cdots\cdots&\cdot&\cdots\cdots\cr
                -{\rm v}_0^i&\vdots&\delta_{ij}\cr }\right].  
\label{eq:acmetr}             
\end{equation}
Note that, although the underlying physics is non-relativistic (one
can safely consider $c\approx\infty$ in this sort of problems), we can
see that the fluctuations are nevertheless experiencing a full
spacetime metric in which exits a group of Lorentz transformations
where the role of the speed of light is played by the speed of sound
$\cs$. On one hand the fluid particles couple {\em only} to the
physical (flat) spacetime metric
$\eta_{\mu\nu}\equiv(\mbox{diag}[-c^2,1,1,1])_{\mu\nu}$. On the other
hand the sound waves do not experience this ``physical metric'' and
instead couple to the acoustic metric $g_{\mu\nu}$.

In spite of this, it has to be stressed that the hydrodynamical system
which we found is far from being equivalent to \GR .  In fact the acoustic
metric depends {\em algebraically} on the distribution of matter (the
density, velocity of the flow and local speed of sound in the fluid)
and it is governed by the fluid equations of motion which constrain
the background geometry. In Einstein gravity, the spacetime metric is
instead related to the distribution of matter by the non-linear
Einstein--Hilbert differential equations.

This difference is also confirmed by looking at the degrees of freedom
of these theories. In a completely general $(3+1)$--dimensional
Lorentzian geometry the metric has six degrees of freedom (it is
described by a $4\times 4$ symmetric matrix which gives $10$
independent components from which one has to subtract the $4$
coordinate conditions). In contrast an acoustic metric is completely
specified by the three scalars
$\rho_{0}(t,\x),\psi_{0}(t,\x),\cs(t,\x)$ so it has at most 3 degrees
of freedom per point in spacetime. Considering also the continuity
equations, these are further reduced to 2 (for $\psi_{0}(t,\x)$ and
$\cs(t,\x)$).
\begin{center}
  \setlength{\fboxsep}{0.5 cm} 
   \framebox{\parbox[t]{14cm}{ 
%------------------------------------------------------------------------      
       {\bf Comment}: We want here to suggest the idea that a more
       complex fluid could actually be built in order to have six
       degrees of freedom per point. The most straightforward way to
       do this might be to define a ``fluid'' whose pressure is a
       tensor $p_{ij}$ so that the barotropic equation has the form:
      \begin{equation}
       \label{eq:genbar}
        p_{ij}=\cs^{2}\delta_{ij}\rho+\pi_{ij} 
      \end{equation}
      where $\pi_{ij}$ is traceless and symmetric.
      
      It is possible to show that also in this case all of the fluid
      perturbations couple to the same acoustic metric which now has
      the form:
      \begin{equation}
       g_{\mu\nu}(t,\x) \equiv 
       \left[ \matrix{-(\cs^2-{\rm v}_{0}^{2})& \vdots & -{\rm v}_0^j\cr
       \cdots\cdots\cdots\cdots&\cdot&\cdots\cdots\cr
       -{\rm v}_0^i&\vdots&\delta_{ij}+\partial_{\rho}\pi_{ij}\cr }\right].
      \end{equation}
      This provides a theory which has actually one degree of freedom
      more than \GR\ because the metric does not depend on $\rho$.
      Indeed it is easily to realize that in this case one ends up
      with a scalar-tensor theory which could be linked to a standard
      Brans--Dicke one via the redefinition
      \begin{equation}
      \label{eq:bd}
       \rho=e^{\phi/\cs^2}
      \end{equation}
      Unfortunately this approach is quite difficult to pursue. The
      equations which one finds for the perturbations are very complicated
      and moreover one requires extra equations (with respect to the
      continuity and Euler equations) to fix the behaviour of
      $\pi_{ij}$. 
%-------------------------------------------------------------------------
} }
\end{center}

From now on we shall discuss mainly the structure of the acoustic
manifold and so we shall always deal with the background field
quantities.  For the sake of simplicity we shall omit the subscript 0
except when there would be a risk of confusion.

%--------------------------------------------------------------------------
\subsection[{Ergoregions and acoustic horizons}]
{Ergoregions and acoustic horizons}
\label{subsec:ergo}
%--------------------------------------------------------------------------

From what we have seen so far, it is not very surprising that acoustic
geometries can show all of the well-known concepts which people are
used to in \GR\ including ergo-regions and trapped surfaces.

We can start by considering the integral curves of the vector field
$K^{\mu}\equiv (\partial/\partial t)^{\mu}=(1,0,0,0)^{\mu}$. In the
case of a ``steady flow'' (what in \GR\ would be called a ``stationary''
solution) this coincides with the time translation Killing vector.
It is easy to see that
\begin{equation}
  \label{eq:kil}
  g_{\mu\nu}(\partial/\partial t)^{\mu}(\partial/\partial t)^{\nu}=g_{tt}=
  -[\cs^2-{\rm v}^2]
\end{equation}
and so the time translation Killing vector becomes spacelike when
$|\v|>\cs$. Any region of supersonic flow is then a part of the
acoustic manifold where it is impossible for any observer to be at
rest with respect to another one at infinity (because it would require
moving 
faster that the speed of sound). This is the sonic analogue of the
ergoregions of~\GR\ generally associated with rotating black holes.
The boundary of these regions, where $|\v|=\cs$, are called
ergo-surfaces~\footnote{
%--------------------------------------------
  In General Relativity the ${\rm v} \to c$ surface would be called an
  ``ergosphere'', however proving that this surface generically has
  the topology of a sphere is a result special to General Relativity
  which depends critically on the imposition of the Einstein
  equations. In the present fluid dynamics context there is no
  particular reason to believe that the ${\rm v}\to c$ surface would
  generically have the topology of a sphere and it is preferable to
  use the more general term ``ergo-surface'' for the boundary of an
  acoustic ergo-region.}.
%------------------------------------------

Another concept which can be translated from \GR\ to acoustic
geometries is that of trapped surfaces. In fact it is easy to see that
an outer-trapped surface can easily be built in an acoustic geometry.
Let us consider a closed two-surface. If the fluid flow is everywhere
inward-pointing and the normal component of the fluid velocity is
everywhere supersonic then a sound wave, irrespective of its direction
of propagation, will be swept inward by the overwhelming fluid flow
and trapped inside the surface. An inner-trapped surface can similarly
be realised by assuming a supersonic outward-pointing flow.

Such simple definitions can be used in this case because we have the
underlying Minkowski geometry which provides a natural definition of
what it means to be ``at rest''. The same concepts in \GR\ require a
much more complicated set of additional machinery such as the notions
of ``expansion'' of bundles of geodesics.

An acoustic trapped region can be similarly defined as a region
containing outer trapped surfaces and the boundary of such a region is
the (acoustic) apparent horizon.  An acoustic (future) event horizon
can also be defined as the boundary of the region from which the null
geodesics (in our case the ``phonons'', in the \bh\ case the photons)
cannot escape. Also in this case the event horizon is a null surface
whose generators are null geodesics.

%--------------------------------------------------------------------------
\subsection[{Acoustic black holes}]{Acoustic black holes} 
\label{subsec:acbh}
%--------------------------------------------------------------------------

In the previous sections we have almost always been discussing
acoustic horizons without directly referring to acoustic black holes.
The main reason for choosing this generic phrasing is that
the above terminology can be misleading when talking about acoustic
manifolds.  In fact, if for acoustic black holes one assumes flows
which have acoustic metrics of the kind generally associated with
\bh s then this is actually equivalent to considering a very small subset
of all of the possible flows which generically show an acoustic horizon.
For a detailed discussion of all these examples see~\cite{Visser:1998yu}.

%-----------------------------------------
\subsubsection{Nozzle}
%-----------------------------------------

A particularly easy example of a geometry which shows an acoustic
horizon is that of a laminar fluid flow across a
nozzle~\cite{Unruh81}. Reducing the radius of the nozzle speeds up the
fluid flow until a velocity is reached which exceeds the local speed of
sound. The
region inside the surface at which $|\v|=\cs$ is an ergo-region and in
the case of purely spherically symmetric flow the ergo-surface is also
an event horizon. 

%---------------------------------------
\subsubsection{Vortex geometry}
%---------------------------------------

Another realization of an acoustic horizon is that obtainable from a
draining bathtub. In fact the swirling flow around the drain can
provide a useful analogy with a rotating (Kerr) \bh .

In the case of a $(2+1)$-dimensional flow, the continuity and Euler
equations, together with the requirements of absence of vorticity and
the conservation of angular momentum are enough for obtaining the
velocity of the fluid flow and the acoustic metric.  The former is
given by:
\begin{equation}
  \label{eq:vortvfl}
   \v=\frac{\left( A \hat{r}+B\hat{\theta}\right)}{r}
\end{equation}
and the metric takes the form
\begin{equation}
  \label{eq:vortex}
  \d s^{2}=-\left(\cs^2-\frac{A^2+B^2}{r^2}\right)\d t^{2}-2\frac{A}{r}\d r\d t
    -2B\d\theta\d t+\d r^2+r^2\d \theta^2
\end{equation}
We can clearly see that there is an ergosurface at 
\begin{equation}
  \label{eq:vorr}
  r_{\mathrm{ergo}}=\frac{\sqrt{A^2+B^2}}{\cs} 
\end{equation}
and from the velocity profile we can deduce that the speed of the
fluid equals that of sound for
\begin{equation}
  \label{eq:vorrho}
   r_{\mathrm{horizon}}=\frac{|A|}{\cs}
\end{equation}
%
%---------------------------------------
\subsubsection{Slab geometry}
%---------------------------------------

Another flow model often used in the literature is the stationary
one-dimensional slab flow.  Here the fluid velocity is just in the $z$
direction and the velocity profile depends just on $z$.

In this case the continuity equation implies
$\rho(z){\rm v}(z)=\mbox{constant}$ and the metric element takes the form:
\begin{equation}
  \label{eq:slab}
  \d s^2\propto \frac{1}{{\rm v}(z)\cs(z)}
       \left[-\cs(z)^2 \d t^2+\left(\d z - {\rm v}(z)\d t\right)^2 
                                            +\d x^2 +\d y^2 \right]
\end{equation}
%
%---------------------------------------
\subsubsection{\Sch -like flow}
%---------------------------------------

Although we have seen that the acoustic metrics associated with
horizons are quite different from the standard \bh\ ones, it can be
interesting to ask what is the flow with a line element like the \Sch\ 
one. If one wants to work with the class of metrics~(\ref{eq:acmetr})
which gives the easy to handle minimally coupled equation~(\ref{E:EOM})
for $\psi_{1}$, then the best that one can obtain is a metric element
which
is conformal to the Painlev\'e--Gullstrand (PG) form of the \Sch\ metric
\cite{Pain21,Gull22,Lema33,Isr300}.

This PG line element can be obtained from the \Sch\ one with a quite
unusual change of coordinates and takes the form:
\begin{equation}
  \label{eq:pg}
 \d s^2=-\cs^2\d t^2+
 \left( \d r \pm \cs \sqrt{\frac{2G_{\mathrm N}M}{r}} \d t \right)^2
         +r^2 \left( \d \theta^2+\sin^2\theta\d \phi^2\right) 
\end{equation}
Although apparently it is enough to assume that $\rho$ and $\cs$ are
constant and to set ${\rm v}=\sqrt{2G_{\mathrm N}M/r}$ to ``translate''
(\ref{eq:acmetr}) into (\ref{eq:pg}), these assumptions are actually not
compatible with the continuity equation (\ref{E:continuity}).

Instead, assuming a constant speed of sound, one can take ${\rm
  v}=\sqrt{2G_{\mathrm N}M/r}$ but then impose the validity of the
continuity equation $\vec{\nabla}\cdot(\rho\v)=0$ to deduce
$\rho\propto r^{-3/2}$.

So the final result is
\begin{equation}
  \label{eq:acpg}
 \d s^2\propto r^{-3/2} 
  \left[-\cs^2\d t^2+
  \left( \d r \pm \cs \sqrt{\frac{2G_{\mathrm N}M}{r}} \d t \right)^2
         +r^2 \left( \d \theta^2+\sin^2\theta\d \phi^2\right)\right] 
\end{equation}
\begin{center}
  \setlength{\fboxsep}{0.5 cm} \framebox{\parbox[t]{14cm}{
%------------------------------------------------------------------------      
      {\bf Comment}: A point which is often glossed over in the 
      literature is the fact that the above fluid flow has to be
      compatible with {\em both} of the fundamental hydrodynamical
      equations and so a constraint has to come also from the Euler
      equation (\ref{E:euler}). In our case this implies that a very
      special external potential $\Phi$ has to be set up in order to
      sustain a PG flow. In $d$ space dimensions this is
       \begin{equation}
        \Phi(r)=\cs^2\left(d-{3\over 2}\right)\ln\left({r\over r_0}
         \right)-{G_{\mathrm N}M\over r}+\mbox{const}.
        \label{E:pgpot}
       \end{equation}
       We shall discuss further in the next section the stability of
       the acoustic horizon in PG flow. We just stress here that such
       a special form for the potential makes the concrete
       realizability of the PG geometries quite difficult.
%-----------------------------------------------------------------------
}}
\end{center}

\subsubsection{Supersonic cavitation}

As a final example of acoustic geometry, we want to discuss the case of
a spherically symmetric flow in a constant density, inviscid fluid.  The
independence on position of $\rho$ together with the continuity
equation implies in this case that ${\rm v}\propto 1/r^2$. The barotropic
equation assures that also the pressure and the speed of sound are
position independent. It is then possible to introduce a
normalization constant $r_{0}$ and set
\begin{equation}
  \label{eq:vsc}
  {\rm v}=\cs \frac{r^{2}_{0}}{r^2}
\end{equation}
The acoustic metric than takes the form:
\begin{equation}
  \label{eq:metrsc}
  \d s^2=-\cs^2 \d t^2+
  \left( \d r \pm \cs \frac{r^{2}_{0}}{r^2} \d t \right)^2
         +r^2 \left( \d \theta^2+\sin^2\theta\d \phi^2\right) 
\end{equation}
This is not at all similar to any \bh -like metric of \GR\ but
nevertheless it has the very interesting property that its
time-dependent generalization has an easy experimental
realization~\cite{Hochberg97}. In fact it corresponds to the acoustic
metric associated with a bubble in a liquid which has an oscillating
radius $R(t)$.  In this case $r_{0}=R\sqrt{\dot{R}/\cs}$.

These bubbles have actually been produced in laboratory experiments
and are the subject of study in relation to the still unexplained
emission of light in {\em Sonoluminescence}. We shall discuss this
phenomenon extensively in the next chapter but we want to stress
already now that the analogy with black holes of sonoluminescent
bubbles certainly does not indicate that the way to explain the observed
light emission is as a type of Hawking radiation. In fact one should
remember that the sonic analogy deals with phonon creation and not
with photons.

It should nevertheless be said that estimates of the temperature of
the Hawking thermal bath of phonons from such solutions indicate that
they could have much better chances of detection than those from the
other acoustic geometries which we considered~\cite{Hochberg97}.

Now that we have seen how concepts like event horizons and black holes
can be ``exported'' into the acoustic manifold framework one may
wonder if also the Hawking radiation, which is generally associated
with horizons, has an acoustic counterpart. The answer to this
question is yes and shall now show how this works.

%--------------------------------------------------------------------------
\subsection[{Phonon Hawking radiation}]{Phonon Hawking radiation} 
\label{subsec:phr}
%--------------------------------------------------------------------------

The existence of phonon Hawking radiation can be deduced easily from
Eq.~(\ref{E:EOM}) if one postulates the presence of an acoustic
horizon.  The only real subtlety in this derivation is related to the
correct identification of the ``surface gravity'' for these horizons.
In fact it is important to keep in mind that in this case that the
surface gravity is linked to the {\em Newtonian} acceleration of
infalling observers at the horizon. This is again related to the fact
that acoustic geometries are able to really mimic \GR\ only in the
kinematical physics of the fields.  The physics that depends on the
dynamical equations of motion is instead rather different (because, as
we said, the hydrodynamic equations are not equivalent to the Einstein
ones).  In particular in the case of acoustic geometries there is a
privileged parameterization of the null geodesics which generate the
acoustic horizon in terms of the Newtonian time coordinate of the
underlying physical metric.  As a consequence of this the surface
gravity can always (even for non-stationary acoustic geometries) be
defined unambiguously.  It is nevertheless much easier to limit the
discussion to stationary cases (where moreover the equivalence with
the definitions of \GR\ is clear).

\subsubsection{Surface gravity}

In his original work, Unruh~\cite{Unruh81} found the surface gravity to
be equal to the acceleration of the fluid as it passes the event
horizon
\begin{equation}
g_{\mathrm h}\equiv\cs\frac{\partial {\rm v}}{\partial n}=a_{\mathrm{fluid}}
\end{equation}
but it easy to see that this result is valid only for a position
independent speed of sound and perpendicular flow with respect to the horizon.

An elegant generalization of the above formula was found by Visser
in~\cite{Visser98} who showed that the surface gravity has two terms,
one coming from acceleration of the fluid, the other coming from
variations in the local speed of sound
\begin{equation}
  \label{eq:acsg}
  g_{\mathrm h}\equiv \half\left| \frac{\partial}{\partial n}
                          \left( \cs^{2}-{\rm v}_{\bot}^{2} \right) 
                  \right|
          = \left| \frac{\partial}{\partial n}
                 \left( \cs-{\rm v}_{\bot} \right)
            \right|
\end{equation}
where $\v_{\bot}=v_{\bot}\hat{n}$ is the component of the flow speed
orthogonal to the horizon and $\partial/\partial n$ is the normal derivative.

As a last remark we just stress that the well known conformal
invariance of the surface gravity and hence of the Hawking
temperature~\cite{JacKan93} prevents any doubt about the validity of
the above results due to the ``conformal ambiguity'' which we previously
pointed out in the determination of the acoustic geometry.  The only
way in which a conformal factor can influence the Hawking radiation is
by backscattering effects on the metric which contribute only in
the grey-body factor.  Notably this also implies that acoustic
metrics which are just conformal to the familiar black hole metrics
(such as the example of the Painlev\'e--Gullstrand geometry discussed
above)
are perfectly suitable for discussing the radiation of phonons from
acoustic horizons.

\subsubsection{The acoustic Hawking radiation and lessons learned from it}

In the first derivation of this effect Unruh~\cite{Unruh81} postulated
just the quantization of the $\psi_{1}$ field in an acoustic geometry
which was, as we said, spherically symmetric, stationary, convergent
and admitting an ergoregion (which with these assumptions has the
event horizon as its boundary).  This is a variant of the nozzle
geometry discussed above.

Under these assumptions, the acoustic metric element can be cast in a
form which is extremely similar to the \Sch\ one close to the horizon
and so the standard Bogoliubov coefficient technique~\cite{Haw75} can
be applied and the Hawking radiation of quanta of $\psi$ derived.

The final result is that for a stationary acoustic geometry the
acoustic horizon will emit phonon radiation characterized by a
quasi-Planckian spectrum with a temperature
\begin{equation}
  \label{eq:THawac}
  k_{\mathrm B}T_{\mathrm H}=\frac{\hbar g_{\mathrm h}}{2\pi\cs}
           =\frac{\hbar}{2\pi}
            \left|\frac{\partial}{\partial n}
             \left(\cs-{\rm v}_{\bot}\right)\right|
\end{equation}
As in the \bh\ case, near to the horizon the spectrum is almost exactly
Planckian but it differs more and more from this as one goes away from
the horizon due to phonon back-scattering by the acoustic geometry.

It should be noted that this investigation has a very interesting didactic
side with respect to the major problems of quantum black holes. In particular
we want to stress here a couple of ``lessons'' which we have learned so
far.

The first issue is related to the problem of trans-Planckian
frequencies in the Hawking effect which we discussed in
Chapter~\ref{chap:1}.  In fact, as the \bh\ event horizon is an
infinite redshift surface for the modes of the quantum field, so the
acoustic horizon is a surface of infinite redshift for the sound
waves. In this sense there is a ``trans-Bohrian'' puzzle. Nevertheless
the acoustic case has the ``advantage'' of being endowed with a natural
rest frame (that of the physical geometry) and a natural short-scale
cutoff (the intermolecular distance below which the fluid-dynamical
description breaks down).  

Apparently if one imposes such a cutoff, all of the ingoing modes with
wavelength less than this, will not be there and so the corresponding
outgoing modes will not be created at all.  How then, for acoustic
horizons, can the presence of thermal distributions of the outgoing
phonons be compatible with a short wavelength cutoff? This has led to
the idea that a sort of ``mode regeneration'' is at
work~\cite{Jacobson91,Unruh94,Jacobson00}.  In fact if the modes
cannot escape from inside the horizon then somehow the high energy
outgoing modes must be generated from the interaction of lower energy
incoming ones.

Since a mode cannot be ``reflected'' at the horizon, which is not a
true boundary, the basic idea is that the mechanism must rely on a
smooth change in the group velocity of the ingoing modes.
Substantially, the dispersion relation of the phonons changes at short
wavelengths in such a way as to generate a reversal of the group
velocity and convert ingoing modes into outgoing ones.

This was proved in the case of a free scalar field in a
two-dimensional \bh\ spacetime by Unruh~\cite{Unruh94} using the sonic
framework. In his theory, the dispersion relation of the field had
large deviations from the massless case only in the short wavelength
limit. He proved that in this treatment the group velocity can reverse
close to the acoustic horizon and that some of the ingoing modes can
be converted into outgoing ones in exactly such a way as to preserve
the Hawking result.  This discussion has been developed and confirmed
in various other works, (see~\cite{Jacobson00} for a comprehensive
review).
\begin{center}
  \setlength{\fboxsep}{0.5 cm} 
   \framebox{\parbox[t]{14cm}{
%------------------------------------------------------------------------      
       {\bf Comment}: In the absence of a definitive theory of quantum
       gravity, it is still unclear how to transport such a framework
       to the case of the ``true'' Hawking radiation. In this case
       one can assume the cut-off to be the typical length at which
       the idea of a continuum spacetime loses its meaning. A natural
       candidate would then be the Planck scale. After this it is quite
       difficult to imagine what the mechanism for changing the
       dispersion relation for the photon would be. In a certain sense
       this would imply a breakdown of Lorentz invariance at high
       energies (in such a way as to change the dispersion relation for
       near Planckian modes). In this regard we can cite the work of
       Nielsen and
       collaborators~\cite{Nielsen78,Nielsen83a,Nielsen83b}, 
       who have shown that Lorentz invariance is often a symmetry in
       the low-energy limit even if the underlying physics explicitly
       breaks it.
%------------------------------------------------------------------------
}}
\end{center}

The second lesson which can be learned from acoustic black holes is
similarly deep and important.  We have just seen how Hawking radiation
can be discussed very precisely in the sonic framework. One can then
wonder whether we might be able to gain further insight about black
hole thermodynamics, and in particular about the origin of black hole
entropy.  In this last case the answer turns out to be clearly
negative but nevertheless instructive~\cite{Visser:1998yu}.

The standard derivation of the Hawking radiation does not rely in any
step on the Einstein equations.  Similarly in Euclidean quantum
gravity the Hawking temperature can be derived from just geometrical
properties (removal of the conical singularity in the $\tau-r$ plane;
see section~\ref{sec:EPI}). This explains the fact that Hawking
radiation can be ``exported'' to acoustic geometries so easily.

On the other hand, the nature of \bh\ thermodynamics is strictly
related to the Einstein equations (see e.g.~\cite{Jacobson:1995ab})
and the \bh\ entropy can be even derived as a general property of
Einstein--Hilbert-like diffeomorphism covariant actions (see
chapter~\ref{chap:2}). Substantially one can say that entropy equals
area (with eventual correction factors) only if the action is the
Einstein--Hilbert action (plus eventual correction
factors)~\cite{Visser:1998yu}.

In summary: although the acoustic geometries cannot fully mimic \GR\ 
dynamics (and hence \bh\ thermodynamics) they do teach us that the
Hawking radiation is a much more general (kinematical) phenomenon
associated with the quantization of fields in the presence of an event
horizon (or just an apparent horizon) and is independent of the
dynamical equations which underlie the geometry.

%----------------------------------------------------------------------------
\section[{Stability of acoustic horizons}]{Stability of acoustic horizons} 
\label{sec:achor}
%----------------------------------------------------------------------------

Apart from the general ``lessons'' about the nature of semiclassical
gravity results which we have just discussed, one of the reasons why
acoustic black holes are so popular is that it seems that prospects
for experimentally building an acoustic horizon are much better than
for a creating a general relativistic event horizon. An early estimate
can be found in \cite{Unruh81}, and related comments are to be found
in \cite{Visser98}. Additionally, there is an impressive body of work
due to Volovik and collaborators, who have extensively studied the
prospects for building such a system using superfluids such as
He${}^3$ and He${}^4$ \cite{Volovik:1995ja,Volovik:1997xi,
  Volovik:1997gg,Kopnin:1998jy,Volovik:1998de,
  Volovik:1997pf,Jacobson:1998ms,Volovik:1998pf,
  Jacobson:1998he,Volovik:1999zs,Volovik:1999fc,Volovik:1999cn}.

More recently, Garay {\etal} have investigated the technical
requirements for implementing an acoustic horizon in Bose--Einstein
condensates~\cite{Garay}, and some of the perils and pitfalls
accompanying acoustic black holes have been discussed in Jacobson's
mini-survey~\cite{Jacobson00}.

Obviously the main reason for studying acoustic horizons is the
possibility of actually detecting the phonon Hawking radiation and
hence indirectly testing semiclassical quantum gravity.

As a first check one can now wonder whether the temperature given in
Eq.~(\ref{eq:THawac}) is detectable or not.  It is easy to see that
\begin{equation}
  \label{eq:Thawnum}
  T_{\mathrm H}=1.2\times10^{-6}K \mm \left[\frac{\cs}{10^3 {\rm m/sec}}\right]
            \left[
                  \frac{1}{\cs}
                  \frac{\partial\left(\cs-{\rm v}_{\bot}\right)} {\partial n}
            \right]
\end{equation}
Dimensional reasons lead one to assume that 
\begin{equation}
  \label{eq:naive}
    \frac{\partial {\rm v}_{\bot}}{\partial r} \approx \frac{\cs}{R}
\end{equation}
where $R$ is a typical length scale associated with the flow (a nozzle
radius, or the radius of curvature of the horizon)~\cite{Unruh81}. Then
for supersonic flow through a $1 mm$ nozzle one has $T_{\mathrm H}\approx
10^{-6}K$.

This result is quite disappointing because this temperature is
just a little below the present conceivable limits of detection.

Nevertheless we shall now see in the next section that this estimate
is quite naive.  Our analysis will suggest that it may in general be
misleading because it does not take into account the information about
the dynamics of the flow. In this case we shall see that the effective
``surface gravity'' could be considerably larger than previously
expected.

The work presented in the following sections has been done in
collaboration with Matt Visser and Sebastiano Sonego and is mainly
contained in ref.~\cite{LSV00}.

%----------------------------------------------------------------------
\subsection[{Regularity conditions at ergo-surfaces}]
{Regularity conditions at ergo-surfaces}
\label{S:ergo-surface}
%----------------------------------------------------------------------

Let us start by establishing a useful mathematical identity.
If we write $\v=v\n$, where $\n$ is a unit vector and $v\geq
0$, then
\begin{equation}
\vec\nabla\cdot\v={\d v\over\d n}+vK,
\label{E:useful}
\end{equation}
where $\d /\d n=\n\cdot\vec\nabla$ and
$K=\vec\nabla\cdot\n$. If the Frobenius condition is
satisfied,\footnote{
%---------------------------------------------------
The Frobenius condition is $\v\cdot\vec\nabla\times\v=0$, or
equivalently $\n\cdot\vec\nabla \times\n=0$. This is
sometimes phrased as the statement that the flow has zero
``helicity''. The Frobenius condition is satisfied whenever
there exist a pair of scalar potentials such that $\v =
\alpha \vec \nabla \beta$, in which case the velocity field
is orthogonal to the surfaces of constant $\beta$. In view
of this fact the velocity field is said to be a ``surface
orthogonal vector field''.}
%----------------------------------------------------
then there exist surfaces which are everywhere orthogonal to the fluid
flow. In this situation, $K$ admits a geometrical interpretation as
the trace of the extrinsic curvature of these surfaces. It should be
noted that, although zero vorticity is a sufficient condition for this
to happen, it is not a necessary one.

We now focus our attention on the component of the fluid
acceleration along the flow direction, $a_n=\mathbf{a}\cdot\n$. This can
be obtained straightforwardly by projecting the
Navier--Stokes equation (\ref{E:euler}) along $\n$:
\begin{equation}
\rho\, a_n
=
- \cs^2\,{\d\rho\over\d n}
-\rho\,{\d\Phi\over\d n}
+ \rho\nu\,\n\cdot
\left(\nabla^2 \v+\frac{1}{3}\vec\nabla (\vec\nabla \cdot \v ) \right),
\label{E:aaa}
\end{equation}
where we have used the barotropic condition.

Next, we rewrite the continuity equation as
\begin{equation}
{\partial\rho\over\partial t}
+v\,{\d\rho\over\d n}
+\rho\left({\d v\over\d n}+vK\right)=0,
\label{E:urca}
\end{equation}
where the identity (\ref{E:useful}) has been used. We can
express $\d v/\d n$ in terms of $a_n$ noticing that, by the
definition (\ref{E:acc}) of $\a$,
\begin{equation}
a_n={\partial v\over\partial t}+v\,{\d v\over\d n}.
\label{E:boh}
\end{equation}
Thus, equation (\ref{E:urca}) can be rewritten as
\begin{equation}
\rho \; a_n
=
- v^2\,{\d\rho\over\d n}
-\rho v^2K
+ \rho\,{\partial v\over\partial t}
- v\,{\partial \rho\over\partial t}\;.
\label{E:bbb}
\end{equation}
Equations (\ref{E:aaa}) and (\ref{E:bbb}) can be solved for
both $a_n$ and $\d\rho/\d n$, obtaining:
\begin{equation}
a_n
=
{1\over \cs^2-v^2}
\left\{
  v^2\left[{\d\Phi\over\d n}- \cs^2K-
\nu\,\n\cdot
\left(\nabla^2 \v+\frac{1}{3}\vec\nabla (\vec\nabla \cdot \v ) \right)\right]
+ \cs^2\left({\partial v\over\partial t}
           -{v\over\rho}\,{\partial\rho\over\partial t}
     \right)
\right\};
\label{E:an}
\end{equation}
\begin{equation}
{\d\rho\over\d n}
=
{1\over \cs^2-v^2}
\left\{
-\rho\left[{\d\Phi\over\d n}
-v^2K
-\nu\,\n\cdot
\left(\nabla^2 \v+\frac{1}{3}\vec\nabla (\vec\nabla \cdot \v ) \right)\right]
-\rho\,{\partial v\over\partial t}
+v\,{\partial\rho\over\partial t}
\right\}.
\label{E:gradrho}
\end{equation}
In general we see that there is risk of a divergence in the
acceleration and the density gradient as $v \to \cs$, which indicates
that the ergo-surfaces must be treated with some delicacy. The fact
that gradients diverge in this limit is our key observation, and we
shall now demonstrate that this has numerous repercussions throughout
the physics of acoustic black holes.

Since $v^2=\cs^2$ at the ergo-surface it is evident that the
acceleration and the density gradient both diverge, unless
the condition
\begin{equation}
\left.{\d\Phi \over \d n}\right|_{\mathrm h}
- c_{\mathrm h}^2 K_{\mathrm h}
- \nu\,\n_{\mathrm h}\cdot
\left(\nabla^2 \v +\frac{1}{3}\vec\nabla (\vec\nabla \cdot \v )\right)_{\mathrm h}
+ \left.{\partial v\over\partial t}\right|_{\mathrm h}
- \left.{\cs \over\rho}\right|_{\mathrm h}
\,\left.{\partial\rho\over\partial t}\right|_{\mathrm h}=0
\label{E:main}
\end{equation}
is satisfied. Equation (\ref{E:main}) is therefore a relationship that
the various quantities must satisfy in order to have a physically
acceptable model. Of course, it is only a {\em necessary\/} condition,
because $a_n$ and $\d\rho/\d n$ may diverge at the ergo-surface even
when (\ref{E:main}) is fulfilled, if the quantities in square brackets
in the right hand sides of (\ref{E:an}) and (\ref{E:gradrho}) tend to
zero more slowly than $c^2-v^2$ as one approaches the ergo-surface.

For a stationary, non-viscous flow, (\ref{E:main}) reduces to
\begin{equation}
\left.{\d\Phi \over \d n}\right|_{\mathrm h}=\left[\cs^2 K\right]_{\mathrm h}.
\label{E:fine-tuning}
\end{equation}
Thus, in this case it seems that a special fine-tuning of the external
forces is needed in order to keep the acceleration and density
gradient finite at the ergo-surface. If the condition
(\ref{E:fine-tuning}) is not fulfilled but still $\nu=0$, the flow
cannot be stationary.  Near the ergo-surface, an instability will make
the time derivatives in (\ref{E:main}) different from zero, so that
they could compensate the mismatch between the two sides of
(\ref{E:fine-tuning}). More realistically, we shall see later that for
a given potential, either no horizon forms, or the flow assumes
a configuration in which (\ref{E:fine-tuning}) is automatically
satisfied.

%----------------------------------------------------------------------
\subsection[{Regularity conditions at horizons}]
{Regularity conditions at horizons}
\label{S:horizon}
%----------------------------------------------------------------------

If we now look at the ``surface gravity'' of an acoustic black hole it
is most convenient to first restrict attention to a stationary flow.
For additional technical simplicity we shall further assume that at
the acoustic horizon (the boundary of the trapped region) the fluid
flow is normal to the horizon. Under these circumstances the technical
distinction between an ergo-surface and an acoustic horizon vanishes
and we can simply define an acoustic horizon by the condition ${\rm v}\to
\cs$. Then the surface gravity is simply given by Eq.(\ref{eq:acsg})
taking ${\rm v}_{\bot}=v$. 

For us it is useful to work with the general quantity
\begin{equation}
g =
\half \left.{\d(\cs^2-v^2)\over\d n}\right.
=
\half \left.{\d \cs^2\over\d n}\right. - a_n.
\end{equation}
which, in our framework, coincides at the acoustic horizon with the
surface gravity.  Now we have, using equations (\ref{E:urca}) and
(\ref{E:boh}),
\begin{eqnarray}
{\d \cs^2\over\d n}
=
{\d \cs^2\over\d \rho} \; {\d\rho\over\d n}
%\nonumber\\
%&=&
&=&
-{\d^2 p\over\d \rho^2}
\left[
\rho\left(K + {1\over v}
{\d v\over\d n}\right)+ {1\over v} \,
{\partial\rho\over\partial t}
\right]
\nonumber\\
&=&
-{\d^2 p\over\d \rho^2}
\left[
\rho
\left(
K + {a_n\over v^2}-
{1\over v^2}\,{\partial\rho\over\partial t}
\right)
+{1\over v}\,{\partial\rho\over\partial t}
\right],
\end{eqnarray}
and so we find, using (\ref{E:an}):
\begin{eqnarray}
g
%&=&
=
\frac{\textstyle \left\{
\left(
\cs^2 + 
\frac{\textstyle \rho}{\textstyle 2}\,
\frac{\textstyle \d^2 p}{\textstyle \d \rho^2} 
\right)
\left(
K v^2-
\frac{\textstyle \partial v}{\textstyle \partial t}+
\frac{\textstyle v}{\textstyle \rho}\,
\frac{\textstyle \partial\rho}{\textstyle\partial t}
\right)
%\right.
%\nonumber\\
%&&
%\left.
-\left(
v^2 + 
\frac{\textstyle \rho}{\textstyle2}\,
\frac{\textstyle \d^2 p}{\textstyle \d \rho^2}
\right)
\left[
\frac{\textstyle \d\Phi}{\textstyle \d n} - \nu\,\n\cdot 
\left(\nabla^2 \v+\frac{1}{3}\vec\nabla (\vec\nabla \cdot \v ) \right)\right]
\right\}}
{\textstyle \cs^2-v^2}.
\end{eqnarray}

Under the present assumptions, time derivatives vanish and $v^2=\cs^2$
at the horizon, so it is now evident that the surface gravity (as well
as the acceleration and the density gradient) diverges unless the
condition
\begin{equation}
\left.{\d\Phi \over \d n}\right|_{\mathrm h}
- \left(\cs^2 K\right)_{\mathrm h}
- \nu\,\n_{\mathrm h}\cdot
\left(\nabla^2 \v+\frac{1}{3}\vec\nabla (\vec\nabla \cdot \v )
\right)_{\mathrm h}
= 0
\label{E:main2}
\end{equation}
is satisfied. For a non-viscous flow (\ref{E:main2}) again reduces to
(\ref{E:fine-tuning}), and the same considerations made about the
acceleration and density gradient apply.

Now all of this discussion is based on the assumption that acoustic
horizons can actually form, and would be of no use if it were to be
shown that something prevents the fluid from reaching the speed of
sound. In order to deal with this possibility we shall now check that
at least in some specific examples it is possible to form acoustic
horizons under the current hypotheses. For analyzing these specific
cases it is useful to consider the stationary, spherically symmetric
case.

%----------------------------------------------------------------------
\section[{Spherically symmetric stationary flow}]
{Spherically symmetric stationary flow}
\label{S:sss}
%----------------------------------------------------------------------

For simplicity, we now deal with the case of a spherically symmetric
stationary flow in $d$ space dimensions.  Additionally, for the time
being we shall assume the absence of viscosity, $\nu=0$.

For a spherically symmetric steady inflow, $\n$ is minus the radial
unit vector. Then $\d\n/\d n=\mathbf{0}$; also
\begin{equation}
{\d\over\d n} = -{\d\over\d r},
\end{equation}
and
\begin{equation}
K = - {d-1\over r}.
\end{equation}
{From} equation (\ref{E:acc}) it follows that $\mathbf{a}$ has only a
radial component, which coincides with $-a_n$ and is
\begin{equation}
a = - v^2 \; { \cs^2 {(d-1)/r} - {\d\Phi/\d r}
\over
\cs^2 - v^2 }.
\label{E:accsph}
\end{equation}
This result could also be obtained directly, without the general
treatment of the previous section. For a steady flow the continuity
equation implies
\begin{equation}
\rho \; v \; r^{d-1} = J=\mbox{constant}.
\label{E:continew}
\end{equation}
Taking the logarithmic derivative of the above equation one easily
gets
\begin{equation}
{\d \rho\over \d r} = - \rho {(d-1)\over r} - {\rho\over v}
\; {\d v\over\d r}.
\label{E:contsph}
\end{equation}
On the other hand, in this case the Euler equation (\ref{E:euler}) takes
the form
\begin{equation}
\rho \; v \; {\d v\over\d r} = -\cs^2 \;
{\d \rho\over \d r} -
\rho {\d\Phi\over\d r},
\label{E:eulsph}
\end{equation}
where we have used the barotropic condition. Equations
(\ref{E:contsph}) and (\ref{E:eulsph}) can be combined to give the
useful result
\begin{equation}
v \; {\d v\over\d r} = \cs^2
\left( {(d-1)\over r} + {1\over v} \;
{\d v\over\d r} \right) - {\d\Phi\over\d r},
\label{E:glu}
\end{equation}
which allows one to easily compute the acceleration $a=v\,\d v/\d r$
of the fluid for this specific case, recovering equation\ 
(\ref{E:accsph}), and to obtain a differential equation for the
velocity profile $v(r)$:
\begin{equation}
{\d v\over\d r} =
- v \;
{
\cs^2 {(d-1)/r} -  (\d\Phi/\d r)
\over
\cs^2 - v^2
}.
\label{E:diffeq}
\end{equation}

When it comes to calculating $g$, the same analysis as previously
developed now yields
\begin{equation}
g=
{1\over \cs^2 - v^2}
\left[
\left(v^2 + {\rho\over 2}\,{\d^2 p\over\d \rho^2}\right)
{\d\Phi\over\d r}
-\left(\cs^2 + {\rho\over 2}\,{\d^2 p\over\d \rho^2}\right)
{v^2 (d-1)\over r}
\right].
\end{equation}
Therefore the acceleration at the acoustic horizon, whose location
$r_{\mathrm h}$ is
the solution of the equation $v(r_{\mathrm h})^2=\cs(r_{\mathrm h})^2$, 
formally goes to
infinity unless the external body force satisfies the condition
\begin{equation}
\left[{\d\Phi\over\d r} - \cs^2 \;
{(d-1)\over r}\right]_{\mathrm h} = 0.
\label{E:sss-fine-tuning}
\end{equation}

Any further analysis requires one to integrate the differential
equation (\ref{E:diffeq}). However, this can be done only by assigning an
equation of state $p=p(\rho)$, and integrating simultaneously equation
(\ref{E:contsph}) in order to get the dependence of $c$ on $r$. We
consider such a specific model in the next section.

%----------------------------------------------------------------------
\subsection[{Constant speed of sound}]
{Constant speed of sound}
\label{S:css}
%---------------------------------------------------------------------

In order to get further insight, let us consider the simple case of a
fluid with a constant speed of sound,
\begin{equation}
{\d^2 p \over \d\rho^2} = 0.
\label{E:c=const}
\end{equation}

It is easy to see that, in this case, the condition
(\ref{E:sss-fine-tuning}) is also sufficient in order to keep the
physical quantities finite on the horizon. Consider equation
(\ref{E:diffeq}) and apply the L'Hospital rule in order to evaluate
$(\d v/\d r)_{\mathrm h}$. One gets
\begin{equation}
\left({\d v\over\d r}\right)^2_{\mathrm h}=
-{1\over 2}\left(\left.{\d^2\Phi\over\d r^2}\right|_{\mathrm h}
+{\cs^2 (d-1)\over r_{\mathrm h}^2}\right),
\label{E:dvdrh}
\end{equation}
and so $(\d v/\d r)_{\mathrm h}$ has a finite value. As a corollary of
(\ref{E:dvdrh}), we see that at the horizon one must have
\begin{equation}
\left.{\d^2\Phi\over\d r^2}\right|_{\mathrm h}\leq
-{\cs^2 (d-1)\over r_{\mathrm h}^2};
\label{E:d2}
\end{equation}
and so, in particular, no potential with a non-negative second derivative
can lead to a horizon on which $\d v/\d r$ is finite.

With the assumption (\ref{E:c=const}), the differential equation
(\ref{E:diffeq}) for the velocity profile can be easily integrated.
Its general solution is
\begin{equation}
\frac{1}{2}
\left[
\cs^2 \;
\ln\left( \frac{v^2}{v^2_0}\right)-v^2+v^2_0
\right]
=
-\cs^2 \;\left(d-1\right)\;\ln{\left(\frac{r}{r_0}\right)}
+ \Phi(r)-\Phi(r_0),
\label{E:d>1Kcos}
\end{equation}
where $r_0$ is arbitrary and $v_0$ is the speed of the fluid at
$r_0$.\footnote{Equation (\ref{E:d>1Kcos}) simply expresses
  Bernoulli's theorem. Indeed, it can be written in the form
  $v^2/2+\Phi(r)+h(v,r)=\mbox{const}$, where $h=\int\d p/\rho$ can be
  found from (\ref{E:contsph}).}

We now come to a crucial point of our analysis. On rewriting
(\ref{E:d>1Kcos}) as
\begin{equation}
F(r,v;r_0,v_0)=0,
\label{E:F}
\end{equation}
we
can represent the location $r_{\mathrm h}$ of the horizon, for a given
potential $\Phi$ and given boundary data $(r_0,v_0)$, as the
solution of the equation
\begin{equation}
F(r_{\mathrm h},c;r_0,v_0)=0.
\label{E:rH}
\end{equation}
On the other hand, differentiating (\ref{E:F}) and comparing with
(\ref{E:diffeq}) we can rewrite the regularity condition
(\ref{E:sss-fine-tuning}) as
\begin{equation}
{\partial F\over\partial r}(r_{\mathrm h},c;r_0,v_0)=0.
\label{E:regcon}
\end{equation}
It is clear that, if we impose the boundary data $(r_0,v_0)$, then
(\ref{E:regcon}) expresses a fine-tuning condition on $\Phi$ in order
to have $\d v/\d r$ finite at the horizon. However, we can reverse the
argument and consider the more realistic case in which one looks for a
physically acceptable flow compatible with an {\em assigned\/}
$\Phi$, {\em without\/} trying to force the boundary condition
$v(r_0)=v_0$. In this case, equations (\ref{E:rH}) and
(\ref{E:regcon}), when solved simultaneously, give the location of the
horizon, $r_{\mathrm h}$, {\em and\/} the value $v_0$ of the fluid speed at
$r_0$.  Thus, requiring regularity of the flow for a given potential
amounts to solving an eigenvalue problem, while if one insists on
assigning a boundary condition for the speed, a careful fine tuning of
$\Phi$ is needed in order to avoid infinite gradients. We stress,
however, that although from a strictly mathematical point of view both
types of problem can be considered, it is the first one that is
relevant in practice.

%----------------------------------------------------------------------
\subsection[{Examples}]{Examples}
%---------------------------------------------------------------------

We now consider some specific choices of $\Phi(r)$, in order to see
what happens.

%----------------------------------------------------------------------
\subsubsection{Constant body force}
%---------------------------------------------------------------------

Let us begin with a constant body force, with the linear potential
\begin{equation}
\Phi(r) = \mbox{\sc k} r,
\end{equation}
where $\mbox{\sc k}$ is a constant. Equation (\ref{E:d>1Kcos})
becomes, in this case,
\begin{equation}
\frac{1}{2}
\left[
\cs^2 \;
\ln\left( \frac{v^2}{v^2_0}\right)-v^2+v^2_0
\right]
=
-\cs^2\;\left(d-1\right)\;\ln{\left(\frac{r}{r_0}\right)}
+ \mbox{\sc k}(r-r_0).
\label{E:d>1Kcos2}
\end{equation}
Following the discussion at the end of section \ref{S:css}, we can
regard (\ref{E:sss-fine-tuning}) as the equation for the locations of
$r_{\mathrm h}$ where $\d v/\d r$ is finite. We have, in this case,
$r_{\mathrm h}=\cs^2(d-1)/\mbox{\sc k}$ so, excluding the uninteresting 
possibility
$r_{\mathrm h}=0$ for $d=1$, we see immediately that there can be no regular
flow with an acoustic horizon when $\mbox{\sc k}\leq 0$. For $\mbox{\sc k}>0$,
replacing the value of $r_{\mathrm h}$ for $r$ in (\ref{E:d>1Kcos2}), one can
see that no real solutions exist for $v_0$. These conclusions are in
agreement with equation (\ref{E:d2}), which implies that $\d v/\d r$
cannot be finite at $r_{\mathrm h}$, because $\d^2\Phi/\d r^2\equiv 0$ in this
case. Thus, either $v(r)\neq c$ for all values of $r$, or $\d v/\d r$
diverges at the horizon. We shall see that the second possibility is
the correct one, by examining some plots of the solution of equation
(\ref{E:d>1Kcos2}) with arbitrarily chosen boundary conditions.

Without loss of generality we can rescale the unit of distance to set
\begin{equation}
\mbox{\sc k} = \left\{
\begin{array}{c}+ \cs^2/r_0,\\0,\\-\cs^2/r_0.\end{array}
\right.
\end{equation}
Let us treat these three cases separately.

%-----------------------------------------------------------------------
%\paragraph{$\mathbf{\mbox{\sc k}>0}$}
%------------------------------------------------------------------------

For $\mbox{\sc k} > 0$ figure \ref{F:d>1K} clearly shows that there is no
obstruction to reaching the acoustic horizon. In addition, if we keep
the distance scale fixed and instead vary $\mbox{\sc k}$ we find the curves
of figure \ref{F:d=3K>1}.

%===========================================================================
\begin{figure}[htbp]
\vbox{
\hfil
\scalebox{0.42}{{\includegraphics{k+1.epsf}}}
\hfil
}
\bigskip
\caption[Plot of the solutions of equation (\ref{E:d>1Kcos}) for
several values of $d$ and $\mbox{\sc k}>0$]{
%-------------------------------------------------------------
Plot of the solutions of equation (\ref{E:d>1Kcos}) for
several values of $d$ and $\mbox{\sc k}>0$. We have first fixed
$\mbox{\sc k}=+\cs^2/r_0$, and then set $c=1$, $r_0=1$, and
$v_0=1/2$.
%-------------------------------------------------------------
}
\label{F:d>1K}
\end{figure}
%===========================================================================
%===========================================================================
\begin{figure}[htbp]
\vbox{
\hfil
\scalebox{0.42}{{\includegraphics{kd3K1.epsf}}}
\hfil
%\hbox to 1.0 in{ }
}
%\vspace{-1.5in}
\bigskip
\caption[Plot of the solutions of equation (\ref{E:d>1Kcos}) for
$\mbox{\sc k}=2,3,4$]{
%------------------------------
Plot of the solutions of equation (\ref{E:d>1Kcos}) for
$\mbox{\sc k}=2,3,4$ with $d=3$, $c=1$, $r_{0}=1$, and $v_{0}=1/2$. 
%------------------------------
}
\label{F:d=3K>1}
\end{figure}
%===========================================================================

%\clearpage

The four things to emphasize here are that:
\begin{enumerate}
\item Velocities equal to the speed of sound are indeed attained;
\item The gradient $\d v/\d r$ is indeed infinite at the acoustic horizon;
\item These particular solutions break down {\em at\/} the acoustic
  horizon and cannot be extended {\em beyond\/} it;
\item The particular solutions which we have obtained all exhibit a
  double-valued behaviour, there is a branch with subsonic flow that
  speeds up and reaches $v=c$ at the acoustic horizon; and there is a
  second supersonic branch, defined on the same spatial region, that
  slows down and reaches $v=c$ at the acoustic horizon.
  Mathematically, this happens because the equation $x=\ln x+k$, with
  $k$ a constant, has two solutions when $k<1$.
\end{enumerate}
%
%----------------------------------------------------------------------
%\paragraph{$\mbox{\sc k}=0$ (no body force)}
%---------------------------------------------------------------------

\begin{figure}[htbp]
\vbox{
\hfil
\scalebox{0.42}{{\includegraphics{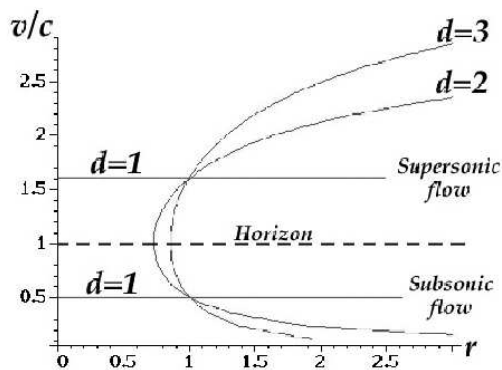}}}
\hfil
%\hbox to 1.0 in{ }
}
%\vspace{-1.5in}
\bigskip
\caption[Plot of the solution of equation (\ref{E:d>1Kcos}) for 
$\mbox{\sc k}=0$]{
%------------------------------
  Plot of the solution of equation (\ref{E:d>1Kcos}) for $\mbox{\sc k}=0$
  with the same initial values. ($c=1$, $r_0=1$, and $v_0=1/2$.) Note
  the triviality of the $d=1$ solution, which exhibits two branches,
  with subsonic and supersonic speeds respectively.
%------------------------------
}
\label{F:d>1}
\end{figure}

If there is no external body force ($\mbox{\sc k}=0$), then $d=1$ is
uninteresting (the velocity is constant). If we now look at $d=2$ and
higher then equation (\ref{E:accsph}) again easily gives us the
acceleration of the fluid
\begin{equation}
a = v \; {\d v\over\d r} =
- v^2 \;
{
\cs^2 {(d-1)/r}
\over
\cs^2 - v^2 }\;,
\end{equation}
%
%so
%
%\begin{equation}
%{\d v\over\d r} =
%- v \;{ \cs^2 {(d-1)/r} \over \cs^2 - v^2}.
%\end{equation}
%
Explicit integration leads us to the solution
\begin{equation}
r=r_0\; \left(\frac{v}{v_0}\right)^{\frac{1}{1-d}}
\exp\left(\frac{v^2-v^2_0}{2(d-1)\cs^2}\right),
\end{equation}
which is equivalent to equation (\ref{E:d>1Kcos}) with $\mbox{\sc k}=0$.
This can easily be plotted for different values of the dimension $d$
as shown in figure \ref{F:d>1}.

%==============================================================================

%\clearpage % flush all figures
%----------------------------------------------------------------------
%\paragraph{$\mbox{\sc k}<0$}
%---------------------------------------------------------------------

\begin{figure}[htbp]
\vbox{
\hfil
\scalebox{0.42}{{\includegraphics{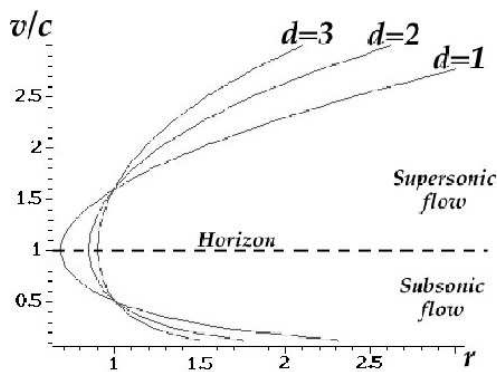}}}
\hfil
%\hbox to 1.0 in{ }
}
%\vspace{-1.5in}
\bigskip
\caption[Plot of the solution of equation (\ref{E:d>1Kcos}) for $d=1,2,3$,
  and $\mbox{\sc k}=-\cs^2/r_0$,]{
%------------------------------
  Plot of the solution of equation (\ref{E:d>1Kcos}) for $d=1,2,3$,
  and $\mbox{\sc k}=-\cs^2/r_0$, where we again set $c=1$, $r_0=1$, and
  $v_0=1/2$.
%------------------------------
}
\label{F:k=-1}
\end{figure}

For $\mbox{\sc k}<0$ the solutions are plotted in figure
\ref{F:k=-1}.

%==============================================================================
%==============================================================================

%\newpage

Finally it is interesting to compare the different behaviour of the
solutions for the different signs of the body force as shown in figure
\ref{F:d2k101}.

In all three cases ($\mbox{\sc k}>0$, $\mbox{\sc k}=0$, $\mbox{\sc
  k}<0$) we see that the acoustic horizon does in fact form as naively
expected, and that the surface gravity and acceleration are indeed
infinite at the acoustic horizon. Naturally, this should be viewed as
evidence that some of the technical assumptions usually made are no
longer valid as the horizon is approached. In particular in the next
section (section \ref{S:viscosity}) we shall discuss the role of
viscosity as a regulator for keeping the surface gravity finite.

%===========================================================================
\begin{figure}[htbp]
\vbox{
\hfil
\scalebox{0.42}{{\includegraphics{kd2K101.epsf}}}
\hfil
%\hbox to 1.0 in{ }
}
%\vspace{-1.5in}
\bigskip
\caption[Plot of the solutions of equation (\ref{E:d>1Kcos}) for
  $\mbox{\sc k}=\pm1,0$]{
%------------------------------
  Plot of the solutions of equation (\ref{E:d>1Kcos}) for
  $\mbox{\sc k}=\pm1,0$ and $d=2$, with $c=1$, $r_0=1$, and $v_0=1/2$.
%------------------------------
}
\label{F:d2k101}
\end{figure}
%===========================================================================

%=============================================================================
%\newpage
%\clearpage % flush all figures
%=============================================================================

%-------------------------------------------------------------
\subsubsection{Schwarzschild geometry}
%-------------------------------------------------------------

So far, the discussion has concerned the attainability
of acoustic horizons in general, without focusing on any particular
acoustic geometry. A more specific, and rather attractive possibility
is to attempt to build a flow with an acoustic metric that is as close
as possible to one of the standard black hole metrics of general
relativity. We have seen in section~\ref{subsec:acbh} that this is
actually conceivable and that the acoustic line element which one obtains
is
conformal to the {\Painleve}--Gullstrand form of the Schwarzschild
geometry.

This possibility has stimulated considerable work concerning the
physical realization of an experimental setup that could actually
produce such a flow (or, more precisely, a two-dimensional version of
it~\cite{Volovik:1999fc}).

We have seen that for this kind of fluid configuration the potential
must be carefully chosen to be of the form given in
Eq.~(\ref{E:pgpot}), which will not be easy to do in a laboratory. If
one does manage to construct such a potential $\Phi(r)$ it will
automatically fulfill the fine-tuning condition (\ref{E:fine-tuning})
at the acoustic horizon, $r_{\mathrm h}=2M/\cs^2$.  This is only to be
expected, because $\d v/\d r=-\sqrt{G_{\mathrm N}M/2r^3}$ diverges
only as $r\to 0$. Also, since we know that the surface gravity of a
Schwarzschild black hole is finite, any fluid flow that reproduces the
Schwarzschild geometry must by definition satisfy the fine tuning
condition for a finite surface gravity.

%===========================================================================
\begin{figure}[htbp]
\vbox{
\hfil
\scalebox{0.45}{{\includegraphics{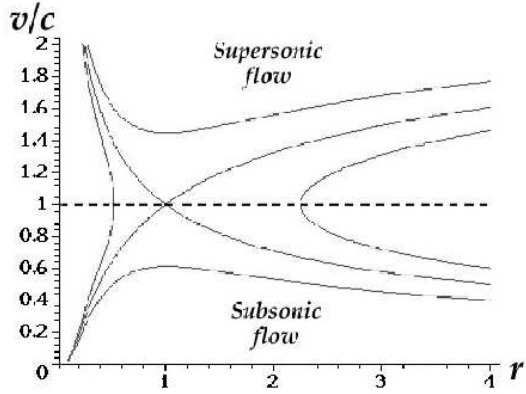}}}
\hfil
%\hbox to 1.0 in{ }
}
%\vspace{-1.5in}
\bigskip
\caption[{\Painleve}--Gullstrand flow]{
%------------------------------
  Plot of the solutions of equation (\ref{E:d>1Kcos}) for the
  potential (\ref{E:pgpot}), with $G_{\mathrm N}M=1/2$, $d=3$, $c=1$, and
  $r_0=4$. The corresponding boundary conditions are $v_0=0.4$,
  $v_0=0.5$, and $v_0=0.6$.
%------------------------------
}
\label{F:pg}
\end{figure}
%===========================================================================

Looking at the issue from the point of view discussed at the end of
section \ref{S:css}, one expects that, given the potential
(\ref{E:pgpot}), the value $v_0=\sqrt{2G_{\mathrm N}M/r_0}$ is the
solution of equations (\ref{E:rH}) and (\ref{E:regcon}), while
$v(r)=\sqrt{2G_{\mathrm N}M/r}$ is the corresponding eigenfunction
that is selected by the requirement of having a regular flow. This is
indeed the case: Equation (\ref{E:rH}) now gives $r_{\mathrm
  h}=2G_{\mathrm N}M/\cs^2$ which, when substituted into (\ref{E:regcon}),
leads to the following equation for $v_0$:
\begin{equation}
{\cs^2\over 2}\,\ln\left({2G_{\mathrm N}M\over r_0v_0^2}\right)+{v_0^2\over
2}={G_{\mathrm N}M\over r_0}.
\label{E:v0pg}
\end{equation}
It is trivial to check that $v_0=\sqrt{2G_{\mathrm N}M/r_0}$ is, in fact,
a solution of (\ref{E:v0pg}).

Considering the same potential (\ref{E:pgpot}), but values of $v_0$
different from $\sqrt{2G_{\mathrm N}M/r_0}$, corresponds to flows either with
no horizon, or in which $(\d v/\d r)_{\mathrm h}$ diverges. This is evident in
figure \ref{F:pg}, which confirms the ``eigenvalue character'' of the
problem of finding a regular flow. Notice that there are two solutions
that are regular at the horizon, with opposite values of $(\d v/\d
r)_{\mathrm h}$, in full agreement with the fact that equation (\ref{E:dvdrh})
only determines the {\em square\/} of $(\d v/\d r)_{\mathrm h}$.

%----------------------------------------------------------------------
\subsection[{Viscosity}]{Viscosity}
\label{S:viscosity}
%----------------------------------------------------------------------

Coming back to the issue of the stability of the acoustic horizon for
constant body force (the easiest sort of potential to set up
experimentally), one can ask if there are alternatives to the breakdown
of stationarity of the flow in order to avoid the divergence of the
gradients of physical quantities at the horizon. We shall now see that
a general, alternative, way for the fluid to avoid these divergences is
to develop a viscosity component at the onset of acoustic
horizons.  This way not only tames the divergences but can in principle
provide the mechanism for the mode regeneration that would allow the
presence of the Hawking radiation.

\subsubsection{General equations for viscid flow}

In the presence of viscosity, the general situation is governed by the
Navier--Stokes equation (\ref{E:euler}), from which we deduce (again
confining attention to stationary spherical symmetry)
\begin{equation}
a = v \; {\d v\over\d r} =
- {v^2 \over \cs^2 - v^2} \; \left[
\frac{\cs^2(d-1)}{r}-\frac{\d\Phi}{\d r} +
\bar{\nu} \left(
\frac{\d^2 v}{\d r^2}+\frac{(d-1)}{r}{\d v\over\d r}-\frac{(d-1)v}{r^2}
\right)\right]
\end{equation}
Here we have rescaled the viscosity coefficient in such a way that
$\bar\nu={4\over 3}\nu$. 

The acceleration is then infinite at the horizon unless
\begin{equation}
\left(1-\frac{\bar{\nu}}{\cs r_{\rm h}} \right)
{\cs^2(d-1)\over r_{\mathrm h}} - \left.{\d\Phi\over\d r}\right|_{\mathrm h} +
\bar\nu \left[{\d^2 v\over dr^2}+\frac{(d-1)}{r_{\rm h}}{\d v\over\d r}
\right]_{\mathrm h} = 0.
\end{equation}
Since this involves higher-order derivatives at the horizon,
it can no longer be regarded as a fine-tuning constraint, or
as an equation for $r_{\mathrm h}$, but merely as a statement about
the shape of the velocity profile near the horizon. 

Although it is possible to find the general solutions for $v$ and
$\d v/\d r$~\cite{LSV00}, it is unfortunately impractical to find an
analytic form for $\d^2 v/\d r^2$ and hence we must resort to the
numerical analysis of special cases.

%-------------------------------------------------------------
\subsubsection{$d=1$, constant body force}
%-------------------------------------------------------------

\begin{figure}[htbp]
\vbox{
\hfil
\scalebox{0.45}{{\includegraphics{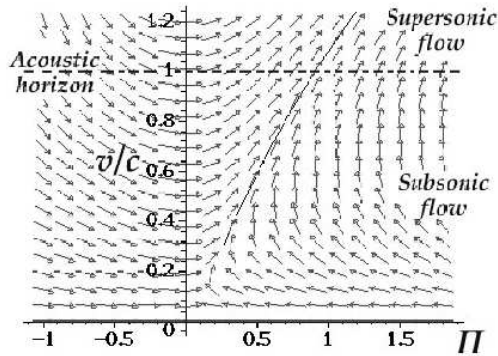}}}
\hfil
%\hbox to 1.0 in{ }
}
%\vspace{-1.5in}
\bigskip
\caption[Plot of the phase space for $d=1$, $\mbox{\sc k}=\mbox{constant} > 0$, and
  nonzero viscosity.]{
%------------------------------
  Plot of the phase space for $d=1$, $\mbox{\sc k}=\mbox{constant} > 0$, and
  nonzero viscosity. The transverse line identifies a separatrix in
  the integral curves. Note that the integral curves can intersect the
  acoustic horizon at arbitrary values of the surface gravity.
%------------------------------
}
\label{F:phase}
\end{figure}

Even the case of constant body force is intractable unless $d=1$, in
which case we get (following the steps above)
\begin{equation}
a = v \; {\d v\over\d r} =
- v^2 \;
{
-\mbox{\sc k} + \bar\nu ({\d^2 v/\d r^2}) \over \cs^2 - v^2
}.
\end{equation}
That is
\begin{equation}
{\d v\over\d r} =
- v \;
{
-\mbox{\sc k} + \bar\nu ({\d^2 v/\d r^2}) \over \cs^2 - v^2
}.
\end{equation}
This single second-order differential equation can be turned into an
{\em autonomous\/} system of first-order equations
\begin{equation}
\left \{ \begin{array}{lll}
\displaystyle{\frac{\d v}{\d r}}&=&\Pi, \\
&&\\
\displaystyle{\frac{\d \Pi}{\d r}}
&=&\displaystyle{
\frac{\mbox{\sc k}}{\bar\nu}+\frac{v}{\bar\nu}\left(1-
\frac{\cs^2}{v^2}\right)\Pi}.
\end{array}
\right.
\end{equation}
We can plot the flow of this autonomous system in the usual way and it
clearly shows that it is possible to cross the acoustic horizon $v=c$
at arbitrary accelerations $a_{\mathrm h}$ and arbitrary surface
gravity $g_{\mathrm h}$ (see figure \ref{F:phase}).

%==============================================================================
%=============================================================================

%-------------------------------------------------------------
\subsubsection{$d=1$, zero body force}
%-------------------------------------------------------------

If $d=1$ and $\mbox{\sc k} =0$ the once-integrated equation for $\d v(r)/\d
r$ reduces to
\begin{equation}
\left.{\d v\over\d r}\right|_r =
\left.{\d v\over\d r}\right|_{r_0}
-
{1\over2\bar\nu} \;
\left[
\cs^2 \;
\ln\left( \frac{v^2}{v^2_0}\right)-v^2+v^2_0
\right].
\end{equation}
In this particular case the analysis is sufficiently simple that we
can say something about the acceleration at the horizon, namely
\begin{equation}
\left.{\d v\over\d r}\right|_{\mathrm h} =
\left.{\d v\over\d r}\right|_{r_0}
-
{1\over2\bar\nu} \;
\left[
\cs^2 \;
\ln\left( \frac{\cs^2}{v^2_0}\right)-\cs^2+v^2_0
\right].
\end{equation}
That is
\begin{equation}
g_{\mathrm h} = a_{\mathrm h} = \cs \left.{\d v\over\d r}\right|_{r_0}
-
{\cs \over2\bar\nu} \;
\left[
\cs^2 \;
\ln\left( \frac{\cs^2}{v^2_0}\right)-\cs^2+v^2_0
\right].
\end{equation}
This is an explicit analytic verification that viscosity regularizes
the surface gravity of the acoustic horizon.

\begin{figure}[htbp]
\vbox{
\hfil
\scalebox{0.45}{{\includegraphics{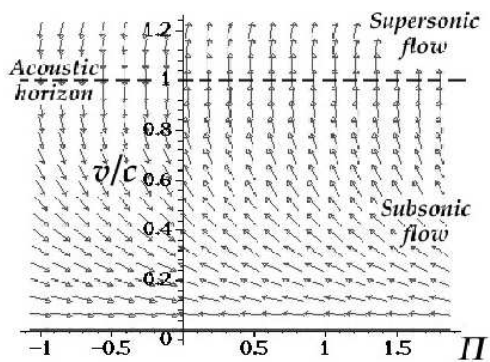}}}
\hfil
%\hbox to 1.0 in{ }
}
%\vspace{-1.5in}
\bigskip
\caption[Plot of the phase space for $d=1$, $\mbox{\sc k}=0$, and nonzero
  viscosity.]{
%------------------------------
  Plot of the phase space for $d=1$, $\mbox{\sc k}=0$, and nonzero
  viscosity. Again note that the integral curves can intersect the
  acoustic horizon at arbitrary values of the surface gravity.
%------------------------------
}
\label{F:phasek0}
\end{figure}

We can plot the flow in the usual way and it again clearly shows that
it is possible to cross the acoustic horizon $v=c$ at arbitrary
accelerations $a_{\mathrm h}$ (see figure \ref{F:phasek0}).

%==============================================================================
%=============================================================================

%-------------------------------------------------------------
\subsubsection{$d>1$, constant body force}
%-------------------------------------------------------------

For the case of $d>1$ an analytic solution cannot be found even for
constant body force. The relevant equation is
The relevant equation is
\begin{equation}
{\d^2 v\over\d r^2} +\left({(d-1)\over r}+{1\over
\bar{\nu}}{c^2-v^2\over v}\right){\d v\over\d r}-{(d-1)v\over
r^2}= {1\over \bar{\nu}}
\left(\mbox{\sc k} - {c^2(d-1)\over r}\right),
\end{equation}
which can be recast as
\begin{equation}
\left \{ \begin{array} {lll}
\displaystyle{{\d v \over \d r}} &=&\Pi \\
&&\\
\displaystyle{\frac{\d \Pi}{\d r}}
&=& {\displaystyle {1\over
\bar{\nu}}\left(\mbox{\sc k}-\frac{c^2(d-1)}{r}\right)+ {(d-1)v\over
r^2} +\left(\frac{v}{\bar{\nu}}\left(1-\frac{c^2}{v^2}\right)-
{(d-1)\over r}\right)\Pi}.
\end{array}
\right.
\label{E:non-autonomous}
\end{equation}
This is no longer an {\em autonomous} system of differential
equations, (there is now an explicit $r$ dependence) and so a flow diagram
is meaningless. Nonetheless the system
can be treated numerically and curves plotted as a function of initial
conditions. As an example we plot some curves in the phase space for
$d=2$, and verify that at least some of these curves imply formation
of an acoustic horizon.

%==============================================================================
\begin{figure}[htbp]
\vbox{
\hfil
\scalebox{0.45}{{\includegraphics{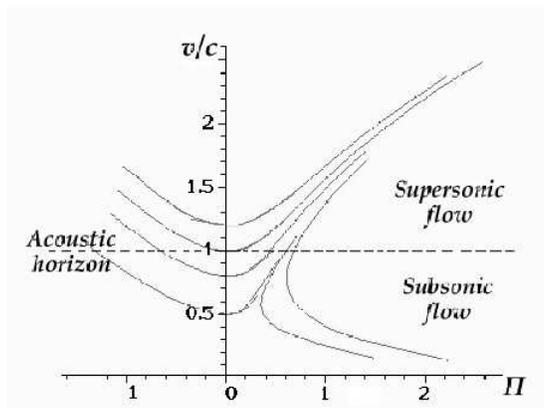}}}
\hfil
%\hbox to 1.0 in{ }
}
%\vspace{-1.5in}
\bigskip
\caption[Plot of some solutions to the non-autonomous system
  (\ref{E:non-autonomous})]{
%------------------------------
    Plot of some solutions of the non-autonomous system
    (\ref{E:non-autonomous}) for various initial conditions for $d=2$,
    $\mbox{\sc k}=\mbox{constant}$, and nonzero viscosity. Note that
    at least some of these curves intersect the acoustic horizon, and
    do so at various finite values of the surface gravity.
%------------------------------
}
\label{phased>1}
\end{figure}
%=============================================================================
%\newpage

As a final remark it can be useful to discuss briefly the effect of
viscosity in relation to the Hawking radiation. It has been shown that
the addition of viscosity to the fluid-dynamical equations is
equivalent to the introduction of an explicit violation of Lorentz
invariance at short scales~\cite{Visser98}. In this case one may
wonder whether such an explicit breakdown would not lead to a
suppression of the Hawking flux as well. In fact the violation of
Lorentz invariance is important for wavelengths of order $k\approx
k_{0}\approx \cs/\bar\nu$~\cite{Visser98}, introducing in this way a
sort of cutoff (via a modified dispersion relation) for short
wavelengths which, as we have discussed, can dramatically affect the
Hawking flux. In this case the same factor which removes the
unphysical divergences at the acoustic horizons would also ``kill''
the phenomenon which we are looking for.
  
This problem has been extensively discussed in the literature (see
e.g.~\cite{Jacobson00}) and it can be shown that such a
violation of Lorentz invariance is not only harmless but even natural.
In particular the viscosity can be shown to
induce~\cite{Visser98} those sorts of modification of the phonon
dispersion relation which are actually required for circumventing the
problem cited above~\cite{Jacobson91,Jacobson93,Unruh94,Jacobson00}.
  
The emergence of viscosity appears then to be indeed a crucial factor
for allowing the formation of acoustic horizons and, at the same time,
for implementing the mechanism of ``mode regeneration'' which should
permit Hawking radiation in the presence of a short-distance cutoff.

%----------------------------------------------------------------------
\subsection[{Discussion}]{Discussion}
%---------------------------------------------------------------------

Let us summarize the results that we have obtained for a fluid
subjected to a given external potential. In the viscosity-free,
stationary case, we have seen that if the flow possesses an acoustic
horizon, the gradients of physical quantities, as well as the surface
gravity and the corresponding Hawking flux, in general exhibit formal
divergences. There are two ways in which a real fluid can circumvent
this physically unpalatable result. For a broad class of potentials,
there is one particular flow which is regular everywhere, even at the
horizon. In this case, it is obvious that the fluid itself will
``choose'' such a configuration.  Mathematically, imposing a
regularity condition at the horizon amounts to formulating an
eigenvalue problem. Unfortunately, there are physically interesting
potentials --- such as a linear one --- for which this is impossible.
In such cases, the divergences will be avoided in reality simply
because one or both of the simplifying hypothesis (stationarity, no
viscosity), become invalid as $v\to c$.

In conclusion, we view the analytic and numerical results obtained in
this section as both a danger and an opportunity. They are a danger
because infinite accelerations are clearly unphysical and indicate
that the idealization of considering an {\em irrotational barotropic
  inviscid perfect fluid\/} (and this idealization underlies the
standard derivations of the notion of the acoustic
metric~\cite{Unruh81,Visser93,Visser98}), is sure to break down in the
neighborhood of any putative acoustic horizon.

On the other hand, this may be viewed as an opportunity: once we
regulate the infinite surface gravity, by adding for instance a finite
viscosity, we find that the surface gravity becomes an extra free
parameter, divorced from naive estimates based on the geometry of the
fluid flow (\ref{eq:naive}). This suggests that detecting the acoustic
Hawking effect could be much easier than expected once one manages to
actually form an acoustic horizon.

%--------------------------------------------------------------------------
\section[{Prospects: condensed matter black holes?}]
{Prospects: condensed matter black holes?}
\label{sec:optbh}
%--------------------------------------------------------------------------

We will end this chapter by discussing a recent development in the
study of possible laboratory realizations of event horizons. The basic
point which makes conceivable their formation in hydrodynamics is that
fluid perturbations in their propagation experience a Lorentzian
structure which has a ``limit'' velocity equal to the sound speed
$\cs$. Since a fluid can in principle flow at supersonic speeds (such
as, for example, in the case of shock waves) then it is possible to
conceive the formation of horizons for phonons.

Another possibility that one can be tempted to explore is to instead
work in a regime where light itself is ``slowed down'' to such an
extent that a fluid flow can be faster than the light velocity. This
slowing down of light is now a concrete reality: in fact, more and
more refined techniques now exist to allow us to make the refractive
index of some materials extremely large. Experimental physicists have
now managed to get refractive indices up to $n \approx 10^7$ giving a
light speed of order $v=c/n\approx 10$meters/second~\cite{Hau}.  This
makes the proposal to build {\em optical hydrodynamical black holes}
extremely appealing.

Recently Leonhardt and Piwnicki~\cite{LP99,LP00a,LP00b} have proposed
a model of a dielectric fluid where the propagation of electromagnetic
waves is described by an effective metric which is actually an
algebraic combination of the refractive index, the fluid velocity
and the background Minkowski metric~\footnote{ 
%-----------------------------------------------------------------------  
  This is actually a development of a formulation of electromagnetism
  in dispersionless media in terms of an effective metric proposed by
  Gordon in 1923~\cite{Gordon}
%----------------------------------------------------------------------
  }.  In particular, in the case of non-dispersive media light
experiences a metric of the form~\cite{LP99}:
\begin{equation}
 g_{\mu\nu}(t,\x) \equiv 
 \left[ \matrix{( \frac{1}{n^2}-
                   \frac{{\rm u}^{2}}{c^2} )&\vdots& 
                +{{\rm u}\over c}^j\cr
                \cdots\cdots\cdots\cdots&\cdot&\cdots\cdots\cr
                +{{\rm u}\over c}^i &
                \vdots &
                -\delta_{ij}\cr }
 \right].  
\label{eq:optmetr}             
\end{equation}
Here $c$ is the speed of light in vacuum and the above form is valid
in the limit of large refractive index $n$ and low local velocities
$\uu$ (with respect to $c$) of the medium.

The analogy of this electromagnetic metric with the acoustic metric
given in Eq.~(\ref{eq:acmetr}) is quite a strict one. It is then easy to
see that most of the structures discussed in acoustic geometry
can have a correspondence in the case of dielectric flows. In the same
way, all of the differences between acoustic manifolds and \GR\ 
similarly apply in this case. 

The most important point is that it is conceivable~\cite{Visser00} to
build up fluid flows with nozzle or vortex geometries which accelerate
a low-compressibility dielectric fluid to velocities which somewhere
have radial components faster that the phase velocity of light. Once
an ``optical event horizon'' is built up then a {\em photon} Hawking
flux should be expected. This will have a near-thermal distribution
characterized by a Hawking temperature proportional to the
acceleration of the fluid as it crosses the horizon. Unfortunately the
large values of the refractive index mentioned above have so far been
obtained for very dispersive media ($n(\omega)$ has a resonance for
some special frequencies). The analogy between these optical black
holes and the acoustic ones is not so strict and so no easy
exportation of the results can be made.

Finally one can wonder whether some analogue of the easily set up
supersonic cavitation configuration which we saw before can be found
in the promising case of a dielectric fluid.  This introduces us to
our discussion in the next chapter where we shall develop a model
based on the dynamics of the refractive index for explaining the
radiation observed in Sonoluminescence which we mentioned before.

Our approach to Sonoluminescence will make use of spatio-temporal
changes in refractive index induced by the fluid dynamics.  Though
there are some weak formal similarities with the phonon Hawking
radiation of this current chapter, it must be emphasised that the
basic physics is rather different.

%%%%%%%%%%%%%%%%%%%%%%%%%%%%%%%%%%%%%%%%%%%%%%%%%%%%%%%%%%%%%%%%%%%%%%%%%%%%
% S.Liberati Ph.D. Chapter 4: Toward experimental semiclassical gravity
%%%%%%%%%%%%%%%%%%%%%%%%%%%%%%%%%%%%%%%%%%%%%%%%%%%%%%%%%%%%%%%%%%%%%%%%%%%%
\chapter[{Sonoluminescence}]
{Toward experimental semiclassical gravity: Sonoluminescence}
\label{chap:3b}
%Signature (+,-,-,-)
%%%%%%%%%%%%%%%%%%%%%%%%%%%%%%%%%%%%%%%%%%%%%%%%%%%%%%%%%%%%%%%%%%%%%%%%%%%%

\vspace*{0.5cm} \rightline{\it Chi disputa allegando l'autorit\`a non
  adopra lo 'ngenio,}\rightline{\it ma pi\`u tosto la
  memoria.\footnote{Who discuss using his authority does not use the
    intelligence but the memory instead.}}  
\vspace*{0.5cm} \rightline{\sf Leonardo da Vinci}

\vspace*{1cm} 
\rightline{\it A belief is like a guillotine,}
\rightline{\it just as heavy, just as light.}
\vspace*{0.5cm} \rightline{\sf Franz Kafka}

\newpage

In this chapter we shall pursue our study of the possible experimental
tests of the dynamical Casimir effect.  In particular we shall focus
our attention on the phenomenon of {\em Sonoluminescence}.  We shall
develop an original model for explaining the origin of the radiation
emitted in this process, based on dynamical particle production
from the quantum vacuum driven by a time varying refractive index
(which acts as an external field perturbing the QED vacuum).  Although
the model developed could be greatly improved if further information
about the realistic dynamics of the refractive index could be gained
from condensed matter studies, it has to be stressed that it is already
able to propose some tests amenable to observations.

The contents of this chapter are original and represent an extract from
a ``corpus'' of works~\cite{Qed0,PRL,SnPr,Qed1,Qed2,2Gamma} dedicated to
the subject. All of these have been done in collaboration with
Francesco Belgiorno, Matt Visser and Dennis Sciama without the
enthusiasm and guidance of whom they would never have appeared.

%%%%%%%%%%%%%%%%%%%%%%%%%%%%%%%%%%%%%%%%%%%%%%%%%%%%%%%%%%%%%%%%%%%%%%%%%
\section[{Sonoluminescence as a dynamical Casimir Effect}]
{Sonoluminescence as a dynamical Casimir Effect}
\label{sec:sonol}
%%%%%%%%%%%%%%%%%%%%%%%%%%%%%%%%%%%%%%%%%%%%%%%%%%%%%%%%%%%%%%%%%%%%%%%%%

Sonoluminescence (SL) is the phenomenon of light emission by a
sound-driven gas bubble in a fluid \cite{Physics-Reports}. In SL
experiments, the intensity of a standing sound wave is increased until
the pulsations of a bubble of gas trapped at a velocity node have
sufficient amplitude to emit brief flashes of light having a
``quasi-thermal'' spectrum with a ``temperature'' of several tens of
thousands of Kelvin.  The basic mechanism of light production in this
phenomenon is still highly controversial. We first present a brief
summary of the main experimental data (as currently understood) and of
their sensitivities to external and internal conditions.  For a more
detailed discussion see~\cite{Physics-Reports}.

SL experiments usually deal with bubbles of air in water, with ambient
radius $R_{\mathrm{ambient}} \approx 4.5 \; \mu {\rm m}$.  The bubble
is driven by a sound wave of frequency of 20--30 kHz.  (Audible
frequencies can also be used, at the cost of inducing deafness in the
experimental staff.)  During the expansion phase, the bubble radius
reaches a maximum of order $R_{\mathrm{max}}\approx \; 45 \;\mu {\rm
  m}$, followed by a rapid collapse down to a minimum radius of order
$R_{\mathrm{min}}\approx 0.5 \; \mu {\rm m}$. 

%%%==========================================================
\begin{figure}[htb]
 \vbox{\hfil
\scalebox{1.00}{\includegraphics{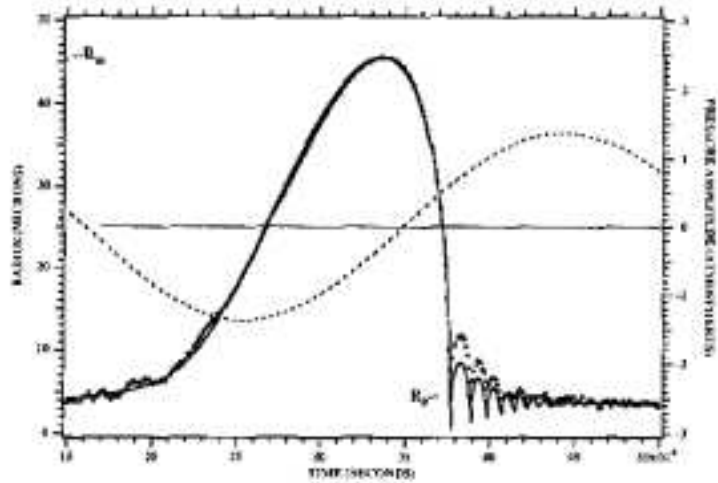}}
\hfil}
 \bigskip
\caption[Radius of sonoluminescent bubble as a function of time]{
%----------------------------------------------------------------------  
  Typical variation of the bubble radius as a function of time and in
  relation to the sound wave cycle (dotted line).  Note the
  peculiar oscillatory behaviour after the collapse. The picture is
  taken from ref.~\cite{Physics-Reports}.
%---------------------------------------------------------------------
}\label{F:bounce}
\end{figure}
%%%==================================================================

\begin{figure}[htb]
 \vbox{\hfil
\scalebox{0.70}{\includegraphics{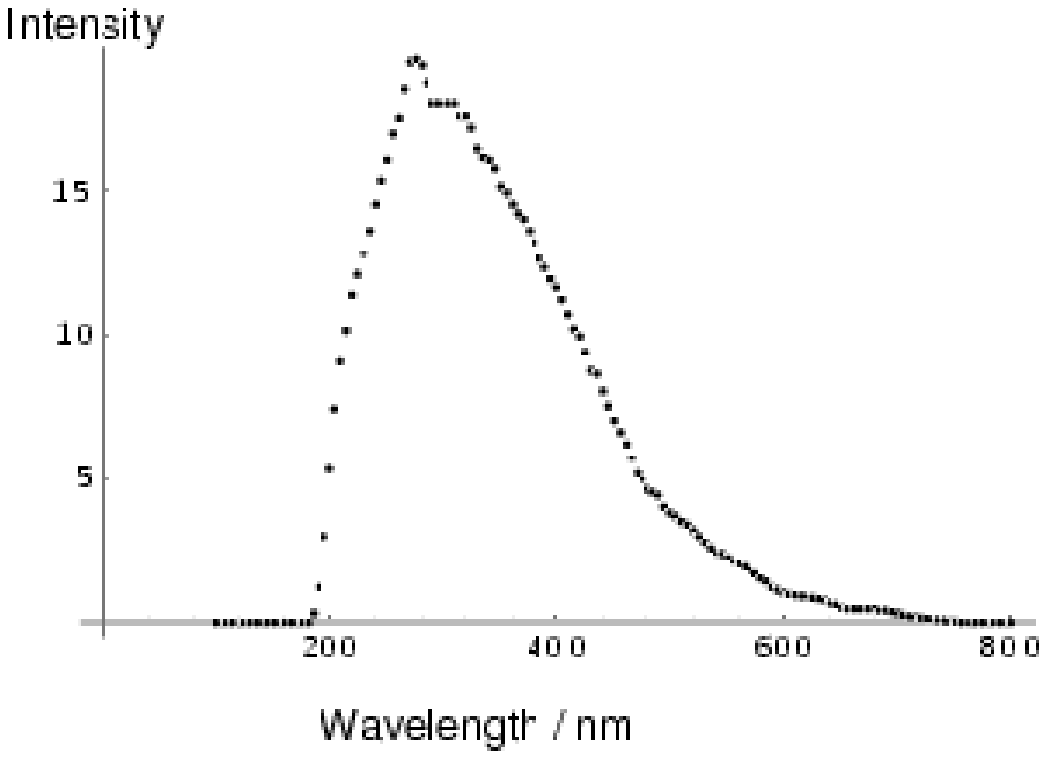}}
\hfil}
 \bigskip
\caption[Sonoluminescence: experimental spectrum in wavelengths]{
%----------------------------------------------------------------------  
  Typical experimental spectrum: The data has been taken from figure
  51 of reference 1, and has here been plotted with intensity (in
  arbitrary units) given as a function of wavelength. Note that no
  data has been taken at frequencies below the visible range. The
  spectrum is a broad-band spectrum without significant structure. The
  physical nature of the cutoff (which occurs in the far ultraviolet)
  is one of the key issues under debate.
%---------------------------------------------------------------------
}\label{F:experiment-1}
\end{figure}

\begin{figure}[htb]
\vbox{\hfil
\scalebox{0.70}{\includegraphics{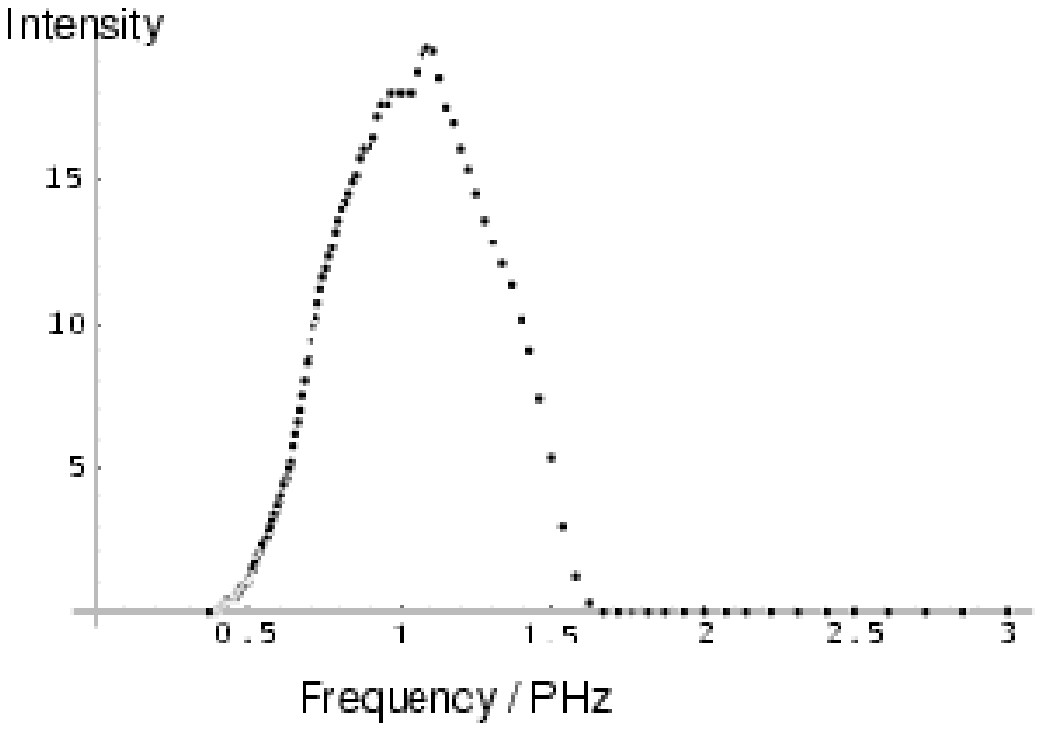}}
\hfil}
\caption[Sonoluminescence: experimental spectrum in frequencies]{ 
%---------------------------------------------------------------------- 
  Typical experimental spectrum: The data has been taken from figure
  51 of reference 1, and has here been plotted with the number
  spectrum given as a function of frequency.
%----------------------------------------------------------------------
}\label{F:experiment-2} 
\end{figure}

The photons emitted by such a pulsating bubble have typical
wavelengths of the order of visible light. The minimum observed
wavelengths range between $200\; {\rm nm}$ and $100\; {\rm nm}$. This
light appears distributed with a broad-band spectrum (no resonance
lines, roughly a power-law spectrum with exponent depending on the gas
admixture entrained in the bubble, and with a cutoff in the extreme
ultraviolet). For a typical example, see figures~\ref{F:experiment-1}
and~\ref{F:experiment-2}. If one fits the data to a Planck black-body
spectrum, the corresponding temperature is several tens of thousands
of Kelvin (typically $70,000\; {\rm K}$, though estimates varying from
$40,000 \; {\rm K}$ to $100,000 \; {\rm K}$ are common). There is
considerable doubt as to whether or not this temperature parameter
corresponds to any real physical temperature.  There are about one
million photons emitted per flash, and the time-averaged total power
emitted is between $30$ and $100\; \hbox{ mW}$.

%%%==========================================================
%%%==================================================================
%%%====================================================================

The photons appear to be emitted from a very small spatio-temporal
region: Estimated flash widths vary from less than $35 \; {\rm ps}$ to
more than $380 \; {\rm ps}$ depending on the gas entrained in the
bubble \cite{Flash1,Flash2}. There are model-dependent (and
controversial) claims that the emission times and flash widths do not
depend on wavelength \cite{Flash2}. As for the spatial scale, there
are various model-dependent estimates but no direct measurement is
available \cite{Flash2}.  Although it is clear that there is a
frequency cutoff at about $1 \; {\rm PHz}$, the physics leading to
this is controversial. Standard explanations are
\begin{enumerate}
\item a thermal cutoff (deprecated because the observed cutoff is much
  sharper than exponential) , or
\item the opacity of water in the UV (deprecated because of the
  observed absence of dissociation effects).  Alternatively, Schwinger
  suggested that the critical issue is that the real part of the
  refractive index of water goes to unity in the UV (so that there is
  no change in the Casimir energy during bubble collapse). We shall
  add another possible contribution to the mix:
\item a rapidly changing refractive index causes photon production
  with an ``adiabatic cutoff' that depends on the timescale over which
  the refractive index changes. (Because the observed falloff above
  the physical cutoff is super-exponential it is clear that this
  adiabatic effect is at most part of the complete picture.)
\end{enumerate}
Any truly successful theory of SL must also explain a whole series of
characteristic sensitivities to different external and internal
conditions. Among these dependencies the main one is certainly the
mysterious catalytic role of a noble gas admixture. (Most often, this
amounts to a few percent in the air entrained in the bubble. One can
obtain SL from air bubbles with a $1\%$ content of argon, and also
from pure noble gas bubbles, but the phenomenon is practically absent
in pure oxygen bubbles.) In fact, it has been suggested that physical
processes concentrate the noble gases inside the bubble to the extent
that the bubble consists of almost pure noble gas~\cite{Argon}, and
some experimental results seem to corroborate this
suggestion~\cite{Matula}.

Other external conditions that influence SL are: 
\begin{itemize}
\item Magnetic Fields --- If the frequency of the driving sound wave
  is kept fixed, SL disappears above a pressure-dependent threshold
  magnetic field: $H \geq H_{0}(p)$. On the other hand, for a fixed
  value of the magnetic field $H_{0}$, there are both upper and lower
  bounds on the applied pressure that bracket the region of SL, and
  these bounds are increasing functions of the applied magnetic
  field~\cite{YSK}.  This is often interpreted as suggesting that the
  primary effect of magnetic fields is to alter the condition for
  stable bubble oscillations. See. also~\cite{Magnetic}.
\item Temperature of the water --- If $T_{H_{2}O}$ decreases then the
  emitted power $W$ increases.  The position of the peak of the
  spectrum depends on $T_{H_{2}O}$. It has been suggested that the
  increased light emission at lower water temperature is associated
  with an increased stability of the bubble, allowing for higher
  driving pressures~\cite{Hilgen}.
\end{itemize}
These are only the most salient features of the SL phenomenon.  In
attempting to explain such detailed and specific behaviour the
dynamical Casimir approach (QED vacuum approach) encounters problems
which are equivalent to those in all of the other approaches. Nevertheless
we
shall argue that SL explanations using a Casimir-like framework are
viable, and merit further investigation.

%-------------------------------------------------------------------------
\subsection[{Casimir models: Schwinger's approach}]
{Casimir models: Schwinger's approach}
\label{subsec:Schw}
%-------------------------------------------------------------------------

The idea of a ``Casimir route'' to SL is due to Schwinger who several
years ago wrote a series of papers~\cite{Sc1,Sc2,Sc3,Sc4,Sc5,Sc6,Sc7}
regarding the so-called dynamical Casimir effect. Considerable
confusion has been caused by Schwinger's choice of the phrase
``dynamical Casimir effect'' to describe his particular model. In
fact, Schwinger's original model is not dynamical but is instead
quasi-static as the heart of the model lies in comparing two static
Casimir energy calculations: one for an expanded bubble and the other
for a collapsed bubble. One key issue in Schwinger's model is thus
simply that of calculating Casimir energies for dielectric spheres ---
and there is already considerable disagreement on this issue. A second
and in some ways more critical question is the extent to which this
difference in Casimir energies may be converted into real photons
during the collapse of the bubble.  The original quasi-static
incarnation of the Schwinger model had no real way of estimating
either photon production efficiency or timing information ({\em when}
does the flash occur?).  In contrast, the model of
Eberlein~\cite{Eberlein1,Eberlein2,Eberlein3} (more fully discussed
below) is truly dynamical but uses a much more specific physical
approximation --- the adiabatic approximation.  The two models should
not be confused. We shall argue that the observed features of SL force
one to make the sudden approximation. We can then estimate the
spectrum of the emitted photons by calculating an appropriate
Bogoliubov coefficient relating two states of the QED vacuum. The
resulting variant of the Schwinger model for SL is then rather tightly
constrained, and should be amenable to experimental verification (or
falsification) in the near future.

In his papers on SL, Schwinger showed that the dominant bulk
contribution to the Casimir energy of a bubble (of dielectric constant
$\epsilon_{\mathrm{inside}}$) in a dielectric background (of dielectric
constant $\epsilon_{\mathrm{outside}}$) is~\cite{Sc4}
\begin{eqnarray}
 E_{\mathrm{cavity}}
 &=&
 +2\frac{4\pi}{3}R^3 \; \int_0^K
 {4 \pi k^2 \d k\over(2\pi)^3}    \; \frac{1}{2}  \hbar c k
 \left(
 {1\over\sqrt{\epsilon_{\mathrm{inside}}}}- 
 {1\over\sqrt{\epsilon_{\mathrm{outside}}}}
 \right) +\cdots
 \nonumber\\
 &=&+\frac{1}{6 \pi} \hbar c R^3 K^4
 \left(
 {1\over\sqrt{\epsilon_{\mathrm{inside}}}}- 
 {1\over\sqrt{\epsilon_{\mathrm{outside}}}}\right) +\cdots
\label{E:schwinger}
\end{eqnarray}
The corresponding number of emitted photons is
\begin{eqnarray}
 N &=&
 +2\frac{4\pi}{3}R^3 \; \int_0^K
 {4 \pi k^2 \d k\over(2\pi)^3}    \; \frac{1}{2}
 \left(
 \frac{\sqrt{\epsilon_{\mathrm{outside}}}}{\sqrt{\epsilon_{\mathrm{inside}}}}
  - 1 \right) +\cdots
 \nonumber\\
 &=&+\frac{2}{9 \pi} (RK)^3
 \left(
 \frac{\sqrt{\epsilon_{\mathrm{outside}}}}{\sqrt{\epsilon_{\mathrm{inside}}}}
  - 1 \right) +\cdots
\label{N:schwinger}
\end{eqnarray}   
Here we have inserted an explicit factor of two with respect to
Schwinger's papers to take into account both photon polarizations.
There are additional sub-dominant finite volume effects discussed
in~\cite{CMMV1,CMMV2,MV}. Schwinger's result can also be rephrased in
a clearer and more general form as~\cite{CMMV1,CMMV2,MV}:
\begin{equation}
 E_{\mathrm{cavity}} = + 2 V \int_0^K \frac{\d^3\vec{k}}{(2 \pi)^3} \; 
 \frac{1}{2}
 \hbar  \left[ \omega_{\mathrm{inside}}(k) - 
 \omega_{\mathrm{outside}}(k)  \right] + \cdots
\end{equation}
\begin{equation}
 N = + 2 V \int_0^K \frac{\d^3\vec{k}}{(2 \pi)^3} \; 
 \frac{1}{2}
 \left[{\omega_{\mathrm{inside}}(k)\over\omega_{\mathrm{outside}}(k)}
 - 1 \right]
 +\cdots
\end{equation}
Here it is evident that the Casimir energy can be interpreted as a
difference in zero point energies due to the different dispersion
relations inside and outside the bubble.  The quantity $K$ appearing
above is a high-wavenumber cutoff that characterizes the wavenumber at
which the (real part of) the refractive indices ($n=\sqrt{\epsilon}$)
drop to their vacuum values. It is important to stress that this
cutoff it is not a regularization artifact to be removed at the end of
the calculation. The cutoff has a true physical meaning in its own
right.

The two main points of strength of models based on zero point
fluctuations ({\em e.g.}, Schwinger's model and its variants) are:
\begin{enumerate}
\item There is no actual production of far ultraviolet photons
  (because the refractive index goes to unity in the far ultraviolet)
  and so one does not expect the dissociation effects in water that
  other models imply. Models based on the quantum vacuum automatically
  provide a cutoff in the far ultraviolet because of the behaviour of
  the refractive index --- this observation goes back to Schwinger's
  first papers on the subject.
\item One naturally gets the right energy budget. For $n_{\mathrm
    {outside}} \approx 1.3$, $n_{\mathrm{inside}} \approx 1$, $K$ in
  the ultra-violet, and $R\approx R_{\mathrm{max}}$, the change in the
  static Casimir energy is approximately equal to the energy emitted
  during each cycle.
\end{enumerate}
This last issue has been the subject of a heated debate.  Milton and
collaborators~\cite{M95,M96,M97,M98}, Brevik {\em et al}~\cite{Brevik}
and Nesterenko and Pirozhenko~\cite{Nesterenko}, have strongly
criticized Schwinger's result claiming that actually the Casimir
energy contains at best a surface term, the bulk term being discarded
via (in our opinion) a physically dubious renormalization argument.
These points have been discussed extensively in~\cite{CMMV1,CMMV2,MV}
where it is emphasized that one has to compare two different physical
configurations of the same system, corresponding to two different
geometrical configurations of matter, and thus must compare {\em two}
different quantum states defined on the {\em same} spacetime\footnote{
%---------------------------------------------------------------------
  This point of view is also in agreement with the bag model results
  of Candelas~\cite{Candelas}.  It is easy to see that in the bag
  model one finds a bulk contribution that happens to be zero only
  because of the particular condition that $\epsilon \mu = 1$
  everywhere. This condition ensures the constancy of the speed of
  light (and so the invariance of the dispersion relation) on all
  space while allowing the dielectric constant to be less than one
  outside the vacuum bag (as the model for quark confinement
  requires).}.
%--------------------------------------------------------------------
In a situation like Schwinger's model for SL one has to subtract from
the zero point energy (ZPE) for a vacuum bubble in water the ZPE for
water filling all space.  It is clear that in this case the bulk term
is physical and {\em must} be taken into account. Surface terms are
also present, and eventually other higher order correction terms, but
they prove not to be dominant for sufficiently large
cavities~\cite{MV}.

While the contentious issues of how to define the
Casimir energy are successfully dealt with in~\cite{CMMV1,CMMV2,MV},
one of the subsidiary aims of the research which we shall present here has
been to side-step this whole argument and provide an independent
calculation demonstrating efficient photon production. After the
former works I think that the issue is now completely fixed and
Schwinger's correctness vindicated.

In contrast to the points of strength outlined above, the main
weakness of the original quasi-static version of Schwinger's idea is
that there is no real way to calculate either timing information or
conversion efficiency.  A naive estimate is to simply and directly
link power produced to the change in volume of the bubble.  As pointed
out by Barber {\em et al.}~\cite{Physics-Reports}, this assumption
would imply that the main production of photons may be expected when
the fractional rate of change of the volume is a maximum, which is
experimentally found to occur near to the maximum radius. However, the
emission of light is experimentally found to occur near the point of
minimum radius, where the rate of change of area is maximum. Everything 
else
being equal, this would seem to indicate a surface dependence and
might be interpreted as a true weakness of the dynamical Casimir
explanation of SL.  In fact we shall show that the situation is
considerably more complex than might naively be thought, but before
this we shall discuss the other main Casimir model for Sonoluminescence.

%-------------------------------------------------------------------------
\subsection[{Eberlein's dynamical model for SL}]
{Eberlein's dynamical model for SL}
\label{subsec:Ebe}
%-------------------------------------------------------------------------

The quantum vacuum approach to SL was developed extensively in the
work of Eberlein~\cite{Eberlein1,Eberlein2,Eberlein3}.  The basic
mechanism in Eberlein's approach is a particular implementation of the
dynamical Casimir effect: Photons are assumed to be produced due to an
{\em adiabatic} change of the refractive index in the region of space
between the minimum and the maximum bubble radius (a related
discussion for time-varying but spatially-constant refractive index
can be found in the discussion by Yablonovitch~\cite{Yablonovitch}).
This physical framework is actually implemented via a boundary between
two dielectric media which accelerates with respect to the rest frame
of the quantum vacuum state. The change in the zero-point modes of the
fields produces a non-zero radiation flux.  Eberlein's contribution
was to take the general phenomenon of photon generation by moving
dielectric boundaries and attempt a specific implementation of these
ideas as a candidate for explaining SL.

It is important to realize that this is a second-order effect. Although
he was unable to provide a calculation to demonstrate it, Schwinger's
original discussion is posited on the direct conversion of zero point
fluctuations in the expanded bubble vacuum state into real photons
plus zero point fluctuations in the collapsed bubble vacuum state.
Eberlein's mechanism is a more subtle (and much weaker) effect
involving the response of the atoms in the dielectric medium to
acceleration through the zero-point fluctuations. The two mechanisms
are quite distinct and considerable confusion has been engendered by
conflating the two mechanisms. Criticisms of the Eberlein mechanism do
not necessarily apply to the Schwinger mechanism, and vice versa.

In the Eberlein analysis the motion of the bubble boundary is taken
into account by introducing a velocity-dependent perturbation to the
usual Hamiltonian of electromagnetism:
\begin{eqnarray}
 H_{\epsilon} &\!=&\! 
 \frac{1}{2} \int{\rm d}^3{\bf r} 
 \left(
 {{\bf D}^2\over\epsilon} + {\bf B}^2 
 \right)\;,
\\
 \Delta H &\!=&\!  
 \beta
 \int{\rm d}^3{\bf r} 
 \frac{\epsilon -1}{\epsilon}\;
 ({\bf D}\wedge{\bf B})\cdot{\bf\hat r}\;.
\end{eqnarray}
This is an approximate low-velocity result coming from a power series
expansion in the speed of the bubble wall $\beta= \dot R/c$.  The
bubble wall is known to collapse with supersonic velocity, values of
Mach 4 are often quoted, but this is still completely non-relativistic
with $\beta\approx 10^{-5}$.

The Eberlein calculation consists of a novel mixture of the standard
adiabatic approximation with perturbation theory. In principle, the
adiabatic approximation requires the knowledge of the complete set of
eigenfunctions of the Hamiltonian for any allowed value of the
parameter $\beta$. In the present case only the eigenfunctions of part
of the Hamiltonian, namely those of $H_\epsilon$, are known (and these
are known explicitly only in the adiabatic approximation where
$\epsilon$ is treated as time-independent.) The calculation consists
of initially invoking the standard application of the adiabatic
approximation to the full Hamiltonian, then formally calculating the
transition coefficients for the vacuum to two photon transition to
first order in $\beta$, and finally in explicitly calculating the
radiated energy and spectral density. In this last step Eberlein used
an approximation valid only in the limit $k R\gg1$ which means the
limit of photon wavelengths much smaller than the bubble radius. This
implies that the calculation will completely miss any resonances that
are present.

Eberlein's final result for the energy radiated over one acoustic
cycle is:
\begin{equation}
 {\cal W} = 1.16\:\frac{(n^2-1)^2}{n^2}\,\frac{1}{480\pi}
 \left[{\hbar\over c^3}\right]
 \int_{0}^{T} {\rm d}\tau\; 
 \frac{\partial^5 R^2(\tau)}{\partial \tau^5}\,R(\tau)\beta(\tau)\;.
\end{equation}
Eberlein approximates $n_{\mathrm{inside}} \approx n_{\mathrm{air}}
\approx 1$ and sets $n_{\mathrm{outside}} = n_{\mathrm{water}} \to n$.
The $1.16$ is the result of an appropriate numerical integration. The
precise nature of the semi-analytic approximations made as a prelude to
performing the numerical integration are far from clear.

By a double integration by parts, the above formula can be re-cast as
\begin{equation}
 {\cal W} = 1.16\:\frac{(n^2-1)^2}{n^2}\,\frac{1}{960\pi}
 \left[{\hbar\over c^4}\right]
 \int_{0}^{T} {\rm d}\tau\; 
 \left(\frac{\partial^3 R^2(\tau)}{\partial \tau^3}\right)^2.
\end{equation}
Then the energy radiated is also seen to be proportional to
\begin{equation}
 \int_0^T (\dot R)^2 (\ddot R)^2 + ...
\end{equation}
explicitly showing that the acceleration of the interface ($\ddot
R$) and the strength of the perturbation ($\dot R$) both contribute
to the radiated energy.

The main advantage of this model over the original Schwinger model
is the ability to provide basic timing information: In this mechanism
the massive burst of photons is produced at and near the turn-around
at the minimum radius of the bubble. There the velocity rapidly
changes sign, from collapse to re-expansion. This means that the
acceleration is peaked at this moment, and so are higher derivatives
of the velocity.  Other main points of strength of the Eberlein model
are the same as previously listed for the Schwinger model plus the
ability to explicitly show that one does {\em not} need to achieve
``real'' temperatures of thousands of Kelvin inside the bubble. 

As discussed in section~\ref{subsec:squeezed}, quasi-thermal
behaviour is generated in quantum vacuum models by the squeezed nature
of the two photon states created \cite{Eberlein2}, and the
``temperature'' parameter is a measure of the squeezing, not a measure
of any real physical temperature~\footnote{
%----------------------------------------------------------------------    
  This ``false thermality'' must not be confused with the very
  specific phenomenon of the Unruh temperature.  In that case, valid only
  for {\em uniformly accelerated} observers in flat spacetime, the
  temperature is proportional to the constant value of the
  acceleration. The temperature of the quasi-Planckian spectrum found
  by Eberlein is not just a function of the first derivative of the
  bubble wall speed.}.
%----------------------------------------------------------------------

Of course, one should remember that the experimental data merely
indicates an approximately power-law spectrum [$N(\omega) \propto
\omega^\alpha$] with some sort of cutoff in the ultraviolet, and with
an exponent that depends on the gases entrained in the bubble; the
much-quoted ``temperature'' of the SL radiation is merely an
indication of the scale of this cutoff $K$.
   
In spite of these points of strength, Eberlein's model exhibits
significant weaknesses:
\begin{enumerate}
\item The calculation is based on an adiabatic approximation which
  does not seem consistent with results.  The adiabatic approximation
  would seem to be justified in the SL case by the fact that the
  frequency $\Omega$ of the driving sound is of the order of tens of
  kHz, while that of the emitted light is of the order of $10^{15}$
  Hz. But if one takes a timescale for bubble collapse of the order of
  milliseconds, or even microseconds, then photon production is
  extremely inefficient, being exponentially suppressed [as we shall
  soon see] by a factor of $\exp(-\omega/\Omega)$.
  
  In order to compare with the experimental data, the model requires,
  as external input, the time dependence of the bubble radius.  This
  is expressed as a function of a parameter $\gamma$ which describes
  the time scale of the collapse and re-expansion process.  In order
  to be compatible with the experimental values for emitted power
  $\cal W$ one has to fix $\gamma \approx 10\; {\rm fs}$. This is far
  too short a time to be compatible with the {\em adiabatic}
  approximation.  Although one may claim that {\em the precise
    numerical value of the timescale} can ultimately be modified by
  the eventual inclusion of resonances it would seem reasonable to
  take this ten femtosecond figure as a first self-consistent
  approximation for the characteristic timescale of the driving system
  (the pulsating bubble).  Unfortunately, the characteristic timescale
  of the collapsing bubble then comes out to be of the same order as
  the characteristic period of the emitted photons, {\em violating the
    adiabatic approximation used in deriving the result}. Attempts at
  bootstrapping the calculation into self-consistency instead bring it
  to a regime where the adiabatic approximation underlying the scheme
  cannot be trusted.

\item The Eberlein calculation cannot deal with any resonances that
  may be present.  Eberlein does consider resonances to be a possible
  important correction to her model, but she is considering
  ``classical'' resonances (scale of the cavity of the same order as
  the wavelength of the photons) instead of the probably more
  interesting possibility of parametric resonances.
\end{enumerate}
Finally one should mention a recent calculation that gives
qualitatively the same results as the Eberlein model although leading
to different formulae. \Schutzhold, Plunien, and Soff~\cite{SPS} adopted
a slightly different decomposition into unperturbed and perturbing
Hamiltonians and their result for the total energy emitted per cycle is given
analytically by
\begin{equation}
 {\cal W} = {n^2(n^2-1)^2 \over 1890\pi}
 \left[{\hbar\over c^6}\right]
 \int_{0}^{T} {\rm d}\tau\; 
 \left(\frac{\partial^4 R^3(\tau)}{\partial \tau^4}\right)^2.
\end{equation}
The key differences are that this formula is analytic (rather than
numerical) and involves fourth derivatives of the volume of the bubble
(rather than third derivatives of the surface area).  The main reason
for the discrepancy between this and Eberlein's result can be seen as
being due to a different choice of the dependence on $r$ of $\beta(r,t)$.
In reference~\cite{SPS} they considered the more physical case of a
localized disturbance that yields significant contributions only over
a bounded volume. (Eberlein makes the simplifying assumption that
$\beta(r,t)$ is a function of $t$ only, which is incompatible with
continuity and the essentially constant density of water. In contrast,
\Schutzhold\ {\em et al.} take the radial velocity of the water
outside the bubble to be $\beta(r,t) = f(t)/r^2$.)

Putting these models aside, and before proposing new routes for
developing further research in SL, I shall give below a more detailed
discussion of some important points of Schwinger's model which seem to
be crucial in order to understand the possibility of a vacuum
explanation for SL.

%-------------------------------------------------------------------------
\subsection[{Timescales: The need for a sudden approximation}]
{Timescales: The need for a sudden approximation}
\label{subsec:ts}
%-------------------------------------------------------------------------

One of the key features of photon production by a space-dependent and
time-dependent refractive index is that for a change occurring on a
timescale $\tau$, the amount of photon production is exponentially
suppressed by an amount $\exp(-\omega\tau)$. 

The importance for SL is that the experimental spectrum is {\em not\,}
exponentially suppressed at least out to the far ultraviolet.
Therefore any mechanism of Casimir-induced photon production based on
an adiabatic approximation is destined to failure: Since the
exponential suppression is not visible out to $\omega \approx 10^{15}
\hbox{ Hz}$, it follows that {\em if\,} SL is to be attributed to
photon production from a time-dependent external field ({\em i.e.},
to a dynamical Casimir effect), {\em then} the timescale for change of
this field should be of order a {\em femtosecond}.  

In the Eberlein model of sonoluminescence the time varying external
field is the bubble refractive index. Its variation (consequential to
the change in the bubble radius) perturbs the QED vacuum and hence
leads to the emission of photons. Unfortunately the bubble radius
changes on a much longer time scale than femtoseconds so any
Casimir--based model has to take into account that {\em the change in
  the refractive index cannot be due just to the change in the bubble
  radius}.

The SL flash is known to occur at or shortly after the point of
maximum compression. The light flash is emitted when the bubble is at
or near minimum radius $R_{\mathrm{min}} \approx 0.5\;\mu \hbox{m} =
500\;\hbox{nm}$.  Note that to get an order femtosecond change in
refractive index over a distance of about $500\;{\rm nm}$, the change
in refractive index has to propagate at relativistic speeds.  To
achieve this, one must adjust basic aspects of the model: We will move
away from the original Schwinger suggestion, in that it is no longer
the collapse from $R_{\mathrm{max}}$ to $R_{\mathrm{min}}$ that is
important.  {\em Instead we will postulate a rapid (order femtosecond)
change in refractive index of the gas bubble when it hits the van der
Waals hard core.}

It is important to stress that, once one takes into account the
refractive index of the final state, this condition can be somewhat
relaxed. We shall ultimately see in section~\ref{subsec:numest} that
we can tolerate a refractive index that changes as slowly as on a
picosecond timescale, but this is still far to rapid to be associated
with physical collapse of the bubble.  For the time being we focus on
the femtosecond timescale (which actually makes things more difficult
for us) to check the physical plausibility of the scenario, but keep
in mind that eventually things can be relaxed by a few orders of
magnitude.

The underlying idea is that there is some physical process that gives
rise to a sudden change of the refractive index inside the bubble when
it reaches maximum compression.  We have to ensure that the velocity
of mechanical perturbations, that is the sound velocity, can be a
significant fraction of the speed of light in this critical regime.
Actually this condition can also be slightly relaxed: one can conceive
of the change in refractive index being driven by a shock wave that
appears at the van der Waals hard core. Now a shock wave is by
definition a supersonic phenomenon. If the velocity of sound is itself
already extremely high then the shock wave velocity may be even higher.

Most of the viable models for the gas dynamics during the collapse
predict the formation of strong shock waves; so we are adapting
physics already envisaged in the literature. At the same time we are
asking for less extreme conditions ({\em e.g.}, we can be much more
relaxed regarding the focusing of these shock waves) in that we just
need a rapid change of the refractive index of the entrained gas, and
do not need to propose any overheating to ``stellar'' temperatures.
Noticeably dramatic changes in the refractive index (but that of the
surrounding water) due to the huge compression generated by shock waves
have already been considered in the literature~\cite{HRB} ({\em cf.}
page 5437).

%-------------------------------------------------------------------------
\subsection[{Bubbles at the van der Waals hard core}]
{Bubble at the van der Waals hard core}
\label{subsec:vWhc}
%-------------------------------------------------------------------------

In support of this proposal I shall first show that the minimum radius
experimentally observed is of the same order as the van der Waals hard
core radius $R_{\mathrm{hc}}$.  The latter can be deduced as follows:
It is known that the van der Waals excluded volume for air is
$b=$0.036 l/mol~\cite{Wu-Roberts}.  The minimum possible value of the
volume is then $V_{\mathrm{hc}} = b\cdot (\rho V_{\mathrm{ambient}})/m$,
where $(\rho V_{\mathrm{ambient}})/m$ gives the number of moles and
$V_{\mathrm{ambient}}$ is the ambient value of the volume. {From}
$R_{\mathrm{hc}} = R_{\mathrm{ambient}} \cdot (b \rho/\mu)^{1/3}$ and
assuming for the density of air $\rho=10^{-3}\; {\rm gr}/{\rm cm}^3$
[$1.3 \times 10^{-3} \; {\rm gr}/{\rm cm}^3$ at STP (standard
temperature and pressure)], one gets $R_{\mathrm hc}\sim 0.48\; \mu
{\mathrm m}$.  This value compares favorably with the experimentally
observed value of $R_{\mathrm{min}}$.  Moreover, the role of the van
der Waals hard core in limiting the collapse of the bubble is
suggested in~\cite{Physics-Reports} (cf.  fig. 10, p.78), and a
careful hydrodynamic analysis for the case of an Argon
bubble~\cite{Fluid} reveals that for Sonoluminescence it is necessary
that the bubble undergoes a so called strongly collapsing phase where
its minimum radius is indeed very near the hard core radius\footnote{
%--------------------------------------------------------- 
  Noticeably, from~\cite{Fluid} it is easy to estimate the van der
  Waals radius for the Argon bubble: $R_{\mathrm{hc}} \approx
  R_{\mathrm{ambient}}/8.86$, with $R_{\mathrm{ambient}}=0.4 \mu {\rm
    m}$.  This again gives $R_{\mathrm{hc}}(\mbox{Argon}) \approx 0.45 \; \mu
  {\rm m} \approx R_{\mathrm{min}}$.}.
%-------------------------------------------------------

It is crucial to realize that a van der Waals gas, when compressed to
near its maximum density, has a speed of sound that goes relativistic.
To see this, write the (non-relativistic) van der Waals equation of
state as
\begin{equation}
p 
= {n k T \over 1-n b} - a n^2 
= {\rho kT/m \over 1- \rho/\rho_{\mathrm{max}}} 
- a {\rho^2\over m^2}.
\end{equation}
Here $n$ is the number density of molecules; $\rho$ is mass density;
$m$ is average molecular weight ($m=28.96\; \hbox{amu/molecule} =
28.96\; \hbox{gr/mol}$ for air).

Now consider the (isothermal) speed of sound for a van der Waals gas
\begin{equation}
v_{\mathrm{sound}}^2 
= \left({\partial p \over \partial \rho}\right)_{T} 
= {(kT/m) \over (1- \rho/\rho_{\mathrm{max}})^2} 
- 2a {\rho\over m^2}.
\end{equation}
Near maximum density this is
\begin{equation}
v_{\mathrm{sound}} 
\approx {\sqrt{kT/m} \over (1- \rho/\rho_{\mathrm{max}}) },
\end{equation}
and so it will go relativistic (formally infinite) for densities close
enough to maximum density~\footnote{
%---------------------------------------------------------------------------
  It should be noted that similar unphysical divergences also affect the
  ``shock wave''--based models~\cite{Physics-Reports}: Indeed, the
  Mach number of the shock formally diverges as the shock implodes
  towards the origin ({\em cf.} page 126 of \cite{Physics-Reports}).
  One way to overcome this type of problem is the suggestion that that
  very near the minimum radius of the bubble there is a breakdown of
  the hydrodynamic description~\cite{Fluid}. If so, the thermodynamic
  description in terms of state equations should probably be
  considered to be on a heuristic footing at best.}.
%-------------------------------------------------------------------------- 

It is {\em only} for the sake of simplicity that it has been considered
the isothermal speed of sound. One does not expect the process of bubble
collapse and core bounce to be isothermal.  Nevertheless, this
calculation is sufficient to demonstrate that in general the sound
velocity becomes formally infinite (and it is reasonable that it goes
relativistic) at the incompressibility limit (where it hits the van
der Waals hard core).  This conclusion is not limited to the van der
Waals equation of state, and is not limited to isothermal (or even
isentropic) sound propagation (see~\cite{Qed1} for further details).
There is something of a puzzle in the fact that hydrodynamic
simulations of bubble collapse do not see these relativistic effects.
Notably, the simulations by Moss {\em et al}~\cite{Moss-et-al} seem to
suggest collapse, shock wave production, and re-expansion all without
ever running into any van der Waals hard core. The fundamental reason
for this is that the model equation of state they choose does not have
any hard core for the bubble to bounce off.  Instead, there are a number
of free parameters in their equation of state which are chosen in such
a way as to make their equation of state stiff at intermediate
densities, even if their equation of state is by construction always
soft at van der Waals hard core densities. If the equation of state is
made sufficiently stiff at intermediate densities then a bounce can be
forced to occur long before hard core densities are encountered.
However, as previously mentioned, the experimental data and
hydrodynamic analysis seem to indicate that hard core densities {\em
  are} achieved at maximum bubble compression.

Given all this, the use of a relativistic sound speed is now
physically justifiable, and the possibility of femtosecond changes in
the refractive index is at least physically plausible (even though we
cannot say that femtosecond changes in refractive index are
guaranteed).

So our new physical picture is this: The ``in state'' is a small
sphere of gas, radius about 500 nm, with some refractive index $\ngi$
embedded in water of refractive index $\nl$. There is a sudden
femtosecond change in refractive index, essentially at constant
radius, so the ``out state'' is gas with refractive index $\ngo$
embedded in water of refractive index $\nl\;$.  In view of this
femtosecond change of refractive index, we would be justified in
making the sudden approximation for frequencies less than about a PHz.

%-------------------------------------------------------------------------
\section[{Sonoluminescence as a dynamical Casimir effect: 
Homogeneous model}]
{Sonoluminescence as a dynamical Casimir effect: Homogeneous model}
\label{sec:homog}
%-------------------------------------------------------------------------

As a first approach to the problem of estimating the spectrum and
efficiency of photon production we can start by studying in detail the
basic mechanism of particle creation and to test the consistency
of the Casimir energy proposals previously described.  With this
aim in mind we shall look at the effect of a changing dielectric constant
in a homogeneous medium.  At this stage of development, we shall not be
concerned with the detailed dynamics of the bubble surface, and
confine attention to the bulk effects, deferring consideration of
finite-volume effects to the next section.

Let us consider two different asymptotic configurations.  An
``in'' configuration with refractive index $\ni$, and an ``out''
configuration with a refractive index $\no$.  These two configurations
will correspond to two different bases for the quantization of the
field.  (For the sake of simplicity we take, as Schwinger did, only
the electric part of QED, reducing the problem to a scalar
electrodynamics).  The two bases will be related by Bogoliubov
coefficients in the usual way.  Once we determine these coefficients
we easily get the number of created particles per mode and from
this the spectrum.  Of course it is evident that such a model cannot
be considered a fully complete and satisfactory model for SL.  This
present calculation must still be viewed as a test calculation in
which basic features of the Casimir approach to SL are investigated.

In the original version of the Schwinger model it was usual to
simplify calculations by using the fact that the dielectric constant
of air is approximately equal 1 at standard temperature and pressure
(STP), and then dealing only with the dielectric constant of water
($n_{\mathrm{liquid}} = \sqrt{\epsilon_{\mathrm{outside}}} \approx
1.3$). One should avoid this temptation on the grounds that the
sonoluminescent flash is known to occur within $500$ picoseconds of
the bubble achieving minimum radius. Under these conditions the gases
trapped in the bubble are close to the absolute maximum density
implied by the hard core repulsion incorporated into the van der Waals
equation of state.  Gas densities are approximately one million times
atmospheric and conditions are nowhere near STP.  For this reason we
shall explicitly keep track of both initial and final refractive indices.

I shall now describe a simple analytically tractable model for the
conversion of zero point fluctuations (Casimir energy) into real
photons.  The model describes the effects of a time-dependent
refractive index in the infinite volume limit. We shall see that
for sudden changes in the refractive index the conversion of
zero-point fluctuations is highly efficient, being limited only by
phase space, whereas adiabatic changes of the refractive index lead
to exponentially suppressed photon production.

%-------------------------------------------------------------------------
\def\curl{\nabla \times}
\def\div{\nabla \cdot}
\def\grad{\nabla}
%-------------------------------------------------------------------------

%-------------------------------------------------------------------------
\subsection[{Defining the model}]
{Defining the model}
%-------------------------------------------------------------------------

Take an infinite homogeneous dielectric with a permittivity
$\epsilon(t)$ that depends only on time, not on space. The homogeneous
($\d F=0$) Maxwell equations are
\begin{equation}
B = \curl A;
\end{equation}
\begin{equation}
E = - \grad \phi - {1\over c} {\partial A\over\partial t};
\end{equation}
while the source-free inhomogeneous ($*\d *F=0$) Maxwell equations
become
\begin{equation}
\div (\epsilon E) = 0;
\end{equation}
\begin{equation}
\curl {B} = 
+{1\over c} {\partial\over\partial t} (\epsilon E). 
\end{equation}
Substituting into this last equation
\begin{equation}
\curl\left(\curl A\right)  =
- {1\over c} {\partial\over\partial t}
\left[  \epsilon \left( 
\grad \phi + {1\over c} {\partial A\over\partial t}
\right) \right].
\end{equation}
Suppose that $\epsilon(t)$ depends on time but not space, then
\begin{equation}
( \grad (\div A) - \nabla^2 A )= 
-  \grad {1\over c} {\partial\over\partial t}
(\epsilon \phi ) -  {1\over c^2 } {\partial\over\partial t} \epsilon 
{\partial A\over\partial t}.
\end{equation}
Now adopt a {\em generalized} Lorentz gauge
\begin{equation}
\div A  + 
{1\over c} {\partial\over\partial t} (\epsilon \phi ) = 0.
\end{equation}
Then the equations of motion reduce to 
\begin{equation}
{1\over c^2 } {\partial\over\partial t} \epsilon 
{\partial A\over\partial t}  = \nabla^2 A.
\end{equation}
We now introduce a ``pseudo-time'' parameter by defining
\begin{equation}
{\partial\over\partial \tau} = \epsilon(t) {\partial \over \partial t}.
\end{equation}
That is
\begin{equation}
\tau(t) = \int {\d t\over\epsilon(t)}.
\end{equation}
In terms of this pseudo-time parameter the equation of motion is 
\begin{equation}
{\partial^2\over\partial\tau^2} A = 
c^2 {\epsilon(\tau)}\nabla^2 A.
\label{eqm}
\end{equation}
Compare this with equation (3.86) of Birrell and
Davies~\cite{Birrell-Davies}. Now pick a convenient profile for
the permittivity and permeability as a function of this pseudo-time.
(This particular choice of time profile for the refractive index
is only to make the problem analytically tractable, with a little
more work it is possible to consider generic monotonic changes of
refractive index and place bounds on the Bogoliubov
coefficients~\cite{Visser}.) Let us take
\begin{eqnarray}
\label{E:profile}
{\epsilon(\tau)} &=& a + b \tanh(\tau/\tau_0)
\\
&=& \half (\ni^2 + \no^2) + \half(\no^2-\ni^2)\;\tanh(\tau/\tau_0).
\end{eqnarray}
Here $\tau_0$ represents the typical timescale of the change of the
refractive index in terms of the pseudo-time we have just defined.  We
are interested in computing the number of particles that can be
created passing from the ``in'' state ($t\to-\infty$, that is,
$\tau\to-\infty$) to the ``out'' state ($t\to+\infty$, that is,
$\tau\to+\infty$). This means we must determine the Bogoliubov
coefficients that relate the ``in'' and ``out'' bases of the quantum
Hilbert space.  Defining the inner product as:
\begin{equation}
(\phi_1,\phi_2) =
\im \int_{\Sigma_\tau} \phi_1^* \;
{\stackrel{\leftrightarrow}{\partial}\over\partial\tau}  \;
\phi_2\: \d^3x,
\label{eq:hoinpro}
\end{equation}  
The Bogoliubov coefficients can now be {\em defined} as\\
\begin{eqnarray}
\alpha_{ij}
&=&
-({A_{i}^\out },{A_{j}^\inn }),
\\
\beta_{ij}
&=&(
{A_{i}^\out }^{*}, {A_{j}^\inn }).
\end{eqnarray}
Where ${A_{i}^\inn }$ and ${A_{j}^\out }$ are solutions
of the wave equation (\ref{eqm}) in the remote past and remote future
respectively.  We shall compute the coefficient $\beta_{ij}$ It is
this quantity that is linked to the spectrum of the ``out'' particles
present in the ``in'' vacuum, and it is this quantity that is related
to the total energy emitted.  With a few minor changes of notation we
can just write down the answers directly from pages 60--62 of Birrell
and Davies~\cite{Birrell-Davies}.  Birrell and Davies were interested
in the problem of particle production engendered by the expansion of
the universe in a cosmological context. Although the physical model is
radically different here the mathematical aspects of the analysis
carry over with some minor translation in the details.  Equations
(3.88) of Birrell--Davies become
\begin{equation}
\omega^\tau_\inn  
= k \sqrt{a-b}
= k \sqrt{\epsilon_\inn } 
= k \; \ni ;
\end{equation}
\begin{equation}
\omega^\tau_\out  
= k \sqrt{a+b} 
= k \sqrt{\epsilon_\out } 
= k \; \no ;
\end{equation}
\begin{equation}
\omega^\tau_{\pm} = 
\half \; k \; \left| \ni  \pm  \no  \right| = 
\half | \omega^\tau_\inn \pm \omega^\tau_\out|.
\end{equation}
(Here it should be emphasized that these frequencies are those
appropriate to the ``pseudo-time'' $\tau$.)  The Bogoliubov $\alpha$
and $\beta$ coefficients can be easily deduced from Birrell--Davies
(3.92)+(3.93)
\begin{eqnarray}
\alpha(\vec k_\inn , \vec k_\out )  
&=& \frac{\sqrt{\omega^\tau_\out \; \omega^\tau_\inn }}{\omega^\tau_{+}} \;\;
{\Gamma(-\im \omega^\tau_\inn  \tau_0) \;
 \Gamma(-\im \omega^\tau_\out  \tau_0) \over
 \Gamma(-\im \omega^\tau_{-} \tau_0)^2 } \;
\delta^3(\vec k_\inn  - \vec k_\out  )\\
\beta(\vec k_\inn , \vec k_\out )  
&=& -
\frac{\sqrt{\omega^\tau_\out \; \omega^\tau_\inn }}{\omega^\tau_{-}} \;\;
{\Gamma(-\im \omega^\tau_\inn  \tau_0) \;
 \Gamma(\im \omega^\tau_\out  \tau_0) \over
 \Gamma(\im \omega^\tau_{-} \tau_0)^2 } \;
\delta^3(\vec k_\inn  + \vec k_\out  ).
\end{eqnarray}
Now square, using Birrell--Davies (3.95). One obtains\footnote{
%-----------------------------------------------
Note that these are the Bogoliubov coefficients for a scalar field
theory. For QED in the infinite volume limit the two photon
polarizations decouple into two independent scalar fields and these
Bogoliubov coefficients can be applied to each polarization state
independently. Finite volume effects are a little trickier.}
%------------------------------------------------
%
\begin{eqnarray}
|\beta(\vec k_\inn , \vec k_\out )|^2 
&=& 
{\sinh^2(\pi\omega^\tau_{-}\tau_0)\over
  \sinh(\pi\omega^\tau_\inn \tau_0) \;
 \sinh(\pi\omega^\tau_\out \tau_0)} \;
{V\over (2\pi)^3 } \; 
\delta^3(\vec k_\inn  + \vec k_\out  ).
\end{eqnarray}
We now need to translate this into physical time, noting that
asymptotically, in either the infinite past or the infinite future,
$t \approx \epsilon \tau + (\mbox{constant})$, so that for physical
frequencies
\begin{equation}
\omega_\inn  = 
{\omega^\tau_\inn \over\epsilon_\inn } = 
{\omega^\tau_\inn \over n_\inn ^2} = 
{k \sqrt{a-b}\over \epsilon_\inn }  = 
k \sqrt{1\over\epsilon_\inn } 
=  {k \over n_\inn };
\end{equation}
\begin{equation}
\omega_\out  = 
{\omega^\tau_\out \over\epsilon_\out } = 
{\omega^\tau_\out \over n_\out ^2} = 
{k \sqrt{a+b}\over \epsilon_\out } = 
k \sqrt{1\over\epsilon_\out } 
= {k \over n_\out }.
\end{equation}
Note that there is a symmetry in the Bogoliubov coefficients under
interchange of ``in'' and ``out''.  

We also need to convert the timescale over which the refractive index
changes from pseudo-time to physical time. To do this we define
\begin{equation}
t_0 \equiv \tau_0 \left.{\d t\over \d\tau}\right|_{\tau=0}.
\end{equation}
For the particular temporal profile we have chosen for analytic
tractability this evaluates to
\begin{equation}
t_0 = \half  \tau_0 \left( \ni^2 + \no^2 \right).
\end{equation}
After these substitutions, the (squared) Bogoliubov coefficient becomes
\begin{eqnarray}
&&|\beta(\vec k_\inn , \vec k_\out )|^2={V\over (2\pi)^3 } \; 
\delta^3(\vec k_\inn  + \vec k_\out)  
\nonumber \\
&&\qquad \times
{
\sinh^2\left(
\pi\; 
{\textstyle |\ni^2 \omega_\inn -\no^2 \omega_\out| \over\textstyle \ni^2+\no^2}
\; t_0
\right)
\over
\sinh\left(
2\pi \; {\textstyle \ni^2 \over \textstyle \ni^2+\no^2} \; \omega_\inn t_0
\right) \;
\sinh\left(
2\pi \; {\textstyle \no^2 \over \textstyle \ni^2+\no^2} \; \omega_\out t_0
\right)
} \;.
\label{bog2}
\end{eqnarray}
We now consider two limits, the adiabatic limit and the sudden limit,
and investigate the physics.

%-----------------------------------------------------------------------
\subsubsection[{Sudden limit}]
{Sudden limit}
%-------------------------------------------------------------------------

Take
\begin{equation}
\max \{\omega^\tau_\inn , \omega^\tau_\out , \omega^\tau_-\} \; \tau_0 \ll 1.
\end{equation}
This corresponds to a rapidly changing refractive index. In terms of
physical time this is equivalent to
\begin{equation}
2\pi\;\max 
\left\{1, {\ni\over\no}, \half\left|{\ni\over\no}-1\right|\right\} \; 
{ \no^2 \over \ni^2+\no^2} \; \omega_\out t_0  \ll 1,
\end{equation}
which can be simplified to yield
\begin{equation}
2\pi\;\max \{\ni,\no\} \; 
{ \no \over \ni^2+\no^2} \; \omega_\out t_0  \ll 1.
\end{equation}
So the sudden approximation is a good approximation for frequencies
{\em less} than $\Omega_{\mathrm{sudden}}$, where we define
\begin{equation}
\label{E:timescale}
\Omega_{\mathrm{sudden}} = {1\over 2\pi t_0} \; 
{\ni^2+\no^2\over \no \; \max\{\ni,\no\}}.
\end{equation}
This shows that the frequency up to which the sudden approximation
holds is not just the reciprocal of the timescale of the change in the
refractive index: there is also a strong dependence on the initial and
final values of the refractive indices. This implies that we can
relax, for some ranges of values of $\ni$ and $\no$, our figure of
$t_0\sim O({\rm fs})$ by up to a few orders of magnitude.
Unfortunately the precise shape of the spectrum is heavily dependent
on all the experimental parameters ($K,\ni,\no,R$). This
discourages us from making any sharp statement regarding the exact
value of the timescale required in order to fit the data.

In the region where the sudden approximation holds the various
$\sinh(x)$ functions in equation (\ref{bog2}) can be replaced by their
arguments $x$. Then
\begin{equation}
|\beta|^2 \propto 
{ (\pi [\ni-\no])^2 \over 
(2\pi \ni) \; (2\pi \no) }. 
\label{E:sb}
\end{equation}
More precisely
\begin{equation}
|\beta(\vec k_\inn, \vec k_\out)|^2 \approx 
{1\over 4} 
{ (\ni - \no)^2 \over 
\ni \; \no }\;
{V\over (2\pi)^3 } \; \delta^3(\vec k_\inn + \vec k_\out ),
\label{E:sbsq}
\end{equation}
For completeness we also give the unsquared Bogoliubov
coefficients evaluated in the sudden approximation:
\begin{eqnarray}
\alpha(\vec k_\inn, \vec k_\out) 
&\approx&
{1\over 2}
{\ni +\no \over \sqrt{\ni \; \no }}\;
\delta^3(\vec k_\inn - \vec k_\out ),\\
\beta(\vec k_\inn, \vec k_\out) 
&\approx&
{1\over 2}
{|\ni -\no| \over \sqrt{\ni \; \no }}\;
\delta^3(\vec k_\inn + \vec k_\out ).
\end{eqnarray}
As expected, for $\ni \rightarrow \no$, we have $\alpha \rightarrow
\delta^3(\vec k_\inn - \vec k_\out )$ and $\beta \rightarrow 0$.

%-------------------------------------------------------------------------
\subsubsection{Adiabatic limit}
%-------------------------------------------------------------------------

Now take
\begin{equation} 
\min\{\omega^\tau_\inn , \omega^\tau_\out , \omega^\tau_-\} \; \tau_0 \gg 1.
\end{equation}
This corresponds to a slowly changing refractive index.  In this limit
the $\sinh(x)$ functions in the exact Bogoliubov coefficient can be
replaced with exponential functions $\exp(x)$.  Then
\begin{eqnarray}
|\beta|^2 &\propto& 
{ \exp(2\pi \omega^\tau_-  \tau_0) \over 
\exp(\pi\omega^\tau_\inn \tau_0) \;
\exp(\pi\omega^\tau_\out \tau_0)}
\\
&=&
{ \exp(\pi \; |\omega^\tau_\inn - \omega^\tau_\out|  \; \tau_0) \over 
\exp(\pi\omega^\tau_\inn \tau_0) \; 
\exp(\pi\omega^\tau_\out \tau_0)}.
\end{eqnarray}
More precisely 
\begin{equation}
|\beta(\vec k_\inn , \vec k_\out )|^2 
\approx 
\exp\left(
-2\pi \; \min\{\omega^\tau_\out ,\omega^\tau_\inn \} \; 
\tau_0
\right) \;
{V\over (2\pi)^3 } \; 
\delta^3(\vec k_\inn  + \vec k_\out  ).
\end{equation}
In terms of physical time the condition defining the adiabatic limit
reads
\begin{equation} 
2\pi\;\min 
\left\{1, {\ni\over\no}, \half\left|{\ni\over\no}-1\right|\right\} 
\; 
{ \no^2 \over \ni^2+\no^2} \; \omega_\out t_0  \gg 1.
\end{equation}
The Bogoliubov coefficient then becomes
\begin{equation}
|\beta(\vec k_\inn , \vec k_\out )|^2 
\approx 
\exp\left(-4\pi  \; {\min\{\ni,\no\} \;
\no \over \ni^2 + \no^2} \; \omega_\out t_0
\right) \;
{V\over (2\pi)^3 } \; 
\delta^3(\vec k_\inn  + \vec k_\out  ),
\label{bogadb}
\end{equation}
This implies exponential suppression of photon production for
frequencies {\em large} compared to 
\begin{equation}
\Omega_{\mathrm{adiabatic}} \equiv 
{1\over 2\pi t_0} 
{\ni^2 + \no^2 \over \no \;\min\{\ni,\no,\half|\ni-\no|\} }.
\end{equation}
Eberlein's model~\cite{Eberlein1,Eberlein2,Eberlein3} for
Sonoluminescence explicitly makes the adiabatic approximation and this
effect is the underlying reason why photon production is so small in
that model; of course the technical calculations of Eberlein's model
also include the finite volume effects due to finite bubble radius
which somewhat obscures the underlying physics of the adiabatic
approximation.

%-------------------------------------------------------------------------
\subsubsection{The transition region}
%-------------------------------------------------------------------------

Generally there will be a transition region between $\Omega_{\mathrm
{sudden}}$ and $\Omega_{\mathrm{adiabatic}}$ over which the Bogoliubov
coefficient has a different structure from either of the asymptotic
limits.  In this transition region the Bogoliubov coefficient is well
approximated by a monomial in $\omega$ multiplied by an exponential
suppression factor, but the e-folding rate in the exponential is
different from that in the adiabatic regime. Fortunately, we will not
need any detailed information about this region, beyond the fact that
there is an exponential suppression.

%-------------------------------------------------------------------------
\subsubsection{Spectrum}
%-------------------------------------------------------------------------

The number spectrum of the emitted photons is
\begin{equation}
\label{E:spectrum0}
{\d N(\vec k_\out )\over \d^3\vec k_\out }
=
\int|\beta(\vec k_\inn ,\vec k_\out )|^{2} 
\d^3\vec k_\inn .
\end{equation}
Taking into account that $\d^3\vec k_\out = 4\pi k_\out^2 
\; \d k_\out $ this easily yields 
 \begin{equation}
\label{E:spectrum1} 
{\d N(\omega_\out )\over \d\omega_\out }
=
{
\sinh^2\left(
{ \textstyle\pi\; |n_\inn -n_\out | \; n_\out  \; \omega_\out  t_0
\over
\textstyle (\ni^2+\no^2)
}
\right)
\over
\sinh\left(
{\textstyle2\pi n_\inn \; n_\out  \omega_\out  t_0
\over 
\textstyle(\ni^2+\no^2)
}
\right) \;
\sinh\left(
{\textstyle 2\pi n_\out ^{2}\;  \omega_\out  t_0
\over
\textstyle (\ni^2+\no^2)
}
\right)
} \;
{2V\over (2\pi)^3 } \; 
4\pi \omega_\out ^2 \; n_\out ^3 . 
\end{equation}
(Here the factor 2 is introduced by hand by taking into account
the 2 photon polarizations).  For low frequencies (where the sudden
approximation is valid) this is a phase-space limited spectrum with
a prefactor that depends only on the overall change of refractive
index. For high frequencies (where the adiabatic approximation
holds sway) the spectrum is cutoff in an exponential manner depending
on the rapidity of the change in refractive index. 

\begin{figure}[htb]
\vbox{\hfil
\scalebox{0.25}{\rotatebox{270}{\includegraphics{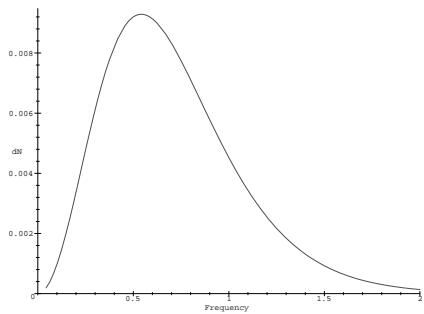}}}
\hfil}
\caption[Sonoluminescence: number spectrum]{
%----------------------------------------------------------------------
  Number spectrum (photons per unit volume) for $ n_\inn =1$, $ n_\out
  =2$.  The horizontal axis is $\omega_\out$ and is expressed in PHz.
  The typical timescale $t_{0}$ is set equal to one ${\rm fs}$.  The
  vertical axis is in arbitrary units.
%----------------------------------------------------------------------
}\label{F:toy-spectrum-1} 
\end{figure}

\begin{figure}[htb]
\vbox{\hfil
 \scalebox{0.25}{\rotatebox{270}{\includegraphics{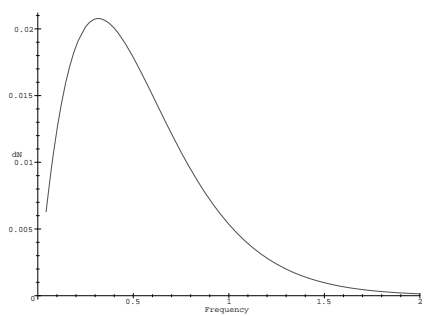}}}
\hfil}
\caption[Black body spectrum]{
%---------------------------------------------------------------------
  Number spectrum for a Planck blackbody curve with $k_{\rm B} T=
  (\ni^2+\no^2)/ (4\pi t_0\; \no \; \min\{\ni,\no,\half|\ni-\no|\} ).
  $ The horizontal axis is $\omega_\out$ and is expressed in PHz. The
  typical timescale $t_{0}$ is set equal to one femtosecond.  The
  vertical axis is in arbitrary units (but with the same normalization
  as figure 1).
%---------------------------------------------------------------------
}\label{F:toy-spectrum-2} 
\end{figure}

\begin{figure}[thb]
\vbox{\hfil
 \scalebox{0.25}{\rotatebox{270}{\includegraphics{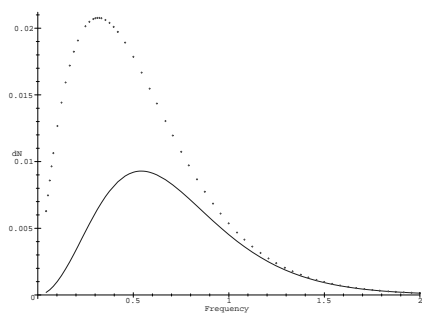}}}
\hfil}
\caption[Superimposed number spectra]{
%---------------------------------------------------------------------
  Superimposed number spectra (the Planck spectrum is the dotted one).
  This figure demonstrates the similar high-frequency behaviour
  although low energy behaviour is different (quadratic versus
  linear).
%---------------------------------------------------------------------
}\label{F:toy-spectrum-3} 
\end{figure}

A sample spectrum is plotted in figure \ref{F:toy-spectrum-1}.  For
comparison figure \ref{F:toy-spectrum-2} shows a Planckian spectrum
with the same exponential falloff at high frequencies, while the two
curves are superimposed in figure \ref{F:toy-spectrum-3}.

%%%==========================================================
%%%==========================================================
%%%==========================================================
%%%==========================================================
%%%==========================================================
\clearpage
%-------------------------------------------------------------------------
\subsubsection{Lessons from this toy model}
%-------------------------------------------------------------------------

\begin{enumerate}
\item This is only a toy model, but it adequately confirms what we
  have seen in chapter~\ref{chap:1}: efficient photon production
  occurs only when the sudden approximation holds, and that photon
  production is suppressed in the adiabatic regime.  The particular
  choice of profile $\epsilon(\tau)$ was merely a convenience, it
  allowed us to get analytic exact results, but it is not a critical
  part of the analysis. One might worry that the results of this toy
  model are specific to the choice of profile (\ref{E:profile}). That
  the results are more general can be established by analyzing general
  bounds on the Bogoliubov coefficients, which is equivalent to
  studying general bounds on one-dimensional potential
  scattering~\cite{Visser}. I shall here quote only the key result
  that for any monotonic change in the dielectric constant the sudden
  approximation provides a strict upper bound on the magnitude of the
  Bogoliubov coefficients~\cite{Visser}.
  
\item Eberlein's model for
  Sonoluminescence~\cite{Eberlein1,Eberlein2,Eberlein3} explicitly
  uses the adiabatic approximation. For arbitrary adiabatic changes we
  expect the exponential suppression to still hold with $\rho$ now
  being some measure of the timescale over which the refractive index
  changes.
  
\item Schwinger's model for
  Sonoluminescence~\cite{Sc1,Sc2,Sc3,Sc4,Sc5,Sc6,Sc7} implicitly uses
  the sudden approximation.  It is only for the sudden approximation
  that we recover Schwinger's phase-space limited spectrum.  For
  arbitrary changes the sudden approximation provides a rigorous upper
  bound on photon production. It is only in the sudden approximation
  that efficient conversion of zero-point fluctuations to real photons
  takes place.  Though this result is derived here only for a
  particularly simple toy model one can expect this part of the analysis to
  be completely generic and that any mechanism for converting
  zero-point fluctuations to real photons will exhibit similar
  effects.
\end{enumerate}

%-------------------------------------------------------------------------
\subsection[{Extension of the model}]{Extension of the model}
%-------------------------------------------------------------------------

The major weaknesses of the toy model are that it currently includes
neither dispersive effects nor finite volume effects. Including
dispersive effects amounts to including condensed matter physics by
letting the refractive index itself be a function of frequency.  To do
this carefully requires a very detailed understanding of the condensed
matter physics, which is quite beyond the scope of the present
investigation. Instead, in this section we shall content ourselves
with making order-of-magnitude estimates using Schwinger's sharp
cutoff for the refractive index and the sudden approximation.

The second issue, that of finite volume effects, is addressed more
carefully in section~\ref{sec:volum}. Finite volume effects are
expected to be significant but not overwhelmingly large.  {From}
estimates of the available Casimir energy developed in~\cite{MV}, the
fractional change in available Casimir energy due to finite volume
effects is expected to be of order $1/(K R) = \hbox{(cutoff
  wavelength)} /(2\pi \hbox{(minimum bubble radius)})$ which is
approximately $(300\; \hbox{nm})/(2\pi \;500\; \hbox{nm}) \approx
10\%\;$\footnote{
%------------------------------------------------------------------------
Here the cutoff wavelength is estimated from the location of
the peak in the SL spectrum. If anything, this causes us to overestimate
the finite volume effects.}.
%-----------------------------------------------------------------------
Returning to dispersive issues: if the refractive indices 
were completely non-dispersive (frequency-independent), then the sudden
approximation would imply infinite energy production. In the real
physical situation $\ni$ is a function of $\omega_\inn $
and $\no$ is a function of $\omega_\out $.  Schwinger's
sharp momentum-space cutoff for the refractive index is equivalent,
in this formalism, to the choice
\begin{equation}
\ni(k) = \ni \; \Theta(K_\inn -k) + 1 \; \Theta(k-K_\inn),
\end{equation}
\begin{equation}
\no(k) = \no \; \Theta(K_\out -k) + 1 \; \Theta(k-K_\out),
\end{equation}
(More complicated models for the cutoff are of course possible at the
cost of obscuring the analytic properties of the model. See
\cite{CMMV2} for further discussion.) Although in general the two
cutoff wavenumbers $K_\inn$ and $K_\out$ can be different, generally
they will both lie in the UV range. This allows us to assume a single
common cutoff $K\equiv \min\{K_\inn,K_\out\}$.  {From} equation
(\ref{E:sbsq}), taking into account the two photon polarizations, one
obtains
\begin{equation}
|\beta(\vec k_\inn ,\vec k_\out )|^{2}
\approx
{1 \over 2} \frac{\left(\no-\ni\right)^2}{\ni \no}
{V\over(2\pi)^3}\; 
\Theta(K - k_\inn ) \;  \Theta(K - k_\out ) \;
\delta^3(\vec k_\inn  + \vec k_\out ).
\label{E:largeb2}
\end{equation}
As a consistency check, expression (\ref{E:largeb2}) has the desirable
property that $\beta\to0$ as $\no\to\ni$: That is, if there is no
change in the refractive index, there is no particle production. 
\begin{center}
  \setlength{\fboxsep}{0.5 cm} \framebox{\parbox[t]{14cm}{
%--------------------------------------------------------------------------
      {\bf Warning}: Equation (\ref{E:largeb2}) is not the complete Bogoliubov
      coefficient. We are missing the contribution coming from the
      case $K_{\inn}>K_{\out}$ or $K_{\inn}<K_{\out}$. Nevertheless
      the above cited physical fact, that one can expect
      $K_{\inn}\approx K_{\out}$, makes this contribution negligible
      with respect to the above one.  In any case this approximation
      amounts to a slight underestimate of the total number of
      produced photons.
%--------------------------------------------------------------------------
}}
\end{center}
The computed Bogoliubov coefficient is directly related to the
physical quantities one is interested in
\begin{equation}
{\d N\over \d\omega_\out}
=4\pi \frac{\no^3 \omega_\out^2}{c^3} \; 
\int|\beta(\omega_\inn ,\omega_\out )|^{2}
\; \d^3 \vec k_\inn,
\end{equation}
\begin{equation} 
 N=\int {\d N\over \d\omega_\out} \; \d\omega_\out ,
\end{equation}
and
\begin{equation}
 E= \hbar \int {\d N(\omega_\out )\over \d\omega_\out }   
 \; \omega_\out  \; \d\omega_\out .
\end{equation}
So we can now compute the spectrum, the number, and the total energy
of the emitted photons.
\begin{eqnarray}
 {\d N(\omega_\out )\over \d\omega_\out }
&=&
 \frac{\no}{c} \; {\d N(\omega_\out )\over \d k_\out }
 =
 \frac{\no}{c} \; 4\pi k^{2}_\out {\d N(\omega_\out )\over \d^3{\vec k}_\out }
\nonumber \\
&\approx&
 {\no \over (2\; c)}\;
 {\left(\no-\ni\right)^2\over\no\,\ni}
 {V\over(2\pi)^3}\; 4\pi \; k_\out ^2 \;
 \Theta(K - k_\out ) \;
\\
&=&
 {1\over (2\; c^3)}\;{\no^2} \;
 {\left(\no-\ni\right)^2\over\ni}
 {V\over(2\pi)^3}\; 4\pi \; \omega_\out ^2 \;
 \Theta\left(K - \frac{\no \omega_\out}{c}\right)\nonumber
\end{eqnarray}
The number of emitted photons is then approximately
\begin{equation}
 N \approx
 {1\over 12\pi^2} \;
 {\left(\no-\ni\right)^2\over\ni\no} \; {V K^3}.
\end{equation}
So that for a spherical bubble  
\begin{equation}
 N \approx
 {1\over 9\pi} \;
 {\left(\no-\ni\right)^2\over\no\ni} \; (R K)^3.
\end{equation}
It is important to note that the wavenumber cutoff $K$ appearing in
the above formula is not equal to the observed wavenumber cutoff
$K_\observed$. The observed wavenumber cutoff is in fact the upper
wavelength measured once the photons have left the bubble and entered
the ambient medium (water), so actually
\begin{equation}
K={\omega_{\mathrm{max}}\no\over c}={\no\over\nl} \; K_\observed.
\end{equation}
Thus
\begin{eqnarray}
N
&\approx&
{1\over 9\pi} \;
{\left(\no-\ni\right)^2\over\no\ni} \;
\left(R \;{\no\over\nl} \; K_\observed\right)^3
\nonumber
\\
&=&
{1\over 9\pi} \;
{\left(\no-\ni\right)^2\over\ni} \; \no^2 \; 
\left({R \omega_{\mathrm{max}} \over c}\right)^3.
\end{eqnarray}
The total emitted energy is approximately
\begin{eqnarray}
E&\approx&
 \frac{\no^2}{2c^3} \;
 {\left(\no-\ni\right)^2\over\ni}
 {V\over(2\pi)^3}\; 4\pi \;
 \int \hbar \omega_\out  \; \omega_\out ^2 \;
 \Theta\left(K - \frac{\no \omega_\out}{c} \right) \d\omega_\out 
\nonumber \\
&=&
 {\hbar\over 2} \; \frac{\no^2}{c^3} \;
 {\left(\no-\ni\right)^2\over\ni} \;
 {V\over(2\pi)^3}\; {4\pi\over4} \; (K\;c/\no)^4
\nonumber \\ 
&=&
{1\over16\pi^2} \;
{\left(\no-\ni\right)^2\over\ni\no^2} \;
\hbar\; c\; K \; {V K^3}
\nonumber \\
&=&
\frac{3}{4}\; N\; \hbar \omega_{\mathrm{max}}.
\label{E:energy}
\end{eqnarray}
Taking into account that the maximum photon energy is $\hbar \,
\omega_{\mathrm{max}} \approx 4\; {\rm eV}$, the average energy per
emitted photon is approximately
\begin{equation}
\langle E \rangle = 
{3\over 4} \hbar c\; K/\no = {3\over 4} \hbar  \, \omega_{\mathrm
{max}}\sim 3 \; {\rm eV}.
\end{equation}

Taking into account this extra factor we can now consider some  numerical 
estimates based on our results.

%-------------------------------------------------------------------------
\subsection[{Some numerical estimates}]
{Some numerical estimates}
\label{subsec:numest}
%-------------------------------------------------------------------------

In Schwinger's original model he took $\ngi\approx 1$, $\ngo \approx
\nl \approx 1.3$, $V= (4\pi/3) R^3$, with $R \approx R_{\mathrm
{max}} \approx 40 \; \mu {\rm m}$ and  $K\approx 2\pi/(360\; {\rm
nm})$ \cite{Sc4}.  Then $KR \approx 698$. Substitution of these
numbers into equation (\ref{E:schwinger}) leads to an energy budget
suitable for about {\em three} million emitted photons.

By direct substitution in equation (\ref{E:energy}) it is easy to
check that Schwinger's results can qualitatively be recovered also in
our formalism: in our case we get about {\em 1.8 million} photons for
the same numbers of Schwinger and about {\em 4 million} photons using
the updated experimental figures $R_{\mathrm{max}} \approx 45 \;
\mu {\rm m}$ and  $K\approx 2\pi/(300\; {\rm nm})$. 

A sudden change in refractive index {\em would} indeed convert the most 
of the energy budget based on static Casimir energy calculations into 
real photons. This may be interpreted as an independent check on Schwinger's
estimate of the Casimir energy of a dielectric sphere.  Unfortunately,
the sudden (femtosecond) change in refractive index required to
get efficient photon production is also the fly in the ointment
that kills Schwinger's original choice of parameters:  The collapse
from $R_{\mathrm{max}}$  to $R_{\mathrm{min}}$ is known to require
approximately $10\; {\rm ns}$, which is far too long a timescale
to allow us to adopt the sudden approximation.

In the new version of the model presented here one has $R\approx R_{\mathrm
{light-emitting-region}} \approx R_{\mathrm{min}} \approx 500 \;
\hbox{nm} $ and take $K_\observed\approx 2\pi/(200\; {\mathrm{nm}})$ so
that $K_\observed R\approx 5 \pi \approx 15$. To get about one million
photons one now needs, for instance, $\ni\approx 1$ and $\no\approx 12$,
or $\ni\approx 2\times 10^4$ and $\no\approx 1$, or even $\no \approx
25$ and $\ni\approx 71$, though many other possibilities could be
envisaged.  In particular, the first set of values could correspond to
a change of the refractive index at the van der Waals hard core due to
a sudden compression {\em e.g.}, generated by a shock wave. In this
framework it is obvious that the most favorable composition for the
gas would be a noble gas since this mechanism would be most effective
if the gas could be enormously compressed without being easily
ionizable.

Note that the estimated values of $\ngo$ and $\ngi$ are extremely
sensitive to the precise choice of cutoff, and the size of the light
emitting region, and that the approximations used in taking the
infinite volume limit underlying the use of our homogeneous dielectric
model are uncontrolled. We should not put to much credence in the
particular numerical value of $\ngo$ estimated by these means, but
should content ourselves with this qualitative message: one needs the
refractive index of the contents of the gas bubble to change
dramatically and rapidly to generate the photons.

As a final remark it is important to stress that equation
(\ref{E:schwinger}) and equation (\ref{E:energy}) are not quite
identical. The volume term for photon production that we have just
derived [equation (\ref{E:energy})] is of second order in $(\ni-\no)$
and not of first order like equation (\ref{E:schwinger}). This is
ultimately due to the fact that the interaction term responsible for
converting the initial energy in photons is a pairwise squeezing
operator (see~\cite{2Gamma}).  Equation (\ref{E:energy}) demonstrates
that any argument that attempts to deny the relevance of volume terms
to Sonoluminescence due to their dependence on $(\ni-\no)$ has to be
carefully reassessed.  In fact what you measure when the refractive
index in a given volume of space changes is {\em not} directly the
change in the static Casimir energy of the ``in'' state, but rather
the fraction of this static Casimir energy that is converted into
photons. We have just seen that once conversion efficiencies are taken
into account, the volume dependence is conserved, but not the power in
the difference of the refractive index.  Indeed the dependence of
$|\beta|^2$ on $(\ni-\no)^2$, and the symmetry of the former under the
interchange of ``in'' and ``out'' states, also proves that it is the
amount of change in the refractive index and not its ``direction''
that governs particle production.  This apparent paradox is easily
solved by taking into account that the main source of energy is the
acoustic field and that the amount of this energy actually converted
in photons during each cycle is a very small fraction of the total
acoustic energy.

%-------------------------------------------------------------------------
\subsubsection{Estimate of the number of photons}
%-------------------------------------------------------------------------

Using the above as a guide to the appropriate starting point, we can
now systematically explore the relationship between the in and out
refractive indexes and the number of photons produced. Using
$K_\observed R \approx 15$ one gets
\begin{equation}
N = {119\over\nl^3}\; (\no-\ni)^2 \; {\no^2 \over \ni}.
\end{equation}

\begin{figure}[htb]
\vbox{\hfil
 \scalebox{0.70}{{\includegraphics{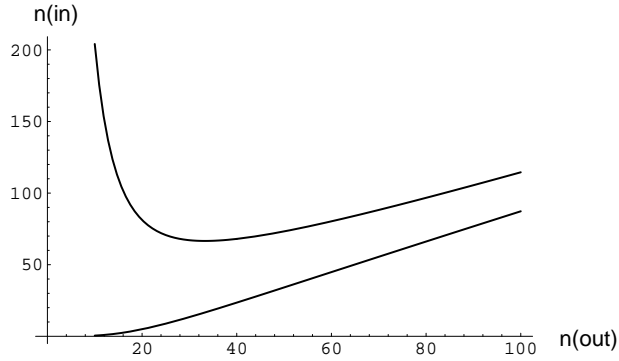}}}
\hfil}
\caption[Values of $\ni$ and $\no$ leading to one million photons]{
%-------------------------------------------------------------------
The initial refractive index $\ni$ plotted as a function of $\no$ when
one million photons are emitted in the sudden approximation.}
%-------------------------------------------------------------------
\label{F:photons-1} 
\end{figure}

This equation can be algebraically solved for $\ni$ as a function of
$\no$ and $N$. (It's a quadratic.) For $N=10^6$ emitted photons the
result is plotted in figure (\ref{F:photons-1}). For any specified
value of $\no$ there are exactly two values of $\ni$ that lead to one
million emitted photons.  To understand the qualitative features of
this diagram we can consider three sub-regions.

%%%==========================================================
%%%==========================================================

\begin{figure}[htb]
\vbox{\hfil
 \scalebox{0.70}{{\includegraphics{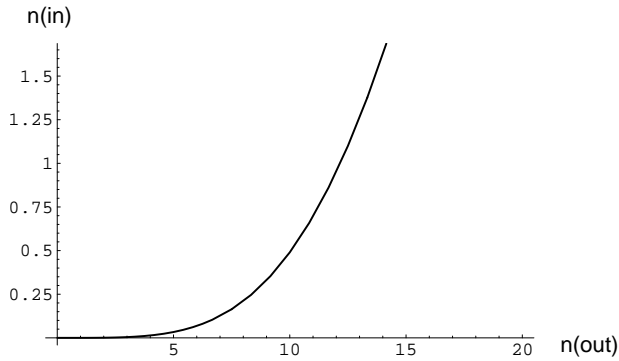}}}
\hfil}
\caption[Values of $\ni$ and $\no$ leading to one million photons:branch 
           that approaches the origin]{
%-------------------------------------------------------------------
  The initial refractive index $\ni$ plotted as a function of $\no$
  when one million photons are emitted in the sudden approximation.
  Here we focus on the branch that approaches the origin.
%--------------------------------------------------------------------
}
\label{F:photons-2} 
\end{figure}

First, if $\ni \ll \no$ then we can approximate
\begin{equation}
\ni \approx {119 \; \no^4 \over \nl^3 \; N}.
\end{equation}
This corresponds to the region near the origin, and one can focus on this
region in figure (\ref{F:photons-2}).

%%%==========================================================
%%%==========================================================

\begin{figure}[htb]
\vbox{\hfil
 \scalebox{0.70}{{\includegraphics{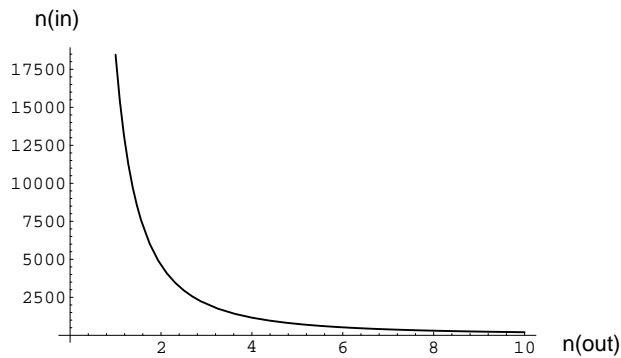}}}
\hfil}
\caption[Values of $\ni$ and $\no$ leading to one million photons:branch 
           that approaches the $y$ axis]{
%-----------------------------------------------------------------------
  The initial refractive index $\ni$ plotted as a function of $\no$
  when one million photons are emitted in the sudden approximation.
  Here we focus on the branch that approaches the $y$ axis.
%---------------------------------------------------------------------
}
\label{F:photons-3} 
\end{figure}

Second, if $\ni \gg \no$ then one can approximate
\begin{equation}
\ni \approx {\nl^3 \; N \over 119 \; \no^2}.
\end{equation}
This corresponds to the region near the $y$ axis, and this
region is illustrated in figure (\ref{F:photons-3}).

%%%==========================================================
%%%==========================================================

Third, if $\ni \approx \no$ then one can approximate
\begin{equation}
N \approx {119\over \nl^3} \; (\ni - \no)^2 \; \no,
\label{nn}
\end{equation}
so that
\begin{equation}
\ni \approx \no \pm \sqrt{N \; \nl^3 \over 119 \; \no}.
\label{nnn}
\end{equation}
This corresponds to the region near the asymptote $\ni=\no$.

Thus to get a million photons emitted from the van der Waals hard core
in the sudden approximation requires a significant (but not enormous)
change in refractive index. There are many possibilities consistent
with the present model and the experimental data.

%-------------------------------------------------------------------------
\subsubsection{Estimate of the timescale}
%-------------------------------------------------------------------------

We have already seen that the actual physical timescale required for
getting a non-adiabatically-suppressed spectrum up to femtosecond
frequencies is given by formula (\ref{E:timescale}). This is indeed
rather different from the simple inverse of $\Omega_{\mathrm{sudden}}$, being a
function of both the sudden cutoff frequency and the refractive
indices: $t_{0}=F(\Omega_{\mathrm{sudden}},\ni,\no)$.

Fixing $\Omega_{\mathrm{sudden}}=1 \;{\rm PHz}$ we can easily plot $t_{0}$ as a
function of $\ni$ and $\no$. From the following graphs it is easy to
see that there is a large range of values for the refractive indices
for which $t_{0}\gg 1\;{\rm fs}$.  In particular the case
$\ni\approx2\;10^4$, $\no\approx1$, would permit a rather relaxed
timescale of just $t_{0}\approx 6.4\;{\rm ps}$ in order to get 1
million photons and a spectrum rising up to 1 PHz.

%%%==========================================================
\begin{figure}[htb]
\vbox{\hfil
 \scalebox{0.80}{\rotatebox{270}{\includegraphics{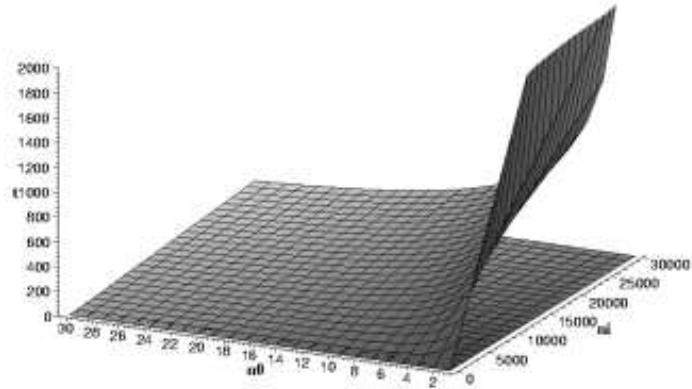}}}
\hfil}
\bigskip
\caption[Required timescales as a function of $\ni$ and $\no$, I]{
%---------------------------------------------------------------------
  Plot of $t_{0}$ as a function of $\ni$ and $\no$ with
  $\Omega_{\mathrm{sudden}}=1$PHz. The units on the $t_{0}$ axis are
  femtoseconds.  The plotted range is range $1<\ni<3\;10^4$ and
  $1<\no<30$. The plane underneath corresponds to $t_{0}\equiv 1\;{\rm
    fs}$.
%---------------------------------------------------------------------
}
\label{time1}
\end{figure}
%%%==========================================================
%%%==========================================================
\begin{figure}[htb]
\vbox{\hfil
 \scalebox{0.80}{\rotatebox{270}{\includegraphics{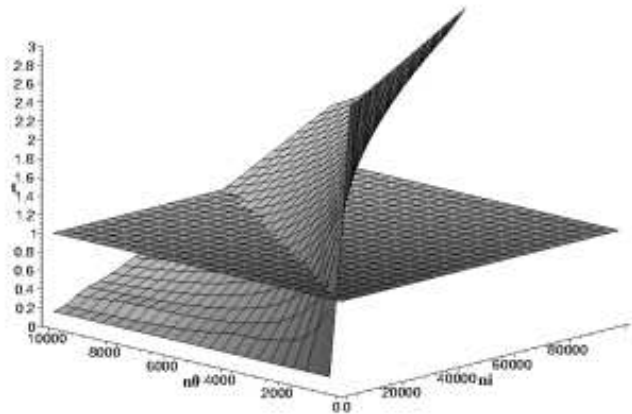}}}
\hfil}
\bigskip
\caption[Required timescales as a function of $\ni$ and $\no$, II]{
%-------------------------------------------------------------------
  Plot of $t_{0}$ as a function of $\ni$ and $\no$ with
  $\Omega_{\rm sudden}=1$PHz. The units on the $t_{0}$ axis are
  femtoseconds.  The plotted range is $1<\ni<10^5$ and $1<\no<10^4$.
  The horizontal plane corresponds to $t_{0}\equiv 1\;{\rm fs}$.
%-------------------------------------------------------------------
}
\label{time2} 
\end{figure}
%%%==========================================================

These results, while encouraging, should be taken with some caution.
Equation (\ref{E:timescale}) is actually dependent on the temporal
profile we choose for the change in the refractive index. Furthermore,
the precise values of the refractive indices depend sensitively on
other aspects of the model.  To fix these uncertainties one requires a
fully developed condensed matter analysis (either by simulation or
direct measurement) able to provide a detailed dynamics for the
refractive index under the rather extreme conditions encountered at
the van der Waals bounce. No such analysis is currently extant, or
even practical.

So far in order to keep that discussion tractable, the technical
computations have been limited to the case of a homogeneous
dielectric medium. We shall now investigate the additional
complications introduced by finite-volume effects. The basic physical
scenario remains the same, but we shall now deal with finite spherical
bubbles. We shall see how to set up the formalism for
calculating Bogoliubov coefficients in the sudden approximation, and
show that one qualitatively retains the results previously obtained
using the homogeneous-dielectric (infinite volume) approximation.
\clearpage
%-------------------------------------------------------------------------
\section[{Sonoluminescence as a dynamical Casimir effect: 
Finite Volume model}]
{Sonoluminescence as a dynamical Casimir effect: Finite Volume model}
\label{sec:volum}
%-------------------------------------------------------------------------

In this case our strategy will be analogous the subtraction procedure
of the static calculations of
Schwinger~\cite{Sc1,Sc2,Sc3,Sc4,Sc5,Sc6,Sc7} or of Carlson {\em et
  al.}~\cite{CMMV1,CMMV2,MV} we shall consider two different
configurations.  An ``in'' configuration with a bubble of refractive
index $\ngi$ in a medium of dielectric constant $\epsilon_{\outside}$,
and an ``out'' configuration with a bubble of refractive index $\ngo$
in a medium of dielectric constant $\epsilon_{\outside}$.  These two
configurations will correspond to two different bases for the
quantization of the field.  (For the sake of simplicity we can take, as
Schwinger did, only the electric part of QED, reducing the problem to
a scalar electrodynamics).  The two bases will be related by Bogoliubov
coefficients in the usual way.  Once we determine these coefficients
we easily get the number of created particles per mode, and from this
the spectrum.  

Let us adopt the Schwinger formalism and consider the equations of the
electric field in spherical coordinates and with a time-independent
dielectric constant. (We can temporarily set $c=1$ for ease of notation,
and shall reintroduce appropriate factors of the speed of light when
needed for clarity.)  Then in the asymptotic future and asymptotic
past, where the refractive index is taken to be time-independent, we
are interested in solving
\begin{equation}
\label{E:static}
 \epsilon(r) \; \partial_{0}(\partial_{0} E)-\nabla^{2} E=0,
\end{equation}
with $\epsilon(r)$ being piecewise constant.  We can look for
solutions of the form
\begin{equation}
 E=\Phi(r,t) \; Y_{lm}(\Omega) \; {1\over r}.
\end{equation}
Then one finds
\begin{equation}
 \epsilon(\partial^{2}_{0}\Phi)-(\partial_{r}^{2} \Phi)+{1\over r^{2}}
 l(l+1) \Phi =0.
\label{E:eqm2}
\end{equation}
For both the ``in'' and ``out'' solution the field equation in $r$
is given by:
\begin{equation}
 \epsilon\partial_{0}^{2} \Phi-\partial_{r}^{2}\Phi +{1\over r^{2}} l(l+1)
 \Phi=0.
\end{equation}
In both asymptotic regimes (past and future) one has a static
situation (a bubble of dielectric $\ngi$ in the dielectric
$\nl$, or a bubble of dielectric $\ngo$ in the dielectric $\nl$)
so one can in this limit factorize the time and radius dependence
of the modes:  $\Phi(r,t)=e^{\im\omega t} f(r)$.  One gets
\begin{equation}
 f^{''}+\left (\epsilon \omega^{2} -{1\over r^{2}} l(l+1)\right) f=0.
\end{equation}
This is a well known differential equation. To handle it more easily
in a standard way we can cast it as an eigenvalues problem
\begin{equation}
 f^{''}-\left( {1\over r^{2}} l(l+1) \right)f=-\varpi^{2}f,
\end{equation}
where $\varpi^2=\epsilon \omega^{2}$.  With the change of variables
$f=r^{1/2} G$, so that $\Phi(r,t)=e^{\im\omega t} r^{1/2} G(r)$, one gets
\begin{equation}
 G^{''}+{1\over r}G^{'}+\left(\varpi^{2}-{\nu^2 \over r^{2}} \right)G=0.
\end{equation}
This is the standard Bessel equation. It admits as solutions the
Bessel and Neumann functions of the first type, $J_{\nu}(\varpi
r)$ and $N_{\nu}(\varpi r)$, with $\nu=l+1/2$.  Remember that for
those solutions which have to be well-defined at the origin, $r=0$,
regularity implies the absence of the Neumann functions.  For both the
``in'' and the ``out'' basis one has to take into account that the
dielectric constant changes at the bubble radius ($R$). In fact one has
\begin{equation}
 \epsilon^\inn=\left\{ 
  \begin{array}{llllll}
   \vphantom{\Bigg|}
    \epsilon_\inside^\inn 
     & = & (\ngi)^2 
     & = & 
     \mbox{\small dielectric constant of air-gas mixture} & 
     \mbox{if $r\leq R$},\\
   \vphantom{\Bigg|}
    \epsilon_\outside^\inn 
     & = & n^2_\liquid 
     & = & 
     \mbox{\small dielectric constant of ambient liquid} & 
     \mbox{if $r > R$}.
  \end{array}
 \right.
\end{equation}
After the change in refractive index, one gets
\begin{equation}
 \epsilon^\out=\left\{ 
  \begin{array}{llllll}
   \vphantom{\Bigg|}
    \epsilon_\inside^\out 
     & = & (\ngo)^2
     & = & 
     \mbox{\small dielectric constant of air-gas mixture} & 
     \mbox{if $r\leq R$},\\
  \vphantom{\Bigg|}
   \epsilon_\outside^\out 
    & = & n^2_\liquid 
    & = & 
    \mbox{\small dielectric constant of ambient liquid} & 
    \mbox{if $r > R$}.
  \end{array}
 \right.
\end{equation}
Defining the ``in'' and ``out'' frequencies, $\omega_\inn$ and
$\omega_\out$ respectively, one has
\begin{equation}
 G^\inn_{\nu}(\ngi,\nl,\omega_\inn,r) =\left \{
  \begin{array}{ll}
   \vphantom{\Bigg|}
    \Xi_\nu^\inn \; A_{\nu}^\inn \; 
     J_{\nu}(\ngi\, \omega_\inn r) & 
      \mbox{if $r\leq R$},\\ 
   \vphantom{\Bigg|}
    \Xi_\nu^\inn \; \left[
     B_{\nu}^\inn \; J_{\nu}(\nl\, \omega_\inn r)+
     C_{\nu}^\inn \; N_{\nu}(\nl\, \omega_\inn r) \right]&
      \mbox{if $r > R$.}
  \end{array}
 \right.
\end{equation}
Here $\Xi_\nu^\inn$ is an overall normalization. The $A_\nu^\inn$,
$B_\nu^\inn$, and $C_\nu^\inn$ coefficients are determined by the
matching conditions at $R$
\begin{equation}
 \begin{array}{lll}
  \vphantom{\Bigg|}
   A_{\nu}^\inn \; J_{\nu}(\ngi\, \omega_\inn  R)&=&
   B_{\nu}^\inn \; J_{\nu}(\nl\, \omega_\inn R)+ 
   C_{\nu}^\inn \; N_{\nu}(\nl\, \omega_\inn R),\\
  \vphantom{\Bigg|}
   A_{\nu}^\inn \; J_{\nu}{'}(\ngi\, \omega_\inn R)&=&
   B_{\nu}^\inn \; J_{\nu}{'}(\nl\, \omega_\inn R)+
   C_{\nu}^\inn \; N_{\nu}{'}(\nl\, \omega_\inn R),
 \end{array}
\label{E:coef}
\end{equation}
(the primes above denote derivatives with respect to $r$),
together with the convention that
\begin{equation}
 |B|^2 + |C|^2 = 1
\end{equation}
The ``out'' basis is easily obtained solving the same equations but
systematically replacing $\ngi$ by $\ngo$. There will be additional
coefficients, $\Xi_\nu^\out$, $A_\nu^\out$, $B_\nu^\out$, and
$C_\nu^\out$, corresponding to the ``out'' basis.

To proceed further we now need to spend some words about the way the
inner product must be generalized in the case of time-dependent and
space-dependent refractive index. This will allow us to impose the
normalization conditions on the $\Phi$ functions and fix the
$\Xi_{\nu}$ coefficients.

%-------------------------------------------------------------
\subsection[{Generalizing the inner product}]
{Generalizing the inner product}
\label{subsec:inpro}
%-------------------------------------------------------------

For the differential equation
\begin{equation}
\label{E:static1}
\epsilon \; \partial_{0}(\partial_{0} E)-\nabla^{2} E=0,
\end{equation}
it is a standard exercise to write down a density and flux,
\begin{equation}
\rho = \epsilon 
\left( E_1^* \; \partial_t E_2 - E_2 \; \partial_t E_1^*\right),
\end{equation}
\begin{equation}
j = E_1^* \nabla E_2 - E_2 \nabla E^*_1,
\end{equation}
and to then show that, by virtue of the differential equation
(\ref{E:static1}), these quantities satisfy a continuity equation
\begin{equation}
\partial_t \rho - \nabla \cdot j = 0.
\end{equation}
Suppose now one has two {\em solutions} of the differential equation
(\ref{E:static1}), one can then define an inner product
\begin{equation}
(E_1, E_2) = -\im \; 
\epsilon \int_t \left( E^*_1 \partial_t E_2 - E_2 \partial_t E_1^*\right),
\end{equation}
where the integral is taken over a constant-time spacelike hypersurface.
By virtue of the above, this inner product is {\em independent} of
the time $t$ at which it is evaluated.

Now what happens if the dielectric is allowed to depend on both
space and time? First the differential equation of interest is
generalized to
\begin{equation}
\label{E:change1}
\partial_{0}(\epsilon(x,t) \partial_{0} E)-\nabla^{2} E=0.
\end{equation}
Second, the density and flux become,
\begin{equation}
\rho = \epsilon(r,t) 
\left( E_1^* \partial_t E_2 - E_2 \partial_t E_1^*\right),
\end{equation}
\begin{equation}
j = E_1^* \nabla E_2 - E_2 \nabla E_1^*.
\end{equation}
By virtue of the differential equation (\ref{E:change1}), 
\begin{eqnarray}
\partial_t \rho &\equiv& 
E_1^* \; \partial_t \left(\epsilon(x,t) \partial_t E_2\right) 
- E_2 \; \partial_t \left(\epsilon(x,t) \partial_t E_1^*\right)
\\
&=&
E_1^* \nabla^2 E_2 - E_2 \nabla^2 E_1^*
\\
&=&
\nabla \cdot \left( E_1^* \nabla E_2 - E_2 \nabla E_1^* \right)
\\
&=& \nabla \cdot j.
\end{eqnarray}
Which implies that these generalized quantities satisfy a continuity
equation
\begin{equation}
\partial_t \rho - \nabla \cdot j = 0.
\end{equation}
This implies that the generalized inner-product [for two solutions
$E_1$ and $E_2$ of the equation (\ref{E:change1}) for time-dependent
and space-dependent dielectric constants] must be
\begin{equation}
(E_1, E_2) = -\im
\int_t \epsilon(x,t) 
\left( E^*_1 \partial_t E_2 - E_2 \partial_t E_1^*\right).
\label{eq:geninpro}
\end{equation}
By the continuity equation this inner product is independent of
the time $t$ at which the integral is evaluated [provided of course,
that $E_1$ and $E_2$ both satisfy (\ref{E:change1})]. This construction
can be made completely relativistic. Define a four-vector $J^\mu$
by
\begin{equation}
J^\mu \equiv ( \rho; j^i).
\end{equation}
Then for any edgeless achronal spacelike hypersurface $\Sigma$ there is
a conserved inner product
\begin{equation}
(E_1, E_2) = -\im \; 
\int_\Sigma \epsilon(x,t) \; J^\mu \; \d\Sigma_\mu.
\end{equation}
%
%------------------------------------------------------------------------
\subsection[{Fixing the normalization constants}]
{Fixing the normalization constants}
%------------------------------------------------------------------------

One can now demand the existence of a normalized scalar product such
that one can define orthonormal eigenfunctions
\begin{equation}
\left(\Phi^{i},\Phi^{j}\right)=\delta^{ij}.
\end{equation}
For the special case $\ng=\nl$, (corresponding to a completely
homogeneous space, in which case $A=1=B$, $C=0$), Eq.~\ref{eq:geninpro}
takes the trivial form (which was implicitly in Eq.~\ref{eq:hoinpro})
\begin{equation} 
\left(\phi_{1},\phi_{2}\right)=\im \, n^2 \int_{\Sigma_t}\phi_{1}^* \stackrel
 {\leftrightarrow}{\partial}_{0}\phi_{2}\: \d^{3}x,
\end{equation}
If we now take the scalar product of two
eigenfunctions, we expect to obtain a normalization condition which
can be written as
\begin{equation}
\left(\Phi^{i}_{[\ng=\nl]},\Phi^{j}_{[\ng=\nl]}\right) = \delta^{ij}.
\end{equation}
Inserting the explicit form of the $\Phi$ functions one gets~\footnote{
%--------------------------------------------------------------------------
  Here we used the inversion formula for Hankel Integral
  transforms~\cite{Bateman,Jackson}, which can be written as
\begin{equation}
\label{E:hankel}
 \int_{0}^{\infty}   r \d r \;
 J_{\nu}(\varpi_1 r) \; J_{\nu}(\varpi_2 r) =
 {\delta(\varpi_1 - \varpi_2) \over\sqrt{\varpi_1\;\varpi_2}}=
 2 \; {\delta(\varpi_1-\varpi_2)\over(\varpi_1+\varpi_2)} =
 2 \;  \delta(\varpi_1^2 - \varpi_2^2),
\end{equation}
  this result being valid for $Re(\nu) > - \half$, and $\varpi_{(1,2)} >0$.  
%----------------------------------------------------------------------------
}
\begin{eqnarray}
\left(\Phi^{i}_{[\ng=\nl]},\Phi^{j}_{[\ng=\nl]}\right)=
%&=&
%\Xi_i^* \; \Xi_j \; \delta_{ll'} \; \delta_{mm'} \; 
%n^2 (\omega_i+\omega_j) 
%\int_{0}^{\infty} r dr \;
%J_{\nu}(n \omega_i r) \; J_{\nu}(n \omega_j r) \;
%e^{i(\omega_i-\omega_j)t}\\
%&=& 
%2 \; \Xi_i^* \; \Xi_j \; \delta_{ll'} \; \delta_{mm'} \; 
%n^2 \; (\omega_i+\omega_j) \;
%\frac{\delta(n \omega_i- n\omega_j)}{n \omega_i + n\omega_j} \;
%e^{i(\omega_i-\omega_j)t}
%\\
%&=& 
2 \; \Xi_i^* \; \Xi_j \; \delta_{ll'} \; \delta_{mm'} \; 
n \; \delta(\varpi_i - \varpi_j),
\end{eqnarray} 
One can now compare this to the behaviour of the three-dimensional delta
function in momentum space
\begin{eqnarray}
\delta^3 (\vec \varpi_i - \vec \varpi_j ) 
&=&
{\delta (\varpi_i - \varpi_j ) \over \varpi_i \; \varpi_j} \; 
\delta^2(\hat \varpi_i - \hat \varpi_j) 
\\
&=&
{\delta (\varpi_i - \varpi_j ) \over \varpi_i \; \varpi_j} \; 
\sum_{lm} Y^*_{lm} (\theta_i,\phi_i) \; Y_{lm}(\theta_j,\phi_j)
\\
&\to&
{\delta (\varpi_i - \varpi_j ) \over \varpi_i \; \varpi_j} \; \delta_{ll'} \; \delta_{mm'},
\end{eqnarray}
to deduce that for homogeneous spaces the most useful normalization is
\begin{equation}
2 \; \Xi_i^* \; \Xi_j \; n  = {1\over \varpi_i \; \varpi_j}.
\end{equation}
This strongly suggests that even for static but non-homogeneous
dielectric configurations it will be advantageous to set
\begin{equation}
\left|\,\Xi^i\right| = {1\over\sqrt{2 n} \; \varpi_i}.
\end{equation}
where $n$ is now the refractive index at spatial infinity~\footnote{
%-----------------------------------------------------------------------
  In our calculations the phase of $\Xi$ is never physically
  important. If desired it can be fixed~\cite{Qed2} by using the
  well-known decomposition of the plane-waves into spherical harmonics
  and Bessel functions. See, {\em e.g.}  Jackson~\cite{Jackson} pages
  767 and 740, equations (16.127) and (16.9).}
%-----------------------------------------------------------------------

To confirm that this is still the most appropriate normalization for
non-homogeneous dielectrics requires a careful discussion which can
be found in~\cite{Qed2}. The central result is that it is indeed the
refractive index at spatial infinity the relevant one for this overall
normalization. So, if one adopts the convention that
\begin{equation}
 |B|^2 + |C|^2 = 1,
\end{equation} 
then proper normalization of the wavefunctions demands
\begin{equation}
 \left|\Xi^i\right| = {1\over\sqrt{2\,\nl} \; \varpi_i},
\end{equation}
%
%---------------------------------------------------------------------
\subsection[{Finite volume calculation}]
{Finite volume calculation}
%--------------------------------------------------------------------

We are now ready to ask what happens if we change the refractive index
by making $\epsilon(r,t)$ a function of both position and time. We are
interested in solving the equation (\ref{E:change1}) and from the
previous discussion it should now be clear that the inner product must
now be modified (in what is now a rather obvious fashion).
\begin{equation} 
\left(\phi_{1},\phi_{2}\right) =
\im\, \int_{\Sigma_t} \epsilon(r,t) \; \phi_{1}^*
\stackrel{\leftrightarrow}{\partial}_{0}\phi_{2}\: \d^{3}x,
\end{equation}
The Bogoliubov coefficients (relative to this inner product) can
now be {\em defined} as
\begin{eqnarray}
\alpha_{ij}
&=&
-\left({E_{i}^\out},{E_{j}^\inn}\right),
\\
\beta_{ij}
&=&\left({E_{i}^\out}^*, {E_{j}^\inn}\right).
\end{eqnarray}
Where $E_{j}^\inn$ now denotes an {\em exact} solution of the
time-dependent equation (\ref{E:change1}) that in the infinite past
approaches a solution of the static equation (\ref{E:static}) with
$\epsilon \to \epsilon_\inn(r)$ and eigen-frequency $\omega_j$.
Similarly $E_{i}^\out$ now denotes an {\em exact} solution of the
time-dependent equation (\ref{E:change1}) that in the infinite future
approaches a solution of the static equation (\ref{E:static}) with
$\epsilon \to \epsilon_\out(r)$ and eigen-frequency $\omega_i$. The
inner product used to define the Bogoliubov coefficients has been
carefully arranged to correspond to a ``conserved charge''. With the
conventions we have in place the absolute values of the Bogoliubov
coefficient are {\em independent} of the choice of time-slice
$\Sigma_t$ on which the spatial integral is evaluated. With minor
modifications, as explained in section~\ref{subsec:inpro}, the inner
product can further be generalized to enable it to be defined for any
arbitrary edgeless achronal spacelike hypersurface, not just the
constant time time-slices.  (This whole formalism is very closely
related to the S-matrix formalism of quantum field theories, where the
S-matrix relates asymptotic ``in'' and ``out'' states.)

Of course, evaluating the Bogoliubov coefficients involves solving the
{\em exact} time-dependent problem (\ref{E:change1}), subject to the
specified boundary conditions, a task that is in general very difficult. It
is at this stage that we shall explicitly invoke the sudden
approximation by choosing the dielectric constant to be
\begin{equation}
\label{E:eps-t}
\epsilon(r,t) = 
\epsilon_\inn(r) \; \Theta(-t) + 
\epsilon_\out(r) \; \Theta(t).
\end{equation}
This is a simple step-function transition from $\epsilon_\inn(r)$ to
$\epsilon_\out(r)$ at time $t=0$. For $t<0$ the exact eigenstates are
given in terms of the static problem with $\epsilon= \epsilon_\inn(r)$,
and for $t>0$ the exact eigenstates are given in terms of the static
problem with $\epsilon=\epsilon_\out(r)$. To evaluate the Bogoliubov
coefficients in the simplest manner, we can chose the spacelike
hypersurface to be the $t=0$ hyperplane. The inner product then
reduces to
\begin{equation} 
(\phi_{1},\phi_{2}) =
\im\; \int_{t=0} 
\epsilon(r,t=0) \;
\phi_{1}^{*} 
\stackrel{\leftrightarrow}{\partial}_{0}
\phi_{2} \: \d^{3}x,
\end{equation}
with the relevant eigenmodes being those of the {\em static} ``in''
and ``out'' problems. (At a fundamental level, this formalism is just
a slight modification of the standard machinery of the sudden
approximation in quantum mechanical perturbation theory.)\\
There is actually a serious ambiguity hiding here: What value are we
to assign to $\epsilon(r,t=0)$? One particularly simple candidate is
\begin{equation}
\epsilon(r,t=0) 
\to 
\half 
\left[ \epsilon_\inn(r) + \epsilon_\out(r) \right]
=
\half 
\left[ n_\inn(r)^2 + n_\out(r)^2 \right],
\end{equation}
but this candidate is far from unique. For instance, we could
rewrite (\ref{E:eps-t}) as
\begin{equation}
\epsilon(r,t) = 
\exp\Big( 
\ln\{\epsilon_\inn(r)\}  \; \Theta(-t) + 
\ln\{\epsilon_\out(r)\} \; \Theta(t)
\Big).
\end{equation}
For $t\neq0$ this is identical to (\ref{E:eps-t}), but for
$t=0$ this would more naturally lead to the prescription
\begin{equation}
\epsilon(r,t=0) 
\to 
\sqrt{\epsilon_\inn(r) \epsilon_\out(r)}
= 
n_\inn(r) n_\out(r).
\end{equation}
By making a comparison with the analytic calculation for homogeneous
media presented in section~\ref{sec:homog} it is in fact possible to
show that this is the correct prescription~\cite{Qed2}. For the moment
it is nevertheless convenient to adopt the notation
\begin{equation}
\epsilon(r,t=0) =
\gamma\left(n_\inn(r);n_\out(r)\right),
\end{equation}
where the only property of $\gamma(n_1;n_2)$ that we really need
to use at this stage is that when $n_1 =  n_2$
\begin{equation}
\gamma(n;n) = n^2.
\end{equation}
(This property follows automatically from considering the
static time-independent case.)

We are mainly interested in the Bogoliubov coefficient $\beta$, since
it is $|\beta|^{2}$ that is linked to the total number of particles
created.  By a direct substitution it is easy to find the expression:
\begin{eqnarray}
 &&\beta_{ll',mm'}(\omega_\inn,\omega_\out)=\nonumber\\
 &&\quad
 =\im \int_{0}^{\infty} 
 \gamma\left(\ni(r),\no(r)\right)
 \left( \Phi_\out(r,t) \; Y_{lm}(\Omega)  \; {1\over r} \right) 
 \stackrel{\leftrightarrow}{\partial}_{0}
 \left( \Phi_\inn(r,t) \; Y_{l^{\prime}m^{\prime}}(\Omega)  
 \;{1\over r}
 \right) \: r^2 \d r \d\Omega,\nonumber\\
 &&\quad
 = -(\omega_\inn-\omega_\out) \; 
 e^{\im(\omega_\out+\omega_\inn)t} 
 \delta_{l l^{\prime}}\; \delta_{m,-m^{\prime}} \; \nonumber \\
 && 
 \quad \quad \times
 \int_{0}^{\infty}
 \gamma\left(\ni(r);\no(r)\right) \;
 G^\out_{l}(\ngo,\nl,\omega_\out,r) \; 
 G^\inn_{l^{\prime}}(\ngi,\nl,\omega_\inn,r) \: r \d r.
\label{Eq:beta1}
\end{eqnarray}
(The $\delta_{m,-m'}$ arises because of the {\em absence } of a
relative complex conjugation in the angular integrals for $\beta$ or
alternatively can be seen as a consequence of the conservation of angular
momentum. On the other hand, the $\alpha$ coefficient will be
proportional to $\delta_{mm'}$.) To compute the radial integral one
needs some ingenuity, let us write the equations of motion for two
different values of the eigenvalues, $\varpi_1$ and $\varpi_2$.
\begin{eqnarray}
G_{\varpi_1}^{''}+{1\over r}G_{\varpi_1}^{'}
+\left (\varpi_1^{2}-
{1\over r^{2}} (l+\half)^2  \right)G_{\varpi_1}
&=&0, \\
G_{\varpi_2}^{''}+{1\over r}G_{\varpi_2}^{'}
+\left (\varpi_2^{2}-
{1\over r^{2}} (l+\half)^2 \right)G_{\varpi_2}&=&0.
\end{eqnarray}
By multiplying the first by $G_{\varpi_2}$ and the second by
$G_{\varpi_1}$ one gets
\begin{eqnarray}
G_{\varpi_1}^{''}G_{\varpi_2}+{1\over r}G_{\varpi_1}^{'}G_{\varpi_2}
+ \left  (\varpi_1^{2}-
{1\over r^{2}} (l+\half)^2\right)G_{\varpi_1}G_{\varpi_2}
&=&0,
\\
G_{\varpi_2}^{''}G_{\varpi_1}+{1\over r}G_{\varpi_2}^{'}G_{\varpi_1}
+ \left (\varpi_2^{2}-
{1\over r^{2}} (l+\half)^2 \right)G_{\varpi_1}G_{\varpi_1}&=&0.
\end{eqnarray}
Subtracting the second from the first we then obtain
\begin{equation}
\left( G_{\varpi_1}^{''}G_{\varpi_2}-G_{\varpi_2}^{''}G_{\varpi_1}\right)+
{1\over r}\left(
G_{\varpi_1}^{'}G_{\varpi_2}-G_{\varpi_2}^{'}G_{\varpi_1}
\right)+
(\varpi_2^{2}-\varpi_1^{2}) G_{\varpi_1}G_{\varpi_2}=0.
\end{equation}
The second term on the left hand side is a pseudo--Wronskian determinant
\begin{equation}
W_{\varpi_1\varpi_2}(r)
=
G_{\varpi_1}^{'}(r) G_{\varpi_2}(r) - G_{\varpi_2}^{'}(r) G_{\varpi_1}(r),
\end{equation}
and the first term is its total derivative $\d W_{\varpi_1\varpi_2}/\d r$.
(This is a pseudo-Wronskian, not a true Wronskian, since the two
functions $G_{\varpi_1}$ and $G_{\varpi_2}$ correspond to different
eigenvalues and so solve different differential equations.) The
derivatives are all with respect to the variable $r$.  Using this
definition we can cast the integral over $r$ of the product of two
given solutions into a simple form. Generically:
\begin{equation}
\left(\varpi_2^{2}-\varpi_1^{2}\right) \int_{a}^{b} r \d r\; 
 G_{\varpi_1}G_{\varpi_2}
=
\int_{a}^{b} r\d r \; \frac{\d W_{\varpi_1\varpi_2}}{\d r}+ \int_{a}^{b} 
\d r\; W_{\varpi_1\varpi_2}.
\end{equation}
That is
\begin{equation}
\int_{a}^{b} r \d r\; G_{\varpi_1}G_{\varpi_2}
=
{1\over
{\varpi_2^{2}-\varpi_1^{2}}}
\left [ 
\left. W_{\varpi_1\varpi_2}\:r \right|_{a}^{b} 
-\int_{a}^{b} \d r \;
W_{\varpi_1\varpi_2}+ \int_{a}^{b} \d r \; W_{\varpi_1\varpi_2}.
\right]
\end{equation}
So the final result is
\begin{equation}
\int_{a}^{b} r \d r \; G_{\varpi_1}G_{\varpi_2}= 
\left.
{1\over {\varpi_2^{2}-\varpi_1^{2}} } \; 
\left( W_{\varpi_1\varpi_2} \; r\right) 
\right|^{b}_{a}.
\end{equation}

This expression can be applied (piecewise) in our specific case
[equation (\ref{Eq:beta1})]. We obtain:
\begin{eqnarray}
&&\int^{\infty}_{0} r  \d r \: 
\gamma\left(\ni(r);\no(r)\right) \; 
G^\out_{\nu}(\ngo,\nl,\omega^\out,r) \; 
G^\inn_{\nu}(\ngi,\nl,\omega^\inn,r)
\nonumber\\
&&\quad =
\int^{R}_{0} r \d r\: 
\gamma\left(\ngi;\ngo\right) \;
G^\out_{\nu}(\ngo\,\omega_\out r)
G^\inn_{\nu}(\ngi\,\omega_\inn r)
\nonumber\\
&&\qquad\qquad
+\int_{R}^{\infty} r \; \d r \: (\nl)^2 \; 
G^\out_{\nu}(\nl\,\omega_\out r) 
G^\inn_{\nu}(\nl\,\omega_\inn r)
\nonumber\\
&&\quad= 
\gamma\left(\ngi;\ngo\right) \; 
\frac{
\left\{ r 
W[G^\out_{\nu}(\ngo\,\omega_\out r),
G^\inn_{\nu}(\ngi\,\omega_\inn r)]
\right\}^{R}_{0}
}
{(\ngo\,\omega_\out)^2-(\ngi\,\omega_\inn)^2} 
\nonumber\\
&& \qquad\qquad
+
(\nl)^2
\frac{
\left\{r 
W[G^\out_{\nu}(\nl\,\omega_\out r),
G^\inn_{\nu}(\nl\,\omega_\inn r)]
\right\}^{\infty}_{R}
}
{(\nl\,\omega_\out)^2-(\nl\,\omega_\inn)^2}
\nonumber\\
&&\quad=
R\,\Bigg[\gamma\left(\ngi;\ngo\right)
\frac{
W[G^\out_{\nu}(\ngo\,\omega_\out r),
G^\inn_{\nu}(\ngi\,\omega_\inn r)]_{R_{-}}
}
{(\ngo\,\omega_\out)^2-(\ngi\,\omega_\inn)^2}
\nonumber\\
&&\qquad\qquad 
-
\frac{
W[G^\out_{\nu}(\nl\,\omega_\out r),
G^\inn_{\nu}(\nl\,\omega_\inn r)]_{R_{+}}
}
{(\omega_\out)^2-(\omega_\inn)^2}
\Bigg],
\end{eqnarray}
where it is used the fact that the above forms are well behaved (and
equal to $0$) for $r=0$. There is an additional delta-function
contribution, proportional to $\delta(\omega_\inn - \omega_\out)$,
arising from spatial infinity $r=\infty$.  In the case of the $\beta$
Bogoliubov coefficient this can quietly be discarded because of the
explicit $ (\omega_\inn - \omega_\out)$ prefactor.  For the $\alpha$
Bogoliubov coefficient we would need to explicitly keep track of this
delta-function contribution, since it is ultimately responsible for
the correct normalization of the eigenmodes if we were to take $\ngo
\to \ngi$.  (Here and henceforth I shall automatically give the same
$l$ value to the ``in'' and ``out'' solutions by using the fact that
equation (\ref{Eq:beta1}) contains a Kronecker delta in $l$ and
$l^{\prime}$.) Finally the two pseudo-Wronskians above are actually
equal (by the junction condition (\ref{E:coef})).  This equality
allows to rewrite integral in equation (\ref{Eq:beta1}) in a more
compact form
\begin{eqnarray}
&&\int^{\infty}_{0} r \; \d r \:
\gamma\left(\ni(r);\no(r)\right) \; 
G^\out_{\nu}(\ngo,\nl,\omega_\out,r) \; 
G^\inn_{\nu}(\ngi,\nl,\omega_\inn,r)
\nonumber\\
&&\qquad = 
\Xi_\inn \; \Xi_\out \; A_{\nu}^\inn \; A_{\nu}^\out R \;
\left[ 
\frac{\gamma\left(\ngi;\ngo\right)}
{(\ngo\,\omega_\out)^2-(\ngi\,\omega_\inn)^2}-
\frac{1}{(\omega_\out)^2-(\omega_\inn)^2}
\right]
\nonumber\\
&&
\qquad\qquad
\times
W[J_{\nu}(\ngo\,\omega_\out r),
  J_{\nu}(\ngi\,\omega_\inn r)]_{R}\nonumber\\
&&\qquad =
\Xi_\inn \; \Xi_\out \;
A_{\nu}^\inn \; A_{\nu}^\out R \;
{
[
\{\gamma\left(\ngi;\ngo\right) - (\ngo)^2 \}
\omega_\out^2 
-
\{\gamma\left(\ngi;\ngo\right) - (\ngi)^2 \}
\omega_\inn^2
]
\over 
[\omega_\out^2-\omega_\inn^2]
}\:
\nonumber\\
&&
\qquad\qquad
\times \frac{
W[J_{\nu}(\ngo\,\omega_\out r),
J_{\nu}(\ngi\,\omega_\inn r)]_{R}
}
{[(\ngo\,\omega_\out)^2-(\ngi\,\omega_\inn)^2]}.
\end{eqnarray}
Inserting this expression into equation (\ref{Eq:beta1}) one gets
\begin{eqnarray}
&&\beta_{lm,l'm'}(\omega_\inn,\omega_\out)=
\Xi_\inn \; \Xi_\out \;
A_\nu^\inn \; A_\nu^\out \; R\:
\delta_{l l^{\prime}} \;
\delta_{m,-m^{\prime}}
\nonumber\\
&& \qquad \times 
\frac{\left [\{\gamma\left(\ngi;\ngo\right) - (\ngo)^2 \}\omega_\out^2 
- \{\gamma\left(\ngi;\ngo\right) - (\ngi)^2 \}
\omega_\inn^2
\right ]
}
{\omega_\out+\omega_\inn} \;
\\
&&
\qquad \qquad \times
\frac{
W[J_{\nu}(\ngo\,\omega_\out r),
J_{\nu}(\ngi\,\omega_\inn r)]_{R}} 
{[(\ngo\,\omega_\out)^2-(\ngi\,\omega_\inn)^2]}
\; 
e^{\im (\omega_\out+\omega_\inn)t}  \nonumber
\end{eqnarray}
As a consistency check, this expression has the desirable property
that $\beta\to0$ as $\ngo\to\ngi$: That is, if there is no change in
the refractive index, there is no particle production.  We are mainly
interested in the square of this coefficient summed over $l$ and
$m$. It is in fact this quantity that is linked to the spectrum of the
``out'' particles present in the ``in'' vacuum, and it is this
quantity that is related to the total energy emitted.  Including all
appropriate dimensional factors ($c$, $\hbar$) we would have (in a
plane wave basis)
\begin{equation}
{\d N(\vec \varpi_\out^{\;\liquid})\over \d^3 \vec \varpi_\out^{\; 
\liquid}} 
=\int|\beta(\vec \varpi_\inn^{\; \liquid},\vec
\varpi_\out^{\; \liquid})|^2
\; \d^3 \vec \varpi_\inn^{\; \liquid}.
\end{equation}
Here, since are are interested in the asymptotic behaviour of the
photons after they escape from the bubble and move to spatial
infinity, we have been careful to express the wave-vectors in terms of
the refractive index of the ambient liquid. This is equivalent
to\footnote{%
%---------------------------------------
Remember that when the photons cross the gas-liquid interface their
frequency, though not their wave-number, is conserved. So we do not
need to distinguish $\omega_\gas$ from $\omega_\liquid$.}
%--------------------------------------
%
\begin{equation}
{\d N(\vec \varpi_\out^{\; \liquid})\over \d \varpi_\out^{\liquid}} 
= \int \left|\beta(\vec \varpi_\inn^{\; \liquid},\vec
\varpi_\out^{\;\liquid})\right|^2
\; \left(\varpi_\inn^\liquid\right)^2 \left(\varpi_\out^\liquid
\right)^2 
\d \varpi_\inn^\liquid \d^2\Omega_\inn \d^2\Omega_\out.
\end{equation}
If we now convert this to a spherical harmonic basis the angular
integrals must be replaced by sums over $l,l'$ and $m,m'$. Furthermore
we can also replace the $\d \varpi_\inn$ and $\d \varpi_\out$ by the
associated frequencies $\d\omega_\inn$ and $\d\omega_\out$ to obtain
\begin{equation}
{\d N(\omega_\out)\over \d\omega_\out} 
=\int \sum_{ll'} \sum_{mm'} 
\left|\beta_{ll',mm'}(\omega_\inn,\omega_\out)\right|^2 \; \nl^2 
\left(\varpi_\inn^\liquid \right)^2 \left(\varpi_\out^\liquid
\right)^2 \; \d \omega_\inn \;.
\end{equation}
In view of our previous definition of the $\Xi$ factors this implies
\begin{equation}
\label{E:N-spectrum}
{\d N(\omega_\out)\over \d\omega_\out} 
= \quarter \int 
{|\beta(\omega_\inn,\omega_\out)|^2 \over |\Xi_\inn|^2 \; |\Xi_\out|^2 } \;
\d \omega_\inn,
\end{equation}
where we have now defined
\begin{equation}
\left|\beta(\omega_\inn,\omega_\out)\right|^2
= \sum_{lm}\sum_{l^{\prime}m^{\prime}}
\left[
\beta_{lm,l^{\prime}m^{\prime}}(\omega_\inn,\omega_\out)
\right]^2.
\end{equation}
Note that the normalization factors $\Xi$ quietly cancel out of the
physically observable number spectrum. Other quantities of physical
interest are
\begin{equation}
\label{E:N}
N = \int {\d N(\omega_\out)\over \d\omega_\out} \; \d\omega_\out,
\end{equation}
and
\begin{equation}
\label{E:E}
E= \hbar \int {\d N(\omega_\out)\over \d\omega_\out} 
\; \omega_\out \; \d\omega_\out.
\end{equation}
Hence we shall concentrate on the computation of:
\begin{eqnarray}
&&\left|\beta(\omega_\inn,\omega_\out)\right|^{2}
= \sum_{lm}\sum_{l^{\prime}m^{\prime}}
\left[
\beta_{lm,l^{\prime}m^{\prime}}(\omega_\inn,\omega_\out)
\right]^2
\\
&&\qquad= 
R^2 
\left({
[
\{\gamma\left(\ngi;\ngo\right) - (\ngo)^2 \}
\omega_\out^2 
-
\{\gamma\left(\ngi;\ngo\right) - (\ngi)^2 \}
\omega_\inn^2
]
\over 
\omega_\out+\omega_\inn
}\right)^2 
\nonumber\\
&&\qquad
\qquad \times
\sum_{l=1}^\infty (2l+1) \;
 |\Xi_\inn|^2  |\Xi_\out|^2 
 \left| A_{\nu}^\inn \right|^{2}  
 \left| A_{\nu}^\out \right|^{2}
 \left[ 
        {W[J_{\nu}(\ngo\,\omega_\out r/c),
           J_{\nu}(\ngi\,\omega_\inn r/c)]_{R}
        \over
        (\ngo\,\omega_\out)^2-(\ngi\,\omega_\inn)^2} 
  \right]^2 \nonumber
\label{E:b2}
\end{eqnarray}
(Note the symmetry under interchange of ``in'' and ``out''; moreover
$l=0$ is excluded since there is no monopole radiation for
electromagnetism.  Also, note that the refractive index of the
liquid in which the bubble is embedded shows up only indirectly: 
in the $A$ and $\Xi$ coefficients.) The above is a general result
applicable
to {\em any} dielectric sphere that undergoes sudden change in
refractive index. However, this expression is far too complex to
allow a practical analytical resolution of the general case.  For
the specific case of Sonoluminescence, using our variant of the
dynamical Casimir effect, we can show that the terms appearing
in it can be suitably approximated in such a way as to obtain a
tractable form that yields useful information about the main
predictions of this model. 

%---------------------------------------------------------------------------
\subsection[{Behaviour for finite radius: Numerical analysis}]
{Behaviour for finite radius: Numerical analysis}
%---------------------------------------------------------------------------

We shall now turn to the study of the predictions of the model in the
case of finite radius. Unfortunately this cannot be done 
analytically due to the wild behaviour of the pseudo--Wronskian
of the Bessel functions.  Nevertheless with some ingenuity,
and a detailed study of the different parts of the Bogoliubov
coefficient, we shall be led to some reasonable approximations
that allow a clear description of the photon spectrum predicted
by the model.

%---------------------------------------------------------------------------
\subsubsection{The A factor}
%---------------------------------------------------------------------------

The $A_{\nu}$, $B_{\nu}$, and $C_{\nu}$ factors can be obtained
by a two step calculation. First one must solve the system
(\ref{E:coef}) by expressing $B$ and $C$ as functions of $A$. Then
one can fix $A$ by requiring $B^{2}+C^{2}=1$, a condition which
comes from the asymptotic behaviour of the Bessel functions.
Following this procedure, and again suppressing factors of $c$ for
notational convenience, one finds that for the ``in'' coefficients
\begin{eqnarray}
A_{\nu}^\inn 
&=& 
\left.
\frac{W[J_{\nu}(\nl\, \omega_\inn r), 
        N_{\nu}(\nl\, \omega_\inn r)]}
{\sqrt{W[ J_{\nu}(\ngi\, \omega_\inn r), 
          N_{\nu}(\nl\, \omega_\inn r)]^2+ 
       W[ J_{\nu}(\ngi\, \omega_\inn r), 
          J_{\nu}(\nl\, \omega_\inn r)]^2}}
\right|_{R},
\nonumber\\ 
B_{\nu}^\inn 
&=& 
\left. A_{\nu}^\inn  
\frac{W[J_{\nu}(\ngi\,\omega_\inn r), 
        N_{\nu}(\nl\, \omega_\inn r)]}
{W[ J_{\nu}(\nl\, \omega_\inn r), 
    N_{\nu}(\nl\, \omega_\inn r) ]}
\right|_{R},
\\
C_{\nu}^\inn 
&=& 
\left.
A_{\nu}^\inn  
\frac{W[J_{\nu}(\nl\, \omega_\inn r), 
      J_{\nu}(\ngi\, \omega_\inn r)]}
{W[J_{\nu}(\nl\, \omega_\inn r), 
   N_{\nu}(\nl\, \omega_\inn r)]}
\right|_{R}.\nonumber
\label{E:coef2}
\end{eqnarray}
We are mostly interested in the coefficient $A_{\nu}$. This can be
simplified by using a well known formula (cf. \cite{as}, page
360 formula 9.1.16) for the (true) Wronskian of Bessel functions of the
first and second kind.
\begin{equation}
W_{\rm true}[J_{\nu}(z), N_{\nu}(z)]=\frac{2}{\pi z}.
\end{equation} 
In our case, taking into account that for our pseudo--Wronskian the
derivatives are with respect to $r$ (not with respect to $z$), one
gets for the numerator of $A_{\nu}$:
\begin{equation}
W[J_{\nu}(\nl\, \omega_\inn r), 
  N_{\nu}(\nl\, \omega_\inn r)]_{R}
= \nl\, \omega_\inn 
\frac{2}{\pi (\nl\, \omega_\inn R)}=\frac{2}{\pi R}.
\end{equation}
Moreover adopting the notation $y = \ngi\, \omega_\inn R/c$ and
$y_\liquid = \nl\, \omega_\inn R/c = (\nl/\ngi) y$ and $\Ni = \nl/\ngi$. 
Then
\begin{equation}
|A_{\nu}^\inn(y,\Ni)|^{2} = 
{4/\pi^2 \over
  \left|
  \begin{array}{rr}
  J_{\nu}(y)&N_{\nu}(\Ni y)\\
  y\;J_{\nu-1}(y) & \Ni y\; N_{\nu-1}(\Ni y)
  \end{array}
  \right|^{2}
+
  \left|
  \begin{array}{rr}
  J_{\nu}(y)&J_{\nu}(\Ni y)\\
  y\;J_{\nu-1}(y) & \Ni y\; J_{\nu-1}(\Ni y)
  \end{array}
  \right|^{2}
}.
\end{equation}
where we have used the standard identities $ x J'_\nu(x) = x
J_{\nu-1}(x) - \nu J_\nu(x)$ and $ x N'_\nu(x) = x N_{\nu-1}(x) - \nu
N_\nu(x)$, and applied properties of the determinant. A similar formula
holds of course for $A_\nu^\out$ in terms of $x$ and $x_\liquid$.

Now, by considering the small argument expansions for the Bessel
functions, it is relatively easy to see that for small $y$ (holding
$\Ni$ fixed)
\begin{equation}
|A_{\nu}^\inn(y\to 0,\Ni)|^{2} \to  (\Ni)^{2\nu} + O(y).
\end{equation}
On the other hand, for large values of the argument $y$ the asymptotic
forms of the Bessel functions can be used to demonstrate that 
\begin{equation}
|A_{\nu}^\inn(y\to \infty,\Ni)|^{2}
\sim
{2\ngi\,\nl\over (\ngi)^2+\nl^2 + 
[\nl^2-(\ngi)^2]\sin(2 y  - \nu\pi)}.
\end{equation} 
Numerical plots of  $|A_\nu|^2$  show that it is an oscillating 
function of $y$ which rapidly reaches this asymptotic form.
The mean value for large arguments is simply:
\begin{equation}
|A_{\nu}^\inn(y\to \infty,\Ni)|^2 \approx {1\over2\pi} 
\int_0^{2\pi} dz 
{2\ngi\,\nl\over (\ngi)^2+\nl^2 + [\nl^2-(\ngi)^2]\sin(z)} = 1.
\label{E:app}
\end{equation}
While these results are general, for our particular application to SL
it is the small $y$ behaviour that is most
relevant. Also, keep in mind that this large $y$ asymptotic formula
holds for $y$ large but ignoring dispersive effects (that is, assuming
a frequency independent index of refraction). If we model dispersive
effects by a Schwinger-like cutoff where the refractive index drops to
unity (see below) then above the cutoff we will have $A_\nu \equiv 1$
holding as an identity.

%---------------------------------------------------------------------------
\subsubsection{The Pseudo--Wronskian}
%---------------------------------------------------------------------------

Use the simplified notation in which $x = \ngo\, \omega_\out R/c$, $y
= \ngi\, \omega_\inn R/c$. In these dimensionless quantities, after
making explicit the dependence on $R$ and $c$, and inserting the
particular choice of $\gamma$ motivated by the large-$R$ limit,
equation (\ref{E:b2}) takes the form:
\begin{equation}
\label{E:b2-2}
\left|\beta(x,y)\right|^{2}=\frac{R^2}{c^2} 
\left(\ngo-\ngi\right)^2 \;
 |\Xi_\inn|^2 \; |\Xi_\out|^2 \;
\left(
{\ngi \; x^2 + \ngo \; y^2
\over
\ngi \; x + \ngo \; y }
\right)^2
F(x,y).
\end{equation}
Here $F(x,y)$ is shorthand for the function
\begin{equation}
F(x,y) = \sum_{l=1}^\infty (2l+1) \; 
|A_l{}^\inn|^2 \; |A_l{}^\out|^2
{ 
  \left|
  \begin{array}{rr}
  J_{\nu}(x)&J_{\nu}(y)\\
  x\;J^{\prime}_{\nu}(x) & y\;J^{\prime}_{\nu}(y)
  \end{array}
  \right|^{2}
\over
(x^2-y^2)^2
},
\label{E:bint}
\end{equation}
where in this equation the primes now signify derivatives with respect
to the full arguments ($x$ or $y$). It is convenient to define a
dimensionless Bogoliubov coefficient, and a dimensionless spectrum, by
taking
\begin{equation}
|\beta(x,y)|^2 = {R^2\over c^2} \; |\beta_0(x,y)|^2,
\end{equation}
so that
\begin{equation}
\label{E:N-spectrum-dimensionless}
{\d N(x)\over \d x} 
= {1\over 4\;\ngi\ngo} \int_0^\infty \d y \;
{|\beta_0(x,y)|^2 \over |\Xi_\inn|^2 \; |\Xi_\out|^2 }.
\end{equation}
The total number of photons is then
\begin{equation}
\label{E:N-dimensionless}
N 
= {1\over 4\;\ngi\ngo} \int _0^\infty \d x \int_0^\infty \d y \;
{|\beta_0(x,y)|^2 \over |\Xi_\inn|^2 \; |\Xi_\out|^2 }.
\end{equation}
The total energy emitted is given by a very similar formula\footnote{%
%-------------------------------------
For a flash occurring at minimum radius $\hbar c/R \approx 0.4 \;
\rm{eV}.$}
%------------------------------------
%
\begin{equation}
\label{E:E-dimensionless}
E
= {\hbar c \over R \ngo} \; {1\over 4\;\ngi\ngo} 
\int _0^\infty \d x \int_0^\infty \d y \; x\;
{|\beta_0(x,y)|^2 \over |\Xi_\inn|^2 \; |\Xi_\out|^2 }.
\end{equation}
In order to proceed in our analysis we need now to perform the
summation over angular momentum.  Although the infinite sum is
analytically intractable, one can easily demonstrate that it is
convergent and can physically argue that the lowest angular momentum
modes will dominate the sum.  Consider the large order expansion ($\nu
\gg x$ at fixed $x$) of the Bessel functions. In this limit one gets
\cite{Jeffrey}:
\begin{equation}
\label{E:asymp}
J_{\nu}(x) 
\sim {\frac{1}{\sqrt{2\pi \nu}}}\left(\frac{e x}{2\nu}\right)^{\nu}
\end{equation}
This can be used to obtain the asymptotic form of the 
pseudo--Wronskian appearing in equation  (\ref{E:bint}).
\begin{eqnarray}
\tilde W_\nu(x,y)&\equiv&\left|
\begin{array}{rr}
J_{\nu}(x)&J_{\nu}(y)\\
x\;J^{\prime}_{\nu}(x) & y\;J^{\prime}_{\nu}(y)
\end{array}
\right|\\
&=&-\left|
\begin{array}{rr}
J_{\nu}(x)&J_{\nu}(y)\\
x\;J_{\nu+1}(x) & y\;J_{\nu+1}(y)
\end{array}
\right|\\
&\sim& \frac{(x^2-y^2)}{2\pi (\nu)^{1/2} (\nu+1)^{3/2}} 
\left(\frac{xy}{\nu(\nu+1)}\right)^{\nu}  
\left(\frac{e}{2}\right)^{2\nu+1}.
\end{eqnarray}
where we have used the standard recursion relation for the Bessel
functions $zJ^{\prime}_{\nu}(z)=\nu J_{\nu}(z)-z J_{\nu+1}(z)$.
This indicates that the sum over $\nu$ is convergent: the terms
for which $(xy/\nu^{2})\leq 1$ are suppressed. Whatever the values
of $x$ and $y$ are, for sufficiently large angular momenta this
asymptotic form guarantees the convergence of the sum over angular
momenta.

%---------------------------------------------------------------------------
\subsection[{Implementation of the cutoff}]{Implementation of the cutoff}
%---------------------------------------------------------------------------

Everything so far has been predicated on the absence of dispersion:
the refractive index is independent of frequency. In real physical
materials the refractive index is known to fall to unity at high
enough energies. (Sufficiently high energy photons ``see'' a vacuum
inhabited by effectively-free isolated charged particles. The manner
in which the refractive index approaches unity is governed by the
plasma frequency, and the location of this physical cutoff is governed
by the resonances present in the atomic structure of the atoms.) This
situation is far too complex to be modelled in detail, but it is easy
to see that an upper bound on emitted photon energies implies an upper
bound on the allowed angular momentum modes: Basically, if one
supposes the photons to be produced inside or at most on the surface
of the light emitting region, then the upper limit for the angular
momentum (as seen at spatial infinity) will be attained by photons
emitted tangentially from the edge of the light emitting region: this
maximal angular momentum is the product of the radius of the light
emitting region times the maximum observed ``out'' momentum.  Then one
gets:
\begin{equation}
l_{\mathrm{max}}^\outside
= { (\hbar K_{\mathrm{observed}}) \times R \over \hbar} 
= R K_{\rm observed}.
\label{E:lmax}
\end{equation}
For Sonoluminescence $K_{\mathrm{observed}}$ is of order $2\pi/(200\; {\rm
nm})$. Since the light emitting region is known to be approximately
$500 \; {\rm nm}$ wide we shall be most interested in the case $K
R\approx 5\pi \approx 15$, with a corresponding maximum angular
momentum $l_{\mathrm{max}}$ approximately $15$.  Under these
conditions, the bulk of the radiation will be into the lowest allowed
angular momentum modes. The precise value of the angular momentum
cutoff $l_{\mathrm{max}}$ is sensitive to the details of both the
frequency cutoff in refractive index, and the size of the light
emitting region. For instance, in some of Schwinger's papers he took
$K \approx 2\pi/(400\; {\rm nm})$ in which case (taking again $R
\approx 400\; {\rm nm}$) $l_{\mathrm{max}} \approx 5\pi/2 \approx 7$.
Whatever ones views as to the precise value of this cutoff it is clear
that the emitted radiation is limited to low angular momenta.

A subtlety is that this is the angular momentum as measured at spatial
infinity (in the ambient liquid---water). This is not the same as the
angular momentum the photons have while they are inside the bubble
(since it is frequency, not wavenumber, that is conserved when photons
cross a timelike interface [spacelike normal]).\footnote{%
%-------------------------------------------------------------------
Contrast this to a spacelike interface (timelike normal; sudden
temporal change in the refractive index) for which it is the
wavenumber, not the frequency, that is conserved across the
interface. During photon {\em production} one is dealing with a
spacelike interface, whereas when the photons {\em escape} from the
gas bubble one is dealing with a timelike interface. }
%-------------------------------------------------------------------
Taking this into account
\begin{equation}
l_{\mathrm{max}}^\inside \approx {\ngo\over\nl} \; l_{\mathrm{max}}^\outside
\approx {\ngo\over\nl}\;  R K_{\rm observed} \approx  {\ngo\over\nl}\; 15.
\label{E:lmax-2}
\end{equation}
If one adopts a Schwinger-like momentum-space cutoff in the refractive
index, then because we have defined the variables $x$ and $y$ partly
in terms of the refractive index, we must carefully assess the meaning
of these variables. In terms of momenta, Schwinger's cutoff is
\begin{equation}
\ni(\varpi) = \ni \; \Theta(K_\inn -\varpi) + 1 \; \Theta(\varpi-K_\inn),
\end{equation}
\begin{equation}
\no(\varpi) = \no \; \Theta(K_\out -\varpi) + 1 \; \Theta(\varpi-K_\out),
\end{equation}
This implies that the photon dispersion relation $\omega(\varpi)$ has a
kink at $\varpi=K$, and that one can write
\begin{equation}
\omega_\inn(\varpi) = {c \varpi\over \ni}\; \Theta(K_\inn -\varpi) + 
\left({c K_\inn\over \ni} + c (\varpi-K_\inn) \right) \; \Theta(\varpi-K_\inn),
\end{equation}
\begin{equation}
\omega_\out(\varpi) = {c \varpi\over \no}\; \Theta(K_\out -\varpi) + 
\left({c K_\out\over \no} + c (\varpi-K_\out) \right) \; \Theta(\varpi-K_\out).
\end{equation}
Finally, the variables $x$ and $y$ generalize (actually, simplify) to
\begin{equation}
x = \varpi_\out R/c; \qquad y = \varpi_\inn R/c,
\end{equation}
so that
\begin{equation}
\label{E:ny}
\ni(y) = \ni \; \Theta(y_* -y) + 1 \; \Theta(y-y_*),
\end{equation}
\begin{equation}
\label{E:nx}
\no(x) = \no \; \Theta(x_*-x) + 1 \; \Theta(x-x_*),
\end{equation}
where $x_* \equiv K_\out R/c;\ y_* \equiv K_\inn R/c$.  Now all these
changes do not affect $F(x,y)$, which is why we defined it the way we
did, but they do affect the prefactors appearing in equation
(\ref{E:b2-2}).  An immediate consequence is that the $(x,y)$ plane
naturally separates into four regions and that $|\beta(x,y)|^2 = 0$ in
the region $x>x_*$ and $y>y_*$. We shall soon see that the two
``tail'' regions $(x<x_*,y>y_*)$ and $(x>x_*,y<y_*)$ are relatively
uninteresting, and that the bulk of the contribution to the emission
spectrum comes from the region $(x<x_*,y<y_*)$.\footnote{%
%--------------------------------
  If one is too enthusiastic about adopting the sudden approximation
  then the integral over these tail regions will be divergent. This,
  however, is not a physical divergence, but is instead a purely
  mathematical artifact of taking the sudden approximation all the way
  out to infinite frequency. The integral over these two tail regions
  is in fact cut off by the fact that for high enough frequency the
  sudden approximation breaks down. As a practical matter we have
  found that the numerical contribution from these tail regions are
  small.}
%------------------------------

Finally, when it comes to choosing specific values for $x_*$ and
$y_*$, we use the fact that the variables $x$ and $y$ are related to
the angular momentum cutoff discussed in the previous subsection to
set
\begin{equation}
x_* = y_* =  {\ngo\over\nl}\; 15.
\end{equation}
%

%----------#---------------------------------------------------------------
\subsection[{Analytical approximations}]{Analytical approximations}
%-------------------------------------------------------------------------

To study in more detail the behaviour of the function $F(x,y)$ when
higher angular momentum modes are retained one can perform a Taylor
expansion of $F(x,y)$ around $x=y$.

It can be shown that, as expected, each term of $F(x,x)$ is 
finite along the diagonal and equal to zero at $x=y=0$.
Moreover 
\begin{eqnarray}
 D(x)\equiv F(x,x)=
\sum_{l=1}^{\infty}(2l+1) 
 {\left\{
 (2l+1) J_{l+1/2}(x) J_{l-1/2}(x) 
 -x \; \left[J_{l+1/2}^{2}(x)+J_{l-1/2}^{2}(x)\right]
 \right\}^2
 \over
 4x^{2}}
%\nonumber 
\end{eqnarray} 
This sum can easily be checked to be convergent for fixed $x$. [Use
equation (\ref{E:asymp}).] With a little more work it can be shown that
\[
\lim_{x\to\infty} D(x) = {1\over2\pi^2}.
\]
The truncated function obtained after summation over the first few
terms (say the first ten or so terms) is a long and messy combination
of trigonometric functions that can however be easily plotted and
approximated in the range of interest.  A semi-analytical study led us
to the approximate form of $D(x)$
\begin{equation}
D(x) \approx \frac{1}{2\pi^{2}}
\frac{x^{6}}{250+x^{6}}.
\end{equation}

To numerically perform the integrals needed to do obtain the spectrum
it is useful to note the approximate factorization property
\begin{equation}
F(x,y) \approx F\left({x+y\over2},{x+y\over2}\right) \; 
G\left(\frac{x-y}{2}\right). 
\end{equation}
That is: to a good approximation $F(x,y)$ is given by its value
along the nearest part of the diagonal, multiplied by a universal
function of the distance away from the diagonal. A little experimental
curve fitting is actually enough to show that to a good approximation
\begin{equation}
F(x,y) \approx  D\left({x+y\over2}\right) \; 
{\sin^2(3[x-y]/4)\over (3[x-y]/4)^2} .
\end{equation}
It is important to stress that this approximation is based on
numerical experimentation, and is not an analytically-driven
approximation.  (In the infinite volume case we know that $F(x,y) \to
(constant) \times \delta(x-y)$. The effect of finite volume is to
``smear out'' the delta function. In this regard, it is interesting to
observe that the combination $\sin^2(x)/(\pi x^2)$ is one of the
standard approximations to the delta function.)  Our approximation is
quite good everywhere except for values of $x$ and $y$ near the origin
(less than 1) where the contribution of the function to the integral
is very small.

%-------------------------------------------------------------------------
\subsection[{The spectrum: numerical evaluation}]
{The spectrum: numerical evaluation}
%-------------------------------------------------------------------------

We have now transformed the function $F(x,y)$ into an easy to handle
product of two functions
\begin{equation}
F(x,y) \approx  
\frac{1}{2\pi^{2}}
\frac{(x+y)^{6}}{16000+(x+y)^{6}}
{\sin^2(3[x-y]/4)\over(3[x-y]/4)^2}. 
\label{fapp}
\end{equation}
One can draw tri-dimensional graphs for both the exact (apart from the
approximation of truncating the sum at a finite $l$) and approximate
forms of the function $F(x,y)$.  It has been chosen the case of $R=500 \;
{\rm nm}$ (corresponding to $y_{*}=15 \;\ngo/\nl$ as previously
explained).
%
%======================================================================
\begin{figure}[htb]
\vbox{
\hfil
\scalebox{0.60}{\rotatebox{270}{\includegraphics{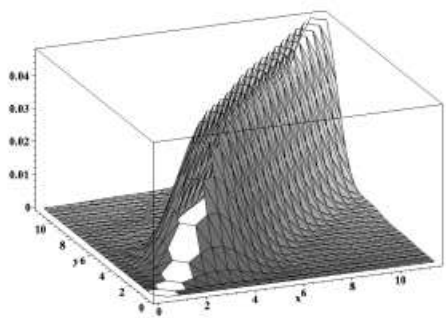}}}
\hfil
}
\bigskip
\caption[Plot of the exact $F(x,y)$]{%
%------------------------------
Plot of the exact $F(x,y)$ in the range $0<x<12$, $0<y<12$. The jagged
behaviour along the diagonal is a numerical artifact, as the function
is known to be smooth there.
%------------------------------
}
\label{F:exact}
\end{figure}
%======================================================================
%======================================================================
\begin{figure}[htb]
\vbox{
\hfil
\scalebox{0.60}{\rotatebox{270}{\includegraphics{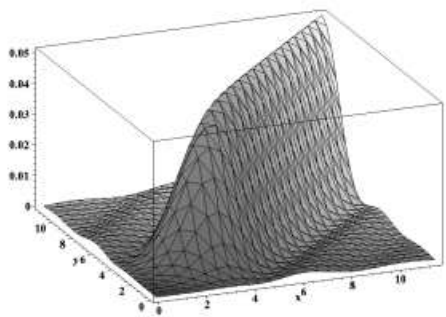}}}
\hfil
}
\bigskip
\caption[Plot of the approximated $F(x,y)$]{%
%------------------------------
Plot of the approximated $F(x,y)$ in the range $0<x<12$,
$0<y<12$
%------------------------------
}
\label{F:factorized}
\end{figure}
%=========================================================================
%\clearpage
%
The dimensionless spectrum, based on equations (\ref{E:b2}) and
(\ref{E:b2-2}), is
\begin{equation}
\label{E:dimensionless-spectrum-2}
{\d N\over \d x} 
=\frac{(\ngi-\ngo)^2}{2 \;\ngi \ngo} 
\int_0^\infty
\left( \frac{\ngi \, x^2+\ngo \, y^2}{\ngi \, x+\ngo \, y} \right)^2 
D\left({x+y\over2}\right) 
{\sin^2(3[x-y]/4)\over(3[x-y]/4)^2} \d y,
\end{equation}
where $\ngo(x)$ and $\ngi(y)$ are now the appropriate functions of $x$
and $y$ (See equations (\ref{E:nx}) and (\ref{E:ny})). A factor $2$ has been
manually inserted to account for the photon polarizations.

As a consistency check, the infinite volume limit is equivalent to
making the formal replacements~\footnote{ 
  This replacement can be
  formally justified as follows. It is known that a sequence of smooth
  functions approximating the delta function is given by
\begin{equation}
f_{s}(x)=\frac{1}{s\; \pi}\; \frac{\sin^2 (s\; x)}{x^2};
\end{equation}
  indeed, one get
\begin{equation}
\lim_{s\to \infty}\; f_{s}(x)=\delta(x).
\end{equation}
  Then, it is straightforward to show that
\begin{eqnarray}
{\sin^2(3[x-y]/4)\over(3[x-y]/4)^2} &=& 
{\sin^2(R \; 3[\ngi \oi-\ngo \oo]/(4 c))
\over(R \; 3[\ngi \oi-\ngo \oo]/(4 c))^2}
\nonumber\\
&\to& \frac{\pi}{R}\; \delta ( \pi[\ngi \oi-\ngo \oo]/(4 c))
\nonumber\\
&=& {4\pi\over3} \; \delta(x-y).
\end{eqnarray}
%-------------------------------------------------------------------------
}
\begin{equation}
{\sin^2(3[x-y]/4)\over(3[x-y]/4)^2} \to {4\pi\over3} \delta(x-y),
\end{equation}
and
\begin{equation}
D\left({x+y\over2}\right) \to {1\over2\pi^2}.
\end{equation}

Doing so, equation (\ref{E:dimensionless-spectrum-2}) reduces to the
spectrum obtained for homogeneous dielectrics in~\cite{Qed1}.
Indeed 
\begin{equation}
{\d N\over \d x} = 
{1\over3\pi} \frac{(\ngi-\ngo)^2}{\ngi \ngo}  \; x^2 \; \Theta(x_* - x).
\end{equation}
With these consistency checks out of the way, it is now possible to
perform the integral with respect to $y$ to estimate the spectrum for
finite volume, and similarly to perform appropriate double integrals
with respect to $x$ and $y$ to estimate both total photon production
and average photon energy.  In the previous section we have seen that
in the infinite volume limit there were two continuous branches of
values for $\ngi$ and $\ngo$ that led to approximately one million
emitted photons with an average photon energy of $3/4$ the cutoff
energy. If we now place the same values of refractive index into the
formula (\ref{E:dimensionless-spectrum-2}) derived above, numerical
integration again yields approximately one million photons with an
average photon energy of $3/4$ times the cutoff energy. The total
number of photons is changed by at worst a few percent, while the
average photon energy is almost unaffected.  (Some specific sample
values are reported in Table I.) The basic result is this: as
expected, finite volume effects do not greatly modify the
results estimated by using the infinite volume limit. Note that $\hbar
\Omega_{\mathrm{max}}$ is approximately $4$ eV, so that average photon
energy in this crude model is about $3$ eV.

%=================================================================
\begin{center}
\bigskip
\begin{tabular}{|c|c|c|c|}
\hline
$\;\ngi\;$ & $\;\ngo\;$  & Number of photons & 
$\langle E \rangle/\hbar \Omega_{\mathrm{max}}$ \\
\hline
\hline
$2\times10^4$ & $1$  & $1.06\times 10^6$ & $0.803$ \\
\hline
$71  $   & $25$ & $1.00\times 10^6$ & $0.750$ \\
\hline
$68  $   & $34$ & $1.06\times 10^6$ & $0.751$ \\
\hline
$9$   & $25$ & $0.955\times 10^6$ & $0.750$ \\
\hline
$1$      & $12$ & $0.98\times 10^6$ & $0.765$ \\
\hline
\end{tabular}
\medskip
\center{Table I: Some typical cases.}
\bigskip
\label{T:table}
\end{center}
%=================================================================

In addition, for the specific case $\ngi=2\times10^4$, $\ngo=1$, we
have calculated and plotted the form of the spectrum. We find that the
major result of including finite volume effects is to smear out the
otherwise sharp cutoff coming from Schwinger's step-function model for
the refractive index. Other choices of refractive index lead to
qualitatively similar spectra.  These results are in reasonable
agreement (given the simplicity of the present model) with
experimental data.
%
%======================================================================
\begin{figure}[htb]
\vbox{
\hfil
\scalebox{0.80}{\rotatebox{270}{\includegraphics{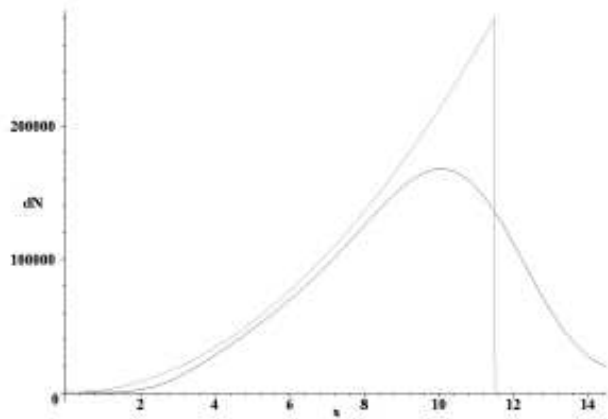}}}
\hfil
%\hbox to 1.0 in{\ } 
}
%\vspace{-1.5in}
\bigskip
\caption[Spectrum $\d N/\d x$ 
 obtained by integrating the approximated Bogoliubov coefficient.]{%
%------------------------------
  Spectrum $\d N/\d x$ obtained by integrating the approximated Bogoliubov
  coefficient.  We integrate from $y=0$ to $y_*=11.5$ and plot the
  resulting spectrum from $x=0$ to $x=14.5$.  For $\no=1$ and
  $R=500$nm the relation between the non-dimensional quantity $x$ and
  the frequency $\nu$ is $x\sim \nu \cdot 10.5 \cdot 10^{-15}$s. So
  $x\approx 11.5$ corresponds to $\nu \approx 1.1$ PHz.  The curve
  with the sharp cutoff is the infinite volume approximation. Finite
  volume effects tend to smear out the sharp discontinuity, but do not
  greatly affect the total number of photons emitted.
%------------------------------
} 
\label{F:spectrum} 
\end{figure}
%========================================================================
\clearpage

%-------------------------------------------------------------------------
\section[{Experimental features and possible tests}]
{Experimental features and possible tests}
%-------------------------------------------------------------------------

Our proposal shares with other proposals based on the dynamical
Casimir effect the main points of strength previously sketched.  On
the other hand it is important to stress that the above model implies
a much more complex and rich collection of physical effects due to the
fact that photon production from vacuum is no longer due to the simple
motion of the bubble boundary. The model indicates that a viable
Casimir route to SL cannot avoid a ``fierce marriage'' with features
related to condensed matter physics.  As a consequence our proposal is
endowed both with general characteristics, coming from its Casimir
nature, and with particular ones coming from the details underlying
the sudden change in the refractive index.

Although the calculation presented above is just a ``probe'', we
can see that it is already able to make some general predictions
that one can expect to see confirmed in a more complete approach.
First of all the photon number spectrum the model predicts is not
a black body. It is polynomial at low frequencies ($\omega^{2}$ in
the infinite volume approximation we used), and in principle
this difference can be experimentally detected.  (The same qualitative
prediction can be found in \Schutzhold\ {\em et al.} \cite{SPS}.)
Moreover the spectrum is expected to be a power law dramatically
ending at frequencies corresponding to the physical wave-number
cutoff $K$ (at which the refractive indices go to 1). This cutoff
implies the absence of hard UV photons and hence, in accordance
with experiments, the absence of dissociation phenomena in the
water surrounding the bubble.

In this type of model, the flash of photons is predicted to occur at
the end of the collapse, the scale of emission zone is of the order of
$500\; \hbox{nm}$, and the timescale of emission is very short, with a
rise-time of the order of femtoseconds, though the flash duration may
conceivably be somewhat longer~\footnote{
%%%%%%%%%%%%%%%%%%%%%%%%%%%%%%%%%%%%%%%%%%%%%%%%%%%%%%%%%%%%%%%%%%%%%%%%
  It would be far too naive to assume that femtosecond changes in the
  refractive index lead to pulse widths limited to the femtosecond
  range. There are many condensed matter processes that can broaden
  the pulse width however rapidly it is generated. Indeed, the very
  experiments that seek to measure the pulse width
  \cite{Flash1,Flash2} also prove that when calibrated with laser
  pulses that are known to be of femtosecond timescale, the SL system
  responds with light pulses on the picosecond timescale.
%%%%%%%%%%%%%%%%%%%%%%%%%%%%%%%%%%%%%%%%%%%%%%%%%%%%%%%%%%%%%%%%%%%%%%%%%
  }.  These are points in substantial agreement with observations. In
the infinite volume limit the photons emerge in strictly back-to-back
fashion. In contrast, for a finite volume bubble we have seen that the
size of the emitting region constrains the model to low angular
momentum for the out states. This is a very sharp prediction that is
in principle testable with a suitable experiment devoted to the study
of the angular momentum decomposition of the outgoing radiation.

Regarding other experimental dependencies, such as the temperature of
the water or the role of noble gases, we can give general arguments
but a truly predictive analysis can be done only after focusing on a
specific mechanism for changing the refractive index.

For instance, the presence of noble gas is likely to change
solubilities of gas in the bubble, and this can vary both bubble
dynamics and the sharpness of the boundary.  Alternatively, a small
percentage of noble gas in air can be very important in the behavior
of its dielectric constant at high pressure. Indeed, while small
admixtures of noble gas will not significantly alter the
zero-frequency refractive index, from the Casimir point of view the
behaviour of the refractive index over the entire frequency range up
to the cutoff is important.

Finally, the temperature of water can instead affect the dynamics of
the bubble boundary by influencing the stability of the bubble,
changing either the solubility of air in water or the surface tension
of the latter. As observed by Schwinger, temperature can also affect
the dielectric cutoff, and so temperature dependence of SL is quite
natural in Schwinger-like approaches.

The above discussion should make clear that an actual experimental
test of the proposed explanation of Sonoluminescence would require as
an unavoidable precondition a much more detailed understanding of
the dynamics of the refractive indices of the gas and of the
surrounding water. Nevertheless one may wonder if a general signature
of the quantum particle creation from the vacuum can be found.

As we said the quasi-thermal nature of the emitted photons can be
explained by the squeezed nature of the photon pairs that are
generically created via the dynamical Casimir effect.  In this case the
core of the bubble is not required to achieve the tremendous physical
temperatures envisaged by other models.

We have seen in chapter~\ref{chap:1} how this apparent thermality can
emerge, in what follow we shall try to investigate this aspect of the
photon emission for our specific models and to find an
univocal signature of the squeezed nature of the photons pair produced
by an eventual dynamical Casimir effect.  

%-----------------------------------------------------------------------
\subsection[{Squeezed states as a test of Sonoluminescence}]
{Squeezed states as a test of Sonoluminescence}
%-----------------------------------------------------------------------

From what we said, thermal characteristics in single photon
measurements in Sonoluminescence can be associated with {\em at least}
two hypotheses: (a) real physical thermalization of the photon
environment; (b) pseudo-thermal single photon statistics due to
tracing over the unobserved member of a photon pair that is actually
produced in a two-mode squeezed state.  We shall call case (a) {\em
  real thermality}; while case (b) will be denoted {\em effective
  thermality}.  Of course, case (b) has no relation with any concept
of thermodynamic temperature, though to any such squeezed state one
may assign a (possibly mode-dependent) {\em effective temperature}.

Our aim is to find a class of measurements able to discriminate
between cases (a) and (b), and to understand the origin of the roughly
thermal spectrum for Sonoluminescence in the visible frequency
range. In principle, the thermal character of the experimental
spectrum could disappear at higher frequencies, but for such
frequencies the water medium is opaque, and it is not clear how we
could detect them.  (Except through heating effects.)

To treat Sonoluminescence, we introduce a quantum field theory
characterized by an infinite set of bosonic oscillators (as in bosonic
Thermofield Dynamics; not just two oscillators as in the case of
``signal-idler'' systems studied in quantum optics).  The simple
two-mode squeezed vacuum discussed in section \ref{subsec:squeezed} is
then replaced by
\begin{eqnarray}
\label{E:general}
%&&
|\Omega[\zeta(k,k')]\rangle \equiv 
%\nonumber\\
%&& \qquad
\exp\left[
-\int \d^3 k \; \d^3k' \;  \zeta(k,k')
\left(a_{k} b_{k'}-a^{\dagger}_{k} b^{\dagger}_{k'}\right)\right] |0\rangle,
\end{eqnarray}
where the function $\zeta(k,k')$ is peaked near $k+k'=0$, and becomes
proportional to a delta function in the case of infinite volume
[$\zeta(k,k') \to \zeta(k) \delta(k+k')$] when the photons are emitted
strictly back-to-back.  To be concrete, let us refer to the
homogeneous dielectric model presented in Sec.\ \ref{sec:homog}. In
this limit there is no ``mixing'' and everything reduces to a sum of
two-mode squeezed-states, where each pair of back-to-back modes is
decoupled from the other. The frequency $\omega$ is the same for each
photon in the couple, in such a way that we are sure to get the same
``temperature'' for both. The two-mode squeezed vacuum then simplifies
to
\begin{equation}
|\Omega (\zeta_k)\rangle \equiv 
\exp\left[-\int \d^3 k\ \zeta_k 
\left(a_k a_{-k}-a^{\dagger}_k a^{\dagger}_{-k}\right)\right] |0\rangle.
\end{equation}
It is interesting to note that, if photons are pair produced in
two-mode squeezed-states by a suitable pair production interaction
term, then $T_{\mathrm{squeezing}}$ is a function of both frequency and
squeezing parameter, and in general only a special ``fine tuning''
would allow us to get the {\em same} effective temperature for all
couples. If we consider the expectation value on the state $|\Omega
(\zeta_k) \rangle $ of $N_k\equiv a^{\dagger}_k a_k$ we get
\begin{equation}
\langle \Omega(\zeta_k)|N_k| \Omega(\zeta_k)\rangle =\sinh^2 (\zeta_k),
\end{equation}
so we again find a ``thermal'' distribution for each value of $k$
with temperature
\begin{equation}
k_{\rm B}T_k\equiv \frac{\hbar\omega_{k}}{2\;\log(\coth(\zeta_k))}.
\end{equation}
The point is that for $k \neq {\bar{k}}$ we generally get $T_k \neq
T_{\bar{k}}$ {\em unless a fine tuning condition holds}.  This
condition is implicitly made in the definition of the thermofield
vacuum and it is possible only if we have
\begin{equation}
\coth (\zeta_k)=\hbox{e}^{ \kappa \omega_k},
\end{equation}
with $\kappa$ some constant, so that the frequency dependence in
$T_k$ is canceled and the same $T_{\rm squeezing}$ is obtained for
all couples.

For models of Sonoluminescence based on the dynamical Casimir effect
({\em i.e.} squeezing the QED vacuum) we cannot rely on a definition
to provide the fine tuning, but must perform an actual calculation.
The model of Sec.\ \ref{sec:homog} is again a useful tool for a
quantitative analysis.  We have (omitting indices for notational
simplicity; our Bogoliubov transformation is diagonal) the following
relation between the squeezing parameter and the Bogoliubov coefficient
$\beta$
\begin{equation}
\langle N \rangle = \sinh^2(\zeta) = |\beta|^2,
\end{equation}
where, by defining $\varphi\equiv \pi\; t_0/ ({\textstyle
\ni^2+\no^2})$, (here $t_0$ is the timescale on which the refractive
index changes) one has from Eq. (\ref{bog2})
\begin{eqnarray}
|\beta|^2
&\propto&
{
\sinh^2\left(
{\textstyle |\ni^2 \omega_\inn -\no^2 \omega_\out|} \; \varphi
\right)
\over
\sinh\left(
2\; {\textstyle \ni^2 } \; \omega_\inn \varphi
\right) \;
\sinh\left(
2 \; {\textstyle \no^2} \; \omega_\out \varphi
\right)
}.
\end{eqnarray}
In the adiabatic limit (large frequencies) we get a Boltzmann
factor (\ref{bogadb})
\begin{equation}
|\beta|^2
\approx
\exp\left(-4  \; \min\{\ni,\no\} \no \;
\omega_\out \varphi
\right).
\end{equation}
Since $|\beta|$ is small, $\sinh(\zeta)\approx\tanh(\zeta)$, so that
in this adiabatic limit
\begin{equation}
|\tanh(\zeta)|^2 \approx
\exp\left(-4 \; \min\{\ni,\no\} \no  \;
\omega_\out \varphi
\right),
\end{equation}
\begin{equation}
k_{\rm B}T_{\mathrm{effective}} \approx 
\frac{\hbar}{8\pi t_{0}}\,
\frac{\ni^2+\no^2}{\no \min\{\ni,\no\}}
\end{equation}
Thus for the entire adiabatic region we can assign a {\em single}
frequency-independent effective temperature which is really a measure
of the speed with which the refractive index changes. Physically, in
Sonoluminescence this observation applies only to the high-frequency
tail of the photon spectrum.

In contrast, in the low frequency region, where the bulk of the
photons emitted in Sonoluminescence are to be found, the sudden
approximation holds and the spectrum is phase space limited (a power
law spectrum), not Planckian. It is nevertheless still
possible to assign a {\em different} effective temperature for each
frequency.

Finite volume effects smear the momentum space delta function so we no
longer get exactly back-to-back photons. This represents a further
problem because we have to return to the general squeezed vacuum of
equation (\ref{E:general}). It is still true that photons are emitted
in pairs, pairs that are now approximately back-to-back and of
approximately equal frequency. We can again define an effective
temperature for each photon in the couple as in the ``signal-idler''
systems of quantum optics. Such temperature is no longer the same for
the two photons belonging to the same couple and no ``special
condition'' for getting the same temperature for all the couples
exists.  We have seen how analyzing these finite volume distortions is not
easy.

In summary: The photons produced in a dynamical Casimir effect are not
truly thermal but can be cast in the framework of ``effective
thermality''.  The spectrum is Planckian or not depending on whether
the adiabatic regime or sudden regime holds sway, but even in the
sudden regime a frequency-dependent effective temperature can be
assigned to each photon mode. Finite volume effects are difficult to
deal with quantitatively, but the qualitative result that in any
dynamic Casimir effect model of Sonoluminescence there should be
strong correlations between approximately back-to-back photons is
robust. It is this last observation that leads us to the following
proposal.

%--------------------------------------------------------------------
\subsubsection{Two photon observables}
%---------------------------------------------------------------------  
Define the observable
\begin{equation}
N_{ab} \equiv N_{a}-N_{b}, 
\end{equation}
and its variance
\begin{equation}
\Delta (N_{ab})^2=
\Delta N_{a}^{2}+\Delta N_{b}^{2}
-2 \langle N_{a} N_{b}\rangle 
+2 \langle N_{a}\rangle \langle N_{b} \rangle.
\end{equation}
These number operators $N_{a},N_{b}$ are intended to be relative to
photons measured, {\em e.g.}, back to back photons.  In the case of true
thermal light we get
\begin{equation}
\Delta N_{a}^{2} = \langle N_{a}\rangle(\langle N_{a} \rangle +1),
\end{equation}
\begin{equation}
\langle N_{a} N_{b}\rangle = \langle N_{a}\rangle \langle N_{b}\rangle,
\end{equation} 
so that 
\begin{equation}
\Delta (N_{ab})^2_{\mathrm{thermal\ light}}
=\langle N_{a}\rangle(\langle N_{a}\rangle+1)
+\langle N_{b}\rangle(\langle N_{b}\rangle+1).
\end{equation}
For a two-mode squeezed-state
\begin{equation}
\Delta (N_{ab})^2_{\mathrm{two\ mode\ squeezed\ light}}=0.
\end{equation}
Due to correlations, $\langle N_{a} N_{b}\rangle \neq \langle
N_{a}\rangle \langle N_{b}\rangle$. Note also, that if you measure
only a single photon in the couple, you get (as expected) a thermal
variance $\Delta N_{a}^{2} = \langle N_{a}\rangle(\langle N_{a}
\rangle +1) $. Therefore a measurement of the variance $\Delta
(N_{ab})^2$ can be decisive in discriminating if the photons are
really thermal or if nonclassical correlations between the photons
occur \cite{bk2}.  If the ``thermality'' in the Sonoluminescence
spectrum is of this squeezed-mode type, one will ultimately desire a
more detailed model of the dynamical Casimir effect involving an
interaction term that produces pairs of photons in two-mode
squeezed-states. 

%-------------------------------------------------------------------------
\section[{Discussion and Conclusions}]{Discussion and Conclusions}
%-------------------------------------------------------------------------

We have verified by explicit computation that photons are produced by
rapid changes in the refractive index, and are in principle able to
explain the observed photons of Sonoluminescence.  

Based largely on the fact that efficient photon production requires
timescales of the order of femtoseconds, we were led to consider rapid
changes in the refractive index as the gas bubble bounces off the van
der Waals hard core. It is important to realize that the speed of
sound in the gas bubble can become relativistic at this stage.

A key lesson learned from this investigation is that in order for the
conversion of zero-point fluctuations to real photons to be relevant
for Sonoluminescence, we would want the sudden approximation to hold
for photons all the way out to the cutoff ($200\;\hbox{nm}$;
corresponding to a period of $0.66 \times 10^{-15} \;
\hbox{seconds}$), a {\em femtosecond} timescale. This implies that if
conversion of zero-point fluctuations to real photons is a significant
part of the physics of Sonoluminescence then the refractive index must
be changing significantly on femtosecond timescales.  {\em Thus the
  changes in refractive index cannot be only due to the motion of the
  bubble wall.}  (The bubble wall is moving at most at Mach
4~\cite{Physics-Reports}; for a $1\;\mu\hbox{m}$ bubble this gives a
collapse timescale of $10^{-10}$ seconds, about 100 picoseconds.)

We have thus suggested that one should not be focusing on the actual
collapse of the bubble, but rather on the way in which the refractive
index changes as a function of space and time. As the bubble collapses
the gases inside are compressed, and although the refractive index for
air (plus noble gas contaminants) is 1 at STP it should be no surprise
to see the refractive index of the trapped gas undergoing major
changes during the collapse process --- especially near the moment of
maximum compression when the molecules in the gas bounce off the van
der Waals' hard core repulsive potential.

Thus attempts at using the dynamical Casimir effect to explain SL are
now much more tightly constrained than previously. We have shown that
any plausible model using the dynamical Casimir effect to explain SL
must use the sudden approximation, and must have very rapid changes in
the refractive index. Our model shows that this timescale is a
non-trivial function of the sudden approximation cut-off frequency,
$\Omega_{\rm sudden}$, and of the initial and final values of the
refractive index $(n_{\inn},n_{\out})$. For this reason a million
photons can be obtained for values of the typical timescale up to some
picoseconds.

A complete theory of SL will need to address {\em much more} specific
timing information and this will require a fully dynamical approach
(from the QFT point of view) and a deeper understanding (from the
condensed matter side) of the precise spatio-temporal dependence of
the refractive index as the bubble collapses. Moreover also
theoretically our model could be improved by taking into account a
multicycle process. In this case the periodicity in the change of the
refractive index should lead to a parametric-resonance behaviour. Such
sorts of regime are very important (as we shall see) in the discussion
of the {\em preheating} phase in inflationary scenarios. In analogy, a
parametric resonance regime could greatly improve the efficiency of the
production of photons.

In the absence of such a more detailed description, the present calculation
is a useful first step. Moreover it allows us to specify certain basic
``signatures'' of the effect that may be amenable to experimental
test.

We have proposed the possibility of discriminating between real
thermal photons and dynamical Casimir photons by means of a careful
analysis of two-photon correlation statistics.  Let us start by making
the ansatz that thermality in the spectrum is either real or
``effective'' in the sense described above. In the former case,
adiabatic heating models are not compatible with some recent
experimental data showing that there is no time delay between
different emitted frequencies~\cite{Flash1,Flash2}. However there
remains the possibility for nonadiabatic heating (Bremsstrahlung or
shock wave models). For thermal light one should find thermal variance
for photon pairs. On the other hand, thermofield-like photons should
show zero variance in appropriate pair correlations.  So we propose to
not deal directly with the issue of thermality by looking at the
spectrum, on the contrary we propose to deal with correlations of
photons approximately emitted back to back.

In conclusion we believe that the present calculation (limited though
it may be) represents an important advance: there can now be no doubt
that changes in the refractive index of the gas inside the bubble lead
to the production of real photons --- the controversial issues now
move to quantitative ones of precise fitting of the observed
experimental data.  Casimir models are now driven into a relatively
small region of parameter space, and one can be hopeful of having
experimental verification (or falsification) of them in the not too
distant future.

%%%%%%%%%%%%%%%%%%%%%%%%%%%%%%%%%%%%%%%%%%%%%%%%%%%%%%%%%%%%%%%%%%%%%%%%%%%%
%S.Liberati Ph.D. Chapter 5: The dynamical Casimir effect in cosmology
%%%%%%%%%%%%%%%%%%%%%%%%%%%%%%%%%%%%%%%%%%%%%%%%%%%%%%%%%%%%%%%%%%%%%%%%%%%%
%%%%%%%%%%%%%%%%%%%%%%%%%%%%%%%%%%%%%%%%%%%%%%%%%%%%%%%%%%%%%%%%%%%%%%%%%%%
\chapter[{Quantum vacuum effects in the early universe}]
{Quantum vacuum effects in the early universe} 
\label{chap:4}
%signature (-,+,+,+)  
%%%%%%%%%%%%%%%%%%%%%%%%%%%%%%%%%%%%%%%%%%%%%%%%%%%%%%%%%%%%%%%%%%%%%%%%%%%
\vspace*{0.5cm}
\rightline{\it As a working hypothesis to explain the riddle of our
  existence,} 
\rightline{\it I propose that our universe is the most interesting of
  all possible universes,} 
\rightline{\it and our fate as human beings is to make it so.}
\vspace*{0.5cm} \rightline{\sf Freeman Dyson}        
%%%%%%%%%%%%%%%%%%%%%%%%%%%%%%%%%%%%%%%%%%%%%%%%%%%%%%%%%%%%%%%%%%%%%%%%%%%
\newpage

In this chapter we shall deal with some quantum vacuum effects in the
very early universe. After discussing the influence of the topological
Casimir effect in cosmology, we shall review the inflationary
paradigm, paying particular attention to the theory of preheating via
parametric resonance.  In this framework we shall propose a new
channel of particle creation which we shall call ``geometric
preheating''. We shall show that this effect can lead to the
production of extremely heavy particles as required by baryogenesis
theory.  After this investigation we shall consider the possibility of
developing an alternative scenario to inflation for solving the
cosmological puzzles.  In particular we shall focus our attention on
the recently proposed variable speed of light cosmologies (VSL). A
critical discussion concerning the foundations of such scenarios will
lead us to propose a new class of such models, the
$\chi$--Variable--Speed--of--Light cosmologies.  Their value as
possible alternatives to inflation or as a ``partner'' for this,
nowadays standard, framework, will be investigated in detail.

%%%%%%%%%%%%%%%%%%%%%%%%%%%%%%%%%%%%%%%%%%%%%%%%%%%%%%%%%%%%%%%%%%%%%%%%%%%
\section[{The role of the quantum vacuum in modern cosmology}]
{The role of the quantum vacuum in modern cosmology}
\label{sec:qveu}
%%%%%%%%%%%%%%%%%%%%%%%%%%%%%%%%%%%%%%%%%%%%%%%%%%%%%%%%%%%%%%%%%%%%%%%%%%%

Although theoretical speculation about the meaning and the
structure of the universe has characterized almost all of the
civilizations which have appeared on earth it is nevertheless true
that cosmology has become a true science only with the advent of \GR .
It is true, in fact, that it is just with the full understanding of the
Einstein equations that the basis for the development of predictive
cosmological models arose.

A quite outstanding fact of this research is that it has led to a
dramatic revolution in our understanding of the place that we have in
the universe and to the final realization that the ``cosmos'' is not a
never-changing place, quiet and peaceful. It appears to us instead as
a dynamical ensemble of matter and energy, a place where huge forces
are in action and can finally lead to unexpected results.

Among the ``surprises'' which \GR\ had in store for us, mention should
be made of the discovery of space-time singularities. In the
particular case of cosmology it was realized that singularities are
unavoidable in \GR\ if matter satisfies some of the energy conditions
which we discussed in chapter \ref{chap:1}. The realization that
quantum vacuum effects in strong fields generally violate most of
these energy conditions, soon led to a great interest in the
application of these effects to cosmology and now this has grown into
a well-established branch of research.

The quantum vacuum can influence cosmological models in two general
ways. The renormalized SET can be such as to strongly modify the
evolution of the universe (via the semiclassical Einstein equations).
Alternatively one can have situations in which the vacuum effects are
manifested via dynamical particle production; in these cases the
gravitational field or a matter field drives the emission of quanta of
some quantum fields, sometimes with relevant effects for the late-time
appearance of the universe.

Two typical examples for the first class of
phenomena are the topological Casimir effect and the inflationary
phase of an FLRW universe. The second class has instead its major
realizations in the generation of perturbations in the inflationary
framework and in the theory of formation and evaporation of primordial
black holes.

We shall not study in detail this second class of phenomena although
we shall discuss the role of the quantum vacuum in the production of
the primordial spectrum of perturbations. Regarding primordial black
holes, they may affect the outcome of Big Bang
nucleosynthesis~\cite{BBNlimits} and be of astrophysical interest
through their evaporative production of very high energy cosmic
rays~\cite{cosmicrays} and as promising candidates for explaining the
Massive Compact Halo Objects (MACHOs) detected in the halo of our
galaxy (these objects would have masses of about $0.2$ to $0.8$
$M_{\odot}$)~\cite{PF00}.  Also, considerations of the possible
formation of primordial black holes during cosmic phase transitions
(e.g., the electroweak and quantum chromodynamic (QCD) transitions)
may be helpful for improving our understanding of the related particle
physics.

In the following we shall concentrate our attention on the topological
Casimir effect in cosmology and its possible role in the evolution of
our universe.

%--------------------------------------------------------------------------
\section[{The topology of the universe and the quantum vacuum}]
{The topology of the universe and the quantum vacuum}
\label{subsec:tCe}
%--------------------------------------------------------------------------

The information which we can get about a cosmological solution by
using the Einstein equations is not complete. In fact these are
partial differential equations and are hence able to describe just the
{\em local} properties of spacetime. These properties are indeed
completely encoded in the infinitesimal distance element $\d
s^2=g_{\mu\nu}\d x^{\mu}\d x^{\nu}$.

The {\em global} structure --- the topology --- of the spacetime is
instead out of the domain of the Einstein equations, actually several
distinct topologies can be associated with the same local metric
element.  This realization has important consequences for modern
cosmology.

According to the standard cosmological framework, the requirement of
spatial homogeneity and isotropy has led to the so-called
Friedmann-Lema\^{\i}tre-Robertson-Walker metrics
\begin{eqnarray}
\label{eq:flrw}
\d s^2  &=& -c^2 \; \d t^2 + a^{2}(t) \; h_{ij} \;\d x^i \; \d x^j,\\
h_{ij}\d x^{i}\d x^{j} &=& \frac{\d r^2}{1-Kr^2}
      +r^2\left(\d\theta^{2}+\sin^{2}\theta\d\varphi^{2}\right)\nonumber
\end{eqnarray}
where $K=0,\pm 1$ denotes the space curvature. For generic $a(t)$, the
above metric describes Friedmann models; for $a(t)$ described by
hyperbolic laws we get the maximally symmetric de~Sitter solution, an
exponentially expanding 3-space. More precisely one gets
\begin{eqnarray}
 && a(t) = H^{-1} \cosh(Ht) \qquad \mbox{for} \quad K=+1\\
 && a(t) = a_{0}   \exp(Ht) \qquad\quad\,\, \mbox{for} \quad K=0 \\
 && a(t) = H^{-1} \sinh(Ht) \qquad \mbox{for} \quad K=-1
\end{eqnarray}
where $H=\dot{a}/a$ is the Hubble parameter and $a_{0}$ is the value
of $a(t)$ at $t=0$. In dealing with FLRW models it is generally
assumed that the global structure of space is that of the infinite Euclidean
space $R^3$, or that of the finite hypersphere $S^3$ or that of the
infinite hyperbolic space $H^3$. The choice depends on whether the
constant spatial curvature is respectively zero, positive or negative
($K=0,\pm 1$).

The above considerations instead lead to the possibility that the
above spatial topologies are not the only ones compatible with the
FLRW metric. Indeed a large number of topologies obtained from the
former ones by opportune identifications of some set of points, are
compatible with (\ref{eq:flrw}). The standard spatial topologies
associated with $R^3$, $S^3$ and $H^3$ describe simply connected
manifolds, that is manifolds for which, at any point $\x$, a loop can
be continuously shrunk to a point. The identifications of some points
of these manifolds leads to their corresponding ``non-simply
connected'' or ``multiply-connected'' versions. In particular, this
implies that even for zero or negative spatial curvatures, the FLRW
models could nevertheless correspond to compact space topologies. For
spaces of constant curvature (such as the FLRW models) these compact
manifolds can be expressed as the quotient ${\cal M}\equiv \tilde{\cal
  M}/{\Gamma}$ where $\tilde{\cal M}$ is one of the three simply
connected topologies mentioned above and $\Gamma$ is a subgroup of
isometries of $\tilde{\cal M}$ acting freely and discontinuously
($\Gamma$ can also be interpreted as the {\em holonomy group} in
$\tilde{\cal M}$, see~\cite{Luminet} for an extensive review of these
issues).

Although some important work has been done in the last few decades on the
possibility that the topology of our universe could be non-simply
connected, cosmologists have often ignored this important possibility
which is moreover apparently favoured by quantum cosmology predictions
(see for example~\cite{Haw84,ZG84}). 

For what concerns the spirit of our research, the possibility of having
non-trivial topologies for the spatial sections of FLRW spacetimes can
certainly be considered as being among the most promising testing fields
for the theory of quantum vacuum effects in strong fields.

We have seen in chapter~\ref{chap:1} how the non-trivial topology of
flat space can lead to a vacuum polarization, a topological Casimir
effect.  In cosmological models with non simply connected spaces, the
same sort of effects can be expected and indeed
predicted~\cite{DB,Ish,GB}. In these cases the Casimir effect will be
calculated using as the reference vacuum state --- with respect to
which the Casimir subtraction is performed --- the appropriate one in
the simply connected manifold ${\cal M}$.

In this way it is possible to obtain a finite Casimir contribution to
the total quantum field vacuum SET, $\langle
T_{\mu\nu}\rangle_{\mathrm{tot}}$, and hence a semiclassical
backreaction effect on the spacetime metric, via the semiclassical
Einstein equations:
\begin{equation}
  \label{eq:semiein}
  G_{\mu\nu}+\Lambda g_{\mu\nu}=
    8\pi G_{\mathrm N}\langle T_{\mu\nu}\rangle_{\mathrm{tot}}
\end{equation}
We note that this is indeed a nice example of how the global
configuration can influence the local structure of the universe via
the quantum vacuum.

In connection with the problem of backreaction, particular attention
should be devoted to the so-called self-consistent models of the
universe. These are cosmological solutions which do not contain any
classical matter field but are instead completely governed by vacuum
quantum effects. Consistency is achieved when the SET for a given
class of metrics gives a metric of the same form if it is re-inserted
into the semiclassical Einstein equations (\ref{eq:semiein}).

\begin{center}
  \setlength{\fboxsep}{0.5 cm} 
   \framebox{\parbox[t]{14cm}{
%------------------------------------------------------------------------      
       {\bf Comment}: Self-consistency is not generic. This can be
       understood from the simple fact that even the choice between
       periodic and anti-periodic conditions on the quantized fields,
       can lead to different signs for the renormalized SET. Moreover
       in~\cite{DB} it is shown that the tensorial structure of the
       SET obtained by the topological Casimir effect in Clifford-Klein
       spaces of the form $\tilde{\cal M}=S^3/\Gamma$ is generally
       different from that of the Einstein tensor
       $G_{\mu\nu}=R_{\mu\nu}-g_{\mu\nu}R/2$ because the former
       contains additional geometrical structure. It is perhaps
       conceivable to impose constraints on the admitted non-simply
       connected realizations of FLRW cosmologies exactly by looking
       at the compatibility of the tensorial structure of the
       renormalized SET with the Einstein equations.
%--------------------------------------------------------------------------
}}
\end{center}
As an example of a self-consistent model we can consider the
topological Casimir effect associated with a non-simply connected
version of the flat FLRW cosmological model, taking $\tilde{\cal
  M}=R^3/\Gamma$ and $K=0$ in Eq.~(\ref{eq:flrw}). For our purposes is
convenient to use Cartesian coordinates $(t,x,y,z)$, and the metric
element~(\ref{eq:flrw}) is then
\begin{equation}
  \label{eq:cartfrwme}
  \d s^2=-\d t^2 +a^2 (t)\left(\d x^2+\d y^2 +\d z^2 \right)  
\end{equation}
All of the possible non-simply connected versions of such a space have
been described by Wolf ~\cite{Wolf} and there are 18 variants
(including the simply connected topology $R^3$).

We shall now focus on the very simple case in which the space sections
have a toroidal topology of characteristic length $L$.  This
topological structure corresponds to some natural periodic conditions
\begin{equation}
  \label{eq:topcond}
  \left\{
    \begin{array}{lll}
       x &=& x+ kL\\
       y &=& y+ mL\\
       z &=& z+ nL
    \end{array}
   \right. \qquad k,m,n=\pm 1,\pm 2,\dots
\end{equation}
In this model, the three-dimensional volume of the universe is always
finite and is given by $V=a^3(t)L^3$. 

The topological Casimir effect in compact three-dimensional spaces with
flat geometry is a somewhat trivial generalization of the one-dimensional  
example discussed in chapter~\ref{chap:1}. In
particular, the renormalized energy density for a massless scalar field
in a torus with typical scales $b,c,d$ can be shown to
be~\cite{MostTru}
\begin{equation}
 \label{eq:bcd}
 \epsilon=-\frac{\pi^2}{90 b^4}-\frac{\zeta(3)}{2\pi bc^3}-\frac{\pi}{6 bcd^2}
\end{equation}
where $\zeta$ is the so called Riemann $\zeta$-function and
$\zeta(3)\approx 1.202$. We stress that we have here lost the symmetry
under permutations of $a$, $b$ and $c$ because some terms are dropped
making the assumption $b\leq c\leq d$.

The pressures along the spatial axes can also be obtained by using
their relation with the total energy $E=bcd\epsilon$ say
\begin{equation}
  \label{eq:presstopo}
  P_{x}=-\frac{1}{cd}\frac{\partial E}{\partial b} \quad
  P_{y}=-\frac{1}{db}\frac{\partial E}{\partial c} \quad
  P_{z}=-\frac{1}{bc}\frac{\partial E}{\partial d} \quad
\end{equation}
In our case $b=c=d=a(t)L$ and so the Casimir SET for a massless,
conformally coupled field, takes the form (taking into account the
terms previously ignored in Eq.(\ref{eq:bcd}))~\cite{MostTru}
\begin{equation}
  \label{eq:settopo}
  \langle T^{0}_{0}\rangle=-\frac{0.8375}{a^4(t)L^4}, \quad
  \langle T^{i}_{i}\rangle=-\frac{1}{3}\langle T^{0}_{0}\rangle
\end{equation}
The inclusion of fields with anti-periodic conditions changes, in this
case, just the numerical coefficient in the above
formula~\cite{MostTru} and for generality we can replace the $0.8375$
by a positive constant $A$. Inserting now the above SET in the
Einstein equations (\ref{eq:semiein}) for the flat FLRW metric, one
gets
\begin{equation}
  \label{eq:tpei}
  3\dot{a}^2-\Lambda a^2=-\frac{8\pi G_{\mathrm N} A}{a^{2}L^4}
\end{equation}
which admits the solution~\cite{ZS84}
\begin{equation}
  \label{eq:tpds}
  a(t)=\frac{1}{L}\left(\frac{8\pi G_{\mathrm N}A}{\Lambda}\right)^{1/4}
      \left[\cosh \left( 2\sqrt{\frac{\Lambda}{3} t}\right)\right]^{1/2}
\end{equation}
This implies that, for the case being considered, the topological
Casimir SET induced by the compactification of flat $R^3$ spatial
sections leads to a non-singular universe with accelerated expansion,
a de~Sitter spacetime. This was the sort of solution to be expected
just because of the presence of a non-zero cosmological constant
$\Lambda$ and hence the result is a nice example of a self-consistent
model~\footnote{
%----------------------------------------------------------------------
  It is nevertheless important to note that the scale factor $a(t)$
  which we found, is the one appropriate for compact spatial sections,
  although we started by assuming that $K=0$. The presence of
  $\Lambda$ in a simply connected flat FLRW spacetime would have led
  to an exponential growth of the scale factor.
%---------------------------------------------------------------------
}.

The property of de~Sitter spacetime which we have just found is a good
starting point for our next topic. In fact this spacetime has some
other peculiar characteristics which play a crucial role in modern
cosmology.

%--------------------------------------------------------------------------
\section[{Inflation}]{Inflation} 
\label{subsec:inflation}
%--------------------------------------------------------------------------

As we said at the start of this chapter, modern cosmology is based on
the equations of \GR\ and on the Copernican principle of isotropy and
homogeneity of our universe.  The SET of matter is assumed to be that
of a perfect fluid~\footnote{
%-------------------------------------------------------------------------  
  Due to the special importance that it will have in the rest of this
  chapter, from here on we shall always make explicit any dependence
  on the speed of light.}.
%-------------------------------------------------------------------------
%
\begin{equation}
 T^{\mu\nu}= 
  \left[ \matrix{\rho c^2 &  0  &  0  &  0  \cr
                       0  &  p  &  0  &  0  \cr                  
                       0  &  0  &  p  &  0  \cr
                       0  &  0  &  0  &  p }\right].  
  \label{eq:tpfset}             
\end{equation}
where $\rho$ and $p$ are respectively the total density and the total
pressure of all of the different kinds of matter considered in the
model.

By these basic elements, it is straightforward to arrive at the FLRW
cosmologies. These are characterized by the class of metrics
determined by (\ref{eq:flrw}) and are governed by the so-called
Friedmann equations
\begin{eqnarray}
{\left({\dot a\over a}\right)}^2
&=&
{8\pi G_{\mathrm N}\over3} \; \rho
-{{Kc^2} \over a^2}+\frac{\Lambda c^2}{3},
\label{eq:fried1}
\\
{\ddot a\over a}
&=&
-{4\pi G_{\mathrm N}\over {3c^2}}\; \left(\varepsilon+3p\right).
\label{eq:fried2}
\end{eqnarray}
where, again, $K=0,\pm1$.

This framework nowadays takes the name of ``the standard model of
cosmology'' (SM), and the wide consensus which it has managed to obtain in
the scientific community is based mainly on three confirmed experimental
predictions:
\begin{enumerate}
\item The Hubble law
\item The Cosmic Microwave Background Radiation (CMBR)
\item The abundances of light elements
\end{enumerate}
The third observation, which is a direct test of the more general
nucleosynthesis theory, represents the earliest-time probe of the SM
to have been developed so far. In fact, the light element production
is supposed to have taken place between a second and a few minutes
after the Big Bang (the initial singularity predicted by the
Hawking--Penrose theorem which we cited in chapter~\ref{chap:1}).  For
times between that one and the Planck time $t_{\pl}=10^{-44} s$ (where
quantum gravity should come onto the stage) we have no direct proof of
the validity of the SM. It is exactly from this ``no-man's land'' that
the main trouble for the standard model arises.

%---------------------------------------------------------------------------
\subsection[{The cosmological Puzzles}]
{The cosmological Puzzles}
\label{subsec:cospu}
%---------------------------------------------------------------------------

In spite of its remarkable success, the SM has proved to be unable to
explain some striking observational features of our universe.  All of
these features seem to have an extremely remote origin (before the
first second after the Big Bang). 

These ``failures'' of the standard model are generally called the
``cosmological puzzles''. These are substantially things that the
framework is unable to justify and about which it cannot give to us
any satisfactory response without appealing to very special initial
conditions for the cosmological evolution. Here we shall limit ourselves
to a rapid survey of these puzzles, directing the reader to standard
cosmology textbooks for further information.

%---------------------------------------------------------------------------
\subsubsection[{The horizon problem}]
{The horizon problem}
%---------------------------------------------------------------------------

The CMBR is isotropic to $1$ part in $10^5$ (after excluding the
dipole due to our peculiar velocity). The horizon problem arises from
the fact that regions at the antipodes of our horizon have never had
the chance to get in causal contact according to the SM. More
concretely, one finds that the particle horizon at the time of photon
decoupling (redshift $z\approx 1200$) subtends just one degree in the
sky and so, at the time when the photons were emitted, two nowadays
antipodal points were separated by very many particle horizons of that
time and were hence causally disconnected. How then is it possible
that regions which had never exchanged information appear to be at the
same temperature with such high precision?

\begin{center}
  \setlength{\fboxsep}{0.5 cm} 
   \framebox{\parbox[t]{14cm}{
%------------------------------------------------------------------------      
       {\bf Comment}: We would like to make clear a point related to
       the definition of horizons in cosmology. Speaking about regions
       in causal contact, the pertinent horizon which one should 
       consider is the so-called \emph{particle horizon}
       \begin{equation}
         \label{eq:prthor}
         \ell_{\mathrm h}(t)=a(t)\int_{0}^{t}{ \frac{c\, \d t^{'}} {a(t^{'})} }
       \end{equation}
       This is telling us the distance covered by light in a time $t$ 
       and hence the largest region which can be in causal contact at
       that time.  As clearly explained in~\cite{Ellis-Rothman}, this
       quantity should not be confused with the length scale given by
       \begin{equation}
         \label{eq:rhub}
         R_{\mathrm H} = {c \over H}.
       \end{equation}
       The above quantity (known as the \emph{Hubble radius}) is often
       mistakenly confused with the particle horizon. The Hubble scale
       evolves in the same way as the particle horizon in simple FLRW
       models and hence measures the domain of future influence of an
       event in these models~\cite{Causal}.  If fields interact only
       through gravity, then the Hubble scale \emph{is} useful as a
       measure of the minimum spatial wavelength of those modes that
       are effectively ``frozen in'' by the expansion of the universe.
       A mode is said to be ``frozen in'' if its frequency is smaller
       than the Hubble parameter, since then there is not enough time
       for it to oscillate before the universe changes substantially
       and so the evolution of that mode is governed by the expansion
       of the universe. Therefore, for modes traveling at the speed
       $c$, if the ``freeze out'' occurs at $\omega<H$, this implies
       that $\lambda> c/H$, as claimed above.
       
       So while the particle horizon is truly the extension of a
       region potentially causally connected via microphysics, the
       Hubble scale has more the meaning of a ``dynamical'' scale. This
       is not surprising if one realizes that the Hubble time
       $H^{-1}\approx 1/\sqrt{G_{\mathrm N}\rho}$ is exactly the dynamical
       time which characterizes all of the gravitational phenomena.
%--------------------------------------------------------------------------
}}
\end{center}

As a final remark it is useful to stress that, even if regions at the
antipodes of our universe would have somehow been in causal contact,
this is still just a necessary, but not sufficient, condition in order
to get isotropization. As shown in~\cite{CH73}, the solutions which
achieve a permanent isotropization represent a subset of measure zero
in the set of all possible SM solutions.

%---------------------------------------------------------------------------
\subsubsection[{The flatness problem}]
{The flatness problem}
%---------------------------------------------------------------------------

The standard model predicts that the universe can be flat, closed or
open depending on its content of matter. The Friedmann equations show
that there is a critical value for the mass density, which determines
which regime the universe is in. Actually, imposing $\Lambda=K=0$ in
Eq.~(\ref{eq:fried1}), one finds that an FLRW model admits Euclidean
spatial sections if
\begin{equation}
  \label{eq:rhocri}
  \rho_{\c}=\frac{3 H^2}{8 \pi G_{\mathrm N}}
\end{equation}
Current observations seem to indicate that the ratio between the
present density $\rho_{0}$ and the critical one is
$\Omega_{0}=\rho_{0}/\rho_{\c}\approx 1$. More precisely the
observations bound $\Omega_{0}$ in the range $[0.1,1]$.

The flatness problem arises from the fact that in FLRW cosmologies the
solutions with $\Omega\approx 1$ are not at all attractors. On the
contrary they represent unstable solutions given the fact that
$\Omega(t)\to 0$ or $\infty$ in time.

%---------------------------------------------------------------------------
\subsubsection[{The entropy problem}]
{The entropy problem}
%---------------------------------------------------------------------------

It is interesting to note that (at least in the usual framework)
the two major cosmological puzzles described above (isotropy/horizon
and flatness) can be reduced to a single problem related to the huge
total amount of entropy that our universe appears to have
today~\cite{Guth81,Guth300,Hu0}. If we define $s\propto T^3$ the entropy
density associated with relativistic particles and $S=a^{3}(t)s$ the
total entropy per comoving volume, then it is easy to see from the
Friedmann equation (\ref{f1}) that
\begin{equation}
  \label{eq:flrwent}
   a^2=\frac{K \cg^2}{H^2 (\Omega-1)},
\end{equation}
and so
\begin{equation}
  \label{eq:sflrw}
  S=\left[ \frac{K \cg^2}{H^2\left(\Omega-1\right)}\right]^{3/2} s.
\end{equation}
The value of the total entropy can be evaluated at the present time
and comes out to be $S>10^{87}$. One can then see that explaining why
$\Omega\approx 1$ (the flatness problem) is equivalent to explaining
why the entropy of our universe is so huge.

In a similar way one can argue (at least in the usual framework) that
the horizon problem can be related to the entropy
problem~\cite{Guth81,Guth300,Hu0}. In order to see how large the causally
connected region of the universe was at the time of decoupling with
respect to our present horizon, we can compare the particle horizon at
time $t$ for a signal emitted at $t=0$, $\ell_{\mathrm h}(t)$, with
the radius at same time, $L(t)$, of the region which now corresponds
to our observed universe of radius $L_{\mathrm{present}}$. The fact
that (assuming insignificant entropy production between decoupling and
the present epoch) $(\ell_{\mathrm h}/L)^{3}
|_{t_{\mathrm{decoupling}}} \ll 1$ is argued to be equivalent to the
horizon problem.  

Once again, a mechanism able to greatly increase $S$
via a non-adiabatic evolution would also automatically lead to the
resolution of the puzzle.

%---------------------------------------------------------------------------
\subsubsection[{The monopole problem}]
{The monopole problem}
%---------------------------------------------------------------------------

Another interesting problem arises when one tries to take into account
particle physics knowledge within the SM framework. It is in fact a
general prediction of Grand Unified Theories (GUTs) --- the set of
theories that try to unify the electro-weak and strong interaction ---
that monopoles should form as a consequence of the $SU(2)$ subgroup
structure inherited from electroweak interactions. These monopoles are
topologically stable knots in the Higgs field expectation value and
are expected to have masses of order $m_{\mathrm{m}}\approx 10^{16}
GeV$.

In particular, the Kibble mechanism~\cite{Kibble76} predicts a number
density of topological defects $n_{\mathrm m}$ which is inversely
proportional to the third power of the correlation length $\xi$ of the
Higgs field
\begin{equation}
  \label{eq:mono}
  n_{\mathrm{m}}\approx {1\over \xi^3}
\end{equation}
In the standard model the natural upper bound for the monopole density
is given by the particle horizon at the moment of their formation.  In
fact causality constrains the correlation length of the Higgs field
to be less that the particle horizon $\ell_{\mathrm h}$. In GUTs, the
phase transition which generates these monopoles is generally
predicted to occur at $T_{\mathrm{GUT}}\approx 10^{14} GeV$ and this
temperature is achieved at approximately $t_{\mathrm{GUT}}\approx
10^{-35}s$ after the big bang. In standard cosmology, one has that in
the radiation dominated era $a(t)\propto \sqrt{t}$ so $\ell_{\mathrm
  h}=2t$ and so
\begin{equation}
  \label{eq:mono2}
  n_{\mathrm{m}}\approx {1\over \xi^3}\geq\frac{1}{8t^{3}_{\mathrm{GUT}}}
\end{equation}
From the fact that the monopole mass density should be
$\rho_{\mathrm{m}}= n_{\mathrm{m}} m_{\mathrm{m}}$, one can derive the
present monopole contribution to the mass density of the universe.
The problem arises from the fact that this turns out to be
\begin{equation}
  \label{eq:omegamono}
  \Omega_{\mathrm{m}}=\frac{\rho_{\mathrm{m}}}{\rho_{\c}}\geq 10^{11}
\end{equation}
which is an outrageously high energy density and is totally
incompatible with the observational limits on the present value of
$\Omega$.

%---------------------------------------------------------------------------
\subsubsection[{The creation of primordial perturbations}]
{The creation of primordial perturbations}
%---------------------------------------------------------------------------

This issue, although it is not strictly a cosmological puzzle,
nevertheless represents an intrinsic lack of predictive power of the
SM for observational facts, in this case the existence of large scale
structures in our universe such as clusters and galaxies.

The standard model does not provide any mechanism for the production
of the primordial fluctuations which should, later on, have evolved into
the present inhomogeneities. Moreover mass perturbations are naturally
gravitationally unstable and tend to grow with time. This implies that
if, for example, one traces back the magnitude of the perturbation
which should have given rise to galaxies, then the result is that at
times of order of the GUT symmetry breaking $t_{\mathrm{GUT}}\approx
10^{-35}s$ the density perturbations should have been extremely
small. Guth estimated in~\cite{Guth300} that
\begin{equation}
  \label{eq:densper}
 \frac{\delta \rho}{\rho}\mbox{(galaxy scale)} \approx 10^{-49}\quad 
  \mbox{at}\quad t=t_{\mathrm{GUT}}
\end{equation}
This is a really tiny figure if one takes into account that it
corresponds to perturbations several orders of magnitude smaller than
the Poissonian fluctuations typical of microscopic systems. Such small
fluctuations are instead typical of strongly correlated (causally
connected) systems (such as a degenerate Fermi gas or a crystal) which
require some form of interaction in order to exist.

Unfortunately, the SM also predicts that any perturbation which now
has a scale of cosmological interest had to be created on a scale
larger than the causal horizon at that time~\cite{Guth300} and so
microphysical interactions are apparently excluded.

%-------------------------------------------------------------------------
\subsection[{General framework}]
{General Framework}
\label{subsec:GFinf}
%---------------------------------------------------------------------------

The inflationary framework resolves (or more correctly mitigates) the
above described puzzles by assuming a period of anomalous, non-adiabatic,
evolution of our universe at a time near to that of the GUT symmetry
breaking.
The basic ``ingredients'' of inflation can be summarized in a few points
\begin{itemize}
\item Some QFT effect led to cosmological evolution equations
  dominated by a vacuum energy which violates the SEC ($\rho+3p<0$)
\item The violation of the strong energy condition leads to
  accelerated expansion for the universe~\cite{Science,Visser:1997tq,
    Visser:1997au}. This can be easily seen from
  Eq.~(\ref{eq:fried2}). In fact for $\rho+3p<0$ one gets $\ddot{a}/a
  >0$. In the case of constant energy density of the vacuum (and in a
  flat FLRW spacetime) one gets an exponential increment in the scale
  factor, $a(t)\propto \exp{(Ht)}$ with $H=\mbox{constant}$ and the
  FLRW metrics acquire a quasi-de~Sitter form (exact de~Sitter is
  invariant under time translations so it cannot have a start or an
  end).
\item The expansion leads to a stretching of the particle horizon due
  to the hyperbolic increment in the scale factor. For a sufficiently long
  inflationary phase one can then solve the horizon and monopole
  problems, in the latter case the huge increase of the particle
  horizon dilutes the density of monopoles.  Similarly
  the flatness problem can be solved. From the Friedmann equation we can
  write
  \begin{equation}
   \label{eq:flateps}
   \epsilon\equiv\Omega-1=
    \frac{K\;c^2}{H^{2}\;a^{2}}
    =\frac{K \;c^{2}}{\dot{a}^2},
  \end{equation}
  and by differentiating the above equation, we can see that on purely
  \emph{kinematic} grounds
  \begin{equation}
  \label{eq:flateps2}
   \dot{\epsilon} = - 2K \; c^{2} \left(\frac{\ddot{a}}{\dot{a}^{3}}\right) 
     = -2\epsilon \left(\frac{\ddot{a}}{\dot{a}}\right).
  \end{equation}
  If during the inflationary phase one has $\ddot{a}/a>0$ then
  $\dot{\epsilon}/\epsilon<0$ and so $\Omega\to 1$. Note that the
  violation of the SEC is not strictly necessary in order to deal with
  the horizon and monopole problems but it is crucial for solving the
  flatness one.
\item The accelerated expansion corresponds to a non-adiabatic change
  of the gravitational fields, this leads to quantum particle
  production and so to causal generation of primordial perturbations.
  In the case of de~Sitter space, the particle spectrum is thermal at
  a ``Hawking temperature'' given by
  $T_{\mathrm{dS}}=H/{2\pi}$~\cite{Lapedes78, Brand85}.  This leads to
  a scale-invariant, Harrison--Zel'dovich (HZ), spectrum of
  perturbations (so far experimentally confirmed over a wide
  wavelength range). In addition, the huge increase in the scale
  factor at almost constant Hubble scale, rapidly makes the
  perturbations larger than the latter scale. As we said, this implies
  them being ``frozen in'' and stopping their evolution. Only after
  inflation, when the Hubble scale starts to grow again and the
  perturbation scale becomes smaller than $H^{-1}$ will they ``freeze
  out'' and start evolving again. In this way much larger initial
  values for the primordial perturbations can be allowed in
  inflationary scenarios.
 
\begin{center}
  \setlength{\fboxsep}{0.5 cm} 
   \framebox{\parbox[t]{14cm}{
%------------------------------------------------------------------------  
  {\bf Warning}: The HZ spectrum is really exact just in the case of
  a precise exponential expansion for an arbitrary long time. In the case
  of quasi-de~Sitter evolution for a finite time it should be
  considered as just a zero-order approximation.
%------------------------------------------------------------------------  
}}
\end{center}

\item The huge expansion ``supercools'' the universe and hence leads to
  a decrement in the entropy density. This implies that inflation
  should admit a graceful exit in which the universe should be
  ``reheated''.  This process breaks the adiabatic evolution of the
  universe and solves the entropy problem by enormously increasing
  the total entropy of the universe $S=s(t)a^{3}(t)$ (at the end of
  inflation the temperature is approximately the same as before
  but the volume is increased by an exponential factor).  This {\em
    reheating} phase is crucial not only for raising up the
  temperature but also because it can lead to further effects on the
  spectrum of primordial perturbations.
\end{itemize}
As a final remark it should be stressed that inflation is generally
able to mitigate the cosmological paradoxes but it is never able to
exactly solve them. The parameter $\epsilon$ is indeed driven towards
zero during the inflationary stage but after that it will still be at
an unstable point in the solution phase space. Similar discussions can
be made for the other puzzles. One can say that inflation enlarges the
region of initial conditions which lead to the observed universe but
it does not make it infinite.

%-------------------------------------------------------------------------
\subsection[{Possible implementations of the framework}]
{Possible implementations of the framework}
%---------------------------------------------------------------------------

~From the above summary of the key features of inflation it should be
clear that the basic requirements for a successful inflationary framework
are quite general (SEC violation, huge increment of the particle
horizon, non adiabatic evolution). Actually what one deals with is a sort
of paradigm which admits a large number of possible implementations.

As we have seen above, the basic requirement which one has to satisfy in
order to get a successful inflationary model, is to have a vacuum
dominated phase of evolution of the universe and to have this for a
sufficient time.

The standard mechanism for obtaining this effect is to assume the
presence of some primordial scalar field $\phi(t,\x)$, generally
called ``the inflaton'', that suddenly finds itself in a false vacuum
state due to the form of its effective potential $V(\phi)$. Generally
the emergence of a false vacuum can be due to temperature induced
spontaneous symmetry breaking but, as we shall see, one can also
assume that the field is just created with $\langle V(\phi)\rangle\neq
0$~\footnote{
%-------------------------------------------------------------------------  
  Although they do not share the same success as the frameworks described  
  above, it is worthwhile to say that in the past several other ways for 
  inducing a de~Sitter phase in the early universe have been
  proposed. Among these proposals a special mention is due to the one by
  Starobinski~\cite{Staro80}. In Starobinski inflation, the
  high-order curvature terms, appearing in the renormalized SET of
  a massless conformally coupled scalar field in FLRW spacetime, can
  remain approximately constant. So once this SET is used in the
  semiclassical Einstein equations (\ref{eq:semiclaseineq}) it can be
  used to start a de~Sitter inflationary phase. This model is now
  abandoned because it was shown in~\cite{Simon92} that when the
  effective action of semiclassical gravity is derived as a
  perturbative approximation to the full (non-perturbative) effective
  action, then the Starobinski inflation is no longer an allowed
  solution.
%----------------------------------------------------------------------
  }.  This means that most of the work on the inflationary model has
been mainly a fine tuning exercise on the inflaton potential. Since
the seminal paper by Alan Guth~\cite{Guth81} several models have been
proposed and the detailed discussion of all of them is out of the
scope of the present work.  In what follows we shall present the basic
characteristics of the most widely applied frameworks.

%-------------------------------------------------------------------------
\subsubsection{Old inflation}
%-------------------------------------------------------------------------

The basic idea in the old inflationary scenario~\cite{Guth81,Guth300}
is the presence of a Higgs field $\phi(t,\x)$ governed by an
effective potential $V(\phi,T)$ (due to some GUT external fields or to gravity
itself) with a pair of minima. Nowadays at $T\approx 0$,
it is assumed that one has a global minimum, say for
$\phi=\phi_{\mathrm{true}},\; V(\phi_{\mathrm{true}},0)\approx 0$, a
local minimum at $\phi=\phi_{\mathrm{false}},\;
V(\phi_{\mathrm{false}},0)\neq 0$.

Now, if the dependence of the effective potential on temperature
is such that, for temperature $T$ greater than some $T_{\mathrm{crit}}$,
$V(\phi_{\mathrm{true}},T)>V(\phi_{\mathrm{false}},T)$ then one gets
that in the early universe the field would have preferably been in the
false vacuum characterized by a constant energy density
$\varepsilon_{0}$ (determined by the shape
of the potential near the false minimum).

In GUTs this constant energy density is proportional to the fourth
power of the mass scale of the theory ($M_{\mathrm{GUT}}\approx
10^{14} GeV$) and hence it is huge.  It is easy to see that if the
Friedmann equation (\ref{eq:fried1}) is dominated by a constant energy
density then one gets a scale factor evolution
\begin{equation}
  \label{eq:evds}
    a(t)\sim a_{\mathrm{init}} e^{H(t-t_{\mathrm{init}})}
      \qquad \mbox{with}\quad H = \sqrt{ \frac{8\pi G_{\mathrm N} }{3}\rho_{0}}
\end{equation}
which is a de~Sitter phase of exponential expansion. 

Unfortunately this model is affected by two main flaws.  The first one
is related to the fact that the de~Sitter phase should end, when $T$
becomes less than $T_{\mathrm{crit}}$, via a first order phase
transition from the false vacuum to the true one. The field has to do
a quantum tunneling from one region to the other.  This implies that
there will be coexistence of regions of true vacuum in an environment
still in the false vacuum state. The problem arises from the fact that
the latter is still expanding exponentially and so the bubble of true
vacuum will never manage to percolate and dominate the universe.

Even in the case that an effective percolation of a true vacuum bubble
happens and it manages to form a bubble able to contain our universe,
one has another problem related to the fact that bubble collisions are
highly non-linear phenomena which would destroy the homogeneity and
isotropy obtained via inflation.

To overcome these problems the new inflation scenario was proposed.

%-------------------------------------------------------------------------
\subsubsection{New inflation}
%-------------------------------------------------------------------------

The so called ``new model'' of inflation (which actually dates back to
1982~\cite{Linde82,Linde82b,AS82}) involves a quite similar scenario
to that described above but with the difference that, in this case,
one has a potential which for some $T<T_{\mathrm{crit}}$, goes from a
shape having a single global minimum, generally assumed to be at
$\phi=0$, (think for example of a paraboloid profile) to a shape in
which the former global minimum becomes a local maximum and a new
global minimum arises (think for example of the standard Mexican hat
shape).

In this case one has a second order phase transition.  The field,
which at $T>T_{\mathrm{crit}}$ was in its global minimum at $\phi=0$,
finds itself at a local maximum on the top of the potential and hence
naturally (by quantum fluctuations) tends to ``roll down'' towards the
true minimum. 

If the potential shape is flat enough, one would get
$V(\phi)\approx\mbox{constant}$ for a sufficient time, and enough
inflationary expansion to solve the cosmological puzzles. The
inflationary epoch ends when the slow rolling phase breaks down and
the field rapidly moves towards its true minimum where it should decay
into lighter fields.

This proposal, although more robust than the previous
one, was also soon found to be plagued by several problems.

Firstly, it is clear that the model requires an extreme fine tuning of
the potential. Potentials with long and flat regions are not generic
in QFT and in the absence of any definitive indication by some GUT,
they appear as a technical artifact.

A second problem is linked to the fact that the initial condition
$\phi=0$ is also non generic in QFT. Moreover in a semiclassical
approach one should deal with $ \langle \phi \rangle $ which, for a
potential invariant under the transformation $ \phi \to -\phi $, is
{\em zero} at all times~\cite{GP85}. This problem can be overcome by
assuming that $ \langle \phi \rangle \neq 0$ in different large regions of
our universe but its value is still zero when one takes a global average.

Finally, it should be stressed that this model of inflation is based on
having a symmetry breaking mechanism which comes into
action when temperature gets low enough. Typically this is equivalent
to roughly saying that inflation can start only for $t \gg
t_{\pl}$. The problem is that FLRW universes are generally
extremely unstable and most of them would have already re-collapsed,
or undergone a large expansion, so that few of them would ``survive''
until well after $t_{\pl}$. We see then that in these models the
fine tuning issues of the SM are mainly moved to the Planck epoch but
are still present.

%-------------------------------------------------------------------------
\subsubsection{Chaotic inflation}
%-------------------------------------------------------------------------

The so called ``chaotic scenario''~\cite{Linde90} was proposed as a
way to overcome the above difficulties. In this class of models one
assumes no spontaneous symmetry breaking for inducing a phase
transition in the inflaton potential. It is instead assumed that the
initial value of the field takes random values in different parts of
the universe and the values of curvature and of the inflaton potential
are of the order expected for them at the end of the Planck epoch
($V(\phi)\leq M^{4}_{\pl}$).

A standard potential considered in this framework is $V(\phi)=\lambda
\phi^4/4$. For reasonable values of $\lambda$ (from the point of view
of QFT) this implies super-Planckian initial values for the mass of
the field. This may appear paradoxical but one can check that the
actual energy stored in the field is ``just'' of Planck order (because
the latter is determined by the gradients of the field and not by its
expectation value). If the field starts at a very high value, it will
then naturally ``roll'' towards the minimum of the potential. For
special forms of the potential, the kinetic energy is constant and
large for a long enough time to be sufficient for starting inflation.
In the end, the field will decay to the global minimum and reheat the
universe.

Although chaotic inflation was more successful than the previous
proposals, it also has several problems among which one should note
that it is necessary to suppose sufficiently homogeneous initial
conditions for the field at the Planck epoch in order to get large
enough regions to go to the inflationary era at the same time.
Finally it is still unclear how quantum corrections to the potential
can influence the framework. The exact form of the curvature of the
effective potential is a crucial feature in order to get sufficient
inflation but at the same time it is extremely sensitive to the
details of the quantum theory being considered.

As we have said, the models discussed above are not the only ones
which have been proposed. For example there are recent, more
realistic, implementations of inflation based on supersymmetric
models.  Nevertheless what is important for us is that they mainly
differ in the way in which they induce the vacuum dominated phase of
evolution of the universe. The basic points which we summarized before
are generally all still valid for whatever framework one chooses.

\begin{center}
  \setlength{\fboxsep}{0.5 cm} 
   \framebox{\parbox[t]{14cm}{
%------------------------------------------------------------------------    
       {\bf Comment}: We have seen at the start of this chapter how
       the topological Casimir effect can sometimes lead to an
       inflationary evolution of the universe. Therefore it can be
       natural to ask whether these sorts of vacuum effect can be used
       as well in order to drive an inflationary scenario. Although
       the example which we showed may appear to be promising, it
       should be emphasized that the vacuum stress energy tensor which
       we got {\em does satisfy} the SEC and hence it cannot by itself
       help in solving the flatness paradox.  The possibility of
       finding an inflationary solution was just due to the presence
       of the positive cosmological constant $\Lambda$, which violates
       the SEC.  It is easy to check that in the case $\Lambda=0$ one
       would get from the SET (\ref{eq:settopo}) just a power law
       expansion for the scale factor.
       
       Although more complex examples of the topological Casimir effect in
       non-trivial spacetimes can be considered it is an open issue
       whether or not they would be able to lead to inflationary scenarios.
%--------------------------------------------------------------------------
}}
\end{center}
%-------------------------------------------------------------------------

%-------------------------------------------------------------------------
\subsection[{Reheating}]
{Reheating}
\label{sec:reheat}
%---------------------------------------------------------------------------

The general foundations of inflation which we have just been
discussing are a typical example of how a vacuum effect can influence
the dynamical evolution of our universe. It is nevertheless an amusing
characteristic of this paradigm that the particle production from
quantum vacuum also comes into action. As we said, the De-Sitter like
phase of the FLRW universe can be shown to drive a production of
particles~\cite{Lapedes78,Brand85} but this is not the end of the
story.

We have seen that the universe underwent a rapid cooling due to the
exponential expansion. As a consequence of this, a very efficient
process is needed for releasing the energy ``stored'' in the inflaton
in order to then reheat the universe sufficiently. Such a process was
only recently developed into a complex theory, the theory of {\em
  preheating}.

The current general theory of reheating can be summarized in three
fundamental steps
\begin{itemize}
\item The classical inflaton field, coherently oscillating at the true
  minimum, decays via parametric resonance into massive bosons.  These
  bosons are generally very far from thermal equilibrium and are
  characterized by very large occupation numbers.  This is the
  ``preheating phase''.
\item After some (model-dependent) time, the backreaction of the
  particle production and the cosmological expansion shuts off the
  inflaton oscillations in such a way that parametric resonance is no
  longer efficient. At this stage the standard channels for the decay
  of the massive bosons start to be relevant. This is what used to be
  the ``old reheating phase''.
\item Finally a ``thermalization phase'' occurs and determines the final
  temperature at which the universe stabilizes after inflation.
\end{itemize}
We shall not deal here with the last two of the above stages but we
shall discuss the basic aspects of parametric resonance in preheating
and propose a possible way to implement this via gravitational effects.

%--------------------------------------------------------------------------
\subsection[{Basic Preheating}]
{Basic Preheating}
\label{subsec:prh}
%--------------------------------------------------------------------------

In this section we shall present some basic features of the
theory of inflationary preheating via parametric resonance.

To start our investigation we can consider the simple case of an
inflation model where an inflaton field has the effective potential
\begin{equation}
  \label{eq:simplpote}
  V(\phi)=\frac{m_{\phi}^{2}\phi^{2}}{2}
\end{equation}
and is coupled to a massive scalar field $\chi(t,\x)$ via the coupling term
\begin{equation}
  \label{eq:simplinte}
  -\frac{1}{2} g^2 \phi^{2}\chi^{2}
\end{equation}
For the moment we shall suppose $\chi(t,\x)$ to be minimally coupled.

The inflaton field is described by the Klein-Gordon equation in FLRW 
\begin{equation}
  \label{eq:kgflrw}
 \left[\partial_{t}^{2}+3\frac{\dot{a}}{a}\partial_{t}
                -\frac{1}{a^{2}}\nabla^{2}\right]\phi(t,\x)=-V^{'}(\phi) 
\end{equation}
We can now restrict our attention to some domain of the universe
where the inflaton can be considered homogeneous. At the end of the
inflationary era the expectation value of the field will start
oscillating around the minimum of the effective potential. Under the
above assumptions, and ignoring the backreaction of quantum
fluctuations, the equation of motion of the field (\ref{eq:kgflrw})
becomes
\begin{equation}
  \label{eq:inflf}
  \ddot{\phi}+3H\dot{\phi}+V^{'}(\phi)=0
\end{equation}
where the prime denotes derivatives with respect to the inflaton field.

The solution of this equation asymptotically approaches the form of
decaying oscillations $\phi(t)=\Phi(t)\sin(m_{\phi}t)$, where $\Phi$
stands for the amplitude of the oscillations and, in the absence of
particle production from the vacuum, behaves like~\cite{KLS94}
\begin{equation}
  \label{eq:phin}
\Phi\sim \frac{M_{\mathrm{pl}}}{m_{\phi} t}\sim   
         \frac{M_{\mathrm{pl}}}{2\pi N}
\end{equation}
where $N$ is the number of oscillations.

If we assume, for the moment, the scalar field $\chi(t,\x)$ to be
minimally coupled to gravity ($\xi=0$), then the equation of motion
for the modes (quantum fluctuations) of the field $\chi$ with physical
momentum $\k/a(t)$ takes the form~\cite{KLS94}
\begin{equation}
 \label{eq:chimod}
  \ddot{\chi}_{k}+3H\dot{\chi}_{k}+\left( \frac{k^{2}}{a^2(t)}+m^{2}_{\chi}+
     g^2\Phi^2\sin^{2}(m_{\phi}t)\right)\chi_{k}=0
\end{equation}
where $k=\sqrt{\k}$. 

As a first approximation we can consider what happens to modes for
which the expansion is an adiabatic effect, that is those modes for
which $\omega H^{-1} \gg 1$. In this frequency range one can safely
neglect the effect of expansion and for the moment set it to zero
($H=0$). In this case Eq.(\ref{eq:chimod}) describes a harmonic
oscillator with a variable frequency
$\Omega^2_{k}(t)=(k/a)^{2}+m^2_{\chi}+g^2\Phi^2\sin^{2}(m_{\phi}t)$. 

It is  now easy to check that, by defining the dimensionless constants
\begin{equation}
  \label{eq:redmat}
 A_{\chi}=\frac{k^{2}}{m^2_{\phi}a^2}+\frac{m^{2}_{\chi}}{m^{2}_{\phi}}+2q\, ,
 \qquad 
 q=\frac{g^2\Phi^2}{4 m^{2}_{\phi}}
\end{equation}
Eq.~(\ref{eq:chimod}) can be cast in the familiar Mathieu form
\begin{equation}
  \label{eq:mathi}
  \chi^{''}_{k}+\left[ A(k)-2q\cos(2z)\right]\chi_{k}=0
\end{equation}
where $z=m_{\phi}t$ and the primes stand for derivatives with respect to $z$.

We met the above equation in chapter~\ref{chap:1} (see
Eq.~(\ref{eq:mathieu})) when we discussed the phenomenon of parametric
resonance. As explained, an important property of the solutions of
Eq.~(\ref{eq:mathi}) is that for some values of the parameters $A$ and
$q$ an exponential instability appears. Generically one should expect
that some modes $\chi_{k}$ will be exponentially amplified
\begin{equation}
  \label{eq:expmod}
  \chi_{k}=p_{k}e^{(\mu_{k}^{(n)}m_{\phi}t)}
\end{equation}
where $p_{k}$ are periodic functions (with the same period as that of
the oscillations of the inflaton field) and $\mu^{(n)}_{k}$ is the
Floquet index corresponding to the $n$-instability band (see
Eq.~(\ref{eq:floq})). This can be interpreted as a very efficient
process of particle production.
% 
%==============================================================================
\begin{figure}[hbt]
  \vbox{\hfil \scalebox{0.60}{{\includegraphics{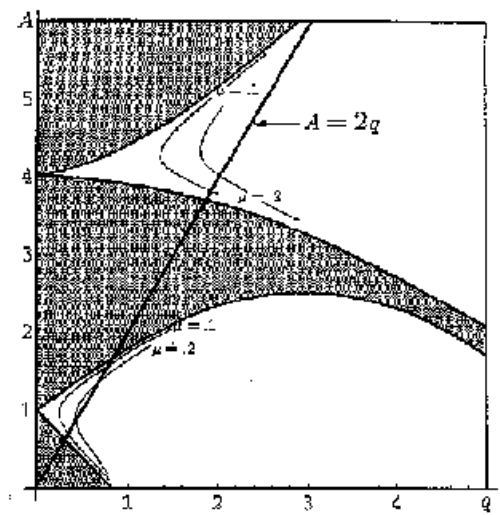}}} \hfil }
  \bigskip
\caption[Band structure of the Mathieu function (\ref{eq:mathi})]{
%----------------------------------------------------------------------------
  Band structure of the Mathieu function (\ref{eq:mathi}).  The white
  and grey bands corresponds respectively to instability and stability
  regions. The line $A=2q$ shows the values of $A$ and $q$ for $k=0$
  and $m_{\chi}/m_{\phi}\approx 0$. This picture is taken
  from~\cite{KLS94}.
%---------------------------------------------------------------------------
} \label{f:mathi}
\end{figure}
%=============================================================================
%
Before proceeding with this introduction to the theory of preheating we
want to stress that we shall deal here only with bosonic
reheating.  Fermionic preheating is also conceivable although in this
case the Pauli exclusion principle puts strong constraints on the
production of particles from the quantum vacuum. Nevertheless it was
soon realized that coherent oscillations of the inflaton can indeed
lead to parametric production of fermions.  We shall not treat this
theory here but direct the reader to some seminal papers in the
subject~\cite{Fermionic}.

%------------------------------------------------------------------------
\subsubsection{Parametric regimes}
%------------------------------------------------------------------------

It should be stressed that very different regimes can be envisaged for
the production of particles at different values of the relevant
parameter $q$. For simplicity we shall describe them in a Minkowski
spacetime ($a(t)=1$) discussing later the effects of expansion. We
shall also consider a massless $\chi$ field.

\paragraph{Narrow resonance}

A first possible regime corresponds to very small values of $q$ ($q\ll
1$ that is $g\Phi<m$) and takes the name of ``narrow resonance''. In
this case the particle production can be treated perturbatively and is
concentrated in the first instability band for modes with $k^2\sim
m^2_{\phi}(1-2q\pm q)$~\cite{KLS97}.  Here the Floquet index takes the
form~\cite{mac}
\begin{equation}
  \label{eq:frb}
  \mu_{k}=\sqrt{\left(\frac{q}{2}\right)^2
          -\left(\frac{2k}{m_{\phi}}-1\right)^{2}}
\end{equation}
The modes $\chi_{k}$ with $k\sim m$ grow as $\exp(qz/2) \sim
\exp(g^2\Phi^2t/8m_{\phi})$ and the number of $\chi$ particles goes
like $n_{k}(t) \sim \exp(2\mu^{(n)}_{k}z) \sim
\exp(g^2\Phi^2t/4m_{\phi})$. This process can be interpreted as the
conversion of two $\phi$ particles into a pair of $\chi$ particles with
momenta $k\sim m$.
% 
%==============================================================================
\begin{figure}[hbt]
  \vbox{\hfil \scalebox{0.40}{{\includegraphics{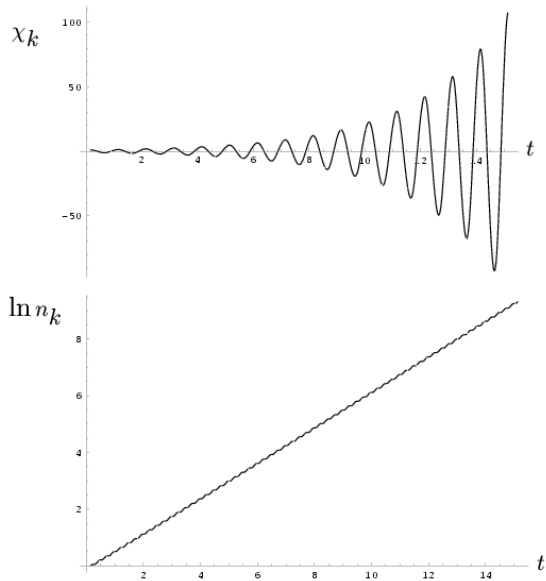}}} \hfil }
  \bigskip
\caption[Mode and particle number amplification for the narrow resonance
  regime]{
%----------------------------------------------------------------------------
    Mode and particle number amplification for the narrow resonance
    regime with $q\approx 1$. The time is in units of $m_{\phi}/2\pi$
    and is hence equal to the number of oscillations of the inflaton
    field $\phi$.  From~\cite{KLS97}.
%---------------------------------------------------------------------------
} \label{f:narrow}
\end{figure}
%=============================================================================
%
\paragraph{Broad resonance}

A second important regime comes into action for wide oscillations of
the inflaton, that is for $q\gg 1$ ($g\Phi\gg m)$. This is the so-called
``broad resonance'' regime. It is a typically non-perturbative
process and leads rapidly to non-negligible backreaction by the
created particles on the inflaton field. The resonance occurs not just
in the first instability band but generically for modes above the
$A=2q$ line in the Mathieu plot, Fig.~\ref{f:mathi}, which have
momenta $k^{2}/m^{2}=A-2q$. In the broad resonance regime the
occupation numbers of the particles produced are extremely large,
typically of order $n_{k}\sim1/g^2$.

\begin{figure}[hbt]
  \vbox{\hfil \scalebox{0.40}{{\includegraphics{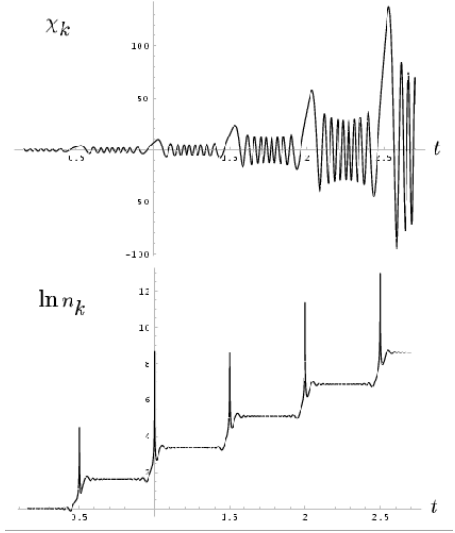}}} \hfil }
  \bigskip
\caption[Mode and particle number amplification for the broad resonance
  regime]{
%----------------------------------------------------------------------------
    Mode and particle number amplification for the broad resonance
    regime with $q\gg 1$. The time is in units of $m_{\phi}/2\pi$.
    Note the rapid oscillations in $\chi_{k}$ and the ``stair-like''
    behaviour of the particle number.  From~\cite{KLS97}.
%---------------------------------------------------------------------------
} \label{f:broad}
\end{figure}

It is interesting to note that, due to the interaction term
(\ref{eq:simplinte}), the $\chi$ field acquires an effective mass
$m_{\chi}(t)=g\phi(t)$ which, in this special regime, can be much
larger that the inflaton mass. As a result, the typical frequency of
oscillation of the $\chi$ field $\omega(t)=
\sqrt{k^{2}+m_{\chi}^{2}(t)}$ becomes larger than that of the
inflaton.  This also implies that during most of the period of
oscillation of $\phi$ the mass of the $\chi$ field changes
adiabatically. Only for $\phi(t)\approx 0$, where the effective mass
of the $\chi$ field is almost zero, is there violation of the adiabatic
condition
\begin{equation}
  \label{eq:adi}
  \frac{\d\omega}{\d t}\leq \omega^2
\end{equation}
and efficient production of particles as shown in
Fig.~\ref{f:broad}.
%
%==============================================================================
%=============================================================================

A final comment is in order. One should not think that arbitrarily high
values for $q$ would lead to higher and higher bursts of particle
creation. In fact, for $q\geq 10^3$, generally non-perturbative
effects (like backreaction and rescattering) soon become crucial for
determining the duration of the resonance regime and the final variance
of both the $\chi$ and $\phi$ fields.  We direct the reader to~\cite{KLS97}
for further insight into this subject.

%-----------------------------------------------------------------------------
\paragraph{Stochastic resonance} 
%----------------------------------------------------------------------------

To conclude this short introduction to the post-inflationary
preheating phase, we shall now briefly discuss the role of expansion
which we have neglected so far. We can start by noticing that if expansion
is included, then the modes suffer a redshift $\k\to\k/a(t)$ and the
amplitude of the inflaton oscillations decreases in time as $t^{-1}$.
This drives the parameters $A$ and $q$ towards zero and hence kills the
parametric amplification.

Moreover it should be noted that Eq.~(\ref{eq:phin}) also tells us
that the amplitude of the $\phi$ oscillations decreases very rapidly
if their number is increased. Since $q\sim \Phi^2 \sim 1/N^2$, only a
few oscillations will be enough to greatly deplete $q$.  This means
that in the case of narrow resonance (when $q \ll 1$ from the
beginning) the amplification is not able to start at all.

The former point is nevertheless an important one. In fact if we assume to
start in a broad resonance regime then one can easily show that
the field $\chi$ moves through several bands after a few oscillations. In
fact in Mathieu theory the band number is given by $n=\sqrt{A}$ and so if
the resonance happens mostly for $A\sim 2q $ one gets $n\sim
\sqrt{2q}\sim g\Phi/(m_{\phi}\sqrt{2})$. If we take
$m_{\phi}\approx 10^{-6} M_{\pl}$, $g\approx 10^{-1}$ then, by
using Eq.~(\ref{eq:phin}) and Eq.~(\ref{eq:redmat}), one finds that
after the first oscillation $n\approx 10^{4}$ but $n\approx
5\cdot 10^{3}$ after the second.

Therefore we see that even during a single oscillation the field does
not remain in one instability band but moves through a thousand of
them.  The standard method for looking at particle production in a
chosen instability band then fails completely here. It is nevertheless
still possible to make a numerical analysis under suitable
approximations. In \cite{KLS97} such an analysis is performed and
shows a behaviour in which the particle number generally increases
(discontinuously) but sometimes even decreases. The $X_{k}$ modes are
defined via rescaling of the $\chi_{k}$ ones,
$X_{k}=\chi_{k}(t/t_{0})$ where $t_{0}$ is the starting time.  The
occupation number of particles per mode can be constructed as
\begin{equation}
  \label{eq:nummof}
  n_{k}=\frac{\omega_{k}}{2}
\left(\frac{|\dot{X}_{k}|^{2}}{\omega_{k}^{2}}+|X_{k}|^2\right)-\frac{1}{2}
\end{equation}
%
%==============================================================================
\begin{figure}[hbt]
  \vbox{\hfil \scalebox{0.40}{{\includegraphics{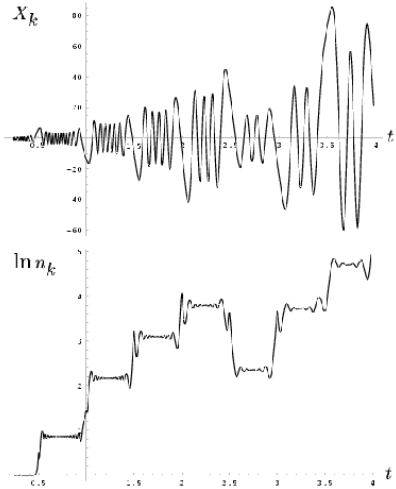}}} \hfil }
  \bigskip
\caption[Mode and particle number amplification for the stochastic resonance
  regime]{
%----------------------------------------------------------------------------
    Mode and particle number amplification for the stochastic
    resonance regime with $q\sim 3\cdot 10^3$. Note the rapid
    oscillations in $\chi_{k}$ and the ``stair-like'' behaviour of the
    particle number but this time including several drops in the
    particle number. This is a purely quantum mechanical effect which
    would be impossible if these particles were in a classical state.
    From~\cite{KLS97}.
%---------------------------------------------------------------------------
} \label{f:stoca}
\end{figure}
%=============================================================================
%

After this brief overview of the basic elements of the the theory of
inflationary preheating we shall now discuss a different ``channel''
for parametric resonance to occur, the so-called ``geometric
preheating''. This original investigation has been made in
collaboration with B.A.~Bassett and most of the material
presented here has been published in reference~\cite{BL98}.

%--------------------------------------------------------------------------
\section[{Geometric preheating}]
{Geometric Reheating}
\label{subsec:grh}
%--------------------------------------------------------------------------

The preheating theory which we have seen so far is based on the direct
coupling between the inflaton and the quantum field to be produced.
Although we have worked with the simple coupling term
(\ref{eq:simplinte}) other forms of coupling can be considered and
other fields (such as fermions or vector bosons) can be coupled to the
inflaton and eventually amplified by its oscillations. Obviously weak
coupling will strongly depress any reheating mechanism (from
(\ref{eq:redmat}) it is easy to see that small values of $g$ end up
depressing the value of $q$).

One can nevertheless wonder whether there is the possibility, for some
generic coupling, for this process to play an important role and
whether gravity should not play a more active part in the process of
preheating. In particular we have considered so far a minimally
coupled scalar field $\chi(t,\x)$ excluding any direct action of
gravity.

Our starting point is that this last assumption is far from being
generic. Renormalization group studies in curved spacetime
\cite{BOS92,RH97,BDH94} have shown that even if the bare coupling
$\xi_0$ is minimal, after renormalization one should generically
expect $\xi \not = 0$.  In particular, while in the infrared limit
fixed points may correspond to a conformally invariant field ($m=0$,
$\xi= {1\over6}$)~\footnote{
%------------------------------------------------------------------
  Note that the relevance of conformal coupling in the low energy
  limit of the theory can be indirectly confirmed by the requirement
  that the equivalence principle holds also at the semiclassical
  level~\cite{SF93}.
%-----------------------------------------------------------------
}, in different GUT models the coupling may also diverge, $|\xi|
\rightarrow \infty$, in the UV limit~\cite{BOS92,PT}.
In both cases the nature of the preheating is very different from
the standard models based on explicit self-interactions or
particle-physics couplings between fields (see e.g. \cite{KLS97}).

Since we are particularly interested in the preheating regime which
occurs when inflation ends, we are interested in the ultra-violet (UV)
fixed points of the renormalization group equations. As a consequence
of this we can assume a general non minimal coupling with possibly
large values of the parameter $\xi$.

To be concrete, consider a generic FLRW universe described by the
metric element (\ref{eq:flrw}) and a minimally coupled scalar field
$\phi$ which is governed by the evolution equation (\ref{eq:inflf}).

Let us restrict ourselves to the quadratic potential
(\ref{eq:simplpote}) which we know to give, for $K=0$, an oscillatory
behaviour of the field: $\phi=\Phi \sin(m_{\phi} t)$, with $\Phi \sim
1/m_{\phi}t$.  In the following we shall try to preserve maximal
generality but when results are derived specifically for the potential
(\ref{eq:simplpote}) we shall denote them with $\doteq$.

The energy density and pressure for a minimally coupled scalar field,
treated as a perfect fluid, are 
\begin{equation}
  \label{eq:endp}
 \mu = \kappa_{\mathrm N}\left({1\over2}\dot{\phi}^2 + V(\phi)\right),\qquad 
   p = \kappa_{\mathrm N}\left({1\over2}\dot{\phi}^2 - V(\phi)\right),
\end{equation}
where $\kappa_{\mathrm N}=8\pi G_{\mathrm N}$.  This breaks down if the field is
non-minimally coupled (an imperfect fluid treatment must be used), if
the effective potential is not adequate~\cite{devega}, or if large
density gradients exist.  The FLRW Ricci tensor is~\cite{KS84}
\begin{eqnarray}
  R^0{}_0 &=& 3\frac{\ddot{a}}{a} \nonumber \\
  R^i{}_j &=& \left[\frac{\ddot{a}}{a} +
   2\left(\frac{\dot{a}}{a}\right)^2 + \frac{2K}{a^2}\right]
 \delta^{i}{}_j \,, 
 \label{eq:ricci}
\end{eqnarray}
where as usual $i,j = 1..3$. The Ricci scalar is~\footnote{Note: we
  are assuming that at the onset of preheating the inflaton $\phi$ is
  the dominant matter field. For this reason we are neglecting the
  contribution of the $\chi$ field.}:
\begin{equation} 
 R = 6\left(\frac{\ddot{a}}{a} +
   \left(\frac{\dot{a}}{a}\right)^2 + \frac{K}{a^2}\right) \,.
 \label{eq:ricciscalar} 
\end{equation} 
We now see that the curvature scalar can be expressed in terms of a
combination of the derivatives of the FLRW scale factor. In order to
express it directly as a function of the inflaton field and of its
effective potential --- we are assuming that at the onset of
preheating the inflaton is still the dominant matter in the universe
--- we can look at the Raychaudhuri equation for the evolution of the
expansion factor $\Theta = 3 \dot{a}/a$~\footnote{
%-------------------------------------------------------------------------  
  The expansion is generally defined as $\Theta \equiv u^a{}_{;a}$
  where $u^a$ is the 4-velocity and $;$ denotes a covariant derivative
  \cite{ellis71}.
%-------------------------------------------------------------------------
}
\begin{equation}
  \dot{\Theta} = -\frac{3\kappa_{\mathrm N}}{2} \dot{\phi}^2+ 
                   \frac{3 K}{a^{2}}\,, 
 \label{eq:ray} 
\end{equation}
We can also get information from the Friedmann equation
(\ref{eq:fried1}) which takes the form
\begin{equation}
 \Theta^2 + {9K\over a^{2}} = 3\kappa_{\mathrm N} \mu 
                            = 3\kappa_{\mathrm N} \left({1\over 2}
                                \dot{\phi}^{2}+ V(\phi)\right) \,. 
\label{eq:fried}
\end{equation}
Using equations (\ref{eq:ray}) and (\ref{eq:fried}), one can
systematically replace in (\ref{eq:ricciscalar}) all factors of
$\dot{a}, \ddot{a}$ with factors of $\dot{\phi}$ and $V(\phi)$.

As an example, taking the usual quadratic potential, $K = 0$ and
$\dot{a}/(a m_{\phi}) \ll 1$, one may solve Eq. (\ref{eq:fried})
perturbatively~\cite{KH96} obtaining
\begin{equation}
 \Theta \doteq \frac{2}{t}
  \left[1 - \frac{\sin 2 m_{\phi} t}{2 m_{\phi}t}\right]\,,
 \label{eq:pertth}
\end{equation}
to first order in $\dot{a}/(a m_{\phi})$. This is only valid after
preheating when $\Phi \ll 1$ but shows that the expansion oscillates
about the mean Einstein-de~Sitter ({\sc eds}) pressure-free solution.
Eq.~(\ref{eq:pertth}) can be integrated to give the scale factor:
\begin{equation}
 a(t) \doteq \overline{a} \exp\left( \frac{\sin 2 m_{\phi}t}{3m_{\phi} t} - 
 \frac{2 \mbox{ci}(2 m_{\phi} t)}{3}\right) 
 \label{eq:scale}
\end{equation}
where $\overline{a} = t^{2/3}$ is the background {\sc eds} evolution,
and $\mbox{ci}(m_{\phi} t) = -\int^{\infty}_{t} \cos(m_{\phi} z)/z
dz$. This example explicitly demonstrates how by just temporal
averaging (which yields $\overline{a}$) one would completely miss the
oscillatory behaviour of the expansion factor. It is instead this
feature that we shall use to get parametric resonance.

The basic idea which we shall follow is that the oscillations in the
inflaton field {\em induce} an oscillatory behaviour in the expansion
factor and hence in the Ricci scalar. It then follows that any field
which is non-minimally coupled to gravity will be potentially subject
to parametric amplification without the need for a direct coupling to
the inflaton. In this sense this preheating channel, which can be called 
{\em geometric reheating} can be considered as
a typical example of the dynamical Casimir effect in a periodic external
field where the latter is the gravitational one.

In the following, we shall apply this simple idea to the case
of scalar, vector and tensor fields.

%--------------------------------------------
\subsection[{Scalar fields}]
{Scalar fields}
%-------------------------------------------

We start by considering the easiest case when $N$ massive scalar fields
$\chi^{(\nu)}(t,\x)$, $\nu=1 \dots N$, are non-minimally coupled to gravity 
\begin{equation}
 \L(x) = \frac{1}{2}\sum_{\nu}^N 
  \left[g^{\alpha\beta}\nabla_{\alpha}\chi^{(\nu)}\nabla_{\beta} \chi^{(\nu)}+ 
   \left(m^{(\nu)}_{\chi}\chi^{(\nu)}\right)^{2} 
         + \xi^{(\nu)} \left(\chi^{(\nu)}\right)^{2} R \right] 
\label{eq:lag} 
\end{equation}
so we now have the inflaton, with potential $V(\phi)$, coupled only
via gravity to the other scalar fields, which have no
self-interactions, masses $m^{(\nu)}_{\chi}$ and non-minimal couplings
$\xi^{(\nu)}$. The equation of motion for modes of the $\nu$-th field is:
\begin{equation}
 \ddot{\chi}^{(\nu)}_k + \Theta \dot{\chi}^{(\nu)}_k + \left(\frac{k^2}{a^2} + 
   \left(m^{(\nu)}_{\chi}\right)^{2} + 
    \xi^{(\nu)} R\right) \chi^{(\nu)}_k = 0\,,
 \label{eq:nuth}
\end{equation}
~From Eq (\ref{eq:ricciscalar},\ref{eq:ray},\ref{eq:fried}) the Ricci 
scalar is given by 
\begin{equation}
  R = - \kappa_{\mathrm N}\dot{\phi}^2 + 4\kappa_{\mathrm N} V(\phi)~~~(K = 0)\,.
\label{eq:ricci2}
\end{equation}
We can now treat separately the cases of minimally and non-minimally
coupled fields and investigate the different behaviour of the parametric 
amplification.

%----------------------------------------------------------------------
\subsubsection{The minimally coupled case}
\label{ssec:mcfield}
%----------------------------------------------------------------------

Consider $\xi_{\nu} = 0$.  We can manipulate Eq.~(\ref{eq:nuth}) by
changing variables from $\chi^{(\nu)}_{k}$ to
$\bar{\chi}^{(\nu)}_{k}=a^{3/2}\chi^{(\nu)}_{k}$ in order to remove
the first derivative term. Then by making use of
equations~(\ref{eq:ray},\ref{eq:fried}), we can reduce it to:
\begin{equation}
 \frac{d^2 \bar{\chi}^{(\nu)}_k }{dt^2} + \left[\frac{k^2}{a^2} + 
  \left(m_{\chi}^{(\nu)}\right)^2 + \kappa_{\mathrm N}{3\over8}\dot{\phi}^2
  - \kappa_{\mathrm N}{3\over4}V(\phi)+{3\over4} {K\over{a^{2}}}\right]
  \bar{\chi}^{(\nu)}_k = 0
\label{eq:vac2} 
\end{equation}
where we have used the useful identity:
$\Theta^{2}+\dot{\Theta}=3 (\dot{a}^2/a^{2}+2\ddot{a}/a)$.

We can now see that there is parametric resonance just because the
expansion $\Theta$ oscillates.  In fact if we take the potential
(\ref{eq:simplpote}) then we can recover the Mathieu equation
(\ref{eq:mathi}) by defining $z=m_{\phi}t+\pi/2$, and introducing the
time-dependent parameters
\begin{eqnarray}
    A(k,t) &\doteq & 
{{4 k^2+3K} \over {4 a^{2} m^{2}_{\phi} } }+ {\left(m^{(\nu)}_{\chi}\right)^2 
   \over m^{2}_{\phi} }\\
         q &\doteq & {3\over16} \kappa_{\mathrm N} \Phi^2
 \label{eq:mcparam} 
\end{eqnarray}
~From this we see that the production of particles is reduced as
$m^{(\nu)}_{\chi}$ increases. Indeed, since $A \rightarrow
m^2_{\chi}/m^2_{\phi}$, $q \rightarrow 0$ due to the expansion,
production of minimally coupled bosons is rather weak and shuts off
quickly due to horizontal motion on the stability chart. We stress
that the production is, however, much stronger than that obtained in
previous studies where the scale factor evolves monotonically
\cite{matacz}. This mild situation changes dramatically when a
non-minimal coupling is introduced.

%-----------------------------------------------------------------------
\subsubsection{Non-minimal preheating}
\label{ssec:nmc}
%-----------------------------------------------------------------------

We now include the {\em arbitrary} non-minimal coupling $\xi_{\nu}$. 
Using Eq. (\ref{eq:ricci2}) one can reduce Eq. (\ref{eq:nuth}) to:
\begin{eqnarray}
 \frac{d^2 \bar{\chi}^{(\nu)}_k}{dt^2}+
  \left[\frac{4k^2+3K}{4a^2} +
   \left(m_{\chi}^{(\nu)}\right)^2 + 
    \kappa_{\mathrm N}\left(\frac{3}{8} - \xi_{\nu}\right)
     \dot{\phi}^2 -
      \kappa_{\mathrm N}\left(\frac{3}{4}-4\xi_{\nu}\right) V(\phi) \right] 
       \bar{\chi}^{(\nu)}_k 
  = 0  
 \label{eq:nmcvac2}
\end{eqnarray}
Adopting again the new variable $z=m_{\phi}t+\pi/2$, we can also
express Eq. (\ref{eq:nmcvac2}) in the Mathieu form (\ref{eq:mathi}).
The time-dependent parameters are now
\begin{eqnarray}
 A(k,t) &\doteq & {{4k^2+3K} \over {4a^{2} m^{2}_{\phi} } } +
  {m^{2}_{\chi} \over m^{2}_{\phi} } + 
   {{\kappa_{\mathrm N} \xi_{\nu}} \over 2}\Phi^{2}\\
    q(t) &\doteq & {3\over4} \kappa_{\mathrm N} 
     \left({1\over 4} - \xi_{\nu} \right) \Phi^2 
 \label{eq:nmcparam} 
\end{eqnarray}
The crucial observation is that since $\xi_{\nu}$ is initially free to take
on any value \footnote{
%----------------------------------------------------------------------  
  The only constraints that one might impose are that the effective
  potential should be bounded from below and that the strong energy
  condition be satisfied. The first is difficult to impose since $R$
  oscillates and the second is difficult since one should use the
  renormalized stress-tensor, $\langle T_{ab}\rangle$.
%---------------------------------------------------------------------
  }, $A(k)$ is not restricted to be either positive or small.  ~From
Eq. (\ref{eq:nmcparam}) it is clear that $A(k) < 0$ for sufficiently
negative $\xi_{\nu}$. The possibility of a negative $A$ was the
suggestion of the work by Greene {\em et al} \cite{GPR97}. However, in
their model, this powerful negative coupling instability was only
partially effective due to the non-zero vacuum expectation value
acquired by the $\chi$ field due to its coupling, $g$, with the
inflaton. Here we only have gravitational couplings and this
constraint is removed.

Actually if we assume that $\xi_{\nu}<0$, then it is easy to see that
for the $k=0$ mode and for $m_{\chi}/m_{\phi}\approx 0$ the relation
between $A$ and $q$ is no longer, as in Eq.~(\ref{eq:redmat}),
$A\approx 2q$. Instead we now have that, in the same limits
\begin{equation}
  A(k,t) \approx -\frac{2}{3}|q|+\frac{\kappa_{\mathrm N}}{8}\Phi^{2}
 \label{eq:nmcparam2} 
\end{equation}
and so a negative $A$ (induced when $\xi_{\nu} < 0$) implies that the
properties of resonance are quite different in the $\xi_{\nu}>0$ case.
%
%==============================================================================
\begin{figure}[hbt]
  \vbox{\hfil \scalebox{0.40}{{\includegraphics{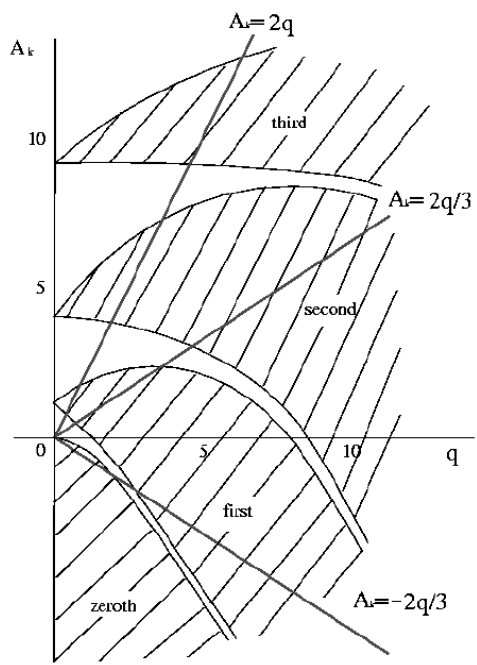}}} \hfil }
  \bigskip
\caption[Mathieu function band structure for $\xi_{\nu}<0$]{
%----------------------------------------------------------------------------
 Band structure for $\xi_{\nu}<0$, from~\cite{Tsujikawa:1999jh}. 
%---------------------------------------------------------------------------
} \label{fig:geom}
\end{figure}
%=============================================================================
%
As shown in Fig.~\ref{fig:geom}, the regions with $A<0$ of the $(A,q)$
plane possess a new instability band (sometimes called the zeroth
instability band) which extends below some critical curve located
approximately at $A\approx -q^2/2$ for small $|q|$.  The line
$A=-2q/3$ crosses this curve for $q\approx 1.4$ and so since $q$ drops
below unity because of the cosmological expansion, there are still
some modes in the zeroth band~\cite{Tsujikawa:1999jh}.

\begin{figure}[hbt]
  \vbox{\hfil \scalebox{0.60}{{\includegraphics{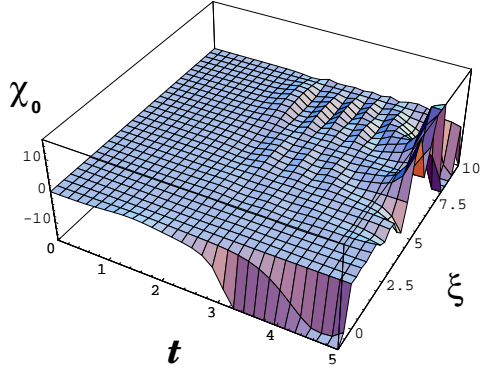}}} \hfil }
  \bigskip
\caption[Geometric preheating: evolution of the $k = 0$ mode]{
%----------------------------------------------------------------------------
  The evolution of the $k = 0$ mode (with $m_{\nu}^2/m_{\phi}^2 \simeq
  1$), as a function of time and of the non-minimal coupling parameter
  $\xi_{\nu}$. For positive $\xi_{\nu}$ the evolution is
  qualitatively that of the standard preheating with resonance bands.
  However, for negative $A$ (negative $\xi_{\nu}$) the solution
  changes qualitatively and there is a negative coupling instability.
  There are generically no stable bands and the Floquet index
  corresponding to $-|\xi_{\nu}|$ is much larger, scaling as $\mu_k
  \sim |\xi_{\nu}|^{1/2}$.
%---------------------------------------------------------------------------
} \label{fig:nmc1}
\end{figure}

\begin{figure}[hbt]
  \vbox{\hfil \scalebox{0.60}{{\includegraphics{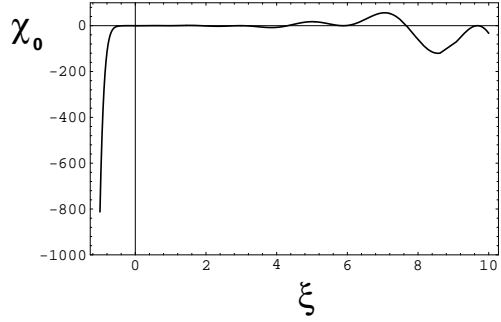}}} \hfil }
  \bigskip
\caption[Geometric preheating: a slice of the spectrum in fig.~\ref{fig:nmc1}]{
%----------------------------------------------------------------------------
  A slice of the spectrum in fig.~\ref{fig:nmc1} at $t = 5$ as a
  function of the non-minimal coupling $\xi_{\nu}$. The qualitative
  differences between $\xi_{\nu} < 0 $ and $\xi_{\nu} > 0$ are clear.
%---------------------------------------------------------------------------
} 
\label{fig:nmc2} 
\end{figure}

When $2|q|/3 > |A| \gg 1$ we have $\mu_k \sim |q|^{1/2} \simeq (6\pi
G_{\mathrm N} |\xi_{\nu}|)^{1/2}\Phi$ along the physical separatrix $A
= \kappa_{\mathrm N}\Phi^{2}/8 - 2|q|/3$.  Since the renormalized
$|\xi_{\nu}|$ may have very large values, this opens the way to
exceptionally efficient reheating - see Figs.
(\ref{fig:nmc1},\ref{fig:nmc2}) - via resonant production of highly
non-minimally coupled fields with important consequences for GUT
baryogenesis \cite{GPR97} and non-thermal symmetry restoration.

For example, let us consider $m_{\phi} \simeq 2\times 10^{13} GeV$ as
required to match CMB anisotropies $\Delta T/T \sim 10^{-5}$. Then GUT
baryogenesis with massive bosons $\chi$ with $m_{\chi} > 10^{14} GeV$
simply requires that geometric preheating with $A<0$ happens with
$\xi_{\nu} < - (\kappa_{\mathrm N} \Phi^2)^{-1} 10^2$.  Instead if one
requires the production of GUT-scale gauge bosons with masses
$m_{\mathrm{gb}} \sim 10^{16} GeV$ this is still possible if the
associated non-minimal coupling is of order $\xi_{\nu} \sim -
(\kappa_{\mathrm N} \Phi^2)^{-1} 10^6$.  Such coupling values have
been considered in, for example~\cite{SBB89}.  The massive bosons with
$m_{\chi} \sim 10^{14} GeV$ can be produced in the usual manner via
parametric resonance if $\xi_{\nu} > 0$, but this process is weaker
(c.f.~\cite{KLR96,Tsujikawa:1999jh}).

%==============================================================================
%=============================================================================
%
%==============================================================================
%=============================================================================
%

Since the coupling between $\phi$ and $\chi$ is purely gravitational,
backreaction effects in the standard sense (see \cite{KLS97,devega})
cannot shut off the resonance (they are generally based on direct
coupling terms like (\ref{eq:simplinte})). The inflaton continues to
oscillate and produce non-minimally coupled particles, receiving no
corrections to $m_{\phi, \eff}^2$ from $\langle \chi^2 \rangle$ (as
is instead the case if one has, for example, a coupling term of the
form $g\chi^{2} \phi^{2} /2$).  Understanding when the parametric
resonance is shut down in geometric preheating is therefore rather
difficult.

The standard method is to establish the time when the resonance is
stopped by the growth of $A(k)$ which pushes the $k = 0$ mode out of
the dominant first resonance band.  In our case we must understand how
$A(k)$ changes as the $\chi$-field gains energy and alters the Ricci
curvature. In fact the basic mechanism which one can expect is that at
some time the energy transferred into the $\chi$ field will be enough
so that it is no longer possible to neglect its backreaction on the
Ricci scalar.

If one assumes that most of the energy goes into the $\chi_0$ mode,
then the change to the Ricci curvature can be expressed as $\delta
R_{\chi} = 8 \pi(E - S)$, where~\cite{marcelo}:
\begin{equation}
 E = \frac{G_{\eff}}{G_{\mathrm N}}\left[\frac{\dot{\chi}_0^2}{2} + 
 \frac{m_{\chi}^2 \chi_0^2}{2} - 12 \xi_{\nu} \chi_0 \dot{\chi}_0 
 \frac{\dot{a}}{a}\right]
 \label{eq:nmce}
\end{equation}
is the $T^{00}$ component of the $\chi$ stress tensor, and
\begin{equation}
 S = \frac{3G_{\eff}}{1 + 192\pi G_{\eff} G_{\mathrm N}\xi_{\nu}^2 \chi_0^2} 
 \left[\frac{\dot{\chi}_0^2}{2} - \frac{m_{\chi}^2 
 \chi^2_0}{2}  
 + 4\xi_{\nu} \left( \frac{\dot{a}^2}{a^2} - \chi_0 
 \dot{\chi}_0 \frac{\dot{a}}{a} -  m_{\chi}^2 \chi_0^2 \right) + 64\pi G_{\mathrm N}  
 \xi_{\nu}^2 \chi_0^2 E \right]
 \label{eq:nmcs} 
\end{equation}
is the spatial trace of the stress tensor $T^i{}_i$ corresponding to
$3p$ in the perfect fluid case.  (In the above formulae $G_{\eff} =
G_{\mathrm N}/(1 + 16\pi G_{\mathrm N}\xi_{\nu} \chi^2) $ is the effective
gravitational constant.)

Since $\chi_0$ is rapidly growing, the major contribution
of $\delta R_{\chi}$ will be to $A(k)$, causing a rapid vertical
movement on the instability chart. Once $\delta A + A > 2|q| +
|q|^{1/2}$, the resonance is shut off. If $\xi_{\nu} < 0$, most of the
decaying $\phi$ energy is pumped into the small $k$ modes (see
Fig.~\ref{fig:nmc2}).  Subsequently we expect the oscillations in
$\chi_0$ to produce a secondary resonance due to the self-interaction
and non-linearity of Eqs (\ref{eq:nmce},\ref{eq:nmcs}).

%---------------------------------------------------------
\subsection[{Vector fields}]
{Vector fields}
%-----------------------------------------------------------

Until now, reheating studies have been limited to minimally-coupled
scalar fields, fermions and gauge bosons~\cite{BHP97}. In the case of
vector fields the minimum one can do to preserve gauge-invariance is
to couple to a complex scalar field via the current since real scalar
fields carry no quantum numbers.  We consider here only vacuum vector
resonances, however.

A massive spin--1 vector field in curved spacetime satisfies the
equations:
\begin{equation}
 (-\nabla_a \nabla^a + m_{\aa}^2)\aa^{b} + R^{b}{}_{a} \aa^a = 0
 \label{eq:vec1}
\end{equation}
These equations are equivalent to the Maxwell--Proca equations for the
vector potential $\aa_a$ only after an appropriate gauge choice which
removes one unphysical polarization.  In our case we shall use the
so-called tridimensional transversal constraint:
\begin{equation}
\aa_0 = 0,~~~ \nabla^i \aa_i = 0
\label{eq:vec2}
\end{equation}
This set is equivalent to the Lorentz gauge, although it does not
conserve the covariant form of the latter.  Nonetheless, in either
case, gauge-invariant quantities such as the radiation energy density,
are unaffected.

In a {\sc FLRW} background, the Ricci tensor is diagonal, which
together with the gauge choice (\ref{eq:vec2}) and expansion over
eigenfunctions, ensures the decoupling of the set of equations
(\ref{eq:vec1}). We can reduce the system to a set of decoupled
Mathieu equations. For $K=0$ the Ricci tensor is (see
Eq.~\ref{eq:ricci})
\begin{equation}
 R^a{}_b = \kappa_{\mathrm N} V(\phi) \delta^a{}_b 
   - \kappa_{\mathrm N} \dot{\phi}^2 \delta^a{}_0 \delta^0{}_b\,, 
 \label{eq:ricci1}
\end{equation}
which leads to the Mathieu parameters for the spatial components
$(a^{3/2} \aa^i)$
\begin{equation}
 A(k) \doteq \frac{k^2}{a^2 m_{\phi}^2} + \frac{m_{\aa}^2}{m_{\phi}^2} +  
 2q~~~,~~ q \doteq \frac{\kappa_{\mathrm N} \Phi^2}{8} 
 \label{eq:spat}
\end{equation}
showing that vector fields are also parametrically amplified (albeit
weakly) during reheating as in the scalar case.

%------------------------------------------
\subsection[{The graviton case}]
{The graviton case}
%-----------------------------------------

It has been shown using the electric and magnetic parts of the Weyl
tensor \cite{bass97} that there exists a formal analogy between the
scalar field and graviton cases during resonant reheating.  Here we
will show that the correspondence also holds in the Bardeen formalism.
The gauge-invariant (at first order) transverse-traceless (TT) metric
perturbations $h_{ij}$ describe gravitational waves in the classical
limit.  In the Heisenberg picture one expands over eigenfunctions,
$Y_{ab}$ of the {\em tensor} Laplace--Beltrami operator with scalar
mode functions $h_k$, which satisfy the equation of motion:
\begin{equation}
 \ddot{h}_k + \Theta \dot{h}_k + \left(\frac{k^2 + 2K}{a^2}\right) h_k = 0\,,
 \label{eq:bard}
\end{equation}
or equivalently
\begin{equation}
 (a^{3/2} h_k)\ddot{} + \left(\frac{k^2 + 2K}{a^2} + 
                          \frac{\textstyle 3}{4}p\right)(a^{3/2} h_k) = 0\,, 
 \label{eq:bard2}
\end{equation}
where $p = \kappa_{\mathrm N}(\dot{\phi}^2/2 - V)$ is the pressure. 
This gives a time-dependent Mathieu equation (c.f. Eq. \ref{eq:mcparam})  
with parameters: 
\begin{equation} 
 A(k) \doteq \frac{k^2 +2K}{a^2 m_{\phi}^2}~~,~~q \doteq 
            - \frac{3\kappa_{\mathrm N} \Phi^2}{16}\,. 
 \label{eq:bardmath}
\end{equation} 
In this case, a negative coupling instability is impossible and only
for $\Phi \sim M_{\pl}$ is there significant graviton production. Note,
however, that if temporal averaging is used, the average equation of
state is that of dust, $\overline{p} = 0$. Eq. (\ref{eq:bard2}) then
predicts (falsely) that there is no resonant amplification of
gravitational waves since the value of $q$ corresponding to the
temporarily averaged evolution vanishes.

%-------------------------------
\subsection[{Discussion}]
{Discussion}
%-------------------------------

We have described a new --- geometric --- reheating channel after
inflation, one which occurs solely due to gravitational couplings.
While this is not very strong in the gravitational wave and minimally
coupled scalar field cases, it can be very powerful in the
non-minimally coupled case, either due to broad-resonance ($\xi \gg
1$) or negative coupling ($\xi < 0$) instabilities. Particularly in
the latter case, it is possible to produce large numbers of bosons
which are significantly more massive than the inflaton, as required
for GUT baryogenesis. It further gives rise to the possibility that
the post-inflationary universe may be dominated by non-minimally
coupled fields. These must be treated as imperfect fluids which would
thus alter both density perturbation and background spacetime
evolution, which are known to be significantly different \cite{HM94}
than in the simple perfect fluid case.  We have further seen a
unified approach to resonant production of vector and tensor fields
during reheating in analogy to the scalar case.

With the study of this nice example of parametric resonance due to a
non-stationary gravitational field, we close our discussion about
inflation. In what follows we shall move our attention to a class of
models that try to present an alternative to this paradigm. In
particular we shall propose a particular implementation of the varying
speed of light cosmologies. This investigation is mainly included
in~\cite{VSL} and was done in collaboration with Bruce A. Bassett,
Carmen Molina--\Paris\ and Matt Visser.

%%%%%%%%%%%%%%%%%%%%%%%%%%%%%%%%%%%%%%%%%%%%%%%%%%%%%%%%%%%%%%%%%%%%%%%%%%%
\section[{Alternatives to inflation?}]
{Alternatives to inflation?}  
\label{sec:altinfl}
%%%%%%%%%%%%%%%%%%%%%%%%%%%%%%%%%%%%%%%%%%%%%%%%%%%%%%%%%%%%%%%%%%%%%%%%%%%

High-energy cosmology is flourishing into a subject of observational
riches but theoretical poverty. Inflation stands as the only
well-explored paradigm for solving the puzzles of the early universe.
This monopoly is reason enough to explore alternative scenarios and
new angles of attack.  Variable--Speed--of--Light (VSL) cosmologies have
recently generated considerable interest as alternatives to
cosmological inflation which serve both to sharpen our ideas regarding
falsifiability of the standard inflationary paradigm, and also to
provide a contrasting scenario that is hopefully amenable to
observational test. The major variants of VSL cosmology under
consideration are those of Clayton--Moffat
\cite{Moffat93a,Moffat93b,Moffat98,Clayton-Moffat-1,Clayton-Moffat-2}
and Albrecht--Barrow--Magueijo
\cite{Albrecht98,Barrow98a,Barrow98b,Barrow98c,Albrecht99}, plus
recent contributions by Avelino--Martins \cite{Avelino-Martins},
Drummond \cite{Drummond}, Kiritsis \cite{Kiritsis}, and Alexander
\cite{Alexander}. The last two are higher-dimensional, brane-inspired
implementations.

Given this wide variety of implementations of the VSL idea it is
difficult to present it as a paradigm. Roughly speaking one can cast
the basic ideas behind these models in the following way.  The
covariance of General Relativity means that the set of cosmological
models consistent with the existence of the apparently universal class
of preferred rest frames defined by the CMBR, is very small and
non-generic.  Inflation alleviates this problem by making the flat
FLRW model an attractor within the set of almost--FLRW models, at the
cost of violating the strong energy condition.  Most of the
above quoted VSL cosmologies, by contrast, sacrifice (or at the very
least, grossly modify) Lorentz invariance at high energies, by
allowing the speed of light to vary in time (and hence setting up a
preferred reference system) in order to mitigate the 
cosmological puzzles discussed above. This is not strictly true for all
of the conceivable VSL models.  We shall see how it is possible to
build up VSL frameworks where Lorentz invariance is not explicitly
violated.

In the following we shall discuss how a general VSL model can be built
up and justified and consider the internal consistency of
cosmological models characterized by a variable speed of light.

%--------------------------------------------------------------------------
\subsection[{Varying--Speed--of--Light Cosmologies}]
{Varying--Speed--of--Light Cosmologies} 
\label{subsec:vsl}
%--------------------------------------------------------------------------

As a starting point for our investigation we want to assess the
internal consistency of the VSL idea and ask to what extent it is
compatible with Einstein gravity. This is not a trivial issue:
Ordinary Einstein gravity has the constancy of the speed of light
built into it at a fundamental level; $c$ is the ``conversion
constant'' that relates time to space.  We need to use $c$ to relate
the zeroth coordinate to time: $\d x^0 = c \; \d t$. Thus, simply
replacing the \emph{constant} $c$ by a position-dependent
\emph{variable} $c(t,\x)$, and writing $\d x^0 = c(t,\x)\; \d
t$ is a suspect proposition.  

Indeed, the choice $\d x^0 = c(t,\x)\; \d t$ is a coordinate
dependent statement. It depends on how one slices up the spacetime
with spacelike hypersurfaces.  Different slicings would lead to
different metrics, and so one has destroyed the coordinate invariance
of the theory right at step one.  This is not a good start for the VSL
programme, as one has performed an act of extreme violence to the
mathematical and logical structure of General Relativistic cosmology.

Another way of viewing the same problem is this. Start with the
ordinary FLRW metric
\begin{equation}
\d s^2 = -c^2 \; \d t^2 + a(t)^2 \; h_{ij} \;\d x^i \; \d x^j,
\end{equation}
and compute the Einstein tensor. In the natural orthonormal basis one
can write
\begin{eqnarray}
G_{\hat t \hat t} 
&=& 
{3 \over a(t)^2} \; \left[ {\dot a(t)^2\over c^2}  + K \right] ,
\\
G_{\hat{\imath}\hat{\jmath}}
&=& 
- \frac{\delta_{ \hat{\imath}\hat{\jmath} }} {a(t)^2}  \;
\left[ 
2 {a(t) \; \ddot a(t)\over c^2} + {\dot a(t)^2 \over c^2} + K 
\right],
\label{einstein}
\end{eqnarray}
with the spatial curvature $K=0,\pm1$.  If one replaces $c \to c(t)$
{\emph{in the metric}}, then the physics does not change since this
particular ``variable speed of light'' can be undone by a coordinate
transformation: $c \; \d t_{\mathrm new} = c(t) \; \d t$.  While a
coordinate change of this type will affect the (coordinate) components
of the metric and the (coordinate) components of the Einstein tensor,
the orthonormal components and (by extension) all physical observables
(which are coordinate invariants) will be unaffected.  Superficially
more attractive, (because it has observable consequences), is the
possibility of replacing $c \to c(t)$ directly \emph{in the Einstein
tensor}.
%----------------------------------------------
(This is the route chosen by Barrow
\etal \cite{Barrow98a,Barrow98b,Barrow98c}, by Albrecht
{\etal} \cite{Albrecht98,Albrecht99}, and by Avelino and Martins
\cite{Avelino-Martins}.)
%----------------------------------------------------------
Then
\begin{eqnarray}
G_{\hat t \hat t}^{\mathrm{modified}} 
&=&
{3\over a(t)^2} \; \left[ {\dot a(t)^2\over c(t)^2}  + K \right] ,
\\
G_{ \hat{\imath}\hat{\jmath} }^{\mathrm{modified}} 
&=&
-{ \delta_{\hat{\imath}\hat{\jmath}} \over a(t)^2} \;
\left[ 
2 {a(t) \; \ddot a(t)\over c(t)^2} + {\dot a(t)^2\over c(t)^2} + K 
\right].
\end{eqnarray}
Unfortunately, if one does so, the modified ``Einstein tensor'' is
\emph{not} covariantly conserved (it does \emph{not} satisfy the
contracted Bianchi identities), and this modified ``Einstein tensor''
is not obtainable from the curvature tensor of \emph{any} spacetime
metric.  Indeed, if we define a timelike vector $V^\mu = (\partial/
\partial t)^\mu = (1,0,0,0)$ a brief computation yields
\begin{equation}
\nabla_\mu G_{\mathrm{modified}}^{\mu\nu} \propto \dot c(t) \; V^\nu.
\end{equation}
Thus, violations of the Bianchi identities for this modified
``Einstein tensor'' are an integral part of this particular way of
trying to make the speed of light variable.
%--------------------------------------------
(In the Barrow {\etal} approach they prefer to define \emph{modified}
Bianchi identities by moving the RHS above over to the LHS. They then
speak of these modified Bianchi identities as being satisfied.
Nevertheless the \emph{usual} Bianchi identities are violated in their
formalism.)
%-------------------------------------------

If one couples this modified ``Einstein tensor'' to the stress-energy
via the Einstein equation
\begin{equation}
G_{\mu\nu} = {8\pi \; G_{\mathrm N}\over c^4} \; T_{\mu\nu},
\label{gmod}
\end{equation}
then the stress-energy tensor divided by $c^4$ cannot be covariantly
conserved either (here we do not need to specify just yet if we are
talking about a variable $c$ or a fixed $c$), and so $T^{\mu\nu}/c^4$
cannot be variationally obtained from \emph{any} action.
%---------------------------------------------------
[The factor of $c^4$ is introduced to make sure that all components of
the stress-energy tensor have the dimensions of energy density,
$\varepsilon$ (the same dimensions as pressure, $p$.)  When needed,
mass density will be represented by $\rho$.]
%---------------------------------------------------
This is an enormous amount of physics to sacrifice before even
properly getting started, and we do \emph{not} wish to pursue this
particular avenue any further --- it is offered as an example of the
sort of fundamental things that can go wrong if one is not careful
when setting up a VSL cosmology.

Apart from the mathematical consistency of the VSL approach, there is
another fundamental issue to be addressed. If $c \to c(t)$
\emph{everywhere} in the theory, how could we tell?  Since our rulers
and clocks are all affected by the change, why does not everything
cancel out, relegating VSL-type approaches to being physically
meaningless changes in the system of units? In \cite{Albrecht98} it is
properly pointed out that it is only through changes in dimensionless
numbers that any physical effect could be detected. It is common to
phrase the discussion in terms of the fine structure constant $\alpha$
\begin{equation}
\alpha = {e^2 \over 4\pi \; \epsilon_0 \;\hbar \; c} \approx {1\over 137}.
\label{fs}
\end{equation}
Unfortunately, if one chooses $\alpha$ as the relevant probe, it is
difficult, if not impossible, to distinguish a variable speed of light
from a variable Planck constant, a variable electric charge, or a
variation in the permeability of the vacuum.

\paragraph{Proposal}

From the former discussion it should now be clear that if one wants to
uniquely specify that it is the speed of light that is varying, then
one should seek a theory that contains two natural speed parameters,
call them $\cp$ and $\cg$ ( $\cg$ is the speed at which small
gravitational disturbances propagate), and then ask that the ratio of
these two speeds is a time dependent quantity.
%----------------
Naturally, once we go beyond idealized FLRW cosmologies, to include
perturbations, we will let this ratio depend on space as well as time.
%----------------
Thus we would focus attention on the dimensionless ratio
\begin{equation}
\zeta \equiv {\cp \over \cg}.
\label{zeta}
\end{equation}
With this idea in mind, it is simplest to take $\cg$ to be fixed and
position-independent and to set up the mathematical structure of
differential geometry needed in implementing Einstein gravity: $\d x^0 =
\cg \; \d t$, the Einstein--Hilbert action, the Einstein tensor,
\etc. One can reserve $\cp$ for photons, and give an objective meaning
to the VSL concept. (Observationally, as recently emphasized by Carlip
\cite{Carlip}, direct experimental evidence tells us that in the
current epoch $\cg\approx\cp$ to within about one percent
tolerance. This limit is perhaps a little more relaxed than one would
have naively expected, but the looseness of this limit is a reflection
of the fact that direct tests of General Relativity are difficult due
to the weakness of the gravitational coupling $G_{\mathrm N}$.)
This approach naturally leads us into the realm of two-metric
theories, and the next section will be devoted to discussing the
origin of our proposal. In brief, we will advocate using at least
\emph{two} metrics: a spacetime metric $g_{\alpha\beta}$ describing
gravity, and a second ``effective metric'' $[g^{\mathrm
em}]_{\alpha\beta}$ describing the propagation of photons. Other
particle species could, depending on the specific details of the model
we envisage, couple either to their own ``effective metric'', to
$g$, or to $g^{\mathrm em}$.

\paragraph{Precursors}

We have seen in chapter~\ref{chap:1} (in section~\ref{sec:VEEFS})
that the basic idea of a quantum-induced effective metric, which
affects only photons and differs from the gravitational metric, is
actually far from radical.  As we said, ``anomalous'' (larger than
$\cg$) photon speeds have been calculated in relation to the
propagation of light in the Casimir vacuum
\cite{Sch90,Bar90,SchBar93}, as well as in gravitational fields
\cite{DH80,DS94,DS96,Shore96}. Moreover we also noted that in recent
papers \cite{DG98,Novello99} it has been stressed that such behaviour
can be described in a geometrical way by the introduction of an
effective metric which is related to the spacetime metric and the
renormalized stress-energy tensor. In curved spaces the natural
generalization of Eq.~(\ref{eq:minkoptmetr}) takes the form
\begin{equation}
[\gem^{-1}]^{\mu\nu}=
A\; g^{\mu\nu} + B\; \langle\psi| T^{\mu\nu} |\psi\rangle, 
\end{equation}
where $A$ and $B$ depend on the detailed form of the
effective (one-loop) Lagrangian for the electromagnetic field.

\begin{center}
  \setlength{\fboxsep}{0.5 cm} 
   \framebox{\parbox[t]{14cm}{
%------------------------------------------------------------------------      
       {\bf Warning}: We will always raise and lower indices using the
       spacetime metric $g$. This has the side-effect that one can no
       longer use index placement to distinguish the matrix $[\gem]$
       from its matrix inverse $[\gem^{-1}]$.  (Since $[\gem]^{\mu\nu}
       \equiv g^{\mu\sigma} \; g^{\nu\rho} \; [\gem]_{\sigma\rho} \neq
       [\gem^{-1}]^{\mu\nu}.$) Accordingly, whenever we deal with the
       EM metric, we will always explicitly distinguish $[\gem]$ from
       its matrix inverse $[\gem^{-1}]$.
%------------------------------------------------------------------------
}}
\end{center} 

It is important to note that such effects can safely be described
without needing to take the gravitational back reaction into account.
The spacetime metric $g$ is only minimally affected by the vacuum
polarization, because the formula determining $[g^{\mathrm {em}}]$ is
governed by the fine structure constant, while backreaction on the
geometry is regulated by Newton's constant. 

Although these deviations from standard propagation are extremely small
for the cases quoted above (black holes and the Casimir vacuum) we can
ask ourselves if a similar sort of physics could have been important
in the early evolution of our universe.
Drummond and Hathrell \cite{DH80} have, for example, computed one-loop
vacuum polarization corrections to QED in the presence of a
gravitational field.  They show that at low momenta the effective
Lagrangian is
%----------------------------------------------------------------
\begin{eqnarray}
{\cal L} =
%&=& 
-\frac{1}{4} \; F_{\mu\nu} \; F^{\mu \nu} 
%\nonumber\\
%&&
- \frac{1}{4m_e^2} (\beta_1 \; R \; F_{\mu\nu} \; F^{\mu \nu} 
+  \beta_2 \; R_{\mu\nu} \; F^{\mu \alpha} \; F^{\nu}{}_{\alpha})
%\nonumber\\
%&& 
- \frac{\beta_3}{4m_e^2} \; R_{\mu\nu\alpha\beta} \; F^{\mu \alpha} \;
  F^{\nu \beta}\,.
\label{dh}
\end{eqnarray}
They were able to compute the low momentum coefficients $\beta_i, i =
1\dots3$, but their results are probably not applicable to the case
$R/m_e \gg 1$ of primary interest here. It is the qualitative
structure of their results that should be compared with our
prescriptions as developed in the next section.

\paragraph{Lorentz invariance}

Actually our approach can be seen as a development of the above
results.  We shall use some classical (external) field to polarize the
vacuum of some quantum fields and hence split the degeneracy (in the sense
described above) between the (effective) null cones of various
species of particles.

It is important to understand that in these two-metric models the
Lorentz symmetry is broken in a ``soft'' manner, rather than in a
``hard'' manner. This ``soft'' breaking of Lorentz invariance, due to
the nature of the ground state or initial conditions, is qualitatively
similar to the notion of spontaneous symmetry breaking in particle
physics, whereas ``hard'' breaking, implemented by explicitly
non-invariant terms in the Lagrangian, is qualitatively similar to the
notion of explicit symmetry breaking in particle physics.

In particular it can be illuminating to consider the Euler--Heisenber
Lagrangian (\ref{eq:EuHei}) that induces the Scharnhorst effect (we
write it again here for the convenience of the reader)
% \label{eq:EuHei}
\begin{equation}
   \L=\frac{\E^2-\B^2}{8\pi}+
     \frac{\alpha^2}{2^3 \cdot 3^2 \cdot 5 \pi^{2} m_{e}^4}
     \left[ \left(\E^2-\B^2\right)^2+7\left(\E\cdot\B \right)\right] 
\end{equation}
It is easy to see that the second term on the right hand side (the one
that induces the non-linearity and the anomalous propagation of the
photons) {\em is}, by itself, Lorentz invariant.  The anisotropic
speed of propagation of light is due to the fact that the Lorentz
symmetry breaking is induced by the external conditions (the parallel
plates) which reflects in a non-Lorentz invariant (Casimir) ground
state. 

As a final remark it has to be pointed out that the VSL
implementations based on two-metric theories, which we shall pursue in
the next section, are certainly closer in spirit to the approaches of
Moffat {\etal}
\cite{Moffat93a,Moffat93b,Moffat98,Clayton-Moffat-1,Clayton-Moffat-2}
and Drummond \cite{Drummond}, than to the Barrow {\etal}
\cite{Barrow98a,Barrow98b,Barrow98c,Albrecht98,Albrecht99} and
Avelino--Martins \cite{Avelino-Martins} prescriptions. In this sense
parts of this investigation can be viewed as an extension (and
criticism) of those models.

%--------------------------------------------------------------------------
\section[{The $\chi$VSL framework}]
{The $\chi$VSL framework} 
\label{sec:chi-vsl}
%--------------------------------------------------------------------------

Based on the preceding discussion, we shall write, as the first step
towards making a VSL cosmology ``geometrically sensible'',
a two-metric theory in the form
\begin{equation}
S_{I} = 
\int \d^4 x \sqrt{-g} \; 
\left\{ R(g) + \L_{\mathrm{matter}}(g) \right\}
+
\int \d^4 x \sqrt{-\gem} \; 
\left\{
[\gem^{-1}]^{\alpha\beta}  \; F_{\beta\gamma} \;
[\gem^{-1}]^{\gamma\delta} \; F_{\delta\alpha}
\right\}
\label{s1}
\end{equation}
Note that here we have just made the first of many \emph{choices}. In fact
we chose the volume element for the electromagnetic Lagrangian
to be $\sqrt{-\gem}$ rather than, say $\sqrt{-g}$. This has been done
to do minimal damage to the electromagnetic sector of the theory. As
long as we confine ourselves to making \emph{only} electromagnetic
measurements this theory is completely equivalent to ordinary curved
space electromagnetism in the spacetime described by the metric
$\gem$. As long as we \emph{only} look at the ``matter'' fields it is
only the ``gravity metric'' $g$ that is relevant.

Since the photons couple to a second, separate metric, distinct from
the spacetime metric that describes the gravitational field, we can
now give a precise physical meaning to VSL. If the two null-cones
(defined by $g$ and $\gem$, respectively) do not coincide one has a
VSL cosmology. Gravitons and all matter except for photons, couple to
$g$. Photons couple to the electromagnetic metric $\gem$.  A more
subtle model is provided by coupling all of the gauge bosons to $\gem$,
but everything else to $g$.
\begin{equation}
S_{II} =
\int \d^4 x \sqrt{-g} \; 
\left\{ R(g) + \L_{\mathrm{ferm}}(g,\psi) \right\} 
+
\int \d^4 x \sqrt{-\gem} \; 
\Tr \left\{
[\gem^{-1}]^{\alpha\beta}  \; F_{\beta\gamma}^{\mathrm{gauge}} \;
[\gem^{-1}]^{\gamma\delta} \; F_{\delta\alpha}^{\mathrm{gauge}} 
\right\}
\label{s2}
\end{equation}
For yet a third possibility: couple \emph{all} the matter fields to
$\gem$, keeping gravity as the only field coupled to $g$.
That is
\begin{eqnarray}
S_{III}=
\int \d^4 x \sqrt{-g} \;  R(g)
&+&
\int \d^4 x \sqrt{-\gem} \; 
\left\{ \L_{\mathrm{ferm}}(\gem,\psi) \right\}\nonumber\\ 
&+&
\int \d^4 x \sqrt{-\gem} \; 
\Tr \left\{
[\gem^{-1}]^{\alpha\beta}  \; F_{\beta\gamma}^{\mathrm{gauge}} \;
[\gem^{-1}]^{\gamma\delta} \; F_{\delta\alpha}^{\mathrm{gauge}}
\right\}
\end{eqnarray}
Note that we have used $\d x^0 = c \; \d t$, with the $c$ in question
being $\cg$. It is this $\cg$ that should be considered fundamental,
as it appears in the local Lorentz transformations that are the
symmetry group of all the non-electromagnetic interactions. It is just
that $\cg$ is no longer the speed of ``light''.

Most of the following discussion will focus on the first model $S_I$,
but it is important to realize that VSL cosmologies can be implemented
in many different ways, of which the models I, II, and III are the
cleanest exemplars.  We will see later that there are good reasons to
suspect that model III is more plausible than models I or II, but we
concentrate on model I for its pedagogical clarity. If one wants a
model with even more complexity, one could give a \emph{different}
effective metric to each particle species. A model of this type would
be so unwieldy as to be almost useless.

If there is no relationship connecting the EM metric to the gravity
metric, then the theory has too much freedom to be useful, and the
equations of motion are under-determined. To have a useful theory we
need to postulate some relationship between $g$ and $\gem$, which in
the interest of simplicity we take to be algebraic.  A particularly
simple electromagnetic (EM) metric which one can consider is
\begin{equation}
[\gem]_{\alpha\beta} = 
g_{\alpha\beta} - 
(A\; M^{-4}) \;
\nabla_\alpha \chi \; \nabla_\beta \chi,
\label{chimodel}
\end{equation}
with the inverse metric
\begin{equation}
[\gem^{-1}]^{\alpha\beta} = 
g^{\alpha\beta} + 
(A\; M^{-4}) \;
{
\nabla^\alpha \chi \; \nabla^\beta \chi
\over
1 - (A\; M^{-4}) \;
(\nabla^\alpha \chi)^2
}.
\end{equation}
Here we have introduced a dimensionless coupling $A$ and taken
$\hbar=\cg=1$, in order to give the scalar field $\chi$ its canonical
dimensions of mass-energy.\footnote{
%--------------------------------------------
Remember that indices are always raised and/or lowered by using the
gravity metric $g$. Similarly, contractions always use the
gravity metric $g$. If we ever need to use the EM metric to contract
indices we will exhibit it explicitly.}
%--------------------------------------------
The normalization energy scale, $M$, is defined in terms of $\hbar$,
$G_{\mathrm N}$, and $\cg$.  The EM lightcones can be much wider than
the standard (gravity) ones without inducing a large backreaction on
the spacetime geometry from the scalar field $\chi$, provided $M$
satisfies $M_{\mathrm{Electroweak}}<M<M_{\pl}$. The presence of this
dimensional coupling constant implies that when viewed as a quantum
field theory, {\VSL} cosmologies will be non-renormalizable.  In this
sense the energy scale $M$ is the energy at which the
non-renormalizability of the $\chi$ field becomes important. (This is
analogous to the Fermi scale in the Fermi model for weak interactions,
although in our case $M$ could be as high as the GUT scale).  Thus,
{\VSL} models should be viewed as ``effective field theories'' valid
for sub-$M$ energies. In this regard {\VSL} models are certainly no
worse behaved than many of the current models of cosmological
inflation and/or particle physics.

The evolution of the scalar field $\chi$ will be assumed to be
governed by some VSL action
\begin{equation}
S_{\mathrm{VSL}} =   \int \d^4 x \sqrt{-g} \; \L_{\mathrm{VSL}}(\chi).
\end{equation}
We can then write the complete action for model I as
\begin{eqnarray}
S_{I} &=&
\int \d^4 x \sqrt{-g} \; \left\{ R(g) + \L_{\mathrm{matter}} \right\} 
+
\int \d^4 x \sqrt{-\gem(\chi)} \;
\left\{
[\gem^{-1}]^{\alpha\beta}(\chi)  \; F_{\beta\gamma} \;
[\gem^{-1}]^{\gamma\delta}(\chi) \; F_{\delta\alpha}
\right\} 
\nonumber\\
&+&
\int \d^4 x \sqrt{-g} \; \L_{\mathrm{VSL}}(\chi)
+ \int \d^4 x \sqrt{-g} \; \L_{\mathrm{NR}}(\chi,\psi),
\end{eqnarray}
where $\L_{\mathrm NR}(\chi,\psi)$ denotes the non-renormalizable
interactions of $\chi$ with the standard model.

Let us suppose the potential in this VSL action has a global minimum,
but the $\chi$ field is displaced from this minimum in the early
universe: either trapped in a metastable state by high-temperature
effects or displaced due to chaotic initial conditions. The transition
to the global minimum may be either of first or second order and
during it $\nabla_\alpha \chi \neq0$, so that $\gem \neq g$. Once the
true global minimum is achieved, $\gem = g$ again. Since one can
arrange $\chi$ today to have settled to the true global minimum,
current laboratory experiments would automatically give $\gem=g$.

It is only via observational cosmology, with the possibility of
observing the region where $\gem \neq g$ that we would expect VSL
effects to manifest themselves.  We will assume the variation of the
speed of light to be confined to very early times, of order of the GUT
scale, and hence none of the low-redshift physics can be directly
affected by this transition. We will see in section
\ref{obst} how indirect tests for the presence of the $\chi$ field are
indeed possible.

Note that in the metastable minimum $V(\chi)\geq 0$, thus the scalar
field $\chi$ can mimic a cosmological constant, as long as the kinetic
terms of the VSL action are negligible when compared to the potential
contribution.  If the lifetime of the metastable state is too long, a
de~Sitter phase of exponential expansion will ensue. Thus, the VSL
scalar has the possibility of driving an inflationary phase in its own
right, over and above anything it does to the causal structure of the
spacetime (by modifying the speed of light). While this direct
connection between VSL and inflation is certainly interesting for its
own sake, we prefer to stress the more interesting possibility that,
by coupling an independent inflaton field $\phi$ to $\gem$, $\chi$VSL
models can be used to improve the inflationary framework by enhancing
its ability to solve the cosmological puzzles. We will discuss this
issue in detail in section \ref{hete}.

During the transition, (adopting FLRW coordinates on the spacetime),
we see
\begin{equation}
[\gem]_{tt} = -1 - (A \; M^{-4}) \; (\partial_t \chi)^2 \leq -1.
\end{equation}
This means that the speed of light for photons will be larger than the
``speed of light'' for everything else --- the photon null cone will
be wider than the null cone for all other forms of matter.\footnote{
%-----------------------------------------------------
For other massless fields the situation depends on whether we use
model I, II, or III. In model I it is \emph{only} the photon that sees
the anomalous light cones, and neutrinos for example are
unaffected. In model II all gauge bosons (photons, $W^\pm$, $Z^0$, and
gluons) see the anomalous light cones. Finally, in model III
everything \emph{except} gravity sees the anomalous light cones.}
%---------------------------------------------------
Actually one has
\begin{equation}
\cp^2 = 
\cg^2 
\left[1 + (A \; M^{-4}) \; (\partial_t \chi)^2\right] 
\geq 
\cg^2 .
\label{cphoton}
\end{equation}
The fact that the photon null cone is wider implies that ``causal
contact'' occurs over a larger region than previously thought --- and
this is what helps to smear out inhomogeneities and solve the horizon
problem.

The most useful feature of this model is that it gives a precise
\emph{geometrical} meaning to VSL cosmologies: something that is
difficult to discern in the existing literature.

Note that this model is by no means unique: (1) the VSL potential is
freely specifiable, (2) one could try to do similar things to the
Fermi fields and/or the non-Abelian gauge fields --- use one metric
for gravity and $\gem$ for the other fields.  We wish to emphasize
some features and pitfalls of two-metric VSL cosmologies:

$\bullet$ The causal structure of spacetime is now ``divorced'' from
the null geodesics of the metric $g$.  Signals (in the form of
photons) can travel at a speed $\cp \geq \cg$.

$\bullet$ We must be extremely careful whenever we need to assign a
specific meaning to the symbol $c$. We are working with a
\emph{variable\,} $\cp$, which has a larger value than the standard
one, and a \emph{constant\,} $\cg$ which describes the speed of
propagation of all the other massless particles. In considering the
cosmological puzzles and other features of our theory (including the
``standard'' physics) we will always have to specify if the quantities
we are dealing with depend on $\cp$ or $\cg$.

$\bullet$ Stable causality: If the gravity metric $g$ is stably
causal~\footnote{
%--------------------------------------------------------------------------  
  Let $t^{\mu}$ be a timelike Killing vector at some point $p$ of a
  manifold $M$. If we define a new metric
  $\tilde{g}_{\mu\nu}=g_{\mu\nu}-t_{\mu}t_{\nu}$ then this will have a
  light cone strictly {\em larger} than that of $g_{\mu\nu}$.
  Therefore if the spacetime $(M,g_{\mu\nu})$ was on the verge of
  allowing closed causal curves, the spacetime
  $(M,\tilde{g}_{\mu\nu})$ could actually admit them.
  
  A spacetime $(M,g_{\mu\nu})$ is said to be {\em stably causal} if
  there exists a continuous non vanishing timelike vector $t^{\mu}$
  such that the spacetime $(M,\tilde{g}_{\mu\nu})$ possesses no closed
  timelike curves.
%--------------------------------------------------------------------------
  }, if the coupling $A \geq 0$, and if $\partial_\mu \chi$ is a
timelike vector with respect to the gravity metric, then the photon
metric is also causally stable. This eliminates the risk of unpleasant
causal problems such as closed timelike loops. This observation is
important since with two metrics (and two sets of null cones), one
must be careful to not introduce causality violations --- and if the
two sets of null cones are completely free to tip over with respect to
each other it is very easy to generate causality paradoxes in the
theory.

$\bullet$ If $\chi$ is displaced from its global minimum we expect it
to oscillate around this minimum, causing $\cp$ to have periodic
oscillations. This would lead to dynamics very similar to that of
preheating in inflationary scenarios~\cite{Preheating}.

$\bullet$ During the phase in which $c_{\rm photon} \gg c_{\rm
gravity}$ one would expect photons to emit gravitons in an analogue of
the Cherenkov radiation.  We will call this effect \emph{Gravitational
Cherenkov Radiation}.  This will cause the frequency of photons to
decrease and will give rise to an additional stochastic background of
gravitons.

$\bullet$ Other particles moving faster than $c_{\rm gravity}$ (\ie,
models II and III) would slow down and become subluminal relative to
$c_{\rm gravity}$ on a characteristic time-scale associated to the
emission rate of gravitons.  There will therefore be a natural
mechanism for slowing down massive particles to below $c_{\rm
gravity}$.

$\bullet$ In analogy to photon Cherenkov emission \cite{cherenkov},
longitudinal graviton modes may be excited due to the non-vacuum
background \cite{vEE}.

%-------------------------------------------------------------------
\subsection
[{Stress-energy tensor, equation of state, and equations of motion}]
{Stress-energy tensor, equation of state, and equations of motion}
%-------------------------------------------------------------------
%\subsubsection{The two stress-energy tensors}
%-------------------------------------------------------------------

The definition of the stress-energy tensor in a VSL cosmology is
somewhat subtle since there are two distinct ways in which one could
think of constructing it. If one takes gravity as being the primary
interaction, it is natural to define
\begin{equation}
T^{\mu\nu} = {2\over\sqrt{-g}} {\delta S\over\delta g_{\mu\nu}},
\end{equation}
where the metric variation has been defined with respect to the
gravity metric. This stress-energy tensor is the one that most
naturally appears in the Einstein equation. One could also think of
defining a different stress-energy tensor for the photon field (or in
fact any form of matter that couples to the photon metric) by varying
with respect to the photon metric, that is
\begin{equation}
\tilde T^{\mu\nu} = {2\over\sqrt{-g_{\mathrm{em}}}}
{\delta S\over\delta g^{\mathrm{em}}_{\mu\nu}}.
\end{equation}
This definition is most natural when one is interested in
non-gravitational features of the physics.

In the formalism which we have set up, by using the chain rule and the
relationship that we have assumed between $g_{\mathrm em}$ and $g$, it
is easy to see that
\begin{equation}
\label{E:two-stress-tensors}
T_{\mathrm{em}}^{\mu\nu} = 
\sqrt{g_{\mathrm{em}}\over g} \; \tilde T_{\mathrm{em}}^{\mu\nu} = 
\sqrt{1-(A \; M^{-4})[(\nabla^\alpha \chi)^2]} \; 
\tilde T_{\mathrm{em}}^{\mu\nu}.
\end{equation}
Thus, these two stress-energy tensors are very closely related. When
considering the way in which the photons couple to gravity, the use of
$T_{\mathrm{em}}^{\mu\nu}$ is strongly recommended. Note that
$T_{\mathrm{em}}^{\mu\nu}$ is covariantly conserved with respect to
$\nabla_g$, whereas $\tilde T_{\mathrm{em}}^{\mu\nu}$ is conserved with
respect to $\nabla_{\gem}$. It should be noted that $\tilde T_{\mathrm
em}^{\mu\nu}$ is most useful when discussing the non-gravitational
behaviour of matter that couples to $\gem$ rather than $g$. (Thus in
type I models this means we should only use it for photons.)  For
matter that couples to $g$ (rather than to $\gem$), we have not found
it to be indispensable, or even useful, and wish to discourage its use
on the grounds that it is dangerously confusing.

An explicit calculation, assuming for definiteness a type I model and
restricting attention to the electromagnetic field, yields
\begin{equation}
T_{\mathrm{em}}^{\mu\nu} =
\sqrt{1-(A \; M^{-4})[(\nabla^\alpha\chi)^2]}
\Bigg\{ 
[\gem^{-1}]^{\mu\sigma} \; F_{\sigma\rho} 
\; [\gem^{-1}]^{\rho\lambda} 
\; F_{\lambda\pi} \; [\gem^{-1}]^{\pi\nu}
- {1\over4} [\gem^{-1}]^{\mu\nu} \; (F^2)
\Bigg\},
\end{equation}
with
\begin{eqnarray}
(F^2) = 
[\gem^{-1}]^{\alpha\beta}  \; F_{\beta\gamma} \;
[\gem^{-1}]^{\gamma\delta} \; F_{\delta\alpha}.
\end{eqnarray} 
(In particular, note that both $\tilde T_{\mathrm{em}}^{\mu\nu}$ and
$T_{\mathrm{em}}^{\mu\nu}$ are traceless with respect to $\gem$, not
with respect to $g$. This observation proves to be very useful.)

%-----------------------------------------------------------------------
\subsubsection{Energy density and pressure: the photon equation--of--state}
%-----------------------------------------------------------------------

In an FLRW universe the high degree of symmetry implies that the
stress-energy tensor is completely defined in terms of energy density
and pressure. We will define \emph{the} physical energy density and
pressure as the appropriate components of the stress-energy tensor
when referred to an orthonormal basis \emph{of the metric that enters
the Einstein equation} (from here on denoted by single-hatted indices)
\begin{eqnarray}
\varepsilon &=&  
 T^{{\hat t} {\hat t}} 
= 
T^{tt}/|g^{tt}| 
= 
|g_{tt}|\; T^{tt},
\label{energyden}\\
p  &=&  
{1\over 3} \delta_{\hat{\imath}\hat{\jmath}} \;  
T^{\hat{\imath}\hat{\jmath}} = {1\over3} \; g_{ij} \; T^{ij}.
\label{E:pressure}
\end{eqnarray}
It is this $\varepsilon$ and this $p$ that will enter the Friedmann
equations governing the expansion and evolution of the universe.

On the other hand, if one defines the stress-energy tensor in terms of
a variational derivative with respect to the electromagnetic metric,
then when viewed from an orthonormal frame adapted to the
{\emph{electromagnetic}} metric (denoted by double hats), one will
naturally define {\emph{different}} quantities for the energy density
$\tilde\varepsilon$ and pressure $\tilde p$. We can then write
\begin{eqnarray}
\tilde\varepsilon &=&  
\tilde T^{\hat{\hat t} \hat{\hat t}} 
= 
\tilde T^{tt}/|g^{tt}_{\mathrm{em}}| 
= 
|g_{tt}^{\mathrm{em}}|\; \tilde T^{tt},
\\
\tilde p  &=&  
{1\over 3} \delta_{\hat{\hat{\imath}}\hat{\hat{\jmath}}} \; 
           \tilde T^{\hat{\hat{\imath}}\hat{\hat{\jmath}}} = 
{1\over3} \; g_{ij}^{\mathrm{em}} \; \tilde T^{ij}.
\end{eqnarray}
{From} our previous discussion [equation (\ref{E:two-stress-tensors})]
we know that the two definitions of stress-energy are related, and
using the symmetry of the FLRW geometry we can write
\begin{equation}
T^{\mu\nu} = {\cp\over\cg} \; \tilde T^{\mu\nu}.
\label{E:two-stress-tensors-FLRW}
\end{equation}
If we combine this equation with the previous definitions, we have
\begin{eqnarray}
\varepsilon &=& {\cg\over\cp} \; \tilde\varepsilon,
\\
p &=& {\cp\over\cg} \; \tilde p.
\label{E:p}
\end{eqnarray}
(Note that the prefactors are \emph{reciprocals} of each other.)  From
a gravitational point of view any matter that couples to the photon
metric has its energy density depressed and its pressure enhanced by a
factor of $\cg/\cp$ relative to the energy density and pressure
determined by ``electromagnetic means''.  This ``leverage'' will
subsequently be seen to have implications for SEC violations and
inflation.

In order to investigate the equation of state for the photon field,
our starting point will be the standard result that the stress-energy
tensor of photons is traceless. By making use of the tracelessness and
symmetry arguments one can (in one-metric theories) deduce the
relationship between the energy density and the pressure $\varepsilon
= 3 p$.  However, in two-metric theories (of the type presented here)
the photon stress-energy tensor is traceless with respect to
$g_{\mathrm{em}}$, but not with respect to $g$.  Thus in this bi-metric
theory we have
\begin{equation}
\tilde \varepsilon = 3 \tilde p.
\end{equation}
When translated into $\varepsilon$ and $p$, (quantities that will
enter the Friedmann equations governing the expansion and evolution of
the universe), this implies
\begin{equation}
p_{\mathrm{photons}} = 
{1\over3} \; \varepsilon_{\mathrm{photons}} \; {\cp^2\over\cn^2}.
\label{eosp}
\end{equation}

As a final remark, it is interesting to consider the speed of sound
encoded in the photon equation of state. If we use the relationship
$\rho_{\mathrm{photons}} = \varepsilon_{\mathrm{photons}}/ \cg^2$, we
can write
\begin{equation}
\rho_{\mathrm{photons}}= 
{3 \; p_{\mathrm{photons}}\over\cp^2}
\label{rho}
\end{equation}
and therefore 
\begin{equation}
(c_{\mathrm{sound}})_{\mathrm{photons}} = 
\sqrt{\partial p_{\mathrm{photons}}\over\partial\rho_{\mathrm{photons}}} = 
{\cp\over\sqrt{3}}.
\end{equation}
That is, oscillations in the density of the photon fluid propagate at
a relativistic speed of sound which is $1/\sqrt{3}$ times the speed of
``light'' \emph{as seen by the photons}.

More generally, for highly relativistic particles we expect
\begin{equation}
\varepsilon_i = 
3 \; p_i \; {\cg^2\over c_i^2},
\end{equation}
and
\begin{equation}
(c_{\mathrm{sound}})_i = 
{c_i\over\sqrt{3}}.
\end{equation}
Note that we could define the mass density (as measured by
electromagnetic means) in terms of $\tilde\rho_{\mathrm{photons}} =
\tilde\varepsilon_{\mathrm{photons}}/\cp^2$.  This definition yields
the following identity
\begin{equation}
\rho_{\mathrm{photons}} = {\cp\over \cn} \;\tilde \rho_{\mathrm{photons}}.
\end{equation}
If the speed of sound is now calculated in terms of $\tilde p_{\mathrm
photons}$ and $\tilde \rho_{\mathrm{photons}}$ we get the same result
as above.

%-------------------------------------------------------------------
\subsubsection{Equations of motion}
%-------------------------------------------------------------------

The general equations of motion based on model I can be written as
\begin{equation}
G_{\mu\nu} = 
{8\pi \; G_{\mathrm N} \over \cg^4} \;
\left(
T_{\mu\nu}^{\mathrm{VSL}} + 
T_{\mu\nu}^{\mathrm{em}} + 
T_{\mu\nu}^{\mathrm{matter}} 
\right).
\end{equation}
All of these stress-energy tensors have been defined with the
``gravity prescription''
\begin{equation}
T_i^{\mu\nu} = {2\over\sqrt{-g}} {\delta S_i\over\delta g_{\mu\nu}}.
\end{equation}

If we make the dependence on the speed of light explicit (and sum over
all particles present), the Friedmann equations
(\ref{eq:fried1},\ref{eq:fried2}) for a {\VSL} cosmology read as follows
\begin{eqnarray}
{\left({\dot a\over a}\right)}^2
&=&
{8\pi G\over3\cg^2} \; \sum_{i} \varepsilon_{i}
-{K \cg^2\over a^2},
\label{f1}
\\
{\ddot a\over a}
&=&
-{4\pi G\over 3\cg^2}\; \sum_{i}{\left(\varepsilon_{i}+3p_{i}\right)}.
\label{f2}
\end{eqnarray}

The constant ``geometric'' speed of light implies that we get from the
Friedmann equation separate conservation equations valid for each
species individually (provided, as is usually assumed for at least
certain portions of the universe's history, that there is no
significant energy exchange between species)
\begin{equation}
\dot\varepsilon_i +
3{\dot a\over a}{\left(\varepsilon_{i}+{p_{i}}\right)}=0.
\end{equation}
In the relativistic limit we have already seen, from equation
(\ref{eosp}), that $p_{i}= {1\over3} \varepsilon_{i} \;
(c_i^2/\cg^2)$.  [We are generalizing slightly to allow each
particle species to possess its own ``speed-of-light''.] So we can
conclude that
\begin{equation}
\dot\varepsilon_i +
\left(3+ {c_i^2\over \cg^2}\right) {\dot a\over a} \; \varepsilon_{i}=0.
\end{equation}
Provided that $c_i$ is slowly changing with respect to the expansion of
the
universe (and it is not at all clear whether such an epoch ever
exists), we can write for each relativistic species
\begin{equation}
\varepsilon_i \; a^{3+ (c_i^2/\cg^2)} \approx {\mathrm{constant}}.
\end{equation}
This is the generalization of the usual equation $(\varepsilon_i a^4
\approx {\mathrm{constant}})$ for relativistic particles in a
constant-speed-of-light model. This implies that energy densities will
fall much more rapidly than naively expected in this bi-metric VSL
formalism, provided $c_i > \cg$.

%-------------------------------------------------------------------
\subsection[{Cosmological puzzles and Primordial Seeds}]
{Cosmological puzzles and Primordial Seeds}
%-------------------------------------------------------------------

In what follows we will deal with the main cosmological puzzles
showing how they are mitigated (if not completely solved) by the
\VSL~models.  

%----------------------------
\subsubsection{Isotropy}
%----------------------------

As discussed in section~\ref{subsec:cospu}, one of the major puzzles of
the standard cosmological model is the apparent conflict between the
isotropy of the CMB and the best estimates of the size of causal
contact at last scattering. In VSL cosmologies the formula for the
(coordinate) size of the particle horizon at the time of last
scattering $t_*$ is
\begin{equation}
R_{\mathrm{particle-horizon}}=\frac{\ell_{H}(t)}{a(t)} 
= \int_0^{t_*} {\cg \; \d t \over a(t)}.
\end{equation}
For photons this should now be modified to
\begin{equation}
R_{\mathrm{photon-horizon}} = \int_0^{t_*} {\cp \; \d t \over a(t)} 
\geq R_{\mathrm{particle-horizon}}.
\end{equation}
The quantity $R_{\mathrm{photon-horizon}}$ sets the distance scale over
which photons can transport energy and thermalize the primordial
fireball.  On the other hand, the coordinate distance to the surface
of last scattering is
\begin{equation}
R_{\mathrm{last-scattering}} = \int_{t_*}^{t_0} {\cp \; \d t \over a(t)}.
\end{equation}
The observed large-scale homogeneity of the CMB implies (without any
artificial fine-tuning)
\begin{equation}
R_{\mathrm{photon-horizon}} \geq R_{\mathrm{last-scattering}},
\label{E:horizon}
\end{equation}
which can be achieved by having $\cp \gg \cg$ early in the expansion.
(In order not to change late-time cosmology too much it is reasonable
to expect $\cp \approx \cg$ between last scattering and the present
epoch.) Instead of viewing our observable universe as an inflated
small portion of the early universe (standard inflationary cosmology),
we can say that in a VSL framework the region of early causal contact
is underestimated by a factor that is roughly approximated by the
ratio of the maximum photon speed to the speed with which
gravitational perturbations propagate.

%-----------------------------------------
\subsubsection[{Flatness}]
{Flatness}
%------------------------------------------

We have seen that the flatness problem is related to the fact that in
FLRW cosmologies the $\Omega=1$ solution appears as an unstable point
in the evolution of the universe. Nevertheless observations seem to be
in favour of such a value. In this section we will show that any
two-metric implementation of the kind given in equation
(\ref{chimodel}) does not by itself solve the flatness problem, let
alone the quasi-flatness problem \cite{Barrow98a}. In spite of this we
shall show that {\VSL} can nevertheless enhance any mild SEC violation
originated by an inflaton field coupled to $\gem$.

%---------------------------------------------------------------
\paragraph{Flatness in ``pure'' {\VSL} cosmologies}
%---------------------------------------------------------------

The question ``Which $c$ are we dealing with?'' arises once more when
we address the flatness problem.  We can start by taking again
Eq.~(\ref{eq:flateps}).
We already know that one cannot simply replace $c \to \cp$ in the
above equation. The Friedmann equation is obtained by varying the
Einstein--Hilbert action.  Therefore, the $c$ appearing here must be
the fixed $\cg$, otherwise the Bianchi identities are violated and
Einstein gravity loses its geometrical interpretation.  Thus, we have
\begin{equation}
\epsilon=\frac{K \;\cg^{2}}{\dot{a}^2}.
\end{equation} 
So in our framework Eq.~(\ref{eq:flateps2}) takes the form
\begin{equation}
\dot{\epsilon}
=
- 2K \;\cg^{2} \left(\frac{\ddot{a}}{\dot{a}^{3}}\right) 
= 
-2\epsilon \left(\frac{\ddot{a}}{\dot{a}}\right).
\end{equation}
{From} the way in which we have implemented VSL cosmology (two-metric
model),
it is easy to see that this equation is independent of the photon
sector; it is unaffected if $\cp\neq\cg$.  The only way that VSL
effects could enter this discussion is indirectly. When $\cp\neq\cg$
the photon contribution to $\rho$ and $p$ is altered.

In particular, if we want to solve the flatness problem by making
$\epsilon=0$ a stable fixed point of the evolution (at least for some
portion of the history of the universe), then we must have $\ddot a >
0$, and the expansion of the universe must be accelerating (for the
same portion in the history of the universe).

We have seen in section~\ref{subsec:GFinf} that the condition $\ddot a
> 0$ leads to violations of the SEC. Namely, violations of the SEC are
directly linked to solving the flatness problem.  By making use of the
Friedmann equations (\ref{f1}, \ref{f2}), this can be rephrased as
\begin{equation}
\dot{\epsilon}
=
2\epsilon 
\left[
\frac{ 
{
4\pi\;G_{\mathrm N}} \; \sum_{i}(\varepsilon_{i}+3p_{i}) 
}{
3H\cg^2
}
\right].
\end{equation}
In our bi-metric formalism the photon energy density $\varepsilon$ and
photon pressure $p$ are both positive, and from equation (\ref{eosp})
it is then clear that also $\varepsilon+3p$ will be positive. This is
enough to guarantee no violations of the SEC.
%---------------------------------
This means that bi-metric VSL theories are no better at solving the
flatness problem than standard cosmological (non-inflationary) FLRW
models. To ``solve'' the flatness problem by making $\epsilon=0$ a
stable fixed point will require some SEC violations and cosmological
inflation from other non-photon sectors of the theory.

As a final remark we stress that Clayton and Moffat, in
their \emph{vector} scenario~\footnote{ 
%----------------------------------------------------------------------
  Moffat \cite{Clayton-Moffat-1} introduces a somewhat similar model
  for an effective metric which in our notation could be written as
  \[
  [\gem]_{\alpha\beta} = 
  g_{\alpha\beta} - 
  (A\; M^{-2}) \;
  V_\alpha \; V_\beta,
  \]
  with the inverse metric
  \[
  [\gem^{-1}]^{\alpha\beta} = 
  g^{\alpha\beta} + 
  (A\; M^{-2}) \;
  { V^\alpha \; V^\beta \over 1 + (A\; M^{-2}) \;(V^\alpha)^2
  }.
  \]
  In the more recent paper \cite{Clayton-Moffat-2} a scalar scenario
  more similar to our own is discussed.
%------------------------------------------------------------------------
  } as discussed in \cite{Clayton-Moffat-1}, claim to be able to solve
the flatness problem. It can be shown that this claim is induced by a 
premature conclusion. Their model does not lead to any SEC violation and no real 
contradiction with our results is indeed present. We direct the reader 
to~\cite{VSL} for a more detailed discussion on this point.

%-------------------------------------------------------------------
\paragraph{Flatness in Heterotic (Inflaton+\VSL) models.}
\label{hete}
%-------------------------------------------------------------------

Although the conclusions above may appear to be disappointing for the
VSL framework, it is interesting to note that nevertheless two-metric
VSL cosmologies \emph{enhance} any inflationary tendencies in the
matter sector.

Let us suppose that we have an inflaton field coupled to the
\emph{electromagnetic} metric. We know that during the inflationary
phase we can write approximately
\begin{equation}
T^{\mu\nu}_{\mathrm{inflaton}} \propto g_{\mathrm{em}}^{\mu \nu}.
\end{equation}
We have repeatedly emphasized that it is important to define
\emph{the} physical energy density and pressure ($\varepsilon, p$) as
the appropriate components of the stress-energy tensor when referred
to an orthonormal basis \emph{of the metric that enters the Einstein
  equation}.  The condition $T^{\mu\nu}_{\mathrm{inflaton}} \propto
g^{\mu\nu}_{\mathrm{em}}$, when expressed in terms of an orthonormal
basis of the metric $g$ gives
\begin{equation}
 p_{\mathrm{inflaton}} = - {\cp^2\over\cg^2} \; 
 \varepsilon_{\mathrm{inflaton}}.
\end{equation}
That is
\begin{equation}
(\varepsilon + 3p)_{\mathrm{inflaton}} = 
\left(1-3 {\cp^2\over\cg^2}\right) \varepsilon_{\mathrm{inflaton}}.
\end{equation}
Thus, any ``normal'' inflation will be amplified during a VSL
epoch. It is in this sense that VSL cosmologies heterotically improve
standard inflationary models.

We can generalize this argument. Suppose that the ``normal'' matter, when
viewed from an orthonormal frame adapted to the \emph{electromagnetic}
metric, has energy density $\tilde\varepsilon$ and pressure $\tilde
p$.  {From} our previous discussion [equations
(\ref{E:two-stress-tensors-FLRW})---(\ref{E:p})] we deduce
\begin{eqnarray}
\varepsilon+3p &=& 
{\cg\over\cp} \; \tilde\varepsilon + 3 \; {\cp\over\cg} \; \tilde p.
\end{eqnarray}
In particular, if $\tilde p$ is slightly negative, VSL effects can
magnify this to the point of violating the SEC (defined with respect
to the gravity metric). It is in this sense that two-metric VSL
cosmologies provide a natural enhancing effect for negative pressures
(possibly leading to SEC violations), even if they do not provide the
seed for a negative pressure.

We point out that this same effect makes it easy to violate \emph{all}
of the energy conditions. If ($\tilde\varepsilon,\tilde p$) satisfy all
of the energy conditions with respect to the photon metric, and provided
$\tilde p$ is only slightly negative, then VSL effects make it easy
for ($\varepsilon,p$) to violate all of the energy conditions with
respect to the gravity metric --- and it is the energy conditions with
respect to the gravity metric which are relevant for the singularity
theorems, the positive mass theorem, and the topological censorship
theorem.
\begin{center}
  \setlength{\fboxsep}{0.5 cm} 
   \framebox{\parbox[t]{14cm}{
%------------------------------------------------------------------------     
       {\bf Comment}: We have shown, in section~\ref{subsec:cospu},
       that one can reformulate both the horizon and flatness
       problems as entropy problems and so it can appear to be a
       contradiction that this equivalence seems to break down in the case
       of \VSL\ cosmologies. To understand how this may happen is
       indeed very instructive.

       First of all, we can try to understand what happens to the entropy per
comoving volume $S=a^{3}(t)s$. In the case of inflation we saw that
the non-adiabatic evolution $\dot{S}\neq 0$ was due to the fact that
although the entropy densities do not significantly change,
$s_{\mathrm{before}} \approx s_{\mathrm{after}}$ thanks to reheating,
nevertheless the enormous change in scale factor
$a(t_{\mathrm{after}}) = \exp[
H(t_{\mathrm{after}}-t_{\mathrm{before}}) ] \cdot
a(t_{\mathrm{before}})$ drives an enormous increase in total entropy
per comoving volume.  (Here ``before'' and ``after'' are intended with
respect to the inflationary phase.)

In our case (bimetric VSL models) the scale factor is unaffected by
the transition in the speed of light  if the $\chi$ field is not the 
dominant energy component of the universe. Instead what changes is the
entropy density $s$.  As we have seen, a sudden phase transition
affecting the speed of light induces particle creation and raises both
the number and the average temperature of relativistic particles.
Therefore one should expect that $s$ grows as $\cp\to\cg$.

~From equation (\ref{E:horizon}) it is clear that the increased speed
of light is enough to ensure a resolution of the horizon problem,
regardless of what happens to the entropy. At the same time one can
instead see that the flatness problem is not solved at all. Equation
(\ref{eq:sflrw}) tells us that it is the {\em ratio} $S\;H^3/s\approx
\dot{a}^3$ which determines the possibility of stretching the
universe. Unfortunately this is not a growing quantity in the standard
model as well as in {\em pure} bi-metric VSL theory. Once again only
violations of the SEC ($\ddot{a}>0$) can lead to a resolution of the
flatness problem.
       
%------------------------------------------------------------------------
}}
\end{center} 
%

%------------------------------------------
\subsubsection{Monopoles and Relics}
%------------------------------------------

The Kibble mechanism predicts topological defect densities that are
inversely proportional to powers of the correlation length $\xi_\Phi$
of the Higgs fields.  We saw that causality constrains this length to
be less than the particle horizon $\ell_{\mathrm{h}}$. In our case we
can use the fact that $\ell_{\mathrm{h}}\approx R_{\mathrm H}$ in FLRW
models to consider the Hubble distance (\ref{eq:rhub}) as an indicator
of the typical correlation scale of the field.

If we now suppose a good thermal coupling between the photons and the
Higgs field to justify using the photon horizon scale in the Kibble
freeze-out argument~\footnote{Alternatively, we could arrange a model
  where both photons and the Higgs field couple directly to $\gem$,
  along the lines of $S_{III}$ above.}
\begin{equation}
R_{\mathrm H} = {\cp \over H}.
\end{equation}
then we can conclude that a transition from a ``fast photon'' regime
to a ``standard photon'' regime corresponds to a reduction of the
Hubble distance.

While inflation solves the relics puzzle by diluting the density of
defects to an acceptable degree, $\chi$VSL models deal with it by
varying $c$ in such a way as to make sure that the Hubble scale is
large when the defects form. Thus, we need the transition in the speed
of light to happen \emph{after} the SSB that leads to monopole
production.

So far the discussion assumes thermal equilibrium, but one should
develop a formalism which takes into account the non-equilibrium
effects and the characteristic time scales (quench and critical
slowing down scales). As a general remark one can note that the larger
is the Higgs correlation length $\xi_\Phi$,\footnote{
%----------------------------------------------------------------------
  This correlation length characterizes the period \emph{before} the
  variation of the speed of light, when we suppose that the creation
  of topological defects has taken place.}
%----------------------------------------------------------------------
the lower the density of defects will be (with respect to the standard
estimates).  For example in the Zurek mechanism
$\rho_{\mathrm{defects}}\sim \xi_{\Phi}^{-n}$ with $n=1,2,$ and $3,$
for domain walls, strings, and monopoles,
respectively~\cite{Zurek}.

%-------------------------------------------------
\subsubsection{$\Lambda$ and the Planck problem}
%-------------------------------------------------

In this $\chi$VSL approach we are not affecting the cosmological
constant $\Lambda$, except indirectly via ${\cal L}_{\mathrm{VSL}}$.
The vacuum energy density is given by
\begin{equation}
\rho_{\Lambda}=\frac{\Lambda \; c^{2}}{8\pi\; G_{\mathrm N}}.
\label{rlam}
\end{equation}
But which is the $c$ appearing here? Is it the speed of light $\cp$?
Or is it the speed of gravitons $\cg$?  In our two-metric approach it
is clear that one should use $\cg$.\footnote{
%-------------------------------------------------------------
  On the other hand, for any contribution to the total cosmological
  constant from quantum zero-point fluctuations (ZPF) the situation is
  more complex. If the quantum field in question couples to the metric
  $\gem$, one would expect $\cp$ in the previous equation, not least
  in the relationship between $\rho_{\mathrm{zpf}}$ and
  $p_{\mathrm{zpf}}$.}
%-------------------------------------------------------------

While we do nothing to mitigate the cosmological constant problem we
also do not encounter the ``Planck problem'' considered by Coule
\cite{Coule}.  He stressed the fact that in earlier VSL formulations
\cite{Moffat98,Albrecht98,Barrow98c} a varying speed of light also
affects the definition of the Planck scale. In fact, in the standard
VSL one gets two different Planck scales (determined by the values of
$c$ before and after the transition). The number of Planck times
separating the two Planck scales turns out to be larger than the
number of Planck times separating us from the standard Planck era.
Therefore, in principle, the standard fine-tuning problems are even worse
in these models.

In contrast, in our two-metric formulation one has to decide from the
start which $c$ is referred to in the definition of the Planck length.
The definition of the Planck epoch is the scale at which the
gravitational action becomes of the order of $\hbar$. This process
involves gravity and does not refer to photons. Therefore, the $c$
appearing there is the speed of propagation of gravitons, which is
unaffected in our model.  Hence we have a VSL cosmology without a
``Planck problem'', simply because we have not made any alterations to
the gravity part of the theory.

%-----------------------------------------------------------------------
\subsubsection{Primordial fluctuations}
%----------------------------------------------------------------------

As previously explained, the inflationary scenario owes its popularity
not just to its ability to solve the main problems of the background
cosmology. It is also important because it provides a plausible,
causal, micro-physics explanation for the origin of the primordial
perturbations which may have seeded large-scale structure.  The phase
of quasi-de~Sitter expansion excites the quantum vacuum and leads to
particle creation in squeezed states. As the expansion is almost
exactly exponential, these particles have an (almost exactly)
scale-invariant spectrum with amplitude given by the ``Hawking
temperature" $H/2\pi$ \cite{Hu}.

In the case of {\VSL} the creation of primordial fluctuations is again
generic. The basic mechanism can be understood by modelling the change
in the speed of light as a changing ``effective refractive index of
the EM vacuum''. In an FLRW background
\begin{equation}
n_{\rm em}=\frac{\cg}{\cp}=
{1\over\sqrt{1+(A\,M^{-4})(\partial_{t}\chi)^2}}.
\label{neff}
\end{equation}
Particle creation from a time-varying refractive index is a well-known
effect \cite{Yablonovitch,Qed1,2Gamma,PRL}~\footnote{
%------------------------------------------------------------------------
  It is important to stress that in the quoted papers the change of
  refractive index happens in a flat static spacetime.  It is
  conceivable and natural that in an FLRW spacetime the expansion rate
  could play an important additional role. The results of
  \cite{Yablonovitch,Qed1,2Gamma,PRL} should then be considered as
  precise in the limit of a rapid (${\dot n}/n \gg {\dot a}/a$)
  transition in the speed of light.}
%------------------------------------------------------------------------
and shares many of the features calculated for its inflationary
counterpart (\eg, the particles are also produced as squeezed
pairs). We point out at this stage that these mechanisms are not
identical. In particular, in {\VSL} cosmologies it is only the fields
coupled to the EM metric that will primarily be excited. Of course, it
is conceivable, and even likely, that perturbations in these fields
will spread to the others whenever some coupling exists. Gravitational
perturbations could be efficiently excited if the $\chi$ field is
non-minimally coupled to gravity.

A second, and perhaps more fundamental, point is that a
scale-invariant spectrum of metric fluctuations on large scales is by
no means guaranteed. The spectrum may have a nearly thermal
distribution over those modes for which the adiabatic limit holds
($\tau \omega>1$, where $\tau$ is the typical time scale of the
transition in the refractive index).  If we assume that
$\tau$ is approximately constant in time during the phase transition,
then it is reasonable to expect an approximately Harrison--Zel'dovich
spectrum over the frequencies for which the adiabatic approximation
holds. Extremely small values of $\tau$, or very rapid changes of
$\tau$ during the transition, would be hard to make compatible with
the present observations.  Since a detailed discussion of the final
spectrum of perturbations in {\VSL} cosmologies would force us to take
into account the precise form of the $\chi$-potential $V(\chi)$,
(being very model dependent), we will not discuss these issues further
here.

As final remarks we want to mention two generic features of
the creation of primordial fluctuations in {\VSL} cosmologies.  Since
we require inflation to solve the flatness problem, the {\VSL}
spectrum must be folded into the inflationary spectrum as occurs in
standard inflation with phase transitions (see \eg, \cite{Julien}).
In addition to this, also a preheating phase is conceivable in {\VSL}
models if $\chi$ oscillates coherently. This would lead to production
of primordial magnetic fields due to the breaking of the conformal
invariance of the Maxwell equations.

%------------------------------------------------------------------------
\subsection[{Observational tests and the low-redshift {\VSL} universe}]
{Observational tests and the low-redshift {\VSL} universe}
\label{obst}
%------------------------------------------------------------------------

At this point, it is important to note that due to the nature of the
interaction (\ref{s1}), the $\chi$ field appears unable to decay
completely. Decay of the $\chi$ field proceeds via $2\chi \rightarrow
2\gamma$ and hence, once the density of $\chi$ bosons drops
considerably, ``freeze-out'' will occur and the $\chi$ field will stop
decaying.  This implies that the $\chi$ field \emph{may} be
dynamically important at low-redshift \emph{if} its potential is such
that its energy density drops less rapidly than that of radiation.

However, the $\chi$ correction to $g_{em}$ corresponds to a dimension
eight operator, which is non-renormalizable.  The vector model of
Moffat \cite{Clayton-Moffat-1} is a dimension six operator.
Nevertheless, for energies below $M$ it is difficult to argue why
either of these operators should not be negligibly small relative to
dimension five operators, which would cause single body decays of the
$\chi$ field. While it is possible that these dimension five operators
are absent through a global symmetry \cite{Carroll-1}, or the lifetime
of the $\chi$ bosons is extremely long, we will see later that such
non-renormalizable interactions with the standard model give rise to
serious constraints. For the time being we neglect single-body decays,
and we can imagine two natural dark-matter candidates, with the added
advantage that they are distinguishable and detectable, at least in
principle.

(i) If $V(\chi)$ has a quadratic minimum, the $\chi$ field will
oscillate about this minimum and its average equation of state will be
that of dust. This implies that the $\chi$ field will behave like
axions or cold dark matter. Similarly if the potential is quartic, the
average equation of state will be that of radiation.

(ii) If $V(\chi)$ has quintessence form, with no local minimum but a
global minimum at $\chi \rightarrow \infty$: a typical candidate is a
potential which decays to zero at large $\chi$ (less rapidly than an
exponential) with $V(\chi) > A e^{-\lambda \chi}$ for $\lambda > 0$.

These two potentials lead to interesting observational implications
for the low-redshift universe which we now proceed to analyze and
constrain.

%------------------------------------------------------------------------
\subsubsection{Clustering and gravitational lensing}
%------------------------------------------------------------------------

It is interesting to note that the effective refractive index which we
introduced in equation (\ref{neff}) may depend, not just on time, but
also on space and have an anisotropic structure.  In particular the
dispersion relation of photons in an anisotropic medium is
\begin{equation}
\omega^{2}= [n^{-2}]^{ij} \; k_{i} \; k_{j},
\end{equation}
and from the above expression it is easy to see that the
generalization of equation (\ref{neff}) then takes the form
\begin{equation}
[n^{-2}]^{ij}=g^{ij}_{\mathrm{em}}/|g^{tt}_{\mathrm{em}}|.
\label{E:aneff}
\end{equation}

Scalar fields do not support small scale density inhomogeneities
(largely irrespective of the potential). This implies that the
transfer function tends to unity at small scales and the scalar field
is locally identical to a cosmological constant.

However, on scales larger than $100 Mpc$, the scalar field can cluster
\cite{Ma}. During such evolution both $\dot\chi \neq 0$ and
$\partial_i \chi \neq 0$ will hold. This would lead to deviations from
equation (\ref{neff}), as the ratio between the two speeds of light
will not only be a function of time.

{For} instance, let us suppose we are in a regime where time
derivatives of $\chi$ can be neglected with respect to spatial
derivatives. Under these conditions the EM metric reduces to
\begin{eqnarray}
g^{\mathrm{em}}_{tt} &=& g_{tt} = - |g_{tt}|, 
\\
g^{\mathrm{em}}_{ij} &=&
g_{ij}-(AM^{-4}) \; \partial_{i}\chi \; \partial_{j}\chi.
\end{eqnarray}  

{From} equation (\ref{E:aneff}) this is equivalent to a tensor refractive
index $n_{ij}$, with
\begin{equation}
[n^{2}]_{ij}=
{g_{ij}-(AM^{-4}) \; \partial_{i}\chi \; \partial_{j}\chi 
\over 
|g_{tt}|}. 
\end{equation}
This tensor refractive index may lead to additional lensing by
large-scale structure, over and above the usual contribution from
gravitational lensing \cite{SEF}.

%------------------------------------------------------------------------
\subsubsection{Quintessence and long-range forces}
%------------------------------------------------------------------------

Another natural application is to attempt to use the $\chi$ field as
the source of the ``dark energy'' of the universe, the putative source
of cosmic acceleration. This is attractive for its potential to unify
a large number of disparate ideas, but is severely constrained as
well.

%------------------------------------------------------------------------
\subsubsection{Constraints arising from variation of the fine-structure
constant} 
%------------------------------------------------------------------------

As noted in the introduction, a change of $c_{\mathrm{photon}}$ will
cause a variation in the fine-structure constant. Such a variation is
very constrained. We point out two particularly interesting
constraints.  The first, arising from nucleosynthesis \cite{KPW}, is
powerful due to the extreme sensitivity of nucleosynthesis to
variations in the proton-neutron mass difference, which in turn is
sensitive to $\alpha$. This places the tight constraint that $
|\dot{\alpha}/\alpha| \leq 10^{-14} {\rm yr}^{-1}$. However, this is only a
constraint on $\dot{c}_{\mathrm{photon}}/c_{\mathrm{photon}}$ if no
other constants appearing in $\alpha$ are allowed to vary. Further we
have assumed $\dot{\alpha}$ was constant through nucleosynthesis.

A similar caveat applies to other constraints which one derives for
variations of $c_{\mathrm photon}$ through variations of
$\alpha$. Other tests are only sensitive to integrated changes in
$\alpha$ over long time scales. At redshifts $z \leq 1$ constraints
exist that $|\Delta{\alpha}/\alpha| < 3 \times 10^{-6}$ (quasar
absorption spectra \cite{QSO}) and $|\Delta{\alpha}/\alpha| < 10^{-7}$
(Oklo natural reactor \cite{oklo}).

%------------------------------------------------------------------------
\subsubsection{Binary pulsar constraints} 
%------------------------------------------------------------------------

Unless we choose the unattractive solution that $\chi$ lies at the
minimum of its potential but has non-zero energy (\ie, an explicit
$\Lambda$ term), we are forced to suggest that $\dot{\chi} \neq 0$
today and $V(\chi)$ is of the form $e^{-\lambda \chi}$ or $\chi^{-n}$
\cite{Quin}. In this case, gravitons and photons do not travel at the
same speed today. The difference between the two velocities is rather
constrained by binary pulsar data to be less than $1\%$ \cite{Carlip};
\ie, $|n_{\rm em} - 1| < 0.01$.

%------------------------------------------------------------------------
\subsubsection{High-energy tests of VSL}
%------------------------------------------------------------------------

Constraints on our various actions $S_I - S_{III}$ also come from high
energy experiments. In model $I$, photons travel faster than any other
fields. This would lead to perturbations in the spectrum of nuclear
energy levels \cite{D96}.

Similarly, high energy phenomena will be sensitive to such speed
differences. For example, if $\cp > c_{e^-}$, the process $\gamma
\rightarrow e^- + e^+$ becomes kinematically possible for sufficiently
energetic photons. The observation of primary cosmic ray photons with
energies up to 20 TeV implies that today $\cp - c_{e^-} <
10^{-15}$~\cite{CG}.  The reverse possibility -- which is impossible
in our model I if $A > 0$ in equation (\ref{cphoton}) -- is less
constrained, but the absence of vacuum Cherenkov radiation with
electrons up to 500 GeV implies that $c_{e^-} - c_{\mathrm{photon}} < 5
\times 10^{-13}$. Similar constraints exist which place upper limits
on the differences in speeds between other charged leptons and hadrons
\cite{CG,cptlorentz}. These will generally allow one to constrain
models I -- III, but we will not consider such constraints further.
 
%------------------------------------------------------------------------
\subsubsection{Non-renormalizable interactions with the standard model}
%------------------------------------------------------------------------

Our $\chi$VSL model is non-renormalizable and hence one expects an
infinite number of $M$-scale suppressed, dimension five and higher,
interactions of the form
\beq
\beta_i \frac{\chi^n}{M^{n}}{\cal L}_i,
\label{nonrenc}
\eeq
where $\beta_i$ are dimensionless couplings of order unity and ${\cal
L}_i$ is any dimension-four operator such as $F^{\mu \nu} F_{\mu
\nu}$.

For sub-Planckian $\chi$-field values, the tightest constraints
typically come from $n =1$ (dimension five operators) and we focus on
this case. The non-renormalizable couplings will cause time variation
of fundamental constants and rotation of the plane of polarization of
distant sources \cite{Carroll}.  For example, with ${\cal L}_{QCD} =
\Tr(G_{\mu\nu} G^{\mu \nu})$, where $G_{\mu \nu}$ is the QCD field
strength, one finds the strict limit~\cite{QCD}
\beq
|\beta_{G^2}| \leq 10^{-4} (M/M_{\mathrm{Planck}})
\eeq
which, importantly, is $\chi$ independent. 

If one expects that $|\beta_i| = O(1)$ on general grounds, then this
already provides as strong a constraint on our model as it does on
general quintessence models.  This constraint is not a problem if
there exist exact or approximate global symmetries
\cite{Carroll-1}. Nevertheless, without good reason for adopting such
symmetries this option seems unappealing.

Another dimension five coupling is given by equation (\ref{nonrenc})
with $ {\cal L}_{F^2} = F_{\mu \nu} F^{\mu \nu}$ which causes
time-variation in $\alpha$. Although there is some evidence for this
\cite{Barrow-alpha}, other tests have been negative as discussed
earlier. These yield the constraint~\cite{Carroll-1}
\beq
|\beta_{F^2}| \leq 10^{-6} (M H/\langle\dot{\chi}\rangle).
\eeq
Clearly this does not provide a constraint on {\VSL} unless we
envisage that $\dot{\chi} \neq 0$ today as required for
quintessence. If $\chi$ has been at the minimum of its effective
potential since around $z < 5$, then neither this, nor the binary
pulsar, constrain {\VSL} models.  The CMB provides a more powerful
probe of variation of fundamental constants and hence provides a test
of {\VSL} if $\chi$ did not reach its minimum before $z \simeq 1100$
\cite{Barrow-cmb}.

Another interesting coupling is ${\cal L}_{F^*F} = F_{\mu \nu}
{}^*F^{\mu\nu}$, where ${}^*F$ is the dual of $F$. As has been
noted~\cite{Carroll-1}, this term is not suppressed by the exact
global symmetry $\chi\rightarrow\chi + {\mathrm constant}$, since it
is proportional to $(\nabla_{\mu}\chi) \; A_{\nu} \; {}^*F^{\mu \nu}$.
A non-zero $\dot{\chi}$ leads to a polarization-dependent ($\pm$)
deformation of the dispersion relation for light
\begin{equation}
\omega^2 = k^2 \pm\beta_{F^*F} (\dot{\chi} k/M)\,.
\end{equation}
If $\dot{\chi} \neq 0$ today, the resulting rotation of the plane of
polarization of light traveling over cosmological distances is
potentially observable.  Indeed claims of such detection
exist~\cite{Ralston}. However, more recent data is consistent with no
rotation~\cite{Leahy,CF}.  Ruling out of this effect by
high-resolution observations of large numbers of sources would be
rather damning for quintessence but would simply restrict the $\chi$
field to lie at its minimum, \ie, $\Delta \chi \simeq 0$ for $z < 2$.

On the other hand, a similar and very interesting effect arises not
from $\dot{\chi}$ but from spatial gradients of $\chi$ at
low redshifts due to the tensor effective refractive index of
spacetime.

%%%%%%%%%%%%%%%%%%%%%%%%%%%%%%%%%%%%%%%%%%%%%%%%%%%%%%%%%%%%%%%%%%%%%%%%%
\section[{Discussion}]
{Discussion} 
\label{sec:astro-persp}
%%%%%%%%%%%%%%%%%%%%%%%%%%%%%%%%%%%%%%%%%%%%%%%%%%%%%%%%%%%%%%%%%%%%%%%%%

In this investigation we have tried to set out a mathematically
consistent and physically coherent formalism for discussing Variable
Speed of Light (VSL) cosmologies. An important observation is that
taking the usual theory and replacing $c \to c(t)$ is not reasonable.
One either ends up with a coordinate change which does not affect the
physics, or a mathematical inconsistency.  In particular, replacing $c
\to c(t)$ in the Einstein tensor of an FLRW universe violates the
Bianchi identities and destroys the geometrical interpretation of
Einstein gravity as arising from spacetime curvature. \footnote{
%-------------------------------------------------
  In the {\VSL} cosmologies presented here a ``geometrical
  interpretation'' is instead preserved given the fact that the
  Bianchi identities are satisfied.}
%--------------------------------------------------

We have instead argued for the usefulness of a two-metric approach. We
have sketched a number of two-metric scenarios that are compatible
with laboratory particle physics, and have indicated how they relate
to the cosmological puzzles. We emphasize that there is considerable
freedom in these models, and that a detailed confrontation with
experimental data will require the development of an equally detailed
VSL model.  In this regard VSL cosmologies are no different from
inflationary cosmologies. Since the models which we have discussed are
non-renormalizable however, there may be interesting implications for
the low-redshift universe through gravitational lensing and
birefringence.

VSL cosmologies should be seen as a general scheme for attacking
cosmological problems. This scheme has some points in common with
inflationary scenarios, but also has some very strange peculiarities
of its own. In particular, once $\cp\neq\cg$ complications may appear
in rather unexpected places. Moreover it should now be clear that in
their ``purest'' form these models cannot induce violations of the SEC
and hence mitigate the flatness paradox. This result can lead to the
development of different strategies
\begin{enumerate}
\item One can consider VSL models as ``auxiliary effects'' with
  respect to inflation. As we have seen, the variation in the speed of
  photons can actually enhance any violation of the SEC in fields
  coupled to $\gem$ and so we can imagine this sort of effect acting
  together with an inflaton field $\phi(t,\x)$ and improving its
  efficiency in solving the cosmological puzzles.
\item Another possibility is that the same field $\chi(t,\x)$, which
  induces the variation in the speed of light, acts as an inflaton.
  This is equivalent to assuming that $V(\chi)$ is such as to lead to
  violations of the SEC at some stage in the evolution of the field.
  It is still unclear whether or not this would weaken the requirement
  of fine tuning of the potential that plagues standard inflationary
  scenarios.
\item Finally one can try to implement a bi-metric framework by
  invoking not a scalar field but rather some special boundary
  conditions (such as, for example, spacetime topology) to get anomalous
  propagation of light {\em and at the same time} violations of the
  SEC.
\end{enumerate}
In any case it should be stressed that these ``hybrid'' models (VSL+SEC
violations) could have several advantages with respect to inflationary
ones. In particular the requirement of violation of the SEC is weaker
than the requirement for a suitably stable quasi-de~Sitter
phase~\footnote{
%-------------------------------------------------------------------------  
  Violations of the SEC enforce accelerated expansions, they are not
  equivalent to getting a quasi-de~Sitter phase.
%-------------------------------------------------------------------------
}.
The latter is generally necessary for ``successful'' inflation --- it
allows the main cosmological puzzles to be solved and also for the
generation of an approximately Harrison--Zel'dovich spectrum of
primordial perturbations --- but it also the source of most of the
fine-tuning problems.

In VSL cosmologies it is not the expansion of the universe which leads
to particle production from the quantum vacuum, and so no special form of
$a(t)$ are required. For this reason it is conceivable that in this
framework (which requires just weak violation of the SEC in order to be
able to deal will {\em all} of the cosmologically puzzles) a less high
degree of fine tuning is necessary. This subject certainly deserves
further investigation.

%%%%%%%%%%%%%%%%%%%%%%%%%%%%%%%%%%%%%%%%%%%%%%%%%%%%%%%%%%%%%%%%%%%%%%%%%%%%
%S.Liberati Ph.D. Chapter 5: Conclusions
%%%%%%%%%%%%%%%%%%%%%%%%%%%%%%%%%%%%%%%%%%%%%%%%%%%%%%%%%%%%%%%%%%%%%%%%%%%%
\chapter*{Conclusions} 
%[\small{Conclusions}]
\pagestyle{myheadings}
\markright{\large \sc Conclusions}
\addcontentsline{toc}{chapter}{Conclusions}
\label{chap:concl}
%%%%%%%%%%%%%%%%%%%%%%%%%%%%%%%%%%%%%%%%%%%%%%%%%%%%%%%%%%%%%%%%%%%%%%%%%%%%
\vspace*{0.5cm}
\rightline{\it There is no intellectual exercise} 
\rightline{\it which is not ultimately useless.}
\vspace*{0.5cm} \rightline{\sf Jorge Luis Borges }  
%%%%%%%%%%%%%%%%%%%%%%%%%%%%%%%%%%%%%%%%%%%%%%%%%%%%%%%%%%%%%%%%%%%%%%%%%%%%%
\vspace*{0.5cm}
\hrulefill
\vspace*{0.5cm}
%%%%%%%%%%%%%%%%%%%%%%%%%%%%%%%%%%%%%%%%%%%%%%%%%%%%%%%%%%%%%%%%%%%%%%%%%%%

This thesis has dealt with a broad area of research developed around
the general theory of vacuum effects in strong external fields. In
particular we have focussed attention on the peculiar role which these
effects have in the presence of gravitational fields.

Perhaps the most remarkable fact is that although these effects can be
described in a similar way to ones in the presence of other external
fields, nevertheless the quantum vacuum appears to have an
unexpectedly central role for gravitational phenomena. Black hole
thermodynamics would be inconsistent without Hawking radiation, and
cosmological puzzles apparently require some sort of quantum vacuum
effects in order to be explained.

This deep link between gravity and the quantum vacuum is the main
reason for pursuing the present research. In fact the study of
semiclassical quantum gravity and of its related phenomena is a
possible way to gain further understanding about the peculiar nature
of gravitational interactions. Hopefully this research will provide a
bridge towards a full quantum gravity theory. In this respect, the
possibility of submitting some of these ideas to direct experimental
test is obviously extremely desirable. For doing this we have at hand
two alternative ways to proceed.

Firstly we can try to observationally search for the strong
gravitational fields necessary for giving relevant vacuum effects
(polarization or particle production). The natural realm of such
extreme regimes is cosmology and astrophysics. The early universe was
probably dominated by quantum vacuum effects and signatures of these
will probably be amenable to observational test in the not too distant
future (e.g. with the forthcoming Planck satellite). Moreover micro
black holes are among the possible by-products of the chaotic youth of
our universe.  Their observation would certainly allow us to test the
prediction of Hawking radiation but, unfortunately, this appears to be
a task far beyond present-day capabilities.

Secondly we can try to reproduce on earth the sort of physics which we
want to study. We are not able (for the moment) to reproduce event
horizons in a laboratory, nor to locally generate inflationary
bubbles. We can nevertheless circumvent this by concentrating on
condensed matter phenomena which share at least some kinematical
aspects with the semiclassical gravity effects which we want to study
(bearing in mind that dynamical aspects cannot be studied in this way
since they are linked to the exact form of the equations of motion).

All of the above problems have been touched on this thesis. In writing
these conclusions we shall summarize the main results, and just at the
very end try to extract some general lessons from our research.

In chapter~\ref{chap:2} we have discussed the thermodynamical
properties of black holes. This is one of the central issues of our
investigation because it clearly shows the special role of vacuum
effects in gravitational fields. We have shown that black hole
thermodynamics is amenable to a Casimir-inspired interpretation, where
the zero-point modes of the quantum fields are the dynamical degrees
of freedom responsible for the gravitational entropy. For consistency,
this approach seems to require an induced gravity framework as a
necessary pre-condition.

This picture is actually compatible with the fact that \bh\ entropy
appears to be induced by the special global properties of spacetime
manifolds with event horizons. We have learned in chapter~\ref{chap:1}
that the quantum vacuum is a globally defined object and that
topological properties of the quantization manifold can indeed
influence the zero-point modes.

All of these investigations seem to converge towards a framework where
gravity emerges as an effective theory, a mesoscopic interaction whose
geometrical interpretation and whose symmetries are probably valid
only at energies which are low compared with the Planck energy.

In particular this interpretation is corroborated by recent results in
string theory and supergravity. The latter is no more that the
supersymmetric generalization of \GR\ and is an effective theory. We
have seen how \bh\ entropy can be interpreted in this framework as a
count on string degrees of freedom. In particular this sort of
interpretation can be taken as a special realization of the general
framework discussed above.

In this investigation, extremal black holes have been shown to play a
special role. They are generally considered to be the zero temperature
states of \bh\ thermodynamics, but they are also problematic objects
for which the the superstring results do not agree with semiclassical
expectations (the gravitational entropy of these objects is expected to
be $A/4$ in superstring-based approaches and $0$ in semiclassical
ones).

The apparently critical role of extremal solutions has been the main
motivation for a study concerning the behaviour of an incipient
extremal black hole. In this case one has a well-defined past history
which allows for a non-ambiguous definition of the vacuum state and a
straightforward discussion. From the apparent paradoxes which we have
encountered in this investigation we can draw two main conclusions.
The first is that extremal solutions require an incredible fine tuning
in order to be generated via gravitational collapse if semiclassical
picture is correct. Secondly we have found that these incipient black
holes do not behave at any time as thermodynamical objects.

The speculative conclusion is that these objects are of a rather
different nature from non-extremal black holes.  It is not
inconceivable that no macroscopic extremal black holes exist at all in
nature, and that such \bh s should be considered only as microscopic
objects, possibly solitons of the microscopic theory which are
preserved as solutions of the semiclassical theory. These aspects
certainly deserve further investigation because they could tell us a
lot about the way in which gravity may emerge as a low-energy theory
and find out what the fundamental properties of its microscopic
counterparts should be.

After this investigation about the very nature of gravity and its
relation to the quantum vacuum we have moved to the exciting
research of earth-based experimental tests of the general class of
phenomena associated with this sort of physics. In particular we have
started by discussing the possibility of reproducing event horizons in
hydrodynamical models and studying their stability under realistic
conditions. 

We have seen that acoustic geometries are promising testing grounds
for studying the kinematical aspects of black hole thermodynamics
(Hawking radiation), but are sufficiently different in their
fundamental equations that they differ substantially in dynamical
aspects (e.g. the gravitational entropy). They are nevertheless
extremely useful tools because they can give us hints about the
mechanisms in action in crucial phenomena like Hawking radiation.

After this study we have moved to the puzzling phenomenon of
Sonoluminescence.  We have proposed a new variant of the dynamical
Casimir effect model for explaining the experimental observations.
This model is based on the dynamical production of particles from the
quantum vacuum due to a rapidly varying external field. In our
approach the latter is identified in the bubble refractive index.

Although simple from a theoretical point of view, this model requires
a complex mathematical analysis (analytical as well as numerical). We
have found strict constraints for our proposal which are amenable to
experimental test. The general picture emerging is nonetheless very
strongly influenced by condensed matter physics issues which deserve
further investigation by experts in the field. This is the main reason
that has pushed us to search for a general ``signature'' of the
Casimir nature of the photons detected in Sonoluminescence.  Such a
signature has been identified in the squeezed nature of the particle
pairs produced. If this property were to be confirmed experimentally,
this would be a definitive proof of the vacuum origin of the
sonoluminescence photons.

Finally we have again shifted our attention back to a general
relativistic context, considering cosmology as another arena for the
general framework of vacuum effects in strong fields. In particular we
have described an application of the general theory of parametric
resonance to the case of reheating after inflation. We have proposed a
new mechanism, that of geometric preheating, where gravity plays a
prominent role.

Parallel to this study we have also searched for applications in
cosmology of some striking features of Casimir-like effects, such as
the Scharnhorst effect. We have found that this branch of research can
be used to improve the recently proposed varying speed of light
scenarios, which have been suggested as possible alternatives to the
inflationary paradigm.

We have seen that these models in their purest form do not
automatically lead to violation of the SEC. This implies that although
they can be very effective in dealing with most of the well-known
cosmological puzzles addressed by inflation, they cannot solve the
flatness problem.  This is not necessarily a deficiency of this
framework however. Violations of the strong energy condition are
actually very easy to obtain via scalar fields or a cosmological
constant and we have seen that variations in the speed of light can
then be very efficient in amplifying them. Our discussion has been so
far very speculative and mainly devoted to the correct implementation
of the VSL idea. It is nevertheless possible that these models will
have further development in the coming years.

As a final remark about the future prospects for this whole area of
research we want to stress a few points that we see as being crucial.
\begin{itemize}
\item Black hole thermodynamics still stands as a major crossroads of
  different branches of physics. Recent developments in string theory
  have shed some light on how these roads meet each other, but we
  are still lacking a clear understanding of how the semiclassical
  limit should be interpreted in this framework. A better
  comprehension of the nature of gravity is a precondition for any
  further development in understanding the theory which should
  substitute it in the high energy regime.
\item The recent achievements in the hydrodynamical description of
  event horizons is a promising development. It hints at the
  possibility that in the not-too-distant future we might be able to
  test at least some form of Hawking radiation. In particular the
  proposal for building optical horizons in dispersive media appears
  to be very promising given the very powerful technology so far
  developed for controlling refractive indices. This research
  certainly deserves further efforts because of its possible
  consequences on the theoretical side as well on the technological
  one.
\item Also the cosmological aspects will probably be an area of
  further advances in the coming years. The advent of more and more
  precise observations (which will hopefully lead to the promised
  ``precision cosmology'') will probably open the way to definitive
  judgments about the plethora of models proposed for the evolution of
  the very early universe. Present theories make predictions but often
  we do not have instruments sensitive enough to test them. It is
  conceivable that this situation will change in the near future.
\item Finally we have seen that semiclassical calculations often
  become analytically intractable. However, problems such as the
  self-consistency of the solutions of the semiclassical Einstein
  equations are of crucial importance for our understanding of
  cosmology, or for example, of the stability of \GR\ structures like
  black holes and wormholes. It is conceivable that the modern
  numerical techniques which are now used for studying classical
  phenomena around black holes and other compact objects will soon be
  applied to the realm of semiclassical quantum gravity as well.
\end{itemize}

In conclusion we can say that this interdisciplinary field of research
is now entering a new era. After a youth based on conjectures and
theoretical speculations, it is now time to confront our knowledge
with direct answers from nature. Whatever these answers will be, they
will help us to make another step on the stairway leading to quantum
gravity.

%%%%%%%%%%%%%%%%%%%%%%%%%%%%%%%%%%%%%%%%%%%%%%%%%%%%%%%%%%%%%%%%%%%%%%%%%%%%
%                    S. Liberati Ph.D. BIBLIOGRAPHY.                       %
%%%%%%%%%%%%%%%%%%%%%%%%%%%%%%%%%%%%%%%%%%%%%%%%%%%%%%%%%%%%%%%%%%%%%%%%%%%%
\pagestyle{headings}

%%%%%%%%%%%%%%%%%%%%%%%%%%%%%%%%%%%%%%%%%%%%%%%%%%%%%%%%%%%%%%%%%%%%%%%

%%%%%%%%%%%%%%%%%%%%%%%%%%%%%%%%%%%%%%%%%%%%%%%%%%%%%%%%%%%%%%%%%%%%%
\end{document}